\documentclass[12pt,a4wide]{book}
\usepackage{mathrsfs}
\usepackage{amsmath,amssymb}
\usepackage{amsfonts}
\DeclareMathSymbol{\mlq}{\mathord}{operators}{``}
\DeclareMathSymbol{\mrq}{\mathord}{operators}{`'}
\usepackage[Sonny]{fncychap}
\usepackage{color}
\usepackage{xcolor}[2007/01/21]
\usepackage{kbordermatrix,blkarray}
\renewcommand{\kbldelim}{(}
\renewcommand{\kbrdelim}{)}
\hypersetup{hyperfootnotes=false,colorlinks=true,pdfborder={0 0 0 0},citecolor=blue,linkcolor=black,urlcolor=black}
\usepackage{url}
\usepackage[
open,
openlevel=3,
atend
]{bookmark}[2011/12/02]

\bookmarksetup{color=blue}

\usepackage{cite}
\usepackage{graphicx}
\usepackage[english]{babel}
\usepackage[T1]{fontenc}
\usepackage[utf8]{inputenc}

\usepackage{float}
\usepackage[font=small,labelfont=bf]{caption}
\usepackage[font=footnotesize,caption=false]{subfig}
\usepackage{etoolbox}
\usepackage{lscape}

\let\emptyset\varnothing
\DeclareSymbolFont{cmcal}{OMS}{cmsy}{m}{n}
\SetSymbolFont{cmcal}{bold}{OMS}{cmsy}{b}{n}
\DeclareSymbolFontAlphabet{\mathcal}{cmcal}

\usepackage{fancyhdr}
\usepackage{enumerate}
\usepackage{enumitem}
\usepackage{pdfpages}
\usepackage{adjustbox}
\usepackage{rotating}
\usepackage{floatpag}
\usepackage{wrapfig}
\usepackage{multicol} 
\usepackage{multirow}
\usepackage{hhline}
\usepackage{floatpag}
\usepackage{tabularx, booktabs} 
\usepackage[title,titletoc,toc]{appendix}
\usepackage{afterpage}

\usepackage[ruled,vlined,algosection]{algorithm2e}
\usepackage{algpseudocode}
\usepackage{framed}
\newlength\marincrease
\makeatletter
\newenvironment{Walgo}[2][htbp]
  {\renewcommand{\@algocf@start}{%
    \setlength\marincrease{#2}
  \@algoskip%
  \begin{lrbox}{\algocf@algobox}%
  \begin{minipage}{\dimexpr\textwidth+2\marincrease\relax}
  \setlength{\algowidth}{\hsize}%
  \vbox\bgroup
  \hbox to\algowidth\bgroup\hbox to \algomargin{\hfill}\vtop\bgroup%
  \ifthenelse{\boolean{algocf@slide}}{\parskip 0.5ex\color{black}}{}%
  \addtolength{\hsize}{-1.5\algomargin}%
  \let\@mathsemicolon=\;\def\;{\ifmmode\@mathsemicolon\else\@endalgoln\fi}%
  \raggedright\AlFnt{}%
  \ifthenelse{\boolean{algocf@slide}}{\IncMargin{\skipalgocfslide}}{}%
  \@algoinsideskip%
  }%
\renewcommand{\@algocf@finish}{%
  \@algoinsideskip%
  \egroup
  \hfill\egroup
  \ifthenelse{\boolean{algocf@slide}}{\DecMargin{\skipalgocfslide}}{}%
  \egroup
  \end{minipage}
  \end{lrbox}%
  \makebox[\linewidth][c]{\algocf@makethealgo}
  \@algoskip%
  \setlength{\hsize}{\algowidth}%
  \lineskip\normallineskip\setlength{\skiptotal}{\@defaultskiptotal}%
  \let\;=\@mathsemicolon%
  \let\]=\@emathdisplay%
}%
  \begin{algorithm}[#1]}
  {\end{algorithm}}
\makeatother

\usepackage{listings}

\definecolor{mygreen}{rgb}{0,0.6,0}
\definecolor{mygray}{rgb}{0.5,0.5,0.5}
\definecolor{mymauve}{rgb}{0.58,0,0.82}

\lstdefinestyle{customc}{
	belowcaptionskip=1\baselineskip,
	breakatwhitespace=false,         
	breaklines=true,                 
	captionpos=b,                    
	commentstyle=\color{mygreen},    
	deletekeywords={...},            
	escapeinside={\%*}{*)},          
	extendedchars=true,      
	frame=single,	                   
	language=C,
	showstringspaces=false,
	basicstyle=\footnotesize\ttfamily,
	keywordstyle=\bfseries\color{green!40!black},
	commentstyle=\itshape\color{purple!40!black},
	identifierstyle=\color{blue},
	stringstyle=\color{orange},
}

\usepackage{longtable}
\usepackage{tabu}
\usepackage{csquotes}
\usepackage{array, booktabs, caption}
\usepackage{stackengine}
\usepackage{makecell}
\renewcommand\theadfont{\bfseries}
\newcommand\blankpage{%
    \null
    \thispagestyle{empty}%
    \addtocounter{page}{-1}%
    \newpage}
    
\makeatletter
\newcommand\fs@norules{\def\@fs@cfont{\bfseries}\let\@fs@capt\floatc@ruled
  \def\@fs@pre{}%
  \def\@fs@post{}%
  \def\@fs@mid{\kern3pt}%
  \let\@fs@iftopcapt\iftrue}

\usepackage{verbatim}
\makeatletter
\newcommand{\verbatimfont}[1]{\def\verbatim@font{#1}}%
\makeatother

\fancyhfoffset[L]{0.5cm}

\makeatother

\usepackage[toc,nonumberlist]{glossaries}
\newglossaryentry{a}{
  name=$\mathbb{N}$,
  description={$\{1,2,3,\cdots\}$, the set of natural numbers},
  symbol={$\mathbb{N}$}
}
\newglossaryentry{b}{
  name=$\mathbb{Z}$,
  description={$\{\cdots,-2,-1,0,1,2,\cdots\}$, the set of integers},
  symbol=$\mathbb{Z}$
}
\newglossaryentry{c}{
  name = CA,
  description = {Cellular Automaton}
}
\newglossaryentry{d}{
  name = CAs,
  description = {Cellular Automata}
}

\newglossaryentry{e}{
  name =$d$,
  description={Number of states per cell of a CA},
  symbol =$d$
}

\newglossaryentry{f}{
  name =$\mathcal{S}$,
  description = {$\{0,1,\cdots,d-1\}$, the set of states of a CA},
  symbol =$\mathcal{S}$
}
\newglossaryentry{g}{
  name =$m$,
  description = {Number of neighbors of a CA},
  symbol=$m$
}

\newglossaryentry{h}{
  name =$D$,
  description = {A positive integer which represents the dimension of a CA},
   symbol=$D$
}

\newglossaryentry{i}{
  name =$\mathscr{L}$,
  description = {The $D$-dimensional cellular space},
  symbol=$\mathscr{L}$
}

\newglossaryentry{k}{
  name =$n$,
  description = {The number of cells of a finite CA},
  symbol=$n$
}
\newglossaryentry{l}{
  name =$R$,
  description = {The local transition function or \emph{rule} of a CA},
  symbol =$R$
}
\newglossaryentry{m}{
  name =$\mathcal{R}$,
  description = {\emph{Rule vector} of a non-uniform CA},
  symbol =$\mathcal{R}$
}

\newglossaryentry{n}{
  name = $\mathbf{\mathscr{R}}$,
  description = {The CA rule $120021120021021120021021210$}, 
  symbol= $\mathbf{\mathscr{R}}$
}

\newglossaryentry{o}{
  name = ECA,
  description = {Elementary Cellular Automaton}
}
\newglossaryentry{p}{
  name =RMT,
  description ={Rule Min Term}
}
\newglossaryentry{q}{
  name = $G$,
  description = {The global transition function of a CA on the set of all infinite configurations},
  symbol = $G$
}

\newglossaryentry{r}{
  name = $G_P$,
  description = {The global transition function of a CA on the set of periodic configurations},
  symbol = $G_P$
}
\newglossaryentry{s}{
  name = $G_F$,
  description = {The global transition function of a CA on the set of all finite configurations},
  symbol = $G_P$
}

\newglossaryentry{t}{
  name = $G_n$,
  description = {The global transition function of a CA over a fixed lattice size $n$},
  symbol = $G_n$
}
\newglossaryentry{u}{
  name =${G_R}^*$,
  description ={The set of global transition functions for $R$, when $n \in \mathbb{N}$},
  symbol=${G_R}^*$
}
\newglossaryentry{v}{
  name = $S_R$,
  description = {The set of CA rules which are reversible for some cell length $n\in \mathbb{N}$},
  symbol = $S_R$
}
\newglossaryentry{w}{
  name = $S_F$,
  description = {The set of CA rules having $G_F$ as bijective},
  symbol = $S_F$
 }

\newglossaryentry{x}{
  name = $S_P$,
  description =  {The set of CA rules having $G_P$ as bijective},
   symbol = $S_P$
 }
 \newglossaryentry{y}{
   name = $S_I$,
   description = {The set of CA rules having $G$ as bijective},
   symbol = $S_P$
  }

\newglossaryentry{z}{
  name =$\sigma^k(x)$,
  description ={$k$-bit left shift of a configuration $x$},
  symbol =$\sigma^k(x)$
}
\newglossaryentry{j}{
  name =$\tilde{x}$,
  description={The RMT sequence corresponding to configuration $x$},
  symbol=$\tilde{x}$
}
\newglossaryentry{c2}{
  name=$Equi_i$,
  description = {A set of RMTs containing RMT $i$ and all of its \emph{equivalents}, where $0 \leq i \leq d^{m-1}-1$},
  symbol=$Equi_i$
}
\newglossaryentry{d2}{
  name =$Sibl_i$,
  description = {A set of RMTs containing RMT $i$ and all of its \emph{siblings}, where $0 \leq i \leq d^{m-1}-1$},
  symbol=$Sibl_i$
}
\newglossaryentry{a2}{
  name = $n_0$,
  description = {The maximum height of the \emph{minimized reachability tree} for $R$},
  symbol = $n_0$
}
\newglossaryentry{i2}{
  name =$N_{i.j}$,
  description ={A node of a reachability tree at level $i$. The $j$ is the node index, where $0 \leq j \leq d^{i}-1$},
  symbol=$N_{i.j}$
}
\newglossaryentry{e2}{
   name =$E_{i.dj+x}$,
   description ={An edge of a reachability tree which is incident to the $x^{th}$ of the $d$ possible children of the node $N_{i.j}$},
   symbol=$E_{i.dj+x}$
  }
\newglossaryentry{f2}{
  name =${\Gamma^{N_{i.j}}_{p}}$,
  description = {The $p^{th}$ set of RMTs of the node $N_{i.j}$ of a reachability tree},
  symbol=${\Gamma^{N_{i.j}}_{p}}$
}
\newglossaryentry{g2}{
  name = $l_{i.dj+x}$,
  description = {The label of an edge $E_{i.dj+x}$ of a reachability tree},
  symbol = $l_{i.dj+x}$
}

\newglossaryentry{h2}{
  name =PRNG,
  description =Pseudo Random Number Generator
}

\makeglossaries
\glsdisablehyper

\newcolumntype{Y}{>{\centering\arraybackslash}X}
\newtheorem{theoremm}{Theorem}[chapter]
\newtheorem{eqed}{Example}[chapter]
\newtheorem {lemmaa}{Lemma}[chapter]
\newtheorem{proposition}{Proposition}

\newtheorem{defnn}{Definition}[chapter]
\newtheorem {corollaryy}{Corollary}[chapter]
\newtheorem {axiomm}{Axiom}[chapter]
\newtheorem {inferencee}{Inference}[chapter]

\newtheorem {hypothesiss}{Hypothesis}[chapter]
\newtheorem {conjecturee}{Conjecture}[chapter]

\newtheorem {prp}{Property}[chapter]
\newenvironment{example}{\begin{eqed} \rm}{\hfill$\Box$ \end{eqed}}
\newenvironment{proof}{\noindent {\bf Proof :\ } }{\hfill$\Box$ }
\newenvironment{infprf}{\noindent {\bf Informal Proof :\ } }{$\Box$ }

\newenvironment{lemma}{\begin{lemmaa} \sl}{\end{lemmaa}}

\newenvironment{Informal Proof}{\begin{prp} \sl}{\end{prp}}
\newenvironment{theorem}{\begin{theoremm}{\bf :}\sl}{\end{theoremm}}

\newenvironment{corollary}{\begin{corollaryy}{\bf :}\sl}{\end{corollaryy}}

\newenvironment{proof of correctness}{\noindent {\bf Proof of Correctness :\ } }{\hfill$\Box$ }
\newtheorem {definition}{Definition}

\providecommand{\floor}[1]{\left \lfloor #1 \right \rfloor }
\DeclareMathVersion{normal2}
\begin{document}
\sloppy

\afterpage{\blankpage}
\pagestyle{empty} 
	\begin{center} 
	\Large{\textbf{{{CELLULAR AUTOMATA: REVERSIBILITY, SEMI-REVERSIBILITY AND RANDOMNESS}}}} \\ 
	\end{center}
	\begin{center} 
	\vspace{4.0cm}
	 \textbf{\normalsize{KAMALIKA BHATTACHARJEE}} \\
	\vspace{4.8cm}
       \begin{figure}[h]
       	\centering
       	\includegraphics[width=1.1in, height = 1.1in]{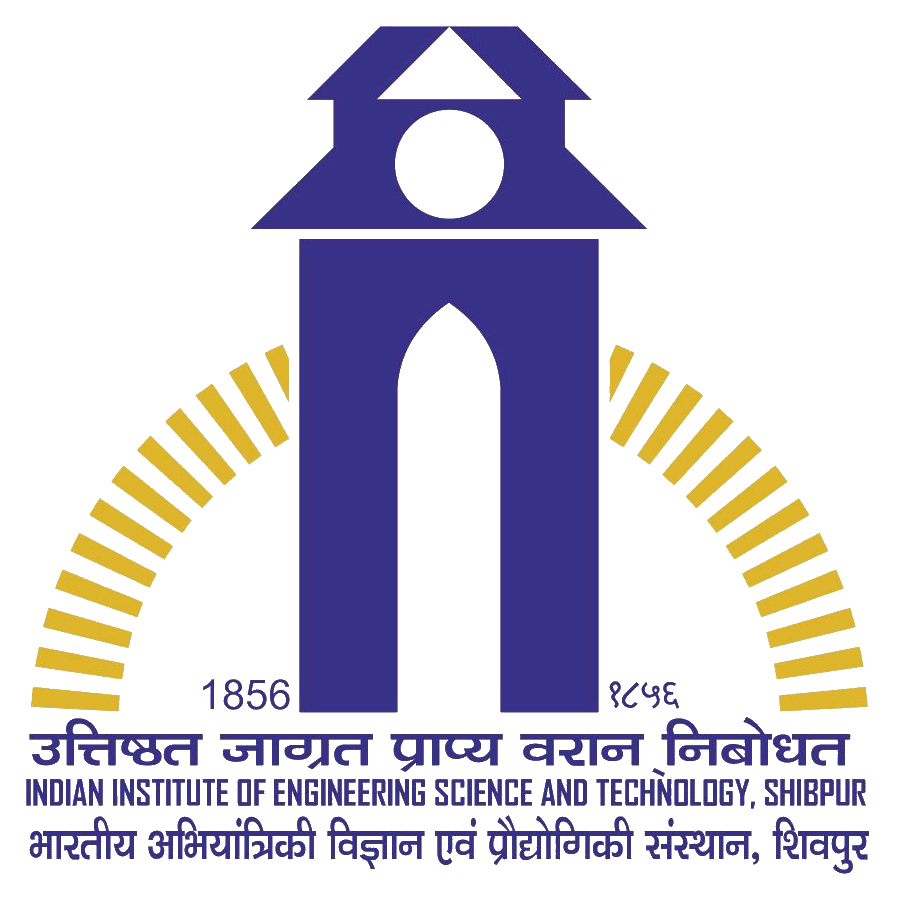}
       	\end{figure}
        \vspace{1.2cm}     
      \small{\textbf{DEPARTMENT OF INFORMATION TECHNOLOGY}}\\ 
      \small\textbf{{INDIAN INSTITUTE OF ENGINEERING SCIENCE \& TECHNOLOGY, SHIBPUR}} \\
     \small{\textbf{HOWRAH, WEST BENGAL, INDIA-711103}}\\
               \vspace{0.8cm} 
   \small{\textbf{2019}} \\  
  	\end{center} 

\newpage

\afterpage{\blankpage}
\pagestyle{empty} 
	\begin{center} 
	\Huge{\textbf{{{Cellular Automata: Reversibility, Semi-reversibility and Randomness}}}}\\ 
        \vspace{1.5cm} 
        \Large{\textbf{Kamalika Bhattacharjee}}\\
        \vspace{2pt}  
         \normalsize{INSPIRE Fellow}\\    
	\vspace{2.0cm}
        \small{\em{A report submitted in partial fulfillment for the degree of}} \\ 
		\vspace{3pt} 
		\textbf{\large{Doctor of Philosophy\\ 
					in \\
			 Engineering}}\\
	\vspace{1.5cm} 
	\normalsize{\em{Under the supervision of}}\\ 
	\vspace{0.2cm} 
	\large{\textbf{Dr. Sukanta Das}}\\ 
	\small{Associate Professor}\\
	\small{Department of Information Technology}\\  
	{Indian Institute of Engineering Science and Technology, Shibpur}\\

	 \vspace{0.2in} 
	\begin{figure}[h]
	\centering
	\includegraphics[width=1.0in, height = 1.0in]{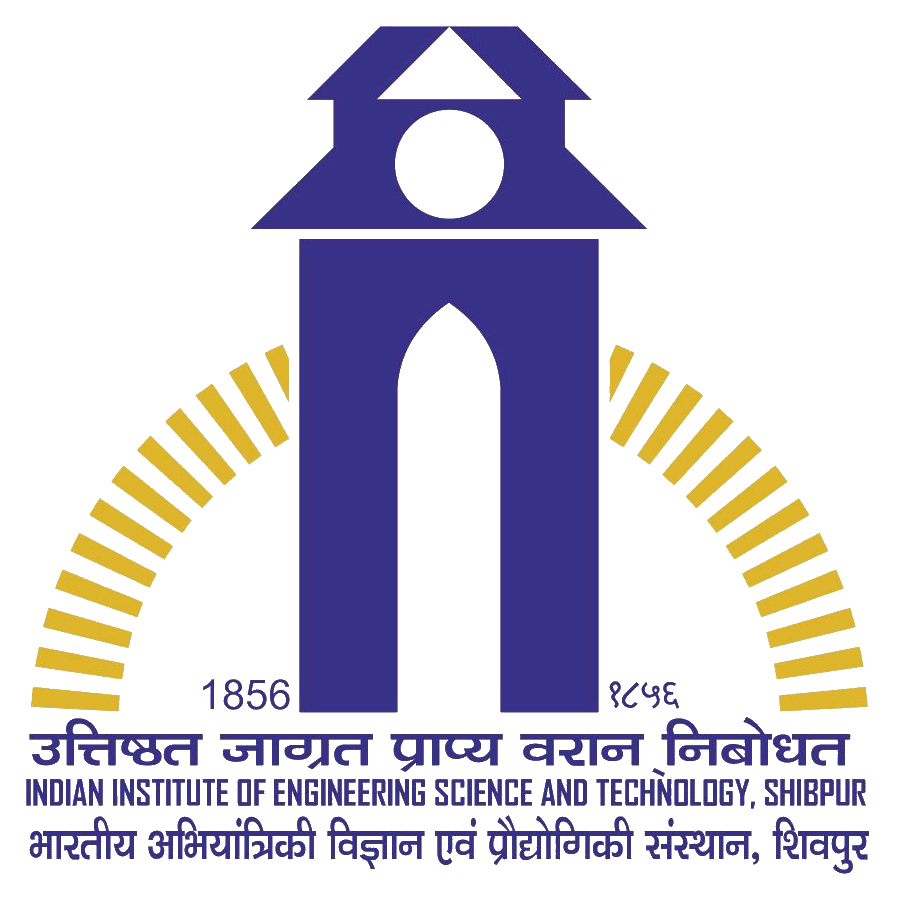}
	\end{figure}

        \vspace{0.2in} 
	{\textbf{Department of Information Technology}}\\ 
{\textbf{Indian Institute of Engineering Science and Technology, Shibpur}}\\
	{\textbf{ West Bengal, India -- 711103}}\\ 
        \vspace{0.5in} 
{\textbf{January, 2019}}\\ 
	\end{center} 
\newpage

\afterpage{\blankpage}
\pagestyle{empty} 
\begin{center} 
\begin{figure}[h]
\centering
	\includegraphics[width=1.1in, height = 1.1in]{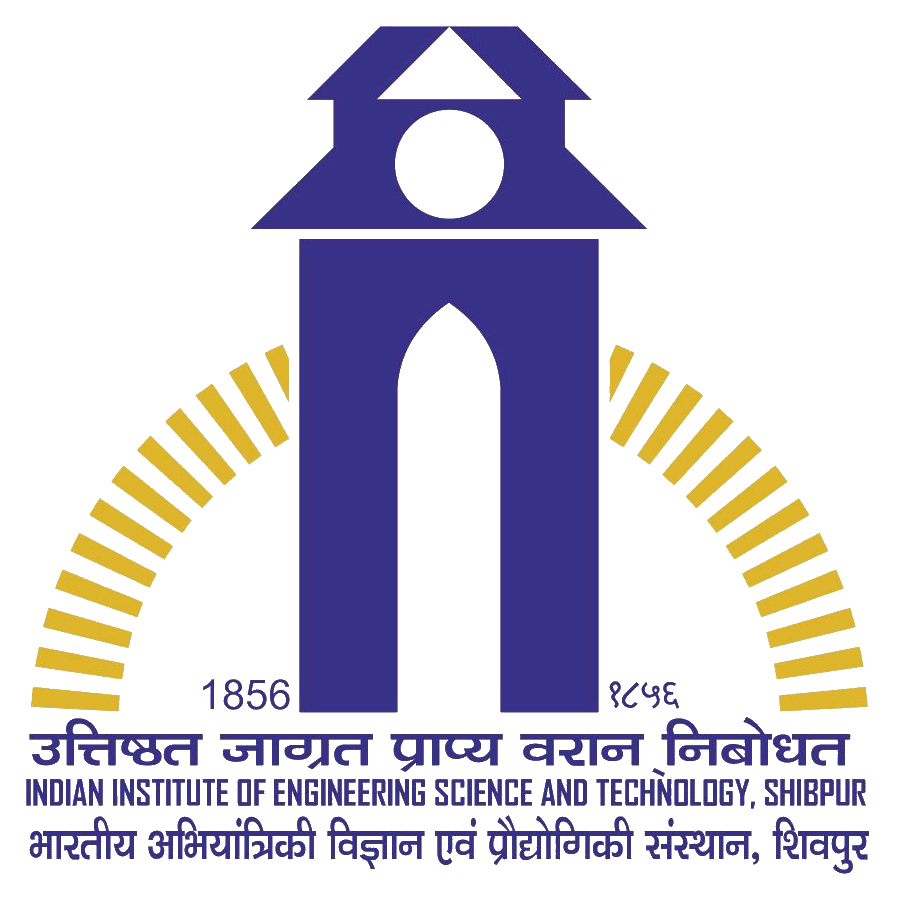}
\end{figure}
\small{\textsc{Department of Information Technology}}\\ 
\small{\textsc{Indian Institute of Engineering Science and Technology, Shibpur}}\\ 
\small{P.O.  Botanic Garden, \\ Howrah--711103}\\ 
\vspace{0.2in} 
 
{\Large\bf CERTIFICATE OF APPROVAL}\\ 
\end{center} 
\vspace{0.3in} 
\normalsize{\par It is certified that, the thesis entitled \emph{\textbf{``Cellular Automata: Reversibility, Semi-reversibility and Randomness''}} is a record of bona fide work carried out under my guidance and supervision by \textbf{Kamalika Bhattacharjee} in the Department of Information Technology of Indian Institute of Engineering Science and Technology, Shibpur.
	
In my opinion, the thesis has fulfilled the requirements for the degree of Doctor of Philosophy in Engineering of Indian Institute of Engineering Science and Technology, Shibpur. The work has reached the standard necessary for submission and, to the best of my knowledge, the results embodied in this thesis have not been submitted for the award of any other degree or diploma.

\vspace{1.0in} 
\begin{tabular}{lr}
 \hspace{3.68in}
	\includegraphics[width=3.0cm, height=1.2cm]{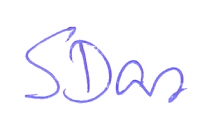}
\\
\hspace{-0.3in}{\textbf{Date:}} 14-JAN-2019 \hspace{2.38in}{\textbf{(Dr. Sukanta Das)}}  \\
\hspace{3.68in}\small{Associate~ Professor} \\
\hspace{3.1in}\small{Department of Information Technology} \\
\hspace{2.5in}\small{Indian Institute of Engineering Science and Technology, } \\
\hspace{2.8in}\small{Shibpur, Howrah, West Bengal, India--711103} \\
\end{tabular}
} 
\newpage

\afterpage{\blankpage}
\vspace*{\fill}  
\begin{center}
	
%

%
	
	\begin{quote}
		{\em ``In questions of science, the authority of a thousand is not worth the humble reasoning of a single individual.'' }
	\end{quote}
	\hspace*{3.55in}{\em -- Galileo Galilei, 1632 }
\end{center}
\vspace*{\fill} 
\newpage

\afterpage{\blankpage}


\vspace*{\fill}  

\begin{quote}
\begin{center}
\textbf{\large{\textit{Dedicated\\to\\ the Hope\\ that\\ every child of\\ the less priviledged genders\\ may fulfil her dreams}}}\\

%
%
%
%
%
%


\end{center}
\end{quote}

\vspace*{\fill} 

\clearpage

\pagestyle{plain}
\pagenumbering{roman}

\cleardoublepage
\phantomsection
\addcontentsline{toc}{chapter}{Acknowledgment}
\begin{center}
\vspace{-5.5cm}
 \textbf{\Large Acknowledgement}
\end{center}
We humans are social creatures who live and learn from the collective experiences and interactions with the fellow living beings, nature and environment. So, each individual achievement is an essence of the cumulative effect of the intentional or unintentional efforts of many known and unknown faces. Although I can not possibly name everyone who knowingly or unknowingly has supported me to reach up to this point, still I feel humbled and honored to get an opportunity to acknowledge with utmost gratefulness the contributions of some of the people who have helped me during my journey.  

First and foremost, I want to mention my mentor and guide Dr. Sukanta Das, Associate Professor, Department of Information Technology, Indian Institute of Engineering Science and Technology (IIEST), Shibpur, who has truly been the personification of an ideal teacher to me: a friend, philosopher and guide in literal sense. I have no words to express the paramount reverence and gratitude I feel towards him. Without his extreme care, ample patience and unconventional ability to understand me, I could not have completed this work.
He has always been a source of blissful shelter and constant support in every hurdle I have faced. 
It is through the enlightened interactions with him that I have learned to be committed in scientific attitude towards life. His incessant encouragement, sincere supervision, valuable advice and perfect guidance have taught me to think independently and deeply about any problem in research.

I would also like to avail this opportunity to extend my sincere admiration and respect to Prof. Mihir K. Chakraborty, Visiting Professor in the Department of Humanities and Social Sciences of this institute for his valuable suggestions and advice which have helped me to be more analytical and rigorous in using scientific methodology. His attitude, sense of beauty and philosophy towards life, society and science have encouraged me to see research and life in a new light. At the same time, I am incalculably grateful to Prof. Subhasis Bandopadhyay, Associate Professor, Department of Humanities and Social Sciences, IIEST, Shibpur and Prof. Biplab K. Sikdar, Professor, Department of Computer Science and Technology, IIEST, Shibpur, whose intellectual associations have helped me in many ways. 
I also want to thank Prof. Abhik Mukherjee, Associate Professor, Department of Computer Science and Technology, IIEST, Shibpur and Prof. Mallika Ghosh Sarbadhikary, Department of Associate Professor, Humanities and Social Sciences, IIEST Shibpur for their support and advice.

I feel special to express my gratitude to the Heads of the Department of Information Technology, IIEST, Shibpur, Prof. Hafizur Rahaman, Prof. Arindam Biswas and Prof. Santi P. Maity who have often taken out of the box way to extend help and support to me. I also want to thank my respected faculty members of this department who have been kind enough to extend their aid whenever required. I am thankful to Amiya sir, Malay Da, Suman Da and Dinu Da and all the other technical and non-technical staffs of this department for their support and service during this research. I am also very grateful to our \emph{Masi}, whose name I have never asked, but who has always taken it as her responsibility to keep our workspace tidy, clean and hygienic.

I want to accept with gratitude the financial support I have received from Innovation in Science Pursuit for Inspired Research (INSPIRE) under Department of Science and Technology, Government of India. Without this funding, it would not have been possible for me to pursue my dream. In the same breath, I also want to thank Santanu da (Santanu Chattopadhyay), my people manager at IBM India Pvt. Ltd., and my dear friend Deenu, who had faith in my abilities and who had encouraged and supported me to leave IBM during bond period and avail this wonderful opportunity.

Many part of this research is result of group work, where others have assisted me in validating my theoretical basis. So, I want to exploit this moment to offer my heartfelt thanks and appreciation to my co-researchers Dipanjyoti, Krishnendu, Nazma di and Souvik for their effort, support and time. 

It has been an immense pleasure to work with the members of our labs - Tuhin da, Srijit da, Subhankar da, Nilanjana di, Priyajit, Chandrakant, Sabyasachi da, Avik and Suman da, who have made this years long journey a joyful one. Specially, I want to thank Nashreen and Somrita di whose insightful conversations have often helped me to get through difficult days. This is also a chance to thank Nazma di and Biswanath da, my seniors in this journey, for their support, encouragement and advice. Further, I would like to take this moment as an opportunity to express my happiness to be part of the self-proclaimed ``Those-Who-Must-Not-Be-Named'' group with Nimisha, Supreeti, Sharmistha, Sankar, Pratima, Mayukh, Sumit and Souvik who have always lightened up my mood and taught me to be sometimes easy on myself. 

Being an introvert, emotional and over-sensitive person, I have often been very difficult to deal with. I am grateful that, I have some friends who have understood and supported me knowing all my vulnerabilities and weaknesses. Here, I feel lack of words to convey my love and gratefulness to my soul-sister Debashri, who has always motivated me to take responsibilities and be independent and confident like her. I also want to admit my deepest gratitude and tenderness to Raju, who has consistently cared, supported and protected me like my own brother in every respect. This is also an opportunity for me to thank Chandan, Pampa and Chhoton for supporting and tolerating me without any expectation. I would also want to express my thankfulness and affection towards Sukanya di, who has always cared and favored me like an elder sister.

I feel privileged to get immense love and support from my family, 
who, because of me, have sacrificed the most. Although no words will be enough to express my intense gratitude and tenderness towards them, still, I take this opportunity to mention them. First, I would like to take this moment to convey my heartfelt respect to my late \emph{dadu} (grandfather) from whom I have learned to be disciplined in life and serious and committed towards study and to my late \emph{thakuma} (grandmother) from whom I have learned to live with dignity and self-respect. Also, I want to express my ardent love and profound regard to my \emph{baba} (father) and \emph{maa} (mother), who have always been my source of inspiration and encouraged me to thrive for excellence and mental gratification with honesty.
I also want to convey my earnest love and gratefulness towards my caring little sister 
who has performed both the roles of me and her in my family so that I can continue my research with peace. I also want to express the same to Ravi, my best friend, who, 
during this journey, has acted as the forbearing receiver of all my irritations, frustrations and resentments, but still calmed and reminded me of my strengths and passions.

Last but by no means least, I want to remember with reverence and extreme gratitude the contributions of my teachers who have shaped my life. I can specially thank Pampa didimoni who has helped me to identify myself and Jiban Sir who has taught me to make sky as the limit to dream. I also would like to thank Jaseem Sir who has influenced me to fall in love with Mathematics and AKM Sir who has always supported and encouraged me to pursue my dreams. Whatever I have achieved, these people are the key shareholders to that success.

It breaks my heart to write that, even in this $21^{st}$ century, I live in a society, where children are abused, deprived and killed everyday because of the social prejudices like gender and caste. Today, while I get a chance to fulfill my passion, I sincerely hope that, I shall see a world, where every child gets equal opportunity and privilege like me and no dream of a child is left unfulfilled because of her gender.
\\ \\

\begin{flushright}
 	\includegraphics[width=5.5cm, height=1.9cm]{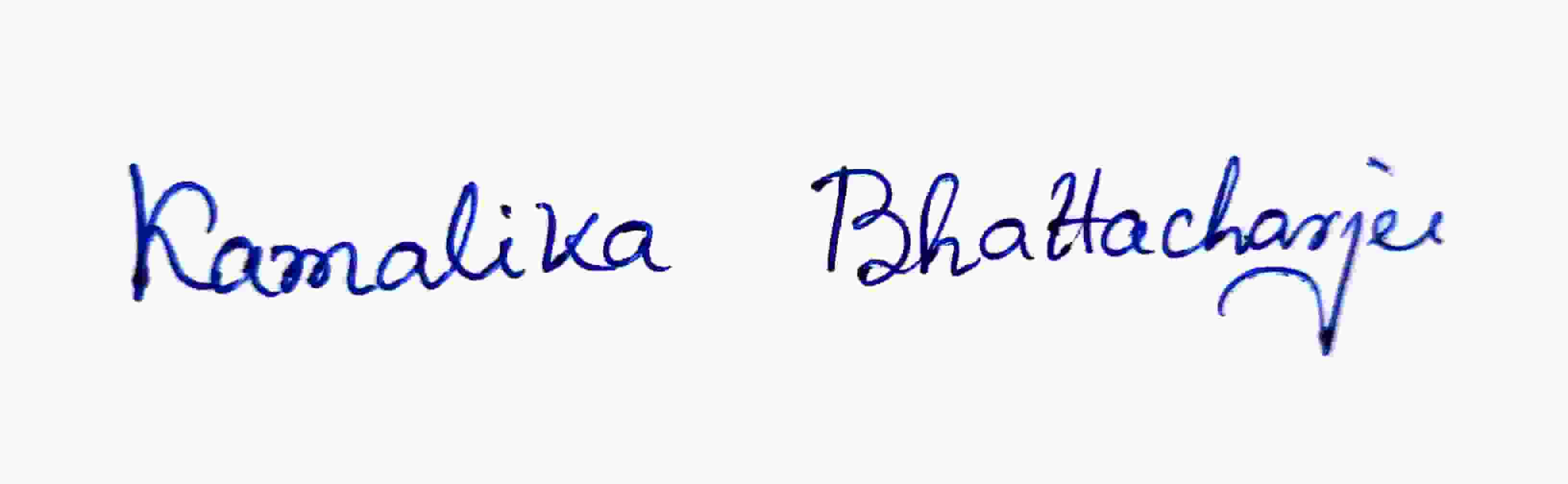}
\end{flushright} 
\vspace{-3.0em}
\noindent{\bf Dated:$~~$}14-JAN-2019  \\
{ Indian Institute of Engineering Science} \hspace*{0.5in} \dotfill\\
{ and Technology, Shibpur}  \hspace*{1.6in} {\bf(Kamalika Bhattacharjee)}\\
{ Howrah, West Bengal, India} \hspace*{1.33in} {[Reg. No: PhD/R/2014/0050]}\\



\cleardoublepage
\phantomsection
\addcontentsline{toc}{chapter}{Abstract}
\chapter*{Abstract}
{\large\textbf{C}}ellular automaton (CA), a discrete dynamical system, has enraptured the attention of researchers to study many of the foundational problems of nature. In this dissertation, we study two of the global properties of cellular automata (CAs), namely, reversibility and randomness. Although classically CAs are defined over infinite lattice, here we concentrate over one dimensional CAs with finite number of cells having periodic boundary condition.

In literature, a number of works are reported dealing with reversibility of finite CAs where the CAs are either linear or have only two states per each cell. However, because of lack in proper characterization tool, the reversibility of finite CAs having $d$ states per cell is mostly untouched. So, this dissertation targets to address this problem by developing a mathematical tool, named \emph{reachability tree}, which can efficiently characterize reversible finite CAs. A number of theorems are reported which signify the features of a reachability tree when it represents a reversible CA of size $n$. Further, we propose a scheme to limit the growth of the reachability tree with $n$ and develop \emph{minimized} reachability tree. For any CA rule, this minimized tree is bounded to a small height. Using this tree, a decision algorithm is developed which takes a CA rule and size $n$ as input and verifies whether the CA is reversible for that $n$. To limit the search space for finding reversible CAs, three greedy strategies are proposed which select a set of CAs candidates to be reversible for $n$.

To decide reversibility of a finite CA, we need to know both the rule and the CA size. However, for infinite CAs, reversibility is decided based on the local rule only. Therefore, apparently, these two cases seem to be divergent. This dissertation targets to construct a bridge between these two cases. To do so, reversibility of CAs is redefined and the notion of \emph{semi-reversible} CAs is introduced. Hence, we get a new classification of finite CAs -- (1) reversible CAs (reversible for every $n \in \mathbb{N}$), (2) semi-reversible CAs (reversible for some $n \in \mathbb{N}$) and (3) strictly irreversible CAs (irreversible for every $n \in \mathbb{N}$). A theorem is developed to identify the strictly irreversible CAs. Then, we again utilize minimized reachability tree and propose an algorithm to decide whether a finite CA is reversible or semi-reversible. Finally, relation between reversibility of finite and infinite CAs is established.

This dissertation also targets to explore CAs as source of randomness and build pseudo-random number generators (PRNGs) based on CAs. We identify a list of properties for a CA to be a good source of randomness. An example window based PRNG is developed using a $3$-state CA which fulfills these properties.
To understand the rank of the proposed PRNG, we conduct a brief survey of existing PRNGs and compare the well known PRNGs with respect to their randomness quality. However, we empirically find that the proposed PRNG along with the existing CAs based PRNGs have average randomness quality. They are not comparable to the elite group of PRNGs like \verb SFMT19937-64, ~\verb MT19937-64, ~etc. ~Nevertheless, we find that the tri-state CA based PRNG that we have developed, offers some inherent benefits. So, we identify a list of $596$ tri-state CAs which have similar qualities. 

It is empirically observed that, the increase of number of states per cell can improve the randomness quality. So, to design a CA based PRNG that can compete with the elite PRNGs in term of randomness quality, we increase the number of states per cell to $10$. Here, we inflict a more reserved property for the CAs to be unpredictable and use two greedy strategies to select CAs satisfying these properties. But the rule space for $3$-neighborhood $10$-state CAs is gigantic. Hence, we propose two heuristic algorithms to synthesize candidate CAs following these strategies. These CAs have great potentiality as PRNGs. We develop two schemes to use these CAs as window-based PRNGs-- (1) as decimal number generators and as (2) binary number generators. These PRNGs perform excellently in all empirical testbeds on the same platform. In fact, in comparison to the best PRNG \verb SFMT19937-64, ~average performance of our proposed PRNGs are slightly better. Hence, our decimal CAs based PRNGs are one of the best PRNGs today. 

\cleardoublepage
\tableofcontents

\cleardoublepage
\phantomsection
\addcontentsline{toc}{chapter}{\listfigurename}
\listoffigures

\cleardoublepage
\phantomsection
\addcontentsline{toc}{chapter}{\listtablename}
\listoftables


\glsaddall
\printglossary[style=long]

\clearpage

\pagestyle{fancy}
\pagenumbering{arabic} 

\chapter{Introduction}\label{Chap:Introduction}
\begin{center}
\begin{quote}
	{\em The real purpose of scientific method is to make sure Nature hasn’t misled you into thinking you know something you don’t actually know.}
\end{quote}
	\hspace*{3.05in}{\em -- Robert M. Pirsig, 1974}
\end{center}

\noindent{\small This dissertation deals with cellular automata and explores some of their behavior. In this context, this chapter provides the background as well as the motivation of undertaking this research. After stating the objective, the chapter briefly points out the significant contributions of the work.}
\section{Background} 
{\large\textbf{H}}uman beings have always been fascinated by nature and its wonders. Their intuitive quest to observe, understand, tame and mimic nature gave birth to Science and its various offspring disciplines $-$ Physics, Mathematics, Chemistry etc. It is through physics by which human discovers and approximates behavior of nature, whereas mathematics is the essential instrument to articulate, verify and predict them. However, often, these two fundamental disciplines unite to understand natural phenomena. In fact, starting with Aristotle until the end of Renaissance in late $17^{th}$ century, the dominant unadulterated study of nature was termed as \emph{Natural Philosophy}, the ancestor of \emph{natural science}. Even Isaac Newton's famous book ``\emph{Philosophiae Naturalis Principia Mathematica}'' (1687), meaning ``Mathematical Principles of Natural Philosophy'', reflects this context. Thanks to the contributions of Copernicus (1473--1543), Galileo Galilei (1564--1642), Kepler (1571--1630), Newton (1642--1727) etc., during $17^{th}$ century, research in science went through a \emph{paradigm shift}, often called the \emph{scientific revolution} and diverged into each of its progeny disciplines. 

Initially, the paramount way of studying nature in Physics was through experiment where a natural phenomenon is observed, measured, and quantified. Whereas, the logical or mathematical explanation of an occurrence in nature is a fundamental theory. However, since the Second World War, with the advance of computer science, nature is often studied through computation. This paradigm, implemented via discrete mathematical model or by computer simulation, plays an elementary role in contemplating situations which can not be directly analyzed or experimented. This modeling of natural systems via computational paradigm is widely used in many diverse fields, like physics, biology, social sciences, arts, economics, engineering etc.

The idea of computation can be traced back to Leibniz's \emph{calculating machine} (1673) which later inspired Hilbert to give his famous \emph{Entscheidungsproblem} \cite{hilbert19281928} (1928) and the concept of \emph{Finitistic Programme}. However, G\"{o}del's \emph{incompleteness theorems} (1931) were the first theoretical proof indicating the flaws in Hilbert's programme. These theorems are the base to the famous \emph{Church-Turing Thesis}
which shows that, solution to the Entscheidungsproblem is, in general, impossible (1936). Consequently, Turing's Turing machine, Church's $\lambda$-calculus and Kleene's \emph{recursion theory} basically convey the same idea, the notion of \emph{effectively calculable}. Today's computer architecture heavily relies on Turing machine as the model of computation, where computation is governed by a central control unit. However, as computer science progressed, study in alternative models of computation prospered; many of these are bio-inspired and offer parallel decentralized computing. Cellular Automata (CAs) emerged as one of the most important developments in this regard.

Cellular Automaton (CA) is a paradigm of uniform fine-grained parallel computation. It is a discrete conceptual model, which helps to understand complex systems by developing its model at the microscopic level. It is also an excellent tool to recognize the core essential elements of a complex system for emergence of a desired masoscopic behavior. Many phenomena of nature can be represented as a CA. Some even claim that nature is a quantum information processing system \cite{PhysRevLett.88.237901} where CA is the way by which nature does this processing \cite{Zuse1982, Wolframbook1}. Its role is similar to mathematical physics and analytical mechanics, and is most naturally and efficiently supported by physics. That is, after the arithmetization of physics (Galilio), the mechanization of physics (Newton, Faraday, Maxwell), the geometrization of physics (Poincare, Einstein), and the informatization of physics at  the statistics level (Boltzman, Gibbs), CA offers the informatization of physics at the fundamental level \cite{Hoekstra:2010:SCS:1855006}.

The notion of cellular automata was conceived in the $1950$s by John von Neumann and Stanislav Ulam in the Los Alamos National Laboratory, USA \cite{Neuma66, ulam1952random}. It is an artificial self-replicating machine capable of self-organization and universal computation. In words of Burks \cite{Burks}, a CA has the following fundamental characteristics: ``Its features are a quantised time and space, a finite number of possible states for each point of space-time, and a computable local transition function or law (not necessarily deterministic or uniform over space) governing the equations of the system through time.'' The operation of a CA is similar to the self-organized systems of nature and human-made world where the unit of computations (cells) perform local interaction among them in different dimensions without any governing central control system.

The signature feature of Cellular Automata is the simple local transition rules that each cell follows to change its state depending on its neighbors. The cells of a CA updates in parallel, thus, the automaton evolutes in discrete time and space. Hence, CAs provide a global framework where microscopic entities (states and rules) are adjusted for implementing parallel complex dynamic systems with emergent behavior.

Although first designed by John von Neumann, his CA (a $29$-state, $2$-dimensional CA capable of self-reproduction and universal computation \cite{Thatcher}) was cumbersome for practical implementation. In fact, CAs become popular when a complex system, named \emph{Game of Life} is implemented in CA by John Conway. The Game of Life represents $2$-state $2$-dimensional CA where extremely complex behavior arises from an exceptionally plain rule. That, even this simple CA is capable of universal computation, is proved by Martin Gardner \cite{Gardner71}. Since then, it has become a prot\'{e}g\'{e} in portraying the potential of CA in enthralling complex behavior of real world in its extraordinary simple grasp. 

In ever increasing field of CA as a computational model, its structure has further been simplified, and still continues to be modified, redefined and explored as per the demand of the system. A notable work in this regard is Stephen Wolfram's $2$-state $3$-neighborhood $1$-dimensional \emph{elementary} CAs \cite{Wolfr83}. In recent years, CA has been applied in engineering fields as technology, for example in VLSI design and test \cite{Horte89a, ppc1, SukantaTH}, cryptography and information security \cite{ppc1, Nandi94a, DBLP:journals/jca/DasR11}, pattern recognition and classification \cite{SMITH1972233, Maji05,maji2007rbffca, jen1986invariant, RAGHAVAN1993145}, pseudo-random number generation \cite{SukantaTH, wolfram86c}, image processing \cite{khan98, Rosin06} etc. Also it has found appeal in medical science \cite{MitraDCN96, DBLP:conf/iicai/GhoshLMC07}, social networking \cite{iet.cp.2013.2283,10.1007/978-3-642-40495-5_3,6108515,ZIMBRES2009157}, sociology \cite{Dabbaghian2012,doi:10.1080/13658810210157769,goldenberg2001using,doi:10.1002, doi:10.1080/136588198241617} etc.

CA is a prototype to observe and predict physical laws and behaviors of nature, and is instrumental in modeling physical and natural phenomenon \cite{Chopard}, such as, traffic systems \cite{nagel1992cellular,PhysRevLett-35}, structural design \cite{Hoekstra:2010:SCS:1855006} etc. CA characterizes some global behaviors of nature, such as reversibility, randomness, chaos and fractals, conservation laws, topology and dynamics, convergence etc. Therefore, scientists and researchers often directly apply CA to understand the global properties at the microscopic level. In this dissertation, we concentrate on two global properties -- reversibility and randomness. Both these are customary problems in Physics and Mathematics. We address these problems through \emph{classical} $1$-dimensional CAs with $d$-states per cell.

Reversibility is a well-known phenomenon in nature. A classical example of a reversible system is the planetary motion in orbits around a star. A deterministic process is called reversible, if its direction of execution in time can be reversed in the sense that, if the process runs back in time, it will return to the initial condition without loosing any of its properties. In mathematics, such a function is bijective or one-to-one, so the inverse of it exists. In thermodynamics, reversibility is related to entropy; there is no increment in entropy, if a process is reversed in the opposite direction by infinitesimally slow changes in the system. Most of the physical laws of nature are \emph{time-symmetric} and can be reversible in ideal situation. In microscopic world of nature, many of the systems and processes are reversible. Therefore, CA, a microscopic model of the complex world, is a natural option to observe this phenomenon.

A CA is reversible, if every configuration of it has exactly one predecessor. This implies, there is no loss of information during the evolution of the CA. This property has direct correspondence with the reversibility of microscopic physical systems, implied by the laws of quantum mechanics. The reversible CAs have been utilized in different domains, like simulation of natural phenomenon \cite{hartman90}, cryptography \cite{Nandi94a}, pattern generations \cite{Kari2012180}, pseudo-random number generation \cite{aspdac04}, recognition of languages \cite{Kutrib20081142} etc. 
Classically, CA is considered as infinite and reversibility in CA depends on CA rule only. However, when CA is defined over finite lattice, the whole scenario changes. For example, a CA which is irreversible when defined over infinite lattice, may become reversible, when its lattice size is a finite value. In this dissertation, we are interested in reversibility of CAs under finite lattice size.

One can observe that, classical reversible CAs are generally very simple. However, the CAs, which are reversible when defined under finite cell length constraint, often exhibit much complexity in their behavior. For instance, often they are unpredictable and can generate configurations which appear to be random. A selection of an object is called random, if each possible outcome has equal chance of getting selected irrespective of the order of the objects. In fact, this notion of randomness is one of the most perplexing and frequently encountered concept in nature and society. In philosophy of science, randomness is often related to indeterminism, unpredictability, probability and chance \cite{Kirschenmann1972,doi:10.1093/bjps/axi138}. It is also often associated with information entropy. Thanks to quantum mechanics and chaos theory, some even consider nature as random at fundamental level \cite{0034-4885-80-12-124001}. This makes the study of randomness a bewitching observation of nature. CA, often considered as a basic model of nature, represents a complex system by its well-defined lattice, local interaction and simple transition rules. Therefore, study of CA is a good approximation of the randomness created by the locally interactive simple systems in nature.

Since $20^{th}$ century, with the development of computer science, there is an increasing demand of apparent randomness generated by a deterministic system, that is generating randomness by computation. This algorithmic randomness or \emph{pseudo}-randomness is required in diverse applications and fields, like probability theory, game theory, information theory, statistics, gambling, computer simulation, cryptography, pattern recognition, VLSI testing etc. All these applications entail numbers, which behave like random ones, but can be reproduced on demand. These deterministic systems are singularly dependent on initial conditions and are preferable to be simple yet unpredictable. CA, as a computational model, in general, aptly qualifies all these criteria; so, we explore CAs to indulge into their randomness.

\section{Motivation and Objectives of the Thesis}
The study of reversibility in cellular automata was started by Hedlund \cite{hedlund69} and Richardson \cite{Richa72}. Since then, it has attracted attention of many researchers \cite{Amoroso72,nasu1977local,Maruoka197947,Maruoka1982269,sato77,di1975reversibility,culik1987invertible,suttner91}. All these works deal with {\em classical reversibility} of CAs. This reversibility is dependent on the local rule only, as here, the CAs are defined over infinite lattice. However, what will happen to its reversibility, if we restrict the size of the CA to be finite? (Strictly speaking, if size of the CA changes, the global transition function also changes, so the CAs are not exactly same; however, here by the same CA, we are implying the local rule to be same.) In fact, finite CAs with a fixed lattice size is desirable for researchers working on practical applications and implementations. Therefore, in this dissertation, we explore the notion of {\em reversibility of a finite CA with fixed size $n$.}

Recently, research on reversibility of finite CA with a fixed size $n$ has gained recognition \cite{zubeyir11,SukantaTH,entcs/DasS09,marti2011reversibility,Ino05}. All these works consider CAs, where the local rule is linear \cite{ppc1,zubeyir11} or the CAs are binary \cite{SukantaTH,Soumya2011,entcs/DasS09,Soumya2010}. The linear CAs is chosen because standard algebraic techniques, like matrix algebra, can be used to characterize them, whereas, for binary CAs, the characterization is simpler. However, reversibility of finite CAs having $d$-states per cell is, in general, an untouched issue. One of the reasons is lack of proper characterization tool for such CAs. Study of reversibility for $d$-state finite CAs is essential for experiencing the reversibility in nature at microscopic level and simulating it on computer. This motivates us to take this research in this dissertation, to fill up the gap and explore reversibility of $1$-dimensional $d$-state CAs with size $n$ by developing a characterization tool.

While studying reversibility of finite CAs, the size of the CA can be any number $n \in \mathbb{N}$. Hence, one can keep on increasing $n$ and check the corresponding reversibility status of the CA. If a CA is reversible for size $n$, it may not remain reversible for size $n+1$. So, what will be its reversibility, when the CA is defined over infinite lattice?
What if, a CA is irreversible when it is infinite, but reversible for some finite cell lengths? What do we call such CAs, where a set (possibly infinite) of lattice sizes exists for which the CA is reversible? All the researches done on classical reversibility are silent about these issues. In the classical notion of reversibility, there is no such concept as reversibility dependent on CA size. So, in this dissertation, we question this classical notion of reversibility and introduce the idea of \emph{semi-reversibility}: reversibility of CAs dependent on its lattice size. We redefine the notion of reversibility into three groups -- \emph{reversible} (reversibility for each $n\in \mathbb{N}$), \emph{semi-reversible} (reversibility for some $n\in \mathbb{N}$), and \emph{strictly-irreversible} (irreversible for each $n\in \mathbb{N}$). Our objective is to connect reversibility of finite CAs with that of the infinite ones and define any relation, if exists, between these two divergent classes.

In general, for an arbitrary CA, there is no known method to predict $t^{th}$ configuration in advance, when $t$ is sufficiently large. By using CA, the level of decadence of local information within the system is also reduced \cite{Hoekstra:2010:SCS:1855006}. Many chaotic CAs exhibit good randomness properties; hence, theoretically, can act as random sequence generator.
In literature, several attempts have been made to use $2$-state finite CAs as pseudo-random number generators (PRNGs) \cite{Horte89a,ppc1,aspdac04,alonso2009elementary,tcad/DasS10,114093,Tomassini96, Marco00,Guan04a,Guan04}. However, none of these CAs are at par with the best PRNGs, like Mersenne Twisters and its variants \cite{Matsumoto:1998:MTE:272991.272995,Saito2008,Saito2009}. We have observed that, due to intrinsic properties of some semi-reversible and strictly irreversible CAs, they can outperform other CAs as source of randomness. This motivates us to take up this research to explore the randomness quality of $d$-state CAs, especially, when the CA is not reversible for all $n \in \mathbb{N}$. Since the search space of $d$-state ($d>2$) CAs is larger than that of $2$-state CAs, we expect to improve randomness by increasing $d$. Our objective is to use these CAs efficiently to build good PRNGs.

Smith proved that, a CA with higher neighborhood dependency can always be emulated by another CA with lesser, say, $3$-neighborhood dependency \cite{Smith71}. So, throughout the thesis, we concentrate mostly on working with $3$-neighborhood CAs (if not otherwise explicitly mentioned) under periodic boundary condition.

\section{Contributions of the Thesis}
This dissertation focuses on two global properties of $1$-dimensional $d$-state CAs under periodic boundary condition -- reversibility and randomness. Research done in these domains can be summarized to the following contributions:
\begin{itemize}
	
	\item A characterization tool, named \emph{Reachability tree} has been extended to characterize $1$-dimensional CAs. This tool is instrumental in developing theories of finite CAs. 
	
	\item A number of theorems are reported which depict the properties of reachability tree when it presents a finite CA, reversible for a lattice size $n$.
	
	\item A decision algorithm to test reversibility of a finite $3$-neighborhood CA with a particular cell length $n$ is developed.
	
	\item Three greedy strategies are identified to synthesize a set of CAs reversible for any CA size. 
	
	\item Identification of \emph{semi-reversible} CAs, which are classically treated as irreversible, but are reversible for a set (possibly infinite) of lattice sizes. Hence, we get a new classification of CAs -- reversible, semi-reversible and \emph{strictly} irreversible.
	
	
	\item A scheme is developed to minimize the reachability tree of a CA with an arbitrary lattice size. An algorithm is reported to find the type of reversibility of a $1$-D CA using the \emph{minimized} reachability tree. 
	
	\item Relation between reversibility of CAs for finite and infinite cases is established.
	
	
	\item A list of desirable properties is recognized for a $1$-dimensional $3$-neighborhood CA to be a good source of randomness.
	
	\item A portable window-based PRNG scheme using a $3$-neighborhood $3$-state CA is developed which offers several benefits, like robustness, exceptionally large cycle length and unpredictability. A list of $3$-neighborhood $3$-state CAs with similar randomness quality is identified.
	
	\item To further improve randomness, decimal CAs with $3$-neighborhood dependency are explored and some greedy strategies are reported which ensures the CAs following these strategies are good source of randomness.
	
	\item A heuristic synthesis algorithm is developed which implements these strategies. A simplified heuristic algorithm is also reported to generate good decimal PRNGs, which can be represented in a decent way. 
	
	
	\item To use these CAs as PRNGs, two schemes are developed -- as decimal number generator and binary number generator. These decimal CAs based PRNGs from the two heuristic algorithms are shown to be at least as good as the best PRNG existing in literature.
\end{itemize}

\section{Organization of the Thesis}
The thesis is organized as follows:

\textbf{Chapter \ref{Chap:surveyOfCA}} reports a tour to the journey of CA starting from its early form to present day. During this journey, different types of automata have been flourished by varying the parameters of CAs. Whereas, several characterization tools have been developed, which are used to understand the behavior of CAs. In this chapter, we briefly survey the types of CAs, the characterization tools and the global behavior of CAs. Then, we focus our discussion on the non-classical CAs - asynchronous CAs, automata networks, and non-uniform CAs. Finally, we briefly describe some of the application areas of CAs.

\textbf{Chapter \ref{Chap:reversibility}} investigates the reversibility of $1$-dimensional $3$-neighborhood $d$-state finite CAs under periodic boundary condition.
To characterize the finite CAs, a discrete tool named {\em reachability tree} is developed which represents all possible reachable configurations of an $n$-cell CA. Then we study the properties of reachability tree when the CA is reversible for size $n$ and develop a set of reversibility conditions. If reachability tree of an $n$-cell CA satisfies these conditions, the CA is reversible for that $n$. However, as reachability tree grows exponentially with $n$, for a large $n$, such approach is difficult to handle. To solve this, we propose \emph{minimized} reachability tree which stores only the unique nodes; hence, maximum size of the tree for any local rule is fixed. Now we report a decision algorithm which develops the minimized reachability tree and checks the conditions of reversibility for an $n$ at addition of each new node in the tree. If at any point, a reversibility condition is violated, the CA is reported as irreversible for that $n$. Complexity of this algorithm is $\mathcal{O}({dM^2})$, where $M$ is the maximum number of unique nodes for the $n$-cell CA. Then we illustrate efficient ways of identifying a set of finite reversible CAs with the help of three greedy strategies.

\textbf{Chapter~\ref{Chap:semireversible}} deals with two questions -- (1) Is it possible to understand the reversibility of a finite CA of size $n\in \mathbb{N}$ by exploring the CA for some small values of $n$? (2) What is the relation between reversibility of finite CA over fixed $n$ to that of infinite CA?
To address these issues, we redefine the notion of reversibility (and irreversibility) of CAs, and classify them into three sets -- (1) reversible (reversibility for each $n \in \mathbb{N}$), (2) strictly irreversible (irreversibility for each $n \in \mathbb{N}$) and (3) semi-reversible (reversibility for some $n \in \mathbb{N}$).
A CA is strictly irreversible if it is irreversible for size $1$. However, to decide whether a CA belongs to any other classes of reversibility, we again take help of minimized reachability tree (extended for $d$-state $m$-neighborhood CAs). For any CA, the minimized tree has finite height, whereas this tree contains all information regarding the reversibility of the CA for any $n$. So, we propose a decision algorithm based on the properties of minimized reachability tree to decide the reversibility class of a finite CA.
Finally, we establish the relation between reversibility of the finite CAs with that of the infinite CAs.

\textbf{Chapter~\ref{Chap:randomness_survey}} explores the concept of pseudo-random number generation and the role of cellular automata in it. After identifying the essential properties of a CA to be a good source of randomness, this chapter develops an example PRNG using a $3$-state CA.
To test the randomness quality of this proposed PRNG, we use Diehard, TestU01 and NIST as the statistical testbeds and lattice test and space-time-diagram for graphical testing. To compare and find the actual ranking of this PRNG among the existing PRNGs, a brief survey of the existing PRNGs is conducted. Among the PRNGs, we choose $28$ well-known PRNGs and rank them according to their randomness quality based on the result of empirical tests on similar platform. It is observed that, among the existing PRNGs, \verb SFMT19937-64 ~outperforms others. Further, we perceive that, the ranking of all CAs based usable PRNGs, including our proposed PRNG, is around average. However, our proposed PRNG offers benefits like ease of implementability, portability etc. The empirical results also hint that, by increasing number of states of a CA, its randomness quality can be improved.


\textbf{Chapter~\ref{Chap:3-stateCA_list}} investigates the prospect of $3$-neighborhood $3$-state CAs as source of randomness.
Here, we first analyze the aspect of using tri-state CA as a PRNG. We observe that, apart from portability and large cycle length, such CAs also offer robustness and possibility of implementation on hardware using ternary logic gates.
Inspired by this result, we further look into the vast rule space ($3^{27}$) of $3$-state CAs to find rules having such qualities. To get the initial set of rules, we apply two greedy strategies developed in Chapter~\ref{Chap:reversibility}.
However, to reduce the rule space while ensuring the selected rules are good source of randomness, we apply theoretical filtering over these rules where inherent structure of the CAs are explored to improve randomness quality.
On theoretically filtered rules, empirical filtering is applied. We use Diehard as the tool for this filtering.
Finally, we report $596$ \emph{good} $3$-state CAs, which qualify as potential PRNGs.

\textbf{Chapter~\ref{Chap:10-stateCA}} covers the potentiality of $3$-neighborhood $10$-state CAs as good PRNGs. Apart from the essential properties, the chapter points out some additional features which, if satisfied by a CA, then it can offer better randomness. Exploring these features, CAs are synthesized using greedy strategies.
While designing PRNGs using these CAs, two schemes are contemplated -- to generate (1) decimal numbers of $w$ digits and (2) binary numbers of $32\times b$ bits. Here, $w$ and $b$ are variables.
Our CA offers flexibility to implement both these schemes. Likewise $3$-state CAs, these PRNGs also have the same benefits -- portability, robustness, very large cycle length and unpredictability. We show that, in terms of the empirical tests, our CAs are at par with \verb SFMT19937-64, ~the best ranked generator in Chapter~\ref{Chap:randomness_survey}. Therefore, if we consider the other strengths, these CAs can beat all other PRNGs in overall performance. This research evidently indicates that, if we increase the number of states of a CA, a finite CA can be an excellent source of randomness.

\textbf{Chapter~\ref{Chap:conclusion}} concludes the thesis with some unsolved questions which may be addressed in future.

\chapter{A Survey of Cellular Automata: Types, Dynamics, Non-uniformity and Applications}\label{Chap:surveyOfCA}

\noindent {\small Cellular automata (CAs) are dynamical systems which exhibit complex global behavior from
simple local interaction and computation. Since the inception of cellular automaton (CA) by von Neumann in $1950s$, it has attracted the attention of several researchers over various backgrounds and fields for modeling different physical, natural as well as real-life phenomena. Classically, CAs are uniform. However, non-uniformity has also been introduced in update pattern, lattice structure, neighborhood dependency and local rule. In this chapter, we survey the various types of CAs introduced till date, the different characterization tools, the global behaviors of CAs, like universality, reversibility, dynamics etc. Special attention is given to non-uniformity in CAs and especially to non-uniform elementary CAs, which have been very useful in solving several real-life problems.}

\section{Introduction}

{\large\textbf{F}}rom the end of the first half of $20^{th}$ century, a new approach has started to come in scientific studies, which after questioning the so called Cartesian analytical approach, says that interconnections among the elements of a system, be it physical, biological, artificial or any other, greatly effect the behavior of the system. In fact, according to this approach, knowing the parts of a system, one can not properly understand the system as a whole. In physics, David Bohm \cite{bohm2002wholeness} is one of the advocates of this approach. This approach is adopted in psychology by Jacob Moreno \cite{moreno1932application}, which later gave birth to a new branch of science, named Network Science. During this time, however, a number of models, respecting this approach, have started to be proposed, see, as an example, \cite{McCulloch1943}. Cellular Automata (CAs) are one of the most important developments in this direction.

The journey of CAs was initiated by John von Neumann \cite{Neuma66} for the modeling of biological \emph{self-reproduction}. A cellular automaton (CA) is a discrete dynamical system comprising of an orderly network of cells, where each cell is a finite state automaton. The next state of the cells are decided by the states of their neighboring cells following a local update rule. John von Neumann's CA is an infinite $2$-dimensional square grid, where each square box, called cell, can be in any of the possible $29$ states. The next state of each cell depends on the state of itself and its four neighbors. This CA can not only model biological self-reproduction, but is also computationally universal. The beauty of a CA is that simple local interaction and computation of cells results in a huge complex behavior when the cells act together.

Since their inception, CAs have captured the attention of a good number of researchers of diverse fields. They have been flourished in different directions -- as universal constructors \cite{Thatcher,banks1970universality}; models of physical systems \cite{Chopard}; parallel computing machine \cite{Toffo87}; etc. In recent years, the CAs are highly utilized as solutions to many real life problems. As a consequence, a number of variations of the CAs have been developed. In this chapter, we conduct a tour to a few aspects of CAs. Our purpose is not to explore any specific aspect of CAs in depth, rather to cover different directions of CAs research along with a good number of references. Interested readers may go to the appropriate references to explore their interests in detail.

We organize this survey with respect to six categories. First, we tour to the various types of CAs, developed since their inception (Section~\ref{Chap:surveyOfCA:sscn_typeCA}). For example, von Neumann's CA and Wolfram's CA are of different types -- first one is defined over $2$-dimensional grid having $29$ states per cell, whereas latter is one-dimensional binary CA. Second, to study the behavior of CAs, few tools and parameters are developed. They are visited in Section~\ref{Chap:surveyOfCA:scn_chrtool}. However, the most interesting property of CAs is probably their complex global behavior due to very simple local computations of the cells. We inspect some of the most explored global behaviors in Section~\ref{Chap:surveyOfCA:sscn_prpCA}.

During last two decades, a new direction in CAs research has been opening up, which introduces non-uniformity in CA structure. Practically, CAs are uniform in all respects - uniformity in cell's update (that is, synchronicity), uniformity in neighborhood dependency, and uniformity in local rule. Though the \emph{uniform} CAs are very good in modeling physical systems, researchers have introduced \emph{non-uniformity} in CA structure to model some physical systems in a better way, and most importantly, to solve some real life problems efficiently. We survey various aspects of non-uniformity in Section~\ref{Chap:surveyOfCA:scn_nunCA}, and put our deeper attention on a special type of non-uniform CAs in Section~\ref{Chap:surveyOfCA:scn_eca}.

Nowadays, CAs are not only modeling tools, but also technologies. Use of CAs in different application domains has established this fact in the last two decades. We explore some of such applications in Section~\ref{Chap:surveyOfCA:scn_appCA}.

\section{Types of Cellular Automata}
\label{Chap:surveyOfCA:sscn_typeCA}
In this section, we survey various types of CAs proposed till date. Before stating these types, let us define cellular automata formally.

\subsection{Basic definition}

A cellular automaton consists of a set of cells which are arranged as a regular network. Each cell of a CA is a finite automaton that uses a finite state set $\mathcal{S}$. The CAs evolve in discrete time and space. During evolution, a cell of a CA changes its state depending on the present states of its neighbors. That is, to update its state, a cell uses a {\em next state function}, also known as {\em local rule}, whose arguments are the present states of the cell's neighbors.
Collection of the states of all cells at a given time is called {\em configuration} of the CA. During evolution, a CA, therefore, hops from one configuration to another. 


\begin{definition}
\label{Def:basic}
A cellular automaton (CA) is a quadruple ($\mathscr{L},\mathcal{S},\mathcal{N},R$), where
\begin{itemize}
\item $\mathscr{L} \subseteq \mathbb{Z}^{D}$ is the $D$-dimensional cellular space. A set of cells are placed in the locations of $\mathscr{L}$ to form a regular network.

\item $\mathcal{S}$ is the finite state set.

\item $\mathcal{N} = (\overrightarrow{v_1}, \overrightarrow{v_2}, \cdots, \overrightarrow{v_m})$ is the neighborhood vector of $m$ distinct elements of $\mathscr{L}$ which associates one cell to its \emph{neighbors}. Generally, the neighbors of a cell are the nearest cells surrounding the cell. However, when the neighborhood vector $\mathcal{N}$ is given, then the neighbors of a cell at location $\overrightarrow{v} \in \mathscr{L}$ are at locations $(\overrightarrow{v} + \overrightarrow{v_i}) \in \mathscr{L}$, for all $i\in \{1, 2, \cdots, m\}$.

\item  $R:\mathcal{S}^{m}\rightarrow \mathcal{S}$ is called the local rule of the automaton. The next state of a cell is given by $R(a_1, a_2, \cdots, a_m)$, where $a_1, a_2, \cdots, a_m$ are the state of its $m$ neighbors.
\end{itemize}
\end{definition}

A cellular automaton is also identified by its {\em global transition function}. Let us consider $\mathcal{C}$ represents $\mathcal{S}^{\mathscr{L}}$, the set of all configurations. Then, a CA is a function $G:\mathcal{C}\rightarrow \mathcal{C}$, which is called global transition function. Classically, CAs are synchronous (that is, all the cells are updated simultaneously) and homogeneous (that is, all the cells use single next sates function). For these cases, if a configuration $y=(y_{\overrightarrow{v}})_{\overrightarrow{v} \in \mathscr{L}}$ is successor of another configuration $x=(x_{\overrightarrow{v}})_{\overrightarrow{v} \in \mathscr{L}}$, that is, $y=G(x)$, then $y$ is the result of the following application: 
For each $\overrightarrow{v} \in \mathscr{L}$
\begin{equation}
y_{\overrightarrow{v}} = G(x)_{\overrightarrow{v}} = G(x_{\overrightarrow{v}})= R(x_{\overrightarrow{v}+\overrightarrow{v_1}}, x_{\overrightarrow{v}+\overrightarrow{v_2}}. \cdots, x_{\overrightarrow{v}+\overrightarrow{v_m}})
\end{equation}

Nevertheless, a CA is described in terms of four parameters -- $\mathscr{L}, \mathcal{S}, \mathcal{N}$ and $R$. Over the years, various types of CAs have been proposed varying the properties of these parameters. Even this basic definition of CA (Definition \ref{Def:basic}) is sometime compromised to get new types of CAs. In this section, we survey different types of CAs, developed till date, under the following headings --

\begin{enumerate}
\item The dimension and neighborhood of a cell.
\item The states of the cell.
\item The lattice size and boundary condition.
\item The local rule.
\end{enumerate}

\subsection{The dimension and neighborhood of a cell}
\label{Chap:surveyOfCA:Sec:DN}
\begin{figure}
     \subfloat[One way CA\label{1way}]{%
       \includegraphics[width=0.20\textwidth, scale = 0.7]{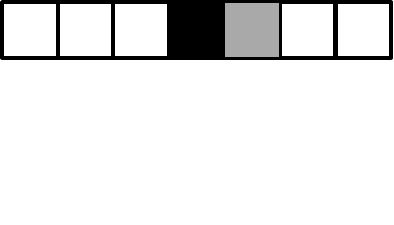}
     }
      \hfill
     \subfloat[Two way CA \label{2way}]{%
       \includegraphics[width=0.20\textwidth, scale = 0.7]{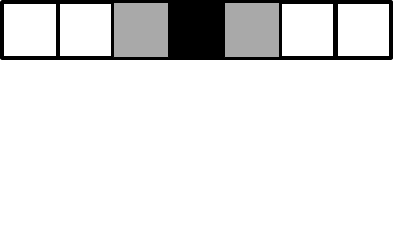}
     }
      \hfill
	\subfloat[von Neumann neighborhood \label{von}]{%
       \includegraphics[width=0.20\textwidth, scale = 0.7]{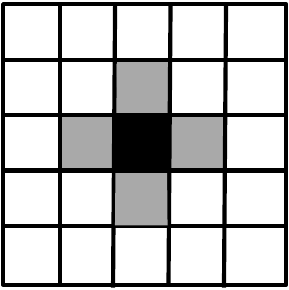}
     }
      \hfill
     \subfloat[Moore neighborhood \label{moore}]{%
            \includegraphics[width=0.20\textwidth, scale = 0.7]{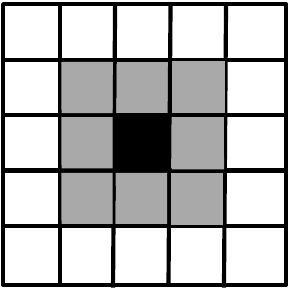}
     }
     \caption{Neighborhood dependencies: \ref{1way} and \ref{2way} are $1$-dimensional CAs, and \ref{von} and \ref{moore} are $2$-dimensional CAs. The black cell is the cell under consideration, and its neighbors are the shaded cells and the black cell itself}
     \label{fig:neighborhood}
 \end{figure}

The neighborhood of CA cells is strongly correlated with the dimension of the CAs. The original CA, proposed by von Neumann, is of $2$-dimension and uses $5$-neighborhood (orthogonal ones and itself) dependency. Figure~\ref{von} shows such a dependency: the neighbors of the black cell is itself and the four shaded cells. This neighborhood dependency is traditionally known as {\em von Neumann neighborhood}.

A natural extension of this neighborhood dependency is the 9-neighborhood dependency, where four non-orthogonal cells are additionally considered as neighbors. Figure~\ref{moore} shows this kind of neighborhood dependency, which was proposed by Edward F. Moore \cite{moore1962machine}, and is traditionally known as {\em Moore neighborhood}.
This neighborhood structure has been utilized to design the famous \emph{Game of Life}, a CA which was introduced by John Conway and popularized by Martin Gardner \cite{Gardner71}. 

\begin{figure}
   \subfloat[Margolus neighborhood\label{Fig:Margolus}]{%
       \includegraphics[scale = 0.3]{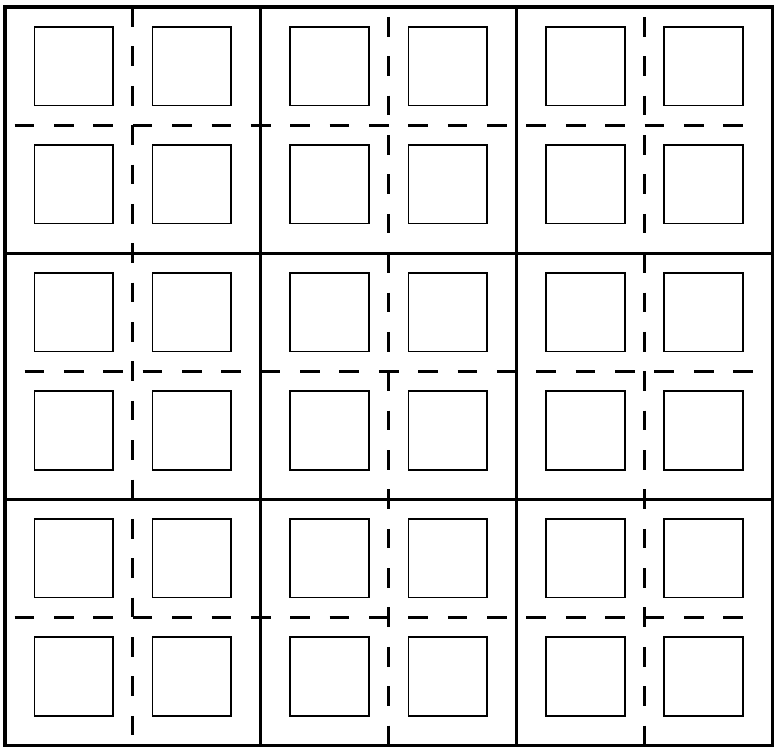}
     }
      \hfill
     \subfloat[Hexagonal CA\label{Fig:Hexagonal}]{%
       \includegraphics[scale = 0.3]{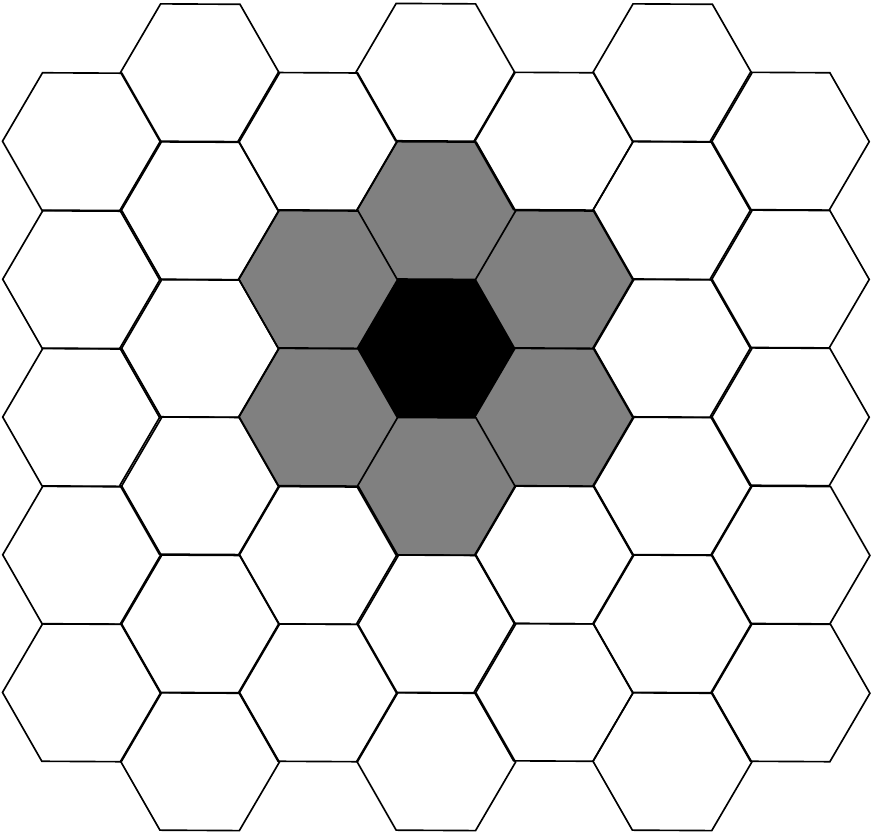}
     }
      \hfill
     \subfloat[Partitioned CA\label{Fig:PartitionedCA}]{%
       \includegraphics[scale = 0.3]{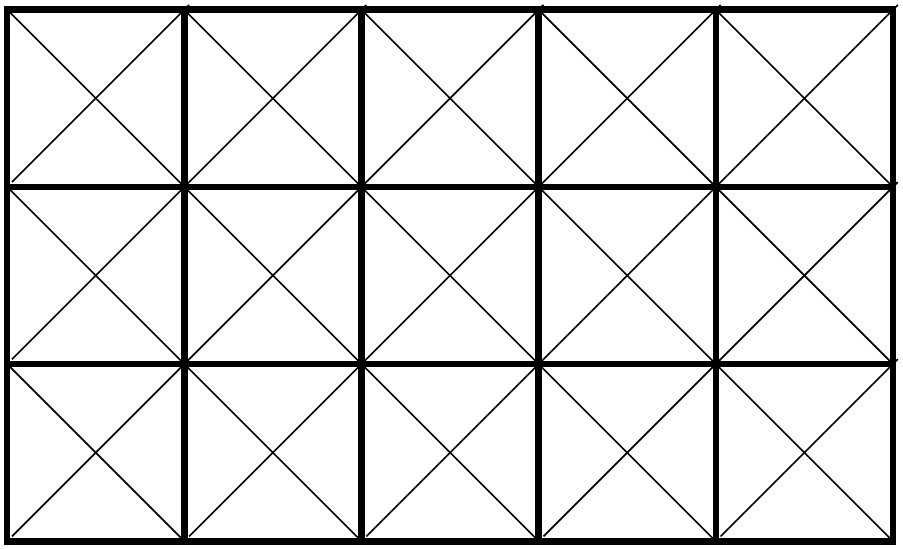}
     }

     \caption{Neighborhood dependencies: \ref{Fig:Margolus} is a block CA, \ref{Fig:Hexagonal} is hexagonal CA, and \ref{Fig:PartitionedCA} is a partitioned CA, where a square cell is partitioned into four triangular parts}
     \label{Fig:Margo+Hexa}
 \end{figure}

Apart from these two popular neighborhood dependencies for $2$-dimensional CAs, some other variations, such as Margolus neighborhood \cite{Toffo87} are also reported. In this neighborhood, the lattice $\mathscr{L}$ is divided into $2 \times 2$ non-overlapping blocks ($2 \times 2 \times 2$ cubes in three dimensions). The partitioning of $\mathscr{L}$ into blocks is applied on different spacial co-ordinates on alternate time steps. Figure~\ref{Fig:Margolus} clarifies the idea: the smaller boxes are the cells, whereas the squares enclosing 4 cells are the blocks. There are two types of overlapping blocks, shown by dark lines and dotted lines. The cells of a block are neighbors of each other, and the type of blocks alternates in alternating time steps. This kind of CAs is also known as {\em block cellular automata} or {\em partitioning cellular automata}. In these CAs, the local rule, or rather {\em block rule}, updates the entire block as a distinct entity rather than any individual cell.

However, in the above types of CAs, the cells are considered as squares. In fact, in most of the CAs, the cells are square in shape. In some of the works, on the other hand, the cells are considered as hexagons over 2-dimensional space, and as a result, we get a different neighborhood dependency; see for example \cite{refId0,Siap2d}. Figure~\ref{Fig:Hexagonal} shows such a neighborhood dependency. Not only hexagonal cells, K. Morita and his co-researchers \cite{IMAI2000181,morita2016universality} have worked with triangular cells also. Further, CAs have been defined in hyperbolic plane in \cite{margenstern1999polynomial,Margenstern200199}. As a result, we have been witnesses of several types of CAs depending on the variations of neighborhood dependencies. As an extension of uniform neighborhood dependencies of cells of CAs, {\em Automata Networks} (see Section~\ref{Chap:surveyOfCA:sscn_NA}) have been defined over arbitrary networks, where the neighborhood dependencies of different cells may be different; see for instance \cite{Marr20,tomassini29,Darabos7}.

{\em Partitioned cellular automata}, which are closely related to block cellular automata, are the CAs where each cell is partitioned into a finite number of parts. Hexagonal CA of \cite{refId0}, triangular CA of \cite{morita2016universality} are the partitioned CAs. Many properties of these CAs have been explored by Morita and his co-researchers. Figure~\ref{Fig:PartitionedCA} shows the partitioning of square cells into four parts. Obviously, neighborhood dependencies of the parts are different than that of the cells. The local rules of partitioned CAs consider the states of parts during the state updates.

Sometimes, neighborhood of a CA is represented by \emph{radius}. By \emph{radius}, we mean the number of consecutive cells in a direction on which a cell depends on. For example, for $1$-dimensional CA, neighborhood $\mathcal{N}$ can be represented as $\mathcal{N}=\lbrace -r, -(r-1), \cdots, -1, 0, 1, \cdots, (r-1), r\rbrace$, where $r$ is the radius of the CA. In Figure\ref{2way}, $r=1$. Classically, the radius is same in every direction.
However, in some works this restriction is removed; see for example \cite{Jump74,boccara-1998-31}. One of such CAs, called \emph{one-way CA}, has been proposed in \cite{Dyer80}, in which communication is allowed only in one direction, that is, in a $1$-dimensional array, the next state of each cell depends on itself and either of its left neighbor(s) or right neighbor(s) (see Figure~\ref{1way}).

Although CAs are originally defined over 2-dimensional space, a large number of researches have been dedicated to explore one-dimensional CAs; see for instance \cite{Amoroso72,Pries86,suttner91,biplabtcad}. Stephen Wolfram has introduced a class of very simple CAs, called \emph{Elementary cellular automata} or ECAs \cite{Wolfr83,Wolframbook}, which are one-dimensional, two-state, and having three-neighborhood dependency (like Figure~\ref{2way}). In last three decades, the major share of cellular automata research has gone to ECAs and their variations.

However, several fundamental properties and parameters of $2$-dimensional CAs are explored \cite{Packa85b,Kari90,Durand93,terrier2004two,deOliveira20061,chr2D}. Higher than $2$-dimensional CAs are also proposed \cite{gandin19973d,Palas1,miller2005two,Mo201431-3DCA}. For instance, in \cite{miller2005two}, two-state three-dimensional reversible CA (RCA) is described and is shown to accomplish universal computation and construction. However, two or higher dimensional CAs sometime behave differently than $1$-dimensional CAs. For example, reversibility, a well addressed problem of $1$-D CA, is undecidable for $2$ or higher dimensional CAs \cite{Kari90}. In fact, most of the decision problems related to two or higher dimensional CAs are undecidable, see for example \cite{DENNUNZIO201440}.


\subsection{The states of the cell} 
A cell can be in any of the states of a finite state set $\mathcal{S}$ at any point of time. The number of states of von Neumann's CA is 29. Initially, many researchers targeted to simplify von Neumann's CA by reducing the number of states. For instance, state count of CAs was reduced to eight in \cite{Codd68}. Thatcher has shown construction and computational universality as well as self-reproducing ability of von Neumann's cellular space \cite{Thatcher}, whereas in \cite{Arbib66}, a simple self-reproducing CA capable of universality is depicted. Then, Banks has proved the universal computability of $2$-state CA and provided a description of self-reproducing CA using only $4$ states \cite{Banks71}. However, all these constructions are for $2$-dimensional CAs, and use $5$-neighborhood dependency (see Figure~\ref{von}). Conway's Game of Life, on the other hand, uses 9-neighborhood dependency (see Figure~\ref{moore}) and two states per cell. Martin Gardner has proved that this binary CA is also computationally universal \cite{Gardner71}. In fact, many CAs, including the ECAs, are binary; that is, the cells use only two states. Smith has shown that neighborhood size and state-set cardinality of a CA are interrelated \cite{Smith71}. A CA with higher neighborhood size can always be emulated by another CA that has lesser neighborhood size but higher number of states per cell, and the contrariwise. 

Generally, the cellular space $\mathscr{L}$ is infinite. Hence, the configurations which are the collections of states of all cells of $\mathscr{L}$ at different time, are also infinite. As is in von Neumann's CA, sometime a restriction is imposed on CAs: a \emph{quiescent state} $q \in \mathcal{S}$ is considered so that $R(q,q,\cdots,q) = q$, where $R$ is the local rule (see Definition~\ref{Def:basic}). 

	\begin{definition}\label{Def:Quiescent}
			Let $R$ be the local rule of a CA. A state $q \in \mathcal{S}$ is called a \textbf{quiescent} state, when $R(q,q,\cdots,q) = q$.
		\end{definition} 

This means, a cell whose neighbors are in quiescent states, remains in quiescent state. As a result, we get a new class of configurations, called {\em finite configurations}.
\begin{definition}
\label{Def:FiniteConf}
Let $q \in \mathcal{S}$ be a quiescent state of a CA ($\mathscr{L},\mathcal{S},\mathcal{N},R$) and the cellular space $\mathscr{L}$ be infinite.
A configuration of the CA is called a \textbf{finite configuration} if all but a finite number of cells are in $q$. 
\end{definition}
Thus, if the initial configuration of the CA is a finite configuration, at every time step the new configuration remains as a finite configuration. Let us consider $\mathcal{C}_F$ represents the set of all finite configurations. Obviously, $\mathcal{C}_F \subset \mathcal{C}$, where $\mathcal{C}=\mathcal{S}^{\mathscr{L}}$, the set of all configurations. Then, the global transition function of the CA over $\mathcal{C}_F$, ${G}_F: \mathcal{C}_F \rightarrow \mathcal{C}_F$ is a restriction of ${G}$. This restriction is at least needed to simulate a CA in computer.

In a number of works, such as \cite{Martin84a,ppc1}, states of a CA are considered as elements of a finite field. B. K. Sikdar and his co-researchers have further considered the states as elements of an extension field \cite{vlsi00d,biplabtcad}. In \cite{ito1983linear,CATTANEO2004249}, the state set has been considered as $\mathbb{Z}_m$ (the integers modulo $m$). 

In almost all cases, however, every cell of a CA uses the same state set. In {\em polygeneous} CAs, on the contrary, a CA may use different state sets \cite{Sarkar00}.

\subsection{The boundary condition}
\begin{figure}
   \subfloat[Null boundary\label{Fig:NullB}]{%
       \includegraphics[scale = 0.45]{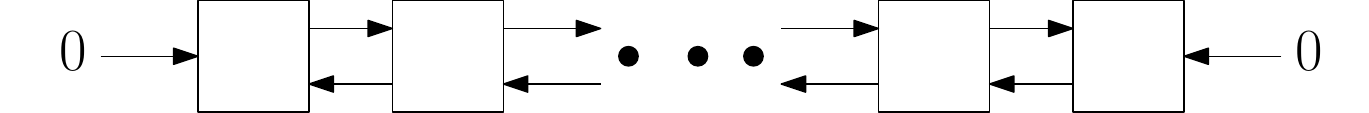}
     }
      \hfill
     \subfloat[Periodic boundary\label{Fig:PeriodicB}]{%
       \includegraphics[scale = 0.35]{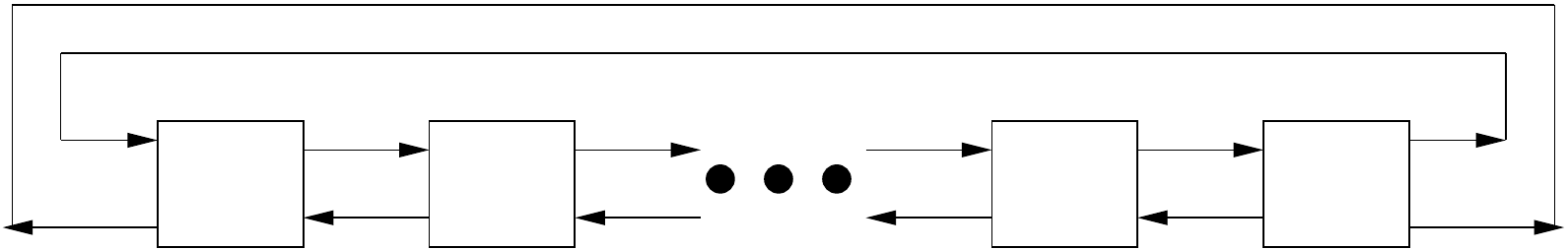}
     }
      \hfill
    \subfloat[Adiabatic boundary\label{Fig:AdiaB}]{%
       \includegraphics[scale = 0.35]{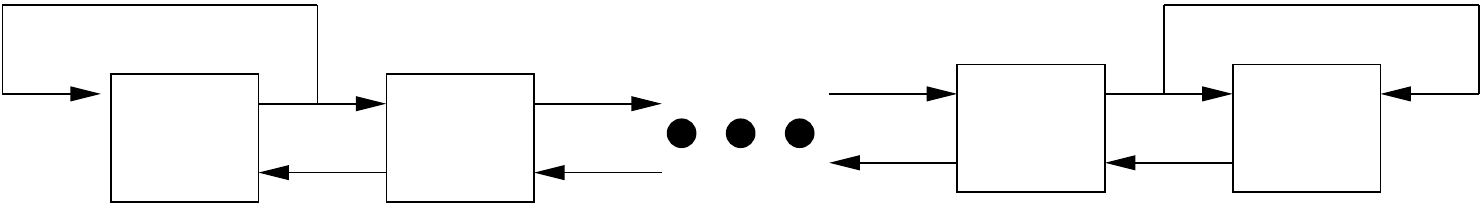}
     }
      \hfill
    \subfloat[Reflexive boundary\label{Fig:ReflB}]{%
       \includegraphics[scale = 0.35]{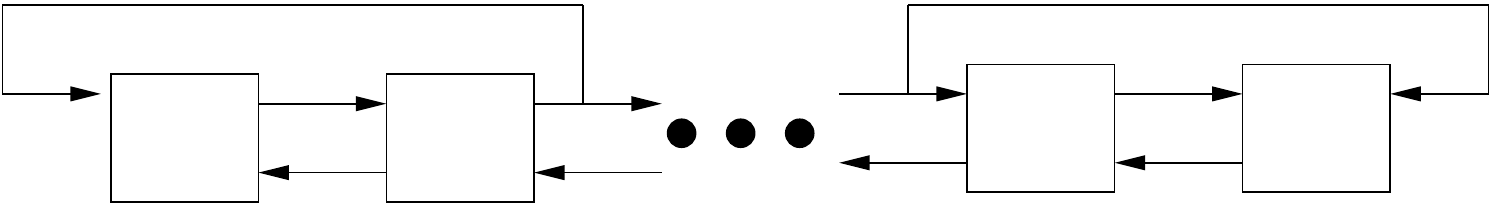}
     } 
           \hfill
    \subfloat[Intermediate boundary\label{Fig:IntmB}]{%
       \includegraphics[scale = 0.35]{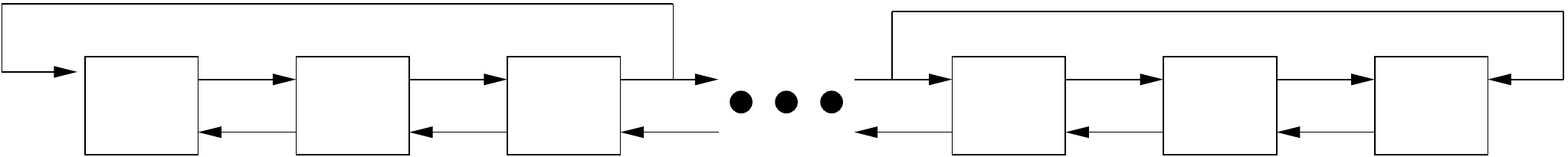}
     }     
     \caption{Boundary conditions of 1-D finite CAs. Arrows pointing to a cell indicate the dependencies of the cell. Here all the CAs use 3-neighborhood dependency}
     \label{Fig:BoundaryC}
 \end{figure}

As the cellular space $\mathscr{L}$ is usually infinite, there is no question of boundary, and boundary condition. However, in a few works, $\mathscr{L}$ is assumed as finite, which is obviously having boundaries. These CAs are finite CAs.
\begin{definition}
A CA is called as a \textbf{Finite Cellular Automaton} if the cellular space $\mathscr{L} \subseteq \mathbb{Z}^{D}$ is finite. Otherwise, it is an \textbf{Infinite Cellular Automaton}.
\end{definition}
Finite CAs are really important if the automata are to be implemented. Two boundary conditions for finite CAs are generally used -- open boundary condition and periodic boundary condition. 
In open (fixed) boundary CAs, the missing neighbors of extreme cells (leftmost and rightmost cells in case of $1$-D CAs) are usually assigned some fixed states. Among the open boundary conditions, the most popular is null boundary, where the missing neighbors of the terminal cells are always in state $0$ (Null). Null boundary CAs have received much attention by the researchers, who have used CAs for VLSI (Very Large Scale Integration) design and test. Some examples of such works are reported in \cite{Horte89a,Tsali91,ppc1,biplabtcad,tcad/DasS10}. Figure~\ref{Fig:NullB} explains null boundary condition of a 1-dimensional CA.

In case of periodic boundary condition, on the other hand, the boundary cells are neighbors of some other boundary cells. For instance, for $1$-D CAs, the rightmost and leftmost cells are neighbors of each other (see Figure~\ref{Fig:PeriodicB}). Higher dimensional CAs under periodic boundary conditions have also been explored by some researchers, such as \cite{Palas1,Jin2012538,uguz2013reversibility} etc. However, the periodic boundaries of a 2-D CA can be visualized by folding right and left sides of a rectangle to form a tube, and then taping top and bottom edges of the tube to form a torus.

Periodic boundary conditions sometime relate finite CAs and infinite CAs through {\em periodic configurations}. 
\begin{definition}\label{Def:periodicConfiguration}
A configuration $x\in \mathcal{C}$ of a D-dimensional infinite CA is \textbf{periodic} if the configuration is invariant under $D$ linearly independent translations. That is, there exist $D$ natural numbers $n_1, n_2,\cdots, n_D$ so that $x$ is same as $n_i$ times shift of $x$ in dimension $i$, for each $i\in\{1, 2, \cdots, D\}$.
\end{definition}
Let $\mathcal{C}_P\subset {\mathcal{C}}$ denote the set of all periodic configurations of a CA. Then, $G_P:\mathcal{C}_P \rightarrow \mathcal{C}_P$ is a restriction of $G$. However, the members of $\mathcal{C}_P$ preserve periodicity of configurations, which is also preserved in periodic boundary condition. So, periodic configurations are sometime called as the periodic boundary conditions of finite CAs.

Some other variations of boundary conditions also exist, such as adiabatic (or reflecting) boundary, reflexive boundary and intermediate boundary. In adiabatic (or reflecting) boundary condition, the boundary cells assume the missing neighbors as their duplicates. Figure~\ref{Fig:AdiaB} shows this boundary condition where the left (right) neighbor of the leftmost (rightmost) cell is the cell itself.
In reflexive boundary condition, shown in Figure~\ref{Fig:ReflB}, the left and right neighbors of boundary cells are the same cell. Finally, Figure~\ref{Fig:IntmB} explains the intermediate boundary condition for a 1-D CA, where a missing neighbor of a boundary cell is the neighbor's neighbor of the boundary cell. In \cite{Nandi96}, the intermediate boundary condition have been used to improve quality of pseudo-random patterns, generated by CAs.

As another special case of open boundary condition, stochastic boundary condition is proposed, where the boundary cells assume some states stochastically. In \cite{PhysRevE-CKS}, this boundary condition have been used to model traffic flow on a one-dimensional lattice with finite number of cells. In this traffic model, the authors have considered that the cars are injected at the leftmost cell with a probability, and the cars are removed from the right boundary with another probability.

\subsection{Local rule} 
\label{Sec:rule}
A cell of a CA changes its state following a next state function $R:\mathcal{S}^{m}\rightarrow \mathcal{S}$ where $\mathcal{S}$ is the set of states and $m$ is the size of the neighborhood. The map $R$, which is generally known as the local rule of the CA, can be conveyed in different ways. As an instance, in Conway's Game of Life, the local rule is $R:\{0,1\}^{9}\rightarrow \{0,1\}$, where state $0$ represents a dead cell and state $1$ represents an alive cell. It is stated as following in \cite{mitcourseware}:

\begin{displayquote}
	\begin{itemize}
\item ``Birth: a cell that is dead at time $t$, will be alive at time $t + 1$, if exactly $3$ of its eight neighbors were alive at time $t$.
\item Death: a cell can die by:
\begin{itemize}
\item Overcrowding: if a cell is alive at time $t$ and $4$ or more of its neighbors are also alive at time $t$, the cell will be dead at time $t + 1$.

\item Exposure: If a live cell at time $t$ has only $1$ live neighbor or no live neighbors, it will be dead at time $t + 1$.
\end{itemize}
\item Survival: a cell survives from time $t$ to time $t + 1$ if and only if $2$ or $3$ of its neighbors are alive at time $t$.''
 \end{itemize}
 \end{displayquote}

However, the local rule $R$ can also be represented in a tabular form. In this form, the table contains entries for the next state values corresponding to each of the possible neighborhood combinations according to the local rule. This representation has been popularized by Wolfram \cite{Wolfr83}, and used to name his Elementary CAs (ECAs). Table~\ref{tab:ruleECA} shows seven rules of ECAs, which are named after the decimal equivalents (last column of Table~\ref{tab:ruleECA}) of 8-bit binary sequence. Obviously, the tabular form is good if size of neighborhood ($m$) and the state set ($\mathcal{S}$) are very small.

\begin{table}[h]
\setlength{\tabcolsep}{1.3pt}
\begin{center}
\caption{Some ECAs rules. Here, PS and NS represent present state and next state respectively}
\label{tab:ruleECA}
\resizebox{0.50\textwidth}{!}{
\begin{tabular}{cccccccccc}
 \toprule
\thead{PS} \hspace{1em}& \thead{111} & \thead{110} & \thead{101} & \thead{100} & \thead{011} & \thead{010} & \thead{001} & \thead{000} & \hspace{1em} \multirow{3}{*}{\thead{Rule}}\\ 

\thead{(RMT)} \hspace{1em}& \thead{(7)} & \thead{(6)} & \thead{(5)} & \thead{(4)} & \thead{(3)} & \thead{(2)} & \thead{(1)} & \thead{(0)} & \\ 
 \midrule
\multirow{3}{*}{}
 &0 & 1 & 0 & 1 & 1 & 0 & 1 & 0 &\hspace{1em} 90\\
 & 1 & 0 & 0 & 1 & 0 & 1 & 1 & 0 &\hspace{1em} 150\\
 & 0& 0& 0& 1& 1& 1& 1& 0&\hspace{1em} 30\\
\thead{NS} & 0& 0& 0& 0& 0& 1& 0& 1&\hspace{1em} 5\\
 & 0& 1& 0& 0& 1& 0& 0& 1&\hspace{1em} 73\\
 & 1& 1& 0& 0& 1& 0& 0& 0&\hspace{1em} 200\\
 & 0& 1& 0& 1& 0& 0& 0& 0&\hspace{1em} 80\\
    \bottomrule
\end{tabular}
}
\end{center}
\end{table} 

Against a binary map $R:\{0,1\}^{m}\rightarrow \{0, 1\}$, an $m$-tuple $(s_1,s_2,\cdots,s_m) \in \mathcal{S}^{m}$ can be viewed as a {\em Min Term} of $m$-variable switching function. So, each of the tuples of the map $R$ is named as {\em Rule Min Term} (RMT). Table~\ref{tab:ruleECA} shows eight RMTs against a rule. Throughout this dissertation, the RMTs of rules have been utilized.
However, all the above works that use RMTs are about 1-dimensional CAs.
\begin{definition}
\label{Def:RMT}
Let $R:\mathcal{S}^{m}\rightarrow \mathcal{S}$ be a local rule of a CA. A tuple $(s_1,s_2,\cdots,s_m) \in \mathcal{S}^{m}$ is called as a {\bf Rule Min Term} (RMT) of the rule $R$, and is usually represented by its decimal equivalent. If $\mathcal{S}=\{0, 1,\cdots, d-1\}$, then the RMT is $r=s_1.d^{m-1}+s_2.d^{m-2}+\cdots+s_{m-1}.d+s_m$. The next state against this RMT of the rule, that is $R(s_1,s_2,\cdots,s_m)$, is also represented as $R[r]$.
\end{definition}

Therefore, a rule can be seen as a collection of RMTs along with their next states. In that sense, RMTs are atomic elements of a rule. The RMTs along with their next state values are sometime called as {\em transitions}. In case of ECAs, where $R:\{0,1\}^{3}\rightarrow \{0, 1\}$, a transition is a quadruplet $(x,y,z,R(x,y,z))$. A transition of an ECA rule is {\em active} if $R(x,y,z)\neq y$; otherwise the transition is {\em passive}. A rule $R$ can also be represented by its active transitions only. To present a rule, in general, at least its active transitions are to be noted down. The rule of Game of Life (see above), for example, shows all active transitions. In case of ECAs, although the numbering style of \cite{Wolfr83} (as shown in Table~\ref{tab:ruleECA}) is widely used, they are alternatively identified by their active transitions only, known as {\em transition codes}. In the works of Nazim Fat{\`{e}}s and his co-researchers \cite{Fates20061,fates00608485}, for example, transition codes of ECAs rules have been used.

When the rule $R$ of a CA is linear, then the CA is also linear. That is, the global transition function $G$ of the CA is a linear function. 

	\begin{definition}
			\label{Def:linearrule}
			A rule $R:\mathcal{S}^m\rightarrow \mathcal{S}$ is called \textbf{linear} if $R$ can be expressed as $R (a_1, a_2, \cdots, a_m) = \sum\limits_{i=1}^m c_i.a_i$, where $c_i$ $\in \mathcal{S}$ is a constant and $\mathcal{S}$ forms a commutative ring with identity; otherwise it is a \textbf{non-linear} rule. 
		\end{definition}
		
There are seven linear ECAs for rules 60, 90, 102, 150, 170, 204 and 240 (excluding rule 0, which is also trivially linear). Apart from these ECAs, there is no additional additive ECAs. So, these rules are presented by some authors as {\em linear/additive} rules.

\begin{figure}
   \subfloat[Triangular Partitioned CA\label{Fig:TPCAr}]{%
       \includegraphics[scale = 0.45]{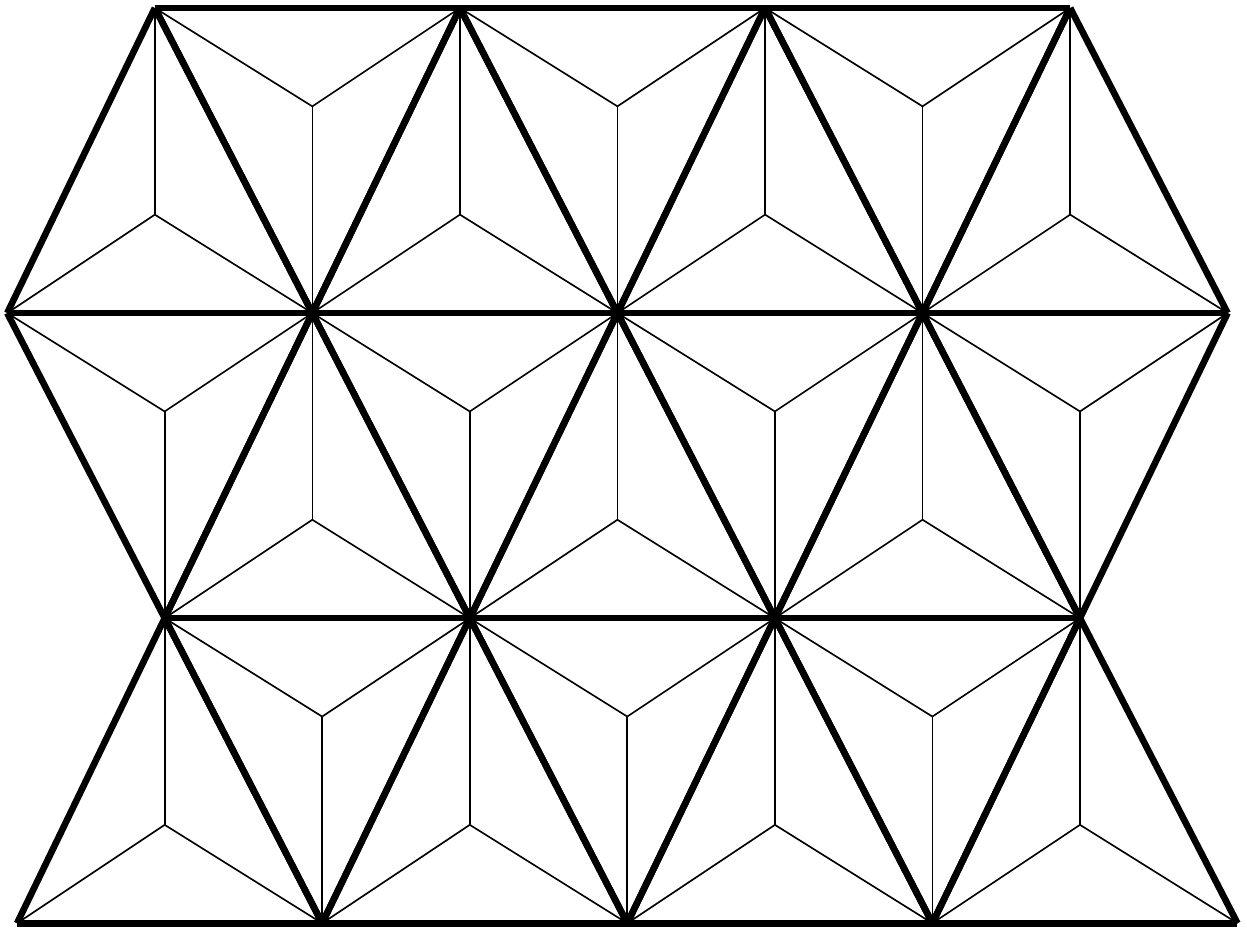}
     }
      \hfill
     \subfloat[A local rule\label{Fig:TPCArule}]{%
       \includegraphics[scale = 0.45]{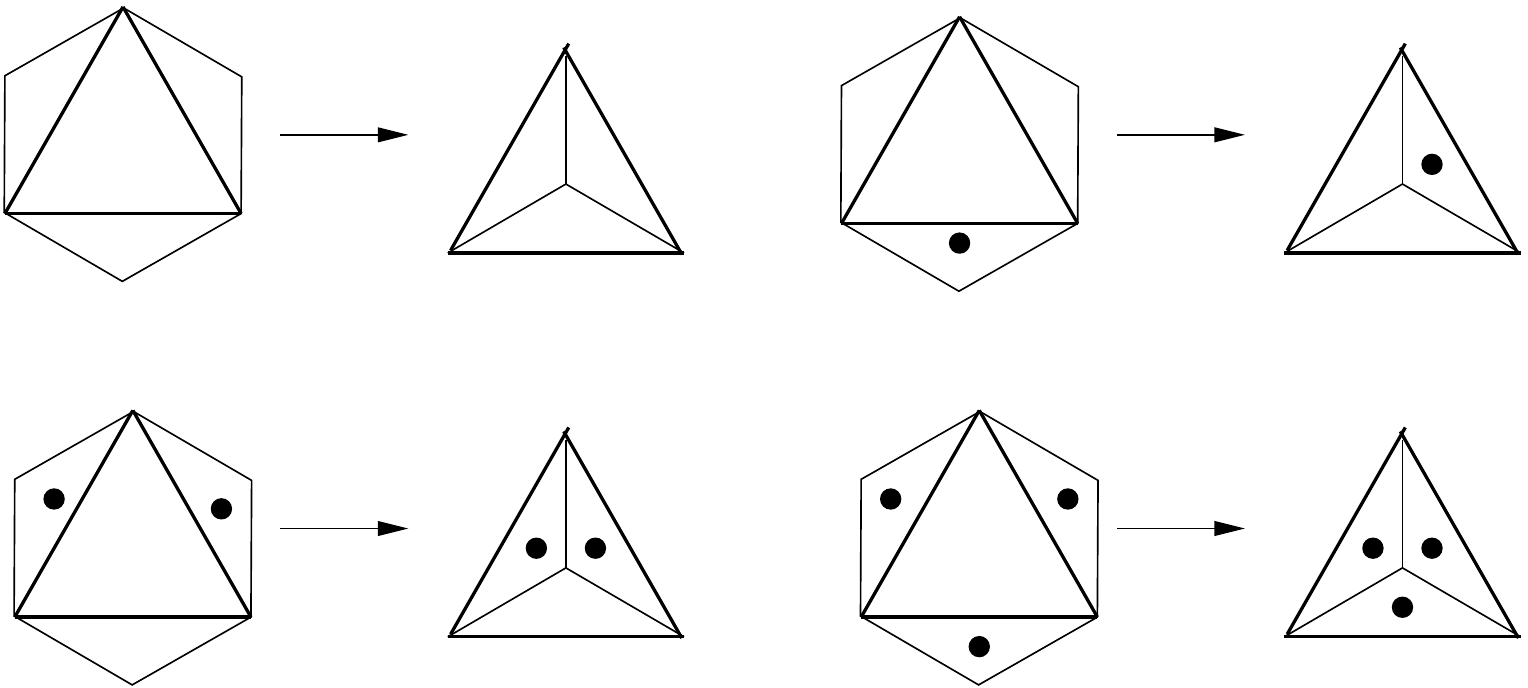}
     }    

     \caption{Triangular partitioned CA and its rule (as given by \cite{morita2016universality}). The rule is rotation symmetric}
     \label{Fig:TPCA+TPCArule}
 \end{figure}

In case of block cellular automata and partitioned cellular automata, which have been briefly discussed in Section~\ref{Chap:surveyOfCA:Sec:DN}, presentation of local rules is altogether different. The rule ({\em block rule}) of block CA does not change individual cells of the CA, rather it looks at the content of the block and updates the whole block. Toffoli and Margolus \cite{Toffo87} have described such rule in their book. For partitioned CAs, on the other hand, each cell is divided into several parts and the next state of each cell is determined by the states of the adjacent parts of the neighbor cells. Figure~\ref{Fig:TPCA+TPCArule} shows a triangular partitioned CA and its rule, which is explored in \cite{morita2016universality}. Like block CA, the local rule of a partitioned CA updates all the parts of a cell. Unlike traditional CA, however, the local rule does not consider the present states of the neighbors of a cell, but adjacent parts of neighboring cells (Figure~\ref{Fig:TPCArule}).

Classically, the CAs are uniform, synchronous and deterministic -- that is, all cells are updated together with the same local rule. A new class of non-classical CAs have been proposed where cells can follow different rules. These CAs are known as {\em hybrid} or {\em non-uniform} CAs. Historically, this class of CAs assume that the rules are the ECAs rules only. Many researchers have explored these CAs, which are one-dimensional and finite, for example \cite{Pries86,Horte89a,ppc1}. In this case, one needs to define the local rules for individual cells; hence she needs a rule vector $\mathcal{R}$.
\begin{definition}
\label{Def:RuleVector}
A {\bf rule vector} of a non-uniform CA over $\mathscr{L}$ is $\mathcal{R}=\langle \mathcal{R}_0, \mathcal{R}_1, \cdots, \mathcal{R}_{n-1}\rangle$, where $\mathcal{R}_i$ is a rule used by cell $i$, $i\in \mathscr{L}$.
\end{definition}
As a proof of concept, let us consider a 5-cell non-uniform CA with rule vector $\mathcal{R}=\langle 90, 150, 90, 5, 150\rangle$. This implies that the first and third cells use ECA rule 90, second and fifth cells use rule 150, and the fourth cell uses rule 5. This class of CAs are specially utilized in VLSI design and test.
Another type of CA, called a programmable CA (with respect to VLSI design), exists, where a cell can choose a distinct local rule at every time instant \cite{Nandi94a}. Nevertheless, a special kind of CAs have also been introduced where a CA updates itself in asynchronous way. In short, non-classical CAs have gained popularity in recent times. Section~\ref{Chap:surveyOfCA:scn_nunCA} and Section~\ref{Chap:surveyOfCA:scn_eca} are dedicated to discuss these non-classical CAs.

\section{Characterization tools of Cellular Automata}
\label{Chap:surveyOfCA:scn_chrtool}

In \emph{A New Kind of Science}, Wolfram has argued that, to find out how a particular CA will behave, one has to observe what is happening just by running the CA \cite{Wolframbook1}. Predicting behavior of a system by means of (mathematical) analysis and without running it is only possible, according to Wolfram, for special systems with simple behavior (page $6$ of \cite{Wolframbook1}). In spite of this observation of some CAs researchers, a few characterization tools and parameters have been proposed in different time to analyze and predict the behavior of some CAs. Needless to say, all kind of dynamic behaviors of a CA may not be analyzed by a tool, but tools are used to discover some specific properties of a CA.

In this section, we survey the characterization tools and parameters, developed till date to analyze the CAs. Tools are mainly developed for one-dimensional CAs, and for two or higher dimensional CAs, ``run and watch'' is the primary technique to study the behavior. Though, few parameters are proposed which can be used to guess the behavior of two or higher dimensional CAs.

\subsection{de Bruijn graph}\label{Chap:surveyOfCA:dbg}
In the different development phases of CAs, graph theory has played an important role. One of the roles is describing the evolution of an automaton, and another is relating local properties to global properties. As an automaton has states which are mapped to another states using overlapping neighborhood sequence, it is very obvious to treat it by shift register. So, de Bruijn graph is considered as an alternative characterization tool.

\begin{definition}\label{Def:dbg}
Let $\Sigma$ be a set of symbols, and $s\ge 1$ be a number. Then, the {\bf de Bruijn graph} is ${B(s, \Sigma)} = (V, E)$, where $V=\Sigma^s$ is the set of vertices, and $E=\{(ax,xb)|a, b\in \Sigma, x\in \Sigma^{s-1} \}$ is the set of edges.
\end{definition}

\begin{figure}
   \subfloat[The de Bruijn graph $B(2,\{0,1\})$\label{Fig:DeBruijn}]{%
       \includegraphics[scale = 0.5]{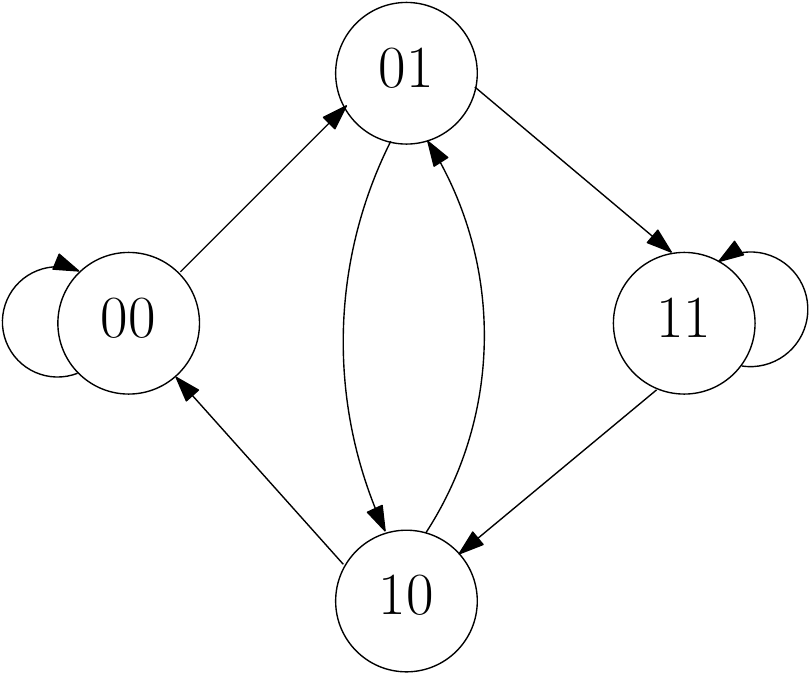}
     }
      \hfill
     \subfloat[de Bruijn graph for rule 90\label{Fig:DB90}]{%
       \includegraphics[scale = 0.5]{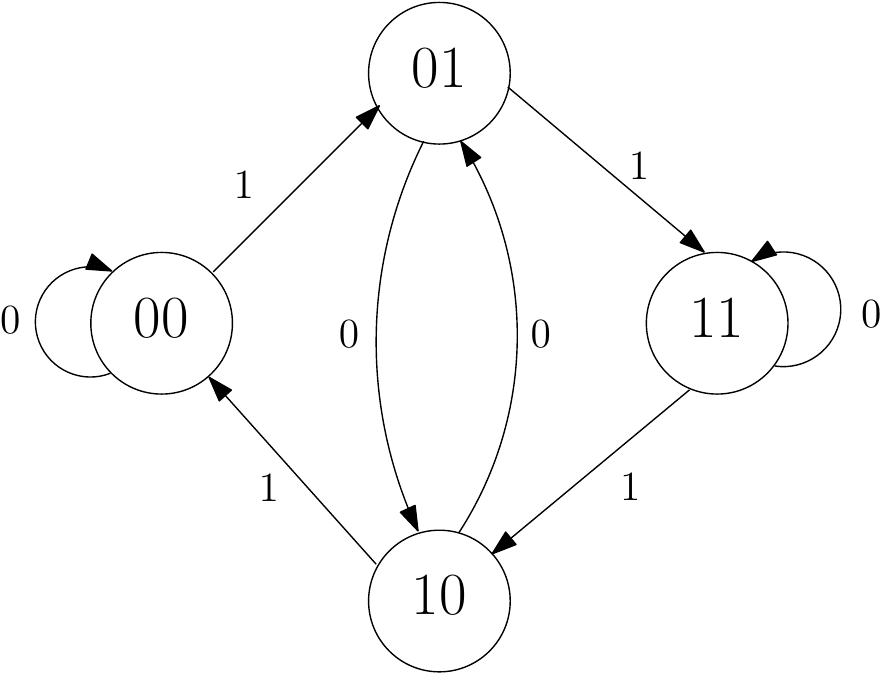}
     }    

     \caption{The de Bruijn Graph of CA with rule $90$ ($2^{nd}$ row of Table~\ref{tab:ruleECA})}
     \label{Fig:DeBruijn90}
 \end{figure}

Figure~\ref{Fig:DeBruijn} shows the de Bruijn graph $B(2,\{0,1\})$. A de Bruijn graph has $k^s$ vertices, consisting of all possible length-$ s $ sequences of $\Sigma$, where $k$ is the number of symbols in $\Sigma$ and $s\ge 1$ is a number. This graph is balanced in the sense that each vertex has both in-degree and out-degree $s$. 

A one-dimensional CA can be represented by a de Bruijn graph. Let us consider $s=m-1$ and $\Sigma=\mathcal{S}$, where $m$ is the size of neighborhood and $\mathcal{S}$ is the state-set of a 1-D CA (see Definition~\ref{Def:basic}). The edges $(ax,xb)$ of $B(m-1,\mathcal{S})$ show the overlap of nodes and $(axb)\in \mathcal{S}^m$. Now label each edge $(ax,xb)$ of $B(m-1,\mathcal{S})$ by $R(axb)\in \mathcal{S}$, where $R$ is the local rule. This labelled graph represents a one-dimensional CA with rule $R$, state set $\mathcal{S}$, and neighborhood size $m$. Since this graph does not relate to the lattice size, de Bruijn graph can be used to study finite and infinite 1-D CAs.

Figure~\ref{Fig:DB90} is the de Bruijn graph for ECA rule 90 (whereas Figure~\ref{Fig:DeBruijn} is a graph for any ECA).  This graph shows that if the left, self and right neighbors of a cell are all $0$s, then next state of the cell (that is, $R(0,0,0)$) is $0$, if the neighbors are $0, 0$ and $1$ respectively, the next state is $1$, and so on. In his work, Sutner has defined {\em $s$-fusion operation} \cite{suttner91}, which is equivalent to labelling an edge in above manner.

Over the years, this graph has been used by various researchers to understand global behavior, like surjectivity and reversibility, number conservation, equicontinuity etc. of 1-dimensional CAs, see for example \cite{suttner91,Soto2008,voorhees2008remarks}. In \cite{Martinez2008}, ECA $110$ has been explored to determine a glider-based regular expressions, whereas, in \cite{de2014complete} ECA $54$ has been studied for similar kind of behavior. Cyclic properties and inverse of a CA are studied using pair diagram of de Bruijn graph in \cite{Mora2008}. 
In \cite{Betel2013}, de Bruijn graph have been used to solve the parity problem. Recently, the periods of pre-images of spatially periodic configurations in surjective CA are studied in \cite{Mariot2017} using de Bruijn graphs.

The de Bruijn graphs are traditionally used to represent and study classical CAs. However, in recent past, Alberto Dennunzio and his co-researchers have shown that de Bruijn graph can also be used to represent non-uniform CAs \cite{DennunzioFP13}. 

\subsection{Matrix algebra}
\label{Chap:surveyOfCA:matrix}

Unlike de Bruijn graph, matrix algebra can represent only finite CAs. In fact, this characterization tool has been developed by \cite{Das90c} to study the behavior of 1-D hybrid or non-uniform CA that uses only linear ECA rules (see Section~\ref{Sec:rule}). A non-uniform CA is {\em linear/additive} if all of its local rules are linear/additive. So non-uniform finite linear/additive CAs can be characterized by matrix algebra. Chaudhuri et al. have provided a good account of works on linear/additive non-uniform finite CAs in their book \cite{ppc1}.

\begin{table}[h]
\begin{center}
\caption{Linear and complemented rules}
\label{CArule}
\resizebox{0.90\textwidth}{!}{
\begin{tabular}{ll|ll}
 \toprule
\thead{Rule} & \thead{Linear} & \thead{Rule} & \thead{Complement} \\
 & \thead{(With $XOR$ logic)} &  & \thead{(With $XNOR$ logic)} \\
 \midrule
					
 60: & $S_i(t+1)~=~S_{i-1}(t) \oplus S_{i}(t) $& 195: & $S_i(t+1)~=\overline{ ~S_{i-1}(t) \oplus S_{i}(t)} $\\
 90: & $S_i(t+1)~=~S_{i-1}(t) \oplus S_{i+1}(t)$& 165: & $S_i(t+1)~=\overline{~S_{i-1}(t) \oplus S_{i+1}(t)}$\\
 102: & $S_i(t+1)~=~S_{i}(t) \oplus S_{i+1}(t)$& 153: & $S_i(t+1)~=\overline{~S_{i}(t) \oplus S_{i+1}(t)}$\\
 150: & $S_i(t+1)~=~S_{i-1}(t) \oplus S_{i}(t) \oplus S_{i+1}(t)$ & 105: & $S_i(t+1)~=\overline{~S_{i-1}(t) \oplus S_{i}(t) \oplus S_{i+1}(t)}$\\
 170: & $S_i(t+1)~=~S_{i+1}(t)$ & 85: & $S_i(t+1)~=\overline{~S_{i+1}(t)}$\\
 204: & $S_i(t+1)~=~S_{i}(t) $ & 51: & $S_i(t+1)~=\overline{~S_{i}(t)} $\\
 240: & $S_i(t+1)~=~S_{i-1}(t) $ & 15: & $S_i(t+1)~=\overline{~S_{i-1}(t)} $\\
 
 \bottomrule
\end{tabular}
}
\end{center}
\end{table}

A binary linear rule $R$ can be expressed as XOR of input variables. For example, ECA rule 90 is a linear rule, and $f_{90}(x,y,z)=x\oplus z$. All the linear ECA rules can similarly be expressed by XOR logic. Table~\ref{CArule} shows such expressions, which is reproduced directly from \cite{ppc1}. However, if we look at the XORed expression of ECA rule 90, it is understood that a cell of ECA 90 depends on left and right neighbors only. When characterized by matrix algebra, a binary finite (uniform/non-uniform) CA is represented by a matrix, and then this neighborhood dependency is taken care of.

Let us consider a binary finite (uniform/non-uniform) CA having $n$ cells. This CA is represented by an $n \times n$ characteristics matrix operating on $GF(2)$. In this matrix, the $i^{th}$ row represents the dependency of the $i^{th}$ cell to its neighbors. The characteristics matrix ($T$), in this case, is formed as:

\begin{equation}
``
  \begin{array}{l}
   \mbox{$T\left[ i, j\right] $ =}
    \left\{
	\begin{array}{ll}
	    \mbox{1 ~if the next state of the $i^{th}$ cell depends on the present state of} ~~~ \text{''}\\ \mbox{~~~the $j^{th}$ cell}   \\
	    \mbox{0 ~otherwise}
	    \end{array}\right.
    \end{array}
 \end{equation}  

The non-uniform CAs can use different rules in different cells, hence we need a {\em rule vector} to specify the rules against cells (see Definition~\ref{Def:RuleVector}). Let us take a $4$-cell non-uniform CA with rule vector $\mathcal{R}=\langle 150, 150, 90, 150 \rangle$  under null boundary condition. Then, the characteristics matrix of the CA is: 
\[
   T
=
\begin{bmatrix}
    1 & 1 & 0 & 0 \\
    1 & 1 & 1 & 0 \\
    0 & 1 & 0 & 1 \\
    0 & 0 & 1 & 1
\end{bmatrix}
\]
Here, third cell's rule is $90$, hence depends on left and right neighbors only (see Table~\ref{CArule}). So, the third row of the $T$ is $\begin{bmatrix}
 0 & 1 & 0 & 1
\end{bmatrix}$. Since boundary condition is null, left neighboring cell of first cell and right neighboring cell of last cell are missing.

The global transition function of such a CA can be expressed by $T$. Let us consider $x$ and $y$ be two length-$n$ configurations of an $n$-cell linear CA, and $x$ be the successor of $y$. Then, $y=T\cdot x$, where $x$ and $y$ are considered as vectors.
The matrix $T$, however, can be used to efficiently study the reversibility properties, convergence etc. of a linear (uniform/non-uniform) CA. We briefly discuss this study in Section~\ref{Chap:surveyOfCA:scn_lnraddCA}.

Originally, the matrix $T$ is defined for 3-neighborhood binary CAs. Later, this tool has been extended in \cite{vlsi00d,biplab} to represent 3-neighborhood, non-binary, finite CAs. However, here the authors have assumed that the states of the cells are the elements of GF($2^p$). These CAs are also linear, and named as {\em hierarchical CAs}.  Further, CAs over GF($2^{p^{q^{r^{\cdots}}}}$) are also designed in \cite{biplabtcad}. 

Apart from 3-neighborhood dependency, 5-neighbor dependent linear CAs are also studied by matrix algebra. For example, in \cite{marti2011reversibility}, reversibility of such binary CAs with null boundary condition have been studied. Whereas, Cinkir et al. have studied reversibility of linear CAs with periodic boundary conditions over $\mathbb{Z}_p$, where $p \geq 2$ is a prime number \cite{zubeyir11}. For $2$-dimensional linear CAs also, a list of works are reported using matrix algebra as characterization tool \cite{Chattopadhyay2d,Siap2d,uguz2013reversibility,chr2D}.

\subsection{Reachability tree} 
\label{Chap:surveyOfCA:tree}
Like matrix algebra, the reachability tree has been proposed to study 1-D non-uniform finite CAs that use only ECAs rules in their rule vectors. Unlike matrix algebra, reachability tree can represent non-uniform non-linear CAs. This tool has first been used in \cite{Acri04} to study non-uniform finite CAs under null boundary condition. In this section, we first define reachability tree for an $n$-cell non-uniform CA with rule vector ${\mathcal{R}}=\langle {\mathcal R}_0, {\mathcal R}_1, \cdots , {\mathcal R}_{n-1}\rangle$ under null boundary condition, where each rule ${\mathcal R}_i$, $i\in \{0, 1, \cdots, n-1\}$ is an ECA rule.

Recall that an ECA rule can be considered as a collection of RMTs (see Definition~\ref{Def:RMT}) along with their next state values. Let us denote the set of RMTs of ${\mathcal R}_i$ as $Z_8^{i}$. That is, $Z_8^{i}=\{0, 1, 2, 3, 4, 5, 6, 7\}$. However, under null boundary condition, RMTs 4, 5, 6 and 7 of ${\mathcal R}_0$ (similarly, RMTs 1, 3, 5 and 7 of ${\mathcal R}_{n-1}$) are {\em invalid}. So, $Z_8^{0}=\{0, 1, 2, 3\}$, and $Z_8^{n-1}=\{0, 2, 4, 6\}$ for null boundary condition.



\begin{definition}
\label{Def:RTNull}
The reachability tree of an $n$-cell ECA with rule vector $\langle {\mathcal R}_0, {\mathcal R}_1, \cdots , {\mathcal R}_i, \cdots, {\mathcal R}_{n-1}\rangle$ under null boundary condition is a rooted and edge-labeled binary tree with $n+1$ levels, where $E_{i.2j} = (N_{i.j}, N_{i+1.2j}, l_{i.2j})$ and $E_{i.2j+1} = (N_{i.j}, N_{i+1.2j+1}, l_{i.2j+1})$ are the edges between nodes $N_{i.j}\subseteq Z_8^{i}$ and $N_{i+1.2j}\subseteq Z_8^{i+1}$ with label $l_{i.2j}\subseteq N_{i.j}$, and between nodes $N_{i.j}$ and $N_{i+1.2j+1}\subseteq Z_8^{i+1}$ with label $l_{i.2j+1}\subseteq N_{i.j}$ respectively  $(0\le i\le n-1$, $0\le j\le 2^i-1)$. Following are the relations which exist in the tree:

\begin{enumerate}

\item \label{rootAtDefNull} {[For root]} $N_{0.0} = Z_8^0 = \{0, 1, 2, 3\}$.

\item \label{edgeAtDefNull} $\forall r\in N_{i.j}$, RMT $r$ of ${\mathcal R}_{i}$  is in $l_{i.2j}$ (resp. $l_{i.2j+1}$), if ${\mathcal R}_{i}[r]$ = 0 (resp. 1). That means,  $l_{i.2j}\cup l_{i.2j+1} = N_{i.j}$ ($0\le i\le n-1$, $0\le j\le 2^i-1 $).

\item \label{nodeAtNodeNull} $\forall r \in l_{i.j}$, RMTs $2r \pmod{8}$ and $2r+1 \pmod{8}$ of ${\mathcal R}_{i+1}$ are in $N_{i+1.j}$ ($0\le i\le n-3$, $0\le j\le 2^{i+1}-1$).

\item \label{n-1AtNodeNull} {[For level $n-1$]} $\forall r \in l_{n-2.j}$, RMT $2r \pmod{8}$ of ${\mathcal R}_{n-1}$ is in $N_{i+1.j}$ ($0\le j\le 2^{n-1}-1$).
	
\item \label{nAtNodeNull} {[For level $n$]} $N_{n,j}=\emptyset$, for any $j$, $0\le j\le 2^n-1$.

\end{enumerate}
\end{definition}

Informally, we can state the reachability tree construction in the following way: Form the root with RMTs 0, 1, 2 and 3 of $\mathcal{R}_0$. Out of the 4 RMTs, the RMTs having next state value 0 are in the label of left edge of the root, and the rest RMTs are in the right edge. Now we can get the nodes of level 1: if an RMT $r$ is in a label of an edge, then put the RMTs $2r\pmod 8$ and $2r+1 \pmod 8$ of $\mathcal{R}_1$ in the node which is connected by the edge. Similarly, we can get next edges and nodes of level 2, and of level 3, etc. However, for the last rule, that is $\mathcal{R}_{n-1}$, all RMTs are not present in null boundary condition. So, only even RMTs of $\mathcal{R}_{n-1}$ can be in the nodes of level $n-1$. Finally, we get the leaves, which are empty.

\begin{figure}
\centering
\includegraphics[height=2.0in, width=4.7in]{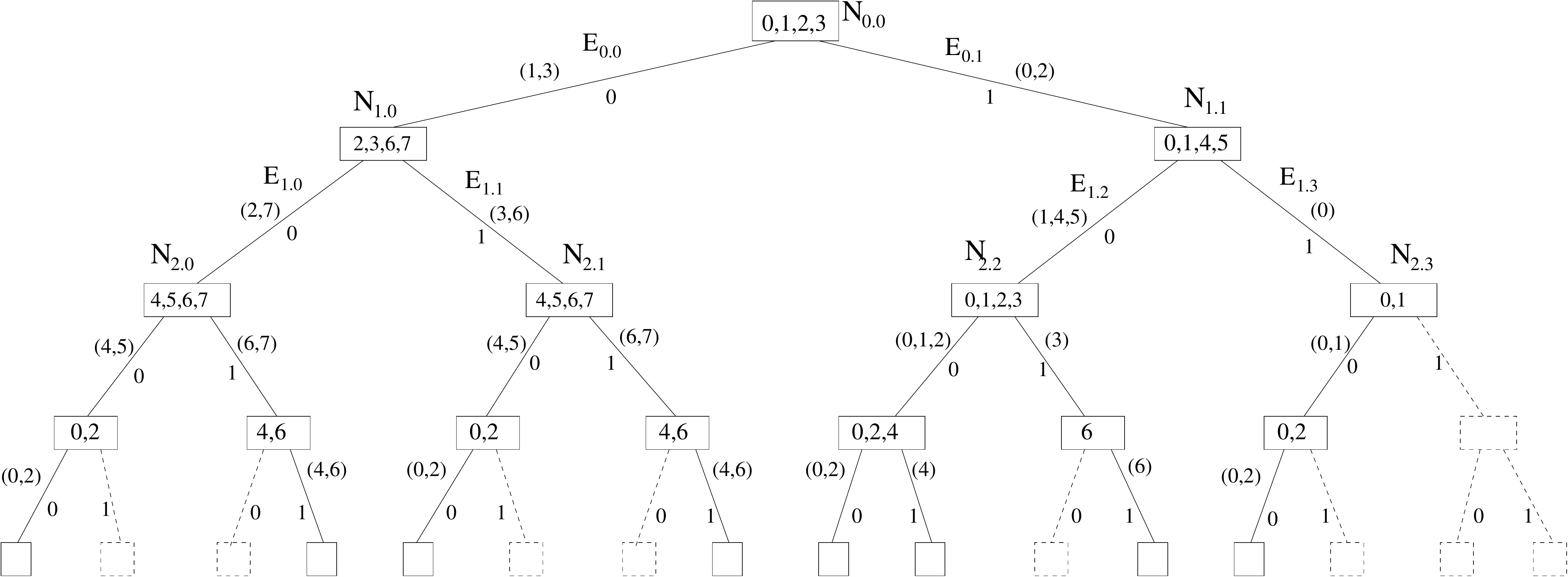}
\caption{Reachability Tree for null boundary $4$-cell non-uniform CA with rule vector $\mathcal{R}=\langle5, 73, 200 ,80\rangle$}
\label{fig: tree}
\end{figure}

Figure~\ref{fig: tree} shows an example of reachability tree of a 4-cell non-uniform CA with rule vector $\mathcal{R}=\langle5, 73, 200 ,80\rangle$ under null boundary condition. The root is a set \{0, 1, 2, 3\}, and the leaves are empty. It can be observed in the figure that the edges between the nodes of two consecutive levels are formed after applying the corresponding rule on the RMTs. However, an edge $E_{i.2j}$ (resp. $E_{i.2j+1}$) is called as {\em 0-edge} (resp. {\em 1-edge}). In Figure~\ref{fig: tree}, 0-edges and 1-edges are marked. 

Reachability tree represents {\em reachable configurations} of a CA. A configuration is {\em reachable} if it has at least one predecessor; otherwise it is a {\em non-reachable} (or Garden-of-Eden) configuration. A sequence of edges from root to a leaf represents a reachable configuration when 0-edges and 1-edges are replaced by 0 and 1 respectively. For example, the configuration 0011 of the CA of Figure~\ref{fig: tree} is reachable. In Figure~\ref{fig: tree}, some edges (and nodes) are shown by dotted lines. These edges do not exist in the tree. So, a configuration 1110, for example, is non-reachable.

Reachability tree has been utilized to discover many properties of above class of non-uniform CAs, such as reversibility, convergence, etc., see for example \cite{SukantaTH,Adak2016OnSO}. We discuss some of these discoveries in Section~\ref{Chap:surveyOfCA:scn_nlCA}.

When boundary condition changes, the structure of reachability tree also changes. This structure is first developed in \cite{entcs/DasS09} to study reversibility of non-uniform CAs under periodic boundary condition.


\subsection{$Z$-Parameter, $\lambda$-Parameter, $\Theta$-Parameter}
\label{Chap:surveyOfCA:Sec:Param}
There are some parameters that can be used to characterize some aspects of one-dimensional CAs, such as the $\lambda$ parameter \cite{langton90}, the $Z$ parameter \cite{WuenscheRePEc,WuenscheI}, and the obstruction ($\Theta$) parameter \cite{voorhees1997some}.
For a $d$-state CA with radius $r$, if $k$ out of the total $d^{2r + 1}$ neighborhood configurations map to a non-quiescent state, then $\lambda$ is defined as: $\lambda = \frac{k}{d^{2r + 1}}$, that is, the percentage of all the entries in a rule table which maps to non-quiescent states. This parameter can be compared with temperature in statistical physics, or the degree of non-linearity in dynamical systems, although these are not equivalent \cite{LangtonII}.

To track behavior of binary CAs, $Z$ parameter is proposed as an alternate. It takes into account the allocation of rule table values to the sub-categories of related neighborhoods and predicts the convergence of global behavior, extremes of local behavior between order and chaos, surjectivity etc. From a given partial pre-image of a CA state, values of the successive cells can be deduced using this parameter. Two probabilities -- $Z_{left}$ and $Z_{right}$ of the next unknown cell being deterministic are calculated from the rule-table. The $Z$ parameter is the greater of these values and varies between $0$ \& $1$. A derivation of the $Z$ parameter in terms of rule table entries is also given in \cite{WuenscheI}. High value of $Z$ implies that number of pre-images of an arbitrary CA state is relatively small.

 In \cite{voorhees1997some}, the $\Theta$ parameter is defined and shown to characterize the degree of non-additivity of a binary CA rule. It is shown that the $\lambda$ parameter and $\Theta$ parameter are equal respectively to the area and volume under certain graphs. 
These parameters prove their utilization in classification of one-dimensional binary CAs.

\subsection{Space-time diagram and Transition diagram}\label{Chap:surveyOfCA:Sec:space-time}
Although neither of space-time diagram and transition diagram (or state-transition diagram) is a characterization tool, but these two diagrams have been used to observe and predict the behavior and dynamics of a CA. These diagrams are finite in size. So, for the CAs with infinite or big sizes, only a part of the whole systems can be viewed through these diagrams.

Space-time diagram is the graphical representation of the configurations of a CA at each time $t$. Here, the configuration lies on $x$-axis and $y$-axis represents time. Each of the CA states are generally depicted by some color (see Figure~\ref{statespace1}). So, the evolution of the CA can be visible from the patterns generated in the state-space diagram. This has been used to study the nature of CA in a set of papers, for example \cite{Wolfr83,wolfram84b,DennunzioFP14}. Some packages are available in public domain which can be used to observe space-time diagrams of CAs - \emph{Fiatlux} \cite{FiatNazim} is one of them.

\begin{figure}[hbtp]
     \subfloat[Space-time diagram \label{statespace1}]{%
            \includegraphics[width=0.38\textwidth, height = 5.8cm]{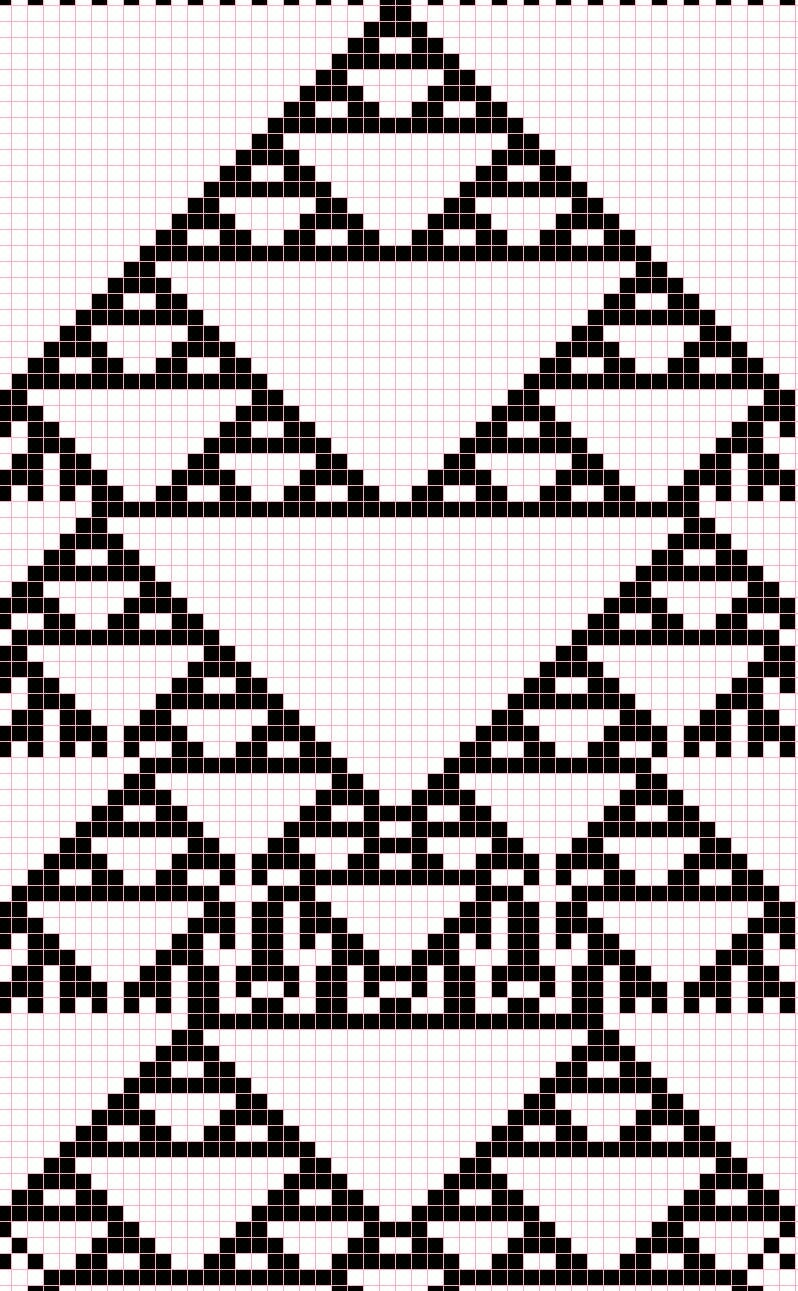}
          }
          \hfill
          \subfloat[Transition diagram \label{transition}]{%
            \includegraphics[width=0.52\textwidth, height = 4.9cm]{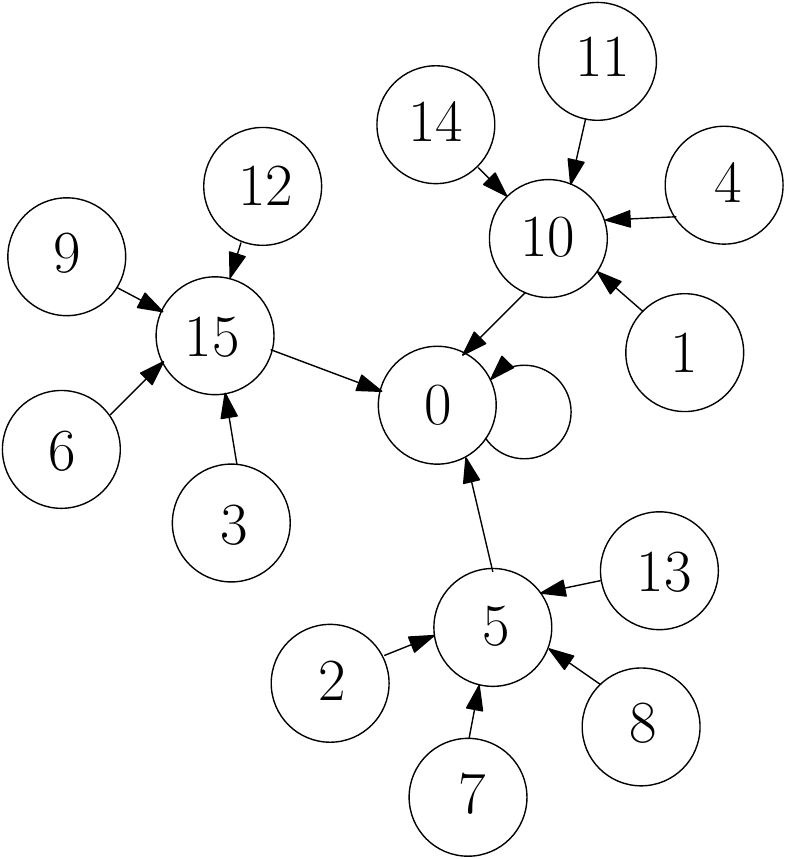}
          }
     \caption{Space-time diagram and transition diagram of $n$-cell ECA $90$ ($2^{nd}$ row of Table~\ref{tab:ruleECA}) under periodic boundary condition. For (a), cell length $n = 50$, and black implies state $1$ whereas white implies state $0$. For (b), $n=4$, and the configurations are in their decimal form}
     \label{fig:evolution diagram}
 \end{figure}

Transition diagram, also called as state transition diagram, of a CA is a directed graph whose vertices are the configurations of the CA, and an edge from a configuration $y$ to another configuration $x$ represents that $x$ is the successor of $y$ (see Figure~\ref{transition}). Sometime the directions are omitted if they can be understood from the context. Transition diagrams are heavily used in the works of Chaudhuri et al. \cite{ppc1} and A. Wuensche et al. \cite{wuensche1992global}. It can be noted that, the dynamic behavior of any CA can also be visualized and studied using its transition diagram. \emph{Discrete Dynamics Lab}, an online laboratory \cite{DDLab}, is a good place to get transition diagrams of CAs.

\section{Global behavior of Cellular Automata}
\label{Chap:surveyOfCA:sscn_prpCA}
The most exciting aspect of CAs is their complex global behavior, which is resulted from simple local interaction and computation.
The CAs have many elementary properties of the real world, such as reversibility and conservation laws, chaotic behavior, etc.
These properties motivate researchers to simulate physical and biological systems by CAs. Vichniac has targeted to simulate physics by CAs \cite{Vichn84}. Lattice gases \cite{Frisc86}, Ising spin models \cite{Grins85}, traffic systems \cite{379-440-26,PhysRevLett-35} etc. are also simulated using CAs. The book of Chopard and Droz \cite{Chopard} shows about these works. Apart from the ability of modeling physical world, the CAs are capable of performing several computational tasks. In this section, we survey some of the global behaviors of CAs such as universality, reversibility, conservation laws, computability, etc.

\subsection{Universality}
The first feature that has attracted the researchers is the capability of CAs in performing universal computation. An arbitrary Turing machine (TM) can be simulated by a CA, so universal computation is possible by CA. The universality of CAs has been first studied in \cite{Thatcher,Arbib66,banks1970universality}. However, the idea of simulating TM does not use the parallelism property of a CA. In \cite{Smith:1971}, the existence of computation-universal cellular spaces with small neighbor-state product is proved.
Later, the procedure of simulating a TM by a reversible CA is shown \cite{Morita89,Dubacq95}. 
Computation universality of one-way CAs and totalistic CAs are also reported \cite{iirgen1987simple}. Computational universality is shown following very simple rules, such as Game of Life \cite{Gardner71}, the billiard ball computer (block CAs) \cite{Durand-Lose98}. The existence of computation-universal one-dimensional CA with $7$-states per cell for radius $r=1$ and $4$-states per cell for $r=2$ is proved in \cite{lindgren90}. A very simple automaton, ECA 110 \cite{wolfram84b} is shown as computationally universal in \cite{cook2004universality}. 

Apart from simulating a Turing machine, a CA can simulate another CA. If a CA can simulate all CAs of the same dimension, it is called intrinsically universal \cite{Kari05}.
The smallest intrinsically universal CA in one-dimension is reported in \cite{ollinger2011four}. For two dimension, nonetheless, the same is a $2$-state and $5$-neighborhood CA \cite{banks1970universality}. Some other notable related works are \cite{Martin94,ollinger2002quest,ollinger2003intrinsic}.

\subsection{Invertibility and Reversibility}
For long, the questions of invertibility have been the major focus of research in CAs. If the global transition function $G$ of a CA is one-to-one, the CA is termed as an {\em injective} CA. However, it is called \emph{surjective} if the function is onto and bijective if $G$ is both onto and one-to-one.
The study of reversibility of CA was started with Hedlund \cite{hedlund69} and Richardson \cite{Richa72}. From \cite{hedlund69}, we get that a function $G:\mathcal{S}^{\mathbb{Z}^D}\rightarrow \mathcal{S}^{\mathbb{Z}^D}$ is the global transition function of a $D$-dimensional (infinite) CA if and only if $G$ is continuous, and it is shift-equivalent (this result is known as {\em Curtis - Hedlund - Lyndon} Theorem). As a consequence of this result, we get that a CA $G$ is {\em reversible} if and only if it is a bijective map. In fact, Hedlund and Richardson independently proved that all injective CAs are reversible. 

In their seminal paper, Amoroso and Patt \cite{Amoroso72} have shown an effective way to decide reversibility of $1$-dimensional infinite CA, on the basis of the local rule. In \cite{di1975reversibility}, a decision algorithm is reported for CAs with finite configurations.
An elegant scheme based on de Bruijn graph for deciding the reversibility of a one dimensional CA is presented in \cite{suttner91}. However, Kari \cite{Kari90} has shown that no algorithm can decide whether or not an arbitrary 2-D CA is reversible. This result can also be extended to higher dimensional CAs. An interesting result has been reported in \cite{toffoli77}, which says, a $(D+1)$-dimensional reversible CA can simulate any $D$-dimensional CA. 
Later, Morita and his co-researchers \cite{Morita89,morita1995reversible} have shown that a $1$-D reversible CA can simulate reversible Turing machines as well as any $1$-D CA with finite configurations. Since reversible Turing machines can be computationally universal, we can find universal 1-D reversible CA. Some other notable works on reversible CAs are \cite{Maruoka197947,culik1987invertible,Durand93,Kari94,Kari2005,moraal2000graph,MoraMM06,Soto2008,Morita2008101}.

However, all the works reported above deal with infinite CAs. In case of finite CAs, things are bit different. For example, the algorithms of \cite{Amoroso72} can not successfully work on finite CAs. 
Recently, reversibility of 1-D linear finite CAs having only two states have been studied \cite{marti2011reversibility,zubeyir11}.
The reversibility of 1-D non-uniform CAs is also deeply studied. We discuss about these studies in Section~\ref{Chap:surveyOfCA:scn_eca}.

\subsection{Garden-of-Eden} 
One of the earliest discovered results on CAs was the \textit{Garden-of-Eden} theorems by Moore \cite{moore1962machine} and Myhill \cite{Myhill63}.
\textit{Injectivity} and \textit{surjectivity} properties of CA are correlated by these theorems. A configuration is named as a Garden-of-Eden configuration, if it does not have a predecessor; that is, if it is a {\em non-reachable} configuration. In 1962, Moore has shown that the existence of mutually erasable configurations in a two-dimensional CA is sufficient for the existence of Garden-of-Eden configurations \cite{moore1962machine}. He has also claimed that existence of a configuration with more than one predecessor ensures existence of another configuration without any predecessor.
However, in \cite{Myhill63}, the reverse was proved. Here, it was shown that the extant of mutually indistinguishable configurations is both necessary and sufficient for the extant of Garden-of-Eden configurations.  

In \cite{amoroso1970garden}, the equivalence between the existence of mutually erasable configurations and mutually indistinguishable configurations has been established. This implies that the converse of Moore's result is true as well. It has also been shown in the paper that, for finite configurations (Definition~\ref{Def:FiniteConf}) both of the above conditions remain sufficient, but neither is then necessary. For finite configurations, a CA is irreversible if and only if a Garden-of-Eden configuration exists. And, a CA is surjective, if and only if it is bijective \cite{amoroso1970garden}. So, the existence of Garden-of Eden configurations violates the injectivity property. The relation between global function of infinite CA and its restriction to finite configurations has been established in \cite{Richa72}. Some other important results are recorded in \cite{maruoka1976condition,sato77,toffoli90,Kari05}.

Garden-of-Eden theorems for CAs have further been extended to Cayley graphs of groups by \cite{doi:10.1137/0406004,ceccherini1999amenable}. In \cite{capobianco2009surjunctivity}, the surjectivity and surjunctivity of CA in Besicovitch topology is recoded. 
Moreover, in \cite{margenstern1999polynomial,Margenstern200199}, CA is defined in hyperbolic plane. In \cite{margenstern2009garden}, it is depicted that, the injectivity and surjectivity properties, proved by Moore and Myhill, are no longer valid for CAs in the hyperbolic plane.

\subsection{Topology, Dynamics and Chaotic behavior of CAs}
The global transition function $G$ of an infinite CA is a continuous map on the compact metric space $\mathcal{S}^{\mathbb{Z}^D}$ according to the topology induced by the metric \textit{d}. This metric \textit{d} is usually considered as the Cantor distance. Therefore, a CA can be viewed as a discrete time dynamical system $\langle \mathcal{S}^{\mathbb{Z}^D}, G\rangle$. So, the (infinite) CAs can be studied by topological dynamics and chaos theory.

Many definitions of chaos, however, use the notion of sensitivity to initial condition. A CA is {\em sensitive to initial condition}, or simply {\em sensitive} if and only if there exists a $\delta >0$ such that, $\forall x \in {\mathcal{S}^{\mathbb{Z}^D}} ~~ \forall \epsilon >0 ~~ \exists y \in \mathcal{S}^{\mathbb{Z}^D} ~~ \exists n \in \mathbb{N}: ~~~ d(x,y) < \epsilon~~  \mbox{and}~~ d(G^n(x),G^n(y))\geq \delta$. That is, in a sensitive CA, a (small) change in initial condition would greatly affect the CA in future. A popular definition of chaos is due to Devaney \cite{Devaney}, which says that a dynamical system is chaotic if and only if it is transitive, has dense periodic points, and sensitive to initial conditions. However, a CA is {\em transitive} if and only if for any two non-empty open subsets $U$ and $V$ of $\mathcal{S}^{\mathbb{Z}^D}$, there exists an $n\in\mathbb{N}$ so that $G^n(U)\cap V \neq \emptyset$. Further, a CA has {\em dense periodic points} if and only if the set $\{x\in \mathcal{S}^{\mathbb{Z}^D}|\exists k\in\mathbb{N}: G^k(x)=x\}$ of all periodic points is a dense subset of $\mathcal{S}^{\mathbb{Z}^D}$. It has been proved in \cite{CM96} that for CAs, transitivity implies sensitivity to initial condition. So, a CA is chaotic in Devaney's sense if and only if it is transitive and has dense periodic points.

Apart from the Devaney's definition, there are more restrictive definitions of chaos, such as Knudsen chaos, positively expansive chaos, etc. A CA is {\em chaotic} according to the definition of Knudsen if and only if it has a dense orbit and sensitive to initial conditions. The CA $G$ has a {\em dense orbit} if and only if there exists a configuration $x$ so that $\forall y\in \mathcal{S}^{\mathbb{Z}^D} ~~ \forall\epsilon >0~~ \exists n\in\mathbb{N}: ~~ d(G^n(x),y)<\epsilon$. On the other hand, a CA is {\em positively expansive chaotic} if and only if it is transitive, has dense periodic points, and positively expansive. Positive expansivity is a stronger form of sensitivity. A CA is {\em positively expansive} CA if and only if there exists a $\delta>0$, such that for any two different configurations $x$ and $y$, $d(G^n(x), G^n(y))\geq \delta$ for some $n\in\mathbb{N}$.  However, these definitions of chaos are related. The chaotic behavior of CAs are studied by a number of works; some examples are \cite{DCTMitchell93,Margara99,Cattaneocht,AcerbiDF09}.

Based on their dynamic behavior, CAs are classified into different classes. One such classification, depending on their degree of {\em equicontinuity}, is due to Kurka \cite{Kurka97}. A configuration $x\in \mathcal{S}^{\mathbb{Z}^D}$ is an {\em equicontinuity} point of the CA if for every $\epsilon>0$, there is a $\delta >0$ such that $\forall y\in \mathcal{S}^{\mathbb{Z}^D}: d(x,y) <\delta ~~\mbox{implies}~~ d(G^n(x),G^n(y))<\epsilon$ for all $n\in\mathbb{N}$.
Now, a CA is called equicontinuous if all configurations are equicontinuity points. There is a connection between sensitivity and equicontinuity points of a CA: a CA $G$ is sensitive to initial conditions, if the CA does not have any equicontinuity points. Following is the classification of Kurka:
\begin{enumerate}
\item[(1)] equicontinuous CAs,
\item[(2)] CAs with some equicontinuity points, 
\item[(3)] sensitive but not positively expansive CAs, and
\item[(4)] positively expansive CAs.
\end{enumerate}
Durand et al. have studied the decision problems to determine whether a given CA belongs to a given class, and they have shown that most of the problems are undecidable \cite{durand2003undecidability}. However, for the additive CAs, Dennunzio et al. \cite{DENNUNZIO20094823} have studied the directional dynamics, and then classified them. Further, the relationships between dynamical complexity and the set of periodic configurations of surjective CAs are investigated in \cite{dennunzio2013periodic}.
 
Apart from the above work, there are many other works, specially for the 1-D case, that target to classify CAs depending on their dynamical behavior. Wolfram has reported a classification of ECAs without considering a precise mathematical definition \cite{wolfram84b}. Culik and Yu have formalized Wolfram's classification \cite{culik88}. In fact, different parameters, discussed in Section~\ref{Chap:surveyOfCA:Sec:Param}, have been developed to classify the CAs depending on their chaotic behavior. For the two-dimensional space, the generalization of the parameters -- sensitivity, neighborhood dominance and activity propagation, is reported in \cite{deOliveira20061}.

Another studied behavior of CAs is {\em nilpotency}. A CA is {\em nilpotent} if for each configuration $x$, $G^n(x)=c$ is a singleton set for sufficiently large $n$. Obviously, $c$ is a \emph{fixed point} (see Section~\ref{Chap:surveyOfCA:sscn_ngrpCA} and Definition~\ref{Def:fixed-point}, Page~\pageref{Def:fixed-point}). In \cite{Culik90}, it is shown that for two or more dimension, the nilpotency of CAs is undecidable. The same result for $1$-D CAs has been proved in \cite{Kari92a}. 
Some more works to study the dynamical properties of CAs, using expansivity, subshift, homomorphism automorphisms and endomorphisms are reported in \cite{blanchard1997dynamical,WARD1994495}. Shereshevsky has defined the left and right Lyapunov exponents for $1$-dimensional CAs \cite{Sheresh}. Finelli et al. generalized the theory of Lyapunov exponents for $D$-dimensional CAs and proved that all expansive CAs have positive Lyapunov exponents for almost all the phase space configurations \cite{Finelli199}.

\subsection{Randomness}\label{Chap:surveyOfCA:sec:randomness}
Stephen Wolfram \cite{Wolfram85c} has introduced CAs as an excellent source of pseudo-randomness. Massive parallelism, simplicity and locality of interactions of CAs, offer many benefits over other techniques, specially in case of hardware implementation. These benefits along with pseudo-randomness and ease of scalability have made CAs as an area of extensive research in VLSI circuit testing \cite{Horte89a,ppc1,tcad/DasS10,Makato98,SukantaTH,Horte89c}, Monte-Carlo simulations \cite{870571}, Field Programmable Gate Arrays \cite{comer2012random}, cryptography \cite{Wolfr86b,DBLP:journals/ccds/DasC13,DBLP:journals/jca/DasR11,Formenti2014,Leporati2014CryptographicPO}) etc. In fact, the most appealing application of pseudo-randomness of CAs is in the domain of cryptography. We briefly discuss these applications in Section~\ref{Chap:surveyOfCA:scn_appCA}.
However, most of the works on pseudo-randomness have been divided in mainly two directions:

-- Most of the research have been going on in the first direction, to generate pseudo-random \emph{numbers} using CAs. Here, generally an integer $X_i$ is generated between zero and some number $m$ (word size of the computer), where the fraction $U_i = \frac{X_i}{m}$ is the real number, uniformly distributed between $0$ and $1$. This number can be generated in several ways. For example, in \cite{wolfram86c}, the sequence is generated by ECA $30$ from the single cell with initial state $1$ among all cells, initiated with state $0$. 
 Sometime whole configurations of a finite CA are also considered as numbers. Some other works of generating pseudo-random numbers are \cite{COMPAGNER1987391,Marco00,alonso2009elementary}. In \cite{Tomassini96,wang2008generating}, some optimization algorithms are applied to CAs, whereas in \cite{Guan03,Guan04}, dynamic behavior is allowed in the cells to generate the pseudo-random numbers.

-- In the second direction, pseudo-random \emph{patterns} are generated using finite CAs. Here, pattern means the configuration of a CA of length $n$, where each cell can take any of the CA states. Note that, in a pattern, individual cell values have significance. For example, a sequence $\langle0110\rangle$ can be a treated as a number $6$, but in case of pattern, it is ``$0110$''. Some notable works on pseudo-random pattern generation using non-uniform CAs are in \cite{ats03,SukantaTH}. Here finding of the minimum cell length is important. For example, in \cite{SukantaTH}, a $45$-cell CA based pseudo-random pattern generator (PRPG) is designed which is shown to beat all existing PRPGs.

\subsection{Conservation law}
Apart from reversibility, there exist other conservation laws, which are equally important in physics. Various direction of conservation (invariants) laws in CAs are reported in \cite{fredkin82,pivato2002-45,Boccara02}. Among them number conserving CAs (NCCAs) are the most studied and used concept. Let us interpret the states of cells as numbers. Then, a CA is an NCCA if the sum of the states remains invariant during evolution of the CA. NCCAs are defined with respect to the spatially periodic configurations and finite configurations. Boccara and Fuk$\acute{s}$ have given the necessary and sufficient conditions of a $1$-D CA to be NCCA, initially for $2$ states per cell in \cite{boccara-1998-31} and then for any arbitrary number of states in \cite{Boccara02}. In \cite{pivato2002-45}, a general treatment to conserved quantities in $1$-D CA is reported. Fuk$\acute{s}$ has shown that motion representations can be constructed by $1$-D binary NCCAs. In \cite{pCA18}, the work has been extended to probabilistic CAs. NCCAs in higher dimensions are also studied by \cite{Durand03}. Further, universality and other dynamics of NCCAs are explored in \cite{Moreira2003711,Formenti2003269}. 

NCCAs have widely appeared as the models of highway traffic. The notable works in this regard are \cite{nagel1992cellular,PhysRevE-40,PhysRevLett-35,379-440-26}. In \cite{Kohyama01011989-27,Kohyama01011989-28}, NCCAs have been used for particle conservation. In $2$-dimension, the Margolus CA is a number conserving CA \cite{Margolus198481}. Apart from NCCAs, additive conserved quantities in CAs are introduced and investigated in \cite{Hattori}. In \cite{takesue1995staggered}, a necessary and sufficient condition for a given CA rule to profess a staggered invariant is studied. In \cite{Morita98,Morita:99}, the computational universality of partitioned number-conserving (and reversible) CA is shown by simulating a universal counter machine. However, these CAs are not exactly the same as NCCAs, because reducing a partitioned CA to a non-partitioned one is not number preserving. The concept of extending conserved quantities to that of monotone quantities was presented in \cite{Kurka:2003}, considering CAs with vanishing particles. Number conserving property of CAs has also been studied by applying the communication complexity approach \cite{GOLES20113616}.

In \cite{das2011characterization}, non-uniform NCCAs are characterized. This paper has reported $O(n)$ time algorithms for verification and synthesis of an $n$-cell non-uniform NCCA. Dennunzio
et al. \cite{DennunzioFP14} have given a generalized definition of number conservation of non-uniform CAs. Further, number conservation property of ECA under asynchronous update has been studied in \cite{hazari14}.

\subsection{Computational tasks}
In the CAs literature, two domains related to computation are mostly studied:
density classification task and synchronization problems. The problem statement of density classification can be summarized as follows - any initial configuration having more $1$s ($0$s) than $0$s ($1$s) must converge to all $1$s ($0$s) configuration. However, in \cite{PhysRevLett.74.5148}, it is proved that it is impossible to solve this problem with $100\%$ accuracy. Because of the impossibility of solving the standard density classification task, research efforts have shifted towards finding the best rule which can solve the problem {\em almost} perfectly -- for one-dimensional CA by \cite{Kari:2012:MTC:2385073.2385086}, for two-dimensional CA by \cite{morales01,deOliveira20061}, for non-uniform CA by \cite{DCTMaiti06} and stochastic CA by \cite{fates00608485}. However, in \cite{Fuk05},  it is shown that the density classification task is solvable by running in sequence the trivial combination of elementary rules $184$ and $232$. A good survey about the problem can be found in \cite{Oliveira13}.

The synchronization problems, like \textit{firing squad}, \textit{queen bee}, \textit{firing mob}, etc. are studied by CAs. The goal of firing squad synchronization problem is designing a CA, which initially has one active cell and evolves to a state, where every cell is simultaneously active. A good number of works on this problem are found in literature; some examples are \cite{FSSPCA2,FSSPCA3,MANZONI2014108}. The firing mob problem, which is a generalization of firing squad problem, is solved by Culik and Dube \cite{Culik91}.
However, the queen bee and leader election problems are considered as the inverse problem of firing squad problem. Here, initially states of all cells are identical and by choosing a proper rule, a cell comes to a special state. Smith has first introduced the problem \cite{Smith76}. Later the problem has been explored by a number of researchers, such as \cite{Mazoyer,BeckersW01,Stratmann154,banda1}.

Another computation problem, named early bird problem has been defined and investigated first by Rosenstiehl et al. \cite{rosen}. Here, any cell in the quiescent state may be excited by the outside world. These excitations result in special ``bird'' states instead of the quiescent states. The task is to give a transition function such that, after a certain time the first excitation(s) can be distinguished from the latter ones. This problem has been studied in \cite{Vollmar2,Legendi,Legendi1}. A variation of this problem is the distributed mutual exclusion problem, where some cells of initial configuration are in a special state (called {\em critical section requesting state}) and during evolution of the CA, these cells will be in another special state (called {\em critical section executing state}) one-by-one. Recently this problem is explored in \cite{SRoy2017}. Some other computation problems, such as French flag problem \cite{Herman01081973}, shortest path problem \cite{short12}, generating discretized circles and parabolae in real time \cite{Delorme1999347} etc. are also discussed in literature.

\section{Non-uniformity in cellular automata}
\label{Chap:surveyOfCA:scn_nunCA}

Conventionally, all the variants of CAs possess basic three properties - \emph{uniformity, synchronicity} and \emph{locality}. The uniformity refers to that each of the CA cells are updated by the identical local rule. The synchronicity implies that all the cells are updated simultaneously; whereas locality refers to that the rules act locally and neighborhood dependencies of each cell is uniform. Note that, the cells perform computation locally, and the global behavior of CA is received due to this local computation only. However, synchronicity is a special type of uniformity, where all the cells are updated simultaneously and uniformly. In fact, uniformity is everywhere in CA, in local rule, cell update and in lattice structure. We can summarize this in the following way:
 \begin{itemize}
 \item Uniformity in update: all cells are updated simultaneously in each discrete time step.
 \item Uniformity in lattice structure and neighborhood dependency:  lattice structure is uniform and each cell follows similar neighborhood dependency.
 \item Uniformity in local rule: each of the cells updates its state following the same rule.
 \end{itemize}
 
Over the years, researchers have successfully been using classical CA as a modeling tool. However, it has become apparent that many phenomena, such as chemical reactions occurring in a living cell, are found in nature which are not uniform. These new modeling requirements led to a new variant of the CAs. As a result, non-uniformity in CAs has been introduced. Following are the main three variants of non-uniformity in CAs, which we get after relaxing above mentioned constraints of uniformity.

\begin{enumerate}
\item Asynchronous cellular automata (ACAs): the cells are not updated at the same (discrete) time step and can be independently updated - breaks the uniform update constraint (discussed in Section~\ref{Chap:surveyOfCA:sscn_ACA}).

\item Automata Network: the CA is on a network and the states of the node evolve with neighborhood defined by the network - breaks uniform neighborhood constraint (discussed in Section~\ref{Chap:surveyOfCA:sscn_NA}).

\item Hybrid or non-uniform cellular automata: cells can assume different local transition functions - breaks uniform local rule constraint (discussed in Section \ref{Chap:surveyOfCA:sscn_HCA}).
\end{enumerate}

\subsection{Asynchronous Cellular Automata ($ACA$s)}
\label{Chap:surveyOfCA:sscn_ACA}
Like other synchronous systems, a CA also assumes a global clock which forces the cells to get updated simultaneously. This assumption of global clock is not very natural, and is relaxed in ACAs. 
The concept of ACAs and their computational ability has first been developed by Nakamura \cite{naka}, it has further been studied in \cite{GOLZE1978176,Nakamura22,Hem82,Ingerson84,LeC89}. ACAs have been developed on two-dimensional grid by Cori et al. \cite{Cor} to report the concurrent situations emerged in distributed systems.

The word `asynchronism' means that the parts of the system do not share the same time. In asynchronous CAs, cells are independent and so, during the evolution of the system, the cells are updated independently. There are several interpretations on the way of applying asynchronism. By simplifying, it can be said that asynchronism is to break the perfect update scheme. The main asynchronous updating schemes found in literature are \emph{fully asynchronous updating} and \emph{$\alpha$-asynchronous updating}.
\begin{itemize}
\item Under \emph{fully asynchronous updating} scheme, a cell is chosen uniformly and randomly at each time step to update. That is, at each step, only one cell is updated.
\item In \emph{$\alpha$-asynchronous updating} scheme, each cell is updated with probability $\alpha$. This implies that a cell does not apply the rule with probability $1-\alpha $, and stays in its old state.
\end{itemize}
The parameter $\alpha$ is known as synchrony rate. Obviously, when $\alpha =1$, the CA becomes synchronous. In that sense, the classical CAs are special case of $\alpha$-asynchronous CAs. Later, other asynchronous updating schemes have been used \cite{probing12}: $\beta$- and $\gamma$-asynchronism. Then, Dennunzio et al. have developed an $m$-asynchronous CA and generalized the various updating methods used so far \cite{Dennunzio13}. A survey on asynchronous update schemes can be found in \cite{Fates14}.

In one of the pointing work, Hemmerling has shown the computation equivalence of synchronous and asynchronous cellular space \cite{Hem82}. He has also shown that, any $d$-state deterministic rule can be simulated by a $3d^2$ state asynchronous rule with same neighborhood dependency. From a different point of view, Golze has shown that, a $(D+1)$-dimensional asynchronous rule can simulate a $D$-dimensional synchronous rule  \cite{GOLZE1978176}. In \cite{Dennunzio16}, it is shown that how fully asynchronous CA can simulate universal Turing machine. 

In an early work, it has been shown that ACA can implement Petri net \cite{Golze1982121}. A Petri net is a directed bipartite graph which is used to model distributed systems. Cori et al. have shown later that ACAs can be used as models of concurrency and distributed systems \cite{Cor}. The computing abilities of these ACAs have further been investigated in \cite{Pighizzini1994179,Droste20001}. As shown in \cite{MANZONI2014108}, ACAs can be used to address firing squad synchronization problem.

In \cite{PhysRevE}, the change that occurs in the Game of Life, when the sites get updated with a given probability, are identified. Ruxton and Saravia have analyzed the sensitivity of ecological system modeled by simple stochastic cellular automata to spatio-temporal ordering \cite{Ruxton}.
In \cite{Tomassini02,fates00608485}, asynchronous rules are used to address the density classification problem. The usage of genetic algorithms for determining the mass-conservative (also called number-conserving) asynchronous models that would generate nontrivial patterns is studied in \cite{Suzudo2004185}. 

It is thought that reversibility and asynchronism are two opposite terms. However, the question of reversibility in ACAs has been studied in \cite{sarkar2012reversibility,SethiD14}. Here, the approach is to study the possibility of returning back to the initial configuration after a finite number of time steps. These so-called reversible ACAs have been used in \cite{ACASir} to generate patterns with specific Hamming distance. These reversible ACAs have further been utilized in \cite{Sethi2016} in symmetric-key cryptography. Mariot has introduced the notion of asynchrony immunity for CAs that could be used for cryptography \cite{Mariot2016}. In \cite{Manzoni2012}, the dynamical properties of CAs, such as injectivity, surjectivity, permutivity, sensitivity, expansivity, transitivity, dense periodic orbits and equicontinuity are re-defined for the asynchronous CAs.

Convergence of ECAs to fixed points under fully asynchronous update has been studied in \cite{Fates20061}. The convergence time of these ACAs have also been explored in this work. Recently, the convergence of these ACAs have been further studied \cite{Sethi2015,CPLX:CPLX21749}, and the convergent ACAs have been used in designing an efficient two class pattern classifier.

\subsection{Automata Network}
\label{Chap:surveyOfCA:sscn_NA}

Traditionally, cellular automata consist of a regular network with local uniform neighborhood dependency. However, in automata network (also called as cellular automata network), this uniform local neighborhood dependency is relaxed. Here, cellular automata rules allow a cell to have an arbitrary number of neighbors, and thus can be set to work on any given network topology, as shown, for example, in \cite{Marr20}. As a matter of fact, the rules of automata networks can not be always local. And, since the non-local and local rules are different, it is, therefore, expected that the non-local rules may lead to different behavior from the conventional local rule-based CAs. Some example works in this regard are \cite{boccara1994some,newman99,yang2007}. 

The earliest version of such non-uniformity in neighborhoods are found in \cite{Jump74,Smith76}.
From the 1990s, networks have been used as an important model for solving different complex problems \cite{Adami199529,watts1998collective}. In fact, after the work of \cite{watts1998collective}, researchers become more interested on automata networks. 
As noted by Tomassini, we get a wider class of generalized automata networks by extending standard lattice cellular automata and random Boolean networks \cite{Tomassini15}. 
It is shown that automata networks with arbitrary topologies perform better than the regular lattice structures for the majority and synchronization problems \cite{tomassini29,Darabos7}. The work of Cori et al. \cite{Cor}, which is a pioneering work on ACAs and which models concurrency and distributed systems, also uses automata network. In \cite{yang2007}, a new type of small-world cellular automata have been developed by combining local updating rules with a probability of long-range short-cuts to simulate the interactions and behavior of a complex system.

Domosi and Nehaniv have investigated automata network as algebraic structures and developed their theory in line with other algebraic theories, such as semi-groups, groups, rings and fields \cite{8718492.ch2}. They have also shown a new method for the emulation of the behavior of any (synchronous) automata network by the corresponding asynchronous one. Kayama and
Imamura have shown the network representation of Game of Life, where the characteristics is like one of Wolfram's class IV rules \cite{Kayama2011}. In \cite{Kayama2012}, the network derived from ECAs and five neighbor totalistic CA rules are further reviewed. However, the studies in this area are still at a very early stage.

\subsection{Non-uniform CAs or Hybrid CAs}
\label{Chap:surveyOfCA:sscn_HCA}

Among the above mentioned models, the most popular and studied model is {\em Hybrid} CA or {\em Non-uniform} CA, where the cells can use different local rules. The study of the non-uniform CA has been started in \cite{Pries86}, where the authors have studied the group properties of $1$-dimensional finite CAs under null and periodic boundary conditions. In that work, a special type of non-uniform CAs have been investigated, where the cells use Wolfram's CAs rules (see Table~\ref{tab:ruleECA}). Since then, however, the major thrust of non-uniform CAs research has been on this class of CAs, see for example \cite{Horte89c,Das91,ppc1,Serra90c,entcs/DasS09}. We dedicate the next section to survey this class of
non-uniform CAs.

In recent years, the generalized definition of non-uniform CAs has been given in \cite{CattaneoDFP09,DennunzioFP12}, where the cells may follow different rules with different neighborhood dependencies. Formenti and his colleagues have been investigating this class of non-uniform CAs, and they have identified various sub-classes of these CAs. Some basic global properties of non-uniform CAs, such as surjectivity, injectivity, equicontinuity, decidability, structural stability etc. have also been explored in  \cite{CattaneoDFP09,DennunzioFP12,DennunzioFP14,salo2014realization}.

\section{Non-uniform ECAs}
\label{Chap:surveyOfCA:scn_eca}

Nowadays, research on non-uniform CAs has gained a popularity. The researchers have been exploring them from various directions, and proposing generalized definition of non-uniform CAs. However, since late $1980$s until today, the primary focus of non-uniform CAs research has been on a special class of $1$-dimensional CAs, where the cells follow Wolfram's rules. The main reason of choosing this class of CA is two-fold -- $(1)$ Wolfram, in early $1980$s, showed the efficacy of $3$-neighborhood binary CAs in modeling physical systems and in producing complex global behavior, and $(2)$ ease of implementing Wolfram's CAs rules in hardware. We call these CAs as non-uniform ECAs, to differentiate them from others. Since the early days, these CAs are explored targeting some hardware-related problems. Obviously, these CAs are finite. In this section, we discuss only about finite non-uniform ECAs.

\begin{figure}
\centering
\includegraphics[width=4.5in,height=1.2in]{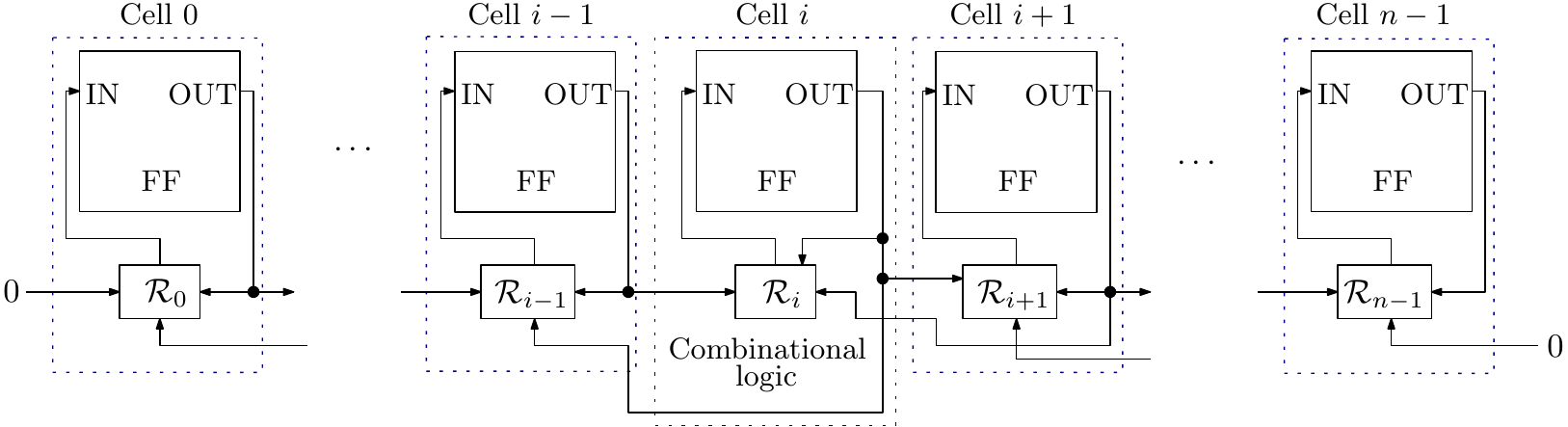}
\caption{Implementation of an $n$-cell non-uniform ECA under null boundary condition}
\label{Fig:ImplNullCA}
\vspace{-1.0em}
\end{figure}

Figure~\ref{Fig:ImplNullCA} shows hardware implementation of an $n$-cell non-uniform ECA under null boundary condition. Here, each cell consists of a flip-flop (FF) to store the state of a cell and a combinational logic circuit to find the next state of the cell. Due to similarity of this structure with finite state machine, many authors, who have been working with finite non-uniform ECAs, consider these CAs as finite state machine (FSM). As a consequence, these authors call the configurations of CAs as states. 

As we have discussed, however, different cells of a non-uniform ECA may use different rules, so we need a rule vector  $\mathcal{R}=\langle \mathcal{R}_0, \mathcal{R}_1, \cdots \mathcal{R}_{n-1} \rangle$ to define these CAs (see Definition~\ref{Def:RuleVector}). Here, cell $i$ uses rule $\mathcal{R}_i$, and $\mathcal{R}_i$ is presented as a decimal number as shown in Table~\ref{tab:ruleECA}. Obviously, uniform or classical CAs are special cases of non-uniform CAs where $\mathcal{R}_0=\mathcal{R}_1= \cdots =\mathcal{R}_{n-1}$.
In this section, we will survey the research organized to understand the behavior of non-uniform ECAs, with respect to two categories: additive/linear CAs and non-linear CAs.

\subsection{Linear/additive CAs}
\label{Chap:surveyOfCA:scn_lnraddCA}
A CA is linear if  $G$, the global transition function of the CA, is linear.
There are seven ECAs having rules $60, 90, 102, 150, 170, 204$ and $240$ which satisfy the linearity conditions (we have excluded rule $0$ from the list, which also satisfies the conditions but a trivial one). These are linear ECAs, and the rules are linear rules (see also Section~\ref{Sec:rule}). However, there is no additive ECAs other than these ECAs. So in literature, the terms ``linear CA'' and ``additive CA'' are used interchangeably.

A (finite) non-uniform ECA is linear/additive if and only if each of the rules in rule vector is linear/additive. Since there are 7 linear/additive rules, the rule vector $\mathcal{R}$ of a linear/additive non-uniform ECA is to be designed with only these 7 rules. The linear ECAs rules can be expressed by $XOR$ logic (see Table~\ref{CArule}, which is reproduced directly from \cite{ppc1}). A linear/additive non-uniform ECA can also be expressed by characteristic matrix, $T$. The matrix $T$ is a tri-diagonal matrix, where all elements except the elements of main, upper and lower diagonals are always zero (see Section~\ref{Chap:surveyOfCA:matrix}). 
In \cite{Das90c,Das91}, the matrix algebraic tool has been used for analyzing state transition behavior of this class of CAs.
From the matrix algebraic tool and characteristic polynomial, several interesting features of the non-uniform ECAs are derived. 

Apart from the seven linear/additive rules, there are another seven rules which are complement of the seven. These rules are named as complemented rules, see Table~\ref{CArule}. The CAs with these complemented rules can also be characterized by matrix algebra. In fact, a CA that uses the fourteen rules, seven linear and seven complemented, can be efficiently characterized by algebraic tools \cite{ppc1,NiloyFI08}\footnote{Some authors call the non-uniform CAs with linear/additive and complemented rules as additive CAs. The reason may be that, these CAs can be characterized using the tools of linear CAs. However, strictly speaking, these CAs are not additive, in general.}. However, for these CAs, we additionally need an {\em Inversion Vector} along with the characteristics matrix. An inversion vector $F$ for such an $n$-cell CA is defined as following:
\begin{equation}
  \begin{array}{l}
   \mbox{$F_i$ =}
    \left\{
	\begin{array}{ll}
	    \mbox{1,  if the $i^{th}$ cell uses a complemented rule}   ~~~~ \\
	    \mbox{0  otherwise}
	    \end{array}\right.
    \end{array}
 \end{equation}  
If a configuration $y$ of the CA is successor of an configuration $x$, then $y = T.x+F$.

Linear/additive CAs are broadly classified as group CAs and non-group CAs, in literature. Next, we briefly discuss about them.

\subsubsection{Group cellular automata}
\label{Chap:surveyOfCA:sscn_grpCA}
A non-uniform ECA is called a group CA if and only if the determinant $det (T) = 1$. The naming of the subclass of linear/additive non-uniform ECAs as group CAs comes from the fact that, these CAs form cyclic group ``under the transformation of operation with $T$'' \cite{ppc1}. Group CAs are reversible CAs. In a group CA, therefore, all the configurations are reachable from some other configurations of the CA. 

In \cite{Pries86}, it is stated that if $\mathcal{R}_i$ is a {\em group rule} then its complement $\overline{ \mathcal{R}_i}$ (that is, $255 - \mathcal{R}_i$) is also a group rule, which is proved in \cite{Das90b}.
A subclass of group CAs is maximal length CAs, in which all non-zero configurations lie in the same cycle. As shown by many authors, such as \cite{Horte89c,Barde90}, these CAs produce pseudo-random patterns having high randomness quality, and are utilized in electronic circuit testing.

In case of maximal length CAs, the characteristics polynomial is primitive. Cattell and
Muzio have given us a scheme of synthesizing a maximal length CA from a given primitive polynomial over $GF(2)$ \cite{cattell1996synthesis}. Rules 90 and 150 are used to construct such non-uniform ECAs, and no single rule can produce a maximal length CA. In \cite{Jetta95}, the maximal length CAs are synthesized up to size 500, where one or two cells follow rule $150$ and the rest follow rule $90$. Boundary condition of these maximal length CAs is null. However, Nandi and Chaudhuri have shown that, for a periodic boundary CA, the characteristic polynomial is factorizable; therefore, there exists no maximal length CA under periodic boundary condition \cite{Nandi96}.

\subsubsection{Non-group cellular automata}
\label{Chap:surveyOfCA:sscn_ngrpCA}

These are irreversible linear/additive non-uniform ECAs. Here, the characteristics matrix $T$ is singular, whereas the $T$ of a group CA is non-singular.
Non-group CAs are explored in different areas, see for example \cite{Bhatt95,Chakr93,santanu00,Rappid}. Any non-group CA is characterized by the following terms - 
\begin{itemize}
\item \emph{attractors}: cyclic states form attractors. If an attractor contains only a single state, it is said to be in graveyard state or a point state attractor or fixed point. 

\item \emph{$\alpha$-basin} or \emph{$\alpha$-tree}: the set of state(s) rooted at any attractor state $\alpha$, is termed as $\alpha$-basin or $\alpha$-tree. 

\item \emph{depth} or \emph{height} of a CA represents the number of single-step evolution, required by the CA to reach to the nearest cyclic state from a non-reachable (that is, Garden-of-Eden) state.
\end{itemize}

Some important findings about number of predecessors of the all-zero configuration and depth of non-uniform ECAs are reported in \cite{ppc1}.
A fraction of all reachable/non-reachable configurations of a uniform CA have been identified in \cite{Martin84a}. However, in general for any additive CA (uniform/hybrid), the fraction of reachable/non-reachable configurations can be numerated from the knowledge of the number of predecessors of a reachable state. Another scheme has been developed in \cite{Acri08b} to identify and count reachable and non-reachable configurations of a finite (uniform/non-uniform) ECA, which can be linear and as well as non-linear.

Some fascinating classes of non-group CAs are also explored, like - multiple attractor CAs (MACAs) \cite{santanu00,maji2003theory}, depth-$1*$ CAs (D1 $*$ CAs) \cite{Chowd92d} and single attractor CAs (SACAs) \cite{Das91}. These CAs have been employed in a broad range of purposes like hashing \cite{adcom00}, classification \cite{NiloyIV,maji2003theory}, designing easy and fully testable FSM \cite{Chowd93a}, etc.

\subsection{Non-Linear CAs}
\label{Chap:surveyOfCA:scn_nlCA}
During the early age of non-uniform ECAs, the non-linear non-uniform ECAs have not been widely explored, due to absence of proper characterization tool. In the past, several attempts have been made to study the characteristics of uniform CAs (linear or nonlinear) qualitatively and quantitatively in terms of parameters, such as $\lambda$ parameter \cite{langton90}, $Z$ parameter \cite{WuenscheRePEc,WuenscheI} etc.
de Bruijn graph is also considered as a characterization tool of CAs (linear/nonlinear) \cite{suttner91,DennunzioFP13}. 

A characterization tool, named {\em reachability tree} (see Section~\ref{Chap:surveyOfCA:tree}), has been proposed to characterize (finite) non-uniform ECAs \cite{SukantaTH,Acri08b,entcs/DasS09}. 
It has been shown that, out of 256 ECAs rules, only 62 rules can design a rule vector of non-uniform reversible ECA under null boundary condition, whereas for periodic boundary condition, additional 8 rules, that is, total 70 rules can take part in a rule vector of reversible automaton. However, the position of these rules in a rule vector is not arbitrary. For the purpose of getting a rule vector of reversible CA, the 62 rules are classified into six classes, which are related to each other. Table~\ref{nextclass} shows the classes and their relations. To get a rule vector of a reversible non-uniform ECA under null boundary condition, one needs to choose a rule from Table~\ref{first}, then the subsequent rules from Table~\ref{nextclass}, and finally the last rule from Table~\ref{last}. It can also be observed that the group CAs under null boundary condition can be decided from the tables. 

In \cite{NazmaTh}, reachability tree has been used to distinguish cyclic/acyclic configurations from the configuration space of non-linear non-uniform ECA. In the same work, $1$-D non-linear non-uniform ECA with point state attractors are also explored using reachability tree. However, in \cite{Soumya2010,Soumya2011}, rule vector graph (RVG) is developed for studying invertibility of $3$-neighborhood CA with null boundary condition.

{\small
\begin{table}[!h]
\caption{First and Last Rule Tables}
\centering
\subfloat[First rule table]{
\label{first}
	\resizebox{0.25\textwidth}{!}{
\begin{tabular}{|c|c|}\hline
Rules for & Class of \\
${\mathcal R}_0$ & ${\mathcal R}_1$ \\\hline
 3, 12 &  I \\
 5, 10 & II\\
 6, 9 & III\\\hline
\end{tabular}}
}
\hspace*{0.75in}
\subfloat[Last rule table]
{
\label{last}
\resizebox{0.30\textwidth}{!}{
\begin{tabular}{|c|c|}\hline
Rule class & Rule set\\
for ${\mathcal R}_{n-1}$ & for ${\mathcal R}_{n-1}$ \\\hline
 I & 17, 20, 65, 68 \\
  II & 5, 20, 65, 80 \\
 III & 5, 17, 68, 80 \\
 IV & 20, 65 \\
 V & 17, 68 \\
 VI & 5, 80 \\\hline
\end{tabular}}
}
\end{table}
}
 
{\small
\begin{table*}[!h]
\caption{Class relationship of ${\mathcal R}_i$ and ${\mathcal R}_{i+1}$}
\begin{center}
\label{nextclass}
	\resizebox{0.60\textwidth}{5.0cm}{
\begin{tabular}{|c|c|c|}\hline
Class of & ${\mathcal R}_i$ & Class of\\
${\mathcal R}_i$ & & ${\mathcal R}_{i+1}$ \\\hline
I & 51,  60,  195, 204 & I\\\cline{2-3}
  &                       85,  90,  165, 170 & II \\\cline{2-3}
  &     102, 105, 150, 153 & III \\\cline{2-3}
  &     53, 58, 83, 92, 163, 172, 197, 202 & IV \\\cline{2-3}
  &     54, 57, 99, 108, 147, 156, 198, 201 & V \\\cline{2-3}
  &     86, 89, 101, 106, 149, 154, 166, 169 & VI \\\hline
II& 15,  30,  45,  60,  75,  90, 105, 120, 135, & I \\
   &                      150, 165, 180, 195, 210, 225, 240 & \\\hline
III& 15,  51, 204, 240 & I \\\cline{2-3}
   &                       85, 105, 150, 170 & II\\\cline{2-3}
   &    90, 102, 153, 165 & III \\\cline{2-3}
   &    23, 43, 77, 113, 142, 178, 212, 232 & IV\\\cline{2-3}
   &    27, 39, 78, 114,  141, 177, 216, 228 & V\\\cline{2-3}
   &    86, 89, 101, 106, 149, 154, 166, 169 & VI \\\hline
IV & 60, 195 & I \\\cline{2-3}
   & 90, 165 & IV \\\cline{2-3}
   & 105, 150 & V \\\hline
V  & 51, 204 & I\\\cline{2-3}
   & 85, 170 & II\\\cline{2-3}
   & 102, 153 & III\\\cline{2-3}
   & 86, 89, 90, 101, 105, 106, 149, 150, & VI\\
   &154, 165,166, 169 & \\\hline
VI & 15, 240 & I\\\cline{2-3}
   & 105, 150 & IV \\\cline{2-3}
   & 90, 165 & V \\\hline
\end{tabular}}
\end{center}
\end{table*}
}

Non-linear non-uniform ECAs are proved to be efficient in various application fields, like VLSI design and test \cite{SukantaTH}, pattern recognition and classification \cite{MajiPhd,DasMNS09}, etc.
Further, designing of pseudo-random pattern generator (PRPG) around reversible non-linear non-uniform ECAs are reported in \cite{ats03,SukantaTH,tcad/DasS10}.

In \cite{NiloyV}, the concept of $GMACA$ (generalized MACA) has been introduced for non-linear non-uniform ECAs. The efficiencies of $MACA$ and $GMACA$ have been compared with respect to pattern recognition. In \cite{MajiPhd}, both linear and non-linear non-uniform ECAs have been used for designing a pattern classifier.

Fuzzy CA, a natural extension of boolean CA, is analyzed and synthesized using matrix algebraic tool \cite{Maji05,DASFAA04}. This CA has also been used to design pattern classifier.

\section{Cellular Automata as Technology}
\label{Chap:surveyOfCA:scn_appCA}

Technology refers to the collection of techniques, methods or processes used to provide some services or solutions to problems, or in the accomplishment of an objective, such as scientific invigilation. The CAs have been historically used as a method for simulating biological and physical systems, and utilized to theoretically study such systems.
Since late $1980$s, however, the CAs have been started to be used as solutions to many real-life problems. In this section, we survey some of such solutions.

\subsection{Electronic circuit design}
The CAs, particularly non-uniform ECAs, have received their popularity as technology in the era of VLSI. The simplicity, modularity and cascadability of CA have enticed the researchers of VLSI domain. Some of these areas are briefly described here.

\subsubsection{Early phase developments:} CAs based machines, CAMs (CA Machines), having high degree of parallelism have been developed by Toffoli and Margolus \cite{Toffo87}, which are ideally suited for simulation of complex systems.
Even before the introduction of CAM, CAs have been utilized as parallel multipliers \cite{Atrub65,Cole69}, parallel processing computers \cite{Manni77}, prime number sieves \cite{Fisch65}, and sorting machine \cite{Nishi81}. Design of fault-tolerant computing machine \cite{Nishio75} and  nanometer-scale classical computer \cite{Benjamin97} are also commendable works. 

\subsubsection{VLSI Design and Test:}
Hortensius et al. have proposed non-uniform ECA based pseudo random pattern generator (PRPG) for built-in self-test (BIST) in VLSI circuits \cite{Horte89a,Horte89c}. Some of the major contributions in the research of PRPGs are reported in \cite{Tsali91,SukantaTH,tcad/DasS10}. The CAs are also proposed as a framework for BIST structures \cite{Tsali90,Das90b,Chowd92d,SukantaTH,DBLP:conf/ats/ChakrabortyC09} and as a deterministic test pattern generator \cite{Albic87a,Das89,Das90b,SukantaTH}. Utilizing the scalability of non-uniform ECAs, a test solution for multi-core chips has been proposed in \cite{tcad/DasS10}.
While testing a CUT (Circuit-Under-Test) with pseudo-random patterns, a set of patterns may be prohibited to the CUT which may adversely affect the circuit. Non-uniform ECAs based solutions to this problem have been proposed in \cite{vlsi02a,ats03,SukantaTH} that can generate pseudo-random-patterns without PPS (Prohibited Pattern Set). Finally, a universal test pattern generator, termed as the $UBIST$ (Universal BIST) has been reported in \cite{Das93,ubist}. This $UBIST$ is able to generate any one of the four types of test patterns - (i) pseudo-random, (ii) pseudo-exhaustive, (iii) pseudo-random without PPS (Prohibited Pattern Set), and (iv) deterministic.

\subsubsection{Synthesis of Finite-State Machine (FSM):}
In \cite{Mitra91b,Misra92b}, the design of testable FSM with CAs have been depicted. In \cite{Chowd93a,Chakr93}, some fascinating properties of a non-group CA and its dual, and their relationship are reported. These papers have also explored a particular class of non-group CAs, named as $D1*$CAs, which have been recommended ``as an ideal test machine which can be efficiently embedded in a finite state machine to enhance the testability of the synthesized design'' \cite{Chakr93}.

\subsubsection{Security and others} A good number of works have been reported in literature that deal with CAs based encryption, cryptography and authentication techniques; see for example \cite{Nandi94a,Bao04,Seredynski2004753,ref1,S0129626409000225,DBLP:journals/jca/DasR11,DBLP:journals/ccds/DasC13,Formenti2014}. Other notable works, related to electronic circuit design are non-uniform ECAs based error correcting codes \cite{Chowd94a,vlsi00b}, signature analysis \cite{Horte90b,Das90e}, etc. More discussions are recorded in \cite{ppc1}.

\subsection{Computer vision and Machine intelligence}
A group of researchers have also explored CAs in the fields of image processing, pattern recognition etc. These fields are roughly named as computer vision and machine intelligence.

\subsubsection{Image processing:} 

CAs, specially $2$-dimensional CAs, are used for image processing. They can address all the significant image processing works like translation, zooming, rotation, segmentation, thinning, compression, edge detection and noise reduction, etc., see for example \cite{Rosin06,Rosin2010790,Okba11}. 
In \cite{khan98}, it is shown that using hybrid CAs, it is possible to rotate any image through an arbitrary angle. In \cite{Paul99}, a new GF($2$) CA based transform coding scheme for gray level and color still images has been proposed. Some works regarding edge detection and noise reduction are reported in \cite{Wongthanavasu03,Sadeghi12}.

\subsubsection{Pattern recognition:} 

CAs have been a popular tool for pattern recognition and classification since long \cite{Jen86,RAGHAVAN1993145,ppc1}. As shown in the book of \cite{ppc1}, MACAs (multiple attractor CAs) can act as a natural classifier.
The correlation between MACA and \emph{Hamming Hash Family} ($HHF$) is shown in \cite{adcom00}. Hamming hash family has inherent capacity to address classification task. This basic framework of linear non-uniform ECA is extended to GMACA (generalized MACA) and utilized for modeling the associative memory \cite{Maji2,MajiPhd}. In \cite{DASFAA04}, fuzzy CAs have been explored as an efficient pattern classifier. Non-linear non-uniform ECAs based pattern classifiers have further been developed in \cite{DasMNS09}. Recently, an asynchronous CA based pattern classifier is reported in \cite{CPLX:CPLX21749}.

\subsubsection{Compression and others:} In \cite{Bhatt95}, some methods are proposed to perform text compression using CA as a technology. In \cite{Lafe}, cellular automata transforms are proposed for digital image compression and data encryption. It is shown that, CAs can generate orthogonal, semi-orthogonal, bi-orthogonal and non-orthogonal bases.
Non-uniform ECA based transforms have been presented in \cite{vlsi00a,ShawSM04,ShawDS06} for developing efficient schemes of image and document compression. Some other related works are reported in \cite{lafe2002method,ye2008novel}. A technology, called \emph{Encompression}, where encryption and compression are married, has been reported in \cite{Chandrama,ShawMSSRC04}.

\subsection{Medical science}

Since the seminal work of von Neumann in $1950$, CA has attracted attention of computer scientists and biologists as an excellent tool to model self-replicating system. The reason of choosing CA for biological modeling is -- ($1$) it is fast and easily implementable and ($2$) the visual result of the simulation provides remarkable resemblance with the original experiment. 
In \cite{Burks1984157}, the research on modeling the evolution and role of DNA sequences within the framework of CA has been initiated. Ermentrout and Edelstein-Keshet have reviewed a number of biologically motivated CAs, such as deterministic or Eurelian automata, lattice gas models and solidification models, heuristic CA etc. and shown the effectiveness of CAs in understanding a physical process \cite{ermentrout1993cellular}.
In \cite{MitraDCN96}, a pioneering work on modeling amino acid using $GF(2^2)$ CA, called as amino acid CA (AACA), has been reported. Some other interesting works are recorded in \cite{zorzenon01,MOREIRA02,JCC:JCC20354,Santos:2013}.

CpG island detection in DNA sequence is a known problem in biological sequence analysis.
Ghosh et al. has addressed this problem \cite{DBLP:conf/iicai/GhoshLMC07}.  In \cite{DBLP:conf/acri/GhoshBMMC10}, a noteworthy class of non-uniform ECAs, known as \emph{Equal Length Cycle CAs}, has been proposed to predict and classify the enzymes. In \cite{Ghosh2012}, another notable class of CAs, termed as \emph{restricted $5$-neighborhood CA} (R5NCA) has been introduced to predict protein structure, and to develop a protein modeling CA machine (PCAM). Here, a R5NCA rule is used to model an amino acid of a protein chain. This PCAM has been used for synthesis of protein structure using an organized knowledge base \cite{DBLP:conf/acri/GhoshMC14}.

\section{Conclusion}
\label{scn_con}

Cellular automaton has gone through a long journey from the early period of von Neumann to elementary form of Wolfram, and to the modern trends of research using this simple yet beautiful model.
In this chapter, various milestones of development regarding CAs are briefly depicted, such as the variety of types, different characterization tools, global behavior and non-uniformity.

During this journey, different types of automata have been developed by varying the parameters of CAs. Based on neighborhood dependencies of cells, various types of CAs have been developed. Similarly, varying dimension of cellular space, states per cell, local rules, we get various types of CAs.

Here, we have discussed about characterization tools, such as de Bruijn graph, matrix algebra, reachability tree, etc, which have been used to understand the behavior of CAs. The global behavior of CAs, which are results of simple local interactions, are very interesting. We have surveyed some of the behaviors, such as universality, reversibility, number conservation, computational ability, etc. Still, there are few more behavior of CAs, such as language recognition, which are not covered in this survey.

In the latter part of the survey, we have focused our discussion on the non-classical CAs - asynchronous CAs, automata networks, and non-uniform CAs. We have noted that, in recent past, a good attention have been put on these non-classical CAs. Among them, the most explored non-classical CAs are (finite) non-uniform CAs, where the local rules are always ECAs rules. Lastly, we have discussed about the potential of CAs as technology.



However, in this dissertation, we particularly concentrate on two global properties of CAs -- reversibility and randomness. In the subsequent chapters, we shall study these two properties in detail.

\chapter{Reversibility of Finite Cellular Automata}\label{Chap:reversibility}

\noindent
{\small This chapter investigates the reversibility of $1$-dimensional $3$-neighborhood $d$-state finite cellular automata (CAs) under periodic boundary condition. A tool named {\em reachability tree} is developed which represents all possible \emph{reachable configurations} of an $n$-cell CA. We identify a large set of CAs, reversible for size $n$, using this tool and some greedy strategies.}

\section{Introduction}
\label{chap:reversibility:sec:intro}

{\large\textbf{T}}he {\em reversibility} property of a cellular automaton (CA) refers to that every configuration of the CA has only one predecessor. That is, the reversible cellular automata (CAs) are injective CAs where the configurations follow one-to-one relationship \cite{hedlund69}. Since late 1960s, the reversibility of CAs has been a point of attraction of many researchers, and a number of works have been carried out in this area, see e.g. \cite{hedlund69,Richa72,Amoroso72,nasu1977local,Maruoka197947,Maruoka1982269,sato77}. The reversible CAs have been utilized in different domains, like simulation of natural phenomenon \cite{hartman90}, cryptography \cite{ppc1}, pattern generations \cite{doi:10.1142/S0218001494000280,Kari2012180}, pseudo-random number generation \cite{aspdac04}, recognition of languages \cite{Kutrib20081142} etc. 

One of the earliest results on CAs is \textit{Garden-of-Eden} theorems by Moore \cite{moore1962machine} and Myhill \cite{Myhill63}, which relate \textit{injectivity} and \textit{surjectivity} of CA with each other. The study on reversibility of CAs was started with Hedlund \cite{hedlund69} and Richardson \cite{Richa72}.  In their seminal paper, Amoroso and Patt provided efficient algorithms to decide whether a one-dimensional CA, defined by a local map $R$, is reversible or not \cite{Amoroso72}. Later, it is shown that it is not possible to design an efficient algorithm that tests reversibility of an arbitrary CA, defined over two or more dimensional lattice \cite{Kari90}. Nevertheless, the research on one-dimensional reversible CAs is continued \cite{di1975reversibility,ito1983linear,morita1995reversible,culik1987invertible,tome1994necessary,MANZINI199860,Soto2008,sato77,Maruoka197947,Aso85,czeizler2007tight,mora4construction,CHANG2016217}. A decision algorithm for CAs with finite configurations is reported in \cite{di1975reversibility}. Some other variants are given in \cite{moraal2000graph,MoraMM06}. In \cite{suttner91}, an elegant scheme based on de Bruijn graph to decide whether a one dimensional CA is reversible is presented. These works, however, deal with infinite lattice. It may be mentioned here that, finite CAs are the interest of researchers, when they are targeted to solve some real-life problems. 

While studying the reversibility (i.e. injectivity) of infinite and finite CAs, one can identify (at least) the following four cases.
\begin{enumerate}\label{reversibility_cases}
\item \label{cs1} An infinite CA whose global function is injective on the set of ``all infinite configurations''.
\item \label{cs2} An infinite CA whose global function is injective on the set of ``all {\em periodic} configurations''. In one-dimension, a configuration $x$ is periodic, or more precisely, spatially periodic if there exists $p \in \mathbb{N}$ such that $x_{i+p}=x_i$ for all $i\in \mathbb{Z}$.
\item \label{cs3} An infinite CA whose global function is injective on the set of ``all finite configurations of length $n$'' for all $n\in \mathbb{N}$.
\item \label{cs4} A finite CA whose global function is injective on the set of ``all configurations of length $n$'' for a fixed $n$.
\end{enumerate}
However, the periodic configurations are often referred to as periodic boundary conditions on a finite CA \cite{Kari05}. Therefore, periodic boundary condition over configuration of length $n$, for all $n \in \mathbb{N}$ evidently implies Case~\ref{cs2}. According to the definitions of periodic configuration (Definition~\ref{Def:periodicConfiguration}, Page~\pageref{Def:periodicConfiguration}) and finite CA (under periodic boundary condition), therefore, Case~\ref{cs2} and Case~\ref{cs3} are equivalent under one dimension. It is also known that Case~\ref{cs1} and Case~\ref{cs2} are equivalent for one-dimensional CAs \cite{Kari05,sato77}. Hence, in one-dimension, cases \ref{cs1}, \ref{cs2} and \ref{cs3} are equivalent, and the Case~\ref{cs4} is different from them. So, the algorithms of \cite{Amoroso72} and \cite{suttner91}, which are the decision procedures for Case \ref{cs1}, can decide the one-dimensional CAs of cases \ref{cs1}, \ref{cs2} and \ref{cs3}. This chapter deals with Case~\ref{cs4}, and reports an algorithm to decide whether a finite one-dimensional CA under periodic boundary condition having a fixed cell length is reversible or not.

Reversibility of finite one-dimensional CAs is also previously tackled \cite{ppc1,entcs/DasS09,Soumya2010,Soumya2011,marti2011reversibility,DOW199767,0305-4470-37-22-006,Ino05,Sato2009,zubeyir11}. For example, in \cite{DOW199767}, reversibility of finite additive CAs under periodic boundary condition is investigated, where the set of states $\mathcal{S}$ is a finite commutative ring. In \cite{marti2011reversibility}, reversibility of binary CAs with null boundary condition is studied. Whereas, in \cite{zubeyir11}, reversibility of linear CAs with periodic boundary conditions over $\mathbb{Z}_p$, where $p \geq 2$ is a prime number, is reported. All these works consider CAs, where the local map $R$ is linear. The reason of choosing the linear CAs is, standard algebraic techniques, like matrix algebra, can be used to characterize them. 

However, reversibility of finite CAs, where the CAs can be non-linear, has also been addressed. For example, in \cite{SukantaTH,entcs/DasS09}, reversibility of $1$-dimensional non-uniform ECAs has been studied under null and periodic boundary conditions. In \cite{Soumya2010}, a linear time algorithm is reported to check invertibility of non-uniform ECAs under null boundary condition. This result is extended for uniform ECAs in \cite{Soumya2011}. Some other works on reversibility of finite CAs, having two states per cell, are observed in \cite{0305-4470-37-22-006,Ino05,Sato2009}. Most of these works are based on binary CAs, where often the neighborhood is considered as $3$. This is because, under this restriction, the CAs are simpler and easy to deal.

Nevertheless, reversibility of finite CAs having the number of states per cell $d\geq 2$ is, in general, an untouched issue. Because, there is lack of proper characterization tool to address this problem. Study of these Case~\ref{cs4} CAs is essential for simulating microscopic reversibility on computer as well as for real-life applications. Hence, in this chapter, we target to address this problem. For doing this, we develop a mathematical tool, named as \emph{Reachability Tree}, which can characterize linear and nonlinear $d$-state finite CAs.

Here, we consider one-dimensional $3$-neighborhood (that is, nearest neighbor) CA with $d$ number of states per cell ($d\geq2$). As is well-known after Smith, a CA with higher neighborhood dependency can always be emulated by another CA with lesser, say $3$-neighborhood dependency \cite{Smith71}. Therefore, in this chapter, if not otherwise specified, by ``CA'', we will mean one-dimensional $3$-neighborhood finite CA having fixed cell length $n$ with $d$ states per cell ($d \geq 2$).

In this chapter and subsequent chapters, many terminologies are used. We first state those definitions and terminologies to be used throughout this dissertation (Section~\ref{chap:reversibility:Sec:CAbasic}). Then, we develop the characterization tool, \emph{Reachability Tree} (Section~\ref{chap:reversibility:Sec:rtree}). This tool is instrumental in developing theories for finite CAs. We identify the properties of reachability tree when it presents a reversible CA (Section~\ref{chap:reversibility:Sec:rev}). Exploring these properties, we develop an algorithm to test reversibility of a finite CA with a particular cell length $n$ (Section~\ref{chap:reversibility:Sec:bij}). We finally report three greedy strategies to get a set of reversible finite CAs (Section~\ref{chap:reversibility:Sec:identify}).

\section{Definitions and Ternimologies}
\label{chap:reversibility:Sec:CAbasic}
In this dissertation, we consider one-dimensional CAs with periodic boundary condition where cells of the CA form a ring $\mathscr{L} = \mathbb{Z}/n\mathbb{Z}$, $n$ is the length of the CA. That is, the CAs are finite. Each cell of such an \emph{$n$-cell} CA can use a set of states $\mathcal{S}$ = $\{0,1, \cdots, d-1\}$, where $d \ge 2$. If not otherwise mentioned, the cells have $3$ neighborhood dependency.

The next state of each cell is determined by a local rule $R: \mathcal{S}^3 \rightarrow \mathcal{S}$. A rule is classically represented by its tabular form. See Table~\ref{chap:reversibility:tab:rule3} for example. This table has an entry for each possible combination of neighborhood $xyz$. Recall that, these neighborhood combinations are referred as \emph{Rule Min Term} or \emph{RMT} (see Definition~\ref{Def:RMT}, Page~\pageref{Def:RMT}). Therefore, there are $d^3$ RMTs in a $3$-neighborhood $d$-state CA. A rule is normally identified by the values of $R[r]$ as a string (with $R[0]$ as the right most digit) or its decimal equivalent.

\begin{table}[h]
	\setlength{\tabcolsep}{1.3pt}
	\begin{center}
		\caption{Rules of $3$-state CAs. Here, PS and NS stands for present state and next state respectively}
		\label{chap:reversibility:tab:rule3}
		\resizebox{1.00\textwidth}{!}{
			\begin{tabular}{cccccccccccccccccccccccccccc}
				\toprule
				\thead{P.S.} & \thead{222} & \thead{221} & \thead{220} & \thead{212} & \thead{211} & \thead{210} & \thead{202} & \thead{201} & \thead{200} & \thead{122} & \thead{121} & \thead{120} & \thead{112} & \thead{111} & \thead{110} & \thead{102} & \thead{101} & \thead{100} & \thead{022} & \thead{021} & \thead{020} & \thead{012} & \thead{011} & \thead{010} & \thead{002} & \thead{001} & \thead{000}\\ 
				
				\thead{RMT} & \thead{(26)} & \thead{(25)} & \thead{(24)} & \thead{(23)} & \thead{(22)} & \thead{(21)} & \thead{(20)} & \thead{(19)} & \thead{(18)} & \thead{(17)} & \thead{(16)} & \thead{(15)} & \thead{(14)} & \thead{(13)} & \thead{(12)} & \thead{(11)} & \thead{(10)} & \thead{(9)} & \thead{(8)} & \thead{(7)} & \thead{(6)} & \thead{(5)} & \thead{(4)} & \thead{(3)} & \thead{(2)} & \thead{(1)} & \thead{(0)}\\ 
				\midrule
				\multirow{3}{*}{}
				&2&0&1&2&1&0&2&1&0&2&0&1&2&1&0&2&1&0&2&0&1&2&1&0&2&1&0\\
				& 2&0&1&0&1&2&2&1&0&2&0&1&0&1&2&2&1&0&2&0&1&0&1&2&2&1&0\\
				\thead{N.S.} & 0&2&0&1&2&0&1&1&2&1&2&2&0&1&0&1&2&0&1&1&0&1&2&2&0&2&0\\
				& 1&0&$\mathbf{2}$&0&$\mathbf{1}$&2&1&$\mathbf{0}$&2&0&1&$\mathbf{2}$&$\mathbf{1}$&0&2&1&$\mathbf{0}$&2&0&$\mathbf{2}$&1&0&2&$\mathbf{1}$&$\mathbf{0}$&1&2 \\
				& 1&2&0&0&2&1&1&2&0&0&2&1&0&2&1&1&2&0&0&2&1&0&2&1&2&1&0\\
				& 1&0&2&2&2&1&0&1&0&1&0&2&2&2&1&0&1&0&1&0&2&2&2&1&0&1&0\\
				& 1&1&2&2&2&1&0&1&0&1&1&2&2&2&1&0&0&0&1&1&2&2&2&1&0&0&0\\
				& 0&1&2&0&1&2&1&2&0&0&1&2&2&1&0&1&0&2&2&0&1&0&2&1&1&0&2 \\
				& 2&2&2&2&1&1&1&1&2&0&0&1&0&0&0&0&0&0&1&1&0&1&2&2&2&2&1 \\
				& 2&1&1&2&1&2&1&1&2&0&2&0&0&0&0&0&2&0&1&0&2&1&2&1&2&0&1\\
				\bottomrule
			\end{tabular}
		}
	\end{center}
\end{table}

To understand global behavior of CAs, a mathematical tool, named de Bruijn graph (see Section~\ref{Chap:surveyOfCA:dbg} of Chapter~\ref{Chap:surveyOfCA}), is also used by various researchers \cite{suttner91,Mora2008,Soto2008}. An $ s $-dimensional de Bruijn graph of $ k $ symbols is a directed and edge-labeled graph representing overlaps between sequences of symbols, where $|\mathcal{S}|=k$ and $s=m-1$, $m$ is the size of the neighborhood of the CA(see Definition~\ref{Def:dbg}). A CA, defined by a local rule $R$, can be expressed as a de Bruijn graph $B(m-1,\mathcal{S})$, where each edge $(ax, xb)$ of $B(m-1, \mathcal{S})$, such that $a, b\in \mathcal{S}, x\in \mathcal{S}^{s-1}$, depicts the overlapping sequence of nodes and $(axb) \in \mathcal{S}^m$. This overlapping sequence $(ax, xb)$ is equivalent to RMT $r$. The graph has $d^{m-1}$ number of nodes and $d^m$ number of edges and is balanced in the sense that each vertex has both in-degree and out-degree $k$. In this chapter, we consider size of neighborhood $m=3$.

\begin{figure*}[hbt]
\centering
\includegraphics[width= 3.8in, height = 3.0in]{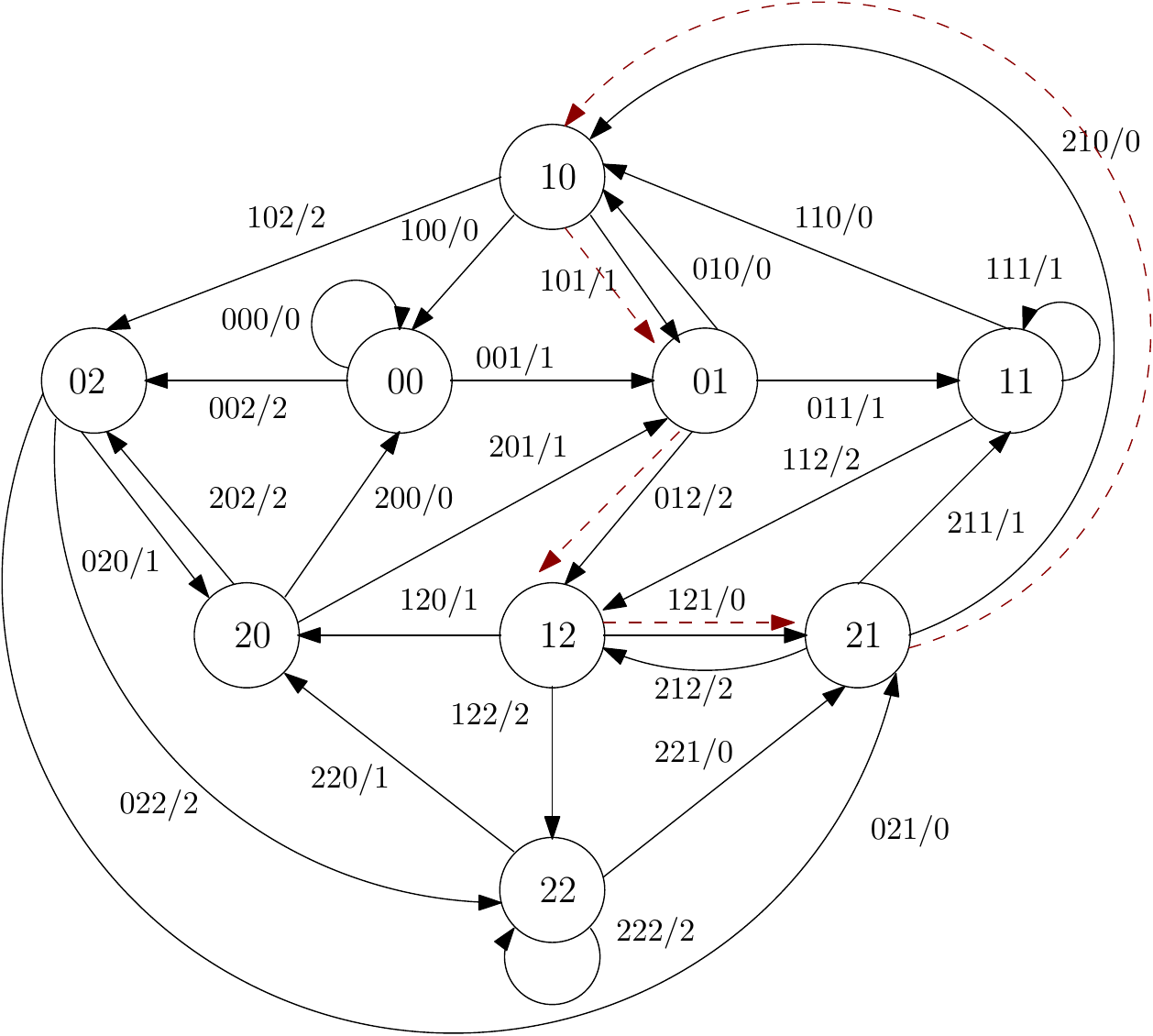}
\caption{The de Bruijn Graph of CA $201210210201210210201210210$}
\label{chap:reversibility:fig:dbg_3_state}
\end{figure*}

Figure~\ref{chap:reversibility:fig:dbg_3_state} represents a $3$-state CA ($3^{rd}$ row of Table~\ref{chap:reversibility:tab:rule3}). The graph shows that if the left, self and right neighbors of a cell are all $0$s, then next state of the cell (that is, $R(0,0,0)$) is $0$, if the neighbors are $0, 0$ and $1$ respectively, the next state is $1$, and so on. 
It can also be observed that, each node has $3$ incoming edges and $3$ outgoing edges. In general, a node of the de Bruijn graph of a $d$-state CA has $d$ incoming edges and $d$ outgoing edges. Therefore, the set of incoming RMTs (resp. outgoing RMTs) are related to each other. In Figure~\ref{chap:reversibility:fig:dbg_3_state}, last (resp. first) $2$ digits of any set of incoming RMTs (resp. outgoing RMTs) are same. We call the set of incoming RMTs as \emph{equivalent} RMTs, and the set of outgoing RMTs as \emph{sibling} RMTs.

\begin{definition}
\label{Def:equivalent}
A set of $d$ RMTs $r_1, r_2, ..., r_d$ of a $d$-state CA rule are said to be \textbf{equivalent} to each other if $r_1 d \equiv r_2 d \equiv ... \equiv r_d d \pmod{ d^3}$.

\end{definition}

\begin{definition}
\label{Def:sibling}
A set of $d$ RMTs $s_1, s_2, ..., s_d$ of a $d$-state CA rule are said to be \textbf{sibling} to each other if $\floor{\frac{s_1}{d}} = \floor{\frac{s_2}{d}} = ... = \floor{\frac{s_d}{d}}$.

\end{definition}

The rationale behind choosing the name \emph{equivalent} is - if one traverses the de Bruijn graph of a $d$-state CA, then a node can be reached through any one of the $d$ incoming edges, hence all edges are equivalent with respect to the reachability of the node. On the other hand, after reaching a node, one can keep on traversing the graph by selecting any of the outgoing edges, to which we name \emph{sibling}, because they are coming out from the same mother node.

We represent $Equi_i$ as a set of RMTs that contains RMT $i$ and all of its equivalent RMTs. That is, $Equi_i = \{i, d^2+i, 2d^2+i, \cdots, (d-1)d^2+i \}$, where $0 \leq i \leq d^2-1$. Similarly, $Sibl_j$ represents a set of sibling RMTs where $Sibl_j = \{d.j, d.j+1, \cdots, d.j+d-1\}$ $(0\leq j \leq d^2-1)$.  However, one can observe an interesting relation among RMTs during traversal of the graph. In Figure~\ref{chap:reversibility:fig:dbg_3_state}, if RMT $1$, (or RMT $10$ or RMT $19$) is used to visit a node, then either RMT $3$ or RMT $4$ or RMT $5$ is to be used to proceed further traversal. Table~\ref{Chap:reversibility:tab:rln} shows the relationship among the RMTs of $3$-state CAs. In general, if RMT $r \in Equi_i$ of a $d$-state CA is used to reach a node, then the next RMT to be chosen is $s \in Sibl_i$.

	\renewcommand{\arraystretch}{0.99}
	\begin{table}[hbtp]
		\begin{center}
			\caption{Relations among the RMTs for $3$-neighborhood $3$-state CAs}\label{Chap:reversibility:tab:rln}
			{
				\resizebox{0.9\textwidth}{!}{
					\begin{tabular}{ccc|ccc}
						\toprule
						\multicolumn{3}{c|}{\thead{Equivalent Set}} & \multicolumn{3}{c}{\thead{Sibling Set}}\\
						\thead{\#Set} & \thead{Equivalent RMTs} & \thead{Decimal Equivalent} & \thead{\#Set} & \thead{Sibling RMTs} & \thead{Decimal Equivalent} \\ 
						\midrule
						
						$Equi_0$ & 000, 100, 200 & 0, 9, 18 & $Sibl_0$ & ~000, 001, 002 & 0, 1, 2 \\ 
						$Equi_1$ & 001, 101, 201 & 1, 10, 19 & $Sibl_1$ & ~010, 011, 012 & 3, 4, 5 \\ 
						$Equi_2$ & 002, 102, 202 & 2, 11, 20 & $Sibl_2$ & ~020, 021, 022 & 6, 7, 8 \\ 
						$Equi_3$ & 010, 110, 210 & 3, 12, 21 & $Sibl_3$ & ~100, 101, 102 & 9, 10, 11 \\ 
						$Equi_4$ & 011, 111, 211 & 4, 13, 22 & $Sibl_4$ & ~110, 111, 112 & 12, 13, 14 \\ 
						$Equi_5$ & 012, 112, 212 & 5, 14, 23 & $Sibl_5$ & ~120, 121, 122 & 15, 16, 17 \\ 
						$Equi_6$ & 020, 120, 220 & 6, 15, 24 & $Sibl_6$ & ~200, 201, 202 & 18, 19, 20 \\ 
						$Equi_7$ & 021, 121, 221 & 7, 16, 25 & $Sibl_7$ & ~210, 211, 212 & 21, 22, 23 \\ 
						$Equi_8$ & 022, 122, 222 & 8, 17, 26 & $Sibl_8$ & ~220, 221, 222 & 24, 25, 26 \\ 
						\bottomrule
					\end{tabular}
				}}
			\end{center}
		\end{table} 

\begin{definition}\label{Def:permutivity}
	Let $R: \mathcal{S}^3 \rightarrow \mathcal{S}$ be the local rule of a CA. Then, the CA is \textbf{left-}(resp. \textbf{right-}) \textbf{permutive} if and only if for any two equivalent (resp. sibling) RMTs $r$ and $s$, $R[r]\ne R[s]$.
\end{definition}
Therefore, in a left-permutive (resp. right-permutive) CA, all RMTs of $Equi_i$ (resp. $Sibl_i$), for all $i$, have distinct next state values. 

\begin{definition}
\label{Def:balancedrule}
A CA rule is \textbf{balanced} if it contains equal number of RMTs for each of the $d$ states possible for that CA; otherwise it is an \textbf{unbalanced} rule.

\end{definition}

\begin{example}
Rule $201210210201210210201210210$ is balanced, because the rule contains nine $0$s, nine $1$s and nine $2$s.
\end{example}

In a CA, the snapshot of the states of all cells at a given time is called a \emph{configuration}, represented by $c:\mathscr{L} \rightarrow \mathcal{S}$. That means, if $x$ is a configuration of an $n$-cell CA, then $x=(x_i)_{i\in\mathscr{L}}$, where $x_i$ is the state of cell $i$. 
During evolution, a CA hops from one configuration to another.
Let $G_n: \mathcal{C}_n \rightarrow \mathcal{C}_n$ be the global transition function for the CA induced by the local rule $R$, where $\mathcal{C}_n=\mathcal{S}^{\mathscr{L}}$ is the set of all configurations of size $n$. Then,  

\begin{equation*}
y=G_n(x) = G_n(x_0x_1\cdots x_{n-1})= (R(x_{i-1}, x_i, x_{i+1}))_{i\in \mathscr{L}}
\end{equation*}
Here, $y$ is the successor of the configuration $x$. However, a configuration can also be depicted by a sequence of RMTs, which we call an \emph{RMT sequence}.

\begin{definition}\label{Def:RMTSequence}
	Let $x=(x_i)_{i\in\mathscr{L}}$ be a configuration of a CA. The {\textbf{RMT sequence}} of $x$, denoted as $\tilde{x}$, is $(r_i)_{i\in\mathscr{L}}$ where $r_i$ is the RMT $(x_{i-1}, x_i, x_{i+1})$.
\end{definition}

\begin{example}
Let $x=1012$ be a configuration for a $4$-cell $3$-neighborhood CA. Then the RMT sequence corresponding to this configuration is $\tilde{x}=\langle210 (21)$, $101 (10)$, $012 (5)$, $121 (16)\rangle$.
\end{example}
Hence, any configuration is a sequence of equivalent and sibling RMTs. However, an arbitrary set of RMTs can not form an RMT sequence. If an RMT is chosen from $Equi_i$, then the next RMT in the sequence must be chosen from $Sibl_i$. Any \emph{cycle} in the de Bruijn graph also represents an RMT sequence corresponding to a configuration. 

 \begin{definition}
 \label{Defi:Cycle}
 A \textbf{\em cycle of length $\mathbf{n}$} in a de Bruijn graph is a sequence of vertices $(v_1, v_2,\cdots, v_n, v_{n+1})$, where $v_{n+1}=v_1$ and edge $e_i = (v_i,v_{i+1})$, $i\in\{1, 2,\cdots,n\}$. We generally represent this sequence as $(e_1,e_2,\cdots,e_n)$.
 \end{definition}

Therefore, a cycle of length $n$ in a de Bruijn graph corresponds to an RMT sequence of length $n$.  In fact, cycles in a de Bruijn graph and RMT sequences (that is, configurations) are synonymous in this context. The next configuration of a given configuration can, therefore, be found by traversing the de Bruijn graph. Following example illustrates this.

\begin{example}
\label{Chap:reversibility:ex_dbg_config}
Let us take the configuration $1012$ of the $4$-cell CA of Figure~\ref{chap:reversibility:fig:dbg_3_state}. To get the next configuration of $1012$, we form a $2$-digit overlapping window and start from node $10$ as the first two digits of $1012$ are $10$. From node $10$, we use edge $101$ and go to node $01$, then from it following edge $012$, we go to node $12$; from node $12$, we go to node $21$ by the edge $121$ and finally, from node $21$, we come back to our starting node $10$ by the edge $210$. For each of the edges we traverse, we get a next state value. By these next states, we get the next configuration as $1200$. 
The traversal of the cycle is shown by dotted arrow in Figure~\ref{chap:reversibility:fig:dbg_3_state}.
\end{example}

\begin{definition}\label{Def:selfreplicating}
	An RMT $r = x \times d^2 + y \times d + z$ is said to be \textbf{self-replicating}, if $R(x,y,z)=y$, where $R: \mathcal{S}^3 \rightarrow \mathcal{S}$ is the rule of the CA.
\end{definition}
\begin{example}
For the CA $102012102012102102021021012$ ($6^{th}$ row of Table~\ref{chap:reversibility:tab:rule3}), the self-replicating RMTs are $2 (002)$, $3 (010)$, $7 (021)$, $10 (101)$, $14 (112)$, $15 (120)$, $19 (201)$, $22 (211)$ and $24 (220)$ [marked in bold face].
\end{example}

\begin{definition}\label{Def:fixed-point}
	A \textbf{fixed-point attractor} is a configuration in CA, for which the next configuration is itself. That means, if a CA reaches a fixed-point attractor, then it remains at that particular configuration forever.                                                                                  
\end{definition}

For example, for the CA $120021120021021120021021210$, $0^{n}$ is a fixed point attractor with $n$ cells, $n$ $\geq$ $3$. If a CA has a quiescent state $x$ (see Definition~\ref{Def:Quiescent}), then there exists a fixed point at $x^{n}$, for any cell length $n$. 	
\begin{example}
For the CA $012012120012210102201021102$ ($10^{th}$ row of Table~\ref{chap:reversibility:tab:rule3}), $1^{n}$ is a fixed point attractor with $n$ cells, $n \in \mathbb{N}$. 
\end{example}

\begin{figure}[hbtp]
\subfloat[$012012120012210102201021102$\label{Chap:reversibility:trans1}]{%
\includegraphics[width=0.47\textwidth, height = 2.7cm]{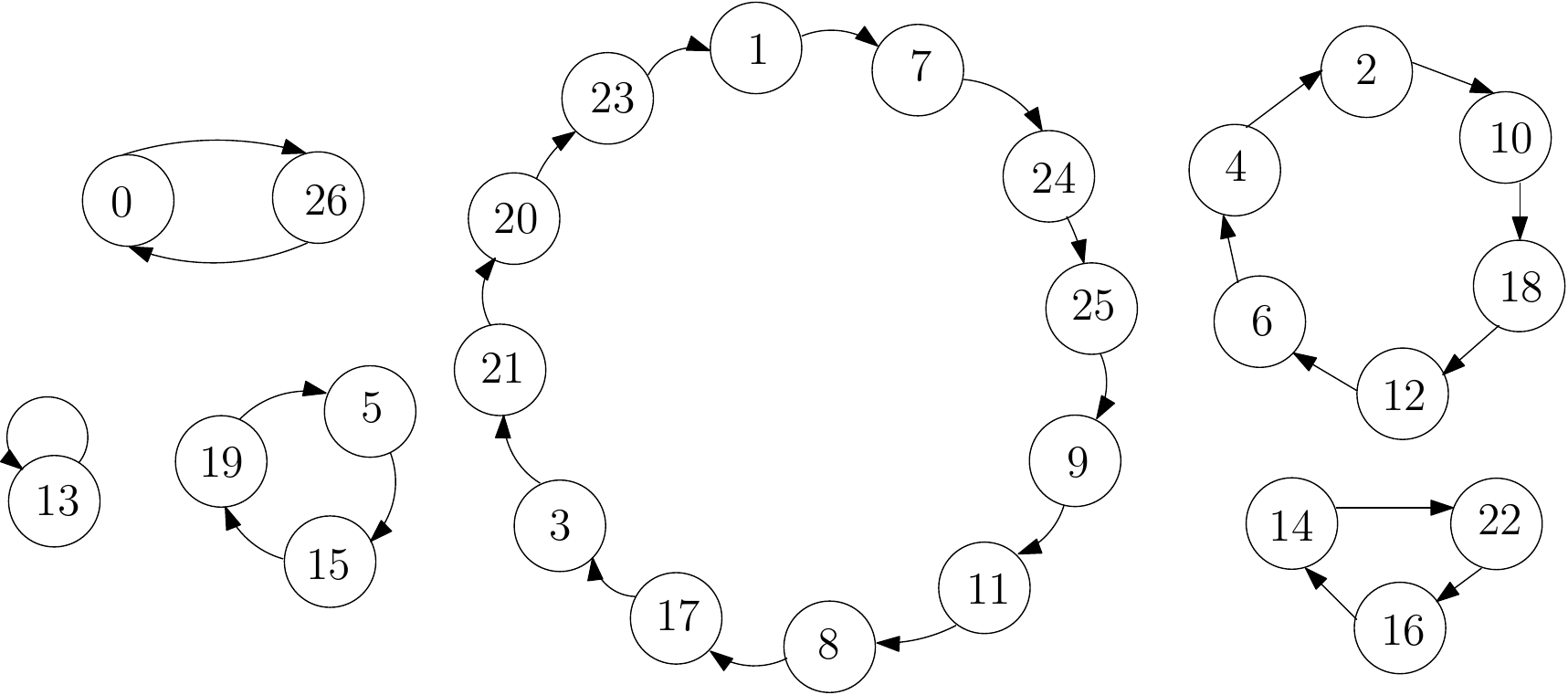}
}
\hfill
\subfloat[$222211112001000000110122221$\label{Chap:reversibility:trans2}]{%
\includegraphics[width=0.47\textwidth, height = 2.7cm]{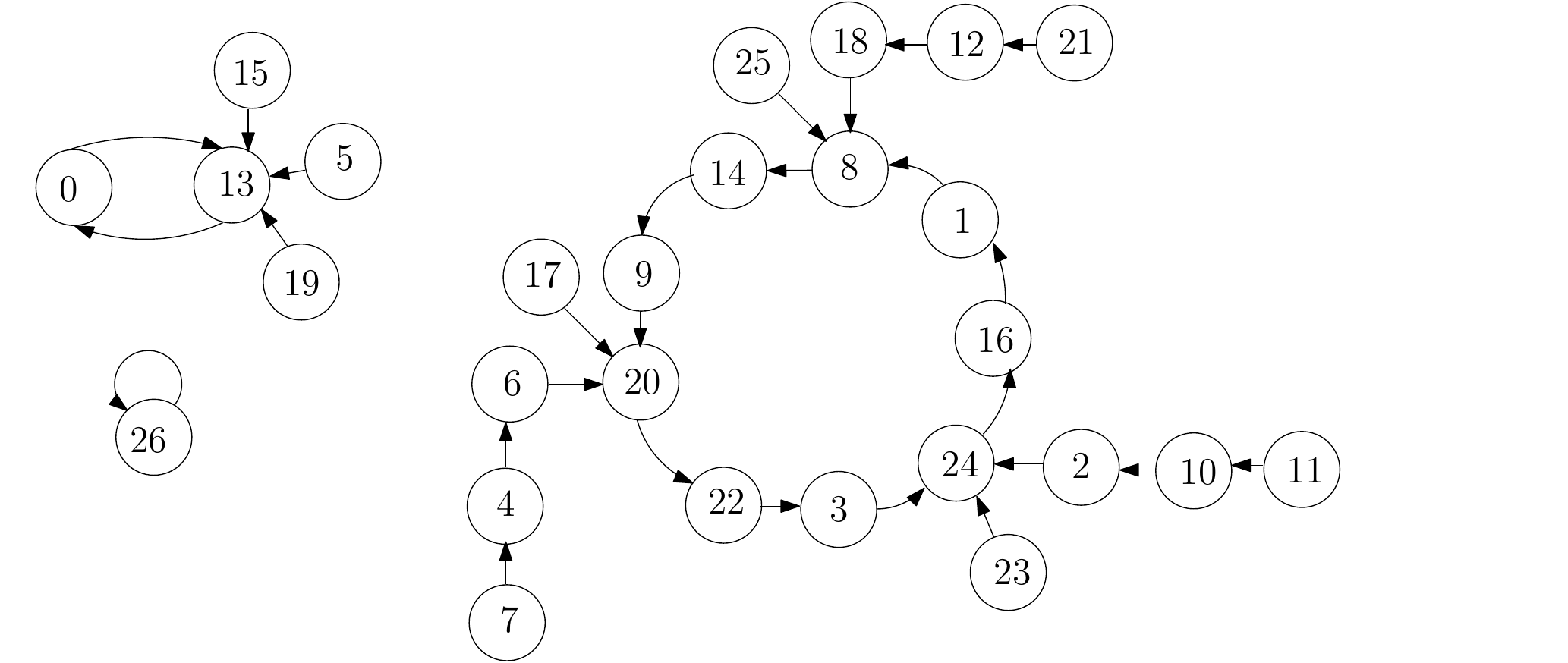}
}
\caption{Configuration transition diagram for two $3$-neighborhood $3$-state CAs with $3$ cells}
\label{fig:config}
\end{figure}
		
\begin{example}\label{Chap:reversibility:ex:trans}
Figure~\ref{fig:config} shows the configuration transition diagram for two $3$-neighborhood $3$-state CAs with $n = 3$. Here, configurations are written in their decimal equivalents. Both of these CAs have fixed points; the configuration $111$ $(13)$ is a fixed-point for Figure~\ref{Chap:reversibility:trans1} and configuration $222$ $(26)$ is a fixed-point for Figure~\ref{Chap:reversibility:trans2}. Moreover, in Figure~\ref{Chap:reversibility:trans1}, for the $3$-cell CA $012012120012210102201021102$ ($10^{th}$ row of Table~\ref{chap:reversibility:tab:rule3}), all configurations are part of some cycle, with the largest cycle length as $12$. However, in Figure~\ref{Chap:reversibility:trans2}, for the CA $222211112001000000110122221$ ($11^{th}$ row of Table~\ref{chap:reversibility:tab:rule3}) with $3$ cells, it is not the case. Here, some configurations are not part of any cycle. The largest cycle of this CA is of length $9$. 
		\end{example}   

%

A configuration is {\emph{reachable}} for a CA, if starting from a distinct initial configuration, that configuration can be reached in the evolution of the CA; otherwise it is {\emph{non-reachable}}.

\begin{definition}\label{Def:reachable}
Let $x \in \mathcal{C}_n$ be a configuration of a CA with global transition function $G_n: \mathcal{C}_n \rightarrow \mathcal{C}_n$. The configuration $x$ is called \textbf{reachable}, if there exists another configuration $y \in \mathcal{C}_n$, such that, $G_n(y)=x$. Otherwise, it is called \textbf{non-reachable}. Here, $\mathcal{C}_n$ is the set of all configurations of length $n$.
\end{definition}

In Example~\ref{Chap:reversibility:ex:trans}, all configurations of Figure~\ref{Chap:reversibility:trans1} are reachable, because every configuration has a predecessor. However, for Figure~\ref{Chap:reversibility:trans2}, the configurations $5,15,19,21,25,17,7,23$ and $11$ are non-reachable, as they can not be reached from any other configurations. 
It can be noted that, a CA can be in a non-reachable configuration, only if it is the initial configuration of the CA.

\begin{definition}\label{Def:reversible_finite}
A finite CA of size $n$ with global transition function $G_n: \mathcal{C}_n \rightarrow \mathcal{C}_n$ is \textbf{reversible}, if for each $x \in \mathcal{C}_n$, where $\mathcal{C}_n$ is the set of all configurations of length $n$, there exists exactly one $y \in \mathcal{C}_n$, such that, $G_n(y)=x$. That is, the CA is reversible if $G_n$ is a bijection. Otherwise, the CA is \textbf{irreversible} for that $n$. 
\end{definition}


\begin{example}
In Figure~\ref{Chap:reversibility:trans1}, the CA $012012120012210102201021102$ is reversible for $n=3$. Whereas, the CA $222211112001000000110122221$ of Figure~\ref{Chap:reversibility:trans2} is irreversible for $n=3$. For this CA, the configurations $8,20,24$ and $13$ have multiple predecessors.
\end{example}

In this chapter, we study the reversibility of $3$-neighborhood finite CAs having $n$ number of cells.
We consider here $n \geq 3$, as $n=1$ and $n=2$ are the trivial cases for $3$-neighborhood CAs.

\section{The Reachability Tree}
\label{chap:reversibility:Sec:rtree}
In this section, we develop a discrete tool, named \emph{Reachability tree}, for an $n$-cell $d$-state CA ($n \geq 3$). The tree enables us to efficiently decide whether a given $n$-cell CA is reversible or not. Moreover, it guides us to identify reversible CAs. 

Reachability tree was initially proposed for studying 1-dimensional non-uniform finite ECAs under null boundary condition \cite{Acri04}. In \cite{entcs/DasS09}, it was extended for non-uniform ECAs having finite size under periodic boundary condition. This mathematical tool has been employed to explore properties like reversibility, convergence, etc. for the above class of non-uniform (linear or non-linear) CAs, see for instance \cite{SukantaTH,Adak2016OnSO,NazmaTh}. For more details, see Section~\ref{Chap:surveyOfCA:tree} (Page~\pageref{Chap:surveyOfCA:tree}) and Section~\ref{Chap:surveyOfCA:scn_nlCA} (Page~\pageref{Chap:surveyOfCA:scn_nlCA}) of Chapter~\ref{Chap:surveyOfCA}. However, in all these works, the CAs are considered non-uniform and binary with $3$-neighborhood dependency. That is, the existing reachability tree does not work on $1$-D $d$-state uniform CAs. Therefore, in this chapter, we generalize reachability tree, so that it can characterize $1$-D $3$-neighborhood $d$-state finite CAs, under periodic boundary condition.

To test reversibility of a CA, de Bruijn graph may be utilized. In \cite{suttner91}, a scheme based on de Bruijn graph was developed to test reversibility of CAs defined over infinite lattice. However, for finite CAs, a straight forward scheme of testing reversibility can be developed - consider each of the possible $d^n$ configurations and find next configuration using de Bruijn graph. If each configuration is reachable and has only one predecessor, declare the CA as reversible for size $n$.

Finding of next configuration of a given configuration using de Bruijn graph is simple, and can be done in $\mathcal{O}(n)$ time, where $n$ is the size of the configuration (see Example~\ref{Chap:reversibility:ex_dbg_config}). However, finding of the next configuration of all possible configurations of an $n$-cell CA is an issue. The de Bruijn graph does not directly give any information about the existence of non-reachable configurations.

Reachability tree, on the other hand, depicts the reachable configurations of an $n$-cell CA. Non-reachable configurations can be directly identified from the tree. The tree has $n+1$ levels, and like de Bruijn graph, edges are labeled. However, here the labels generally contain more than one RMT. A sequence of edges from root to leaf represents a reachable configuration, where each edge represents a cell's state.

\begin{definition} \label{Chap:reversibility:def:tree_3}
Reachability tree of an $n$-cell $d$-state CA is a rooted and edge-labeled $d$-ary tree with $(n+1)$ levels where 
each node $ N_{i.j} ~ (0 \leq i \leq n,~ 0 \leq j \leq d^{i}-1)$ is an ordered list of $d^2$ sets of RMTs, and the root $N_{0.0}$ is the ordered list of all sets of sibling RMTs. We denote the edges between $N_{i.j} ~ (0 \leq i \leq n-1,~ 0 \leq j \leq d^{i}-1)$ and its possible $d$ children as $E_{i.dj+x} = ( N_{i.j}, N_{i+1.dj+x}, l_{i.dj+x} )$ where $l_{i.dj+x}$ is the label of the edge and $0 \leq x \leq d-1$. Like nodes, the labels are also ordered list of $d^2$ sets of RMTs. Let us consider that ${\Gamma_{p}}^{N_{i.j}}$ is the $p^{th}$ set of the node $N_{i.j}$, and ${\Gamma_{q}}^{E_{i.dj+x}}$ is the $q^{th}$ set of the label on edge $E_{i.dj+x}$ $(0 \leq p,q \leq d^2 -1)$, So,  $N_{i.j} = ( {\Gamma_{p}}^{N_{i.j}})_{0 \leq p \leq d^2-1}$ and $l_{i.dj+x} =  ( {\Gamma_{q}}^{E_{i.dj+x}})_{0 \leq q \leq d^2-1}$. Following are the relations which exist in the tree:

\begin{enumerate}
\item \label{Chap:reversibility:def:tree_3:rtd1} [For root] $N_{0.0} = ({\Gamma_{k}}^{N_{0.0}})_{0 \leq k \leq d^2-1}$, where ${\Gamma_{k}}^{N_{0.0}} = Sibl_k$.

\item \label{Chap:reversibility:def:tree_3:rtd2} $\forall r \in {\Gamma_{k}}^{N_{i.j}}, ~ r$ is included in ${\Gamma_{k}}^{E_{i.dj +x}}$, if $R[r] = x, x \in \lbrace 0, 1, \cdots, d-1 \rbrace$, where $R$ is the rule of the CA. That means, $ {\Gamma_{k}}^{N_{i.j}} = \bigcup\limits_{x} {\Gamma_{k}}^{E_{i.dj+x}}$ $(0 \leq k \leq d^2-1, 0 \leq i \leq n-1,~ 0 \leq j \leq d^{i}-1)$.

\item \label{Chap:reversibility:def:tree_3:rtd4}$\forall r$, if $r \in {\Gamma_{k}}^{E_{i.dj+x}}$, then RMTs of $Sibl_p$, that is $d.r \pmod{d^3}, d.r+1 \pmod{d^3}, \cdots, d.r+(d-1) \pmod{d^3} $ are in ${\Gamma_{k}}^{N_{i+1.dj+x}}$, where $0\leq x \leq d-1$ and $r \equiv p \pmod{d^{2}}$.

\item \label{Chap:reversibility:def:tree_3:rtd5} [For level $n-2$] ${\Gamma_{k}^{N_{n-2.j}}} = \lbrace s ~|~$ if $ r \in {\Gamma_{k}^{E_{n-3.j}}} $ then $ s \in \lbrace d.r \pmod{d^3}, d.r+1 \pmod{d^3}, \cdots, d.r+(d-1) \pmod{d^3}\rbrace \cap \lbrace i, i+d, \cdots, i+(d^2-1)d \rbrace \rbrace$ $(i= \floor{\frac{k}{d}}, 0 \leq k \leq d^2-1, 0 \leq j \leq d^{n-2}-1 )$.

\item \label{Chap:reversibility:def:tree_3:rtd6} [For level $n-1$] ${\Gamma_{k}}^{N_{n-1.j}} = \lbrace s ~|~$ if $ r \in {\Gamma_k^{E_{n-2.j}}} $ then $ s \in \lbrace d.r \pmod{d^3}, d.r+1 \pmod{d^3}, \cdots, d.r+(d-1) \pmod{d^3}\rbrace \cap \lbrace k + i \times d^2 ~|~ 0 \leq i \leq d-1 \rbrace \rbrace, { 0 \leq k \leq d^2-1}$.

\end{enumerate}

\end{definition}

Note that, the nodes of level $n-2$ and $n-1$ are different from other intermediate nodes (Points~\ref{Chap:reversibility:def:tree_3:rtd5} and \ref{Chap:reversibility:def:tree_3:rtd6} of Definition~\ref{Chap:reversibility:def:tree_3}). Only a subset of selective RMTs can play as $ {\Gamma_{k}}^{N_{i.j}}$ in a node $ {N_{i.j}}, (0\leq k \leq d^2-1, 0 \leq j \leq d^i-1)$ when $i = n-2$ or $n-1$. In fact, only $\frac{1}{d}$ of the RMTs that are possible for a node following Point~\ref{Chap:reversibility:def:tree_3:rtd4} of Definition~\ref{Chap:reversibility:def:tree_3}, can exist for any node at level $n-2$ or level $n-1$, if we apply Points~\ref{Chap:reversibility:def:tree_3:rtd5} and \ref{Chap:reversibility:def:tree_3:rtd6} of Definition~\ref{Chap:reversibility:def:tree_3} on the node. Finally, we get the leaves with ${\Gamma_{k}}^{N_{n.j}}$, where ${\Gamma_{k}}^{N_{n.j}}$ is either empty or a set of sibling RMTs. The root ${\Gamma_{k}}^{N_{0.0}}$ is a set of sibling RMTs (Point~\ref{Chap:reversibility:def:tree_3:rtd1} of Definition~\ref{Chap:reversibility:def:tree_3}) and ${\Gamma_{k}}^{N_{0.0}} = \bigcup\limits_{j} \Gamma_k^{N_{n.j}} $. However, in our further discussion we shall not explicitly define $i$ and $j$ of node $N_{i.j}$ or edge $E_{i.j}$ if they are clear from the context.

\begin{example}
Reachability tree of a $4$-cell CA with $3$ states per cell is shown in Figure~\ref{Chap:reversibility:fig:rt2}. Table~\ref{Chap:reversibility:Tab:ruleRTree} shows the edges and corresponding nodes at each level. As it is of $3$ states, a node $ N_{i,j} $ can have at most $3$ children - $ N_{i+1.3j} $, $ N_{i+1.3j+1} $ and $ N_{i+1.3j+2} $. Hence, maximum number of nodes possible in the tree for a $4$-cell $3$-state CA is $ \frac{3^{4+1} - 1}{3-1} = 121$. Figure~\ref{Chap:reversibility:fig:rt2}, however, contains $105$ nodes. This is because, in Table~\ref{Chap:reversibility:Tab:ruleRTree}, level of nodes like $N_{4.1}, N_{4.2}, N_{4.3}, N_{4.6}$ etc. are empty; so, these nodes does not exist in the reachability tree (see Figure~\ref{Chap:reversibility:fig:rt2}). 
The root of the tree is $N_{0.0} = (\Gamma_k^{N_{0.0}})_{0 \leq k \leq 8}$, where $\Gamma_k^{N_{0.0}}$ is a set of sibling RMTs $(Sibl_k)$. That means, $N_{0.0}$ = $(\{0,1,2\}$, $\{3,4,5\}$, $\{6,7,8\}$, $\{9,10,11\}$, $\{12,13,14\}$, $\{15,16,17\}$, $\{18,19,20\}$, $\{21,22,23\}$, $\{24,25,26\})$ (see $2^{nd}$ row of Table~\ref{Chap:reversibility:Tab:ruleRTree}). In a reachability tree, the root is independent of CA rule (it depends only on $d$, the number of states per cell), whereas other nodes are rule dependent (see Table~\ref{Chap:reversibility:Tab:ruleRTree} for details). 
 
\begin{scriptsize}\setlength\tabcolsep{5pt}
	\centering
\begin{longtable}{|c|p{6.5cm}|p{6cm}|}
\caption{Reachability Tree for $3$-state CA $201012210201012210201012210$ (Figure~\ref{Chap:reversibility:fig:rt2}) with $n = 4$}
\label{Chap:reversibility:Tab:ruleRTree}\\
\hline
\textbf{Level} & \textbf{Nodes} & \textbf{Edges} \\
\hline
\endfirsthead
\multicolumn{3}{c}%
{\tablename\ \thetable\ -- \textit{Continued from previous page}} \\
\hline
\textbf{Level} & \textbf{Nodes} & \textbf{Edges} \\
\hline
\endhead
\hline \multicolumn{3}{r}{\textit{Continued on next page}} \\
\endfoot
\hline
\endlastfoot
$0$ & $N_{0.0} = (\{0-2\}$, $\{3-5\}$, $\{6-8\}$, $\{9-11\}$, $\{12-14\}$, $\{15-17\}$, $\{18-20\}$, $\{21-23\}$, $\{24-26\})$ & NA \\ 
\hline 

\multirow{3}{*}{$1$} & $N_{1.0}$ = $(\{0-2 \}$, $\{15-17 \}$, $\{21-23 \}$, $\{0-2 \}$, $\{15-17 \}$, $\{21-23 \}$, $\{0-2 \}$, $\{15-17 \}$, $\{21-23 \})$ & $E_{0.0}$ = $(\{0\}$, $\{5\}$, $\{7\}$, $\{9\}$, $\{14\}$, $\{16\}$, $\{18\}$, $\{23\}$, $\{25\})$ \\
 \hhline{~--} 
& $N_{1.1}$ = $(\{3-5 \}$, $\{12-14 \}$, $\{18-20 \}$, $\{3-5 \}$, $\{12-14 \}$, $\{18-20 \}$, $\{3-5 \}$, $\{12-14 \}$, $\{18-20 \})$ & $E_{0.1}$ = $(\{1\}$, $\{4\}$, $\{6\}$, $\{10\}$, $\{13\}$, $\{15\}$, $\{19\}$, $\{22\}$, $\{24\})$ \\
 \hhline{~--} 
& $N_{1.2}$ = $(\{6-8 \}$, $\{9-11 \}$, $\{24-26 \}$, $\{6-8 \}$, $\{9-11 \}$, $\{24-26 \}$, $\{6-8 \}$, $\{9-11 \}$, $\{24-26 \})$ & $E_{0.2}$ = $(\{2\}$, $\{3\}$, $\{8\}$, $\{11\}$, $\{12\}$, $\{17\}$, $\{20\}$, $\{21\}$, $\{26\})$ \\
\hline

\multirow{9}{*}{$2$} & $N_{2.0}$ = $(\{0 \}$, $\{21 \}$, $\{15 \}$, $\{1 \}$, $\{22 \}$, $\{16 \}$, $\{2 \}$, $\{23 \}$, $\{17 \})$ & $E_{1.0}$ = $(\{0\}$, $\{16\}$, $\{23\}$, $\{0\}$, $\{16\}$, $\{23\}$, $\{0\}$, $\{16\}$, $\{23\})$ \\
 \hhline{~--} 
& $N_{2.1}$ = $(\{3 \}$, $\{18 \}$, $\{12 \}$, $\{4 \}$, $\{19 \}$, $\{13 \}$, $\{5 \}$, $\{20 \}$, $\{14 \})$ & $E_{1.1}$ = $(\{1\}$, $\{15\}$, $\{22\}$, $\{1\}$, $\{15\}$, $\{22\}$, $\{1\}$, $\{15\}$, $\{22\})$ \\
 \hhline{~--} 
& $N_{2.2}$ = $(\{6 \}$, $\{24 \}$, $\{9 \}$, $\{7 \}$, $\{25 \}$, $\{10 \}$, $\{8 \}$, $\{26 \}$, $\{11 \})$ & $E_{1.2}$ = $(\{2\}$, $\{17\}$, $\{21\}$, $\{2\}$, $\{17\}$, $\{21\}$, $\{2\}$, $\{17\}$, $\{21\})$ \\
 \hhline{~--} 

& $N_{2.3}$ = $(\{15 \}$, $\{15 \}$, $\{0 \}$, $\{16 \}$, $\{16 \}$, $\{1 \}$, $\{17 \}$, $\{17 \}$, $\{2 \})$ & $E_{1.3}$ = $(\{5\}$, $\{14\}$, $\{18\}$, $\{5\}$, $\{14\}$, $\{18\}$, $\{5\}$, $\{14\}$, $\{18\})$ \\
\hline
\end{longtable}
\end{scriptsize}
\begin{sidewaysfigure}[hbtp]
\thisfloatpagestyle{empty}
   \centering
    \includegraphics[width= 8.5in, height = 3.5in]{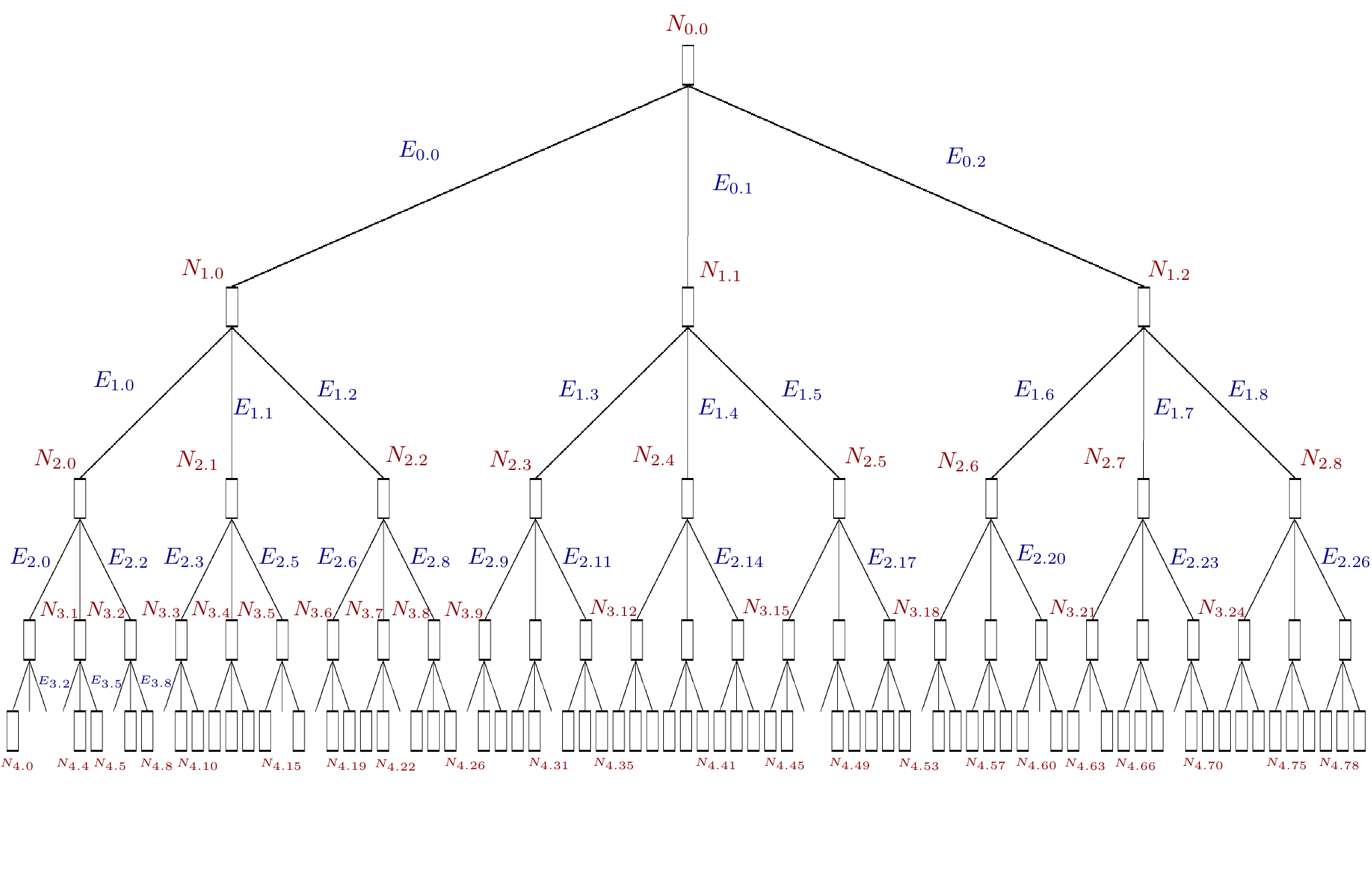} 
   \caption{Reachability tree of $4$-cell $3$-state CA with rule $201012210201012210201012210$}
    \label{Chap:reversibility:fig:rt2}
\end{sidewaysfigure}
However, an arbitrary RMT can not be a part of nodes of level $n-2$ and $n-1$. For example, ${l_{1.1}} = (\{1\}$, $\{15\}$, $\{22\}$, $\{1\}$, $\{15\}$, $\{22\}$, $\{1\}$, $\{15\}$, $\{22\})$, but the corresponding node ${N_{2.1}}$ (that is, $N_{n-2.1}$, since $n=4$) is $(\{3\}$, $\{18\}$, $\{12\}$, $\{4\}$, $\{19\}$, $\{13\}$, $\{5\}$, $\{20\}$, $\{14\})$ ($7^{th}$ row of Table~\ref{Chap:reversibility:Tab:ruleRTree}). Observe that, $\Gamma_1^{N_{2.1}}$ of the node ${N_{2.1}}$ is $\{18\}$. If we follow point~\ref{Chap:reversibility:def:tree_3:rtd4} of Definition~\ref{Chap:reversibility:def:tree_3}, then RMTs $19$ and $ 20$ should also be part of $\Gamma_1^{N_{2.1}}$. But, they could not, because the node is at level $n-2$ (Point~\ref{Chap:reversibility:def:tree_3:rtd5} of Definition~\ref{Chap:reversibility:def:tree_3}). Similarly at level $3$, $l_{2.3} = (\emptyset, \{18\}, \emptyset, \emptyset, \emptyset, \emptyset, \{5\}, \emptyset, \{14\})$ and $N_{3.3}= (\emptyset, \{1\}, \emptyset, \emptyset, \emptyset, \emptyset, \{15\}, \emptyset, \{17\})$ (Point~\ref{Chap:reversibility:def:tree_3:rtd6}). That means, for node $N_{3.3}$, only RMT $1$ is present in the set ${\Gamma_{1}}^{N_{3.3}}$. So, only $\frac{1}{3}$ of the possible RMTs can be part of any set at level $n-2$ or $n-1$. 
\end{example}
Reachability tree gives us information about reachable configurations of the CA. However, some nodes in a reachability tree may not be present, which we call non-reachable nodes, and the corresponding missing edges as non-reachable edges. No RMT is present in a non-reachable node or in the label of a non-reachable edge.

\begin{definition}
A node $N_{i.j}$ is \textbf{non-reachable} if $\bigcup\limits_{0 \leq k \leq d^2 - 1 } \Gamma_k^{N_{i.j}} = \emptyset$. Similarly, an edge $E_{i.j}$ is \textbf{non-reachable} if $\bigcup\limits_{0 \leq k \leq d^2 - 1 } \Gamma_k^{E_{i.j}} = \emptyset$.
\end{definition}

The edges of the tree associate the states of CA cells, and a sequence of edges from the root to a leaf represents a reachable configuration. Since $d$ number of edges can come out from a parent node, we call the left most edge as $0$-edge which represents state $0$, second left most edge as $1$-edge which represents state $1$, and so on. The right most edge represents state $d-1$.

\begin{definition}
An edge $E_{i.j}$ is called $\mathbf{s}$-\textbf{edge} if $R[r] = s $, $ r \in \bigcup\limits_{0 \leq k \leq d^2 - 1 } \Gamma_k^{E_{i.j}}$, where $ 0\leq s \leq d-1$, and $R$ is the CA rule.
\end{definition}

Therefore, the sequence of edges $\langle E_{0.j_1}, E_{1.j_2}, ...,  E_{n-1.j_n}\rangle$, where $0 \leq j_i \leq d^i -1, 1 \leq i \leq n$, represents a reachable configuration.

\begin{example}
In Figure~\ref{Chap:reversibility:fig:rt2}, there are some non-reachable edges - $E_{3.1}, E_{3.2}, E_{3.3}, E_{3.6}$ etc. for which the labels are empty (see Table~\ref{Chap:reversibility:Tab:ruleRTree}). Corresponding nodes $N_{4.1}, N_{4.2}, N_{4.3}$, $N_{4.6}$ etc. are non-reachable nodes. The edge sequence $\langle E_{0.0}, E_{1.0}, E_{2.0}$, $ E_{3.1} \rangle$ represents a non-reachable configuration ($0001$), as the sequence includes a non-reachable edge. On the other hand, the sequence $\langle E_{0.0}, E_{1.1}, E_{2.3}$, $ E_{3.11} \rangle$, represents the reachable configuration $0102$. 
\end{example}

\section{Reachability Tree and Reversible CA}
\label{chap:reversibility:Sec:rev}
This section studies the reachability tree of CAs which are reversible for a fixed lattice size $n$. These studies are utilized in Section~\ref{chap:reversibility:Sec:bij} and Section~\ref{chap:reversibility:Sec:identify}.


\begin{theorem}
\label{Chap:reversibility:revth1}
The reachability tree of a finite reversible CA of length $n$ $(n \geq 3)$ is complete.
\end{theorem} 

\begin{proof}Since all the configurations of a reversible CA are reachable, the number of leaves in the reachability tree of an $n$-cell $d$-state CA is $d^n$ (number of configurations). Therefore, the tree is complete as it is a $d$-ary tree of $(n + 1)$ levels.
\end{proof}

The above theorem points to the fact that the identification of a reversible CA can be done by constructing the reachability tree of the CA. If there is no non-reachable edge in the reachability tree, then the CA is reversible. Following theorem further characterizes the reachability tree of a reversible CA.

\begin{theorem}
\label{Chap:reversibility:revth2}
The reachability tree of a $d$-state finite CA of length $n$ $ (n \geq 3)$ is complete if and only if
\begin{enumerate}

\item \label{Chap:reversibility:c1} The label $l_{n-1.j}$, for any $j$, contains only one RMT, that is,\\ $\mid \bigcup\limits_{0 \leq k \leq d^2 -1} {\Gamma_{k}}^{E_{n-1.j}}\mid = 1$.

\item \label{Chap:reversibility:c2}  The label $l_{n-2.j}$, for any $j$, contains only $d$ RMTs, that is,\\ $ \mid \bigcup\limits_{0 \leq k \leq d^2 -1} {\Gamma_{k}}^{E_{n-2.j}}\mid = d$.

\item \label{Chap:reversibility:c3} Each other label $l_{i.j}$ contains $d^2$ RMTs, that is, $ \mid \bigcup\limits_{0 \leq k \leq d^2 -1} {\Gamma_{k}}^{E_{i.j}}\mid = d^2$, where $ 0 \leq i \leq n-3$.
\end{enumerate}
\end{theorem}

\begin{proof}
\begin{description} [leftmargin=1ex]
\item\noindent\underline{\textit{If Part:}}
 Let us consider, the number of RMTs in the label of an edge is less than that is mentioned in (\ref{Chap:reversibility:c1}) to (\ref{Chap:reversibility:c3}). That means,

(i)\label{Chap:reversibility:i1} There is no RMT in the label $l_{n-1.j}$, for some $j$. That is, $ \bigcup\limits_{0 \leq k \leq d^2 -1} {\Gamma_{k}}^{E_{n-1.j}}= \emptyset $. It implies, the tree has a non-reachable edge and so, it is incomplete.

(ii) \label{Chap:reversibility:i2} The label $l_{n-2.j}$ contains less than $d$ RMTs, for some $j$. That is, \\$ \mid \bigcup\limits_{0 \leq k \leq d^2 -1} {\Gamma_{k}}^{E_{n-2.j}}\mid \leq d-1$. Then, the number of RMTs in the node $N_{n-1.j} \leq d(d-1)$.
Since the node is at level $(n - 1)$, only $\frac{1}{d}$ of the RMTs are valid according to the Definition~\ref{Chap:reversibility:def:tree_3}. So, the number of valid RMTs is $ \leq \frac{d(d-1)}{d} = (d-1)$, which implies that the maximum number of possible edges from the node is $d-1$. Hence, there is at least one (non-reachable) edge ${E_{n-1.b}}$ for which $\bigcup\limits_{0 \leq k \leq d^2 -1} {\Gamma_{k}}^{E_{n-1.b}}= \emptyset $.

(iii) \label{Chap:reversibility:i3} Say, each other label $l_{i.j}$ contains less than $d^2$ RMTs, that is, $ \mid \bigcup\limits_{0 \leq k \leq d^2 -1} {\Gamma_{k}}^{E_{i.j}}\mid < d^2$, $(0 \leq i \leq n-3)$. Then, $ \mid \bigcup\limits_{0 \leq k \leq d^2 -1} {\Gamma_{k}}^{N_{i+1.j}}\mid < d^3$. Here, the node $N_{i+1.j}$ may have $d$ number of edges. In best case, the tree may not have any non-reachable edge up to level $(n - 2)$. Then there exists at least one edge $E_{n-3.p}$, for which $ \mid \bigcup\limits_{0 \leq k \leq d^2 -1} {\Gamma_{k}}^{E_{n-3.p}}\mid < d^2$, which makes a node $N_{n-2.p}$ with $ \mid \bigcup\limits_{0 \leq k \leq d^2 -1} {\Gamma_{k}}^{N_{n-2.p}}\mid < d^3$. Since the node is at level $(n - 2)$, it has maximum $\frac{d(d^2-1)}{d} = (d^2-1)$ valid RMTs. This implies, there exists at least one edge, incident to $N_{n-2.p}$, where $ \mid \bigcup\limits_{0 \leq k \leq d^2 -1} {\Gamma_{k}}^{E_{n-2.q}}\mid < d$, which makes the tree an incomplete one by (ii).

On the other hand, if for any intermediate edge ${E_{i.j_1}}$, $ \mid \bigcup\limits_{0 \leq k \leq d^2 -1} {\Gamma_{k}}^{E_{i.j_1}}\mid$ $\geq d^2$, then an edge $ E_{i.j_2}$ can be found at the same label $i$ for which $ \mid \bigcup\limits_{0 \leq k \leq d^2 -1} {\Gamma_{k}}^{E_{i.j_2}}\mid < d^2$, where $ 0 \leq i \leq n-3$, and $j_1 \neq j_2$. Then, by (iii), the tree is incomplete. Now, if for any $p$, label $l_{n-2.p}$ contains more than $d$ RMTs, then also there exists an edge $E_{n-2.q}$ for which $ \mid \bigcup\limits_{0 \leq k \leq d^2 -1} {\Gamma_{k}}^{E_{n-2.q}}\mid$ $< d$. Hence, the tree is incomplete (by (ii)). Similarly, if for an edge $E_{n-1.x}$, $ \mid \bigcup\limits_{0 \leq k \leq d^2 -1} {\Gamma_{k}}^{E_{n-1.x}} \mid > 1 $, then also the tree is incomplete. Therefore, if the number of RMTs for any label is not same as mentioned in (i) to (iii), the reachability tree is incomplete.

\item\noindent\underline{\textit{Only if Part:}} Now, let us consider that, the reachability tree is complete. The root $N_{0.0}$ has $d^3$ number of RMTs. Now, these RMTs have to be distributed so that the tree remains complete. Let us take that, an edge $E_{0.j_1}$ has less than $d^2$ RMTs, another edge $E_{0.j_2}$ has greater than $d^2$ RMTs and other edges $E_{0.j} (0 \leq j,j_1,j_2 \leq d-1$ and $j_1 \neq j_2 \neq j)$ has $d^2$ RMTs. Then node $N_{1.j_1}$ has less than $d^3$ RMTs, $N_{1.j_2}$ has greater than $d^3$ RMTs and other edges $N_{1.j}$ has $d^3$ RMTs. The tree has no non-reachable edge at level $1$. Now, this situation may continue upto level $n-2$. At level $(n-2)$, only $\frac{1}{d}$ of the RMTs are valid (see Definition~\ref{Chap:reversibility:def:tree_3}). So, the nodes with less than $d^3$ RMTs has at maximum $d^2-1$ valid RMTs and so on. For such nodes at level $n-2$, there exists at least one edge $E_{n-2.p}$, such that $ \mid \bigcup\limits_{0 \leq k \leq d^2 -1} {\Gamma_{k}}^{E_{n-2.p}}\mid < d$ for which the tree will have non-reachable edge (by (ii)). The situation will be similar if any number of intermediate edges have less than $d^2$ RMTs implying some other edges at the same level have more than $d^2$ RMTs. Hence the tree will be incomplete which contradicts our initial assumption. So, for all intermediate edges $E_{i.j}$, $ \mid \bigcup\limits_{0 \leq k \leq d^2 -1} {\Gamma_{k}}^{E_{i.j}}\mid = d^2$, where $ 0 \leq i \leq n-3$.

Now, if this is true, then at level $n-2$, the nodes have $d^3$ RMTs out of which $d^2$ are valid. If an edge $E_{n-2.p}$ has less than $d$ RMTs, then the node $N_{n-1.p}$ has at maximum $d(d-1)$ RMTs out of which only $d-1$ are valid. Hence, at least one edge, incident to $N_{n-1.p}$, is non-reachable making the tree incomplete. Similar thing happens if there exist more edges like $E_{n-2.p}$. So, each edge label $l_{n-2.j}$ must have $d$ RMTs. In the same way, each edge label $l_{n-1.j}$, for any $j$, is to be made with a single RMT to make the tree complete. Hence the proof.
\end{description}\end{proof}

\begin{corollary}
\label{Chap:reversibility:revcor1}
The nodes of a reachability tree of a reversible CA of length $n$ $(n \geq 3)$ contains

\begin{enumerate}

\item $d$ RMTs, if the node is in level $n$ or $n-1$, i.e. $ \mid \bigcup\limits_{0 \leq k \leq d^2 -1}{\Gamma_{k}}^{N_{i.j}} \mid = d$ for any $j$, when $i = n$ or $n-1$.

\item $d^2 $ RMTs, if the node is at level $n-2$ i.e, $ \mid \bigcup\limits_{0 \leq k \leq d^2 -1}{\Gamma_{k}}^{N_{n-2.j}} \mid = d^2$ for any $j$.

\item $d^3$ RMTs for all other nodes $N_{i.j}$, $ \mid \bigcup\limits_{0 \leq k \leq d^2 -1}{\Gamma_{k}}^{N_{i.j}} \mid = d^3$ where ${ 0 \leq i \leq n-3}$.
\end{enumerate}
\end{corollary}

\begin{proof}This is directly followed from Theorem~\ref{Chap:reversibility:revth2}, because for each RMT on an edge $E_{i.j}$, $d$ number of sibling RMTs are contributed to $N_{i+1.j}$.
\end{proof}


Like CA rules, we classify the nodes of a reachability tree as \emph{balanced} and \emph{unbalanced}.

\begin{definition}
\label{Chap:reversibility:def:balancednode}
A node is called \textbf{balanced} if it has equal number of RMTs corresponding to each of the $d$ states possible; otherwise it is \textbf{unbalanced}. So, for a balanced node, number of RMTs with next state value $0$s = number of RMTs with next state value $1$s = $\cdots$ =  number of RMTs with next state value $(d-1)$s.
\end{definition}

Therefore, the root of the reachability tree of a balanced rule is balanced, because number of RMTs associated with each of the $d$ states is $d^2$.

\begin{lemma} 
\label{Chap:reversibility:revcor2}
The nodes of the reachability tree of an $n$-cell $(n \geq 3)$ reversible CA are balanced.

\end{lemma}

\begin{proof}Since the reachability tree of a reversible $d$-state CA is complete, each node has $d$ number of edges i.e $d$ number of children. Since a node $N_{i.j}$ contains $d^3$ RMTs when $i < n-2$ (Corollary ~\ref{Chap:reversibility:revcor1}) and an edge $E_{i+1.k}$, for any $k$, contains $d^2$ RMTs (Theorem~\ref{Chap:reversibility:revth2}), the node $N_{i.j}$ is balanced. Here, number of RMTs, associated with same next state values, is $d^2$. Similarly, the nodes of level $n-2$ and $n-1$ are balanced. 
\end{proof}

\begin{theorem}
\label{Chap:reversibility:revth4}
A finite CA of length $n (n \geq 3)$ with unbalanced rule is irreversible.
\end{theorem}

\begin{proof}If the rule is unbalanced, then it has unequal number of RMTs corresponding to each state. That means, the root node $N_{0.0}$ is unbalanced. Therefore, there exists an edge $E_{0.j}$ where $\mid\bigcup\limits_{0\leq k \leq d^2-1} \Gamma_k^{E_{0.j}}\mid < d^2$ $(0 \leq j \leq d^2-1)$. Hence the CA is irreversible by Theorem~\ref{Chap:reversibility:revth2}.
\end{proof}

However, the CAs with balanced rules can not always be reversible. Following example illustrates this.

\begin{example}
Consider the CA $201012210201012210201012210$ of Figure~\ref{Chap:reversibility:fig:rt2} and Table~\ref{Chap:reversibility:Tab:ruleRTree}. The rule has $9$ RMTs for each of the states $0, 1,$ and $2$, so it is balanced. Each node $N_{i.j}$, when $i \leq n-3 = 1$ contains $27$ RMTs, each node at level $n-2$ i.e $N_{2.j}$ contains $9$ RMTs and each node of level $n-1$ i.e. $N_{3.j}$ contains $3$ RMTs. However, all the nodes of level $n$, i.e. $N_{4.j}$, $0\leq j \leq 3^4-1$ do not contain $3$ RMTs; for example, the nodes $N_{4.5}, N_{4.7}, N_{4.20}$ etc. consist of $6$ RMTs, but the nodes $N_{4.1}, N_{4.6}, N_{4.9}, N_{4.16}$ etc. are empty. 
So, the CA does not satisfy Corollary~\ref{Chap:reversibility:revcor1} and it is irreversible for length $4$.
\end{example}

Depending on the theoretical background developed in this section, we now test reversibility of $d$-state CAs in the next section.

 \section{Decision Algorithm for Testing Reversibility}
\label{chap:reversibility:Sec:bij}

The simplest approach of testing reversibility of an $n$-cell CA is, develop the reachability tree of the CA starting from root, and observe whether the reversibility conditions given by the theorems \ref{Chap:reversibility:revth1} and \ref{Chap:reversibility:revth2} are satisfied for the given rule or not. If there is any such node / edge that does not satisfy any of these conditions, then the CA is \textit{irreversible}, otherwise it is a \textit{reversible} CA. The problem of this approach is that if the CA is reversible then the tree grows exponentially, so when $n$ is not very small, it is difficult to handle the CA with this approach. However, we have following two observations - 

\begin{enumerate}

\item \label{pt1} If $N_{i.j} = N_{i.k}$ when $j \neq k$ for any $i$, then both the nodes are roots of two similar sub-trees. So, we can proceed with only one node. Similarly, if $l_{i.j} = l_{i.k} ~(j \neq k)$, then also we can proceed with only one edge.
\item \label{pt2} If $N_{i.j} = N_{i'.k}$ when $i>i' (0\leq i,i' \leq n-3)$, then the nodes that follow $N_{i'.k}$ are similar with the nodes followed by $N_{i.j}$. Therefore, we need not to explicitly develop the sub-tree of $N_{i.j}$. It is observed that after few levels, no unique node is generated. So, for arbitrary large $n$, it is not required to develop the whole tree.
\end{enumerate}

Following above two observations, we can develop \emph{minimized} reachability tree which does not grow exponentially. In fact, very few nodes are generated in such minimized reachability tree. To develop minimized reachability tree with only unique nodes, we need to put some extra links. For observation~\ref{pt1}, we exclude $N_{i.k}$ and add a link from the parent of $N_{i.k}$ to $N_{i.j}$. For observation~\ref{pt2}, we exclude $N_{i.j}$ and then form a link from the parent of $N_{i.j}$ to $N_{i'.k}$. In this case, a \emph{loop} is embodied between levels $i$ and $i'$. Presence of this loop can be signified by updating the level of the unique node; instead of a single level for each node, a node, now, may have a set of levels. For example, for observation~\ref{pt2}, we update the set of levels associated with node $N_{i'.k}$ as $\{i',i\}$. This loop implies that the node reappears at levels $i + (i - i')$, $i + 2(i-i')$, $i + 3(i-i')$, etc. (Strictly speaking, the minimized reachability tree is not a tree. In our further discussion, however, we call it as tree.) This minimized reachability tree is a directed graph, and the directions are necessary to reconstruct the original tree.

In a minimized reachability tree, a node, say $\mathcal{N}$ can be part of more than one loop, which implies that, $\mathcal{N}$ can appear at levels implied by each of the loops. However, if we observe in more detail, we can find that, although every loop confirms presence of $\mathcal{N}$, but all loops are not significant in the tree. For example, if $\mathcal{N}$ is part of a loop of length $2$ as well as a loop of length $4$, then for the loop of length $4$, the node will not appear in any extra levels than the loop of length $2$; that is, the loop of length $2$ is sufficient for affirming the levels in which $\mathcal{N}$ will appear. Similarly, if $\mathcal{N}$ appears in a loop of length $1$ (self-loop), then it will appear in every successive levels; that means, all other loops for this node will be irrelevant. In the same way, if a node has one loop of length $2$, and another loop whose length is an odd number, then from the last level of the second loop onwards, the node will be present in every level, that is, will behave as having a self-loop. Nonetheless, if we get two loops of length $l_1, l_2$ for a node with lengths of the loops $> 2$ and the lengths are mutually prime (that is, \emph{GCD($l_1,l_2$)} $= 1$), then both these loops are important for the presence of the node at certain levels; but if \emph{GCD($l_1,l_2$)} $> 1$, then none of the lengths will remain relevant and new loop length will be the \emph{GCD} value. In this way, we can find some loops which are important for a node and some loops which are not; the loops that are not important for a node can be discarded. We can also observe that, if $\mathcal{N}$ is in a loop and present in level $i$, then the children of $\mathcal{N}$ are also involved in the loop and always be present in level $i+1$. This implies, whenever $\mathcal{N}$ appears in more than one level, then all the nodes of the sub-tree rooted at $\mathcal{N}$ also appear in levels updated according to levels of $\mathcal{N}$. Similarly, if a node is in self-loop, then the whole sub-tree with that node as root will also have self-loop, that is, will be generated in every level.

\begin{example}	
	The minimized reachability tree of $2$-state CA with rule $01001011$ is shown in Figure~\ref{Chap:reversibility:fig:rt3}. In this figure, the tree has $21$ nodes and last unique node is at level $5$. Every node has $2$ edges, labeled by $0$ and $1$ respectively and a set of levels from which the node was referred. For example, $\{1,3\}$ associated to $N_{1.0}$ implies that, this node has been referred in levels $1$ and $3$ respectively and is part of a loop of length $2$. Directed line (link) from one node to another implies, child of the first node is a duplicate node equivalent to the second node.

	\begin{figure}[hbtp]
		 \centering
			\includegraphics[width= 4.5in, height = 3.7in]{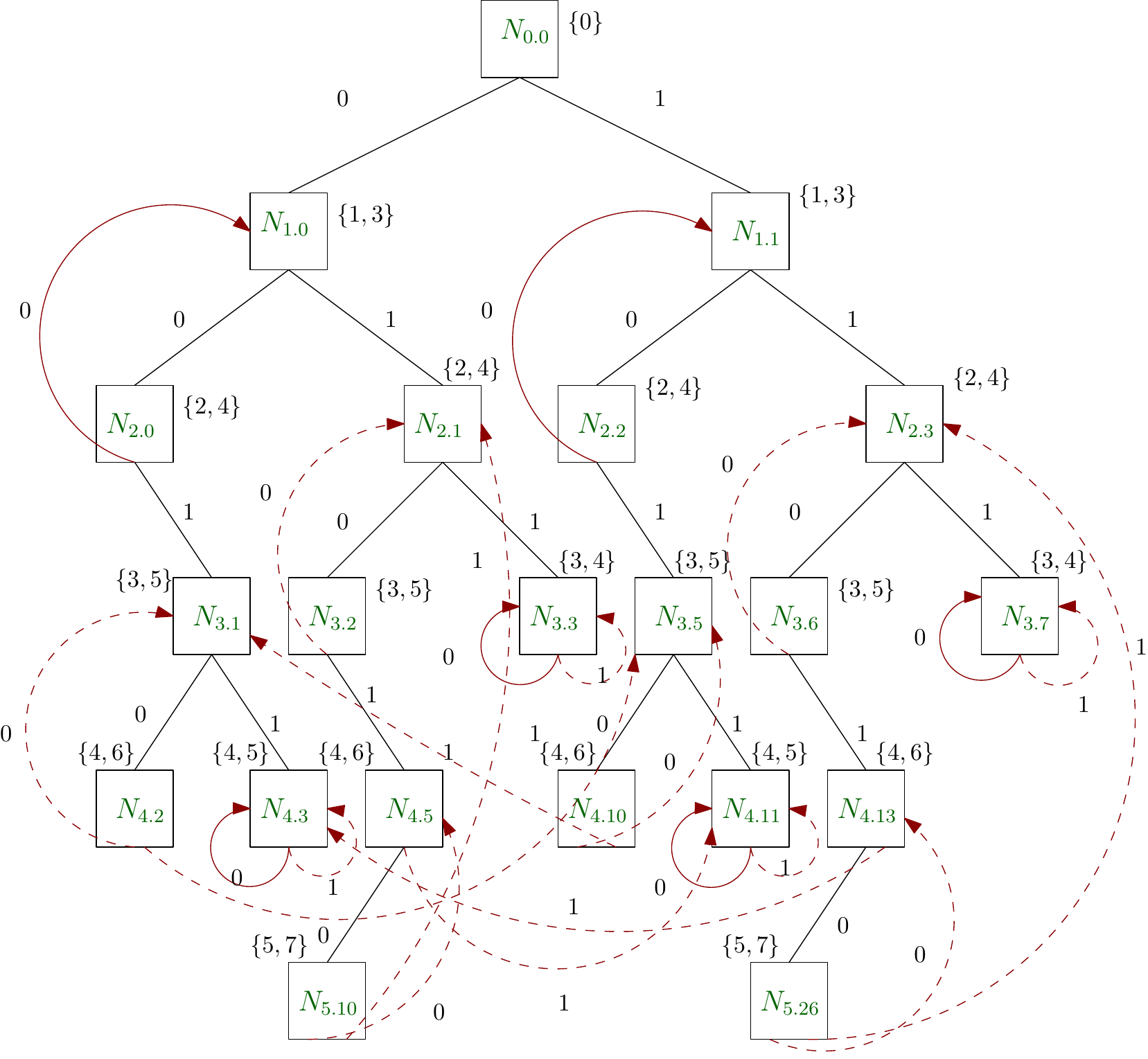}
			\caption{Minimized Reachability tree of $2$-state CA with rule $01001011$}
			\label{Chap:reversibility:fig:rt3}
	\end{figure}

It can be noticed that, although a node is connected with several loops, many of them are not important. For instance, the node $N_{2.1}$ is child of $N_{1.0}$, whose set of levels is updated as $\{1,3\}$ by the link from node $N_{2.0}$. So, as a child, set of levels of $N_{2.1}$ is also updated as $\{2,4\}$. But, this node is also part of two other loops, one from node $N_{3.2}$ and another from node $N_{5.10}$, which want to update its level by $4$ and $6$ respectively. As, $4$ is already present in its set of levels, so the loop from node $N_{3.2}$ is not relevant. Similarly, as the length of the loop from node $N_{5.10}$ is $4$ and length of previous loop is $2$, so, the new loop becomes insignificant and level $6$ is also not added in set of levels of $N_{2.1}$. Observe that, the set of levels for a node is updated only for the relevant loops in the tree. The loops which are not important, are shown in dashed line in Figure~\ref{Chap:reversibility:fig:rt3}. 
	
For any node, self loops always get priority over other loops. For example, set of levels of the node $N_{3.3}$ was $\{3,5\}$ as a child of node $N_{2.1}$. But, when this node gets its self-loop, the set of levels is updated as $\{3,4\}$, that is, previous loop of length $2$ is dominated by the self-loop of length $1$. It can also be noticed that, for many of the nodes, first loop is prevailed and other loops become insignificant.
\end{example}

We can find the possible nodes of an arbitrary level ($p$) from the minimized reachability tree. If a node appears only in level $i$, then the node can not appear in level $p$ $(p > i)$. On the other hand, if a node of the minimized reachability tree appears in level $i$, as well as in level $i'$ (that is, length of the loop is $i-i'$), and if $p - i' \equiv 0 \pmod{(i-i')}$ $(p > i > i')$, then the node is present at level $p$. Since the nodes of level $n-2$ (also of level $n-1$) are special in the reachability tree, we can find the possible nodes of level $n-3$ using this technique, and can then get the nodes of level $n-2$.

In fact, we can verify whether a node belongs to level $n-2$ or level $n-1$ directly in advance - whenever the set of levels of a node, say $\mathcal{N}$ is updated and has multiple elements, using the above technique check whether the node is part of level $n-2$ or level $n-1$. If $\mathcal{N}$ is part of level $n-2$, then use the following set operation:
$\Gamma_{k}^{\mathcal{N'}} \leftarrow \Gamma_{k }^{\mathcal{N}} \cap \lbrace i, i+d, i+2d, \cdots, i+(d^2-1)d \rbrace$, where $ i = \floor{\frac{k}{d}}$ and $ 0 \leq k \leq d^2-1$ (see Point~\ref{Chap:reversibility:def:tree_3:rtd5} of Definition~\ref{Chap:reversibility:def:tree_3}) and if $\mathcal{N}$ is part of level $n-1$, then use the following set operation: 
$\Gamma_{k}^{\mathcal{N''}} \leftarrow \Gamma_{k }^{\mathcal{N}} \cap \lbrace  k + i \times d^2 ~|~ 0 \leq i \leq d-1 \rbrace, { 0 \leq k \leq d^2-1}$ (see Point~\ref{Chap:reversibility:def:tree_3:rtd6} of Definition~\ref{Chap:reversibility:def:tree_3}). Now, we can verify whether the nodes ${(\Gamma_{k}^{\mathcal{N'}})}_{0\leq k \leq d^2-1}$ and ${(\Gamma_{k}^{\mathcal{N''}})}_{0\leq k \leq d^2-1}$ obey the conditions of reversibility given by Corollary~\ref{Chap:reversibility:revcor1} and Lemma~\ref{Chap:reversibility:revcor2}.
Advantage of this procedure is, we can detect many balanced rules, which violate reversibility property for nodes of level $n-2$ or level $n-1$, at the first occurrence of such node.
However, if $n$ is too small, then, we need to have the remaining nodes of level $n-2$ from the unique nodes generated from level $n-3$.

The proposed algorithm (\emph{CheckReversible}) develops the minimized reachability tree and stores the unique nodes of the tree. If any of the nodes is unbalanced (Lemma~\ref{Chap:reversibility:revcor2}) or does not follow the conditions of Corollary~\ref{Chap:reversibility:revcor1}, the CA is reported as irreversible. The algorithm uses two data structures - \emph{NodeList} to store the unique nodes and \emph{NodeLevel} to store the level number(s) of the nodes. Each of the nodes of \emph{NodeList} is also associated with a flag - \emph{selfLoop}, which is set when the node has self loop. The algorithm also uses some variables-- $uId$ as index of \emph{NodeList}, $i$ as the current level of the tree and $p$ as the parent node. As input, it (Algorithm~\ref{chap:reversibility:algo:rev_algo}) takes a $d$-state CA rule and  $n \geq 3$ as the number of cells and outputs $``$Irreversible'' if the $n$-cell CA is irreversible, and $``$Reversible'' otherwise. The algorithm uses following two procedures. The steps of these procedures are not shown in detail in this presentation. 
\begin{description}
	\item[	\emph{verifyLastLevels()}] As argument it accepts a node, and then checks whether it can exist at level $n-2$ or $n-1$. If yes, based on the above mentioned logic the procedure decides whether this presence can make the CA irreversible.
	
	\item[	\emph{updateSubTree()}] This procedure updates a sub-tree when a loop is formed. The update includes the modification of \emph{NodeLevel} of each node of the sub-tree based on the logic presented above. During the update of \emph{NodeLevel} of each node, the procedure \emph{verifyLastLevels()} is called to see if the node can be present at level $n-2$ or $n-1$. As argument, the procedure takes the $uId$ of the node which is the root of the sub-tree.
\end{description}

In the beginning, the algorithm checks whether the input CA is balanced. If not, it decides the CA as irreversible (\ref{Chap:reversibility:algo:st1} of Algorithm~\ref{chap:reversibility:algo:rev_algo}). Otherwise, the root of the reachability tree is formed, and we set \emph{NodeList}[$0$] $\gets$ root, \emph{NodeLevel}[$0$] $ \gets \lbrace 0 \rbrace$ (\ref{Chap:reversibility:algo:st2}).  Then we find the nodes of the next level. If the nodes are unique, they are added to \emph{NodeList}. Otherwise, \emph{NodeLevel} of each node in the sub-tree rooted at the matched node of \emph{NodeList} is updated (\ref{Chap:reversibility:algo:st3}).

In \ref{Chap:reversibility:algo:st3}, the main step in this algorithm, first $d$ nodes of level $1$ are formed. If any of the nodes is similar to the root, the node is dropped and it is checked whether the new loop is valid or not. As \emph{NodeLevel}[$0$] has no existing loop ($\mid NodeLevel[0]\mid = 1$), so, it is set to $\lbrace 0, 1 \rbrace$, and \emph{selfLoop} flag associated with \emph{NodeList}[$0$] is set to $true$. This means, \emph{NodeList}[$0$] appears in level $0$, and level $1$ as well. 
Now, the procedure \emph{updateSubtree()} is called with argument $0$.

Here, the existing sub-tree of \emph{NodeList}[$0$], say, \emph{NodeList}[$1$] in this case, is updated according to levels of its parent \emph{NodeList}[$0$]. That means, \emph{NodeLevel}[$1$] is updated as  $\lbrace 1, 2 \rbrace$ and \emph{selfLoop} flag of \emph{NodeList}[$1$] is set to $true$. If both \emph{NodeList}[$0$] and \emph{NodeList}[$1$] satisfy the conditions of Corollary~\ref{Chap:reversibility:revcor1} and Lemma~\ref{Chap:reversibility:revcor2} for levels $n-2$ and $n-1$, we proceed to find the nodes of next level. To get them, we use the unique nodes of the previous level (\ref{Chap:reversibility:algo:st3}).

At any point of time we get a duplicate node, it is first decided whether the new loop is a relevant one; if it is not relevant, no action is required, otherwise \emph{updateSubTree()} is called. It may be observed that, a new unique node can also be part of a loop, if its parent has loop(s). So, for each new unique node, its \emph{NodeLevel} is updated by its parent's \emph{NodeLevel} and if it has loop,  \emph{verifyLastLevels()} is called with the new $uId$ to ensure early detection of irreversibility. If no unique node is found to add in the \emph{NodeList} in a level, we conclude that the minimized reachability tree is formed (\ref{Chap:reversibility:algo:st4}). The number of unique nodes is stored in \emph{uId}. As, for the minimized reachability tree, reversibility conditions are already asserted, the CA is declared as $``$Reversible'' (\ref{st8}).

However, for small $n$, the tree may not be completely minimized in \ref{Chap:reversibility:algo:st4}, i.e unique nodes may be generated up to level $n-2$. So, to get the remaining nodes of level $n-2$, we first find the unique nodes $(\mathcal{N})$ of level $n-2$ from the minimized reachability tree, and then use the operation $\Gamma_{k}^{\mathcal{N'}} \gets \Gamma_{k }^{\mathcal{N}} \cap \lbrace i, i+d, i+2d, \cdots, i+(d^2-1)d \rbrace$ ($ i = \floor{\frac{k}{d}}$ and $ 0 \leq k \leq d^2-1$) to get the actual nodes for level $n-2$ (\ref{Chap:reversibility:algo:st6}). Finally, we find the nodes of level $n-1$ from these nodes (\ref{Chap:reversibility:algo:st7}).

\begin{Walgo}[hbtp]{1.75cm}

	\BlankLine
	\scriptsize
	\SetKw{Fn}{Procedure}
	\SetKwFunction{FindGCD}{findGCD}
	\SetKwFunction{verifyForLastLevels}{verifyLastLevels}
	\SetKwFunction{updateSubTree}{updateSubTree}
	\SetKwInOut{Input}{Input}
	\SetKwInOut{Output}{Output}

	\Input{A $d$-state CA rule, $n$ (Number of cells)}
	\Output{\textit{Reversible} or \textit{Irreversible}}
	
	\rule[4pt]{0.95\textwidth}{0.95pt}\\
		\hspace{0.04\textwidth} \nlset{Step 1} Check whether the CA rule is balanced or not \; \label{Chap:reversibility:algo:st1} 
		\hspace{0.04\textwidth}	\lIf{ CA is not balanced }{
			Report $``$\textit{Irreversible}'' and
			$return$
		}
		
		\hspace{0.04\textwidth} \nlset{Step 2}	\label{Chap:reversibility:algo:st2} Form the root of the reachability tree \; 
		\hspace{0.04\textwidth}	$NodeList[0] \gets$ root, $NodeLevel[0] \gets \lbrace 0 \rbrace$ \;
		\hspace{0.04\textwidth}	Set $i \gets 1$, $uId \gets 0$, $s \gets 0$, $j \gets 0$,  $tuId \gets 0$ \;		
		\hspace{0.04\textwidth}	\nlset{Step 3}\label{Chap:reversibility:algo:st3} \For {$p = s$ to $j$}{ 
			Get the children of $NodeList[p]$ \;
			\For {each child $\mathcal{N}$ of $NodeList[p]$}{
				\If{$i \neq n-2$}{
					\If{ ($\mathcal{N}$ is not balanced) OR ($\mid \bigcup\limits_{0\leq x \leq d^2-1} {\Gamma_x}^{\mathcal{N}}\mid \neq d^3 $)}{
						Report $``$\textit{Irreversible}'' and
						$return$ \;
					}
				}
			\eIf{ $\mathcal{N}$ matches with $NodeList[k]$}{
				\eIf{$\mid NodeLevel[k] \mid=1$ AND $i \notin NodeLevel[k]$ \tcp{node is referred for the first time}} 
				{
					Set	$NodeLevel[k] \leftarrow NodeLevel[k] \cup \lbrace i \rbrace$ \;
					\lIf{loop is self-loop}{
						Set	$NodeList[k].selfLoop \leftarrow true$
						}
						\updateSubTree{$k$} ; \tcp{update sub-tree adding the new loop}
				}
				{
					Set $loopFlag \leftarrow false$; \tcp{checks whether old loop value remains important}
					\If{$NodeList[k].selfLoop = false$ AND $i \notin NodeLevel[k]$}{
						Set $newLoop \gets i-\min(NodeLevel[k])$ \;
						\eIf{$newLoop = 1$ \tcp{new self-loop detected}}{
							Set $NodeLevel[k] \leftarrow \lbrace (i-1), i \rbrace$,
							$loopFlag \leftarrow true$ \;	
							Set	$NodeList[k].selfLoop \leftarrow true$ \;
							\updateSubTree{$k$} ; \tcp{update sub-tree by the new loop}
						}{
						\ForEach{$ l \in NodeLevel[k]$}{
							Set $oldLoop \gets l-\min(NodeLevel[k])$ \;
							Set $gcd \leftarrow$ \textit{GCD}($oldLoop,newLoop$) \;
							\If{$gcd = oldLoop$  \tcp{that is, old loop value prevailed}}{
								Set $loopFlag \leftarrow true$ and \textbf{break} ; \tcp{new loop is not relevant} 				
							}
							\ElseIf{$oldLoop = 2$ AND $gcd = 1$ \tcp{that is, valid loop length $ = 1$}}{
								Set $NodeLevel[k] \leftarrow \lbrace (i-1), i \rbrace$,
								$loopFlag \leftarrow true$ \;	
								Set $NodeList[k].selfLoop \leftarrow true$ \;
								\updateSubTree{$k$} and \textbf{break} \;
							}
							\ElseIf{$gcd >1$ \tcp{that is, updated valid loop length $ = gcd$}}{
								Set	$NodeLevel[k] \leftarrow \lbrace (i-gcd), i \rbrace$,
								$loopFlag \leftarrow true$ \;	
								\updateSubTree{$k$} and \textbf{break} \;  
							}
						}
						\If{$loopFlag = false$ \tcp{new loop is relevant}}{
							Set	$NodeLevel[k] \leftarrow NodeLevel[k] \cup \lbrace i \rbrace$ \;
							\updateSubTree{$k$} ; \tcp{update sub-tree adding the new loop}
						}
					}
				}
			}
		}	
		{
			Set	$uId \leftarrow uId + 1$, 
			$NodeList[uId] \leftarrow \mathcal{N}$ ; \tcp{add the unique node in the $NodeList$}
			\ForEach{$l \in NodeLevel[p]$}{
				Set	$NodeLevel[uId] \leftarrow \lbrace l+1 \rbrace$ ; \tcp{update child's level by parent's level}
			}
			\lIf{$NodeList[p].selfLoop = true$}{
				Set $NodeList[uId].selfloop \leftarrow true $
			}
			\If{$\mid NodeLevel[uId] \mid > 1$ \tcp{that is, the newly added unique node has a loop}}{
				\verifyForLastLevels{$uId$} \;
			}
		   }   
		 }
		}	
		\caption{\emph{CheckReversible}}
	\label{chap:reversibility:algo:rev_algo}
\end{Walgo}

\setcounter{algocf}{0}

\begin{Walgo}[hbtp]{1.75cm}
	\BlankLine
	\SetKw{Fn}{Procedure}
	\SetKwFunction{FindGCD}{findGCD}
	\SetKwFunction{verifyReversibilityOfLastLevels}{verifyLastLevels}
	\SetKwFunction{updateSubTree}{updateSubTree}
	\scriptsize
			\hspace{0.04\textwidth} \nlset{Step 4}	\label{Chap:reversibility:algo:st4} If $j = uId$, that is, no unique node is generated in \ref{Chap:reversibility:algo:st3}, go to \ref{st8} \; 
			\hspace{0.04\textwidth} \nlset{Step 5}	\label{Chap:reversibility:algo:st5} \If{ $i <n-2$}{ 
				$s \gets j + 1 $; $j \gets uId$;  $i \gets i+1$ \;
				go to \ref{Chap:reversibility:algo:st3} \;}

			\hspace{0.04\textwidth} \nlset{Step 6}	\label{Chap:reversibility:algo:st6} \For {$p = j+1$ to $uId$}	{
				$\mathcal{N'} \gets NodeList[p]$ \;
				$\Gamma_{x}^{\mathcal{N'}} \gets \Gamma_{x }^{\mathcal{N'}} \cap \lbrace i, i+d, i+2d, \cdots, i+(d^2-1)d \rbrace$, where $ i = \floor{\frac{x}{d}}$ and $ 0 \leq x \leq d^2-1$ \; 
				\If{($\mathcal{N'}$ is not balanced) OR ($\mid \bigcup\limits_{0\leq x \leq d^2-1} {\Gamma_x}^{\mathcal{N'}}\mid \neq d^2 $)}{
					Report $``$\textit{Irreversible}'' and $return$ \;
				}
				$NodeListT[tuId] \gets \mathcal{N'}$ \;	  
				$tuId \gets tuId + 1$ \;
			}
			
			\hspace{0.04\textwidth}	\nlset{Step 7} \label{Chap:reversibility:algo:st7} Get the nodes of level $n-1$ (Point~\ref{Chap:reversibility:def:tree_3:rtd6} of Definition~\ref{Chap:reversibility:def:tree_3}) \; 
			\For{ any node $\mathcal{N''}$ of level $n-1$}{ 
				\If{($\mathcal{N''}$ is not balanced) OR ($\mid \bigcup\limits_{0\leq x \leq d^2-1} {\Gamma_x}^{\mathcal{N''}}\mid \neq d $)}{
					Report $``$\textit{Irreversible}'' and
					$return$ \;
				} 
			} 
			
			\hspace{0.04\textwidth}	\nlset{Step 8}  \label{st8} 			                 
			Report $``$\textit{Reversible}'' and
			$return$ \;
			\caption{\emph{CheckReversible} contd...}
			\label{chap:reversibility:algo:rev_algo_1}
			\end{Walgo}
Following examples illustrate the execution of Algorithm~\ref{chap:reversibility:algo:rev_algo}.
\begin{example}
Let us consider a $2$-state CA $01001011$ with $n = 1001$ as input. The CA is balanced, so the root $N_{0.0}$ is added to \emph{NodeList} and $0$ is added to \emph{NodeLevel}[$0$]. Following our algorithm, we get $2$ nodes $N_{1.0}$ and $N_{1.1}$ at level $1$ (see Figure~\ref{Chap:reversibility:fig:rt3}), where $\bigcup\limits_{0 \leq k \leq 3} {\Gamma_k}^{N_{1.0}}$ = $\{0-8\}$ ($ {\Gamma_1}^{N_{1.0}}$ = $\{4,5\}, {\Gamma_2}^{N_{1.0}}$ = $\{0-3\}, {\Gamma_3}^{N_{1.0}}$ = $\{6,7\}$) and $\bigcup\limits_{0 \leq k \leq 3} {\Gamma_k}^{N_{1.1}}$ = $\{0-8\}$ ($ {\Gamma_0}^{N_{1.1}}$ = $\{0-3\}, {\Gamma_1}^{N_{1.1}}$ = $\{6,7\}, {\Gamma_3}^{N_{1.1}}$ = $\{4,5\}$). For each node, the sets whose contents are not mentioned, are empty. These nodes are unique and added to \emph{NodeList} and level $1$ is added to the corresponding \emph{NodeLevel}. $uId$, that is, index of \emph{NodeList} is now increased to $2$. The execution of the algorithm for this CA is shown in Table~\ref{ruleEx1}. In this table, first column represents level $i$, second column the current $uId$, third column content of $NodeList[uId]$ and the fourth column represents the $NodeLevel[uId]$. Other three columns are related to the loop; if $NodeLevel[uId]$ is associated with a new loop, fifth column is set to \emph{yes} and the nodes which are affected by this loop are listed in the sixth column. However, if $NodeLevel$ of a node gets a new loop because of its parent node, the last column of Table~\ref{ruleEx1} represents the parent $uId$.

\begin{table}[!h]
\renewcommand{\arraystretch}{1.1}
\centering
\caption{Execution of Algorithm~\ref{chap:reversibility:algo:rev_algo} for $2$-state CA $01001011$ with $n = 1001$}
\label{ruleEx1}
\resizebox{1.0\textwidth}{9.1cm}{
\vspace{-\topsep} 
\begin{tabular}{|c|c|c@{\hspace{1em}}|c@{\hspace{-0.1em}}|c@{\hspace{-0.1em}}|c@{\hspace{-0.1em}}|c@{\hspace{-0.1em}}|}
\hline 
$i$ & $uId$ & $NodeList[uId]$ &\begin{tabular}{c}$NodeLevel$\\ $[uId]$ \end{tabular} & \begin{tabular}{c}Loop\\ Updated?\end{tabular} &  \begin{tabular}{c} Affected\\ $uId$(s) \end{tabular}  & \begin{tabular}{c} Affected\\ by $uId$ \end{tabular}\\ 
\hline 
$0$ & $0$ & $N_{0.0} = (\{0,1\}$, $\{2,3\}$, $\{4,5\}$, $\{6,7\})$ & $\lbrace 0 \rbrace$ & NA & NA & NA \\ 
\hline 

\multirow{2}{*}{$1$} & $1$ &  $N_{1.0}$ = $(\emptyset, \{4,5\}$, $\{0,1,2,3\}$, $\{6,7\})$ & $\lbrace1\rbrace$ & NA & NA & NA \\ 
\hhline{~------} 
& $2$ &  $N_{1.1} = (\{0,1,2,3\}$, $\{6,7\}, \emptyset, \{4,5\})$ & $\lbrace1\rbrace$ & NA & NA & NA \\ 
\hline 

\multirow{4}{*}{$2$} & $3$ & $N_{2.0}$ = $(\emptyset, \{0,1,2,3\}$, $\{4,5\}$, $\{6,7\})$  & $\lbrace2\rbrace$ & NA & NA & NA \\ 
\hhline{~------} 
& $4$ & $N_{2.1}$ = $(\emptyset, \emptyset, \{0,1,2,3,6,7\}$, $\{4,5\})$ & $\lbrace2\rbrace$ & NA & NA & NA \\ 
\hhline{~------} 
 & $5$ &  $N_{2.2} = (\{4,5\}$, $\{6,7\}, \emptyset, \{0,1,2,3\})$  & $\lbrace2\rbrace$ & NA & NA & NA \\ 
\hhline{~------} 
 & $6$ &  $N_{2.3} = (\{0,1,2,3,6,7\}$, $\{4,5\}, \emptyset, \emptyset)$  & $\lbrace2\rbrace$ & NA & NA & NA \\ 
\hline 

\multirow{12}{*}{$3$} & $1$ & $N_{3.0} \equiv N_{1.0} =(\emptyset, \{4,5\}$, $\{0,1,2,3\}$, $\{6,7\})$ & $\lbrace1, 3\rbrace$ & Yes & $3$, $4$ & NA\\ 
& $3$ & $N_{2.0}$ = $(\emptyset, \{0,1,2,3\}$, $\{4,5\}$, $\{6,7\})$  & $\lbrace2, 4\rbrace$ & Yes & NA & $1$ \\ 
& $4$ & $N_{2.1}$ = $(\emptyset, \emptyset, \{0,1,2,3,6,7\}$, $\{4,5\})$ & $\lbrace2, 4\rbrace$ & Yes & NA & $1$ \\ 
\hhline{~------} 
 & $7$ & $N_{3.1}$ = $(\emptyset, \{0,1,2,3,6,7\}, \emptyset, \{4,5\})$  & $\lbrace3, 5\rbrace$ & NA & NA & $3$ \\ 
\hhline{~------} 
 & $8$ &  $N_{3.2}$ = $(\emptyset, \emptyset, \{4,5,6,7\}$, $\{0,1,2,3\})$ & $\lbrace3, 5\rbrace$ & NA & NA & $4$ \\ 
\hhline{~------} 
 & $9$ &  $N_{3.3}$ = $(\emptyset, \emptyset, \{0,1,2,3,4,5,6,7\}, \emptyset)$  & $\lbrace3, 5\rbrace$ & NA & NA & $4$ \\ 
\hhline{~------} 
 & $2$ & $N_{3.4} \equiv N_{1.1} = (\{0,1,2,3\}$, $\{6,7\}, \emptyset, \{4,5\})$ & $\lbrace1, 3\rbrace$ & Yes & $5$, $6$ & NA\\ 
 & $5$ &  $N_{2.2} = (\{4,5\}$, $\{6,7\}, \emptyset, \{0,1,2,3\})$  & $\lbrace2, 4\rbrace$ & Yes & NA & $2$ \\ 
 & $6$ &  $N_{2.3} = (\{0,1,2,3,6,7\}$, $\{4,5\}, \emptyset, \emptyset)$  & $\lbrace2, 4\rbrace$ & Yes & NA & $2$ \\ 
\hhline{~------} 
 & $10$ & $N_{3.5}$ = $(\emptyset, \{4,5\}, \emptyset,\{0,1,2,3,6,7\})$ & $\lbrace3, 5\rbrace$ & NA & NA & $5$ \\ 
\hhline{~------} 
 & $11$ & $N_{3.6} = (\{4,5,6,7\}$, $\{0,1,2,3\}, \emptyset, \emptyset)$ & $\lbrace3, 5\rbrace$ & NA & NA & $6$\\ 
\hhline{~------} 
 & $12$ & $N_{3.7} = (\{0,1,2,3,4,5,6,7\}, \emptyset, \emptyset, \emptyset)$ & $\lbrace 3, 5\rbrace$ & NA & NA & $6$\\ 
\hline 

\multirow{12}{*}{$4$} & $13$ & $N_{4.2}$ = $(\emptyset, \{4,5,6,7\}, \emptyset, \{0,1,2,3\})$ & $\lbrace4, 6\rbrace$ & NA & NA & $7$\\ 
\hhline{~------} 
 & $14$ & $N_{4.3}$ = $(\emptyset, \{0,1,2,3,4,5,6,7\}, \emptyset, \emptyset)$ & $\lbrace4, 6\rbrace$ & NA & NA & $7$\\ 
\hhline{~------} 
 & $4$ & $N_{4.4} \equiv N_{2.1}$ = $(\emptyset, \emptyset, \{0,1,2,3,6,7\}$, $\{4,5\})$ & $\lbrace2,4\rbrace$ & No & NA & NA\\ 
\hhline{~------} 
 & $15$ & $N_{4.5}$ = $(\emptyset, \emptyset, \{4,5\}$, $\{0,1,2,3,6,7\})$ & $\lbrace4, 6\rbrace$ & NA & NA & $8$\\ 
\hhline{~------} 
 & $9$ & $N_{4.6} \equiv N_{3.3}$ = $(\emptyset, \emptyset, \{0,1,2,3,4,5,6,7\}, \emptyset)$ & $\lbrace3,4\rbrace$ & Yes & $9$ & $9$\\ 
 & $9$ &  $N_{4.7} \equiv N_{3.3}$ = $(\emptyset, \emptyset, \{0,1,2,3,4,5,6,7\}, \emptyset)$ & $\lbrace3,4 \rbrace$ & No & NA & NA\\ 
\hhline{~------} 
 & $16$ &  $N_{4.10}$ = $(\emptyset, \{0,1,2,3\}, \emptyset, \{4,5,6,7\})$ & $\lbrace 4, 6\rbrace$ & NA & NA & $10$\\ 
\hhline{~------} 
 & $17$ & $N_{4.11}$ = $(\emptyset, \emptyset, \emptyset, \{0,1,2,3,4,5,6,7\})$ & $\lbrace4, 6\rbrace$ & NA & NA & $10$\\ 
\hhline{~------} 
 & $6$ &  $N_{4.12} \equiv N_{2.3} = (\{0,1,2,3,6,7\}$, $\{4,5\}, \emptyset, \emptyset)$ & $\lbrace2,4\rbrace$ & No & NA & NA\\ 
\hhline{~------} 
 & $18$ & $N_{4.13} = (\{4,5\}$, $\{0,1,2,3,6,7\}, \emptyset, \emptyset)$ & $\lbrace4, 6\rbrace$ & NA & NA & $11$\\ 
\hhline{~------} 
 & $12$ & $N_{4.14} \equiv N_{3.7} = (\{0,1,2,3,4,5,6,7\}, \emptyset, \emptyset, \emptyset)$ & $\lbrace3,4\rbrace$ & Yes & $12$ & $12$\\ 
 & $12$ & $N_{4.15} \equiv N_{3.7} = (\{0,1,2,3,4,5,6,7\}, \emptyset, \emptyset, \emptyset)$ & $\lbrace3,4\rbrace$ & No & NA & NA \\ 
\hline 

\multirow{12}{*}{$5$} & $7$  &  $N_{5.4} \equiv N_{3.1}$ = $(\emptyset, \{0,1,2,3,6,7\}, \emptyset, \{4,5\})$  & $\lbrace3,5\rbrace$  & No & NA & NA\\ 
\hhline{~------}  
 & $10$  &  $N_{5.5} \equiv N_{3.5}$ = $(\emptyset, \{4,5\}, \emptyset,\{0,1,2,3,6,7\})$ & $\lbrace3,5\rbrace$ & No & NA & NA \\ 
\hhline{~------} 
 & $14$  & $N_{5.6} \equiv N_{4.3}$ = $(\emptyset, \{0,1,2,3,4,5,6,7\}, \emptyset, \emptyset)$ & $\lbrace4,5\rbrace$ & Yes & $14$ & $14$ \\ 
 & $14$  & $N_{5.7} \equiv N_{4.3}$ = $(\emptyset, \{0,1,2,3,4,5,6,7\}, \emptyset, \emptyset)$ & $\lbrace4,5\rbrace$  & No & NA & NA\\ 
\hhline{~------} 
 & $19$  & $N_{5.10}$ = $(\emptyset, \emptyset, \{0,1,2,3\}$, $\{4,5,6,7\})$ & $\lbrace5, 7\rbrace$ & NA & NA & $15$ \\ 
\hhline{~------} 
 & $17$  & $N_{5.11} \equiv N_{4.11}$ = $(\emptyset, \emptyset, \emptyset, \{0,1,2,3,4,5,6,7\})$ & $\lbrace4,5\rbrace$ & Yes & NA & NA \\ 
\hhline{~------} 
 & $10$  & $N_{5.20} \equiv N_{3.5}$ = $(\emptyset, \{4,5\}, \emptyset,\{0,1,2,3,6,7\})$ & $\lbrace3,5\rbrace$ & No & NA & NA\\ 
\hhline{~------} 
 & $7$  & $N_{5.21} \equiv N_{3.1}$ = $(\emptyset, \{0,1,2,3,6,7\}, \emptyset, \{4,5\})$ & $\lbrace 3,5\rbrace$ & No & NA & NA \\ 
\hhline{~------} 
 & $17$  & $N_{5.22} \equiv N_{4.11}$ = $(\emptyset, \emptyset, \emptyset, \{0,1,2,3,4,5,6,7\})$ & $\lbrace4,5\rbrace$ & No & NA & NA \\ 
\hhline{~------} 
 & $17$  & $N_{5.23} \equiv N_{4.11}$ = $(\emptyset, \emptyset, \emptyset, \{0,1,2,3,4,5,6,7\})$ & $\lbrace4,5\rbrace$ & No & NA & NA \\ 
\hhline{~------} 
 & $20$  & $N_{5.26} = (\{0,1,2,3\}$, $\{4,5,6,7\}, \emptyset, \emptyset)$  & $\lbrace5, 7\rbrace$ & NA & NA & $18$ \\ 
\hhline{~------} 
 & $14$  & $N_{5.27} \equiv N_{4.3}$ = $(\emptyset, \{0,1,2,3,4,5,6,7\}, \emptyset, \emptyset)$ & $\lbrace4,5\rbrace$ & No & NA & NA \\ 
\hline 
 
\multirow{4}{*}{$6$}  &  $15$ &  $N_{6.20} \equiv N_{4.5}$ = $(\emptyset, \emptyset, \{4,5\}$, $\{0,1,2,3,6,7\})$  & $\lbrace4,6\rbrace$ & No & NA & NA\\ 
\hhline{~------}  
 &  $4$ & $N_{6.21} \equiv N_{2.1}$ = $(\emptyset, \emptyset, \{0,1,2,3,6,7\}$, $\{4,5\})$ & $\{2,4\}$ & No & NA & NA \\ 
\hhline{~------}  
 &  $18$ & $N_{6.52} \equiv N_{4.13} = (\{4,5\}$, $\{0,1,2,3,6,7\}, \emptyset, \emptyset)$ & $\{4,6\}$ & No & NA & NA \\ 
\hhline{~------}   
 &  $6$ & $N_{6.53} \equiv N_{2.3} = (\{0,1,2,3,6,7\}$, $\{4,5\}, \emptyset, \emptyset)$ & $\{2,4\}$ & No & NA & NA \\ 
\hline 
\end{tabular}
}
\end{table}

From Table~\ref{ruleEx1}, it can be seen that, at level $2$, all nodes are unique and added to \emph{NodeList}. At level $3$, however, $N_{3.0} \equiv N_{1.0}$ and $N_{3.4} \equiv N_{1.1}$; these two loops are valid and accordingly \emph{NodeLevel} of $6$ existing nodes are updated. Moreover, $6$ new unique nodes are also added in this level. As reversibility conditions are sustained for all these nodes, so, the algorithm proceeds to the next level.

At level $4$ also, $6$ unique nodes are added to \emph{NodeList}. As, each of these nodes has multiple levels in their \emph{NodeLevel}, so, each is verified for the reversibility conditions at levels $n-2$ and $n-1$. Among the duplicate nodes, new loops for nodes $N_{2.1}$ and $N_{2.3}$ are not relevant, so, \emph{NodeLevel}[$4$] and \emph{NodeLevel}[$6$] remain unchanged. But, \emph{NodeLevel}[$9$] and \emph{NodeLevel}[$12$] are updated with levels of their new loop value. These nodes have no sub-tree to update. $N_{3.3}$ and $N_{3.7}$ also assert reversibility conditions, so, the algorithm continues to move forward.

At the next level, only $2$ unique nodes are added to \emph{NodeList}. However, among the duplicate nodes only $N_{4.3}$ and $N_{4.11}$ have updated their \emph{NodeLevel}. 

At level $6$, no new unique node is generated, as well as, no new relevant loop is found for the duplicate nodes. So, the algorithm jumps to \ref{st8}. The minimized tree for the CA is shown in Figure~\ref{Chap:reversibility:fig:rt3}. The tree has only $21$ nodes, that is, number of unique nodes ($M$) generated by the algorithm is $21$. For every loop of Figure~\ref{Chap:reversibility:fig:rt3}, the corresponding nodes satisfy reversibility conditions, so, the CA is declared as reversible for $n=1001$.
\end{example}

\begin{example}
Let us take a $3$-state CA $102012120012102120102102120$ with $n = 555$ as input. This CA is also balanced, so the root $N_{0.0}$ is added to \emph{NodeList} and $0$ is added to \emph{NodeLevel}[$0$]. Execution of Algorithm~\ref{chap:reversibility:algo:rev_algo} for this CA is shown in Table~\ref{ruleEx2}.

\begin{table}[!h]
\renewcommand{\arraystretch}{1.45}
\caption{Execution of Algorithm~\ref{chap:reversibility:algo:rev_algo} for $3$-state CA $102012120012102120102102120$ with $n = 555$}
\label{ruleEx2}
\centering
\resizebox{1.0\textwidth}{7.5cm}{
\small{
\begin{tabular}{|c|c|p{11cm}|c|}
	\hline
	\multicolumn{1}{|c|}{$i$}&
	\multicolumn{1}{c|}{$uId$}&
	\multicolumn{1}{c|}{$NodeList[uId]$}&
	\multicolumn{1}{c|}{$NodeLevel$}\\\hline
$0$ & $0$ & $N_{0.0} = (\{0-2\}$, $\{3-5\}$, $\{6-8\}$, $\{9-11\}$, $\{12-14\}$, $\{15-17\}$, $\{18-20\}$, $\{21-23\}$, $\{24-26\})$ & $\lbrace 0 \rbrace$ \\ 
\hline 

\multirow{3}{*}{$1$} & $1$ &  $N_{1.0} =  (\{0-2\}$, $\{ 12-14\}$, $\{21-23 \}$, $\{ 0-2 \}$, $\{12-14\}$, $\{24-26\}$, $\{0-2\}$, $\{15-17\}$, $\{21-23\})$ & $\lbrace 1 \rbrace$ \\ 
\hhline{~---} 
& $2$ &  $N_{1.1} =  (\{6-8\}$, $\{15-17\}$, $\{24-26\}$, $\{6-8\}$, $\{15-17\}$, $\{21-23\}$, $\{6-8\}$, $\{12-14\}$, $\{24-26\})$ & $\lbrace1\rbrace$ \\ 
\hhline{~---} 
& $3$ &  $N_{1.2} =  (\{3-5\}$, $\{9-11\}$, $\{18-20\}$, $\{3-5\}$, $\{9-11\}$, $\{18-20\}$, $\{3-5\}$, $\{9-11\}$, $\{18-20\})$ & $\lbrace1\rbrace$ \\ 
\hline 

\multirow{9}{*}{$2$} & $4$ &  $N_{2.0} =  (\{0-2\}$, $\{ 12-14\}$, $\{15-17 \}$, $\{ 0-2 \}$, $\{12-14\}$, $\{21-23\}$, $\{0-2\}$, $\{24-26\}$, $\{15-17\})$ & $\lbrace 2 \rbrace$ \\ 
\hhline{~---} 
 &  $5$ &  $N_{2.1} =  (\{6-8\}$, $\{15-17\}$, $\{ 12-14\}$, $\{6-8\}$, $\{15-17\}$, $\{24-26\}$, $\{6-8\}$, $\{21-23\}$, $\{12-14\})$ & $\lbrace 2 \rbrace$ \\  
\hhline{~---} 
 &  $6$ &  $N_{2.2} =  (\{3-5\}$, $\{9-11\}$, $\{9-11\}$, $\{3-5\}$, $\{9-11\}$, $\{18-20\}$, $\{3-5\}$, $\{18-20\}$, $\{9-11\})$ & $\lbrace 2 \rbrace$ \\ 
\hhline{~---} 
 &  $7$ &  $N_{2.3} =  (\{21-23\}$, $\{24-26\}$, $\{21-23\}$, $\{21-23\}$, $\{24-26\}$, $\{15-17\}$, $\{21-23 \},  \{12-14\}$, $\{21-23 \})$ & $\lbrace 2 \rbrace$ \\  
\hhline{~---} 
 &  $8$ &  $N_{2.4} =  (\{24-26\}$, $\{21-23\}$, $\{24-26\}$, $\{24-26\}$, $\{21-23\}$, $\{12-14\}$, $\{24-26\}$, $\{15-17\}$, $\{24-26\})$ & $\lbrace 2 \rbrace$ \\  
\hhline{~---} 
 &  $9$ &  $N_{2.5} =  (\{18-20\}$, $\{18-20\}$, $\{18-20\}$, $\{18-20\}$, $\{18-20\}$, $\{9-11\}$, $\{18-20\}$, $\{9-11\}$, $\{18-20\})$ & $\lbrace 2 \rbrace$ \\  
\hhline{~---} 
 &  $10$ &  $N_{2.6} =  (\{ 12-14\}$, $\{0-2\}$, $\{ 0-2 \}$, $\{ 12-14\}$, $\{0-2\}$, $\{ 0-2 \}$, $\{ 12-14\}$, $\{0-2\}$, $\{ 0-2 \})$ & $\lbrace 2 \rbrace$ \\  
\hhline{~---}  
 &  $11$ &  $N_{2.7} =  (\{15-17\}$, $\{6-8\}$, $\{ 6-8\}$, $\{15-17\}$, $\{6-8\}$, $\{ 6-8\}$, $\{15-17\}$, $\{6-8\}$, $\{ 6-8\})$ & $\lbrace 2 \rbrace$ \\ 
\hhline{~---} 
 &  $12$ &  $N_{2.8} =  (\{9-11\}$, $\{ 3-5\}$, $\{3-5\}$, $\{9-11\}$, $\{ 3-5\}$, $\{3-5\}$, $\{9-11\}$, $\{ 3-5\}$, $\{3-5\})$ & $\lbrace 2 \rbrace$ \\ 
\hline 

\multirow{7}{*}{$3$} & $13$ &  $N_{3.0} =  (\{0-2\}$, $\{ 12-14\}$, $\{24-26\}$, $\{ 0-2 \}$, $\{12-14\}$, $\{15-17\}$, $\{0-2\}$, $\{21-23\}$, $\{24-26\})$ & $\lbrace 3 \rbrace$ \\ 
\hhline{~---} 
& $14$ &  $N_{3.1} =  (\{6-8\}$, $\{15-17 \}$, $\{21-23\}$, $\{6-8\}$, $\{15-17 \}$, $\{ 12-14\}$, $\{ 6-8\}$, $\{24-26\}$, $\{21-23\})$ & $\lbrace 3 \rbrace$ \\ 
\hhline{~---} 
& $15$ &  $N_{3.2} =  (\{3-5\}$, $\{9-11\}$, $\{18-20 \}$, $\{3-5\}$, $\{9-11\}$, $\{9-11\}$, $\{3-5\}$, $\{18-20 \}$, $\{18-20 \})$ & $\lbrace 3 \rbrace$ \\ 
\hhline{~---} 
 & $16$ &  $N_{3.3} =  (\{21-23\}$, $\{24-26\}$, $\{12-14\}$, $\{21-23\}$, $\{24-26\}$, $\{21-23\}$, $\{21-23\}$, $\{15-17\}$, $\{12-14\})$ & $\lbrace 3 \rbrace$ \\ 
\hhline{~---} 
& $17$ &  $N_{3.4} =  (\{24-26\}$, $\{21-23\}$, $\{15-17 \}$, $\{24-26\}$, $\{21-23\}$, $\{24-26\}$, $\{24-26\}$, $\{12-14\}$, $\{15-17\})$ & $\lbrace 3 \rbrace$ \\ 
\hhline{~---}  
& $18$ &  $N_{3.5} =  (\{18-20 \}$, $\{18-20 \}$, $\{9-11 \}$, $\{18-20 \}$, $\{18-20 \}$, $\{18-20 \}$, $\{18-20 \}$, $\{9-11 \}$, $\{9-11 \})$ & $\lbrace 3 \rbrace$ \\ 
\hhline{~---} 
 &$10$ &  $N_{3.6} \equiv N_{2.6} =  (\{ 12-14\}$, $\{0-2\}$, $\{ 0-2 \}$, $\{ 12-14\}$, $\{0-2\}$, $\{ 0-2 \}$, $\{ 12-14\}$, $\{0-2\}$, $\{ 0-2 \})$ & $\lbrace 2,3 \rbrace$ \\
\hline 
\end{tabular} 
}}
\end{table}

Following our algorithm, we get that, at level $1$, three nodes $N_{1.0}$, $N_{1.1}$ and $N_{1.2}$ are added to \emph{NodeList} and level $1$ is added to their corresponding \emph{NodeLevel}. At level $2$ also, all $9$ nodes are unique and added to \emph{NodeList}. $uId$ is increased to $12$.

At the next level, $6$ consecutive nodes, from $N_{3.0}$ to $N_{3.5}$ are unique and added to \emph{NodeList} by increasing $uId$ to $18$. However, $N_{3.6} \equiv N_{2.6}$, so, level $3$ is added to \emph{NodeLevel}[$10$], making a loop of length $1$. That means, this node is part of both the levels $n-2$ and $n-1$. But, after applying operation $\Gamma_{k}^{\mathcal{N'}} \leftarrow \Gamma_{k }^{\mathcal{N}} \cap \lbrace  \forall i, k + i \times d^2 ~|~ 0 \leq i \leq d-1 \rbrace, { 0 \leq k \leq d^2-1}$ (see Point~\ref{Chap:reversibility:def:tree_3:rtd6} of Definition~\ref{chap:reversibility:Sec:rtree}), the node $\mathcal{N'}$ does not remain balanced; which implies, it fails to satisfy reversibility conditions for level $n-1$. The algorithm, therefore, stops further processing and declares the CA as irreversible for $n= 555$. Number of unique nodes generated by the algorithm for this CA is $M=19$.
\end{example}

\noindent \textbf{Complexity:} Although Algorithm~\ref{chap:reversibility:algo:rev_algo} takes the cell length $n$ as input, its running time depends only on the unique nodes generated in the reachability tree (stored in \emph{NodeList}), which is a rule specific value. Let us consider the maximum number of unique nodes for the CA with number of cells $n$ is $M$. It may be mentioned here that, when $n$ is very small, $M$ increases with $n$. But, after a certain value of $n$, say $n_0$, the maximum number of unique nodes ($M$) possible in the reachability tree of a CA is independent of $n$, that is, when $n$ is not very small ($n>n_0)$, then $M$ does not depend on $n$. 

 So, execution time of the algorithm depends on \ref{Chap:reversibility:algo:st3}, where, for each node generated in the tree, first, it is checked whether the node is already present in \emph{NodeList} or not. If the node is already present, that is, a duplicate node, and the corresponding loop is a valid one, then, the level information of the loop is added to \emph{NodeLevel} of the matched node and levels of the whole sub-tree of that node are updated. The complexity of the algorithm depends on the total number of nodes visited / processed.

 According to the algorithm, total number of nodes generated for the construction of minimized tree is $d \times M$, as, for each node, $d$ number of children are generated. Now, for each node, the existing \emph{NodeList} is checked to find whether it is already present or not. If the node is unique, all the nodes of the \emph{Nodelist} are visited. But if it is duplicate, we stop at the matched index $k$ of \emph{NodeList} and update \emph{NodeLevel} of the nodes of the sub-tree of \emph{NodeList}[$k$], which obviously is stored from index $k$ onwards in the \emph{NodeList}. So, at maximum, for this duplicate node, the total \emph{NodeList} is visited. 
 However, to check whether a node already exists in the \emph{NodeList}, first node can be visited $dM-1$ times, the second node $dM-2$ times and so on.  As all loops are not relevant and the \emph{NodeList} is updated gradually, total cost of visiting nodes of the \emph{NodeList} 
$< (dM-1) + (dM-2) + \cdots + (dM-M)= dM^2 - \frac{M^2+M}{2} $.
Hence, complexity of Algorithm~\ref{chap:reversibility:algo:rev_algo} is $\mathcal{O}({dM^2})$.

\noindent\textbf{Remark:} Complexity is generally measured in terms of input parameters. Here, the maximum number of possible nodes $N_{i.j}$ is bounded by $({2^{d^2}})^{d^2} = 2^{d^4}$ (any number of sibling RMT sets out of total $d^2$ number of sibling RMT sets to be selected and placed in any number of sets out of total $d^2$ sets). 
Again, for a specific $d$ and cell length $n$, the reachability tree can have at most $d^{n+1}$ number of nodes. Hence, we have the relation, $M < \min ({2^{d^4}, d^{n+1}})$. Here, if $n$ is small, then $M$ is bounded by $d^{n+1}$, and if $n$ is large, then it is bounded by $2^{d^4}$. However, this is not a tight upper bound. Practically, $M$ is much less than $2^{d^4}$. We have showed the values of $M$ for different $d$-state rules in the tables \ref{chap:reversibility:tab:tstg1}, \ref{chap:reversibility:tab:tstg2} and \ref{chap:reversibility:tab:tstg3}. For example, in Table~\ref{chap:reversibility:tab:tstg1}, for the $3$-state CA rule $011101111102012000220220222$ with $n = 100001$, we observe $M = 910$ only, which is very very less than $2^{d^4} = 2^{81}$. It can also be observed that $M$ is rule dependent, and for a specific $d$, there is a sufficient lattice size $n_0$, after which no unique node is added for any $d$-state rule in the tree.
However, finding this tight upper bound of the sufficient lattice size $n_0$ is a future research problem.

\section{Identification of Reversible Cellular Automata}
\label{chap:reversibility:Sec:identify}

This section reports efficient ways of identifying a set of finite CAs reversible for some $n$. 
One can, however, intuitively design the following straight forward approach to get a set of reversible finite CAs of length $n$ - consider a set of CAs and then use our algorithm to select reversible CAs from the set. This trial-and-error approach is not practical, because total number of rules for $d$-state CAs is $d^{d^3}$ and most of the CAs are irreversible. So, it is very difficult to identify a number of reversible CAs.

Instead of considering  a set of arbitrary CAs, one can repeat the above procedure with balanced rules only, because unbalanced rules are always irreversible CAs (Theorem~\ref{Chap:reversibility:revth4}). However, the number of balanced $d$-state CA rules is $ \frac{d^3!}{(d^2!)^d } $ (the total number of arrangements of $d^3$ RMTs where $d$ groups of RMTs have same next state value with each group size $d^2$ (Definition~\ref{Def:balancedrule})), and the ratio of the balanced rules to total number of rules is $\frac{d^3!}{{{(d^2!)}^d}\times{d^{d^3}}}$. This ratio is quite little - for $3$-state CAs, it is $\approx 3\%$, for $4$-state CAs $\approx 0.2\%$ and for $5$-state CAs, it is $\approx 0.009\%$.
Even if we take only balanced rules, we find that most of the balanced rules are irreversible! 
To get a feel about how many balanced rules are reversible, we have arranged an experiment where we have randomly generated one hundred million balanced rules for $3$-state CAs and tested reversibility of those CAs by Algorithm~\ref{chap:reversibility:algo:rev_algo} with random cell length $n$.
And, we have observed that there are only three reversible CAs! A sample result of this experiment is given in Table~\ref{randomRules}. In this table, first column shows the cell length $n$ and the second column shows the rule. Here, both are generated randomly. Column $3$ of Table~\ref{randomRules} notes the number of unique nodes generated before deciding the CA as reversible/irreversible; whereas column $4$ shows the level of the last unique node.
Therefore, arbitrary choosing of balanced rules for testing reversibility is not very helpful. In this scenario, we take greedy approach to choose the balanced rules which are potential candidates to be reversible.

\begin{table}[hbtp]
\centering
		\caption{Sample of randomly generated balanced rules for $3$-state CAs}
		\label{randomRules}
		\resizebox{0.90\textwidth}{10.0cm}{
			\begin{tabular}{|c|c|c|c|c|}
				\hline 
				 $n$ & Rule & $M$ & Last Level & Reversible? \\ 
				\hline
				$180$ & $000102212200012012112121201$ & $1$ & $0$ & No \\
				\hline 
				$583$ & $011220101212120121202201000$ & $2$ & $1$ & No \\
				\hline 
				$636$ & $201212101021200020010212112$ & $2$ & $1$ & No \\
				\hline 
				$966$ & $120201201201120021012210210$ & $3280$ & $7$ & No \\
				\hline 
				$669$ & $102201121210021102202010021$ & $2$ & $1$ & No \\
				\hline 
				$888$ & $102200220002122010110211121$ & $2$ & $1$ & No \\
				\hline 
				$563$ & $001002120120210201221112021$ & $7$ & $2$ & No \\
				\hline
				$387$ & $102002202110121010102012221$ & $2$ & $1$ & No \\
				\hline 
				$13$ & $021022210022012201110110102$ & $4$ & $1$ & No \\
				\hline 
				$36$& $120210201102021210021102120 $ & $3273$ & $7$ & No\\
				\hline
				$946$ & $201022222121010111100202001$ & $1$ & $0$ & No \\
				\hline 
				$264$ & $000222202112020110112001121$ & $7$ & $2$ & No \\
				\hline 
				$467$ & $122010020120002012111221201 $ & $1$ & $0$ & No\\
				\hline
				$837$ & $210001111021121200222202010$ & $1$ & $0$ & No \\
				\hline 
				$162$ & $111212010002002122210210120$ & $4$ & $1$ & No \\
				\hline 
				$247$ & $102012201021020210121120120$ & $5$ & $2$ & No \\
				\hline 
				$931$ & $122011222011100101200222010 $ & $103$ & $4$ & No\\
				\hline
				$932$ & $010101212210201122012201020$ & $1$ & $0$ & No \\
				\hline 
				$277$ & $100001212012122010101102222$ & $1$ & $0$ & No \\
				\hline
				$221$ & $110202112021021220202110001 $ & $242$ & $5$ & No\\
				\hline
				$953$ & $012012120100201221210120201$ & $7$ & $2$ & No \\
				\hline
				$467$ & $212122221120210112001001000$ & $1041$ & $15$ & Yes\\
				\hline
				$939$ & $210102210120120120012201210$ & $109$ & $4$ & No \\
				\hline
				$753$ & $212201110121000102222020110$ & $1$ & $0$ & No \\
				\hline
				$282$ & $012222110210120100002210211$ & $1$ & $0$ & No \\
				\hline 
				$413$ & $012210101222011120120102002$ & $4$ & $1$ & No \\
				\hline
				$56$ & $112201202210012122001010102$ & $4$ & $1$ & No \\
				\hline
				$533$ & $111201011222022220000110102$ & $109$ & $4$ & No \\
				\hline
				$493$ & $120001012212021220010211021$ & $2$ & $1$ & No \\
				\hline
				$222$ & $110022220211121012000112020$ & $1$ & $0$ & No\\ 
				\hline 
				$251$ & $220101112021212100202200101 $ & $2$ & $1$ & No \\
				\hline	
				$991$ & $112020220210102010111020221$ & $1$ & $0$ & No \\
				\hline 
				$152$ & $102200212210001212211100012$ & $2$ & $1$ & No \\
				\hline
				$906$ & $110210101022022202112001102$ & $1$ & $0$ & No \\
				\hline 
				$641$ & $201200102101020222122011011$ & $1$ & $0$ & No \\
				\hline 
				$444$ & $000211122001210122001110222$ & $44$ & $4$ & No \\
				\hline 
				$927$ & $021211112020012011202220010$ & $2$ & $1$ & No \\
				\hline 
				$177$ & $020120101210122111222021000$ & $1$ & $0$ & No \\
				\hline 
				$297$ & $120012021102102210021120102$ & $364$ & $5$ & No \\
				\hline 
				$96$ & $100022110201211022201022110$ & $20$ & $3$ & No \\
				\hline 
				$728$ & $211011002111222022012200001$ & $1$ & $0$ & No \\
				\hline 
				$956$ & $112122000000122112222111000$ & $47$ & $4$ & No \\
				\hline 
				$59$ & $121202101120200010021221102$ & $2$ & $1$ & No \\
				\hline 
				$505$ & $011101021021200212120212002$ & $5$ & $2$ & No \\
				\hline 
				$299$ & $110011110021200201022221220$ & $4$ & $1$ & No \\
				\hline 
				$476$ & $120221022012101120020001112$ & $2$ & $1$ & No \\
				\hline
				$816$ & $210212110101012200002101222$ & $1$ & $0$ & No \\
				\hline 
				$17$ & $120012012012120102201201210$ & $9837$ & $8$ & No \\
				\hline 
				$898$ & $010102102020121021111022220 $ & $1$ & $0$ & No \\
				\hline 
				$611$ & $122210112212201021000011020 $ & $1$ & $0$ & No \\
				\hline
				$183$ & $021101012210201202210012210$ & $8$ & $2$ & No \\
				\hline
			\end{tabular}
			 }
	\end{table}

It is pointed out in Section~\ref{chap:reversibility:Sec:rev} that nodes of a reachability tree of a reversible CA are balanced (see Definition~\ref{Chap:reversibility:def:balancednode} and Lemma~\ref{Chap:reversibility:revcor2}). If a rule is balanced, the root which contains all RMTs of the rule, is also balanced. Our greedy approach is, choose the balanced rules in such a way that all the nodes up to level $n-3$ also remain balanced. Then, use Algorithm~\ref{chap:reversibility:algo:rev_algo} to test reversibility of the selected balanced rules. Success of this scheme, however, remains on how efficiently we are choosing the balanced rules.

We observe that the equivalent RMTs result in a same set of (sibling) RMTs at next level (see Section~\ref{chap:reversibility:Sec:CAbasic}). For example, in a $3$-state CA, RMT $0$ and RMT $9$ are equivalent to each other and both of them produce RMTs $0, 1$ and $2$ in next level (see Table~\ref{Chap:reversibility:tab:rln}). We exploit this property to develop our first greedy strategy. 
Let us recall that, $Equi_i = \{i, d^2+i, 2d^2+i,..., (d-1)d^2+i\}$ is a set of equivalent RMTs where $0 \leq i \leq d^2-1$. However, our first strategy of rule selection is -

\noindent \textbf{STRATEGY I:} \label{stg1}\textit{Pick up the balanced rules in which equivalent RMTs have different next state values, that is, no two RMTs of $Equi_i ~(0 \leq i \leq d^2-1)$ have same next state value.}

If we follow STRATEGY $I$, the label of an edge incident to the root, contains exactly one RMT from $Equi_i$, for any $i$ $(0 \leq i \leq d^2-1)$. This implies, the nodes of level $1$ contain all the $d^3$ RMTs of the rule. 
Hence, the nodes of level $2$ also contain all the $d^3$ RMTs of the rule. This scenario continues until level $n-3$. However, we use our algorithm to test whether this scenario continues further for levels $n-2$ and $n-1$, that is, whether the CA is reversible or not. 
These types of CAs are vibrant candidates to be reversible. 

To observe the effectiveness of this strategy, we have randomly generated $d$-state CA rules applying STRATEGY $I$. The cell lengths are also chosen arbitrarily. Now we get a good number of reversible CAs. Table~\ref{chap:reversibility:tab:tstg1} gives a few examples of this experiment.
In Table~\ref{chap:reversibility:tab:tstg1}, first column represents number of states ($d$), second column the number of cells ($n$), third column the CA rule, fourth column represents the number of unique nodes ($M$) generated by Algorithm~\ref{chap:reversibility:algo:rev_algo} for the CA with $n$ cells and the fifth column represents the level of the last unique node. The result of reversibility test by the algorithm is shown in the sixth column. 
In a sample run, out of one hundred million randomly generated balanced rules following this strategy for $3$-state CAs, we have found more than $1.5\times {10}^{5}$ rules which are reversible by Algorithm~\ref{chap:reversibility:algo:rev_algo} for arbitrary cell length $n$.

It can also be noted that, for each of the rules of Table~\ref{chap:reversibility:tab:tstg1}, although the number of cells $n$ given as an input is a large number, but the number of levels up to which unique nodes are generated for that tree is relatively very small and is independent of $n$. For example, for the $3$-state CA rule $222122122001001000110210211$, $M= 585$ and although $n$ was taken as $10005$, last unique is added in level $11$ and the CA is reported as reversible. However, when $n$ is taken as $1000000$, for the same CA the algorithm generates the tree for level $3$ only with $39$ unique nodes and detects it as an irreversible CA. Even if we change the value of $n$, such that $n \geq 11$, $M$ becomes either $585$ or $39$ depending on whether the CA is reversible or irreversible. So, although $n$ can be very large, but Algorithm~\ref{chap:reversibility:algo:rev_algo} generates the reachability tree for this CA up to maximum level $11$ only.
It is observed through experiment that, for $3$-state reversible CAs, maximum number of unique nodes generated by the algorithm is $1371$ and the last unique node is generated in level $i = 19$.  

\begin{table}[!h]
\centering
\caption{Sample rules of STRATEGY I}
\label{chap:reversibility:tab:tstg1}
{\small
\resizebox{0.99\textwidth}{5.0cm}{
\begin{tabular}{|c|c|c|c|c|c|}
\hline 
$d$ & $n$ & Rule & $M$ & Last Level & Reversible? \\ 
\hline 
$2$ & $ 1001 $ & $ 01001011 $ & $ 21 $ & $ 5 $ & Yes \\ 
\hline 
$ 2 $ & $ 2000 $ & $ 01111000 $ & $ 7 $ & $ 2 $ & No \\ 
\hline 
\multirow{2}{*}{$ 3 $} & $ 2090 $ &\multirow{2}{*}{$ 001101211110010120222222002 $}  & $ 39 $ & $  3 $ & No \\ 
 \hhline{~-~---}
 & $ 2091 $ & & $ 282 $ & $ 9 $ & Yes \\ 
\hline 
\multirow{2}{*}{$ 3 $} & $ 10005 $ & \multirow{2}{*}{$ 222122122001001000110210211 $} & $ 585 $ & $ 11 $& Yes \\ 
\hhline{~-~---}
 & $ 1000000 $ &  & $ 39 $ & $ 3 $ & No \\ 
\hline 
\multirow{2}{*}{$ 3 $} & $ 20000 $ & \multirow{2}{*} {$ 010211101020111202020222102 $} & $ 72 $ & $ 5 $ & No \\ 
\hhline{~-~---}
 & $ 100001 $ & & $ 92 $ & $ 8 $ & Yes \\ 
\hline 
$ 3 $ & $ 100001 $ & $ 222220222110111111001002000 $ & $ 88 $ & $ 6 $ & Yes \\ 
\hline 
\multirow{2}{*}{$ 3 $} & $ 25 $ & \multirow{2}{*}{$ 211212112020000020102121201 $} & $ 1371 $ & $ 19 $ & Yes \\ 
\hhline{~-~---}
 & $ 300 $ & & $ 163 $ & $ 5 $ & No \\ 
\hline
\multirow{2}{*}{$ 3 $} & $ 100001 $ & \multirow{2}{*}{$ 011101111102012000220220222 $} & $ 910 $ & $ 18 $ & Yes \\
\hhline{~-~---}
 & $ 101001 $ &  & $ 104 $ & $ 5 $ & No \\
\hline 
$ 3 $ & $ 101 $ & $ 210020002121111121002202210 $ & $ 1345 $ & $ 19 $ & Yes \\
\hline 
$ 3 $ & $ 25 $ & $ 112222111020000000201111222 $ & $ 114 $ & $ 7 $ & Yes \\
\hline
$ 3 $ & $ 25 $ & $ 201020222020201000112112111 $ & $ 580 $ & $ 14 $ & Yes \\
\hline
\multirow{2}{*}{$ 3 $} & $ 330 $ & \multirow{2}{*}{$ 121000212012122121200211000 $} & $ 122 $ & $ 5 $ & No\\
\hhline{~-~---}
 & $ 334 $ & & $ 269 $ & $ 8 $& Yes\\
\hline 
$ 3 $ & $ 101 $ & $ 122210111211121222000002000  $ & $ 194 $ & $ 10 $ & Yes \\
\hline
$ 3 $ & $ 103 $ & $ 021101110202222202110010021  $ & $ 1345 $ & $ 19 $ & Yes \\
\hline
$ 3 $ & $ 1000 $ & $ 201112201120020012012201120 $ & $ 1335 $ & $ 7 $ & No \\
\hline
$ 3 $ & $ 2551 $ & $ 111011011222220122000102200  $ & $ 252 $ & $ 9 $ & Yes \\
\hline
$ 3 $ & $ 2555 $ & $ 202121202020000020111212111 $ & $ 75 $ & $ 5 $ & Yes \\
\hline
$ 3 $ & $ 105 $ & $ 111211111202000020020122202 $ & $ 339 $ & $ 9 $ & Yes \\
\hline
$ 3 $ & $ 101 $ & $ 111111211222220002000002120  $ & $ 196 $ & $ 9 $ & Yes \\
\hline
\end{tabular} }
}
\end{table}

There are ${(d!)}^{d^2}$ balanced rules that can be selected as candidates following STRATEGY $I$. However, more rules can be selected as candidates if we look into sibling RMTs in similar fashion. Recall that, $Sibl_i = \{d.i, d.i+1, d.i+2, ..., d.i+d-1\}$ is a set of sibling RMTs, where $0 \leq i \leq d^2-1$.
It is directly followed from the definition of the reachability tree that, for any $i$, either all the RMTs of $Sibl_i$ are present in a node or none of the RMTs is present. That is, no node in the tree (except the nodes of levels $n-2$ and $n-1$) partially contains the elements of any sibling set. Keeping in mind this property, we develop our next greedy strategy of rule selection -

\noindent\textbf{STRATEGY II:} \label{stg2} \textit{Pick up the balanced rules in which the RMTs of a sibling set have the different next state values, that is, no two RMTs of $Sibl_i ~(0 \leq i \leq d^2-1)$ have same next state value.}

If a rule is picked up following STRATEGY $II$, all the nodes except the nodes of level $n-2$ and $n-1$ are always balanced. There are ${(d!)}^{d^2}$ number of such balanced rules. These rules are also good candidates to be reversible CAs. Like previous, we use Algorithm~\ref{chap:reversibility:algo:rev_algo} to finally decide which of these rules are reversible. 

Here also, we have experimented in the same way with randomly generated $d$-state rules and arbitrary cell length $n$. Some of the rules of this experiment following STRATEGY $II$ are shown in Table~\ref{chap:reversibility:tab:tstg2}. The columns of this table are defined likewise the columns of Table~\ref{chap:reversibility:tab:tstg1}. For this strategy, in a sample run of one hundred million randomly generated balanced $3$-state CA rules we have got more than $1.6 \times {10}^5$ reversible rules by applying Algorithm~\ref{chap:reversibility:algo:rev_algo}.

We can observe that, here too, for each of the rules of Table~\ref{chap:reversibility:tab:tstg2}, although the input $n$, that is the number of cells, is large, but the number of levels up to which unique nodes are generated for that tree is relatively very small and is independent of $n$. For example, for the $3$-state CA rule $102012102012102102021021012$ with $n= 10001$, $M= 1371$ and last unique is added in level $19$. The CA is reported as reversible. Even if we change the value of $n$, such that $n \geq 19$, the Algorithm~\ref{chap:reversibility:algo:rev_algo} generates the reachability tree for this CA up to maximum level $19$ only with $M \leq 1371$.
For STRATEGY $II$ also, it is observed that maximum number of unique nodes generated by the algorithm for $3$-state reversible CAs is $1371$ and the last unique node is generated in level $i = 19$.  

\begin{table}[!h]
\centering
\caption{Sample rules of STRATEGY II}
\label{chap:reversibility:tab:tstg2}
\resizebox{0.99\textwidth}{5.0cm}{
\begin{tabular}{|c|c|c|c|c|c|}
\hline 
$d$ & $n$ & Rule & $M$ & Last level & Reversible? \\ 
\hline 
$ 2 $ & $ 1001 $ & $ 01011001 $ & $ 21 $ &  $ 5 $ & Yes \\ 
\hline 
$ 2 $ & $ 2000 $ & $ 10010101 $ & $ 13 $ &  $ 3 $ & No \\ 
\hline 
$ 3 $ & $ 25 $ & $ 021021021210210210012012012 $ & $ 229 $ & $ 9 $ & Yes \\ 
\hline 
\multirow{2}{*}{$ 3 $} & $ 25 $ & \multirow{2}{*}{$ 210210012201210021012210210 $} & $ 1315 $ & $ 17 $ & Yes \\ 
\hhline{~-~---} 
 & $ 30 $ &  & $ 138 $ & $ 5 $ & No \\ 
\hline
$ 3 $ & $ 10001 $ & $ 201120102201201201102120201 $ & $ 166 $ & $ 8 $ & Yes \\ 
\hline 
$ 3 $ & $ 25001 $ & $ 210012012120210021012012210 $ & $ 1345 $ & $ 19 $ & Yes \\ 
\hline 
\multirow{2}{*}{$ 3 $} & $ 25 $ & \multirow{2}{*}{$ 021120210021012021021021012 $} & $ 1039 $ & $ 17 $& Yes \\ 
\hhline{~-~---} 
& $ 300 $ & & $ 158 $ & $ 5 $ & No \\ 
\hline
$ 3 $ & $ 1001 $ & $ 021120120210120210120120021 $ & $ 910 $ & $ 18 $ & Yes \\
\hline
$ 3 $ & $ 100001 $ & $ 210012012201201210210201201 $ & $ 592 $ & $ 11 $ & Yes \\
\hline 
\multirow{2}{*}{$ 3 $} & $ 25 $ & \multirow{2}{*}{$ 201021012201201102021021201 $} & $ 382 $ & $ 9 $ & Yes \\ 
\hhline{~-~---}
& $ 20 $ &  & $ 49 $ & $ 4 $ & No \\ 
\hline 
$ 3 $ & $ 10001 $ & $ 102012102012102102021021012 $ & $ 1371 $ & $ 19 $ & Yes \\ 
\hline
$ 3 $ & $ 3333 $ & $ 120021120012021012021021120 $ & $ 192 $ & $ 10 $ & Yes \\
\hline 
$ 3 $ & $ 1000 $ & $ 210102102120201210102021021 $ & $ 716 $ & $ 6 $ & No \\
\hline
$ 3 $ & $ 25 $ & $ 012102012120210120210021102 $ & $ 1252 $ & $ 7 $ & No \\
\hline 
$ 3 $ & $ 3333 $ & $ 210210012201012102210210210 $ & $ 332 $ & $ 9 $ & Yes \\
\hline 
$ 3 $ & $ 101 $ & $ 201021201021201201012210201 $ & $ 985 $ & $ 17 $ & Yes \\
\hline 
$ 3 $ & $ 103 $ & $ 102102201210102012201102102 $ & $ 910 $ & $ 18 $ & Yes \\
\hline 
$ 3 $ & $ 331 $ & $ 102201012102102120102120102 $ & $ 1315 $ & $ 17 $ & Yes \\
\hline 
$ 3 $ & $ 3333 $ & $ 210012210012012210012012120 $ & $ 128 $ & $ 8 $ & Yes \\
\hline 
$ 3 $ & $ 103 $ & $ 012201021012012021012021012 $ & $ 196 $ & $ 9 $ & Yes \\
\hline 
$ 3 $ & $ 103 $ & $ 210210012201210102012210210 $ & $ 910 $ & $ 18 $ & Yes \\
\hline
$ 3 $ & $ 105 $ & $ 210120120102210012021120201  $ & $ 730 $ & $ 6 $ & No \\
\hline
\end{tabular} }
\end{table}

Therefore, if we select rules following STRATEGY $I$ and STRATEGY $II$, we will be able to identify a large set of $n$-cell reversible CAs. However, the set of rules, selected out of STRATEGY $I$ and the set of rules, selected out of STRATEGY $II$ are not disjoint.
We now report our $3^{rd}$ strategy of rule selection where the members of $Sibl_i ~(0\leq i \leq d^2-1)$ have same next state value.

\noindent\textbf{STRATEGY III:} \label{stg3} \textit{Pick up the balanced rules in which - $(1)$ the RMTs of $Sibl_i$ for each $i$, have same next state value, and either $(2)$ the RMTs of $Sibl_{k.d}, Sibl_{k.d + 1}, ..., Sibl_{k.d + d-1}$ have either same next state value, where $~0 \leq k \leq d-1$, or RMTs of those $d$ sets have different next state values; or $(3)$ RMTs of $Sibl_k$ and its equivalent RMTs have different next state values or RMTs of those $d$ sets have same next state value.}

There are $2(d! + (d!)^{d})$ number of balanced rules that can be selected following STRATEGY $III$ as candidates to be reversible CAs. One can easily verify that in this case also, all but nodes of level $n-2$ and level $n-1$ are balanced. 

For this strategy too, we have followed similar experiment on randomly generated $d$-state CA rules with arbitrary $n$. Some examples of such rules are shown in Table~\ref{chap:reversibility:tab:tstg3}. Here also, the columns of Table~\ref{chap:reversibility:tab:tstg3} are defined similar to the columns of Table~\ref{chap:reversibility:tab:tstg1} and Table~\ref{chap:reversibility:tab:tstg2}.

The CAs of STRATEGY $III$ are very simple CAs (actually of $2$-neighborhood dependency) and whatever be the input $n$, for any CA, Algorithm~\ref{chap:reversibility:algo:rev_algo} generates the tree for a small length to detect whether the CA is reversible or irreversible. 
Number of rules following STRATEGY $III$ is less. However, in a sample run of one hundred randomly generated rules following this strategy for $3$-state CAs, we have found $22$ reversible rules by Algorithm~\ref{chap:reversibility:algo:rev_algo}, where maximum number of unique nodes generated by the algorithm for these CAs is $28$ and the last unique node is generated in level $i = 4$.  

\begin{table}[!h]
\centering
\caption{Sample rules of STRATEGY III}
\label{chap:reversibility:tab:tstg3}
\resizebox{0.99\textwidth}{3.5cm}{
\begin{tabular}{|c|c|c|c|c|c|}
\hline 
$d$ & $n$ & Rule & $M$ & Last Level & Reversible? \\ 
\hline
$ 2 $ & $ 100 $ & $ 11001100 $ & $ 5 $ & $ 2 $ & Yes\\
\hline
\multirow{2}{*}{$ 3 $}& $ 199 $ &\multirow{2}{*}{$ 111222000222000222000111111 $}&$ 13 $&$ 2 $ & No\\
\hhline{~-~---}
&$ 200 $&&$ 13 $ & $ 2 $ & No\\
\hline
$ 3 $& $ 101 $ &$ 000111222111222000222000111 $& $ 19 $& $3$ & Yes\\
\hline
\multirow{2}{*}{$ 3 $}&$105$&\multirow{2}{*}{$000111222000222111222000111$}&$28$& $4$ & Yes\\
\hhline{~-~---}
&$1004$&&$15$& $3$ &No\\
\hline
$ 3 $&$ 25 $&$111222000000222111222111000$&$28$& $4$&Yes\\
\hline
$ 3 $ & $1000$ &$000222111000111222111000222$& $16$& $ 3 $& No \\
\hline
$ 3 $ & $103$ &$111222000222000111111000222$& $28$ & $4$ & Yes\\
\hline
$ 3 $&$ 1111 $ &$222111000111000222222000111$&$28$& $ 4 $ & Yes\\
\hline
$ 3 $&$10001$&$111222000222111000111222000$&$7$& $2$& No\\
\hline
$ 3 $&$ 5555 $&$111000222000222111222111000$&$19$&$ 3 $ & Yes\\
\hline
$ 3 $&$ 5555 $&$111222000111000222000111222$&$28$&$ 4 $ & Yes\\
\hline
\multirow{2}{*}{$ 3 $}&$10001$&\multirow{2}{*}{$222111000111000222000222111$}&$19$& $3$ &Yes\\
\hhline{~-~---}
&$ 10000 $&&$ 13 $&  $ 2 $& No\\
\hline
$ 3 $&$ 1111 $ &$222000111000111222111222000$&$19$& $ 3 $ & Yes\\
\hline
$ 3 $&$ 1001 $ &$000111222111222000000222111$&$28$& $ 4 $ & Yes\\
\hline
\end{tabular}} 
\end{table}

Therefore, we can identify a large set of reversible CAs after using above strategies and Algorithm~\ref{chap:reversibility:algo:rev_algo}.

\section{Conclusion}
\label{conclusion}

In this work, we have developed reachability tree to test reversibility of $3$-neighborhood $d$-state periodic boundary finite CAs of length $n$. We have developed an algorithm which tests reversibility of a finite CA with a given length $n$. We have also reported three greedy strategies for finding a set of reversible $d$-state CAs of length $n$. 

In the next chapter, we further explore reversibility of one-dimensional CAs, and relate the different cases of reversibility as mentioned in the beginning of this chapter.

%
%
%
%
%

\chapter{Reversibility and Semi-reversibility}\label{Chap:semireversible}

\noindent{\small Reversibility of a one-dimensional finite cellular automaton (CA) is dependent on lattice size. The decision algorithm, as stated in Chapter~\ref{Chap:reversibility} (see Algorithm~\ref{chap:reversibility:algo:rev_algo} of Page~\pageref{chap:reversibility:algo:rev_algo}), decides reversibility by taking CA rule and lattice size as input. A CA, however, can be reversible for a set of lattice sizes. On the other hand, reversibility of an infinite CA, which is decided by exploring the rule only, is different in its kind from that of finite CA. As pointed out in Chapter~\ref{Chap:reversibility}, the reversibility of finite and infinite cases are to be studied separately. Can we, however, link the reversibility of finite CA to that of infinite CA? This is the primary issue which we investigate in this chapter. In order to address this issue, we introduce a new notion, named \emph{semi-reversibility}.} 

\section{Introduction}
\label{Chap:semireversible:sec:intro}

\noindent {\large\textbf{C}}lassically cellular automata (CAs) are defined over infinite lattice. In classical literature, therefore, reversibility of a CA simply implies the reversibility of the CA, defined over infinite lattice. On the other hand, the issue of reversibility  of finite CAs has gained the interest of researchers in last few decades due to various reasons. For finite CAs, lattice size along with a CA rule is a parameter to decide its reversibility. Hence, there is an ambiguity in the term ``reversibility of a CA'', because the CA can be defined over finite as well as infinite lattice.

Additionally, Chapter~\ref{Chap:reversibility} has discussed that reversibility of CAs under finite and infinite cases are two different issues which are not to be mixed up. There are at least three different cases (see Section~\ref{chap:reversibility:sec:intro} of Chapter~\ref{Chap:reversibility}, Page~\pageref{reversibility_cases}) of reversibility of $1$-dimensional infinite CAs, for which we can use any classical algorithms (e.g. \cite{Amoroso72,suttner91}) to decide their reversibility. However, these classical algorithms do not work for finite CAs. Therefore, there is no apparent bridge between finite and infinite cases. The primary motivation of this chapter is to inquire of a relation between these two cases.

Nevertheless, to search for any such relation, we need to first remove the ambiguity associated with the reversibility of a CA. To do so, we redefine the reversibility, and introduce a new notion named \emph{semi-reversibility} in this chapter. Before understanding the utility of this new definition and notion in the present study, let us examine the existing results and relations among various types of reversibility.


\subsection{Some Existing Results on Reversibility}
As noted in Section~\ref{chap:reversibility:sec:intro} of Chapter~\ref{Chap:reversibility}, there are at least four cases of reversibility of CAs under finite and infinite lattice size. Let us reproduce these four cases:
\begin{description}[itemsep=0ex]
\item[Case $1$:]\label{case1} An infinite CA whose global function is injective on the set of ``all infinite configurations''.
\item[Case $2$:]\label{case2} An infinite CA whose global function is injective on the set of ``all {\em periodic} configurations''.
\item[Case $3$:]\label{case3} An infinite CA whose global function is injective on the set of ``all finite configurations of length $n$'' for all $n\in \mathbb{N}$.
\item[Case $4$:]\label{case4} A finite CA whose global function is injective on the set of ``all configurations of length $n$'' for a fixed $n$.
\end{description} 

Therefore, for a CA with a given local map $R$, there exists at least four types of global transition functions depending on the above four cases -- let, $G$ be the global transition function on the set of all infinite configurations, $G_P$ be the global transition function on the set of periodic configurations, $G_F$ be the global transition function on the set of finite configurations for all lattice size $n \in \mathbb{N}$ and $G_n$ be the global transition function over a fixed lattice size $n$. Recall that, for one-dimensional CAs, periodic boundary condition over configuration of length $n$, for all $n \in \mathbb{N}$ evidently implies periodic configurations (see Definition~\ref{Def:periodicConfiguration} of Page~\pageref{Def:periodicConfiguration}). So, in this chapter, by the global transition function on the set of periodic configurations, we mean both the notions and use result on any one of these to portray the other. 

There are many interesting results regarding reversibility of $1$-dimensional CAs, over a given local map $R$, for the first three cases: 
\begin{theorem}
The following statements hold for $1$-dimensional infinite CAs:
\begin{enumerate}[topsep=0pt,itemsep=0ex,partopsep=2ex,parsep=1ex]
\item A CA, defined over infinite lattice, is reversible, if inverse of its global transition function $G$, $G^{-1}$ is the global transition function of some (infinite) CA \cite{hedlund69}.

\item An infinite CA is reversible, if $G$ is injective \cite{hedlund69,Richa72}.

\item If $G$ is injective, then $G_F$ is surjective \cite{Richa72}.

\item $G_F$ is injective if and only if $G$ is surjective \cite{moore1962machine,Myhill63}.

\item $G_F$ is surjective if and only if it is bijective \cite{amoroso1970garden}.

\item If $G_F$ is surjective, then $G_F$ is injective; but the converse is not true \cite{Richa72}.

\item  $G$ is injective, if and only if $G_P$ is injective \cite{sato77}.


\item If $G_P$ is injective, then $G$ is surjective \cite{sato77}.

\item $G_P$ is surjective, if and only if $G$ is surjective \cite{sato77}.

\item If $G_P$ is injective, then $G_P$ is surjective \cite{sato77}.

\item If $G_P$ or $G_F$ is surjective, then $G$ is surjective \cite{Kari05}.

\item If $G$ is injective, then $G_P$ and $G_F$ are injective \cite{Kari05}.

\item If $G_P$ is injective, then $G$ is injective \cite{Kari05}.

\item If $G$ is surjective, then $G_P$ is surjective \cite{Kari05}.
\end{enumerate}
\end{theorem}

\begin{figure}[!h]
\centering
\includegraphics[width= 3.0in, height = 1.0in]{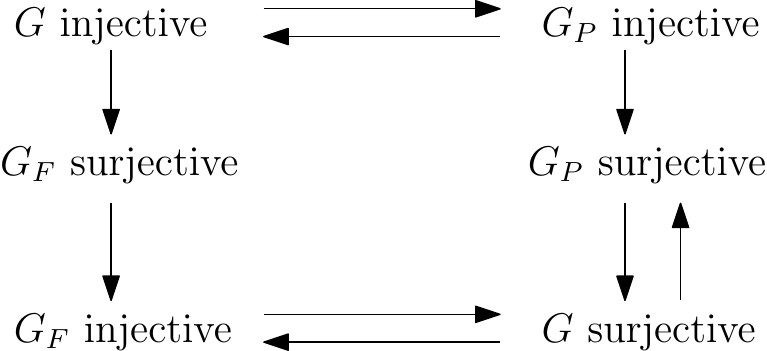}
\caption{Relationship between Injectivity and Surjectivity properties of $1$-dimensional CAs}
\label{Chap:semireversible:fig:rev_rel}
\end{figure}

\noindent The relation among $G$, $G_F$ and $G_P$ is shown in Figure~\ref{Chap:semireversible:fig:rev_rel}. Therefore, for a one dimensional CA with local rule $R$, injectivity of $G$ implies injectivity of $G_F$ and $G_P$. However, Case~\ref{cs4}, that is, reversibility for finite CAs, where $G_n$ is injective over the set of all configurations of length $n$ for a fixed $n$, is different from these three cases. In Chapter~\ref{Chap:reversibility}, an algorithm is presented which can decide reversibility of a $d$-state $3$-neighborhood finite CA for a given lattice size $n$.  

\subsection{The Challenges}
One can observe that, there is a both way implication between the injectivity of $G_P$ and $G$ (see Figure~\ref{Chap:semireversible:fig:rev_rel}). That is, if a CA is injective over the set of all periodic configurations, it is also injective over the set of all infinite configurations and vice versa. Further, injectivity of an infinite CA implies reversibility of it. Hence, one can exploit the injectivity (or, bijectivity) of $G_P$ over periodic configurations to deduce about injectivity (or, bijectivity) of $G$.

Moreover, injectivity of $G_P$ implies injectivity of the CA over configurations of length $n$ under periodic boundary condition for all $n \in \mathbb{N}$. But, injectivity of CAs over configurations of length $n$ under periodic boundary condition is the Case \ref{cs4}, where the global transition function is $G_n$. Hence, by using injectivity of $G_n$ for all $n \in \mathbb{N}$, we may extrapolate about injectivity of $G_P$. That is, by exploiting reversibility of finite CAs with discrete $n$, we are targeting to infer about reversibility of CAs defined over infinite lattice. 

But problem with this approach is, injectivity of $G_n$ (Case~\ref{cs4}) is dependent on the lattice size $n$. That is, for a Case~\ref{cs4} CA where number of states per cell is $d$ and size of the neighborhood is $m$, we need to additionally know the lattice size to decide its reversibility. See for example, Table~\ref{chap:reversibility:tab:tstg1} (Page~\pageref{chap:reversibility:tab:tstg1}), \ref{chap:reversibility:tab:tstg2} (Page~\pageref{chap:reversibility:tab:tstg2}) and \ref{chap:reversibility:tab:tstg3} (Page~\pageref{chap:reversibility:tab:tstg3}) of Chapter~\ref{Chap:reversibility}. In fact, for many of these Case~\ref{cs4} CAs, the set of discrete lattice sizes for which the CA is reversible is an infinite set. For instance, elementary CA (ECA) $101$ is reversible if $n$ is odd, but ECA $51$ is reversible for any $n \in \mathbb{N}$. The existence of ECAs and $2$-neighborhood $3$-state CAs which are reversible for infinite number of lattice sizes under periodic boundary condition is proved in \cite{Ino05,Sato2009, 0305-4470-37-22-006}. 
However, classical reversibility of the infinite CAs referred by Case \ref{cs1} as well as of Case \ref{cs2} and \ref{cs3} is independent of lattice size. So, to conclude about injectivity of $G_P$, and correspondingly about $G$ and $G_F$, $G_n$ has to be injective for every $n \in \mathbb{N}$.

There are many finite CAs (Case~\ref{cs4}) which are reversible for some $n$; that is, for a given local rule $R$, $G_n$ is bijective for those $n$.
However, as $G_n$ is not bijective for all $n \in \mathbb{N}$, $G_P$ is not bijective for those CAs. So, in classical sense, these CAs are irreversible when defined over infinite lattice, but reversible when its lattice size is finite. Therefore, when CA is defined over finite lattice, we get a new class of reversibility -- reversible for some $n \in \mathbb{N}$. To deal with this class of CAs, we introduce the notion of \emph{semi-reversibility}. Hence, there are three types of finite CAs -- (1) CAs for which $G_n$ is injective for each $n \in \mathbb{N}$, (2) CAs for which $G_n$ is injective for some $n \in \mathbb{N}$, and (3) CAs for which $G_n$ is not injective for each $n \in \mathbb{N}$.

The Algorithm~\ref{chap:reversibility:algo:rev_algo} of Chapter~\ref{Chap:reversibility} (see Page~\pageref{chap:reversibility:algo:rev_algo}) can decide reversibility of a finite CA with a local rule $R$ for a given lattice size $n$. However, for a finite CA with a set of (possible infinite) lattice sizes, one needs to check every lattice size $n$, to see whether the CA is reversible/irreversible for each $n$ or some of these $n\in \mathbb{N}$. Nonetheless, if the set is not very small, it is very difficult to do such test.
In this scenario, the following two questions arise -- 
\begin{enumerate}
\item Is it possible to understand the reversibility behavior of a finite CA by exploring the CA for some small, but sufficient lattice size $n$? 

\item What is the relation between reversibility of finite CA over fixed $n$ (Case~\ref{cs4}) to that of the other three cases?
\end{enumerate}

%
%
This chapter targets to answer these two questions for $1$-dimensional $d$-state $m$-neighborhood CAs under periodic boundary condition. Our contributions are mainly the following:
\begin{enumerate}
\item Identification of \emph{semi-reversible} CAs, which are classically treated as irreversible, but are reversible for a set (possibly infinite) of lattice sizes. Hence, we get a new classification of CAs -- reversible, semi-reversible and \emph{strictly} irreversible (Section~\ref{Chap:semireversible:sec:rev_class}).
\item Reachability tree is extended to find the class of reversibility of a $1$-dimensional CA. Hence, we decide the reversibility behavior of a CA for any $n \in \mathbb{N}$ from a finite size (Section~\ref{Chap:semireversible:sec:rtree}).
\item A scheme to decide (semi-)reversibility of a CA along with an expression to find the infinite set of sizes for which it is reversible (Section~\ref{Chap:semireversibility:sec:semi_tree}).
\item Relation among the four cases of reversibility (Section~\ref{Chap:semireversible:sec:remark}).
\end{enumerate}

\section{Primary RMT Sets and Homogeneous Configurations}\label{Chap:semireversible:sec:prim}
\noindent This section introduces another mathematical tool, named \emph{primary RMT sets} which are elemental in representing a configuration of a CA. Recall that, if $\mathcal{C}_n$ be the set of all configurations of an $n$-cell CA and $G_n:\mathcal{C}_n\rightarrow \mathcal{C}_n$ be its global transition function induced by the local rule $R$, then, 
\begin{equation*}
y=G_n(x) = G_n(x_0x_1\cdots x_{n-1})= (R(x_{i-l_r},\cdots, x_i, \cdots, x_{i+r_r}))_{i\in \mathscr{L}}
\end{equation*}
where $l_r$ and $r_r$ are left and right radii of the CA and $m= l_r+r_r +1$. Here, $y$ is the successor of the configuration $x$. As RMT sequence is another representation of a configuration, it can be redefined as:
 \begin{definition}\label{Chap:semireversible:def:RMTseq}
Let $x=(x_i)_{i\in\mathscr{L}}$ be a configuration of a CA. The {RMT sequence} of $x$, denoted as $\tilde{x}$, is $(r_i)_{i\in\mathscr{L}}$ where $r_i$ is the RMT $(x_{i-l_r},\cdots x_i, \cdots, x_{i+r_r})$.
\end{definition}

Hence, number of configurations of an $n$-cell CA = number of possible RMT sequences = number of cycles of length $n$ in the de Bruijn graph = $d^n$. Two configurations are said to be {\em shift-equivalent} to each other if one is obtained by the left shifts of the other. We write $k$-bit left shift of $x$ as $\sigma^k(x)$. Two RMT sequences which are shift-equivalent to each other, essentially refer to the same cycle of length $n$ (see Definition~\ref{Defi:Cycle} of Page~\pageref{Defi:Cycle}). Since their edge sequences are different, we treat them as two cycles of length $n$. We now rewrite following two definitions (first defined in Section~\ref{chap:reversibility:Sec:CAbasic}, Page~\pageref{chap:reversibility:Sec:CAbasic}) considering neighborhood dependency as $m$.

%
%


\begin{definition}
\label{Chap:semireversible:def:equi}
A set of $d$ RMTs $r_1, r_2, ..., r_d$ of a $d$-state CA rule are said to be equivalent to each other if $r_1 d \equiv r_2 d \equiv ... \equiv r_d d \pmod{ d^m}$.
\end{definition}

\begin{definition}
\label{Chap:semireversible:def:sibl}
A set of $d$ RMTs $r'_1, r'_2, ..., r'_d$ of a $d$-state CA rule are said to be sibling to each other if $\floor{\frac{r'_1}{d}} = \floor{\frac{r'_2}{d}} = ... = \floor{\frac{r'_d}{d}}$.
\end{definition}

Therefore, there are $d^{m-1}$ sets of equivalent and sibling RMTs, which can be written respectively as $Equi_i = \{i, d^{m-1}+i, 2d^{m-1}+i, \cdots, (d-1)d^{m-1}+i \}$ and $Sibl_j = \{d.j, d.j+1, \cdots, d.j+d-1\}$, where $0\leq i,j \leq d^{m-1}-1$.
 An interesting relation is followed in the RMT sequence $(r_i)_{i\in\mathscr{L}}$: if $r_i\in Equi_j$, then $r_{i+1}\in Sibl_j$ ($0\le j\le d^{m-1}-1$).

Hence, an arbitrary arrangement of RMTs does not form an RMT sequence. If RMT $r$ is part of any RMT sequence, then the next RMT in the sequence has to be from $Sibl_k = \{d.r \pmod{d^m}, d.r+1 \pmod{d^m}, \cdots, d.r+(d-1) \pmod{d^m}\}$, where $r \equiv k \pmod {d^{m-1}}$. However, as any cycle in a de Bruijn graph $B(m-1,\mathcal{S})$ corresponding to a CA represents an RMT sequence, this graph can be used to get the smallest RMT sequences possible for any CA. 

 For instance, in Figure~\ref{chap:reversibility:fig:dbg_3_state} (Page~\pageref{chap:reversibility:fig:dbg_3_state}), a cycle can be traced by: $200\rightarrow002\rightarrow021\rightarrow210\rightarrow101\rightarrow010\rightarrow102\rightarrow022\rightarrow222\rightarrow221\rightarrow211\rightarrow112\rightarrow120$, which is the RMT sequence $\langle 18,2,7,21,10,3,11,8,26,25,22,14,15\rangle$ constituting the configuration $0021010222112$. However, another RMT sequence $\langle 2,8,25,22,14,15,19,3,9\rangle$ is traced by the elementary cycle $002\rightarrow022\rightarrow221\rightarrow211\rightarrow112\rightarrow120\rightarrow201\rightarrow010\rightarrow100$ representing the configuration ${022112010}$. Note that, none of these RMT sequences can be split into any smaller RMT sequences, but can be used to form a larger RMT sequence. We name the RMT sets corresponding to the RMTs of each elementary cycle as \emph{primary RMT sets} \cite{hazari14, hazariJCA17}.

 \begin{definition}\label{Chap:semireversible:def:primaryRMT} A set of RMTs $P=\{r_1, r_2,\cdots, r_p\}$ is said to be a primary RMT set, if following two conditions are satisfied:
 \begin{enumerate}
 \item If $r \in P$, then $r' \in P$, where $r' \in Sibl_k$, $r \equiv k \pmod {d^{m-1}}$. 
 
 \item No proper subset of $P$ can form an RMT sequence.
 \end{enumerate}
\end{definition}

\begin{figure*}[hbt]
	\centering
	\includegraphics[width= 2.0in, height = 1.0in]{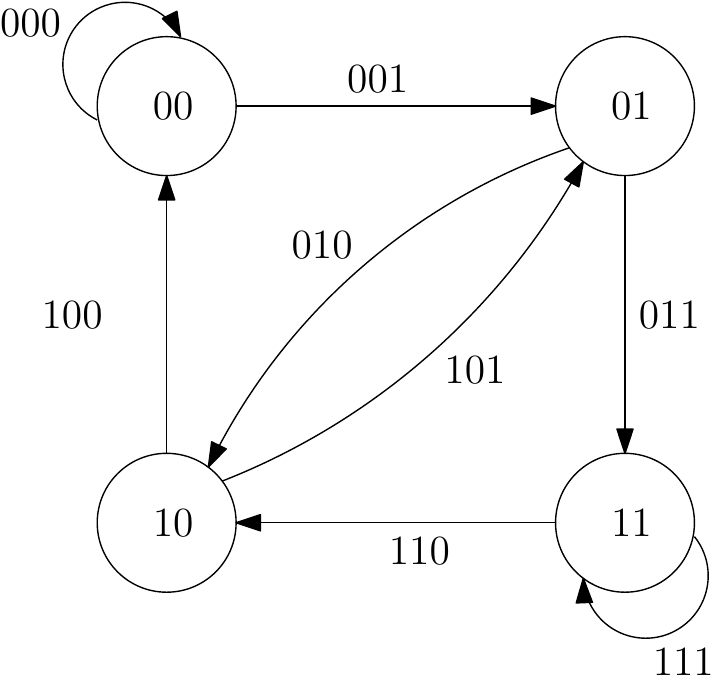}
	\caption{The de Bruijn graph $B(2,\{0,1\})$}
	\label{Chap:semireversible:fig:dbg_eca}
	\vspace{-1.0em}
\end{figure*}

The number of elementary cycles in the de Bruijn graph is the number of primary RMT sets, whereas, the union of all primary RMT sets is the set of all $d^m$ RMTs. Since de Bruijn graph is a Hamiltonian graph, maximum possible cardinality of $P$ is $d^{m-1}$, that is, the number of nodes in the graph. Minimum cardinality of a primary RMT set is $1$, and number of such singleton primary RMT sets are $d$, associated with the RMTs $r = s\times d^{m-1}+s\times d^{m-2}+\cdots+s\times d+s$, $0\leq s \leq d-1$. For ECAs, there are only $6$ primary RMT sets -- $\{0\}$, $\{7\}$, $\{2,5\}$, $\{1,2,4\}$, $\{3,6,5\}$ and $\{1,3,6,4\}$ corresponding to the six elementary cycles, shown in Figure~\ref{Chap:semireversible:fig:dbg_eca}. Note that, as any elementary cycle in the de Bruijn graph corresponds to a primary RMT set, no two RMTs of $Sibl_i$, $0\leq i\leq d^{m-1}$, can belong to the same primary RMT set.
Moreover, for each $l \in \{1, 2, \cdots, d^{m-1}\}$, there exist one or more primary RMT sets with cardinality $l$. In fact, for any $n\in \mathbb{N}$, we can get cycle(s) of length $n$ in de Bruijn graph.

Given a CA, however, finding of its primary RMT sets is a task. From the given CA rule, we can construct the de Bruijn graph (In fact, from only $m$ and $d$ values, we can construct de Bruijn graph). Once we get the de Bruijn graph, we can find the elementary cycles of the graph by using any well-known algorithm \cite{tiernan1970efficient,tarjan1973enumeration,ehrenfeucht1973algorithm,johnson1975finding}. As said before, these elementary cycles correspond to the possible primary RMT sets of the CA.

\begin{definition} 
A configuration, RMT sequence of which is formed with the RMTs of a primary RMT set only, is called  \textbf{homogeneous configuration}.
\end{definition}

Length of a homogeneous configuration, however, depends on the size of the CA. We
represent a homogeneous configuration in terms of its minimal length, that is, the cardinality of corresponding primary RMT set. For example,
take the primary RMT set $\{000\}$. The homogeneous configuration formed by RMT $000~(0)$
is $0^n$, where $n \in \mathbb{N}$. Similarly, for $\{202, 020\}$, a primary RMT set of cardinality $2$, the homogeneous configuration is $(02)^n$, $n \in \mathbb{N}$. In case of ECAs, there are six homogeneous configurations -- ${0}^n$, ${1}^n$, $(01)^n$, $(001)^n$, $(011)^n$ and $(0011)^n$. Obviously, all the homogeneous configurations may not be observed simultaneously in an $n$-cell CA. For example, if $n \notin 3\mathbb{N}$, $(001)^{\frac{n}{3}}$, $(011)^{\frac{n}{3}}$ cannot be observed in an $n$-cell ECA.

In general, an arbitrary RMT sequence $\tilde{x}$ is always formed by using one or more primary RMT sets. In other words, an arbitrary configuration $x$ of a CA is followed from a set of primary RMT sets. Therefore, there is a binary relation from the set of primary RMT sets to the configurations. Let us call this relation ``follows'' and denote it by `$\vdash$'. If there are $k$ primary RMT sets $P_1, P_2, \cdots, P_k$, then $\vdash\subseteq\mathcal{P}(\{P_1, P_2, \cdots, P_k\})\times \mathcal{C}_n$, where $\mathcal{P}(.)$ denotes the power set. Following is the definition of this relation.

\begin{definition}
\label{Chap:semireversible:Defi:TurnStyle}
Let $X$ be a non-empty set of primary RMT sets and $x$ be a configuration of a CA. Then, $X \vdash x$ holds if and only if the RMT sequence $\tilde{x}$ uses all the RMTs of each primary RMT set $P\in X$.
\end{definition}
In case of ECAs, for example, $\{\{2, 5\}, \{4, 1, 3, 6\}\}\vdash 0010101011$.
This configuration, however, contains the {\em signature} of each of the primary RMT sets.

\begin{definition}
A primary RMT set $P$ has a \textbf{\em signature} in a configuration $x$ if $\tilde{x}$ has a subsequence $\tilde{y}=(s_i)_{i\in\{0,1,\cdots,\mathbf{card}(P)-1\}}$ where $\{P\}\vdash y$.
\end{definition}

For a configuration $x=0010101011$ of an ECA, we have $\tilde{x}=4125252536$. Here, $52$ is the subsequence of $\tilde{x}$, and $\{\{2,5\}\}\vdash 01$. Hence, the set $\{2, 5\}$ has a signature in $x$. Similarly, $4136$ is another subsequence of $\tilde{x}$, so the set $\{4, 1, 3, 6\}$ has also a signature in $x$. 
Let us now define following notations, related to a configuration $x$ and a primary RMT set $P$, which are used throughout the chapter.
\begin{itemize}
\item[] $|x|$ is the length of the configuration $x$.
\item[] $|x|_P$ is the number of signatures of $P$ in $x$.
\end{itemize}
In the above example, $|x|_{\{4, 1, 3, 6\}}=1$, $|x|_{\{2,5\}}=3$,
whereas $|x|=10$.

However, an RMT may be part of more than one primary RMT set. Because, an edge of de Bruijn graph may be used by two or more elementary cycles. This implies, if an RMT $r\in P$ is used in $\tilde{x}$, then it may happen that $|x|_P = 0$.

If $x$ and $y$ are two configurations of a CA where $X \vdash x$ and $X \vdash y$, then it can not be said in general that $y$ is shift equivalent to $x$. Even, if  $|x|_P = |y|_P$ for each $P\in X$, then they may not be shift equivalent to each other. Similarly, same RMT sequence may be followed from two different primary RMT sets. For example, for the RMT sequence $\langle4,1,2,5,2,5,2,5,3,6\rangle$, $\{\{2, 5\},\{4, 1, 2\}, \{5, 3, 6\}\}\vdash 0010101011$. Now, we can state the following result. 

\begin{proposition}\label{Chap:semireversible:Proposition:basic} Let $X=\{P_1,P_2,\cdots,P_l\}$ where $l\ge 1$ and $P_1,P_2,\cdots, P_l $ are primary RMT sets of a CA. Now, if $X\vdash x$ then the following hold:
\begin{enumerate}
\item\label{Proposition:condition1}
$X\vdash x^k$ for any $k\in \mathbb{N}$.
\item\label{Proposition:condition2}
$X \vdash \sigma^k(x)$ for any $k\in \mathbb{N}$.
\item \label{Proposition:Condition3}
There exist positive integers $k_1, k_2, \cdots, k_l$ such that $k_1.\mathbf{card}(P_1)+k_2.\mathbf{card}(P_2)+\cdots+k_l.\mathbf{card}(P_l)=|x|$.
\end{enumerate}
\end{proposition}

\begin{proof} Let $X=\{P_1,P_2,\cdots,P_l\}$ where $l\ge 1$ and $X\vdash x$. Then, proofs of conditions \ref{Proposition:condition1} and \ref{Proposition:condition2} are trivial. Hence omitted. Here, we give proof of only condition~\ref{Proposition:Condition3}.

In this case, we get a cycle of length $|x|$ against $\tilde{x}$ in the de Bruijn graph. This cycle can be decomposed into $l$ elementary cycles which correspond to $l$ primary RMT sets of $X$. Lengths of elementary cycles are $\mathbf{card}(P_1), \mathbf{card}(P_2), \cdots, \mathbf{card}(P_l)$. However, an elementary cycle may be repeated more than once in the cycle of length $|x|$. If the elementary cycles are repeated $k_1, k_2,\cdots, k_l$ times, then obviously, $k_1.\mathbf{card}(P_1)+k_2.\mathbf{card}(P_2)+\cdots+k_l.\mathbf{card}(P_l)=|x|$.
\end{proof}

\begin{example}
In case of ECAs, $\{\{4, 1, 2\}\}\vdash 001$ and $\{\{5, 3, 6\}\}\vdash 011$. The RMT sequences of ${001}$ and $011$ are $\langle 412 \rangle$ and $\langle 536 \rangle$ respectively. Here, RMTs $4,5 \in Sibl_2$. Therefore, we get $\{\{4, 1, 2\},\{5, 3, 6\}\}\vdash 001011$, that is, an RMT sequence $\langle 412536412 \rangle$. However, two primary RMT sets $\{0\}$ and $\{2, 5\}$ cannot form any RMT sequence. That is, $\{\{0\},\{2, 5\}\}\nvdash x$ for any $x\in \mathcal{C}_n$ and any $n\in\mathbb{N}$, whereas, $\{\{4, 1, 2\},\{2, 5\}\}\nvdash x$ for any $x\in \mathcal{C}_6$.
\end{example}


\section{Reversibility and Semi-reversibility}\label{Chap:semireversible:sec:rev_class}
\noindent For a finite CA of size $n$, reversibility is dependent on $n$. 
A CA is reversible for length $n$, if $G_n$ is bijective, and irreversible, if $G_n$ is not bijective over $\mathcal{C}_n$. Therefore, injectivity and surjectivity of $G_n$ are not equivalent to that of $G_P$. Recall that, $G_P$ is the global transition function on the set of all periodic configurations.
If $G_P$ is bijective for periodic configurations of length $n$, for all $n \in \mathbb{N}$, $G_n$ is also bijective for an $n$, and if $G_n$ is not bijective for a size $n$, then $G_P$ is not bijective. However, if $G_n$ is bijective for only some $n$, then, $G_P$ is not a bijection. We are interested in inquiring about injectivity of $G_P$ from injectivity of $G_n$. So, in order to relate $G_n$ with $G_P$, we redefine the notion of reversibility and irreversibility of the CA.

\begin{definition}
\label{chap:semireversibility:def:rev}
A CA with rule $R$ is called \textbf{reversible}, if $G_n$ is bijective on the set of all configurations of length $n$, for each $n\in \mathbb{N}$.
\end{definition}

\begin{definition}
\label{chap:semireversibility:def:irrev}
A CA with rule $R$ is called \textbf{ strictly irreversible}, if $G_n$ is not bijective on the set of all configurations of length $n$, for each $n\in \mathbb{N}$.
\end{definition}

\begin{definition}\label{chap:semireversibility:def:semirev} A CA with rule $R$ is called \textbf{semi-reversible}, if $G_n$ is bijective on the set of all configurations of length $n$, for some $n\in \mathbb{N}$.
\end{definition}

The algorithms by Amoroso and Patt \cite{Amoroso72} as well as of Sutner \cite{suttner91}, identify the reversible CAs of Definition~\ref{chap:semireversibility:def:rev} as reversible for infinite configurations and the strictly irreversible CAs of Definition~\ref{chap:semireversibility:def:irrev} as irreversible for infinite configurations.
These also identify semi-reversible CAs of Definition~\ref{chap:semireversibility:def:semirev}, as irreversible for infinite configurations. Therefore, if a CA with rule $R$ is \emph{reversible}, it is actually reversible for all cases (Case \ref{cs1} to \ref{cs4}), that is, $G_P$, $G$ and $G_F$ as well as $G_n$ for each $n \in \mathbb{N}$ are injective. Similarly, if a CA with local map $R$ is \emph{strictly irreversible}, it is irreversible for all cases, implying each of $G_n$, $G_P$, $G$ and $G_F$ is not injective.

Therefore, finite CAs can be classified into three sets -- (1) reversible, (2) strictly irreversible and (3) semi-reversible.
Note that, by the term \emph{reversible}, we mean a CA which is reversible for all sizes $n \in \mathbb{N}$ (Definition~\ref{chap:semireversibility:def:rev}) and by the term \emph{reversible for length $n$}, we mean a CA whose global function $G_n$ is bijective for that length $n$. These two notions are used throughout this dissertation.

\begin{proposition}\label{Chap:semireversible:Proposition:basicReversibility}
Let $\{P\} \vdash x$ and $G_{|x|}$ be bijective for length $|x|$.
\begin{enumerate}
\item Let $y$ be a configuration where $|x| = |y|$. Then  $G_{|x|}(x)= G_{|x|}(y)$ if and only if $x=y$.

\item For any $k\in \mathbb{N}$, $G_{|x|}(x)=G_{|x|}(\sigma^k(x))$ if and only if $x = \sigma^k(x)$;

\end{enumerate}
\end{proposition}
The proof is trivial. Hence omitted.

\begin{proposition}\label{Chap:semireversible:Proposition:blockP}
Let $\{P\} \vdash x$ where $|x|_P=1$ and $Y \vdash y$. If $ G_{|x|}(x) = y^k$, where $k>1$, then CA is irreversible for length $|y|$, or for length $|x|$.
\end{proposition}
\begin{proof}
Let us consider that $\{P\} \vdash x$ where $|x|_P=1$, $Y \vdash y$ and $ G_{|x|}(x) = y^k$ for $k>1$. Now, we can get following two cases - ($i$) the configuration $y$ of the CA of size $|y|$ has no predecessor, or ($ii$) $y$ has at least one predecessor, say $z$. For the first case, the CA is irreversible for the size $|y|$. And for the second case, the configuration $y^k$ has at least two predecessors -- $x$ and $z^k$. Hence irreversible for size $|x|$. 
\end{proof}

%
%
%

\begin{proposition}\label{Chap:semireversible:Theorem:irreversibility}
Let $R$ be a rule of a CA of length $n$. For the CA,
\begin{enumerate}
\item\label{irreversibility_c1} let $\{P\}=\{\{r\}\} \vdash p$ and $\{Q\}=\{ \{s\}\}\vdash q$. If $R[r]=R[s]$, then $G_{|p|}(p)=G_{|q|}(q)$, where $|p|=|q|$;

\item\label{irreversibility_c2} let $\{P\}=\{\{r_1,r_2,\cdots,r_l\}\}\vdash p$ and $l>1$. If $R[r_1]=R[r_2]=\cdots=R[r_l]$, then, $G_{|p|}(p)=G_{|p|}(\sigma^i(p))$, for any $i,k \in \mathbb{N}$ where $|p|=k\times l$.


\end{enumerate}
\end{proposition}

\begin{proof}
Let $\{P\}=\{\{r\}\} \vdash p$ and $\{Q\}=\{ \{s\}\} \vdash q$, where $|p|=|q|$. So, $\tilde{p}=r^n$  and $\tilde{q}=s^n$, $n \in \mathbb{N}$. If $R[r]=R[s]$, then obviously, $G_{|p|}(p)=(R(r))^n=(R(s))^n=G_{|q|}(q)$, for any $n\in \mathbb{N}$.

Let $\{P\}=\{\{r_1,r_2,\cdots,r_l\}\}\vdash p$ and $l>1$. Therefore, $|P|=l$ and $|p| = k\times l$, $k \in \mathbb{N}$. If $R[r_1]=R[r_2]=\cdots=R[r_l]$, then, $p = \sigma^i(p)$ for all $i \in \mathbb{N}$. Hence, $G_{|p|}(p)=G_{|p|}(\sigma^i(p))$ for any $|p|=k\times l$, where $k,i\in \mathbb{N}$,
\end{proof}

%
%
%
%
%
%
%
%
%

However, if a CA rule satisfies condition~\ref{irreversibility_c1} of Proposition~\ref{Chap:semireversible:Theorem:irreversibility}, it essentially means that, the CA is strictly irreversible, and if it satisfies condition~\ref{irreversibility_c2} of Proposition~\ref{Chap:semireversible:Theorem:irreversibility}, then the CA is irreversible for $n=k\times i$, $k\in \mathbb{N}$. 

\begin{example}
For any $n$-cell ECA, the singleton primary RMT sets are $\{0\}$ and $\{7\}$. So, if RMT $0$ and RMT $7$ have same next state values, that is, $R[0]=R[7]$, the CA is strictly irreversible. 
For instance, the ECAs $90(01011010)$ and $30(00011110)$ are strictly irreversible.
However, if RMT $2$ and RMT $5$ have same next state value, that means, $G_n((01)^n) = G_n((10)^n)$, where $\{\{2,5\}\} \vdash (01)^n$ and $\{\{2,5\}\} \vdash (10)^n$. Therefore, the CA is irreversible for $n=2k$, $k\in \mathbb{N}$. For example, the ECAs $45(00101101)$ and $75(01001011)$ are irreversible for $n=2k$, $k \in \mathbb{N}$. Similarly, if $R[1]=R[2]=R[4]$ or $R[3]=R[5]=R[6]$, 
the CA is irreversible for $n=3k$, $k\in \mathbb{N}$. If $R[1]=R[3]=R[4]=R[6]$,
the CA is irreversible for $n=4k$, $k\in \mathbb{N}$.
\end{example}

\begin{theorem}\label{Chap:semi-reversibility:Th:strictirreversibility}
A CA is strictly irreversible if and only if it is irreversible for $n=1$.
\end{theorem}

\begin{proof} \begin{description}[leftmargin=1ex]
\item\noindent\underline{\textit{If Part:}} Let us consider that a CA with rule $R$ is irreversible for $n=1$.
That means, for the CA, there exists at least two configurations $x$ and $y$ such that $G_{|x|}(x) = G_{|y|}(y)$. As $|x|=|y|=1$, these configurations are homogeneous and followed from singleton primary RMT sets. Let, $\{\{r\}\}\vdash x$ and $\{\{s\}\}\vdash y$. 
Now, if $\{\{r\}\}\vdash p$ and $\{ \{s\}\}\vdash q$, then, $p=x^k$ and $q=y^k$ where $|x|=|y|=1$ and $k \in \mathbb{N}$. Therefore, $G_{|p|}(p)=G_{|x^k|}(x^k) = (G_{|x|}(x))^k = (G_{|y|}(y))^k = G_{|y^k|}(y^k)= G_{|q|}(q)$. So, for any CA size $n\in \mathbb{N}$, there exists at least two configurations $p$ and $q$ which have same successor. Hence, the CA is irreversible for each $n \in \mathbb{N}$; that is, it is strictly irreversible.



\item\noindent\underline{\textit{Only if Part:}} Assume that a CA is strictly irreversible. Obviously, it is irreversible for $n=1$.
\end{description}\end{proof}

Table~\ref{Chap:semireversible:tab:ECA_rev_rules} gives the minimal rules (for the minimal ECA rules and their equivalents, see pages 485--557 of \cite{wolfram86}) for reversible, strictly irreversible and semi-reversible ECAs. There are only $6$ ECA rules, which are reversible according to Definition~\ref{chap:semireversibility:def:rev} and $128$ rules which are strictly irreversible (irreversible for each $n \in \mathbb{N}$). Other $122$ rules are semi-reversible (reversible for some $n \in \mathbb{N}$).
  
\begin{table}[h]
  \setlength{\tabcolsep}{1.5pt}
  \begin{center}
  \caption{The reversible, strictly irreversible and semi-reversible rules of ECAs. Here, only minimal representation \cite{wolfram86} of the rules are written and non-trivial semi-reversible rules are represented in bold face}
  \label{Chap:semireversible:tab:ECA_rev_rules}
  \resizebox{0.8\textwidth}{1.5cm}{
 \fbox{
  \begin{tabular}{m{3cm}|m{11.0cm}}
Reversible ECAs & $15$  $51$  $170$  $204$ \\
\hline
\hline
Strictly Irreversible ECAs & $0$ $2$ $4$ $6$ $8$ $10$ $12$ $14$ $18$ $22$ $24$ $26$ $28$ $30$ $32$ $34$ $36$ $38$ $40$ $42$ $44$ $46$ $50$ $54$ $56$ $58$ $60$ $62$ $72$ $74$ $76$ $78$ $90$ $94$ $104$ $106$ $108$ $110$ $122$ $126$ $130$ \\
  \hline
  \hline
   Semi-reversible ECAs & $1$ $3$ $5$  $7$ $9$ $11$ $13$ $19$ $25$ $33$ $35$ $37$ $41$ $73$         $23$ $27$ $29$ $43$ $\mathbf{45}$ $57$ $77$ $\mathbf{105}$ $128$ $132$ $134$ $136$ $138$ $140$  $142$ $146$ $\mathbf{150}$ $152$ $\mathbf{154}$ $156$ $160$ $162$ $164$ $168$ $172$ $178$ $184$ $200$ $232$\\
  \end{tabular}
 } }
  \end{center}
  \vspace{-1.0em}
  \end{table}
 
If a CA is irreversible for $n=1$, it is strictly irreversible (Theorem~\ref{Chap:semi-reversibility:Th:strictirreversibility}). Otherwise, it is either reversible for each $n \in \mathbb{N}$, or semi-reversible. However, among these semi-reversible CAs, some CAs may be present, which are reversible only for a (small) finite set of sizes, and irreversible for an infinite set of sizes. We name such CAs as \textit{trivially} semi-reversible CAs.

\begin{definition}\label{Chap:semireversible:def:triv_semi}
A CA is \textbf{trivially} semi-reversible, if the CA is semi-reversible for a finite set of sizes.
\end{definition}

In Table~\ref{Chap:semireversible:tab:ECA_rev_rules}, the semi-reversible rules written in plain face are trivial semi-revrsible rules. Many of these rules are unbalanced and all of the CAs are not surjective when defined over infinite lattice. However, some CAs exist which are not trivial semi-reversible. For them, an infinite number of sizes exist for which the CA is reversible. In Table~\ref{Chap:semireversible:tab:ECA_rev_rules}, such CAs are marked by bold face. We call these CAs \emph{non-trivial} semi-reversible CAs. 

\begin{definition}\label{Chap:semireversible:def:nontriv_semi}
A semi-reversible CA is \textbf{non-trivial}, if the CA is semi-reversible for an infinite set of sizes.
\end{definition}

%
%

\begin{example}
The ECA rules $45~(00101101)$ and $154~(10011010)$ are reversible for $n =1,3,5,7,\cdots$, that is, when $n$ is odd. The rules $105~(01101001)$ and $150~(10010110)$ are reversible for $n=1,2,4,5,7,8,\cdots $, that is, when $n\ne 3k$, $k\in \mathbb{N}$. Hence, these are \emph{non-trivial} semi-reversible CAs. However, the trivial semi-reversible ECA rules $23~(00010111)$ and $232~(11101000)$ are reversible for $n$ = $1$ and $2$ only, rule $27~(00011011)$ is reversible for $n$ = $1$ and $3$ only and rules $29~(00011101) $, $43~(00101011)$, $57~(00111001)$, $77~(01001101)$, $142~(10001110)$, $156~(10011100)$, $172~(10101100)$, $178~(10110010)$ and $184~(10111000)$ are reversible only for $n$ =$1$, $2$ and $3$.

Let us consider two $4$-neighborhood $2$-state CAs $0101101010101001$ and $1011010011110000$. The first one is trivially semi-reversible, as it is reversible for only $n \in \{1,2\}$, whereas the other one is non-trivially semi-reversible because it is reversible for each odd $n$.
\end{example}


From the discussion of this section, it is evident that, strictly irreversible CAs can be identified beforehand by observing the  singleton primary RMT sets of the CAs. If RMTs of two singleton primary RMT sets have same next state value, the CA is strictly irreversible. However, the following questions need to be addressed: 
\begin{enumerate}
\item How can we identify whether a finite CA is reversible for each $n \in \mathbb{N}$ from bijectivity of $G_n$ for some $n$?
\item Can we identify non-trivial semi-reversibility of a CA from a $G_n$ and list the $n$ for which the CA is reversible? These non-trivial semi-reversible CAs are of interest to us. Obviously, for each of these CAs, infinite number of sizes exist for which the CA is reversible. So, our target is to find this infinite set of sizes by some expression(s).
\item How can we identify a trivially semi-reversible CA?
\end{enumerate}
To address these questions, we again take help of the mathematical tool, named reachability tree, as discussed in the next section.
%

\section{The Reachability Tree and Reversibility}
\label{Chap:semireversible:sec:rtree}
\noindent In this section, we use the discrete mathematical tool, named \emph{reachability tree}, to analyze the reversibility behavior of any $1$-D finite CA. As already discussed in Section~\ref{chap:reversibility:Sec:rtree} of Chapter~\ref{Chap:reversibility} (Page~\pageref{Chap:reversibility:def:tree_3}), reachability tree depicts all reachable configurations of an $n$-cell CA and helps to efficiently decide whether a given $n$-cell CA is reversible or not. We now redefine the reachability tree considering neighborhood size as $m$ (which was $3$ in Definition~\ref{Chap:reversibility:def:tree_3}).

\begin{definition} \label{chap:semireversibility:def:tree}
Reachability tree of an $n$-cell $m$-neighborhood $d$-state CA is a rooted and edge-labeled $d$-ary tree with $(n+1)$ levels where 
each node $ N_{i.j} ~ (0 \leq i \leq n,~ 0 \leq j \leq d^{i}-1)$ is an ordered list of $d^{m-1}$ sets of RMTs, and the root $N_{0.0}$ is the ordered list of all sets of sibling RMTs. We denote the edges between $N_{i.j} ~ (0 \leq i \leq n-1,~ 0 \leq j \leq d^{i}-1)$ and its possible $d$ children as $E_{i.dj+x} = ( N_{i.j}, N_{i+1.dj+x}, l_{i.dj+x} )$ where $l_{i.dj+x}$ is the label of the edge and $0 \leq x \leq d-1$. Like nodes, the labels are also ordered list of $d^{m-1}$ sets of RMTs. Let us consider that ${\Gamma_{p}}^{N_{i.j}}$ is the $p^{th}$ set of the node $N_{i.j}$, and ${\Gamma_{q}}^{E_{i.dj+x}}$ is the $q^{th}$ set of the label on edge $E_{i.dj+x}$ $(0 \leq p,q \leq d^{m-1} -1)$. So,  $N_{i.j} = ( {\Gamma_{p}}^{N_{i.j}})_{0 \leq p \leq d^{m-1}-1}$ and $l_{i.dj+x} =  ( {\Gamma_{q}}^{E_{i.dj+x}})_{0 \leq q \leq d^{m-1}-1}$. Following are the relations which exist in the tree:

\begin{enumerate}
\item \label{rtd1} [For root] $N_{0.0} = ({\Gamma_{k}^{N_{0.0}}})_{0 \leq k \leq d^{m-1}-1}$, where ${\Gamma_{k}^{N_{0.0}}} = Sibl_k$.

\item \label{rtd2} $\forall r \in {\Gamma_{k}^{N_{i.j}}}, ~ r$ is included in ${\Gamma_{k}^{E_{i.dj +x}}}$, if $R[r] = x, (0 \leq x \leq d-1)$, where $R$ is the rule of the CA. That means, $ {\Gamma_{k}^{N_{i.j}}} = \bigcup\limits_{x} {\Gamma_{k}^{E_{i.dj+x}}}$, $(0 \leq k \leq d^{m-1}-1, 0 \leq i \leq n-1,~ 0 \leq j \leq d^{i}-1)$.

\item \label{rtd4}$\forall r$, if $r \in {\Gamma_{k}^{E_{i.dj+x}}}$, then RMTs of $Sibl_p$, that is $\lbrace d.r \pmod{d^{m}}, d.r+1 \pmod{d^m}, \cdots, d.r+(d-1) \pmod{d^m} \rbrace$ are in ${\Gamma_{k}^{N_{i+1.dj+x}}}$, where $0\leq x \leq d-1$, $0 \leq i \leq n-1$, $0 \leq j \leq d^{i}-1$ and $r \equiv p \pmod{d^{m-1}}$.

\item \label{rtd5} [For level $n-\iota$, $1 \leq \iota \leq m-1$ ] ${\Gamma_{k}^{N_{n-\iota.j}}} = \lbrace y ~|~$ if $ r \in {\Gamma_{k}^{E_{n-\iota-1.j}}} $ then $ y \in \lbrace d.r \pmod{d^m}, d.r+1 \pmod{d^m}, \cdots, d.r+(d-1) \pmod{d^m}\rbrace \cap \lbrace i, i+d^{m-\iota}, i+2d^{m-\iota}, \cdots, i+(d^\iota-1)d^{m-\iota} \rbrace \rbrace$, where $i= \floor{\frac{k}{d^{\iota-1}}}$, $0 \leq k \leq d^{m-1}-1$ and $0 \leq j \leq d^{n-\iota}-1 $.
\end{enumerate}
\end{definition}

Note that, the nodes of level $n-\iota$, $1 \leq \iota \leq m-1$, are different from other intermediate nodes (Points~\ref{rtd5} of Definition~\ref{chap:semireversibility:def:tree}). At level $n-\iota$, $1 \leq \iota \leq m-1$, only some specific RMTs can be present, these RMTs are called \emph{valid} RMTs for this level. Recall that, in a reachability tree, the root is at level $0$ and the leaves are at level $n$. A sequence of edges from root to leaf, $\langle E_{0.j_1}, E_{1.j_2}, ...,  E_{n-1.j_n}\rangle$, where $0 \leq j_i \leq d^i -1, 1 \leq i \leq n$, constitutes an RMT sequence, where each edge represents a cell's state. If an edge (resp. node) has no RMT, then it is non-reachable edge (resp. node) and represents a non-reachable configuration.

\begin{example}\label{Chap:semireversibility:ex:tree}
The reachability tree for a $5$-cell $4$-neighborhood $2$-state CA with rule $1010101010101010$ is shown in Figure~\ref{Chap:semireversible:fig:rt2}. The details of the nodes and edges for this reachability tree is shown in Table~\ref{Chap:semireversible:Tab:ruleRTree}. This tree contains no non-reachable nodes or edges, so the tree is complete. One reachable configuration $11010$ is represented by the sequence $\langle E_{0.1}, E_{1.3}, E_{2.6}, E_{3.13}, E_{4.26} \rangle$. 

\begin{figure}[hbtp]
  \centering
    \includegraphics[width= 6.0in, height = 3.0in]{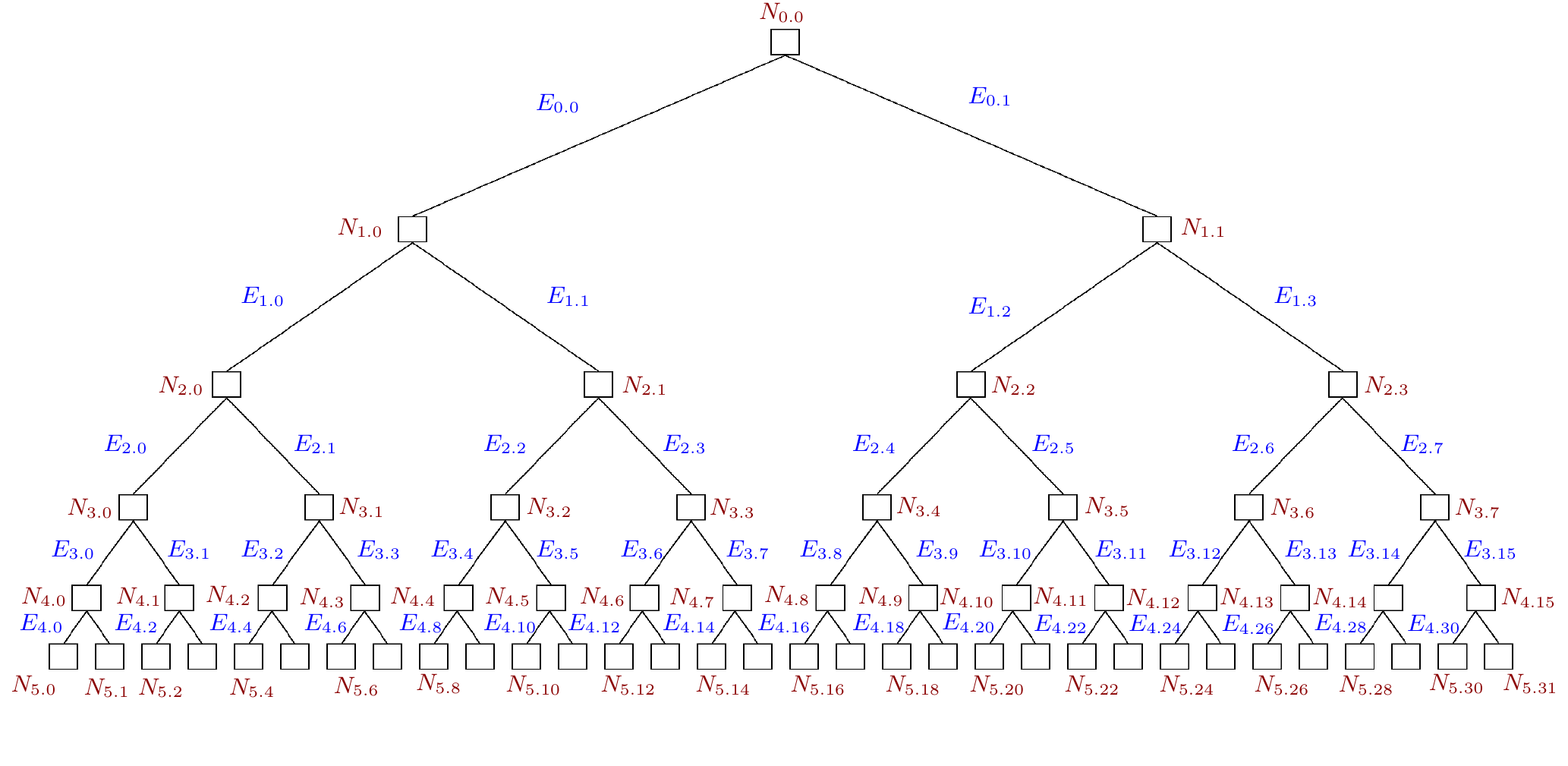} 
   \caption{Reachability tree for $4$-neighborhood $2$-state $5$-cell CA $1010101010101010$}
    \label{Chap:semireversible:fig:rt2}
\end{figure}

\begin{scriptsize}\setlength\tabcolsep{5pt}
	\centering
\begin{longtable}{|c|p{6.5cm}|p{6cm}|}
\caption{Reachability Tree for CA $4$-neighborhood $2$-state CA $1010101010101010$ (Figure~\ref{Chap:semireversible:fig:rt2}) with $n = 5$}
\label{Chap:semireversible:Tab:ruleRTree}\\
\hline
\textbf{Level} & \textbf{Nodes} & \textbf{Edges} \\
\hline
\endfirsthead
\multicolumn{3}{c}%
{\tablename\ \thetable\ -- \textit{Continued from previous page}} \\
\hline
\textbf{Level} & \textbf{Nodes} & \textbf{Edges} \\
\hline
\endhead
\hline \multicolumn{3}{r}{\textit{Continued on next page}} \\
\endfoot
\hline
\endlastfoot
$0$ & $N_{0.0} = (\{0,1\}$, $\{2,3\}$, $\{4,5\}$, $\{6,7\}$, $\{8,9\}$, $\{10,11\}$, $\{12,13\}$, $\{14,15\})$ & NA \\ 
\hline 
\multirow{2}{*}{$1$} & $N_{1.0}$ = $(\{0,1\}$, $\{4,5\}$, $\{8,9\}$, $\{12,13\}$, $\{0,1\}$, $\{4,5\}$, $\{8,9\}$, $\{12,13\})$ & $E_{0.0}$ = $(\{0\}$, $\{2\}$, $\{4\}$, $\{6\}$, $\{8\}$, $\{10\}$, $\{12\}$, $\{14\})$ \\
 \hhline{~--} 
& $N_{1.1}$ = $(\{2,3\}$, $\{6,7\}$, $\{10,11\}$, $\{14,15\}$, $\{2,3\}$, $\{6,7\}$, $\{10,11\}$, $\{14,15\})$& $E_{0.1}$ = $(\{1\}$, $\{3\}$, $\{5\}$, $\{7\}$, $\{9\}$, $\{11\}$, $\{13\}$, $\{15\})$ \\
\hline
 \multirow{4}{*}{$2$} & $N_{2.0}$ = $(\{0\}$, $\{8\}$, $\{0\}$, $\{8\}$, $\{1\}$, $\{9\}$, $\{1\}$, $\{9\})$ & $E_{1.0}$ = $(\{0\}$, $\{4\}$, $\{8\}$, $\{12\}$, $\{0\}$, $\{4\}$, $\{8\}$, $\{12\})$ \\
  \hhline{~--} 
 & $N_{2.1}$ = $(\{2\}$, $\{10\}$, $\{2\}$, $\{10\}$, $\{3\}$, $\{11\}$, $\{3\}$, $\{11\})$ & $E_{1.1}$ = $(\{1\}$, $\{5\}$, $\{9\}$, $\{13\}$, $\{1\}$, $\{5\}$, $\{9\}$, $\{13\})$ \\
  \hhline{~--} 
 & $N_{2.2}$ = $(\{4\}$, $\{12\}$, $\{4\}$, $\{12\}$, $\{5\}$, $\{13\}$, $\{5\}$, $\{13\})$ & $E_{1.2}$ = $(\{2\}$, $\{6\}$, $\{10\}$, $\{14\}$, $\{2\}$, $\{6\}$, $\{10\}$, $\{14\})$ \\
  \hhline{~--} 
 & $N_{2.3}$ = $(\{6\}$, $\{14\}$, $\{6\}$, $\{14\}$, $\{7\}$, $\{15\}$, $\{7\}$, $\{15\})$ & $E_{1.3}$ = $(\{3\}$, $\{7\}$, $\{11\}$, $\{15\}$, $\{3\}$, $\{7\}$, $\{11\}$, $\{15\})$ \\
 \hline
\multirow{8}{*}{$3$} & $N_{3.0}$ = $(\{0\}$, $\{0\}$, $\{1\}$, $\{1\}$, $\emptyset,\emptyset, \emptyset,\emptyset)$ & $E_{2.0}$ = $(\{0\}$, $\{8\}$, $\{0\}$, $\{8\}$, $\emptyset,\emptyset,\emptyset,\emptyset)$\\
 \hhline{~--} 
& $N_{3.1}$ = $(\emptyset,\emptyset, \emptyset,\emptyset, \{ 2\}$, $\{2\}$, $\{3\}$, $\{3\})$ & $E_{2.1}$ = $(\emptyset,\emptyset, \emptyset,\emptyset$,$\{ 1\}$, $\{9\}$, $\{1\}$, $\{9\})$ \\
 \hhline{~--} 
& $N_{3.2}$ = $(\{4\}$, $\{4\}$, $\{5\}$, $\{5\}$, $\emptyset,\emptyset, \emptyset,\emptyset)$ &  $E_{2.2}$ = $(\{2\}$, $\{10\}$, $\{2\}$, $\{10\}$, $\emptyset,\emptyset, \emptyset,\emptyset)$\\
 \hhline{~--} 
& $N_{3.3}$ = $(\emptyset,\emptyset, \emptyset,\emptyset, \{ 6\}$, $\{6\}$, $\{7\}$, $\{7\})$ & $E_{2.3}$ = $(\emptyset,\emptyset, \emptyset,\emptyset$, $\{3\}$, $\{11\}$, $\{3\}$, $\{11\}$ \\
 \hhline{~--} 
& $N_{3.4}$ = $(\{8\}$, $\{8\}$, $\{9\}$, $\{9\}$, $\emptyset,\emptyset, \emptyset,\emptyset)$ & $E_{2.4}$ = $(\{4\}$, $\{12\}$, $\{4\}$, $\{12\}$, $\emptyset,\emptyset, \emptyset,\emptyset)$\\
 \hhline{~--} 
& $N_{3.5}$ = $(\emptyset,\emptyset, \emptyset,\emptyset, \{ 10\}$, $\{10\}$, $\{11\}$, $\{11\})$ & $N_{3.5}$ = $(\emptyset,\emptyset, \emptyset,\emptyset, \{5\}$, $\{13\}$, $\{5\}$, $\{13\})$ \\
 \hhline{~--} 
& $N_{3.6}$ = $(\{12\}$, $\{12\}$, $\{13\}$, $\{13\}$, $\emptyset,\emptyset, \emptyset,\emptyset)$ & $E_{2.6}$ = $(\{6\}$, $\{14\}$, $\{6\}$, $\{14\}$, $\emptyset,\emptyset, \emptyset,\emptyset)$\\
 \hhline{~--} 
& $N_{3.7}$ = $(\emptyset,\emptyset, \emptyset,\emptyset, \{ 14\}$, $\{14\}$, $\{15\}$, $\{15\})$ & $E_{2.7}$ = $(\emptyset,\emptyset, \emptyset,\emptyset,\{7\}$, $\{15\}$, $\{7\}$, $\{15\})$ \\
\hline
\multirow{16}{*}{$4$} & $N_{4.0}$ = $(\{0\}$, $\{1\}$, $\emptyset,\emptyset, \emptyset,\emptyset, \emptyset,\emptyset)$ &  $E_{3.0}$ = $(\{0\}$, $\{0\}$, $\emptyset,\emptyset, \emptyset,\emptyset, \emptyset,\emptyset)$\\
 \hhline{~--} 
& $N_{4.1}$ = $(\emptyset,\emptyset, \{ 2\}$, $\{3\}$, $\emptyset,\emptyset, \emptyset,\emptyset)$ & $E_{3.1}$ = $(\emptyset,\emptyset, \{1\}$, $\{1\}$, $\emptyset,\emptyset, \emptyset,\emptyset)$\\
 \hhline{~--} 
& $N_{4.2}$ = $(\emptyset,\emptyset, \emptyset,\emptyset, \{ 4\}$, $\{5\}$, $\emptyset,\emptyset)$ & $E_{3.2}$ = $(\emptyset,\emptyset, \emptyset,\emptyset, \{ 2\}$, $\{2\}$, $\emptyset,\emptyset)$\\
 \hhline{~--} 
& $N_{4.3}$ = $(\emptyset,\emptyset, \emptyset,\emptyset, \emptyset,\emptyset, \{ 6\}$, $\{7\})$ & $E_{3.3}$ = $(\emptyset,\emptyset, \emptyset,\emptyset, \emptyset,\emptyset, \{ 3\}$, $\{3\})$ \\
 \hhline{~--} 
& $N_{4.4}$ = $(\{8\}$, $\{9\}$, $\emptyset,\emptyset, \emptyset,\emptyset, \emptyset,\emptyset)$ &  $E_{3.4}$ = $(\{4\}$, $\{4\}$, $\emptyset,\emptyset, \emptyset,\emptyset, \emptyset,\emptyset)$\\
 \hhline{~--} 
& $N_{4.5}$ = $(\emptyset,\emptyset, \{ 10\}$, $\{11\}$, $\emptyset,\emptyset, \emptyset,\emptyset)$ & $E_{3.5}$ = $(\emptyset,\emptyset, \{ 5\}$, $\{5\}$, $\emptyset,\emptyset, \emptyset,\emptyset)$\\
 \hhline{~--} 
& $N_{4.6}$ = $(\emptyset,\emptyset, \emptyset,\emptyset, \{ 12\}$, $\{13\}$, $\emptyset,\emptyset)$ & $E_{3.6}$ = $(\emptyset,\emptyset, \emptyset,\emptyset, \{ 6\}$, $\{6\}$, $\emptyset,\emptyset)$ \\
 \hhline{~--} 
& $N_{4.7}$ = $(\emptyset,\emptyset, \emptyset,\emptyset, \emptyset,\emptyset, \{ 14\}$, $\{15\})$ & $E_{3.7}$ = $(\emptyset,\emptyset, \emptyset,\emptyset, \emptyset,\emptyset, \{ 7\}$, $\{7\})$ \\
 \hhline{~--} 
& $N_{4.8}$ = $(\{0\}$, $\{1\}$, $\emptyset,\emptyset, \emptyset,\emptyset, \emptyset,\emptyset)$ & $E_{3.8}$ = $(\{8\}$, $\{8\}$, $\emptyset,\emptyset, \emptyset,\emptyset, \emptyset,\emptyset)$\\
 \hhline{~--} 
& $N_{4.9}$ = $(\emptyset,\emptyset, \{ 2\}$, $\{3\}$, $\emptyset,\emptyset, \emptyset,\emptyset)$ & $E_{3.9}$ = $(\emptyset,\emptyset, \{ 9\}$, $\{9\}$, $\emptyset,\emptyset, \emptyset,\emptyset)$\\
 \hhline{~--} 
& $N_{4.10}$ = $(\emptyset,\emptyset, \emptyset,\emptyset, \{ 4\}$, $\{5\}$, $\emptyset,\emptyset)$ & $E_{3.10}$ = $(\emptyset,\emptyset, \emptyset,\emptyset, \{ 10\}$, $\{10\}$, $\emptyset,\emptyset)$\\
 \hhline{~--} 
& $N_{4.11}$ = $(\emptyset,\emptyset, \emptyset,\emptyset, \emptyset,\emptyset, \{ 6\}$, $\{7\})$ & $E_{3.11}$ = $(\emptyset,\emptyset, \emptyset,\emptyset, \emptyset,\emptyset, \{ 11\}$, $\{11\})$ \\
 \hhline{~--} 
& $N_{4.12}$ = $(\{8\}$, $\{9\}$, $\emptyset,\emptyset, \emptyset,\emptyset, \emptyset,\emptyset)$ & $E_{3.12}$ = $(\{12\}$, $\{12\}$, $\emptyset,\emptyset, \emptyset,\emptyset, \emptyset,\emptyset)$\\
 \hhline{~--} 
& $N_{4.13}$ = $(\emptyset,\emptyset, \{ 10\}$, $\{11\}$, $\emptyset,\emptyset, \emptyset,\emptyset)$ & $E_{3.13}$ = $(\emptyset,\emptyset, \{ 13\}$, $\{13\}$, $\emptyset,\emptyset, \emptyset,\emptyset)$ \\
 \hhline{~--} 
& $N_{4.14}$ = $(\emptyset,\emptyset, \emptyset,\emptyset, \{ 12\}$, $\{13\}$, $\emptyset,\emptyset)$ & $E_{3.14}$ = $(\emptyset,\emptyset, \emptyset,\emptyset, \{ 14\}$, $\{14\}$, $\emptyset,\emptyset)$\\
 \hhline{~--} 
& $N_{4.15}$ = $(\emptyset,\emptyset, \emptyset,\emptyset, \emptyset,\emptyset, \{ 14\}$, $\{15\})$ & $E_{3.15}$ = $(\emptyset,\emptyset, \emptyset,\emptyset, \emptyset,\emptyset, \{ 15\}$, $\{15\})$ \\
\hline

\multirow{32}{*}{$5$} & $N_{5.0}$ = $(\{0,1\}$, $\emptyset$, $\emptyset,\emptyset, \emptyset,\emptyset, \emptyset,\emptyset)$ & $E_{4.0}$ = $(\{0\}$, $\emptyset$, $\emptyset,\emptyset, \emptyset,\emptyset, \emptyset,\emptyset)$ \\
 \hhline{~--} 
& $N_{5.1}$ = $(\emptyset$, $\{2,3\}$, $\emptyset,\emptyset, \emptyset,\emptyset, \emptyset,\emptyset)$ & $E_{4.1}$ = $(\emptyset$, $\{1\}$, $\emptyset,\emptyset, \emptyset,\emptyset, \emptyset,\emptyset)$ \\
 \hhline{~--} 
& $N_{5.2}$ = $(\emptyset,\emptyset, \{ 4,5\}$, $\emptyset$, $\emptyset,\emptyset, \emptyset,\emptyset)$ & $E_{4.2}$ = $(\emptyset,\emptyset, \{ 2\}$, $\emptyset$, $\emptyset,\emptyset, \emptyset,\emptyset)$ \\
 \hhline{~--} 
& $N_{5.3}$ = $(\emptyset,\emptyset, \emptyset$, $\{6,7\}$, $\emptyset,\emptyset, \emptyset,\emptyset)$ & $E_{4.3}$ = $(\emptyset,\emptyset, \emptyset$, $\{3\}$, $\emptyset,\emptyset, \emptyset,\emptyset)$\\
 \hhline{~--} 
& $N_{5.4}$= $(\emptyset,\emptyset, \emptyset,\emptyset, \{ 8,9\}$, $\emptyset$, $\emptyset,\emptyset)$ & $E_{4.4}$= $(\emptyset,\emptyset, \emptyset,\emptyset, \{ 4\}$, $\emptyset$, $\emptyset,\emptyset)$\\
 \hhline{~--} 
& $N_{5.5}$ = $(\emptyset,\emptyset, \emptyset,\emptyset, \emptyset$, $\{10,11\}$, $\emptyset,\emptyset)$ & $E_{4.5}$ = $(\emptyset,\emptyset, \emptyset,\emptyset, \emptyset$, $\{5\}$, $\emptyset,\emptyset)$\\
 \hhline{~--} 
& $N_{5.6}$ =$(\emptyset,\emptyset, \emptyset,\emptyset, \emptyset,\emptyset, \{ 12,13\}$, $\emptyset)$ & $E_{4.6}$ =$(\emptyset,\emptyset, \emptyset,\emptyset, \emptyset,\emptyset, \{6\}$, $\emptyset)$\\
 \hhline{~--} 
& $N_{5.7}$ =$(\emptyset,\emptyset, \emptyset,\emptyset, \emptyset,\emptyset, \emptyset$, $\{14,15\})$ & $E_{4.7}$ =$(\emptyset,\emptyset, \emptyset,\emptyset, \emptyset,\emptyset, \emptyset$, $\{7\})$\\
\hhline{~--} 
& $N_{5.8}$ = $(\{0,1\}$, $\emptyset$, $\emptyset,\emptyset, \emptyset,\emptyset, \emptyset,\emptyset)$ & $E_{4.8}$ = $(\{8\}$, $\emptyset$, $\emptyset,\emptyset, \emptyset,\emptyset, \emptyset,\emptyset)$  \\
 \hhline{~--} 
& $N_{5.9}$ = $(\emptyset$, $\{2,3\}$, $\emptyset,\emptyset, \emptyset,\emptyset, \emptyset,\emptyset)$ & $E_{4.9}$ = $(\emptyset$, $\{9\}$, $\emptyset,\emptyset, \emptyset,\emptyset, \emptyset,\emptyset)$ \\
 \hhline{~--} 
& $N_{5.10}$ = $(\emptyset,\emptyset, \{ 4,5\}$, $\emptyset$, $\emptyset,\emptyset, \emptyset,\emptyset)$ & $E_{4.10}$ = $(\emptyset,\emptyset, \{10\}$, $\emptyset$, $\emptyset,\emptyset, \emptyset,\emptyset)$\\
 \hhline{~--} 
& $N_{5.11}$ = $(\emptyset,\emptyset, \emptyset$, $\{6,7\}$, $\emptyset,\emptyset, \emptyset,\emptyset)$ & $E_{4.11}$ = $(\emptyset,\emptyset, \emptyset$, $\{11\}$, $\emptyset,\emptyset, \emptyset,\emptyset)$ \\
 \hhline{~--} 
& $N_{5.12}$= $(\emptyset,\emptyset, \emptyset,\emptyset, \{ 8,9\}$, $\emptyset$, $\emptyset,\emptyset)$ & $E_{4.12}$= $(\emptyset,\emptyset, \emptyset,\emptyset, \{ 12\}$, $\emptyset$, $\emptyset,\emptyset)$\\
 \hhline{~--} 
& $N_{5.13}$ = $(\emptyset,\emptyset, \emptyset,\emptyset, \emptyset$, $\{10,11\}$, $\emptyset,\emptyset)$ & $E_{4.13}$ = $(\emptyset,\emptyset, \emptyset,\emptyset, \emptyset$, $\{13\}$, $\emptyset,\emptyset)$\\
 \hhline{~--} 
& $N_{5.14}$ =$(\emptyset,\emptyset, \emptyset,\emptyset, \emptyset,\emptyset, \{ 12,13\}$, $\emptyset)$ & $E_{4.14}$ =$(\emptyset,\emptyset, \emptyset,\emptyset, \emptyset,\emptyset, \{14\}$, $\emptyset)$\\
 \hhline{~--} 
& $N_{5.15}$ = =$(\emptyset,\emptyset, \emptyset,\emptyset, \emptyset,\emptyset, \emptyset$, $\{14,15\})$ & $E_{4.15}$ =$(\emptyset,\emptyset, \emptyset,\emptyset, \emptyset,\emptyset, \emptyset$, $\{15\})$\\
 \hhline{~--} 
& $N_{5.16}$ = $(\{0,1\}$, $\emptyset$, $\emptyset,\emptyset, \emptyset,\emptyset, \emptyset,\emptyset)$ & $E_{4.16}$ = $(\{0\}$, $\emptyset$, $\emptyset,\emptyset, \emptyset,\emptyset, \emptyset,\emptyset)$  \\
 \hhline{~--} 
& $N_{5.17}$ = $(\emptyset$, $\{2,3\}$, $\emptyset,\emptyset, \emptyset,\emptyset, \emptyset,\emptyset)$ & $E_{4.17}$ = $(\emptyset$, $\{1\}$, $\emptyset,\emptyset, \emptyset,\emptyset, \emptyset,\emptyset)$ \\
 \hhline{~--} 
& $N_{5.18}$ = $(\emptyset,\emptyset, \{ 4,5\}$, $\emptyset$, $\emptyset,\emptyset, \emptyset,\emptyset)$ & $E_{4.18}$ = $(\emptyset,\emptyset, \{ 2\}$, $\emptyset$, $\emptyset,\emptyset, \emptyset,\emptyset)$\\
& $N_{5.19}$ = $(\emptyset,\emptyset, \emptyset$, $\{6,7\}$, $\emptyset,\emptyset, \emptyset,\emptyset)$ & $E_{4.19}$ = $(\emptyset,\emptyset, \emptyset$, $\{3\}$, $\emptyset,\emptyset, \emptyset,\emptyset)$\\
 \hhline{~--} 
& $N_{5.20}$= $(\emptyset,\emptyset, \emptyset,\emptyset, \{ 8,9\}$, $\emptyset$, $\emptyset,\emptyset)$ & $E_{4.20}$= $(\emptyset,\emptyset, \emptyset,\emptyset, \{ 4\}$, $\emptyset$, $\emptyset,\emptyset)$\\
 \hhline{~--} 
& $N_{5.21}$ = $(\emptyset,\emptyset, \emptyset,\emptyset, \emptyset$, $\{10,11\}$, $\emptyset,\emptyset)$ & $E_{4.21}$ = $(\emptyset,\emptyset, \emptyset,\emptyset, \emptyset$, $\{5\}$, $\emptyset,\emptyset)$\\
 \hhline{~--} 
& $N_{5.22}$ =$(\emptyset,\emptyset, \emptyset,\emptyset, \emptyset,\emptyset, \{ 12,13\}$, $\emptyset)$ & $E_{4.22}$ =$(\emptyset,\emptyset, \emptyset,\emptyset, \emptyset,\emptyset, \{6\}$, $\emptyset)$\\
 \hhline{~--} 
& $N_{5.23}$ = $(\emptyset,\emptyset, \emptyset,\emptyset, \emptyset,\emptyset, \emptyset$, $\{14,15\})$ & $E_{4.23}$ =$(\emptyset,\emptyset, \emptyset,\emptyset, \emptyset,\emptyset, \emptyset$, $\{7\})$\\
 \hhline{~--} 
& $N_{5.24}$ = $(\{0,1\}$, $\emptyset$, $\emptyset,\emptyset, \emptyset,\emptyset, \emptyset,\emptyset)$ & $E_{4.24}$ = $(\{8\}$, $\emptyset$, $\emptyset,\emptyset, \emptyset,\emptyset, \emptyset,\emptyset)$  \\
 \hhline{~--} 
& $N_{5.25}$ = $(\emptyset$, $\{2,3\}$, $\emptyset,\emptyset, \emptyset,\emptyset, \emptyset,\emptyset)$ & $E_{4.25}$ = $(\emptyset$, $\{9\}$, $\emptyset,\emptyset, \emptyset,\emptyset, \emptyset,\emptyset)$\\
 \hhline{~--} 
& $N_{5.26}$ = $(\emptyset,\emptyset, \{ 4,5\}$, $\emptyset$, $\emptyset,\emptyset, \emptyset,\emptyset)$ & $E_{4.26}$ = $(\emptyset,\emptyset, \{ 10\}$, $\emptyset$, $\emptyset,\emptyset, \emptyset,\emptyset)$\\
 \hhline{~--} 
& $N_{5.27}$ = $(\emptyset,\emptyset, \emptyset$, $\{6,7\}$, $\emptyset,\emptyset, \emptyset,\emptyset)$ & $E_{4.27}$ = $(\emptyset,\emptyset, \emptyset$, $\{11\}$, $\emptyset,\emptyset, \emptyset,\emptyset)$\\
 \hhline{~--} 
& $N_{5.28}$= $(\emptyset,\emptyset, \emptyset,\emptyset, \{ 8,9\}$, $\emptyset$, $\emptyset,\emptyset)$ & $E_{4.28}$= $(\emptyset,\emptyset, \emptyset,\emptyset, \{ 12\}$, $\emptyset$, $\emptyset,\emptyset)$\\
 \hhline{~--} 
& $N_{5.29}$ = $(\emptyset,\emptyset, \emptyset,\emptyset, \emptyset$, $\{10,11\}$, $\emptyset,\emptyset)$ & $E_{4.29}$ = $(\emptyset,\emptyset, \emptyset,\emptyset, \emptyset$, $\{13\}$, $\emptyset,\emptyset)$\\
 \hhline{~--} 
& $N_{5.30}$ =$(\emptyset,\emptyset, \emptyset,\emptyset, \emptyset,\emptyset, \{ 12,13\}$, $\emptyset)$ & $E_{4.30}$ =$(\emptyset,\emptyset, \emptyset,\emptyset, \emptyset,\emptyset, \{14\}$, $\emptyset)$\\
 \hhline{~--} 
& $N_{5.31}$ = $(\emptyset,\emptyset, \emptyset,\emptyset, \emptyset,\emptyset, \emptyset$, $\{14,15\})$ & $E_{4.31}$ =$(\emptyset,\emptyset, \emptyset,\emptyset, \emptyset,\emptyset, \emptyset$, $\{15\})$\\
\end{longtable}
\end{scriptsize}
\end{example}

In a reachability tree, some nodes are \emph{balanced} (see Definition~\ref{Chap:reversibility:def:balancednode} of Page~\pageref{Chap:reversibility:def:balancednode}), some are not. 
For instance, all nodes of the balanced rule (see Definition~\ref{Def:balancedrule} of Page~\pageref{Def:balancedrule}) $1010101010101010$ of Example~\ref{Chap:semireversibility:ex:tree} are balanced.

Reachability tree helps to identify whether a CA is reversible for a size $n$. If the reachability tree of an $n$-cell CA has all possible edges/nodes, then this implies that, all configurations are reachable for the CA. That is, if the tree is complete, the CA is reversible for that cell length $n$ (Theorem \ref{Chap:reversibility:revth1} of Chapter~\ref{Chap:reversibility}).
However, for a CA to be reversible for an $n$, $(n \geq m)$, the reachability tree needs to satisfy the following two conditions --
\begin{enumerate}[topsep=0pt,itemsep=0ex,partopsep=2ex,parsep=1ex]
\item The nodes of the reachability tree are to be balanced.
\item Each node of level $n$ is to have $d$ RMTs, each node of level $n-\iota$, $1 \leq \iota \leq m-1$, is to contain $d^\iota $ RMTs, whereas all other nodes are to be made with $d^m$ RMTs.
\end{enumerate}

These conditions are derived from the reversibility theorems for $3$-neighborhood CA (see Section~\ref{chap:reversibility:Sec:rev} of Chapter~\ref{Chap:reversibility} for details), which are extended here for $m$-neighborhood CA. Proofs of these theorems are similar to the theorems of Section~\ref{chap:reversibility:Sec:rev} (Page~\pageref{Chap:reversibility:revth1} to \pageref{Chap:reversibility:revth4}).
\subsection{Reversibility Theorems for Reachability Tree}

\begin{theorem}
\label{chap:semireversibility:th:revth1}
The reachability tree of a finite reversible CA of length $n$ $(n \geq m)$ is complete.
\end{theorem} 

\begin{proof} If a CA is reversible for length $n$, every configuration of it is reachable. So, in the corresponding reachability tree of $n+1$ levels, the number of leaves is $d^n$. Hence, the tree is complete.
\end{proof}

\begin{theorem}
\label{chap:semireversibility:th:revth2}
The reachability tree of a finite CA of length $n$ $ (n \geq m)$ is complete if and only if
\begin{enumerate}[topsep=0pt,itemsep=0ex,partopsep=2ex,parsep=1ex]
\item \label{c2}  The label $l_{n-\iota.j}$, for any $j$, contains only $d^{\iota-1}$ RMTs, where $1 \leq \iota \leq m-1$; that is, \[\mid \bigcup\limits_{0 \leq k \leq d^{m-1} -1} {\Gamma_{k}^{E_{n-\iota.j}}}\mid = d^{\iota-1}\]

\item \label{c3} Each other label $l_{i.j}$ contains $d^{m-1}$ RMTs, where $ 0 \leq i \leq n-m$; that is, \[ \mid \bigcup\limits_{0 \leq k \leq d^{m-1} -1} {\Gamma_{k}}^{E_{i.j}}\mid = d^{m-1}\]
\end{enumerate}
\end{theorem}

\begin{proof}
\begin{description}[leftmargin=0pt]
\item\noindent\underline{\textit{If Part:}} Let us consider, the number of RMTs in the label of an edge is less than that is mentioned in (\ref{c2}) and (\ref{c3}). This implies,

(i)\label{i1} The label $l_{n-1.j}$ is empty, for some $j$. That is, $ \bigcup\limits_{0 \leq k \leq d^2 -1} {\Gamma_{k}^{E_{n-1.j}}}= \emptyset $. So, the tree has a non-reachable edge and it is incomplete.

(ii) \label{i2} In general, for some $j$, the label $l_{n-\iota.j}$ contains less than $d^{\iota-1}$ RMTs, where $1\leq \iota \leq m-1$. That is, $ \mid \bigcup\limits_{0 \leq k \leq d^{m-1} -1} {\Gamma_{k}^{E_{n-\iota.j}}}\mid \leq d^{\iota-1}-1$. Then, the number of RMTs in the node $N_{n-\iota+1.j} \leq d(d^{\iota-1}-1)$. However, according to Point~\ref{rtd5} of the Definition~\ref{chap:semireversibility:def:tree}, only $\frac{1}{d}$ of the RMTs of these levels are valid. So, the number of valid RMTs is $ \leq \frac{d(d^{\iota-1}-1)}{d} = (d^{\iota-1}-1)$. If this scenario is continued for all $\iota$, then in the best case, the tree may not have any non-reachable node up to level $(n - 1)$. But, at level $n-1$, there will be at least one node for which $ \mid \bigcup\limits_{0 \leq k \leq d^{m-1} -1} {\Gamma_{k}^{N_{n-1.g}}}\mid \leq d-1$, that is, the maximum number of possible edges from this node is $d-1$. Hence, at least one (non-reachable) edge ${E_{n-1.b}}$ exist in the tree for which $\bigcup\limits_{0 \leq k \leq d^{m-1} -1} {\Gamma_{k}^{E_{n-1.b}}}= \emptyset $.

(iii) \label{i3} Let, every other label $l_{i.j}$ contains less than $d^{m-1}$ RMTs, that is, \\$ \mid \bigcup\limits_{0 \leq k \leq d^{m-1} -1} {\Gamma_{k}^{E_{i.j}}}\mid < d^{m-1}$, $(0 \leq i \leq n-m)$. Then, $ \mid \bigcup\limits_{0 \leq k \leq d^{m-1} -1} {\Gamma_{k}^{N_{i+1.j}}}\mid < d^m$. The corresponding node $N_{i+1.j}$ may have $d$ number of edges. Here also, in best case, the tree may not have any non-reachable edge up to level $(n - 1)$. Then there exists at least one
 node $N_{n-1.p}$ where $ \mid \bigcup\limits_{0 \leq k \leq d^{m-1} -1} {\Gamma_{k}^{N_{n-1.p}}}\mid < d^2$. Since the node is at level $(n - 1)$, it has maximum $\frac{(d^2-1)}{d} < d$ valid RMTs (see Point~\ref{rtd5} of the Definition~\ref{chap:semireversibility:def:tree}). Therefore, there exists at least one edge from the node $N_{n-1.p}$, where $ \mid \bigcup\limits_{0 \leq k \leq d^{m-1} -1} {\Gamma_{k}^{E_{n-1.q}}}\mid =\emptyset$, which makes the tree an incomplete one by (i).

On the other hand, any intermediate edge ${E_{i.j_1}}$ has, say, more than $d^{m-1}$ RMTs, that is, $ \mid \bigcup\limits_{0 \leq k \leq d^{m-1} -1} {\Gamma_{k}^{E_{i.j_1}}}\mid$ $\geq d^{m-1}$. Then there exists an edge $ E_{i.j_2}$ in the same label $i$ for which \\$ \mid \bigcup\limits_{0 \leq k \leq d^{m-1} -1} {\Gamma_{k}^{E_{i.j_2}}}\mid < d^{m-1}$, where $ 0 \leq i \leq n-m$, and $j_1 \neq j_2$. So, by (iii), the tree is incomplete. Now, if for any $p$, label $l_{n-\iota.p}$ contains more than $d^{\iota-1}$ RMTs, then also there exists an edge $E_{n-\iota.q}$ for which $ \mid \bigcup\limits_{0 \leq k \leq d^{m-1} -1} {\Gamma_{k}^{E_{n-\iota.q}}}\mid$ $< d^{\iota-1}$. Hence, the tree is incomplete (by (ii)).
Hence, if the number of RMTs for any label is not same as mentioned in (\ref{c2}) and (\ref{c3}), the reachability tree is incomplete. 

\item\noindent\underline{\textit{Only if Part:}} Now, consider that, the reachability tree is complete. The root $N_{0.0}$ has $d^m$ number of RMTs. In the next level, these RMTs are to be distributed such that the tree remains complete. Say, an edge $E_{0.j_1}$ has less than $d^{m-1}$ RMTs, another edge $E_{0.j_2}$ has greater than $d^{m-1}$ RMTs and other edges $E_{0.j} (0 \leq j,j_1,j_2 \leq d-1$ and $j_1 \neq j_2 \neq j)$ has $d^{m-1}$ RMTs. Then node $N_{1.j_1}$ has less than $d^m$ RMTs, $N_{1.j_2}$ has greater than $d^m$ RMTs and other edges $N_{1.j}$ has $d^m$ RMTs. There is no non-reachable edge at level $1$. 
Now, this situation may continue up to level $n-m$. However, at level $(n-\iota)$, $1\leq \iota \leq m-1$, only $\frac{1}{d}$ of the RMTs are valid (see Definition~\ref{chap:semireversibility:def:tree}). So, the nodes at level $n-m+1$ with less than $d^m$ RMTs has at maximum $d^{m-1}-1$ valid RMTs, nodes at level $n-m+2$ has $<d^{m-2}$ valid RMTs and so on. So, at level $n-1$, there exists at least one node $N_{n-1.p}$, such that $ \mid \bigcup\limits_{0 \leq k \leq d^{m-1} -1} {\Gamma_{k}^{N_{n-1.p}}}\mid < d$ for which the tree will have non-reachable edge (by (ii)). Similar situation will arise, if any number of intermediate edges have less than $d^{m-1}$ RMTs. Because, this means, some other edges at the same level have more than $d^{m-1}$ RMTs. Hence, the tree will be incomplete which contradicts our initial assumption. So, for all intermediate edges $E_{i.j}$, $ \mid \bigcup\limits_{0 \leq k \leq d^{m-1} -1} {\Gamma_{k}^{E_{i.j}}}\mid = d^{m-1}$, where $ 0 \leq i \leq n-m$.

Now, if this is true, then in general, at level $n-\iota$, the nodes have $d^{\iota}$ valid RMTs and every edge has $d^{\iota-1}$ valid RMTs. If an edge $E_{n-\iota.p}$ has less than $d^{\iota-1}$ RMTs, then the node $N_{n-\iota.p}$ has at maximum $d(d^{\iota-1}-1)$ RMTs out of which only $d^{\iota-1}-1$ are valid. Hence, in this way, at level $n-1$, we can get at least one edge from $N_{n-1.q}$, which is non-reachable making the tree incomplete. So, for the tree to be complete, each edge label $l_{n-\iota.j}$, for any $j$, is to have $d^{\iota-1}$ RMTs. Hence the proof.
\end{description}\end{proof}

\begin{corollary}
\label{chap:semireversibility:th:revcor1}
The nodes of a reachability tree of a reversible CA of length $n$ $(n \geq m)$ contains
\begin{enumerate}[topsep=0pt,itemsep=0ex,partopsep=2ex,parsep=1ex]
\item $d$ RMTs, if the node is in level $n$, i.e. $ \mid \bigcup\limits_{0 \leq k \leq d^{m-1} -1}{\Gamma_{k}^{N_{n.j}}} \mid = d$ for any $j$.

\item $d^\iota $ RMTs, if the node is at level $n-\iota$ i.e, $ \mid \bigcup\limits_{0 \leq k \leq d^{m-1} -1}{\Gamma_{k}^{N_{n-\iota.j}}} \mid = d^\iota$ for any $j$, where $1 \leq \iota \leq m-1$.

\item $d^m$ RMTs for all other nodes $N_{i.j}$, $ \mid \bigcup\limits_{0 \leq k \leq d^{m-1} -1}{\Gamma_{k}^{N_{i.j}}} \mid = d^m$ where ${ 0 \leq i \leq n-m}$.
\end{enumerate}
\end{corollary}

\begin{proof}This can be directly followed from Theorem~\ref{chap:semireversibility:th:revth2}.
\end{proof}

\begin{corollary} 
\label{chap:semireversibility:th:revcor2}
The nodes of the reachability tree of an $n$-cell $(n \geq m)$ reversible CA are balanced.
\end{corollary}

\begin{proof}For an $n$-cell $(n \geq m)$ reversible CA, the reachability tree is complete. So, each node has $d$ number of edges. Each node $N_{i.j}$ contains $d^m$ RMTs when $i < n-\iota$, $1 \leq \iota \leq m-1$ (Corollary ~\ref{chap:semireversibility:th:revcor1}) and each edge $E_{i+1.k}$, for any $k$, contains $d^{m-1}$ RMTs (Theorem~\ref{chap:semireversibility:th:revth2}). So, the node $N_{i.j}$ is balanced and the number of RMTs per each state is $d^{m-1}$. Similarly, the nodes of level $n-\iota$ are balanced. 
\end{proof}

\begin{corollary}
\label{chap:semireversibility:th:revcor3}
Let $R$ be the rule of a CA. Now, the CA is trivially semi-reversible (irreversible for each $n\geq m$) if the following conditions are satisfied--
\begin{enumerate}[topsep=0pt,itemsep=0ex,partopsep=2ex,parsep=1ex]
\item $R$ is unbalanced and
\item no two RMTs $r$ and $s$ exist, such that $\{r\}$, $\{s\}$ are two primary RMT sets and $R[r]=R[s]$.
\end{enumerate}
\end{corollary}

\begin{proof}If the rule $R$ is unbalanced, number of RMTs per each state is unequal. So, for every $n\ge m$, the root node $N_{0.0}$ of the corresponding reachability tree is unbalanced. Hence, there exists at least one edge $E_{0.j}$ where $\mid\bigcup\limits_{0\leq k \leq d^{m-1}-1} \Gamma_k^{E_{0.j}}\mid < d^{m-1}$ $(0 \leq j \leq d^{m-1}-1)$. Therefore, by Theorem~\ref{chap:semireversibility:th:revth2}, the CA is irreversible for every $n\ge m$.

However, the second condition ensures that, the CA does not violate condition~\ref{irreversibility_c1} of Proposition~\ref{Chap:semireversible:Theorem:irreversibility}. So, it is not irreversible for $n=1$. 

Therefore, the CA is reversible for a finite number of lattice sizes (for at least $n=1$) and irreversible for an infinite set of sizes (for each $n \ge m$). Hence, by Definition~\ref{Chap:semireversible:def:triv_semi}, the CA is trivially semi-reversible.
\end{proof}

We can observe that, the tree of Figure~\ref{Chap:semireversible:fig:rt2} is complete, so, by Theorem~\ref{chap:semireversibility:th:revth1}, the CA is reversible for $n=3$. 
However, for very large $n$, the tree grows exponentially with large number of nodes, so it becomes impossible to deal with. We, therefore, construct the \textit{minimized} reachability tree which does not grow exponentially.

\subsection{Construction of Minimized Reachability Tree}\label{Chap:semireversible:sec:tree-cons}
When $n$ grows, number of nodes in reachability tree grows exponentially. Therefore, to deal with this situation, minimized reachability tree is developed which stores only the unique nodes generated in the tree. Section~\ref{chap:reversibility:Sec:bij} of Chapter~\ref{Chap:reversibility} discusses in details the construction of minimized reachability tree for a $3$-neighborhood $d$-state CA. Here, those steps of forming the minimized tree (consisting of all possible unique nodes for the CA) are enlisted briefly.
\begin{description}[topsep=0pt,itemsep=0ex,partopsep=2ex,parsep=1ex]
\item[Step 1:]\label{minimizedTree:step_1} Form the root $N_{0.0}$ of the tree. Set index of level, $i\leftarrow 0$ and set of levels of node $N_{0.0} \leftarrow \{0\}$. As root is always unique, move to the next level.

\item[Step 2:]\label{minimizedTree:step_2} For each node $\mathcal{N}$ of level $i$, 
\begin{enumerate}[leftmargin=0pt,topsep=0pt,itemsep=0ex,partopsep=2ex,parsep=1ex]
\item Find the children of $\mathcal{N}$ for level $i+1$

\item For each child node, say, $N_{i+1.j}$, check whether it is unique or not --
\begin{enumerate}[topsep=0pt,itemsep=0ex,partopsep=2ex,parsep=1ex]
\item If $N_{i+1.j} = N_{i+1.k}$ when $j > k$, then both the nodes are roots of two similar sub-trees. So, we discard $N_{i+1.j}$ and add a link from the parent node $\mathcal{N}$ to $N_{i+1.k}$.

\item If $N_{i+1.j} = N_{i'.k}$ when $i+1>i'$, 
then the nodes that follow $N_{i'.k}$ are similar with the nodes followed by $N_{i+1.j}$. Therefore, we discard $N_{i+1.j}$ and add a link from $\mathcal{N}$ to $N_{i'.k}$. The level of node $N_{i'.k}$ is updated by adding $i+1$ with the existing set of levels. For example, say, $N_{i'.k}$ was present only in level $i'$, then the previous set of levels of $N_{i'.k}$ is $\{i'\}$. This set is now updated as $\{i',i+1\}$. In this case, a \emph{loop} is embodied between levels $i+1$ and $i'$. Presence of this loop is indicated by this set of levels of node $N_{i'.k}$.

\item Otherwise, the node $N_{i+1.j}$ is unique. Add this node in the minimized reachability tree and update its set of levels according to the set of levels of the parent node $\mathcal{N}$.
\end{enumerate}

\item Whenever, a new level is added to the set of levels of an existing node $\mathcal{N}'$ of the minimized tree, the set of levels for each node of the subtree of $\mathcal{N}'$ are updated according to the new set of levels of $\mathcal{N}'$. 
\end{enumerate} 
\item[Step 3:]\label{minimizedTree:step_3} If a new node is added in the tree, update $i \leftarrow i+1$ and go to Step~$2$.

\item[Step 4:]\label{minimizedTree:step_4} If no new node is added in the tree, construction of the minimized reachability tree is complete.
\end{description}



It is observed that after few levels, no unique node is generated. So, for arbitrary large $n$, we need not to develop the whole reachability tree, rather, the minimized reachability tree  can be drawn out of only the \emph{unique} nodes. In fact, if $n_0$ is the level when the last unique node is added in the minimized reachability tree corresponding to a CA, then, for any $n\ge n_0$, the minimized tree for the CA does not change. However, if $n<n_0$, then we need to develop the minimized reachability tree up to level $n-m$ and generate nodes of $n-\iota$, $1 \leq \iota \leq m-1$ according to Point~\ref{rtd4} of Definition~\ref{chap:semireversibility:def:tree}.


Recall that, in the minimized tree, if a node $\mathcal{N}$ is generated in level $i$ as well as level $i'$, $i'\geq i$, then the unique node of level $i$ gets two node levels $\lbrace i, i'\rbrace$ and a loop of length $(i'-i)$ is associated with the node. This loop implies that the node reappears at levels on the arithmetic series $i + (i - i')$, $i + 2(i-i')$, $i + 3(i-i')$, etc. Moreover, a node $\mathcal{N}$ can be part of more than one loop, which implies that, $\mathcal{N}$ can appear at levels implied by each of the loops. However, if we observe in more detail, we can find that, although every loop confirms presence of $\mathcal{N}$, but all loops are not significant in the tree.
We need to keep only the relevant loops in the tree and discard the other loops.



\begin{example} Consider the minimized reachability tree of ECA $75 (01001011)$ shown in Figure~\ref{Chap:reversibility:fig:rt3} (Page~\pageref{Chap:reversibility:fig:rt3}). The minimized tree has only $21$ nodes and the height of the tree is $5$. Therefore, for each $n\ge 5$, the minimized tree remains the same and conveys all information regarding reversibility of the CA for that $n$.
\end{example} 
%
%
%

\subsection{Minimized Tree and Reversibility for all ${n \in \mathbb{N}}$}
Number of possible unique nodes for a CA is fixed and finite. So, the minimized reachability tree of a CA with any $n\ge n_0$, $n \in \mathbb{N}$, is always of finite height. (Here, while constructing minimized tree for a CA, our target is to draw the tree with all possible unique nodes in the CA, without considering the \emph{special} levels $n-\iota$, $1 \leq \iota \leq m-1$. This is because, these special levels can be identified from this minimized tree, as discussed below.) Therefore, a CA with any number of cells can also be represented by its minimized reachability tree of fixed height having only the unique nodes. This minimized tree is instrumental in relating reversibility of infinite CAs with that of finite CAs.

For a finite CA with fixed size $n$, the reversibility of the CA for that $n$ can be decided by using the minimized reachability tree (see Section~\ref{chap:reversibility:Sec:bij} for more details). To decide this, we check whether any node at any level $l$, $1\leq l \leq n$, violates any reversibility conditions. To find the nodes of a level $p$, 
we take help of the loops in the tree.
For instance, if a node $\mathcal{N}$ has three levels associated with it-- $\lbrace i, i', i'' \rbrace$, $i''>i'>i$, then, it can be present in level $p$, if $(p-i) \equiv 0 \pmod {(i'-i)}$, or $(p-i) \equiv 0 \pmod {(i''-i)}$. Therefore, by this procedure, we can identify the nodes of level $n-\iota$, $1 \leq \iota \leq m-1$. A node can be present in level $n-\iota$, if $(n-\iota-i) \equiv 0 \pmod {(i'-i)}$, where $i' > i$ is any level associated with $\mathcal{N}$ and $1 \leq \iota \leq m-1$. However, if a node can exist in level $n-\iota$ ($1 \leq \iota \leq m-1$), Point~\ref{rtd5} of Definition~\ref{chap:semireversibility:def:tree} is applied to this node. So, to be reversible, these (modified) nodes also need to satisfy the reversibility conditions. Hence, we can verify the conditions of reversibility for any $n$.

However, as minimized reachability tree is same for any $n\ge n_0$, this tree contains the same information regarding reachability of the configurations for every $n \in \mathbb{N}$. So, for each $n\in \mathbb{N}$, one can check the minimized tree to find whether the CA is reversible or not. Nevertheless, this process of checking reversibility for every $n$ is not feasible. In this scenario, we exploit the property of the loops in the minimized reachability tree once again. Because, each loop invariably indicates the appearances of a specific node repeatedly at different levels in the tree, that is, for different $n$ in arithmetic series.

For a CA to be reversible (according to Definition~\ref{chap:semireversibility:def:rev}), every node of the minimized reachability tree has to satisfy the reversibility conditions of Corollary~\ref{chap:semireversibility:th:revcor1} and Corollary~\ref{chap:semireversibility:th:revcor2} for each $n \in \mathbb{N}$. In the minimized tree, if a node is not associated with any loop, this node can not appear in any other level except its present level. So, if all such nodes satisfy the reversibility conditions, we can proceed to other nodes associated with loop(s). Note that, to understand reversibility for any $n\in \mathbb{N}$, we need to observe these loops in the minimized tree. As we are to consider all $n$, $n\in \mathbb{N}$, each node associated with a loop can appear in each of the levels $n-\iota$, $1 \leq \iota \leq m-1$ for some $n$. So, every node associated with loop(s) is verified whether it violates any reversibility conditions of Corollary~\ref{chap:semireversibility:th:revcor1} and Corollary~\ref{chap:semireversibility:th:revcor2} for any of levels $n-\iota$, $1 \leq \iota \leq m-1$ by applying Point~\ref{rtd5} of Definition~\ref{chap:semireversibility:def:tree} on that node. If no such node exists in the minimized reachability tree, that means, for every $n \in \mathbb{N}$, the tree satisfies all reversibility conditions. That is, the CA is reversible for every $n \in \mathbb{N}$ (Definition~\ref{chap:semireversibility:def:rev}).
Therefore, by using minimized reachability tree for a CA, we can decide whether a CA is reversible (according to Definition~\ref{chap:semireversibility:def:rev}).

\begin{example}
Consider the $2$-state $4$-neighborhood CA $1010101010101010$. The minimized tree for this CA is shown in Figure~\ref{Chap:semireversible:fig:min_rt_rev}. The tree contains only $15$ unique nodes and the height of the minimized tree is $3$. Hence, for any $n\geq 3$, the minimized tree remains the same.
\begin{figure}[!h]
		 \centering
			\includegraphics[width= 5.5in, height = 2.5in]{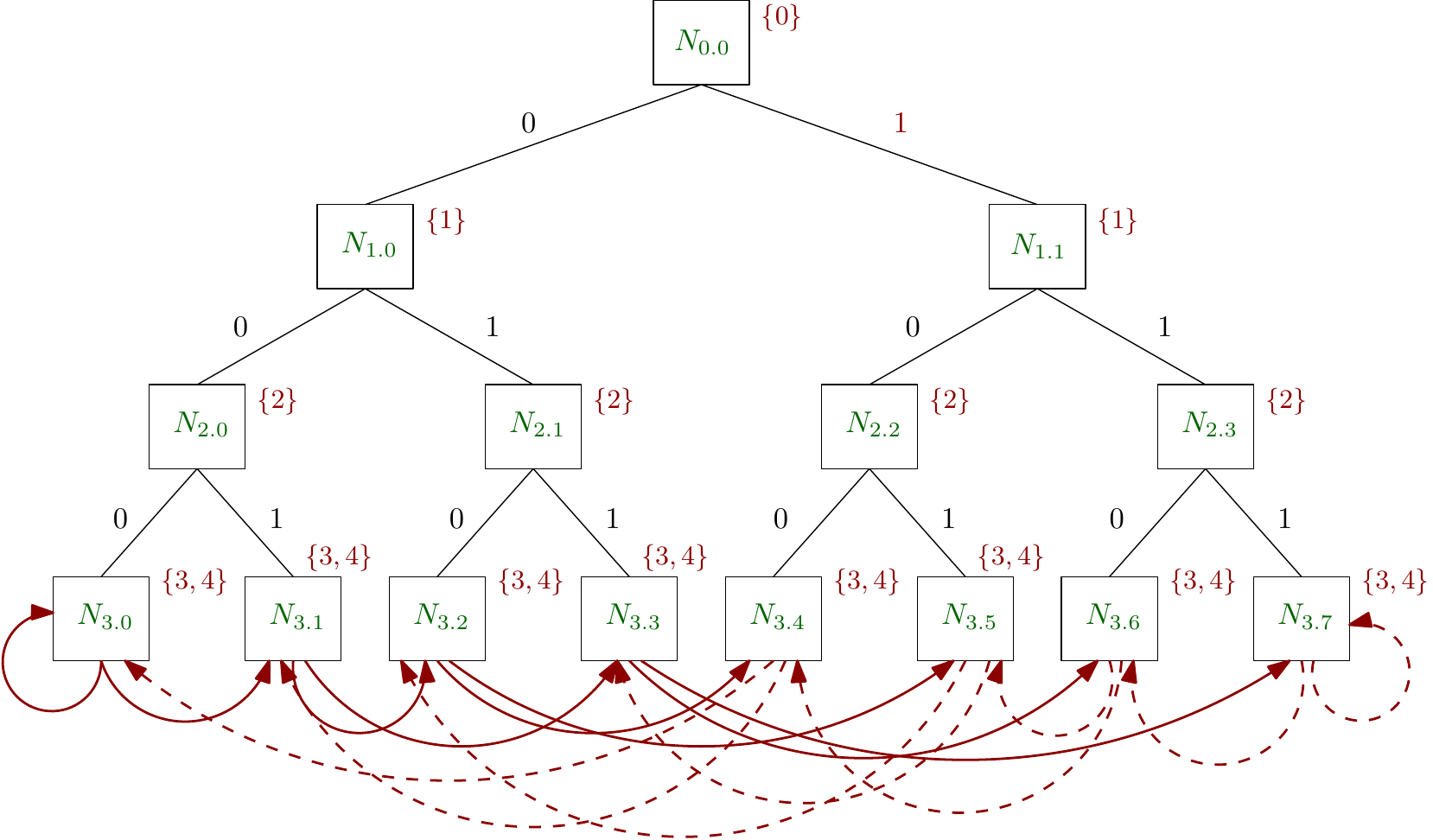}
			\caption{Minimized Reachability tree of $2$-state $4$-neighborhood CA with rule $1010101010101010$}
			\vspace{-1.0em}
			\label{Chap:semireversible:fig:min_rt_rev}
	\end{figure}
For this CA, the nodes of the minimized tree satisfy every reversibility condition. Further, each of the nodes of level $3$ is associated with self-loop. But, even after applying Point~\ref{rtd5} of Definition~\ref{chap:semireversibility:def:tree} on those nodes, none of the modified nodes violate any reversibility conditions for any $n \in \mathbb{N}$. Hence, the CA is reversible for each $n \in \mathbb{N}$.
\end{example}

Moreover, the tree can help us to identify in which class of reversibility, the CA belongs to. The next section discusses about this.

\section{Minimized Reachability Tree and Semi-reversibility}\label{Chap:semireversibility:sec:semi_tree}
Using the minimized tree for a CA, we can understand the reversibility class of the CA. If a CA is strictly irreversible, it can be identified easily by using condition~\ref{irreversibility_c1} of Proposition~\ref{Chap:semireversible:Theorem:irreversibility}. For such a CA, two RMTs exist corresponding to two singleton primary RMT sets for which the next state values are same. For instance, the $3$-neighborhood $3$-state CA $102012120012102120102102120$ has $R[13]=R[0]=0$, so the CA is strictly irreversible and there is no need to construct minimized reachability tree for the CA. Similarly, if a CA rule is unbalanced, but the CA is not strictly irreversible, it is trivially semi-reversible (by Corollary~\ref{chap:semireversibility:th:revcor3}). Take as example the $2$-neighborhood $3$-state CA $021122011$. The rule is unbalanced, so the CA is trivially semi-reversible (irreversible for all $n\ge 2$). Here also, we do not have to construct the minimized tree. However, for other cases, we need to construct the minimized reachability tree for that CA.

During the construction of the minimized  tree, we test the reversibility conditions of Corollary~\ref{chap:semireversibility:th:revcor1} and Corollary~\ref{chap:semireversibility:th:revcor2} at addition of each new node in the tree. If a reversibility condition is violated before completion of the construction, we conclude that the CA is trivially semi-reversible. For example, some nodes may be found, which are not part of any loop, but for which the CA does not satisfy the reversibility conditions. If a CA has such node(s) in the minimized tree at level $i$, that means, for every $n\geq i$ ($i\ge m$), the CA is irreversible. So, such CAs are trivially semi-reversible. (If we draw minimized reachability tree for a strictly irreversible CA, then also, reversibility conditions are violated before completion of the tree.) 
For example, for a $2$-neighborhood $3$-state CA $012122001$, the minimized tree has a unbalanced node $N_{1.0}=(\{3, 4, 5, 6, 7, 8\} ,\emptyset, \{6, 7, 8\})$ at level $1$. So, the CA is irreversible for all $n>1$, that is, it is a trivial semi-reversible CA.

Similarly, if a node has a self-loop at level $i$, that means, it will be present in the reachability tree for each $n\geq i$. If such a node violates conditions of reversibility after applying Point~\ref{rtd5} of Definition~\ref{chap:semireversibility:def:tree}, the CA is trivially semi-reversible. Whenever a CA is detected to be trivial semi-reversible, there is no need to continue construction of the minimized reachability tree. However, in case of non-trivial semi-reversible or reversible CA, we can complete construction of the minimized reachability tree. For the reversible CAs, no node of the minimized reachability tree violates the reversibility conditions. Nevertheless, we next discuss the way of identifying a non-trivial semi-reversible CA with the infinite set of sizes for which the CA is reversible.


\subsection{Identifying Non-trivial Semi-reversible CAs using Minimized Tree}
A CA is non-trivially semi-reversible, if there exists infinite number of sizes for which it is reversible. In the minimized reachability tree, reversibility for any size $n$ can be confirmed if no node in the tree violates reversibility conditions for that $n$. However, if a CA is irreversible for some $n$, the minimized tree has node(s) which violate(s) reversibility conditions for those $n$. If a node at level $i$ violates any reversibility condition, that means the CA is irreversible for all $n>i$. Hence, it is trivially semi-reversible. Nonetheless, a node associated with loop(s) may violate these conditions after applying Point~\ref{rtd5} of Definition~\ref{chap:semireversibility:def:tree} to that node for some loop. This means, the CA is irreversible for all those $n$ in arithmetic series implied by the loop.

For example, say, according to the values of a loop, a node can exist in level $n-\iota$ ($1 \leq \iota \leq m-1$). So, by Point~\ref{rtd5} of Definition~\ref{chap:semireversibility:def:tree}, only $\frac{1}{d}$ of the RMTs of the original node of the loop remains valid for this node. If this (modified) node does not satisfy the reversibility conditions for levels $n-\iota$ ($1 \leq \iota \leq m-1$) (using Corollary~\ref{chap:semireversibility:th:revcor1} and \ref{chap:semireversibility:th:revcor2}), the CA is irreversible for those values of $n$. Therefore, if a node does not satisfy these conditions, we can get an expression over cell length $n=k(i'-i)+i+\iota$, where $k \ge 0$ and $1 \leq \iota \leq m-1$, for which the CA is irreversible. This expression is termed as \textit{irreversibility expression}. For each loop violating the reversibility conditions for some values of $n$, we get an irreversibility expression. By using these irreversibility expressions, we can find the set of sizes for which the CA is irreversible. Let $\mathcal{I}$ be the set of CA sizes from the irreversibility expressions for which the CA is irreversible. Then, $\mathbb{N}\setminus\mathcal{I}$ denotes the set of lattice sizes for which the CA is reversible.


If the minimized reachability tree for a CA has some loops, where for some specific values of $n$, the nodes of the loops violate reversibility condition(s), then the CA is irreversible for only those specific values of $n$. For other $n$, the CA is reversible. Such a CA is non-trivially semi-reversible. Because, for such CAs, the set $\mathbb{N}\setminus\mathcal{I}$ denotes an infinite set of $n$ for which the CA is reversible. However, for trivial semi-reversible CAs, this set indicates only finitely many sizes for which the CA is reversible. Similarly, a CA is declared reversible, when
there is no $n$, for which an irreversibility expression can be found in the minimized tree. That is, the set $\mathcal{I} = \emptyset$.
\begin{example}
	Consider the minimized reachability tree of the ECA $75 (01001011)$ shown in Figure~\ref{Chap:reversibility:fig:rt3} of Page~\pageref{Chap:reversibility:fig:rt3}. Here, the CA violates reversibility conditions only when the lattice size is an even number. So, the irreversibility expressions are $n=2j+2$, $n=2j+4$ and $n=2j+6$, where $j \ge 0$. That is, the final expression of irreversibility is $n = 2j +2$, $j \ge 0$ and the set $\mathcal{I} = \{k ~|~ k\equiv 0\pmod{2}\}$. Therefore, ECA $75(01001011)$ is reversible for the set $\mathbb{N}\setminus\mathcal{I}=\{n ~|~ n \equiv 1 \pmod{2}\}$, that is, when $n$ is an odd number.
	
	Similarly, for the $2$-state $4$-neighborhood CA $1010101010101010$ of Figure~\ref{Chap:semireversible:fig:min_rt_rev}, the set $\mathcal{I} = \emptyset$. So, the CA is reversible for each $n\in \mathbb{N}$.
\end{example}

Therefore, the minimized reachability tree is sufficient to segregate the non-trivial semi-reversible CAs from the trivial semi-reversible CAs and reversible CAs. In other words, given a CA, it can efficiently detect its reversibility class. In the following section, we develop a decision algorithm to decide the reversibility class of a CA.

\subsection{Algorithm for Finding Type of Semi-reversibility using Minimized Tree}\label{Chap:semireversible:sec:Algo}
\noindent The minimized reachability tree is formed with the unique nodes of the reachability tree, where each of the unique nodes can have a set of levels associated with it. 
It provides complete information on reversibility of a CA. 
However, while constructing the minimized reachability tree to decide reversibility of a CA, one may observe the following:
\begin{enumerate}
\item If the CA follows condition~\ref{irreversibility_c1} of proposition~\ref{Chap:semireversible:Theorem:irreversibility}, there is no need to construct the minimized tree, as the CA is strictly irreversible for all cell length $n$ (Theorem~\ref{Chap:semi-reversibility:Th:strictirreversibility}).

\item If the CA is unbalanced, but not strictly irreversible, then also, there is no need to construct the minimized tree, as the CA is trivially semi-reversible for all cell length $n\geq m$ (Corollary~\ref{chap:semireversibility:th:revcor3}).

\item If at any level $i>0$, the minimized tree has a non-reachable node, the CA is irreversible for all $n \geq i$, $i \geq m$. That is, the CA is trivially semi-reversible.

\item If for any level $i\geq 0$, the minimized tree has an unbalanced node (Corollary~\ref{chap:semireversibility:th:revcor2}), or does not follow the conditions of Theorem~\ref{chap:semireversibility:th:revth2}, the CA is irreversible for all $n \geq i$, $i \geq m$.  Here also, the CA is trivially semi-reversible.
\end{enumerate}

For other CAs, we construct the minimized tree. Here, instead of first completing construction of the minimized reachability tree with all unique nodes, every time a loop is detected, we can directly apply the logic for generating irreversibility expression. However, if a node has self-loop at level $i$, that implies, it will be present in every level of the tree. So, if such a node violates Corollary~\ref{chap:semireversibility:th:revcor1} or Corollary~\ref{chap:semireversibility:th:revcor2} for any of the levels $n-\iota$, $1\leq \iota \leq m-1$, the CA is irreversible for every $n>i, (i \geq m)$; that is, it is a trivial semi-reversible CA. Similarly, say, there are $k$ number of irreversibility expressions for a CA of form $n=k.j+i$, $b\leq i \leq k+b$, where $k$ is the length of the loop and $b$ is any number. This implies, the CA is irreversible for each $n \geq b$, $(b \geq m)$, that is, it is a trivial semi-reversible CA.


Now, we design Algorithm~\ref{Chap:semireversible:algo:rev_exp_algo}, which mainly deals with the CAs whose minimized reachability tree does not violate Theorem~\ref{Chap:semi-reversibility:Th:strictirreversibility} and Corollary~\ref{chap:semireversibility:th:revcor3}.
This algorithm \emph{CheckReversibilityForAllCell()} uses the minimized reachability tree of a CA rule, and reports \emph{``Reversible''}, \emph{``Irreversible''}, \emph{``Trivially semi-reversible''} or \emph{``Non-trivially semi-reversible''} with expression of irreversibility.
It uses two data structures -- \emph{NodeList} for storing the unique nodes and \emph{NodeLevel} to store the set of levels associated with each node along with an array \emph{ExpressionArray} to store the irreversibility expressions. Each of the nodes of \emph{NodeList} is associated with a flag - \emph{selfLoop}, which is set when the node has loop of length $1$. 
It also uses some variables, like $uId$ as index of \emph{NodeList}, $i$ as the current level of the tree and $p$ as the parent node. Moreover, two global boolean variables \emph{conditional} and \emph{missing}, initialized as \emph{false} and two procedures \emph{updateSubTreeLevels()} and \emph{verifyReversibility()} 
are used for its operation:

\begin{description} [leftmargin=0pt]
\item[\emph{verifyReversibility()}] 
The procedure takes an index $k$ of \emph{NodeList} as its argument and applies Point~\ref{rtd5} of Definition~\ref{chap:semireversibility:def:tree} on that node. If the modified node violates reversibility conditions for any level $n-\iota$, $1 \leq \iota \leq m-1$, by Corollary~\ref{chap:semireversibility:th:revcor1} or Corollary~\ref{chap:semireversibility:th:revcor2}, the procedure sets \emph{missing} as \emph{true}. It also generates the irreversibility expression for the node with index $k$ and stores in \emph{ExpressionArray}. If the node has a self-loop, or expressions of \emph{ExpressionArray} denote irreversibility for all $n\geq b$, $b$ is any level, \emph{conditional} is set to \textit{false} and the CA is declared as ``Trivially semi-reversible''. Otherwise, the flag \emph{conditional} is set \textit{true}. 

\item[\emph{updateSubTreeLevels()}] This procedure updates a sub-tree when a relevant loop is formed. The update includes modification of \emph{NodeLevel} of each node of the sub-tree based on the logic presented before. During the update of \emph{NodeLevel} of each node, the procedure \emph{verifyReversibility()} is called to see if the node can be present at level $n-\iota$ ($1\leq \iota \leq m-1$) and violates Corollary~\ref{chap:semireversibility:th:revcor1} and Corollary~\ref{chap:semireversibility:th:revcor2}.
As argument, it takes the $uId$ of the node which is the root of the sub-tree.
\end{description}
This algorithm also uses some other variables and flags. For example, for any level $i$, the variables $q$ and $j$ keep track of the unique nodes added in the previous level $i-1$. Similarly, the flag \emph{loopFlag} checks the validity of each new loop. If the new loop value is not relevant, \emph{loopFlag} is set to $true$ and the corresponding \emph{NodeLevel} value is not updated. Likewise Section~\ref{chap:reversibility:Sec:bij} of Chapter~\ref{Chap:reversibility} (see Page~\pageref{chap:reversibility:algo:rev_algo}), old loop value(s) and new loop value are indicated by the variables \emph{oldLoop} and \emph{newLoop} respectively. 
\begin{Walgo}[hbtp]{1.75cm}
	\scriptsize
	\SetKw{Fn}{Procedure}
	\SetKwFunction{verifyReversibility}{verifyReversibility}
	\SetKwFunction{FindGCD}{findGCD}
	\SetKwFunction{updateSubTreeLevels}{updateSubTreeLevels}
	\SetKwFunction{printIrreversibilityExpression}{printIrreversibilityExpression}
	\SetKwInOut{Input}{Input}
	\SetKwInOut{Output}{Output}
	\Input{A CA rule $R$} 
	\Output{\textit{Reversible} or \textit{Irreversible} or \textit{Trivially semi-reversible} or \textit{Non-trivially semi-reversible} with \textit{Expression of Irreversibility}}
		
	\rule[4pt]{0.95\textwidth}{0.85pt}\\
		\hspace{0.04\textwidth} \nlset{Step 1} \label{Chap:semireversible:algo:st1}
		\lIf{ $\exists x, y\in \mathcal{S}, x\neq y$ so that $R[x\times d^{m-1} + \cdots + x \times d +x] =R[y\times d^{m-1} + \cdots + y \times d +y]$}{
					Report $``$\textit{Strictly Irreversible}'' and
					$return$ }
				
		\hspace{0.04\textwidth} \nlset{Step 2}\label{Chap:semireversible:algo:st2} 
		\lIf{ $R$ is unbalanced }{
					Report $``$\textit{Trivial Semi-reversible. Irreversible for each $n\geq m$}'' and	$return$ }
		\hspace{0.04\textwidth} \nlset{Step 3} Form the root of the reachability tree \; 	\label{Chap:semireversible:algo:st3}
		\hspace{0.04\textwidth}	Set $NodeList[0] \gets$ root, $NodeLevel[0] \gets \lbrace 0 \rbrace$, $i \gets 1$, $uId \gets 0$, $q \gets 0$, $j \gets 0$ \;	
		\hspace{0.04\textwidth}	Set $ExpressionArray \gets null$, $missing \gets false$, $conditional \gets false$ \;	
			
		\hspace{0.04\textwidth}	\nlset{Step 4}\label{Chap:semireversible:algo:st4} \For {$p = q$ to $j$}{ 
			Get the children of $NodeList[p]$ \;
			\For {each child $\mathcal{N}$ of $NodeList[p]$}{
			 \If{ ($\mathcal{N}$ is not balanced) OR ($\mid \bigcup\limits_{0\leq q \leq d^{m-1}-1} {\Gamma_q^{\mathcal{N}}}\mid \neq d^m $)}{
						Report $``$\textit{Trivial Semi-reversible. Irreversible for each $n \geq i, (i \geq m)$}'' and
						$return$ \;
			  }
			
			\eIf{ $\mathcal{N}$ matches with $NodeList[k]$, $0 \leq k \leq uId$}{
				\eIf{$\mid NodeLevel[k] \mid=1$ AND $i \notin NodeLevel[k]$ \tcp{node is referred for the first time}} 
				{
					Set	$NodeLevel[k] \leftarrow NodeLevel[k] \cup \lbrace i \rbrace$ ;
					\tcp{Update node level of the unique node $NodeList[k]$}
					\lIf{the directed edge makes a loop of length $1$ (self-loop)}{
							Set	$NodeList[k].selfLoop \leftarrow true$ }
					 \updateSubTreeLevels{$k$} ; \tcp{update sub-tree adding the new loop}
				}
				{	\tcc{If node is referred more than once, check whether any new directed edge is required}
					Set $loopFlag \leftarrow false$; 
					\uIf{$NodeList[k].selfLoop = false$ AND $i \notin NodeLevel[k]$}
					{
						Set $newLoop \gets i-\min(NodeLevel[k])$ ; \tcp{Value of Loop is the difference of levels}
						\uIf{$newLoop = 1$ \tcp{new self-loop detected}}{
							Set $NodeLevel[k] \leftarrow \lbrace (i-1), i \rbrace$,
							$loopFlag \leftarrow true$ ; \tcp{Overwrite existing node levels}	
							Set	$NodeList[k].selfLoop \leftarrow true$ \;
							\updateSubTreeLevels{$k$} ; \tcp{update sub-tree by the new loop}
						}
						\uElse{
							\ForEach{$ l \in NodeLevel[k]$}{
								Set $oldLoop \gets l-\min(NodeLevel[k])$, $gcd \leftarrow$ \textit{GCD}($oldLoop,newLoop$) \;
								\uIf{$gcd = oldLoop$  \tcp{that is, old loop value prevailed}}{
										Set $loopFlag \leftarrow true$ and \textbf{break} ; \tcp{new loop is not relevant} 				
									}
								\uElseIf{$gcd >1$ \tcp{that is, updated valid loop length $ = gcd$}}
								{
									Set	$NodeLevel[k] \leftarrow \lbrace (i-gcd), i \rbrace$,
									$loopFlag \leftarrow true$ ; \tcp{Overwrite node levels}	
									\updateSubTreeLevels{$k$} and \textbf{break} \;  
								}
							}
							\uIf{$loopFlag = false$ \tcp{new loop is relevant}}
							{
								Set	$NodeLevel[k] \leftarrow NodeLevel[k] \cup \lbrace i \rbrace$ \;
								\updateSubTreeLevels{$k$} ; \tcp{update sub-tree adding the new loop}
							}
						}
				 }
			}
		}	
		{
			Set	$uId \leftarrow uId + 1$, 
			$NodeList[uId] \leftarrow \mathcal{N}$ ; \tcp{add the unique node in the $NodeList$}
			\ForEach{$l \in NodeLevel[p]$}{
				Set	$NodeLevel[uId] \leftarrow NodeLevel[uId]\cup \lbrace l+1 \rbrace$ \tcp{update child's level by parent's level}
			}
			\lIf{$NodeList[p].selfLoop = true$}{
						Set $NodeList[uId].selfloop \leftarrow true $ }
			\uIf{$\mid NodeLevel[uId] \mid > 1$ \tcp{that is, the newly added unique node has a loop}}{
				\verifyReversibility{$uId$} \;
			}
		   }   
		 }
		}
		\hspace{0.04\textwidth} \nlset{Step 5}	\label{Chap:semireversible:algo:st5} \lIf{ $j = uId$ \tcp{that is, no unique node is generated in \ref{Chap:semireversible:algo:st4}}}{ go to \ref{Chap:semireversible:algo:st7} } 
			\uElse{ 	$q \gets j + 1 $; $j \gets uId$;  $i \gets i+1$ ;
				go to \ref{Chap:semireversible:algo:st4} \;
			}
			\hspace{0.04\textwidth}	\nlset{Step 6}  \label{Chap:semireversible:algo:st7} 
			\lIf{$missing = false $ AND $conditional=false$}{
				Report $``$\textit{Reversible}'' and
				$return$ }
			\ElseIf{$conditional = true$}{
				Report $``$\textit{Non-trivial Semi-reversible}'' \; 
				Print the final expression of irreversibility from $ExpressionArray$ by merging all expressions and $return$ \;}	
		\caption{\emph{CheckReversibilityForAllCell}}
	\label{Chap:semireversible:algo:rev_exp_algo}
\end{Walgo}

\setcounter{algocf}{1}

\begin{Walgo}[hbtp]{1.70cm}
	\BlankLine
	\SetKw{Fn}{Procedure}
	\SetKwFunction{FindGCD}{findGCD}
	\SetKwFunction{verifyReversibility}{verifyReversibility}
	\SetKwFunction{updateSubTreeLevels}{updateSubTreeLevels}
	\SetKwFunction{printIrreversibilityExpression}{printIrreversibilityExpression}
	\scriptsize
\Fn{\verifyReversibility{$element$}{\\
	\Begin{
	Set $loop \gets 0$, $missing \gets false$\;
	\For{$1 \leq \iota \leq m-1$}{
			$N' \leftarrow NodeList[element] $\;
			$\Gamma_{k}^{N'} \leftarrow \Gamma_{k }^{N'} \cap \lbrace i, i+d^{m-\iota}, i+2d^{m-\iota}, \cdots, i+(d^\iota-1)d^{m-\iota} \rbrace$, where $ i = \floor{\frac{k}{d^{\iota-1}}}$ and $ 0 \leq k \leq d^{m-1}-1$\; 
			\If{ ($N'$ is not balanced) OR ($\mid \bigcup\limits_{0\leq k \leq d^{m-1}-1} {\Gamma_k^{N'}}\mid \neq d^\iota $)}{
				Set $missing \gets true$ \;
					Set $q \gets \min(NodeLevel[element])$ ; \tcp{get the original level of $NodeList[element]$}
					\uIf{$NodeList[element].selfLoop=true$}{
						Set $conditional \gets false$\;
						Report $``$\textit{Trivial Semi-reversible. Irreversible for $n \geq q, (q\geq m)$}'' and $exit$ \;}
					\Else{
						 \ForEach{levels $p > q$ of $NodeLevel[element]$}{
								Set $exp \gets (n = j(p - q) + q + \iota)$, where $j \ge 0$ \;
								\If{$!ExpressionArray.contains(exp)$}{
									Add $exp$ to $ExpressionArray$\;
									\If{$ExpressionArray$ has expressions $n=j\times loop+ i $, $loop=p-q, b\leq i \leq b+loop$, $b$ any constant \tcp{That is, Irreversible for every $n \geq b$}}
									{
									Set $conditional \gets false$\;
									Report $``$\textit{Trivial Semi-reversible. Irreversible for $n \geq b, (b\geq m)$}'' and $exit$ \;
									}
									}
								}
						\lIf{$missing = true$}
						{ Set $conditional \gets true$ }
						}
					}
				}
	 	 	  }
			}
		}

\caption{\emph{CheckReversibilityForAllCell( ) contd...}}
\label{Chap:semireversible:algo:rev_exp_algo_1}
\end{Walgo}


The Algorithm~\ref{Chap:semireversible:algo:rev_exp_algo} takes a CA rule $R$ as input. In \ref{Chap:semireversible:algo:st1}, it checks whether the primary RMT sets of cardinality $1$ violate condition~\ref{irreversibility_c1} of Proposition~\ref{Chap:semireversible:Theorem:irreversibility}. If it is violated, the algorithm reports the CA as ``\emph{Strictly irreversible}'' following Theorem~\ref{Chap:semi-reversibility:Th:strictirreversibility} and returns. Otherwise, it moves to \ref{Chap:semireversible:algo:st2}. In \ref{Chap:semireversible:algo:st2}, it is checked, whether the CA rule $R$ is balanced or not. If unbalanced, the CA is declared as ``\emph{Trivially semi-reversible.}'' following Corollary~\ref{chap:semireversibility:th:revcor3} and exits. As minimum length of the tree to validate Corollary~\ref{chap:semireversibility:th:revcor3} is $m$, the CA with unbalanced rule is \emph{irreversible for all $n\geq m$}. Otherwise, in \ref{Chap:semireversible:algo:st3}, the root of the minimized reachability tree is formed and added to $NodeList$. $NodeLevel[0]$ is set to $\{0\}$. The other variables and data structures are also initialized. Recall that, the variables $q$ and $j$ keep track of the first and last unique nodes generated at previous level respectively and $i$ denotes the current level.

In \ref{Chap:semireversible:algo:st4}, the minimized tree is constructed as long as new unique nodes are added in $NodeList$. In this step, the children nodes of the unique nodes generated at the previous level are calculated. Each child is checked to find whether it violates Corollary~\ref{chap:semireversibility:th:revcor1} and Corollary~\ref{chap:semireversibility:th:revcor2}. If any such node is detected, the CA is declared as ``\emph{Trivially semi-reversible}'' as it is irreversible for all $n \geq i$, $i \ge m$, where $i$ is the current level. Otherwise, it is checked whether the node is duplicate or not. Each time a new node is found to be duplicate of an existing node with index $k$ of $NodeList$, the validity of the new loop is checked. If the new loop is valid or relevant, \emph{updateSubTreeLevels()} is called with the index of the unique node ($k$) to update the levels of the subtree. The procedure \emph{updateSubTreeLevels()} in turn calls \emph{verifyReversibility()} to verify whether any node of the subtree can violate reversibility conditions of Corollary~\ref{chap:semireversibility:th:revcor1} and Corollary~\ref{chap:semireversibility:th:revcor2}. If any such node is detected, the flag $missing$ is set to \emph{true}. If the node has self-loop, the CA is declared as ``\emph{Trivially semi-reversible}'' for all $n\geq q$ ($q$ is the minimum level of the node and $q\ge m$) and exits. Otherwise, the \emph{irreversibility expression} is calculated for that node and added to $ExpressionArray$. Moreover, if expressions of $ExpressionArray$ denote irreversibility for all $n\geq n'$, $n'$ is a level of the node, the algorithm reports the CA as ``\emph{Trivially semi-reversible}'' for all $n \geq n'$ and exits; otherwise the flag $conditional$ is set to \emph{true}. If the new node is found to be an unique node, it is added to $NodeList$ with its $NodeLevel$ updated by its parent node. If the new node had any loop associated with it, again \emph{verifyReversibility()} is called. 

In \ref{Chap:semireversible:algo:st5}, it is checked whether any new unique node is added in $NodeList$. If $NodeList$ has new entry in the last level, the algorithm updates the variables $q,j$ and $i$ and goes back to \ref{Chap:semireversible:algo:st4}, otherwise, it moves to \ref{Chap:semireversible:algo:st7}. As, a node is formed by the combinations of RMTs according to the rule, there can be only a finite number of unique nodes possible in the minimized reachability tree. The construction of minimized tree is complete, when no more unique node is generated in the tree (that is, $uId=j$). Hence, after a finite number of levels, the algorithm always moves to \ref{Chap:semireversible:algo:st7}. In \ref{Chap:semireversible:algo:st7}, the CA is declared ``\emph{Reversible}'', if the flag $missing$ and $conditional$ remained \emph{false}, that is, no node is detected in the tree which violates any reversibility condition. Otherwise, if $conditional$ is set to $true$, the CA is declared as ``\emph{Non-trivially semi-reversible}'' and the algorithm terminates. The final expression of irreversibility is found by combining all the irreversibility expressions.
%

\begin{example}

Let us consider ECA $75 (01001011)$ as input. The minimized reachability tree for this CA is to be drawn. Note that, the CA is balanced and RMTs $0$ and $7$ have different next state values, so the root $N_{0.0}$ is added to \emph{NodeList} and $0$ is added to $NodeLevel[0]$. Following our algorithm, we get $2$ nodes $N_{1.0}$ and $N_{1.1}$ at level $1$ (see Figure~\ref{Chap:reversibility:fig:rt3} of Page~\pageref{Chap:reversibility:fig:rt3}), where $ {\Gamma_1^{N_{1.0}}} = \{4,5\}, {\Gamma_2^{N_{1.0}}} = \{0,1,2,3\}, {\Gamma_3^{N_{1.0}}} = \{6,7\}$ and $ {\Gamma_0^{N_{1.1}}} = \{0,1,2,3\}, {\Gamma_1^{N_{1.1}}} = \{6,7\}, {\Gamma_3^{N_{1.1}}} = \{4,5\}$).
(For each node, the sets whose contents are not mentioned, are empty.)
These nodes are unique and added to \emph{NodeList} and level $1$ is added to the corresponding \emph{NodeLevel}. $uId$, that is, index of \emph{NodeList} is now increased to $2$. Note that, none of these nodes has any loop yet. The execution of the algorithm for this CA is shown in Table~\ref{Chap:semireversible:ex:AlgoEx1}. In this table, first column represents level $i$, second column the current $uId$, third column the content of \emph{NodeList}[$uId$] and the fourth column represents the \emph{NodeLevel}[$uId$]. Other five columns are related to the loop; if \emph{NodeLevel}[$uId$] is associated with a new loop, fifth column is set to \emph{yes} and the nodes which are affected by this loop are listed in the sixth column. However, for the nodes whose \emph{NodeLevel} gets a new loop for their parent node, the seventh column of Table~\ref{Chap:semireversible:ex:AlgoEx1} represents the parent $uId$. If a node has a new loop associated with it (that is, the corresponding fifth column has \emph{yes}), the eighth column is set to the level $n-2$ or $n-1$ for which the CA can be irreversible and the last column stores the corresponding irreversibility expression; otherwise, the eighth column is set to \emph{No}.

\begin{table}[!h]
\renewcommand{\arraystretch}{1.1}
\centering
\caption{Execution of Algorithm~\ref{Chap:semireversible:algo:rev_exp_algo} for ECA rule $75 (01001011)$}
\vspace{-0.5em}
\label{Chap:semireversible:ex:AlgoEx1}
\resizebox{1.0\textwidth}{9.1cm}{
\vspace{-\topsep} 
\begin{tabular}{|c|c|c|c@{\hspace{-0.1em}}|c@{\hspace{-0.1em}}|c@{\hspace{-0.1em}}|c@{\hspace{-0.1em}}|c|c|}
\hline 
$i$ & $uId$ & $NodeList[uId]$ &\begin{tabular}{c}$NodeLevel$\\ $[uId]$ \end{tabular} & \begin{tabular}{c}Loop\\ Updated?\end{tabular} &  \begin{tabular}{c} Affected\\ $uId$(s) \end{tabular}  & \begin{tabular}{c} Affected\\ by $uId$ \end{tabular} & \multicolumn{1}{c|}{\begin{tabular}{c}Unbalanced\\ for level?\end{tabular}}&
	\multicolumn{1}{c|}{ \begin{tabular}{c} Irreversibility \\ Expression \end{tabular}}\\ 
\hline 
$0$ & $0$ & $N_{0.0} = (\{0,1\}$, $\{2,3\}$, $\{4,5\}$, $\{6,7\})$ & $\lbrace 0 \rbrace$ & NA & NA & NA & NA & NA\\ 
\hline 

\multirow{2}{*}{$1$} & $1$ &  $N_{1.0} = (\emptyset, \{4,5\}$, $\{0,1,2,3\}$, $\{6,7\})$ & $\lbrace1\rbrace$ & NA & NA & NA & NA & NA\\ 
\hhline{~--------} 
& $2$ &  $N_{1.1} = (\{0,1,2,3\}$, $\{6,7\}, \emptyset, \{4,5\})$ & $\lbrace1\rbrace$ & NA & NA & NA & NA & NA\\ 
\hline 

\multirow{4}{*}{$2$} & $3$ & $N_{2.0} = (\emptyset, \{0,1,2,3\}$, $\{4,5\}$, $\{6,7\})$  & $\lbrace2\rbrace$ & NA & NA & NA & NA & NA\\ 
\hhline{~--------} 
& $4$ & $N_{2.1} = (\emptyset, \emptyset, \{0,1,2,3,6,7\}$, $\{4,5\})$ & $\lbrace2\rbrace$ & NA & NA & NA & NA & NA\\ 
\hhline{~--------} 
 & $5$ &  $N_{2.2} = (\{4,5\}$, $\{6,7\}, \emptyset, \{0,1,2,3\})$  & $\lbrace2\rbrace$ & NA & NA & NA & NA & NA\\ 
\hhline{~--------} 
 & $6$ &  $N_{2.3} = (\{0,1,2,3,6,7\}$, $\{4,5\}, \emptyset, \emptyset)$  & $\lbrace2\rbrace$ & NA & NA & NA & NA & NA\\ 
\hline 

\multirow{12}{*}{$3$} & $1$ & $N_{3.0} \equiv N_{1.0} =(\emptyset, \{4,5\}$, $\{0,1,2,3\}$, $\{6,7\})$ & $\lbrace1, 3\rbrace$ & Yes & $3$, $4$ & NA & $n-1$ & $n = 2j+2$\\ 
& $3$ & $N_{2.0} = (\emptyset, \{0,1,2,3\}$, $\{4,5\}$, $\{6,7\})$  & $\lbrace2, 4\rbrace$ & Yes & NA & $1$ & $n-2$ & $n=2j+4$\\ 
& $4$ & $N_{2.1} = (\emptyset, \emptyset, \{0,1,2,3,6,7\}$, $\{4,5\})$ & $\lbrace2, 4\rbrace$ & Yes & NA & $1$ & No & NA\\ 
\hhline{~--------} 
 & $7$ & $N_{3.1} = (\emptyset, \{0,1,2,3,6,7\}, \emptyset, \{4,5\})$  & $\lbrace3, 5\rbrace$ & NA & NA & $3$ & $n-1$ & $n=2j+4$\\ 
\hhline{~--------} 
 & $8$ &  $N_{3.2} = (\emptyset, \emptyset, \{4,5,6,7\}$, $\{0,1,2,3\})$ & $\lbrace3, 5\rbrace$ & NA & NA & $4$ & $n-1$ & $n=2j+4$\\ 
\hhline{~--------} 
 & $9$ &  $N_{3.3} = (\emptyset, \emptyset, \{0,1,2,3,4,5,6,7\}, \emptyset)$  & $\lbrace3, 5\rbrace$ & NA & NA & $4$ & No & NA\\ 
\hhline{~--------} 
 & $2$ & $N_{3.4} \equiv N_{1.1} = (\{0,1,2,3\}$, $\{6,7\}, \emptyset, \{4,5\})$ & $\lbrace1, 3\rbrace$ & Yes & $5$, $6$ & NA & $n-1$ & $n=2j+2$\\ 
 & $5$ &  $N_{2.2} = (\{4,5\}$, $\{6,7\}, \emptyset, \{0,1,2,3\})$  & $\lbrace2, 4\rbrace$ & Yes & NA & $2$ & $n-2$ & $n=2j+4$\\ 
 & $6$ &  $N_{2.3} = (\{0,1,2,3,6,7\}$, $\{4,5\}, \emptyset, \emptyset)$  & $\lbrace2, 4\rbrace$ & Yes & NA & $2$ & $n-2$ & $n=2j+4$\\ 
\hhline{~--------} 
 & $10$ & $N_{3.5} = (\emptyset, \{4,5\}, \emptyset,\{0,1,2,3,6,7\})$ & $\lbrace3, 5\rbrace$ & NA & NA & $5$ & $n-1$ & $n=2j+4$\\ 
\hhline{~--------} 
 & $11$ & $N_{3.6} = (\{4,5,6,7\}$, $\{0,1,2,3\}, \emptyset, \emptyset)$ & $\lbrace3, 5\rbrace$ & NA & NA & $6$ & No & NA \\ 
\hhline{~--------} 
 & $12$ & $N_{3.7} = (\{0,1,2,3,4,5,6,7\}, \emptyset, \emptyset, \emptyset)$ & $\lbrace 3, 5\rbrace$ & NA & NA & $6$ & No & NA\\ 
\hline 

\multirow{12}{*}{$4$} & $13$ & $N_{4.2} = (\emptyset, \{4,5,6,7\}, \emptyset, \{0,1,2,3\})$ & $\lbrace4, 6\rbrace$ & NA & NA & $7$ & $n-2$ & $n=2j+6$\\ 
\hhline{~--------} 
 & $14$ & $N_{4.3} = (\emptyset, \{0,1,2,3,4,5,6,7\}, \emptyset, \emptyset)$ & $\lbrace4, 6\rbrace$ & NA & NA & $7$ & No & NA\\ 
\hhline{~--------} 
 & $4$ & $N_{4.4} \equiv N_{2.1} = (\emptyset, \emptyset, \{0,1,2,3,6,7\}$, $\{4,5\})$ & $\lbrace2,4\rbrace$ & No & NA & NA & NA & NA\\ 
\hhline{~--------} 
 & $15$ & $N_{4.5} = (\emptyset, \emptyset, \{4,5\}$, $\{0,1,2,3,6,7\})$ & $\lbrace4, 6\rbrace$ & NA & NA & $8$ & No & NA\\ 
\hhline{~--------} 
 & $9$ & $N_{4.6} \equiv N_{3.3} = (\emptyset, \emptyset, \{0,1,2,3,4,5,6,7\}, \emptyset)$ & $\lbrace3,4\rbrace$ & Yes & $9$ & $9$ & No & NA\\ 
 & $9$ &  $N_{4.7} \equiv N_{3.3} = (\emptyset, \emptyset, \{0,1,2,3,4,5,6,7\}, \emptyset)$ & $\lbrace3,4 \rbrace$ & No & NA & NA & NA & NA\\ 
\hhline{~--------} 
 & $16$ &  $N_{4.10} = (\emptyset, \{0,1,2,3\}, \emptyset, \{4,5,6,7\})$ & $\lbrace 4, 6\rbrace$ & NA & NA & $10$ & No & NA\\ 
\hhline{~--------} 
 & $17$ & $N_{4.11} = (\emptyset, \emptyset, \emptyset, \{0,1,2,3,4,5,6,7\})$ & $\lbrace4, 6\rbrace$ & NA & NA & $10$ & No & NA\\ 
\hhline{~--------} 
 & $6$ &  $N_{4.12} \equiv N_{2.3} = (\{0,1,2,3,6,7\}$, $\{4,5\}, \emptyset, \emptyset)$ & $\lbrace2,4\rbrace$ & No & NA & NA & NA & NA\\ 
\hhline{~--------} 
 & $18$ & $N_{4.13} = (\{4,5\}$, $\{0,1,2,3,6,7\}, \emptyset, \emptyset)$ & $\lbrace4, 6\rbrace$ & NA & NA & $11$ & No & NA\\ 
\hhline{~--------} 
 & $12$ & $N_{4.14} \equiv N_{3.7} = (\{0,1,2,3,4,5,6,7\}, \emptyset, \emptyset, \emptyset)$ & $\lbrace3,4\rbrace$ & Yes & $12$ & $12$ & No & NA\\ 
 & $12$ & $N_{4.15} \equiv N_{3.7} = (\{0,1,2,3,4,5,6,7\}, \emptyset, \emptyset, \emptyset)$ & $\lbrace3,4\rbrace$ & No & NA & NA & NA & NA\\ 
\hline 

\multirow{12}{*}{$5$} & $7$  &  $N_{5.4} \equiv N_{3.1} = (\emptyset, \{0,1,2,3,6,7\}, \emptyset, \{4,5\})$  & $\lbrace3,5\rbrace$  & No & NA & NA & NA & NA\\ 
\hhline{~--------}  
 & $10$  &  $N_{5.5} \equiv N_{3.5} = (\emptyset, \{4,5\}, \emptyset,\{0,1,2,3,6,7\})$ & $\lbrace3,5\rbrace$ & No & NA & NA & NA & NA\\ 
\hhline{~--------} 
 & $14$  & $N_{5.6} \equiv N_{4.3} = (\emptyset, \{0,1,2,3,4,5,6,7\}, \emptyset, \emptyset)$ & $\lbrace4,5\rbrace$ & Yes & $14$ & $14$ & No & NA\\ 
 & $14$  & $N_{5.7} \equiv N_{4.3} = (\emptyset, \{0,1,2,3,4,5,6,7\}, \emptyset, \emptyset)$ & $\lbrace4,5\rbrace$  & No & NA & NA & NA & NA\\ 
\hhline{~--------} 
 & $19$  & $N_{5.10} = (\emptyset, \emptyset, \{0,1,2,3\}$, $\{4,5,6,7\})$ & $\lbrace5, 7\rbrace$ & NA & NA & $15$ & $n-1$ & $n=2j+6$\\ 
\hhline{~--------} 
 & $17$  & $N_{5.11} \equiv N_{4.11} = (\emptyset, \emptyset, \emptyset, \{0,1,2,3,4,5,6,7\})$ & $\lbrace4,5\rbrace$ & Yes & NA & NA & No & NA\\ 
\hhline{~--------} 
 & $10$  & $N_{5.20} \equiv N_{3.5} = (\emptyset, \{4,5\}, \emptyset,\{0,1,2,3,6,7\})$ & $\lbrace3,5\rbrace$ & No & NA & NA & NA & NA\\ 
\hhline{~--------} 
 & $7$  & $N_{5.21} \equiv N_{3.1} = (\emptyset, \{0,1,2,3,6,7\}, \emptyset, \{4,5\})$ & $\lbrace 3,5\rbrace$ & No & NA & NA& NA & NA \\ 
\hhline{~--------} 
 & $17$  & $N_{5.22} \equiv N_{4.11} = (\emptyset, \emptyset, \emptyset, \{0,1,2,3,4,5,6,7\})$ & $\lbrace4,5\rbrace$ & No & NA & NA & NA & NA\\ 
\hhline{~--------} 
 & $17$  & $N_{5.23} \equiv N_{4.11} = (\emptyset, \emptyset, \emptyset, \{0,1,2,3,4,5,6,7\})$ & $\lbrace4,5\rbrace$ & No & NA & NA & NA & NA\\ 
\hhline{~--------} 
 & $20$  & $N_{5.26} = (\{0,1,2,3\}$, $\{4,5,6,7\}, \emptyset, \emptyset)$  & $\lbrace5, 7\rbrace$ & NA & NA & $18$ & No & NA\\ 
\hhline{~--------} 
 & $14$  & $N_{5.27} \equiv N_{4.3} = (\emptyset, \{0,1,2,3,4,5,6,7\}, \emptyset, \emptyset)$ & $\lbrace4,5\rbrace$ & No & NA & NA & NA & NA\\ 
\hline 
 
\multirow{4}{*}{$6$}  &  $15$ &  $N_{6.20} \equiv N_{4.5} = (\emptyset, \emptyset, \{4,5\}$, $\{0,1,2,3,6,7\})$  & $\lbrace4,6\rbrace$ & No & NA & NA & NA & NA\\ 
\hhline{~--------}  
 &  $4$ & $N_{6.21} \equiv N_{2.1} = (\emptyset, \emptyset, \{0,1,2,3,6,7\}$, $\{4,5\})$ & $\{2,4\}$ & No & NA & NA & NA & NA\\ 
\hhline{~--------}  
 &  $18$ & $N_{6.52} \equiv N_{4.13} = (\{4,5\}$, $\{0,1,2,3,6,7\}, \emptyset, \emptyset)$ & $\{4,6\}$ & No & NA & NA & NA & NA\\ 
\hhline{~--------}   
 &  $6$ & $N_{6.53} \equiv N_{2.3} = (\{0,1,2,3,6,7\}$, $\{4,5\}, \emptyset, \emptyset)$ & $\{2,4\}$ & No & NA & NA & NA & NA\\ 
\hline 
\end{tabular}
}
\end{table}

From Table~\ref{Chap:semireversible:ex:AlgoEx1}, it can be seen that, at level $2$ ($i=2$), all nodes are unique and added to \emph{NodeList}. At level $3$, however, $N_{3.0} \equiv N_{1.0}$ and $N_{3.4} \equiv N_{1.1}$; these two loops are valid and accordingly \emph{NodeLevel} of $6$ existing nodes are updated. By the procedure \emph{verifyReversibility}, we get that $NodeList[1]=N_{3.0} \equiv N_{1.0}$ and $NodeList[2]=N_{3.4} \equiv N_{1.1}$ can be unbalanced for level $n-1$, so, we get one irreversibility expression as $n = j(3-1)+1+1=2j+2$ to be added to $ExpressionArray$. Similarly, $NodeList[3]$, $NodeList[5]$ and $NodeList[6]$ can be unbalanced for level $n-2$; another expression of irreversibility as $n = j(4-2)+2+2=2j+4$ is added to $ExpressionArray$. Moreover, $6$ new unique nodes are also added in this level among which $3$ nodes $NodeList[7]$, $NodeList[8]$ and $NodeList[10]$ can be unbalanced for level $n-1$. 

At level $4$ also, $6$ unique nodes are added to \emph{NodeList}. As, each of these nodes has multiple levels in their \emph{NodeLevel}, so, each is verified using \emph{verifyReversibility}; only $NodeList[13]$ is found to unbalanced for level $n-2$ and another expression of irreversibility, $n = 2j+6$ is added to $ExpressionArray$. Among the duplicate nodes, new loops for nodes $N_{2.1}$ and $N_{2.3}$ are not relevant, so, $NodeLevel[4]$ and $NodeLevel[6]$ remain unchanged. But, $NodeLevel[9]$ and $NodeLevel[12]$ are updated with levels of their new loop value. These nodes have no sub-tree to update and does not violate the reversibility conditions for levels $n-2$ and $n-1$.

At the next level, only $2$ unique nodes are added to \emph{NodeList}. However, among these nodes, only $NodeList[19]$ can be unbalanced for level $n-1$, which yields one existing irreversibility expression $n = 2j+6$. Among the duplicate nodes, only $N_{4.3}$ and $N_{4.11}$ have updated their \emph{NodeLevel}, but does not violate reversibility conditions. 

At level $6$, no new unique node is generated, as well as, no new relevant loop is found for the duplicate nodes. So, the algorithm jumps to \ref{Chap:semireversible:algo:st7}. The minimized tree for the CA is shown in Figure~\ref{Chap:reversibility:fig:rt3} (Page~\pageref{Chap:reversibility:fig:rt3}). Here, dashed lines represent the invalid loops which are discarded. The tree has only $21$ nodes and last unique node is generated at level $5$. Here, the flag \emph{missing = true} and neither any of the missing nodes has self-loops nor expressions of $ExpressionArray$ implies trivial semi-reversibility. So, \emph{conditional = true}. Hence, the ECA with rule $75$ is declared \emph{non-trivial semi-reversible} with final expression of irreversibility as $n = 2j +2$, $j \ge 0$, that is, reversible when lattice size $n$ is odd.
\end{example}

\begin{example}
Consider a $2$-state $4$-neighborhood CA $0110010101010101$. Note that this CA is balanced and RMTs $0$ and $15$ have different next state values, so Corollary~\ref{chap:semireversibility:th:revcor3} and Theorem~\ref{Chap:semi-reversibility:Th:strictirreversibility} are not violated. Therefore, the minimized reachability tree for the CA is constructed. 

\begin{scriptsize}\setlength\tabcolsep{5pt}
	\centering
\begin{longtable}{|c|c|p{6.5cm}|c@{\hspace{-0.3em}}|c@{\hspace{-0.5em}}|c@{\hspace{-0.5em}}|}
\caption{Execution of Algorithm~\ref{Chap:semireversible:algo:rev_exp_algo} for $2$-state $4$-neighborhood CA $0110010101010101$}
\vspace{-0.5em}
\label{Chap:semireversible:ex:AlgoEx2}\\
\hline
$i$ & $uId$ & $NodeList[uId]$ &\begin{tabular}{c}$NodeLevel$\\ $[uId]$ \end{tabular} &   \multicolumn{1}{c|}{\begin{tabular}{c}Unbalanced\\ for level?\end{tabular}}&
\multicolumn{1}{c|}{ \begin{tabular}{c} Irreversibility \\ Expression \end{tabular}}\\ 
\hline 
\endfirsthead
\multicolumn{6}{c}%
{\tablename\ \thetable\ -- \textit{Continued from previous page}} \\
\hline
$i$ & $uId$ & $NodeList[uId]$ &\begin{tabular}{c}$NodeLevel$\\ $[uId]$ \end{tabular} &   \multicolumn{1}{c|}{\begin{tabular}{c}Unbalanced\\ for level?\end{tabular}}&
\multicolumn{1}{c|}{ \begin{tabular}{c} Irreversibility \\ Expression \end{tabular}}\\ 
\hline
\endhead
\hline \multicolumn{6}{r}{\textit{Continued on next page}} \\
\endfoot
\hline
\endlastfoot

$0$ & $0$ & $N_{0.0} =(\{0, 1\}$, $\{2, 3\}$, $\{4, 5\}$, $\{6, 7\}$, $\{8, 9\}$, $\{10, 11\}$, $\{12, 13\}$, $\{14, 15\})$ & $\lbrace 0 \rbrace$ & NA & NA \\ 
\hline 

\multirow{2}{*}{$1$} & $1$ &  $N_{1.0} = (\{2, 3\}$, $\{6, 7\}$, $\{10, 11\}$, $\{14, 15\}$, $\{2, 3\}$, $\{6, 7\}$, $\{8, 9\}$, $\{14, 15\})$ & $\lbrace1\rbrace$ & NA & NA \\ 
 \hhline{~-----} 
& $2$ &  $N_{1.1} = (\{0, 1\}$, $\{4, 5\}$, $\{8, 9\}$, $\{12, 13\}$, $\{0, 1\}$, $\{4, 5\}$, $\{10, 11\}$, $\{12, 13\})$ & $\lbrace1\rbrace$ & NA  & NA\\ 
\hline 

\multirow{4}{*}{$2$} & $3$ & $N_{2.0} = (\{6, 7\}$, $\{14, 15\}$, $\{6, 7\}$, $\{14, 15\}$, $\{6, 7\}$, $\{14, 15\}$, $\{2, 3\}$, $\{14, 15\})$  & $\lbrace2\rbrace$ & NA & NA  \\ 
 \hhline{~-----} 
& $4$ & $N_{2.1} = (\{4, 5\}$, $\{12, 13\}$, $\{4, 5\}$, $\{12, 13\}$, $\{4, 5\}$, $\{12, 13\}$, $\{0, 1\}$, $\{12, 13\})$ & $\lbrace2\rbrace$ & NA & NA  \\ 
 \hhline{~-----} 
 & $5$ &  $N_{2.2} = (\{2, 3\}$, $\{10, 11\}$, $\{2, 3\}$, $\{8, 9\}$, $\{2, 3\}$, $\{10, 11\}$, $\{6, 7\}$, $\{8, 9\})$  & $\lbrace2\rbrace$ & NA & NA  \\ 
 \hhline{~-----} 
 & $6$ &  $N_{2.3} = (\{0, 1\}$, $\{8, 9\}$, $\{0, 1\}$, $\{10, 11\}$, $\{0, 1\}$, $\{8, 9\}$, $\{4, 5\}$, $\{10, 11\})$  & $\lbrace2\rbrace$ & NA & NA  \\ 
\hline 

\multirow{8}{*}{$3$} & $7$ & $N_{3.0} \equiv N_{6.12} = (\{14, 15\}$, $\{14, 15\}$, $\{14, 15\}$, $\{14, 15\}$, $\{14, 15\}$, $\{14, 15\}$, $\{6, 7\}$, $\{14, 15\})$ & $\lbrace 3,6 \rbrace$ & No & NA  \\ 
 \hhline{~-----} 
& $8$ & $N_{3.1} \equiv N_{6.13} = (\{12, 13\}$, $\{12, 13\}$, $\{12, 13\}$, $\{12, 13\}$, $\{12, 13\}$, $\{12, 13\}$, $\{4, 5\}$, $\{12, 13\})$ & $\lbrace3,6\rbrace$ & $n-3$ & $n=3j+6$  \\ 
 \hhline{~-----} 
 & $9$ &  $N_{3.2} = (\{10, 11\}$, $\{8, 9\}$, $\{10, 11\}$, $\{8, 9\}$, $\{10, 11\}$, $\{8, 9\}$, $\{2, 3\}$, $\{8, 9\})$  & $\lbrace3\rbrace$ & NA & NA  \\ 
 \hhline{~-----} 
 & $10$ &  $N_{3.3} = (\{8, 9\}$, $\{10, 11\}$, $\{8, 9\}$, $\{10, 11\}$, $\{8, 9\}$, $\{10, 11\}$, $\{0, 1\}$, $\{10, 11\})$  & $\lbrace3\rbrace$ & NA & NA  \\ 
  \hhline{~-----} 
 & $11$ & $N_{3.4} = (\{6, 7\}$, $\{6, 7\}$, $\{6, 7\}$, $\{2, 3\}$, $\{6, 7\}$, $\{6, 7\}$, $\{14, 15\}$, $\{2, 3\})$ & $\lbrace3\rbrace$ & NA & NA  \\ 
  \hhline{~-----} 
  & $12$ &  $N_{3.5} = (\{4, 5\}$, $\{4, 5\}$, $\{4, 5\}$, $\{0, 1\}$, $\{4, 5\}$, $\{4, 5\}$, $\{12, 13\}$, $\{0, 1\})$  & $\lbrace3\rbrace$ & NA & NA  \\ 
  \hhline{~-----} 
 & $13$ & $N_{3.6} = (\{2, 3\}$, $\{2, 3\}$, $\{2, 3\}$, $\{6, 7\}$, $\{2, 3\}$, $\{2, 3\}$, $\{10, 11\}$, $\{6, 7\})$ & $\lbrace3\rbrace$ & NA & NA  \\ 
 \hhline{~-----} 
 & $14$ & $N_{3.7} = (\{0, 1\}$, $\{0, 1\}$, $\{0, 1\}$, $\{4, 5\}$, $\{0, 1\}$, $\{0, 1\}$, $\{8, 9\}$, $\{4, 5\})$ & $\lbrace 3\rbrace$ & NA & NA  \\ 
\hline 

\multirow{16}{*}{$4$} & $15$ & $N_{4.0}\equiv N_{5.0} \equiv N_{5.16} \equiv N_{5.24} \equiv N_{6.16} \equiv N_{6.24} \equiv N_{6.56} \equiv N_{7.16}= (\{14, 15\}$, $\{14, 15\}$, $\{14, 15\}$, $\{14, 15\}$, $\{14, 15\}$, $\{14, 15\}$, $\{14, 15\}$, $\{14, 15\})$ & $\lbrace4,5\rbrace$  & No & NA \\ 
 \hhline{~-----} 
 & $16$ & $N_{4.1} \equiv N_{5.1} \equiv N_{5.17} \equiv N_{5.25} \equiv N_{6.17} \equiv N_{6.25} \equiv N_{6.57} \equiv N_{7.17}= (\{12, 13\}$, $\{12, 13\}$, $\{12, 13\}$, $\{12, 13\}$, $\{12, 13\}$, $\{12, 13\}$, $\{12, 13\}$, $\{12, 13\})$ & $\lbrace4,5\rbrace$  & No & NA \\ 
 \hhline{~-----} 
 & $17$ & $N_{4.2} \equiv N_{7.19} = (\{8, 9\}$, $\{8, 9\}$, $\{8, 9\}$, $\{8, 9\}$, $\{8, 9\}$, $\{8, 9\}$, $\{10, 11\}$, $\{8, 9\})$ & $\lbrace4,7\rbrace$ & $n-2$ & $n=3j+6$  \\ 
 \hhline{~-----} 
 & $18$ & $N_{4.3} \equiv N_{7.18} = (\{10, 11\}$, $\{10, 11\}$, $\{10, 11\}$, $\{10, 11\}$, $\{10, 11\}$, $\{10, 11\}$, $\{8, 9\}$, $\{10, 11\})$ & $\lbrace4,7\rbrace$ & $n-2$ & $n=3j+6$  \\ 
 \hhline{~-----} 
 & $19$ & $N_{4.4} \equiv N_{7.38}= (\{6, 7\}$, $\{2, 3\}$, $\{6, 7\}$, $\{2, 3\}$, $\{6, 7\}$, $\{2, 3\}$, $\{6, 7\}$, $\{2, 3\})$ & $\lbrace4,7\rbrace$ & No & NA  \\ 
 \hhline{~-----} 
 & $20$ &  $N_{4.5} \equiv N_{7.39}= (\{4, 5\}$, $\{0, 1\}$, $\{4, 5\}$, $\{0, 1\}$, $\{4, 5\}$, $\{0, 1\}$, $\{4, 5\}$, $\{0, 1\})$ & $\lbrace4,7\rbrace$ & No & NA  \\ 
 \hhline{~-----} 
 & $21$ & $N_{4.6} \equiv N_{7.36} = (\{2, 3\}$, $\{6, 7\}$, $\{2, 3\}$, $\{6, 7\}$, $\{2, 3\}$, $\{6, 7\}$, $\{2, 3\}$, $\{6, 7\})$ & $\lbrace4,7\rbrace$ & No & NA  \\ 
 \hhline{~-----} 
 & $22$ &  $N_{4.7} \equiv N_{7.37}= (\{0, 1\}$, $\{4, 5\}$, $\{0, 1\}$, $\{4, 5\}$, $\{0, 1\}$, $\{4, 5\}$, $\{0, 1\}$, $\{4, 5\})$ & $\lbrace4,7\rbrace$ & No & NA  \\ 
 & $23$ & $N_{4.8} \equiv N_{7.76}= (\{14, 15\}$, $\{14, 15\}$, $\{14, 15\}$, $\{6, 7\}$, $\{14, 15\}$, $\{14, 15\}$, $\{14, 15\}$, $\{6, 7\})$ & $\lbrace4,7\rbrace$ & No & NA  \\ 
 \hhline{~-----} 
 & $24$ & $N_{4.9} \equiv N_{7.77} = (\{12, 13\}$, $\{12, 13\}$, $\{12, 13\}$, $\{4, 5\}$, $\{12, 13\}$, $\{12, 13\}$, $\{12, 13\}$, $\{4, 5\})$ & $\lbrace4,7\rbrace$  & $n-2$ & $n=3j+6$ \\ 
  \hhline{~-----} 
  & $25$ & $N_{4.10} = (\{10, 11\}$, $\{10, 11\}$, $\{10, 11\}$, $\{2, 3\}$, $\{10, 11\}$, $\{10, 11\}$, $\{8, 9\}$, $\{2, 3\})$ & $\lbrace4\rbrace$ & NA & NA  \\ 
  \hhline{~-----} 
  & $26$ & $N_{4.11} = (\{8, 9\}$, $\{8, 9\}$, $\{8, 9\}$, $\{0, 1\}$, $\{8, 9\}$, $\{8, 9\}$, $\{10, 11\}$, $\{0, 1\})$ & $\lbrace4\rbrace$ & NA & NA  \\ 
  \hhline{~-----} 
  & $27$ & $N_{4.12} \equiv N_{7.72}= (\{6, 7\}$, $\{6, 7\}$, $\{6, 7\}$, $\{14, 15\}$, $\{6, 7\}$, $\{6, 7\}$, $\{6, 7\}$, $\{14, 15\})$ & $\lbrace4,7\rbrace$ & No & NA  \\ 
  \hhline{~-----} 
  & $28$ & $N_{4.13} \equiv N_{7.73}= (\{4, 5\}$, $\{4, 5\}$, $\{4, 5\}$, $\{12, 13\}$, $\{4, 5\}$, $\{4, 5\}$, $\{4, 5\}$, $\{12, 13\})$ & $\lbrace4,7\rbrace$ & $n-2$ & $n=3j+6$  \\ 
  \hhline{~-----} 
  & $29$ &  $N_{4.14} \equiv N_{7.74}= (\{2, 3\}$, $\{2, 3\}$, $\{2, 3\}$, $\{10, 11\}$, $\{2, 3\}$, $\{2, 3\}$, $\{2, 3\}$, $\{10, 11\})$ & $\lbrace4,7\rbrace$ & No & NA  \\ 
  \hhline{~-----} 
  & $30$ & $N_{4.15} \equiv N_{7.75} = (\{0, 1\}$, $\{0, 1\}$, $\{0, 1\}$, $\{8, 9\}$, $\{0, 1\}$, $\{0, 1\}$, $\{0, 1\}$, $\{8, 9\})$ & $\lbrace4,7\rbrace$ & No & NA  \\ 
\hline 

\multirow{20}{*}{$5$} & $31$  & $N_{5.2} \equiv N_{6.59}= (\{8, 9\}$, $\{8, 9\}$, $\{8, 9\}$, $\{8, 9\}$, $\{8, 9\}$, $\{8, 9\}$, $\{8, 9\}$, $\{8, 9\})$ & $\lbrace5,6\rbrace$  & No & NA \\ 
  \hhline{~-----} 
 & $32$  & $N_{5.3} \equiv N_{6.58}= (\{10, 11\}$, $\{10, 11\}$, $\{10, 11\}$, $\{10, 11\}$, $\{10, 11\}$, $\{10, 11\}$, $\{10, 11\}$, $\{10, 11\})$ & $\lbrace5,6\rbrace$   & No & NA \\ 
 \hhline{~-----} 
 & $33$  & $N_{5.4} \equiv N_{5.22}= (\{2, 3\}$, $\{2, 3\}$, $\{2, 3\}$, $\{2, 3\}$, $\{2, 3\}$, $\{2, 3\}$, $\{6, 7\}$, $\{2, 3\})$ & $\lbrace5,8\rbrace$ & $n-1$ & $n=3j+6$  \\ 
 \hhline{~-----} 
 & $34$  & $N_{5.5}\equiv N_{5.23} = (\{0, 1\}$, $\{0, 1\}$, $\{0, 1\}$, $\{0, 1\}$, $\{0, 1\}$, $\{0, 1\}$, $\{4, 5\}$, $\{0, 1\})$ & $\lbrace5,8\rbrace$ & No  & NA \\ 
 \hhline{~-----} 
 & $35$  & $N_{5.6} \equiv N_{5.20} = (\{6, 7\}$, $\{6, 7\}$, $\{6, 7\}$, $\{6, 7\}$, $\{6, 7\}$, $\{6, 7\}$, $\{2, 3\}$, $\{6, 7\})$ & $\lbrace5,8\rbrace$ & $n-1$ & $n=3j+6$  \\ 
 \hhline{~-----} 
 & $36$  & $N_{5.7} \equiv  N_{5.21} = (\{4, 5\}$, $\{4, 5\}$, $\{4, 5\}$, $\{4, 5\}$, $\{4, 5\}$, $\{4, 5\}$, $\{0, 1\}$, $\{4, 5\}) $ & $\lbrace5,8\rbrace$ & No & NA \\ 
 \hhline{~-----} 
 & $37$  & $N_{5.8} = (\{14, 15\}$, $\{6, 7\}$, $\{14, 15\}$, $\{6, 7\}$, $\{14, 15\}$, $\{6, 7\}$, $\{14, 15\}$, $\{6, 7\})$ & $\lbrace5,8\rbrace$ & No & NA  \\ 
 \hhline{~-----} 
 & $38$  & $N_{5.9} = (\{12, 13\}$, $\{4, 5\}$, $\{12, 13\}$, $\{4, 5\}$, $\{12, 13\}$, $\{4, 5\}$, $\{12, 13\}$, $\{4, 5\})$ & $\lbrace5,8\rbrace$ & $n-1$ & $n=3j+6$  \\ 
 \hhline{~-----} 
 & $39$  & $N_{5.10} = (\{10, 11\}$, $\{2, 3\}$, $\{10, 11\}$, $\{2, 3\}$, $\{10, 11\}$, $\{2, 3\}$, $\{10, 11\}$, $\{2, 3\})$  & $\lbrace5,8\rbrace$  & No & NA \\ 
 \hhline{~-----} 
 & $40$  & $N_{5.11} = (\{8, 9\}$, $\{0, 1\}$, $\{8, 9\}$, $\{0, 1\}$, $\{8, 9\}$, $\{0, 1\}$, $\{8, 9\}$, $\{0, 1\})$ & $\lbrace5,8\rbrace$ & No & NA  \\ 
  \hhline{~-----}  
 &  $41$ & $N_{5.12} = (\{6, 7\}$, $\{14, 15\}$, $\{6, 7\}$, $\{14, 15\}$, $\{6, 7\}$, $\{14, 15\}$, $\{6, 7\}$, $\{14, 15\})$ & $\lbrace5,8\rbrace$ & No & NA  \\ 
  \hhline{~-----}  
  &  $42$ & $N_{5.13} = (\{4, 5\}$, $\{12, 13\}$, $\{4, 5\}$, $\{12, 13\}$, $\{4, 5\}$, $\{12, 13\}$, $\{4, 5\}$, $\{12, 13\})$ & $\lbrace5,8\rbrace$ & $n-1$ & $n=3j+6$  \\ 
  \hhline{~-----}   
  &  $43$ & $N_{5.14} = (\{2, 3\}$, $\{10, 11\}$, $\{2, 3\}$, $\{10, 11\}$, $\{2, 3\}$, $\{10, 11\}$, $\{2, 3\}$, $\{10, 11\}) $ & $\lbrace5,8\rbrace$ & No & NA  \\ 
    \hhline{~-----} 
   & $44$  & $N_{5.15} = (\{0, 1\}$, $\{8, 9\}$, $\{0, 1\}$, $\{8, 9\}$, $\{0, 1\}$, $\{8, 9\}$, $\{0, 1\}$, $\{8, 9\})$ & $\lbrace5,8\rbrace$   & No & NA \\ 
   \hhline{~-----} 
   & $45$  & $N_{5.18} \equiv N_{5.27} = (\{8, 9\}$, $\{8, 9\}$, $\{8, 9\}$, $\{10, 11\}$, $\{8, 9\}$, $\{8, 9\}$, $\{8, 9\}$, $\{10, 11\})$ & $\lbrace5,8\rbrace$ & $n-1$ & $n=3j+6$  \\ 
   \hhline{~-----} 
   & $46$  & $N_{5.19} \equiv N_{5.26} = (\{10, 11\}$, $\{10, 11\}$, $\{10, 11\}$, $\{8, 9\}$, $\{10, 11\}$, $\{10, 11\}$, $\{10, 11\}$, $\{8, 9\}) $ & $\lbrace5,8\rbrace$ & $n-1$ & $n=3j+6$ \\ 
   \hhline{~-----} 
   & $47$  & $N_{5.28} \equiv N_{6.7} \equiv N_{6.20} \equiv N_{6.28} \equiv N_{6.60} \equiv N_{7.20} \equiv N_{7.28} \equiv N_{8.156}= (\{6, 7\}$, $\{6, 7\}$, $\{6, 7\}$, $\{6, 7\}$, $\{6, 7\}$, $\{6, 7\}$, $\{6, 7\}$, $\{6, 7\}) $ & $\lbrace5,6\rbrace$ & No  & NA \\ 
   \hhline{~-----} 
   & $48$  & $N_{5.29} \equiv N_{6.8} \equiv N_{6.21} \equiv N_{6.29} \equiv N_{6.61} \equiv N_{7.21} \equiv N_{7.29} \equiv N_{8.157}= (\{4, 5\}$, $\{4, 5\}$, $\{4, 5\}$, $\{4, 5\}$, $\{4, 5\}$, $\{4, 5\}$, $\{4, 5\}$, $\{4, 5\}) $ & $\lbrace 5,6\rbrace$ & No & NA \\ 
   \hhline{~-----} 
   & $49$  & $N_{5.30} \equiv N_{6.4} \equiv N_{6.22} \equiv N_{6.30} \equiv N_{6.62} \equiv N_{7.22} \equiv N_{7.30} \equiv N_{8.158}= (\{2, 3\}$, $\{2, 3\}$, $\{2, 3\}$, $\{2, 3\}$, $\{2, 3\}$, $\{2, 3\}$, $\{2, 3\}$, $\{2, 3\})$ & $\lbrace5,6\rbrace$ & No  & NA \\ 
   \hhline{~-----} 
   & $50$  & $N_{5.31} \equiv N_{6.5} \equiv N_{6.23} \equiv N_{6.31} \equiv N_{6.63} \equiv N_{7.23} \equiv N_{7.31} \equiv N_{8.159}= (\{0, 1\}$, $\{0, 1\}$, $\{0, 1\}$, $\{0, 1\}$, $\{0, 1\}$, $\{0, 1\}$, $\{0, 1\}$, $\{0, 1\})$ & $\lbrace5,6\rbrace$ & No  & NA \\ 
\hline 
 
\multirow{12}{*}{$6$}  & $51$  & $N_{6.8} = (\{6, 7\}$, $\{6, 7\}$, $\{6, 7\}$, $\{6, 7\}$, $\{6, 7\}$, $\{6, 7\}$, $\{14, 15\}$, $\{6, 7\})$  & $\lbrace6,9\rbrace$  & No & NA \\ 
   \hhline{~-----} 
& $52$  & $N_{6.9} = (\{4, 5\}$, $\{4, 5\}$, $\{4, 5\}$, $\{4, 5\}$, $\{4, 5\}$, $\{4, 5\}$, $\{12, 13\}$, $\{4, 5\}) $ & $\lbrace6,9\rbrace$ & $n-3$  & $n=3j+9$ \\ 
    \hhline{~-----}  
&  $53$ & $N_{6.10} = (\{2, 3\}$, $\{2, 3\}$, $\{2, 3\}$, $\{2, 3\}$, $\{2, 3\}$, $\{2, 3\}$, $\{10, 11\}$, $\{2, 3\})$ & $\{6,9\}$ & No  & NA \\ 
    \hhline{~-----}  
&  $54$ & $N_{6.11} = (\{0, 1\}$, $\{0, 1\}$, $\{0, 1\}$, $\{0, 1\}$, $\{0, 1\}$, $\{0, 1\}$, $\{8, 9\}$, $\{0, 1\}) $ & $\{6,9\}$ & No  & NA \\ 
    \hhline{~-----}   
&  $55$ & $N_{6.14} = (\{10, 11\}$, $\{10, 11\}$, $\{10, 11\}$, $\{10, 11\}$, $\{10, 11\}$, $\{10, 11\}$, $\{2, 3\}$, $\{10, 11\}) $ & $\{6,9\}$ & No  & NA \\ 
  \hhline{~-----}   
 &  $56$ & $N_{6.15} = (\{8, 9\}$, $\{8, 9\}$, $\{8, 9\}$, $\{8, 9\}$, $\{8, 9\}$, $\{8, 9\}$, $\{0, 1\}$, $\{8, 9\})$ & $\{6,9\}$ & No  & NA \\ 
 \hhline{~-----}  
 &  $57$ & $N_{6.18} \equiv N_{6.27}= (\{8, 9\}$, $\{10, 11\}$, $\{8, 9\}$, $\{10, 11\}$, $\{8, 9\}$, $\{10, 11\}$, $\{8, 9\}$, $\{10, 11\})$ & $\{6,9\}$ & No  & NA \\ 
 \hhline{~-----}   
 &  $58$ & $N_{6.19} \equiv N_{6.26} = (\{10, 11\}$, $\{8, 9\}$, $\{10, 11\}$, $\{8, 9\}$, $\{10, 11\}$, $\{8, 9\}$, $\{10, 11\}$, $\{8, 9\})$ & $\{6,9\}$ & No  & NA \\ 
     \hhline{~-----}   
 &  $59$ & $N_{6.36} = (\{2, 3\}$, $\{2, 3\}$, $\{2, 3\}$, $\{6, 7\}$, $\{2, 3\}$, $\{2, 3\}$, $\{2, 3\}$, $\{6, 7\}) $ & $\{6,9\}$ & No  & NA \\ 
   \hhline{~-----}   
  &  $60$ & $N_{6.37} = (\{0, 1\}$, $\{0, 1\}$, $\{0, 1\}$, $\{4, 5\}$, $\{0, 1\}$, $\{0, 1\}$, $\{0, 1\}$, $\{4, 5\}) $ & $\{6,9\}$ & No  & NA \\ 
  \hhline{~-----}  
  &  $61$ & $N_{6.38} = (\{6, 7\}$, $\{6, 7\}$, $\{6, 7\}$, $\{2, 3\}$, $\{6, 7\}$, $\{6, 7\}$, $\{6, 7\}$, $\{2, 3\})$ & $\{6,9\}$ & No  & NA \\ 
  \hhline{~-----}   
  &  $62$ & $N_{6.39} = (\{4, 5\}$, $\{4, 5\}$, $\{4, 5\}$, $\{0, 1\}$, $\{4, 5\}$, $\{4, 5\}$, $\{4, 5\}$, $\{0, 1\})$ & $\{6,9\}$ & No  & NA \\ 
\hline 

\multirow{2}{*}{$7$}  & $63$  & $N_{7.78} = (\{10, 11\}$, $\{10, 11\}$, $\{10, 11\}$, $\{2, 3\}$, $\{10, 11\}$, $\{10, 11\}$, $\{10, 11\}$, $\{2, 3\})$  & $\lbrace7,10\rbrace$  & No & NA \\ 
   \hhline{~-----} 
   & $64$  & $N_{7.79} = (\{8, 9\}$, $\{8, 9\}$, $\{8, 9\}$, $\{0, 1\}$, $\{8, 9\}$, $\{8, 9\}$, $\{8, 9\}$, $\{0, 1\}) $ & $\lbrace7, 10\rbrace$ & No  & NA \\ 
   \end{longtable}
   \end{scriptsize}

Table~\ref{Chap:semireversible:ex:AlgoEx2} gives the structure of execution of Algorithm~\ref{Chap:semireversible:algo:rev_exp_algo} for this CA. Here, up to level $4$, all nodes are unique and added to $NodeList$. However, at level $5$, out of $2^5=32$ nodes, $20$ nodes are unique and added to $NodeList$. The duplicate nodes  $ N_{5.0} \equiv N_{5.16} \equiv N_{5.24}\equiv N_{4.0}$ and $N_{5.1} \equiv N_{5.17} \equiv N_{5.25}\equiv N_{4.1} $, creating self-loops for nodes $N_{4.0}$ and $N_{4.1} $. But, none of these nodes or the affected children nodes $N_{5.2} $, $N_{5.3} $ violate reversibility conditions. 

At the next level, only $12$ nodes are unique and added to $NodeList$. Among the duplicate nodes,    $ N_{6.7} \equiv N_{6.20} \equiv N_{6.28} \equiv N_{6.60} \equiv N_{5.28}$, $N_{6.8} \equiv N_{6.21} \equiv N_{6.29} \equiv N_{6.61} \equiv N_{5.29}$, $N_{6.4} \equiv N_{6.22} \equiv N_{6.30} \equiv N_{6.62} \equiv N_{5.30}$ and $N_{6.5} \equiv N_{6.23} \equiv N_{6.31} \equiv N_{6.63} \equiv N_{5.31}$, creating self-loops. But like previous, these self-loops do not violate reversibility conditions. However, $N_{6.12} \equiv N_{3.0} $ and $N_{6.13} \equiv N_{3.1} $. Among the nodes of the subtrees of these two nodes, $N_{3.1}$, $N_{4.2}$, $N_{4.3}$, $N_{5.4}$, $N_{5.6}$ and $N_{6.9}$ do not satisfy reversibility conditions. So, we get irreversibility expressions $n=3j+6$ and $n=3j+9$ to be added to $ExpressionArray$.

At level $7$, only two unique nodes are added to $NodeList$. As $N_{4.4} \equiv N_{7.38}$, $N_{4.5} \equiv N_{7.39}$, $N_{4.6} \equiv N_{7.36}$, $N_{4.7} \equiv N_{7.37}$, $N_{4.8} \equiv N_{7.76}$, $N_{4.9} \equiv N_{7.77}$, $N_{4.12} \equiv N_{7.72}$, $N_{4.13} \equiv N_{7.73}$, $N_{4.14} \equiv N_{7.74}$ and $N_{4.15} \equiv N_{7.75} $, $10$ new loops are also added in the minimized tree. Among these duplicate nodes and subtree of the nodes, $N_{4.9}$, $N_{4.13}$, $N_{5.9}$, $N_{5.13}$, $N_{5.18}$ and $N_{5.19}$ violate irreversibility conditions, but no new expression of irreversibility is formed. 

As, on level $8$, no new unique node is added in the minimized tree, the algorithm concludes. The CA is declared as \textit{non-trivial semi-reversible} with final expression of irreversibility as $n =3j+6$, where $j \ge 0$. 
\end{example}

Therefore, by using the minimized reachability tree, we can deduce the expression of irreversibility for non-trivial semi-reversible CAs. The reversible, trivial semi-reversible and strictly irreversible CAs are also identified using Algorithm~\ref{Chap:semireversible:algo:rev_exp_algo}.

By experimenting on all possible rules for any particular $d $ and $m$, we have observed that, for ECAs ($d=2$, $m=3$), maximum height of the minimized reachability tree according to Algorithm~\ref{Chap:semireversible:algo:rev_exp_algo} is $5$, for $2$-state $4$-neighborhood CAs, the height is $9$ and for $3$-state $3$-neighborhood CAs the maximum height of the minimized tree for deciding non-trivial semi-reversibility is $19 $. Table~\ref{Chap:semireversibility:tab:results} shows some sample results of this experiment. In this table, the first column represents the number of states per cell ($d$), second column represents the number of neighbors ($m$) and the third column shows a CA rule for that $d$ and $m$. The number of unique nodes generated by  Algorithm~\ref{Chap:semireversible:algo:rev_exp_algo} for that CA is reported in the fourth column along with the height of the minimized reachability tree (Column $5$); whereas, the reversibility class of the CA with an expression of irreversibility (Column $7$) is depicted in the sixth column.


\begin{table}[!h]
\begin{center}
\caption{Classification of some sample rules using Algorithm~\ref{Chap:semireversible:algo:rev_exp_algo}}
\label{Chap:semireversibility:tab:results}
\resizebox{1.0\textwidth}{!}{
\begin{tabular}{|c|c|c|c|c|c|c|}
\hline 
$d$ & $m$ & Rule & $M$ & Height of tree & Reversibility Class? & Expression of irreversibility\\ 
\hline 
$2$ & $ 3 $ & $ 01010101 $ & $ 7 $ & $ 2 $ & Reversible & $\emptyset$\\ 
\hline 
$ 2 $ & $ 3 $ & $ 00011110 $ & NA & NA & Strictly irreversible & $\forall n \in \mathbb{N}$\\ 
\hline 
$ 2 $ & $ 3 $ & $ 00101101$ & $ 21 $ &  $ 5 $ & Non-trivial semi-reversible & $n=2j+2$, $j \ge 0$\\ 
\hline 
$ 2 $ & $ 3 $ & $ 10010110 $ & $ 12 $ &  $ 4 $ & Non-trivial semi-reversible & $n=3j+3$, $j \ge 0$ \\ 
\hline 
$ 2 $ & $ 3 $ & $ 00101011 $ & $ 4 $ & $ 2 $ & Trivial semi-reversible & $n\ge 3$\\
\hline
$3$ & $ 3 $ & $ 012012012012012210012102012 $ & $ 84 $ & $ 6 $ & Non-trivial semi-reversible & $n=4j+4$, $j \ge 0$\\ 
\hline 
$ 3 $ & $ 3 $ & $ 012210210102012102210210012 $ & $1371$ & $19$ & Non-trivial semi-reversible &  $n=2j+4$, $n=3j+3$, $j \ge 0$\\ 
\hline 
$ 3 $ & $ 3 $ & $ 012012012012012012012012012 $ & $ 13 $ &  $ 2 $ & Reversible & $\emptyset$ \\ 
\hline 
$ 3 $ & $ 3 $ & $ 210201102201120210201012102 $ & NA & NA & Strictly irreversible & $\forall n \in \mathbb{N}$\\
\hline
$ 2 $ & $ 3 $ & $ 021101110202222202110010021 $ & $ 1345 $ & $ 19 $ & Non-trivial semi-reversible & $n=2j+4$, $n=3j+3$, $j \ge 0$\\
\hline
$3$ & $ 3 $ & $ 111011011222220122000102200 $ & $ 252 $ & $ 9 $ & Non-trivial semi-reversible &  $n=2j+4$, $j \ge 0$\\
\hline 
$3$ & $ 3 $ & $ 102120120102120021120120120 $ & $ 104 $ & $7 $ & Non-trivial semi-reversible &  $n=2j+4$, $j \ge 0$\\
\hline 
$ 2 $ & $ 4 $ & $ 0001000011101111 $ & $65$ & $7$ & Non-trivial semi-reversible & $n=3j+3$, $j \ge 0$\\ 
\hline 
$ 2 $ & $ 4 $ & $ 0101101010100101$ & $ 56 $ &  $ 9 $ & Non-trivial semi-reversible & $n=7j+7$, $j\ge 0$\\ 
\hline 
$ 2 $ & $ 4 $ & $ 0000111101001011 $ & $ 32 $ &  $ 5 $ & Reversible & $\emptyset$ \\ 
\hline 
$ 2 $ & $ 4 $ & $ 0000111101001110 $ & NA & NA & Strictly irreversible & $\forall n \in \mathbb{N}$\\
\hline
$ 2 $ & $ 4 $ & $ 0000111101010101 $ & $ 4 $ & $ 2 $ & Trivial semi-reversible & $n\ge 4$\\
\hline 
$ 2 $ & $ 4 $ & $ 0000111101001011 $ & $ 32 $ &  $ 5 $ & Reversible & $\emptyset$ \\ 
\hline 
$ 2 $ & $ 4 $ & $ 1101110010001101 $ & NA & NA & Strictly irreversible & $\forall n \in \mathbb{N}$\\
\hline
$ 2 $ & $ 4 $ & $ 1101110000110010 $ & $ 2 $ & $ 1 $ & Trivial semi-reversible & $n\ge 4$\\
\hline
\end{tabular} }
\end{center}
\end{table}

Therefore, minimized reachability tree of size $5$, $9$ and $19$ are sufficient to decide reversibility class of ECAs, $4$-neighborhood $2$-state CAs and $3$-neighborhood $3$-state CAs respectively. That is, from a small set of sizes, reversibility class of the CAs can be identified.

\section{New Relation about Reversibility}\label{Chap:semireversible:sec:remark}
\noindent Theorem~\ref{Chap:semi-reversibility:Th:strictirreversibility} and minimized reachability tree guide us to decide the nature of reversibility of any $1$-D CA. By construction, height of the minimized tree for a CA is finite, say $n_0$. So, we can deduce about the reversibility class of the CA, if we construct 
the reachability tree for each $n$ up to length $n_0+m-1$. Otherwise, we can construct 
the minimized reachability tree for an $n$-cell CA, where $n \ge n_0$ and detect the reversibility class of the CA for any $n \in \mathbb{N}$. Therefore, the sufficient length to decide the reversibility class of CA is $n_0+m-1$ for which the minimized reachability tree has height $n_0$ and contains all possible unique nodes for the CA. Now we can relate the different cases of reversibility using the global transition functions.

If a finite CA with rule $R$ is found reversible by minimized reachability tree using Algorithm~\ref{Chap:semireversible:algo:rev_exp_algo} (resp. strictly irreversible by Theorem~\ref{Chap:semi-reversibility:Th:strictirreversibility}), then the CA is also reversible (resp. strictly irreversible) for configurations of length $n$, for all $n \in \mathbb{N}$, under periodic boundary condition; that is, for the set of all periodic configurations. That means, for a CA with local rule $R$, injectivity of $G_n$ (Case $4$) for all $n \in \mathbb{N}$, implies injectivity of $G_P$ (Case $2$), and hence, injectivity of $G$ (Case $1$) and $G_F$ (Case $3$). We can further note the following:
\begin{itemize}

\item For a finite CA with a set of periodic configurations of length $n$ for a fixed $n$, $G_n$ injective $\Rightarrow$ $G_n$ surjective.

\item If a CA is bijective for periodic configurations of length $n$, for all $n$, it is also bijective for periodic configurations of length $n$, for a fixed $n$, but the converse is not true. That is, 
$G_P$ injective \stackunder{$\rightarrow$}{$\nleftarrow$}
$G_n$ injective.

\item If a CA is not bijective for periodic configurations of length $n$, for a fixed $n$, then it is not bijective for periodic configurations of length $n$, for all $n$, but the converse is not true. That is, $G_n$ not bijective \stackunder{$\rightarrow$}{$\nleftarrow$} 
$G_P$ not bijective.

\item If $G$ is injective, then $G_n$ is also injective for any $n\in \mathbb{N}$, but the converse is not true. However, if $G_n$ is not injective, then $G$ is also not injective.

\item The algorithms of Amoroso and Patt \cite{Amoroso72} as well as of Sutner \cite{suttner91}, report the trivial semi-reversible CAs as neither surjective nor injective, but the non-trivial semi-reversible CAs as not injective but surjective for infinite configurations.
\end{itemize}

\noindent We are mainly interested in the \emph{non-trivially} semi-reversible CAs.
By Definition~\ref{Chap:semireversible:def:nontriv_semi}, the finite CAs which are reversible for length $n$, for all $n$, are a special case of non-trivial semi-reversible CAs. However, for a finite CA with a local rule $R$, we can get different $G_n$ by varying the lattice size $n$. To relate the injectivity of $G_P$ with $G_n$, let us define the following:
\begin{definition}
Let, ${G_R}^* = \{G_n ~|~ n \in \mathbb{N}\}$ be the set of global transition functions for rule $R$, when $n \in \mathbb{N}$. ${G_R}^*$ is called \textbf{reversible} if $G_n$ is reversible for each $n \in \mathbb{N}$, \textbf{strictly irreversible} if $G_n$ is irreversible for each $n \in \mathbb{N}$ and \textbf{semi-reversible} if $G_n$ is reversible for some $n \in \mathbb{N}$.
\end{definition}

Target of our work is to relate ${G_R}^*$ with $G$, $G_P$ and $G_F$. That is, deduce and classify reversibility of CAs from studying $G_n$. We can understand that, when ${G_R}^*$ is reversible, that means, the finite CA is reversible for every $n \in \mathbb{N}$, hence $G_P$ is injective. Similarly, if ${G_R}^*$ is strictly irreversible, $G_P$ is not injective.

\begin{figure*}[hbt]
\centering
\includegraphics[width= 4.9in, height = 1.9in]{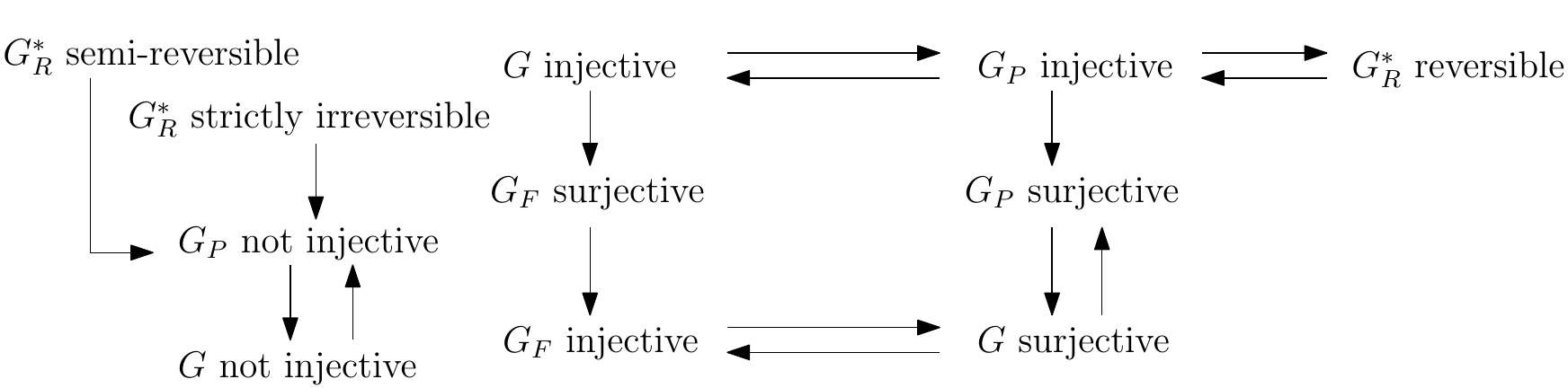}
\caption{Updated relationship among various kinds of reversibility of $1$-dimensional CAs}
\label{Chap:semireversible:fig:rev_rel_2}
\end{figure*}

 Let us define $S_R$ to be the set of CA rules which are reversible for some cell length $n\in \mathbb{N}$. So, all the semi-reversible and reversible CAs belong to this set $S_R$, that is $S_R = \{R ~|~ {G_R}^*$ is reversible $\lor ~{G_R}^*$ is semi-reversible\}. Let us further define $S_I$ to be the set of CA rules whose $G$ is bijective over the set of infinite configurations, $S_F$ to be the set of CA rules whose $G_F$ is bijective over the set of all finite configurations, and $S_P$ to be the set of CA rules whose $G_P$ is bijective over the set of periodic configurations. From the above discussions, it is evident that, $S_P$ is a subset of the set $S_R$.
 As, for one dimensional CAs, $S_F$, $S_I$ and $S_P$ are equivalent, so, $S_R$ is superset of all these sets. That is, $S_P\equiv S_I \equiv S_F \subset S_R$. 
Hence, Figure~\ref{Chap:semireversible:fig:rev_rel} is updated to depict the relation among the cases of reversibility in Figure~\ref{Chap:semireversible:fig:rev_rel_2}.

\noindent The above discourse points to the fact that, the cell length $n$ plays a salient role to determine the reversibility of the CA. Depending on the value of $n$, a CA can be reversible or irreversible. For each CA, we can get an $n_0$, height of the minimized tree, which is necessary to identify its reversibility class. However, is there any sufficient value of $n$, which is agreed by every CA, to predict the reversibility behavior of any CA? This question remains open for future research.\\

\noindent\textbf{Remark:} We have observed that, when defined over infinite lattice, the CAs of STRATEGY I, II and III (see Section~\ref{chap:reversibility:Sec:identify} of Chapter~\ref{Chap:reversibility}) are always surjective according to the global surjectivity testing procedure of Amoroso and Patt \cite{Amoroso72} as well as by the algorithm of Sutner \cite{suttner91}. However, many of them are not injective by these algorithms.
%

\section{Conclusion}
\label{Chap:semireversible:sec:con}
\noindent In this work, we have classified CAs into three sets -- reversible, semi-reversible and strictly irreversible CAs. Semi-reversible CAs are further classified into two sub-classes -- trivially semi-reversible and non-trivially semi-reversible. Reachability tree has been used to identify the appropriate class of a CA and find the expression of irreversibility for any non-trivially semi-reversible CA. It is observed that, from the reversibility for a some small set of sizes, we can confer about the reversibility of the CA defined over periodic configurations, and consequently for infinite CAs. Finally, we have related the four cases of reversibility: infinite configurations with $G$, periodic configurations with $G_P$, finite configurations for all $n \in \mathbb{N}$ with $G_F$ and configurations for fixed $n$ with $G_n$ under periodic boundary condition.

\chapter{Pseudo-random Number Generation and Cellular Automata}\label{Chap:randomness_survey}
\begin{center}
\begin{quote}
\emph{Any one who considers arithmetical methods of producing random digits is, of course, in a state of sin.}
\end{quote}
\hspace*{2.65in}{\em -- John von Neumann, 1951}
\end{center}

\noindent{\small In today's world, several applications demand numbers which appear random but are generated by a background algorithm; that is, the numbers are \emph{pseudo-random}. Since late $19^{th}$ century, researchers have been working on pseudo-random number generators (PRNGs). As a result, a number of PRNGs have been developed using different techniques. Cellular automata (CAs) have also been explored to design PRNGs. In this chapter, we identify the essential properties of a cellular automaton (CA) to be a potential candidate as PRNG, and then propose an example PRNG using a tri-state CA. The randomness quality of a PRNG is tested using some standard empirical tests, which are also used here to verify the quality of the example PRNG. To understand the position of the newly developed PRNG, we rank the existing PRNGs with respect to their randomness quality, and then compare our PRNG with the extant ones. For the comparison purpose, we have chosen $28$ PRNGs, which are widely used and respected as good PRNGs.

\section{Introduction}\label{chap:randomness_survey:sec:Introduction}
\large\textbf{S}}ociety, arts, culture, science and even daily life is engulfed by the concept of randomness. One of the major usage of randomness is in generating numbers for diverse fields of practices. History of human race gives evidence that, since the ancient times, people has generated random numbers for various purposes. As an example, for them, the output of rolling a dice was a sermon of God! However, in the modern times, researchers and scientists have discovered diverse applications and fields, like probability theory, game theory, information theory, statistics, gambling, computer simulation, cryptography, pattern recognition, VLSI testing etc., which require random numbers. Most of these applications entail numbers, which appear to be random, but which can be reproduced on demand. Such numbers, which are generated by a background algorithm, are called pseudo-random numbers and the implementation of the algorithms as pseudo-random number generators (PRNGs). In this dissertation, however, by random number, we will mean pseudo-random numbers only.

Although, PRNGs have a long history of development -- its modern journey starting in late $19^{th}$ century \cite{galton1890dice} to early $20^{th}$ century \cite{tippett1927random,2980655,2983623,von195113} and evolving and getting more powerful ever since \cite{l1996maximally,Lewis:1973:GFS:321765.321777,Matsumoto:1998:MTE:272991.272995,Panneton:2005:XRN:1113316.1113319,Panneton:2006:ILG:1132973.1132974,Saito2008,Saito2009,Tausworthe,Tezuka:1987:DGP:31846.31848,wolfram86c,Marco99,Vigna:2016:EEM:2956571.2845077}.
Initially, two dominating categories of PRNGs existed -- Linear Congruential Generator (LCG) based and Linear Feedback Shift Register (LFSR) based. In $1985$, due to inherent randomness quality of some CAs, cellular automata have also been introduced as a source of randomness \cite{Wolfram85c}. Since then, it has bewitched many researchers to use it as PRNG, see as example \cite{Horte89a,Horte89c,114093,COMPAGNER1987391,Tomassini96, Marco99,122655,alonso2009elementary,tcad/DasS10,SukantaTH, Guan03, Guan04, Marco00,Guan04a,HOSSEINI2014149}.
 
In this chapter, however, we identify the necessary properties of a CA to be a good source of randomness (Section~\ref{chap:randomness_survey:sec:CA_source_of_randomness}). It is shown that, having these attributes essentially mean effectuating the desirable properties of a PRNG. As an example, we select one $3$-neighborhood $3$-state CA distinguished by the string $120021120021021120021021210$ which accomplishes these criteria and propose it as a window-based PRNG (Section~\ref{Chap:randomness_survey:sec:prng_R}). To understand the potentiality of the proposed CA as a random number generator, we test the randomness quality of the CA. For this, we use some empirical testbeds (Section~\ref{chap:randomness_survey:sec:empirical}) to measure the caliber of randomness for the CA (Section~\ref{Chap:randomness_survey:sec:empirical_result_R}). 

Now, we find its rank as PRNG in a set of \emph{good} PRNGs. 
To get an idea about the existing PRNGs in literature, Section~\ref{chap:randomness_survey:sec:class} surveys the journey of the PRNGs through three technologies -- Linear Congruential Generators (LCGs), Linear Feedback Shift Registers (LFSRs) based and CAs based. Total $28$ currently used well-known PRNGs are selected for comparison with the proposed PRNG. Each of these PRNGs claims itself as good. So, to verify their assertions, we test these PRNGs on a common platform.
Section~\ref{chap:randomness_survey:sec:facts} reports the test results of the selected PRNGs. We observe that, for many PRNGs, the claim and actual independent result do not tally. Section~\ref{chap:randomness_survey:sec:final_rank} depicts a relative ranking for these $28$ PRNGs with respect to the overall performance in the empirical testbeds. Finally, in Section~\ref{Chap:randomness_survey:sec:comparison}, we compare our proposed PRNG with these $28$ well-known PRNGs and find its rank with respect to randomness quality.

\section{PRNGs and their Properties} \label{chap:randomness_survey:sec:property}
\noindent Pseudo-random number generators are simple deterministic algorithms which produce deterministic sequence of numbers that appear random. 
In general, a PRNG produces uniformly distributed, independent and uncorrelated real numbers in the interval $[0,1)$. However, generation of numbers in other probability distribution is also possible. Mathematically, a PRNG is defined as the following \cite{LEcuyer1990}:

\begin{definition}\label{Chap:randomness_survey:def:prng}
	A pseudo-random number generator $G$ is a structure $(\mathscr{S},\mu, f,\\ \mathscr{U}, g)$, where $\mathscr{S}$ is a finite set of states, $\mu$ is the probability distribution on $\mathscr{S}$ for the initial state called seed, $f: \mathscr{S}\rightarrow \mathscr{S}$ is the transition function, $\mathscr{U}$ is the output space and $g: \mathscr{S} \rightarrow \mathscr{U}$ is the output function. The generator $G$ generates the numbers in the following way.
	\begin{enumerate}
		\item Select the seed $s_0 \in \mathscr{S}$ based on $\mu$. The first number is $u_0 = g(s_0)$.
		\item At each step $i\geq 1$, the state of the PRNG is $s_i = f(s_{i-1})$ and output is $u_i = g(s_i)$. These outputs of the PRNG are the pseudo-random numbers, and the sequence ${(u_i)}_{i\ge 0}$ is the pseudo-random sequence.
	\end{enumerate}
\end{definition}

Since a PRNG is a finite state machine with a finite number of states, after a finite number of steps, eventually it will come back to an old state and the sequence will be repeated. This property is common to all sequences where the function $f$ transforms a finite set into itself.
This repeating cycle is known as \emph{period}. The length of a period is the smallest positive integer $\rho$, such that, $\forall n\geq k, s_{\rho+n}=s_n$, here $k\geq 0$ is an integer.
 If $k=0$, the sequence is purely periodic. Preferably, $\rho\approx|\mathscr{S}|$, or, $\rho \approx 2^{b}$, if $b$ bits represent each state.

\begin{figure}[hbtp]
\centering
	\vspace{-1.0em}
	\subfloat[Maximum-period PRNG\label{cycle1}]{%
		\resizebox{0.40\textwidth}{!}{
			\includegraphics[width=3.5in, height = 1.3in]{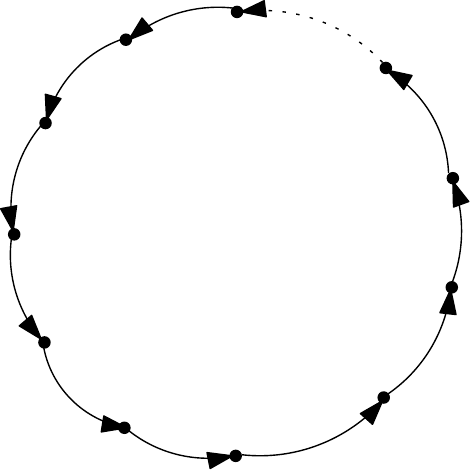}}
	}
	\hfill
	\subfloat[PRNG with non-maximum periods \label{carry}]{%
		\resizebox{0.42\textwidth}{!}{
			\includegraphics[width=3.5in, height = 1.3in]{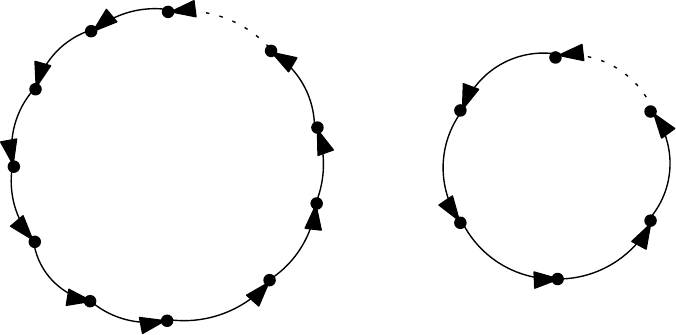}
		}
	}
	\caption{Cycle structure of PRNGs}
	\label{Chap:randomness_survey:fig:cycle} %
\end{figure}

Ideally, a PRNG has only one period, that is, all unique numbers of the output space are part of the same cycle. In that case, the PRNG is \emph{maximum-period generator}. However, many PRNGs exist, which have more than one cycle. So, depending on the seeds, completely different sequence of numbers from distinct cycles may be generated. This situation is shown in Figure~\ref{Chap:randomness_survey:fig:cycle}.

Every PRNG is classified by the functions $ f$ and $g$, and the seed $s_0$.
 Therefore, when a PRNG is observed for its randomness quality, it is considered that the algorithm is not known to the adversary. Following are the desirable properties, which are to be observed in a good PRNG.

\begin{description}[leftmargin=1pt]
	\item[1. Uniformity:] This property implies that, if we divide the set of possible numbers generated by the PRNG (that is, the range of the PRNG) into $K$ equal subintervals, then expected number of samples $(e_i)$ in each subinterval $i, 1\leq i \leq K$, is equal; that is, $e_i = \frac{N}{K}$, where $N$ is the range of the numbers. This ensures that, the generated numbers are equally probable in every part of the number space. 
	
	
	\item[2. Independence:] The generated numbers are to be independent of each other; that is, there should not be any serial correlation between numbers generated in succession. So, any subsequence of numbers have no correlation with any other subsequences. This means, given any length of previous numbers, one can not predict the next number in the sequence by observing the given numbers.
	
	\item[3. Large Period:] Every PRNG has a period after which the sequence is repeated. A PRNG is considered good if it has a very large period. Otherwise, if one can exhaust the period of a PRNG, the sequence of numbers become completely predictable.
	
	\item[4. Reproducibility:] One of the prominent reason of developing a PRNG is its property of reproducibility. This ensures that given the same seed $s_0$, the same sequence of numbers is to be generated. This is very useful in simulation, debugging and testing purposes.
	
	\item[5. Consistency:] The above properties of the PRNGs are to be independent of the seed. That is, all these properties are to be maintained for every seed value.
	
	\item[6. Disjoint subsequences:] There is to be little or no correlation between subsequences generated by different seeds. However, this criterion is difficult to achieve in an algorithmic PRNG. 
	
	\item[7. Permutations:] Every permutation of numbers generated by a PRNG is expected to be equally likely. Otherwise, the numbers can be biased and may help to predict successive numbers. 
	
	\item[8. Portability:] A PRNG is to be portable; that is, the same algorithm can work on every system. Given the same seed, different machines with varied configuration are to give the same output sequence.
	
	\item[9. Efficiency:] The PRNG is to be very fast; which means, generation of a random number takes insignificant time. Moreover, a PRNG should not use much storage or computational overhead. This is to make certain that, the use of PRNGs in an application is not a hindrance to its efficiency.

	
	\item[10. Coverage:] This implies whether the PRNG covers the output space for any seed. Many PRNGs have less coverage. In case the PRNG has more than one cycle, then it may happen that, although it covers the whole output space, but only a part of it is covered by a particular seed. 
	
	\item[11. Spectral Characteristics:] A good PRNG does not generate numbers of one frequency higher than any other. If we plot the consecutive numbers, there is not to be any pattern visible for any length of the sequence.
	
	\item[12. Cryptographically Secure:] To be used in cryptographic applications, the generated numbers have to be cryptographically secure. This is a desirable property often missing in most of the algorithmic PRNGs. 
\end{description}

\noindent Many of these properties are inter-related. For example, if the numbers are not uniform, they are correlated and have identifiable patterns. Ideally, the numbers of a good PRNG are to satisfy all these properties. However, practically, most of the PRNGs do not possess all these properties; for example, the properties $6,7$ and $12$ are often missing in the existing PRNGs. Still, in terms of usage in the applications for which they are intended, many PRNGs are considered good in today's standard. 

In the next section, we explore the desirable properties to be exhibited by a CA to act as a PRNG.

\section{Cellular Automata as Source of Randomness}\label{chap:randomness_survey:sec:CA_source_of_randomness}
One of the most important characteristics of cellular automata is their complex global behavior arising from a simple rule with local interaction and computation and massive parallelism. Behavior of any CA is always consistent, same initial configuration results in the same sequence of next configurations. Further, the intricate chaotic behavior originated by simple functions with local interaction of a CA can be exploited to develop random numbers. Hence, CA was introduced as a source of randomness \cite{Wolfram85c}. However, not all CAs can be used as a source of randomness; for this the CA needs to be \emph{unpredictable}.

\subsection{Unpredictability in CAs}\label{Chap:randomness_survey:sec:unpredictability}
Like every PRNG, CAs are also completely deterministic. By applying the rule on the initial configuration, next configurations can be deduced. However, in a PRNG based on CA, the random numbers are generated out of the configurations of the CA. So, to be used as a PRNG, the configurations of the CA has to satisfy the desirable properties of a good PRNG. For example, the numbers have to be independent of each other, have uniform distribution, the average cycle length of the CA has to be large to confer large period, etc.

However, if one observes the dynamics of $1$-dimensional CAs, it is seen that, a large number of the CAs are \emph{convergent}. That is, starting from any arbitrary configuration, the CAs have tendency to converge to a particular configuration (fixed-point attractor), or to a small cycle of configurations. According to Wolfram's classification \cite{wolfram84b}, these CAs form Class $1$ and Class $2$ respectively. Such CAs are highly predictable; by observing the CA for a small number of time steps, its next configurations can be easily predicted. Hence, they can not be a source of randomness. 

Similarly, there are some CAs, for which the behavior of the CA can change from regular to irregular in an indefinite manner. Sometimes the dynamics of the CAs shows stable or periodic structures, whereas, sometimes the dynamics is complex and chaotic. Wolfram's Class $4$ CA is an example of such CAs. These CAs are not used as candidates to be PRNGs.

Nevertheless, there exist some CAs which are \emph{divergent} in nature. These CAs, defined over infinite lattice, show chaotic behavior, which implies that, knowing the present configuration of such a CA, it is not possible to predict the future configurations of it. This unpredictability has a correspondence to the independence property.
These CAs form the Class $3$ in Wolfram's classification. A tiny perturbation in initial configuration of these CAs greatly affect the future configurations of them. Therefore, slightly different initial configurations of the CA produce different sequences of configurations which are having little or no correlation. These CAs (defined over infinite lattice) are surjective (see Ref. \cite{Kurka97}) which implies that all the configurations are reachable from some initial configurations.

Although the above properties are very appropriate for a PRNG, however, we need finite CAs to design a PRNG. Many properties of a chaotic system disappear when the lattice is restricted to a finite size. Nonetheless, as a borderline, chaotic CAs rules can only be chosen with an expectation that the above properties will be reflected, to some extent, to the finite CAs.

\begin{figure}[hbtp]
\centering
	\subfloat[ECA $30$ \label{statespace_30}]{%
		\includegraphics[width=0.3\textwidth, height = 5.7cm]{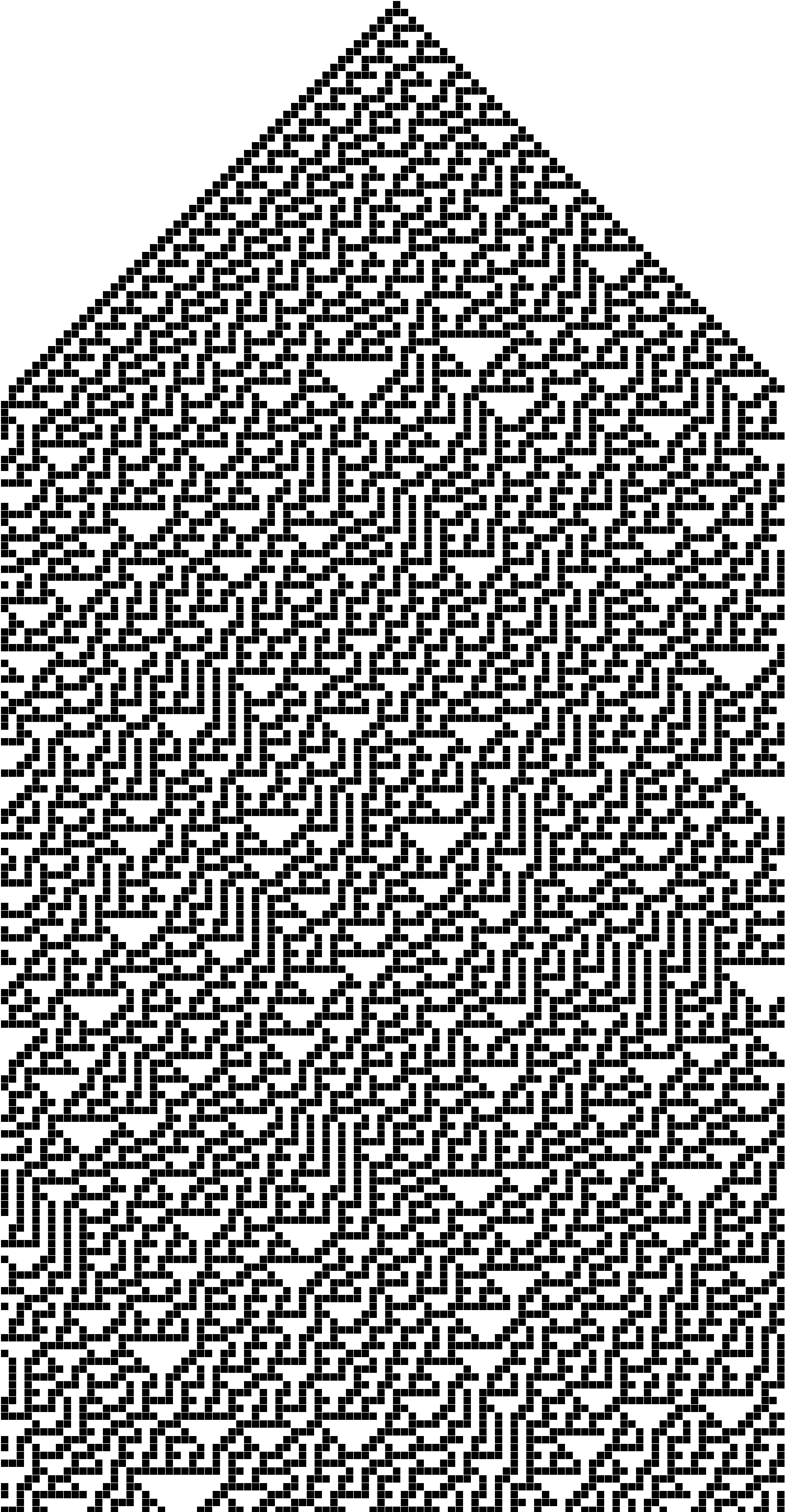}
	}
	\hfill
	\subfloat[ECA $45$ \label{statespace_45}]{%
		\includegraphics[width=0.3\textwidth, height = 5.7cm]{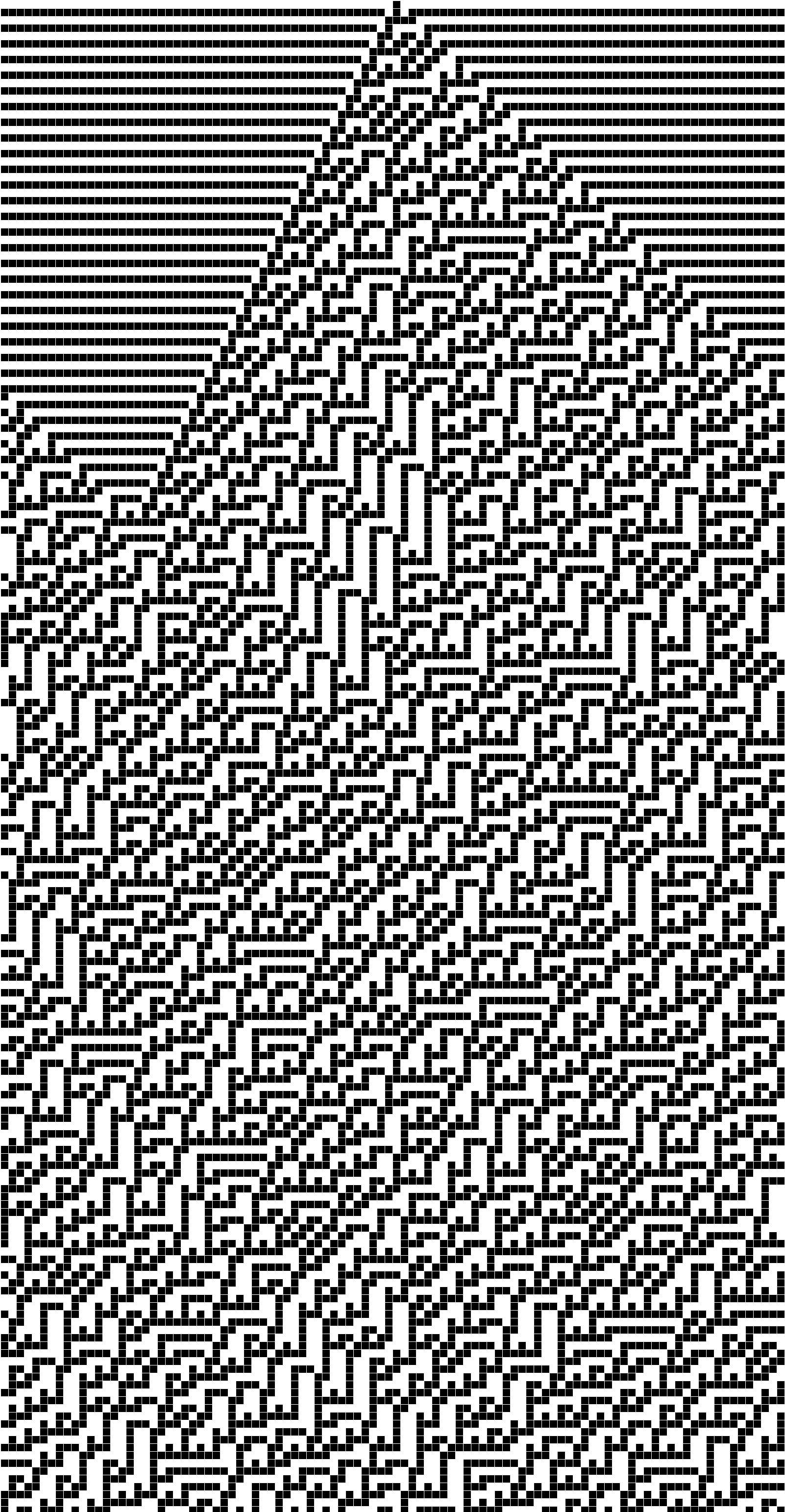}
	}
	\hfill
	\subfloat[ECA $90$ \label{statespace_90}]{%
		\includegraphics[width=0.3\textwidth, height = 5.7cm]{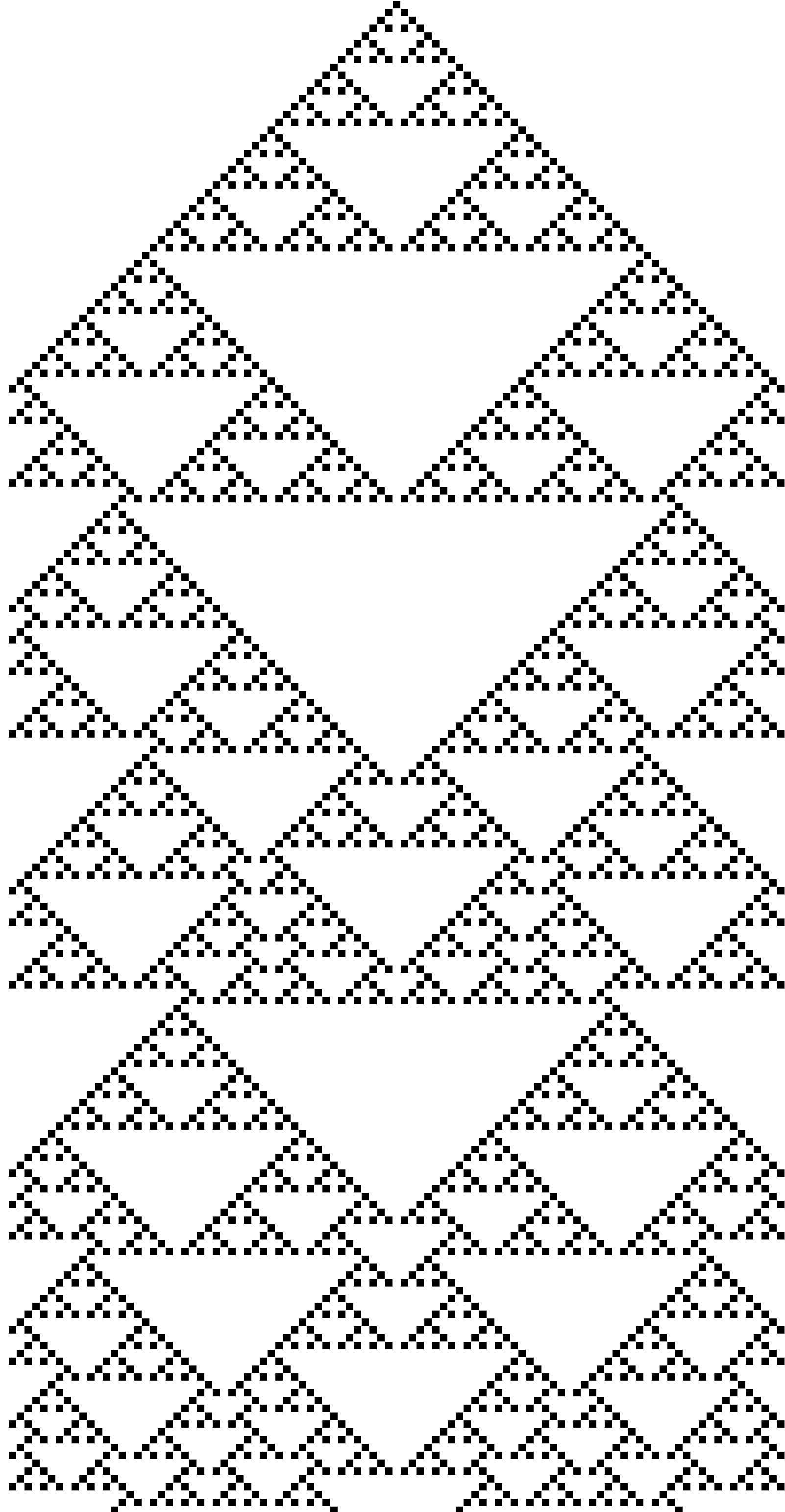}
	}
	\caption{Space-time diagram of three Class $3$ elementary CAs $30~(00011110)$, $45~(00101101)$ and $90~(01011010)$ for $n=100$ with seed $0^{49}10^{50}$}
	\label{Chap:randomness_survey:fig:space-time diagramECA} %
\end{figure}

The chaotic CAs, however, have a tendency to generate \emph{self-similar} patterns. See, for example, Figure~\ref{Chap:randomness_survey:fig:space-time diagramECA} for three Class $3$ ECAs considering lattice sizes as finite. Here, ECA $90$ has regular repeating pattern. If the numbers are generated out of these patterns, they can not show randomness property.
 So, not all chaotic CAs qualify as source of randomness.
In a recent work \cite{Supreeti_2018_chaos}, it is argued that, to be unpredictable, the CA has to \emph{destroy} the information of initial configuration, step by step, during its evolution. If this property is observed in finite case, there is a good chance that the generated configurations show independence property. Further, the CA which wants to act as a PRNG, should have no affinity towards any particular state and its configuration space is to be devoid of any \emph{self-similar} pattern.

\subsection{Necessary Properties of Cellular Automata}\label{Chap:randomness_survey:sec:propertiesOfCA} 
Let us now write down the necessary attributes of a finite CA which can be explored as a good source of randomness. These properties help to make the CA chaotic (and unpredictable) under infinite lattice size. In infinite case, however, a CA may show chaotic behavior without some of the following criteria. But for finite lattice size, these are the essential qualities of a CA which wants to behave as a PRNG. Observe that, a finite CA has finite number of configurations, so, may be seen as a finite state machine (as is seen in many works, e.g. \cite{ppc1}). Recall that, a PRNG is also a finite state machine by Definition~\ref{Chap:randomness_survey:def:prng}.

\subsubsection{Balancedness} The first property, a candidate CA needs to fulfill is, its local rule is to be balanced (see Definition~\ref{Def:balancedrule} of Page~\pageref{Def:balancedrule}). This is because, if a CA rule is unbalanced, then at least one of its state $s\in \mathcal{S}$ has more presence in the rule than the other state(s). So, for an arbitrary configuration, the next configuration of it has a preference to generate a particular state (having more presence in the unbalanced rule) more than the others. This is unacceptable for a random system.


%
%
%
\subsubsection{Flow of Information}\label{Chap:randomness_survey:sec:info_flow}
In a good PRNG, the generated numbers should appear as independent. In case of CAs, only balancedness may not ensure this quality. As the numbers are generated from the configurations, to have good randomness, the configurations of a CA should be independent of each other (as far as possible). That is, it should be practically impossible to infer the future configurations of the CA by using only the current configuration without any knowledge of the rule. Therefore, it is desirable that, even the smallest addition of information to the initial configuration, should propagate in the CA and largely affect the future sequence of configurations.

Moreover, in a CA, the next states of the cells are changed following a rule which takes into account the states of the neighbors. Therefore, the rule of the CA decides whether a cell is \emph{dependent} or \emph{independent} of (some of) its neighbors. If all cells of a CA are independent of their neighbors, any change in the states of the neighbors of a cell does not affect the cell. This CA is \emph{stable} and \emph{predictable} -- by observing the current configuration of the CA, the future configurations can be predicted. Moreover, if only a small number of cells are dependent on their neighbors, then also, for a change in seed, there can be less change in the next configurations.
For all these CAs, a small change in the initial configuration can not largely impact the future configurations. Hence, these CAs are undesirable for random number generation.

However, if the cells are dependent on their neighbors, then change in the neighbors' states induces update of the cells' states. 
That means, even a tiny change in the initial configuration, eventually propagates throughout the cells in the course of evolution of the CA. Therefore, the information of a localized change does not get absorbed in the cells, rather it flows (and spreads) through the configurations. This situation makes the CAs unpredictable and desirable candidates as PRNGs. Hence, for a CA to be good source of randomness, it is expected that most of the cells are dependent on their neighbors, so, more is the information transmission and more independent the next configurations are.

For example, let us consider $3$-neighborhood dependency, where, left and right cells of each cell are neighbors of that cell. So, change in state of the left neighbor (respectively right neighbor) can affect the next state of the cell. Therefore, if we contemplate any change in state of a cell as an information, then this information can flow in two directions -- left and right.

\begin{itemize}[leftmargin=0pt]
\item \textbf{Information flow on right side:}
A $3$-neighborhood (left, self and right) CA can have information flow on right side, if any change in its left neighbor affects the cell itself. This can be measured by observing the change of states in the equivalent RMT set (see Definition~\ref{Def:equivalent} of Page~\pageref{Def:equivalent}). Recall that, each equivalent RMT set is represented by the set $\lbrace 0yz,1yz, \cdots, (d-1)yz \rbrace$, $y,z \in \mathcal{S}$ (see Table~\ref{Chap:reversibility:tab:rln} of Page~\pageref{Chap:reversibility:tab:rln}). Hence, if $R[x'yz] \ne R[x''yz]$, where $x',x'' \in \mathcal{S}$, then, it means that, change in the left neighbor affects the cell. However, if the RMT corresponding to a cell is self-replicating (see Definition~\ref{Def:selfreplicating} of Page~\pageref{Def:selfreplicating}), then practically, it is independent of its neighbors. So, these RMTs are not contributories. Therefore, the rate of information flow on right side is the cumulative sum of the change of states in each equivalent RMT set (except the self-replicating RMTs), divided by the total number of RMTs. Table~\ref{Chap:randomness_survey:tab:equi_R} illustrates this for a $3$-state CA $120021120021021120021021210$ (see Table~\ref{Chap:randomness:survey:tab:rule_R}). 

\renewcommand{\arraystretch}{1.2}
\begin{table}[hbtp]
	\begin{center}
		\caption{RMTs of $Equi_{i} $, $0 \leq i \leq 8$ for CA $\mathbf{\mathscr{R}}=120021120021021120021021210$}
		{
			\resizebox{0.9\textwidth}{2.5cm}{
				\begin{tabular}{c|ccc|ccc|ccc|c}
					\toprule	
					&	\multicolumn{9}{c|}{\thead{Equivalent RMTs}} & \\
					\thead{\#Set~~} & \thead{P.S.} & \thead{RMT $i$} & \thead{$\mathbf{\mathscr{R}}[i]$~~} & \thead{P.S.} & \thead{RMT $i$} & \thead{$\mathbf{\mathscr{R}}[i]$~~}  & \thead{P.S.} & \thead{RMT $i$} & \thead{$\mathbf{\mathscr{R}}[i]$~~} & \thead{Change of States} \\
					\midrule
					$Equi_0$ & $000$ & $(0)$ & $0$ & $100$ & $(9)$ & $0$ & $200$ & $(18) $ & $ 0$ & $0$ \\ 
					$Equi_1$ & $001$ & $(1)$ & $ 1 $ & $101$ & $(10) $ & $ 2$ & $201$ & $(19) $ & $ 2$ & $2$ \\ 
					$Equi_2$ & $002$ & $ (2) $ & $ 2$ & $102$ & $(11) $ & $ 1$ & $202$ & $(20)$ & $1$ & $2$ \\ 
					$Equi_3$ & $010$ & $(3)$ & $1$ & $110$ & $(12)$ & $1$ & $210$ & $(21)$ & $1$ & $0$ \\ 
					$Equi_4$ & $011$ & $(4)$ & $2$ & $111$ & $(13)$ & $2$ & $211$ & $(22)$ & $2$ & $1$ \\ 
					$Equi_5$ & $012$ & $(5)$ & $0$ & $112$ & $(14)$ & $0$ & $212$ & $(23)$ & $0$ & $1$ \\ 
					$Equi_6$ & $020$ & $(6)$ & $1$ & $120$ & $(15)$ & $1$ & $220$ & $(24)$ & $0$ & $2$ \\ 
					$Equi_7$ & $021$ & $(7)$ & $2$ & $121$ & $(16)$ & $2$ & $221$ & $(25) $ & $2$ & $0$ \\ 
					$Equi_8$ & $022$ & $(8)$ & $0$ & $122$ & $(17)$ & $0$ & $222$ & $(26)$ & $1$ & $2$ \\ 
					\midrule
					\multicolumn{10}{c}{\thead{\emph{Total change of $\mathbf{\mathscr{R}}[i]$} depending on $Equi_i$ = }} & \thead{$10$}\\
					\bottomrule
				\end{tabular}
			}\label{Chap:randomness_survey:tab:equi_R}}
	\end{center}
	\vspace{-1.5em}	
\end{table} 

In this table, the RMTs of equivalent RMT set $Equi_0$, that is, RMTs $0$, $9$ and $18$, are all self-replicating, so there is no change of state for this set of RMTs. Similarly, in the $4^{th}$ row of Table~\ref{Chap:randomness_survey:tab:equi_R}, RMT $10$ and RMT $19$ have same next state value $2$ and RMT $1$ have different next state value $1$, so, number of next state changes for this equivalent RMT set $Equi_1$ is $2$. In this way, total number of effective state change of RMTs is $10$ and the rate of information transmission on right side for this CA is $\frac{10}{27} = 37.037 \%$.

\item \textbf{Information flow on left side:}
Similarly, a CA can have information flow on left direction, if the information of right neighbor of a cell passes to the cell itself. As sibling RMT sets (see Definition~\ref{Def:sibling} of Page~\pageref{Def:sibling}) signify the possible combinations of right neighbors, hence, difference in states of sibling RMTs implies the affect of the right neighbor. Recall that, each sibling RMT set is represented by the set $\lbrace xy0,xy1,\cdots,xy(d-1) \rbrace$, $x,y \in \mathcal{S}$ (see Table~\ref{Chap:reversibility:tab:rln}). Therefore, if $R[x,y,z'] \ne R[x,y,z'']$, where $z',z'' \in \mathcal{S}$, then, it means that, change in the right neighbor affects the cell's state. Here also, the self-replicating RMTs do not contribute to the information transmission. Hence, the rate of information flow on left side is the cumulative sum of the change of states in each sibling RMT set (excluding the self-replicating RMTs), divided by the total number of RMTs. Table~\ref{Chap:randomness_survey:tab:sibl_R} shows the information flow on left side for the CA $120021120021021120021021210$ (Table~\ref{Chap:randomness:survey:tab:rule_R}).
\renewcommand{\arraystretch}{1.2}
\begin{table}[hbtp]
	\begin{center}
		\caption{RMTs of $Sibl_{i} $, $0 \leq i \leq 8$ for CA $\mathbf{\mathscr{R}}=120021120021021120021021210$}
		{
			\resizebox{0.9\textwidth}{2.5cm}{
				\begin{tabular}{c|ccc|ccc|ccc|c}
					\toprule	
					&	\multicolumn{9}{c|}{\thead{Sibling RMTs}} & \\
					\thead{\#Set~~} & \thead{P.S.} & \thead{RMT $i$} & \thead{$\mathbf{\mathscr{R}}[i]$~~} & \thead{P.S.} & \thead{RMT $i$} & \thead{$\mathbf{\mathscr{R}}[i]$~~}  & \thead{P.S.} & \thead{RMT $i$} & \thead{$\mathbf{\mathscr{R}}[i]$~~} & \thead{Change of States} \\
					\midrule
					$Sibl_0$ & $000$ & ($0$) & $0$ & $001$ & ($1$) & $1$ & $002$ & ($2$) & $ 2 $ & $2$ \\ 
					$Sibl_1$ &  $010$ & ($3$) & $1$ & $011$ & ($4$) & $2$ & $012$ & ($5$) & $0$ & $2$\\ 
					$Sibl_2$ & $020$ & ($6$) & $1$ & $021$ & ($7$) & $2$ & $022$ & ($8$) & $0$ & $2$ \\ 
					$Sibl_3$ & $100$ & ($9$) & $0$ & $101$ & ($10$) & $2$ & $102$ & ($11$) & $1$ & $2$ \\ 
					$Sibl_4$ & $110$ & ($12$) & $1$ & $111$ & ($13$) & $2$ & $112$ & ($14$) & $0$ & $2$ \\ 
					$Sibl_5$ & $120$ & ($15$) & $1$ & $121$ & ($16$) & $2$ & $122$ & ($17$) & $0$ & $2$ \\ 
					$Sibl_6$ & $200$ & ($18$) & $0$ & $201$ & ($19$) & $2$ & $202$ & ($20$) & $1$ & $2$ \\ 
					$Sibl_7$ & $210$ & ($21$) & $1$ & $211$ & ($22$) & $2$ & $212$ & ($23$) & $0$ & $2$ \\ 
					$Sibl_8$ & $220$ & ($24$) & $0$ & $221$ & ($25$) & $2$ & $222$ & ($26$) & $1$ & $2$ \\ 
					\midrule
					\multicolumn{10}{c}{\thead{\emph{Total change of $\mathbf{\mathscr{R}}[i]$} depending on $Sibl_i$ = }} & \thead{$18$}\\
					\bottomrule
				\end{tabular}
			}\label{Chap:randomness_survey:tab:sibl_R}}
	\end{center}
	\vspace{-1.5em}	
\end{table} 

In this CA, for each $i$, RMTs of $Sibl_i$ have distinct next state values, whereas, only one RMT of each $Sibl_i$ is self-replicating. Hence, $18$ out of $27$ RMTs are affected by the information of their right neighbors. So, the CA has $\frac{18}{27} = 66.667 \%$, that is, the maximum possible information flow on left direction.
\end{itemize}

\subsubsection{Large Cycle Length}\label{Chap:randomness_survey:sec:cycleLength}
In classical PRNGs, the numbers are generated in a cycle.
In a good PRNG, the \emph{period} of the cycle is very large. Because, larger period length implies lesser probability of repeating the same number for a given length of sequence. However, for a CA under periodic boundary condition, the configuration space of the CA is partitioned into a number of subspaces. If the number of subspaces is less, then it is expected that the lengths of cycles of the CA are high, where a cycle may be reached via many chains of configurations. In other words, if number of cycles is large, then expected length of each cycle is less.

Therefore, for a CA to be a good candidate for PRNG, it is expected that, its configuration space is not partitioned into many sub-spaces resulting existence of large cycles. As, for the CAs based PRNGs, the numbers are obtained from individual configurations, average cycle length of the CA should be very large, and it should grow exponentially with CA size $n$. 
\subsubsection{Non-linearity}
To be a good PRNG which could be used in several applications including cryptography, it has to be a non-linear system \cite{Meier1990}. Recall that, a CA is non-linear if and only if its rule  is non-linear (see Definition~\ref{Def:linearrule} of Page~\pageref{Def:linearrule}). For $m$-neighborhood $d$-state CAs, there are $d^m-1$ linear rules, which are balanced. So, these rules can be avoided as PRNGs.
		
\section{Design of an Example PRNG}\label{Chap:randomness_survey:sec:PRNG_design}
In this section, we select a CA -- a $3$-neighborhood $3$-state CA of Table~\ref{Chap:randomness:survey:tab:rule_R}, which satisfies the above properties, and propose a design of PRNG using it. Hereafter, this CA is represented as CA $\mathbf{\mathscr{R}}$.	
\begin{table}[t]
	\setlength{\tabcolsep}{1.3pt}
\centering
		\caption{The cellular automaton $\mathbf{\mathscr{R}}$ 
		}\label{Chap:randomness:survey:tab:rule_R}
		{
			\resizebox{1.00\textwidth}{!}{
				\begin{tabular}{cccccccccccccccccccccccccccc}
					\toprule
					\thead{P.S.} & \thead{222} & \thead{221} & \thead{220} & \thead{212} & \thead{211} & \thead{210} & \thead{202} & \thead{201} & \thead{200} & \thead{122} & \thead{121} & \thead{120} & \thead{112} & \thead{111} & \thead{110} & \thead{102} & \thead{101} & \thead{100} & \thead{022} & \thead{021} & \thead{020} & \thead{012} & \thead{011} & \thead{010} & \thead{002} & \thead{001} & \thead{000}\\ 
					
					\thead{RMT} & \thead{(26)} & \thead{(25)} & \thead{(24)} & \thead{(23)} & \thead{(22)} & \thead{(21)} & \thead{(20)} & \thead{(19)} & \thead{(18)} & \thead{(17)} & \thead{(16)} & \thead{(15)} & \thead{(14)} & \thead{(13)} & \thead{(12)} & \thead{(11)} & \thead{(10)} & \thead{(9)} & \thead{(8)} & \thead{(7)} & \thead{(6)} & \thead{(5)} & \thead{(4)} & \thead{(3)} & \thead{(2)} & \thead{(1)} & \thead{(0)}\\ 
					\midrule
					\thead{N.S.} & 1&2&0&0&2&1&1&2&0&0&2&1&0&2&1&1&2&0&0&2&1&0&2&1&2&1&0\\
					\bottomrule
				\end{tabular}
			} }
\end{table}

\subsection{Properties of the Cellular Automaton $\mathbf{\mathscr{R}}$}\label{Chap:randomness_survey:sec:prop_R}
This CA is balanced.
Further, the rule $\mathbf{\mathscr{R}}$ is non-linear which can easily be verified from Table~\ref{Chap:randomness:survey:tab:rule_R}. For example, for RMTs $4$ and $5$, $\mathbf{\mathscr{R}}[4+5]\ne \mathbf{\mathscr{R}}[4]+\mathbf{\mathscr{R}}[5]$ under modulo $3$ addition. This CA also has following properties --

\subsubsection{Flow of Information}\label{Chap:randomness_survey:sec:info_flow_R}
The CA $\mathbf{\mathscr{R}}$ has information flow on both sides as shown in \ref{Chap:randomness_survey:tab:equi_R} and Table~\ref{Chap:randomness_survey:tab:sibl_R}. It has maximum possible information flow ($66.67 \%$) on left direction and $ 37.037 \%$ information flow on right direction.

\subsubsection{Semi-Reversibility}
The CA $\mathbf{\mathscr{R}}$ is a non-trivial semi-reversible CA. That is, it is not reversible for all $n \in \mathbb{N}$, but reversible for an infinite set of sizes. We find that, it is reversible when $n$ is odd. So, if an odd $n$ is chosen, every configuration of the CA falls into some cycle.

\subsubsection{Space-time Diagram}\label{Chap:randomness_survey:sec:space-time_R}
Space-time diagram (see Section~\ref{Chap:surveyOfCA:Sec:space-time} of Chapter~\ref{Chap:surveyOfCA} for details) is a useful tool to observe the flow of information in evolution of the corresponding CA. It is a graphical representation of the configurations (on $x$-axis) at each time $t$ (on $y$-axis). Each of the CA states are depicted by some color. So, the evolution of the CA can be visible from the patterns generated in the state-space diagram.

\begin{figure}[!hbtp]
	 \vspace{-1.0em}
	\centering
\subfloat[seed $0^{100}1$ \label{statespace2_R}]{%
		\includegraphics[width=0.33\textwidth, height = 8.7cm]{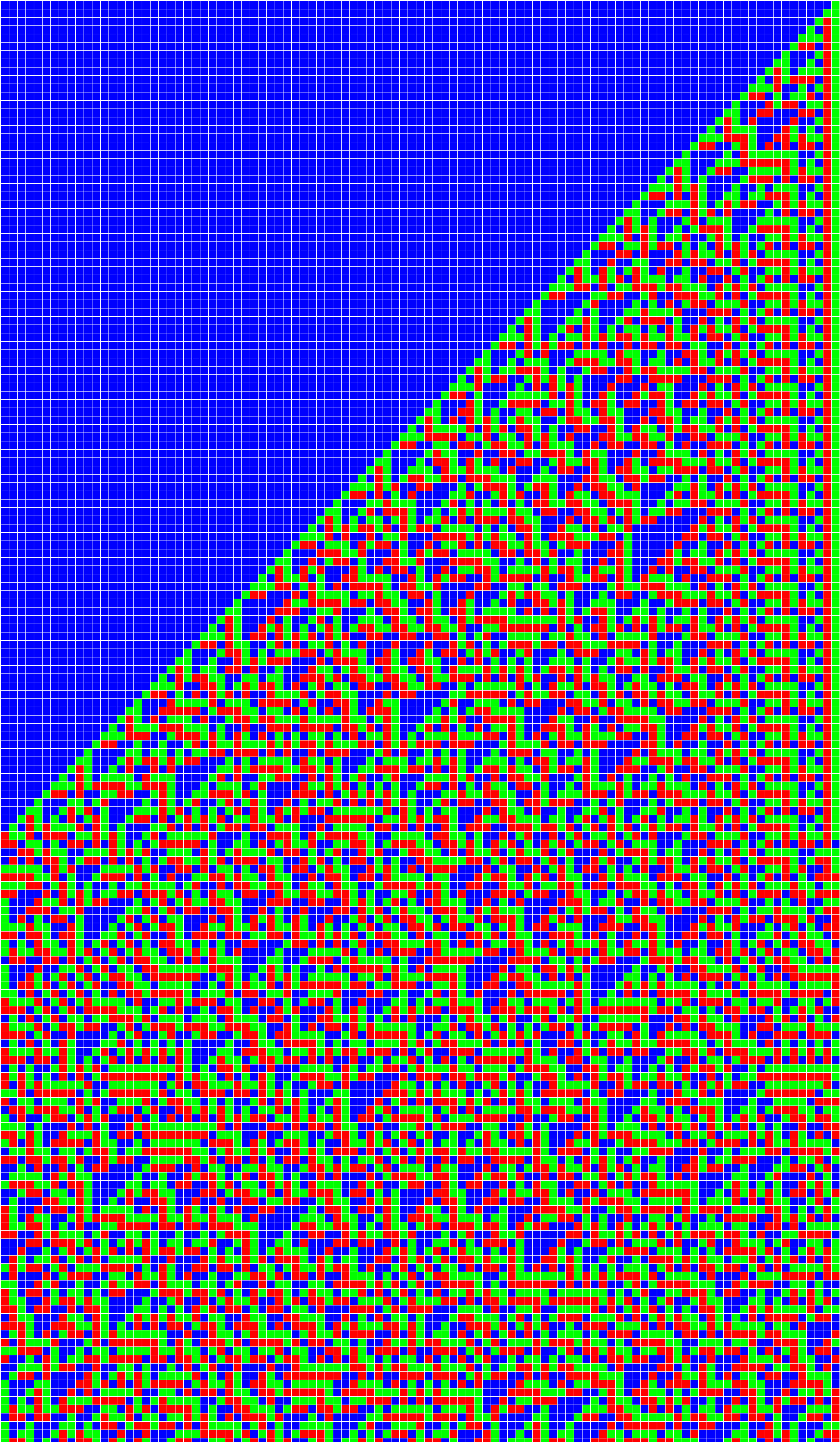}
	}
	\subfloat[seed $1^{100}0$\label{statespace7_R}]{%
		\includegraphics[width=0.33\textwidth, height = 8.7cm]{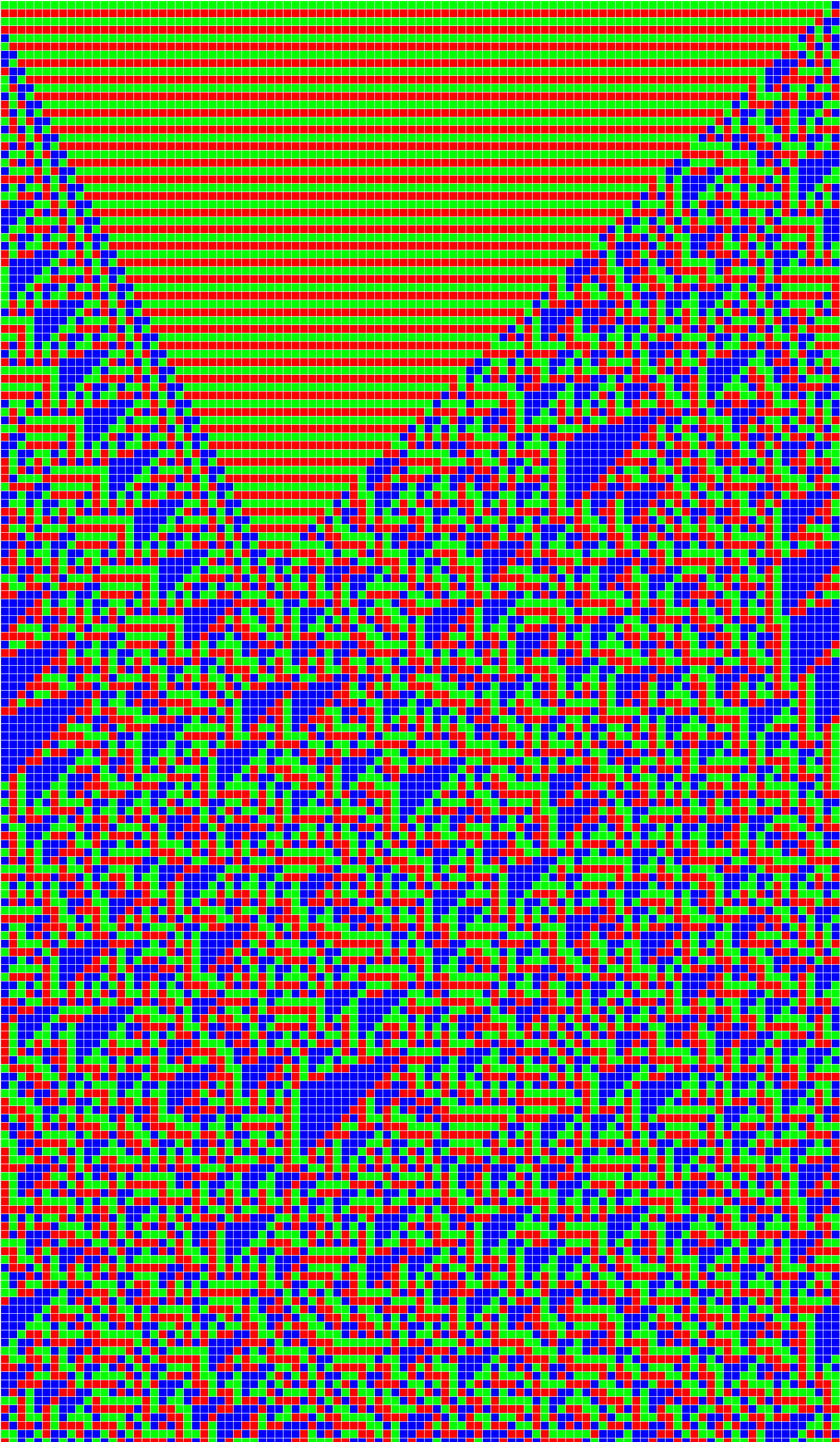}
	}
	\subfloat[seed $0^{100}2$ \label{statespace8_R}]{%
		\includegraphics[width=0.33\textwidth, height = 8.7cm]{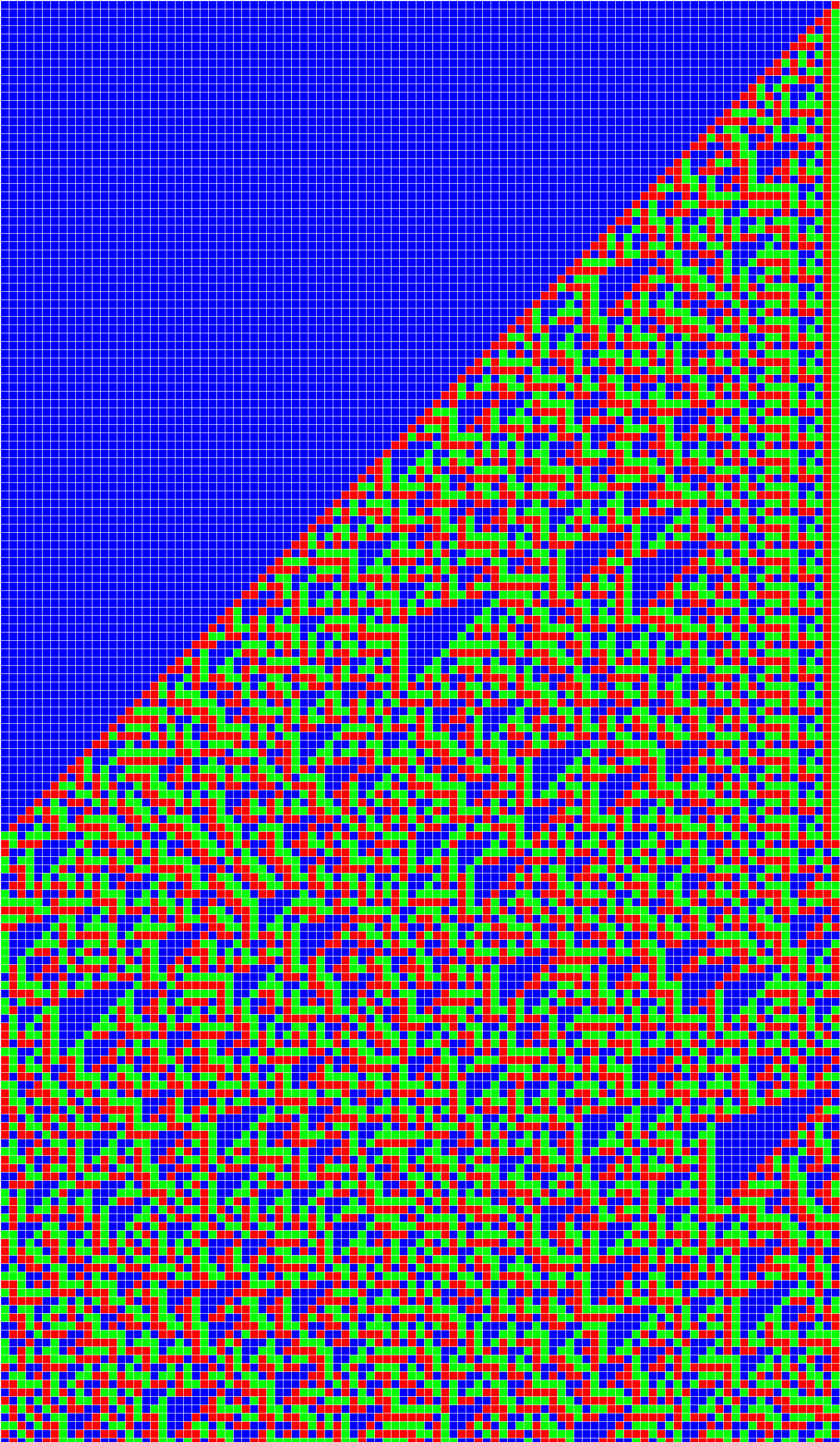}
	}
	\caption{Space-time diagram for the CA $\mathbf{\mathscr{R}}$ with $n=101$. Here, red implies state $2$, green $1$ and blue implies state $0$}
	\label{Chap:randomness_survey:fig:space-time diagram1} %
	       \vspace{-0.5em}
\end{figure}

The space-time diagrams of the CA $\mathbf{\mathscr{R}}$ with $n=101$ are shown in Figure~\ref{Chap:randomness_survey:fig:space-time diagram1}. Here, the figures depict evolution from an almost homogeneous seed with one cell having disparate value (like a tiny addition of noise to the stable configuration).
It is obvious from this figure that, there is information flow to both directions for the initial configuration $1^{n-1}0$. However, for the configurations $0^{n-1}1$ and $0^{n-1}2$, initial information flow is on left side only and after $n^{th}$ configuration, because of change in states at the terminal positions, information flows from both directions. 

\subsubsection{Large Cycle Length of the CA}
Large cycle length is an essential property of a good PRNG. However, apart from evolution, there is no tool to calculate the cycle length(s) of an arbitrary CA. So, we experiment our CA over small $n$ ($5\leq n\leq 14$) and note the maximum cycle length of the CA for that $n$. The ratio between maximum cycle length [maximum of all the cycles for a particular CA] to maximum possible number of configurations [for  $ d $-state, $ n $-length CA MPC is $d^{n}$] is also observed. Table~\ref{Chap:randomness_survey:tab:Tcells} shows the result of this experiment. In this table, the first column records the size of the CA ($n$) and the fourth column shows the maximum possible number of configurations (MPC) for that $n$. However, the actual maximum cycle length (MCL) of CA $\mathbf{\mathscr{R}}$ for that $n$ is recorded in the second column along with the particular seed for which we have got this cycle (Column $3$). The last column shows the ratio between MCL and MPC in percentage.

\begin{table}[hbpt]
	\scriptsize
	\begin{center}
		\caption{Maximum cycle length of the CA with corresponding seed }\label{Chap:randomness_survey:tab:Tcells}
			\resizebox{0.90\textwidth}{2.0cm}{
			\begin{tabular}{|c|c|c|c|c|}
				\hline
				\begin{tabular}{c}\#Cell\\ ($n$)\\ \end{tabular} & \begin{tabular}{c} Maximum\\ Cycle Length\\(MCL))\\\end{tabular} &\begin{tabular}{c} Seed \end{tabular} & 
				\begin{tabular}{c} Max Possible Number\\ of Configurations\\ (MPC)\\ \end{tabular}
				& $ \frac{MCL}{MPC}\% $ \\
				\hline
				
				$ 5 $ & $ 170 $ & $ 00001 $ & $243$ & $ 69.959 $ \\
				\hline
				$ 6 $ & $ 156 $ & $ 000012 $ & $729$ & $ 21.399 $   \\
				\hline
				$ 7 $ & $ 1967 $ & $ 0000001 $ & $2187$ &$ 89.941 $   \\
				\hline
				$ 8 $ & $584$ & $ 00000001 $ & $6561$ & $ 8.901 $   \\
				\hline
				$ 9 $ & $19296$ & $ 000000001 $ & $19683$ & $ 98.034 $   \\
				\hline
				$ 10 $ & $8970$ & $0111111002 $ & $59049$ & $ 15.191 $   \\
				\hline
				$ 11 $ & $121275$ & $ 00000000001 $ & $177147$ & $ 68.460 $   \\
				\hline
				$ 12 $ & $20544$ & $ 000000000001 $ & $531441$ & $ 3.866 $   \\
				\hline
				$ 13 $ & $1073397$ & $ 0000000000001 $ & $1594323$ & $ 67.326 $   \\
				\hline
				$ 14 $ & $59626$ & $ 11011100101010 $ & $4782969$ & $ 1.247 $   \\
				\hline   
				
			\end{tabular}
			}
	\end{center}
	\vspace{-1.5em}
\end{table}
We can pronounce that, the selected CA has the ratio $ \frac{MCL}{MPC} $ as more than $ 50 $  percent for all odd length cell, that is, when the CA is reversible. Also, note that, in Table~\ref{Chap:randomness_survey:tab:Tcells}, the reversible CAs have maximum cycle length when the initial configuration is $0^{n-1}1$. Now, to get an idea about average cycle length of the CA for an odd $n$, a greedy approach is taken. In our experiment, an arbitrary number of cells ($k$) are initialized with random values, whereas, the remaining cells are set with $0^{n-k-1}1$. Next configurations from the seed are generated until either the cycle is completed or some fixed amount of time has elapsed on a $64$-bit Intel Core $i5-3570$ CPU with $3.40GHz \times 4$ frequency and $4$ GB RAM. Some of the results of this experiment are shown in Table~\ref{Chap:randomness_survey:tab:Tcount}.
\begin{table}[!h]
\centering
		\caption{Number of unique configurations in a sample run for different $n$ with corresponding seeds}\label{Chap:randomness_survey:tab:Tcount}
		{
			\resizebox{1.00\textwidth}{!}{
				\begin{tabular}{|c|c|c|c|c|}
					\hline
					\#Cell & Seed & \#Unique Configurations & Running time & Cycle completed?  \\ 
					\hline
					\multirow{4}{*}{}
					& $101211001212101$ & $3292305$ & $15$ min & yes \\
					$15$ & $210011220022201$ & $3564780$ & $15$ min & yes\\
					& $210022002210001$ & $3808665$ & $15$ min & yes\\
					& $121001020201001$ & $3808665$ & $15$ min & yes\\
					\hline
					\multirow{5}{*}{}
					& $2002120000000000000000001$ & $3215366900$ & $20$ min & no\\
					& $2210220000000000000000001$ & $3140559423$ & $20$ min & no\\
					$25$ & $1002210000000000000000001$ & $3079962998$ & $20$ min & no\\
					& $1221220000000000000000001$ & $3073522004$ & $20$ min & no\\
					& $2101220000000000000000001$ & $3078828468$ & $20$ min & no\\
					\hline
					\multirow{5}{*}{}
					& $1012110012000000000000000000001$ & $3278533719$ & $25$ min & no\\
					& $1212100112000000000000000000001$ & $3284478677$ & $25$ min & no\\
					$31$ & $0002011110000000000000000000001$ & $3199265211$ & $25$ min & no\\
					& $0201110020000000000000000000001$ & $3190422305$ & $25$ min & no\\
					& $0022002210000000000000000000001$ & $3011307567$ & $25$ min & no\\
					\hline
					\multirow{5}{*}{}
					& $212022001212210200000000000000000000000000000000001$ & $3003102079$ & $35$ min & no\\
					& $100221201221221200000000000000000000000000000000001$ & $3019073812$ & $35$ min & no\\
					$51$ & $112101220021011200000000000000000000000000000000001$ & $3056772401$ & $35$ min & no\\
					& $002111222111000000000000000000000000000000000000001$ & $3056722864$ & $35$ min & no\\
					& $112101101202100200000000000000000000000000000000001$ & $3139757291$ & $35$ min & no\\
					\hline
					\multirow{5}{*}{}
					& $10220020022(0)^{89}1$ & $2261971435$ & $45$ min & no\\
					& $2220120001(0)^{90}1$ & $2262013800$ & $45$ min & no\\
					$101$ & $01022110022(0)^{89}1$ & $2262024986$ & $45$ min & no\\
					& $0200101121(0)^{90}1$ & $2261970226$ & $45$ min & no\\
					& $21021122222(0)^{89}1$ & $2261869246$ & $45$ min & no\\
					\hline
				\end{tabular}			} }
	\vspace{-1.0em}
\end{table}
It can be observed from this table that, the average cycle length for the CA $\mathbf{\mathscr{R}}$, where most of the cells are initialized with $0$ followed by a single cell with $1$, grows exponentially with $n$ ($n$ is taken as odd length only).

\subsection{The Cellular Automaton $\mathbf{\mathscr{R}}$ as PRNG}
\label{Chap:randomness_survey:sec:prng_R}
To design a PRNG using this CA, the cell length $n$ has to be taken as odd. However, even the longest cycle does not contain all possible configurations. Therefore, from one initial configuration, all possible configurations can not be reached. However, we take here a small window of length, say, $w (w<n)$ and generate the numbers out of the window.
That means, the decimal number corresponding to the ternary string of length $w$ of the window represents a random number. The length of the window can be adjusted as per requirement. For example, to generate a $32$-bit number, the necessary window length is of $20$ (ternary number). Similarly, for generating $64$-bit and $128$-bit numbers, required window sizes are $40$ and $80$ respectively.

To incorporate all possible combinations of the number out of window, the CA size $n$ is taken as a large number. The cells of the window are initialized arbitrarily, however, the remaining $n-w$ cells are initialized with $0^{n-w-1}1$. We have observed that, for the initial configuration $0^{n-1}1$, there is a left directional information flow for the first $n$ configurations (see Figure~\ref{Chap:randomness_survey:fig:space-time diagram1}). So, to avoid the possibility of predicting numbers if seed is given as $0^w$, we produce random numbers from ${(n+1)}^{th}$ configuration onwards.

It is perceived that, the cycle length for the CA is sufficiently large for $20$-length window when $n\ge 51$. So, we choose $n=51$ as the minimum cell length for a $20$-length window, that is, for a $32$-bit PRNG. Here, the range of the generated numbers is $0$ to $3^{20}-1$, which is slightly less than $2^{32}-1$.

\section{Performance of the PRNG}
Now, our target is to measure the randomness quality of the proposed PRNG. To do this we use some well-known empirical testbeds. Next we briefly discuss about the empirical testbeds. These testbeds are singularly used in the rest part of this dissertation.

\subsection{Empirical Tests}\label{chap:randomness_survey:sec:empirical}
Empirical tests target to find some pattern in the generated numbers of a PRNG to prove its non-randomness. These tests aim to check the local randomness property, that is, randomness of the numbers are approximated over a minimum sequence length, rather than the whole period \cite{Knuth2}. Note that, for empirical tests, numbers of a complete period are not necessary. Innumerable such tests can be developed which aim to find any violation of the desirable properties (described in Section~\ref{chap:randomness_survey:sec:property}), if exists, in a PRNG. If a PRNG passes all \emph{relevant} empirical tests, it is declared as a good PRNG. However, usually, there is no known method to find which tests are pertinent for a PRNG to certify its randomness quality. Therefore, the common practice is to use empirical testbeds to identify non-randomness in the generated numbers.

In general, empirical tests can be classified into two groups -- blind tests and graphical tests. In case of blind tests, the tests are based on statistics and computation, so, no human intervention is required in taking a decision. On the other hand, in case of graphical tests, the performance is measured by finding visible patterns in the generated image; so here decision is taken by the coordinating person(s). In the next subsections, the tests used for our purpose are described in more details.

\subsubsection{Blind (Statistical) Tests}\label{sec:BTest}
\noindent In blind tests or statistical tests, the properties of a random sequence are considered to be probabilistic. So, when applied on a random sequence, the likely outcome of these tests is believed to be known a priori and measured in terms of probability. The target of these tests is to find evidence against a specific null hypothesis $(\mathscr{H}_0)$. Usually, this $\mathscr{H}_0$ is, ``the sequence to be tested is random''. For each test, based on the random sequence produced by a PRNG, a decision is taken either to reject or not to reject the null hypothesis $\mathscr{H}_0$. To do this, a suitable randomness statistic, having a distribution of possible values, has to choose which determines the rejection of $\mathscr{H}_0$. Under $\mathscr{H}_0$, the reference theoretical distribution (usually standard normal or chi-square distribution) of this statistic is calculated. A \emph{critical value} $(t)$ is computed for this reference distribution. During a statistical test, the relevant statistic is calculated on the generated random sequence and compared to the critical value. If the test statistic value is greater than the critical value, $\mathscr{H}_0$ is rejected, otherwise it is not rejected. However, the $p$-value of a test measures the strength of evidence against the null hypothesis. It is defined as 
\[p=\Pr[X\geq t|\mathscr{H}_0 ~~{ is~ true }]\]
 where $X$ denotes the test statistics and $t$ the critical value \cite{L'Ecuyer:2007:TCL:1268776.1268777}. If $p$-value is very close to $0$ or $1$, it indicates that the sequence generated by the PRNG is not random. Normally, if $p \geq \alpha$, then the tested sequence is considered random, and $\mathscr{H}_0$ is not rejected, otherwise it is rejected. This $\alpha$ is called \emph{level of significance} of a test and it is set prior to the test. However, if the test statistic has a discrete distribution, the $p$-value is redefined as:
\begin{equation*}
p={\begin{cases}
	p_R,  &\text{if $p_R > p_L$ }\\
	1-p_L, &\text{if $p_R \geq p_L$ and $p_L < 0.5$}\\
	0.5, &\text{otherwise}
	\end{cases}}
\end{equation*} 
where $p_R= \Pr(X \geq t | \mathscr{H}_0 \text{ is true })$ and $p_L = \Pr(X \leq t | \mathscr{H}_0 \text{ is true })$.

There are many statistical battery of tests available which target to find non-randomness based on a collection of statistical tests. The first known statistical battery of tests was offered by Donald Knuth in $1969$ in his book ``The Art of Computer Programming, Vol. $2$''\cite{Knuth2}. Later, other batteries have been developed improving the testing procedures of Knuth. In this dissertation, we have selected tests from three well-known test suites, namely \emph{Diehard}, \emph{TestU01} and \emph{NIST}.

\begin{itemize}[leftmargin=0pt]
\item{\textbf{Diehard battery of Tests:}}\label{sec:diehard}
George Marsaglia in $1996$ provided this battery \cite{diehard}, which is the basic testbed for PRNGs. It consists of $15$ different tests - 
\begin{small}
\begin{framed}
\vspace{-1.0em}
\label{diehardtests}
\begin{center}
\underline{\textbf{Diehard Battery of Tests}}
\end{center}
1. Birthday spacings, 2. Overlapping permutations, 3. Ranks of $31\times31$ and $32\times32$ matrices, 4. Ranks of $6\times8$ matrices, 5. Monkey tests on $20$-bit Words, 6. Monkey tests: OPSO (Overlapping-Pairs-Sparse-Occupancy), OQSO (Overlapping-Quadruples-Sparse-Occupancy) and DNA tests, 7.  Count the $1$'s in a stream of bytes, 8. Count the $1$'s in specific bytes, 9. Parking lot test, 10. Minimum distance test, 11. Random spheres test, 12. The squeeze test, 13. Overlapping sums test, 14. Runs up and runs down test, and 15. The craps test (number of wins and throws/game). 
\end{framed}
\end{small}
To test a PRNG on Diehard for a particular seed, a binary file of size $10-12$ MB is created using the generated numbers of the PRNG with that seed. In our case, we have taken the file size as $11.5$ MB. For each test, one or multiple $p$-values are derived. A test is called \emph{passed}, if every $p$-value of the test is within $0.025$ to $0.975$ \cite{diehard}. A PRNG is supposed to be a good PRNG, if it passes all tests of every testbed for any seed. 

\item{\textbf{TestU01 library of Tests:}}\label{sec:testu01} This library offers implementations of many stringent tests -- the classical ones as well as many recent ones. It was developed by Pierre L'Ecuyer and Richard Simard \cite{L'Ecuyer:2007:TCL:1268776.1268777} to remove the limitations of existing testbeds -- like inability to modify the test parameters (such as, the input file type, $p$-values etc.) as well as to encompass new updated tests. It comprises of several battery of tests, including most of the tests in Diehard and many more with more flexibility
to select the test parameters than in Diehard. However, we have selected the battery \emph{rabbit} to test the PRNGs. The reason for choosing this test-suite is -- this battery is specifically designed to test a sequence of random bits produced by a generator. 
It contains the following $26$ tests from different modules (mentioned in parenthesis for each test):
\begin{small}
\begin{framed}
\vspace{-1.0em}
\label{rabbittests}
\begin{center}
\underline{\textbf{Battery Rabbit of TestU01}}
\end{center}
1. MultinomialBitsOver test (smultin), 2. ClosePairsBitMatch in $t = 2$ dimensions (snpair) and 3. ClosePairsBitMatch in $t = 4$ dimensions (snpair), 4. AppearanceSpacings test (svaria), 5. LinearComplexity test (scomp), 6. LempelZiv test (scomp), 7. Spectral test of Fourier1 (sspectral), and 8. Spectral test of Fourier3 (sspectral), 9. LongestHeadRun test (sstring), 10. PeriodsInStrings test (sstring), 11. HammingWeight with blocks of $L = 32$ bits test (sstring), 12. HammingCorrelation test with blocks of $L = 32$ bits (sstring), 13. HammingCorrelation test with blocks of $L = 64$ bits (sstring) and 14. HammingCorrelation test with blocks of $L = 128$ bits (sstring), 15. HammingIndependence with blocks of $L = 16$ bits (sstring), 16. HammingIndependence with blocks of $L = 32$ bits (sstring) and 17. HammingIndependence with blocks of $L = 64$ bits (sstring), 18.AutoCorrelation test with a lag $d = 1$ (sstring)and 19. AutoCorrelation test with a lag $d = 2$ (sstring), 20. Run test (sstring), 21. MatrixRank test with $32 \times 32$ matrices (smarsa) and 22. MatrixRank test with $320 \times 320$ matrices (smarsa), 23. RandomWalk1 test with walks of length $L = 128$ (swalk), 24. RandomWalk1 test with walks of length $L = 1024$ (swalk), and 25. RandomWalk1 test with walks of length $L = 10016$ (swalk).
\end{framed}
\end{small}


The battery \emph{rabbit} takes two arguments -- a filename and number of bits $(nb)$. The first $nb$ bits of the binary file, filled by the random numbers generated by the PRNG, is tested. For each test, the parameters are a function of $nb$, to make it dynamic. Here, to test a PRNG using \emph{rabbit}, we have set $nb=10^7$ and the file size is taken as $10.4$ MB. A test is declared to be passed for a seed, if each of the $p$-values of the test is within $0.001$ to $0.999$ \cite{L'Ecuyer:2007:TCL:1268776.1268777}.


\item{\textbf{NIST Statistical Test suite:}}\label{sec:nist}
\noindent The NIST Statistical Test Suite is developed to test a PRNG for cryptographic properties \cite{rukhin2001statistical}. It has mainly three tasks -- (1) investigate the distribution of $0$s and $1$s, (2) using spectral methods, analyze the harmonics of bit stream and (3) detect patterns based on information theory and probability theory. This test suite consists of $15$ tests --
\begin{small}
\begin{framed}
\vspace{-1.0em}
\label{nisttests}
\begin{center}
\underline{\textbf{NIST Test Suite}}
\end{center}
1. The Frequency (Monobit) Test,
2. Frequency Test within a Block,
3. The Runs Test,
4. Tests for the Longest-Run-of-Ones in a Block,
5. The Binary Matrix Rank Test,
6. The Discrete Fourier Transform (Spectral) Test,
7. The Non-overlapping Template Matching Test,
8. The Overlapping Template Matching Test,
9. Maurer's ``Universal Statistical'' Test,
10. The Linear Complexity Test,
11. The Serial Test,
12. The Approximate Entropy Test,
13. The Cumulative Sums (Cusums) Test,
14. The Random Excursions Test, and
15. The Random Excursions Variant Test.
\end{framed}
\end{small}

For this test suite, the level of significance $\alpha = 0.01$. So, for a sample size $m$ generated by a PRNG with a particular seed, if $x$ is the minimum pass rate, then for the PRNG to pass the test, minimum number of sequences with $p$-values $\geq 0.01$ has to be $x$.
This minimum pass rate is approximately $615$ for a sample size of $629$ for the random excursion (variant) test and $980$ for sample size $1000$ for each of the rest tests. In this dissertation, to test a PRNG using NIST test suite, sample size is taken as $10^3$ with sequence length = $10^6$ and a binary file of size $125$ MB is given as the input. 
\end{itemize}

\subsubsection{Graphical Test}\label{sec:Gtest}
In graphical tests, the numbers are plotted in a graph to see whether any visible pattern exists.
We have mainly used two graphical tests -- (1) Lattice tests, (2) Space-time diagram (see Section~\ref{Chap:surveyOfCA:Sec:space-time} of Page~\pageref{Chap:surveyOfCA:Sec:space-time} and Section~\ref{Chap:randomness_survey:sec:space-time_R} for details). 

In case of lattice tests, it is checked whether the random numbers form some patterns.
To test this, the consecutive random numbers (in normalized form), generated from a seed, are paired and plotted. Two types of lattice tests are executed on these normalized numbers, namely $2$-D lattice test (takes two consecutive numbers as a point) and $3$-D lattice tests (three consecutive numbers form a point). If the random numbers are correlated, the plots show patterns. Otherwise, the PRNG is considered to be good.

\subsection{Empirical Facts}\label{Chap:randomness_survey:sec:empirical_result_R}
From Section~\ref{Chap:randomness_survey:sec:prop_R}, it is obvious that, our window-based PRNG using the CA  $\mathbf{\mathscr{R}}$ possesses the theoretical properties to be a good source of randomness. In this section, we test its randomness quality in terms of performance in empirical testbeds.

\subsubsection{Uniformity Test} 
Uniformity is one of the main aspect of random number generation. However, in a CA based PRNG, no cycle contains all possible configurations. 
So, to test our PRNG for uniformity, the random numbers generated out of $20$ length window for the $51$ cell CA are stored and frequency of each number is calculated.
We have run our experiment for a couple of hours and generated $129824111$ numbers. It is observed that, among these data, $125819427$ numbers are generated only once, $3922425$ numbers twice, $81048$ numbers thrice, $1192$ numbers  four times and only $19$ numbers are generated five times. That means, about $97\%$ of the generated data are not repeated and the differences among the number of times, each number is generated by the CA, is negligible.

However, this result also indicates that, although very less, but still there exists a probability that a number, once generated, can again be generated in the cycle. But, the beauty is in the fact that, appearance of the same number multiple times does not make the sequence to be generated, same or predictable. This is the essence of randomness in this scheme.

Next, we create the normalized numbers equivalent to the $129824111$ numbers and calculate the frequency of numbers in a fixed sub-interval of $[0,1]$. It is seen that, frequency of the generated numbers in each sub-interval is uniform.

\subsubsection{Independence Test}\label{inde}
For testing independence, we use Diehard battery of tests and battery \emph{rabbit\_file()} of TestU01 library. Some of the results for $32$-bit PRNG (window size $20$, CA size $51$) are shown in Table~\ref{battery}. In this table, only those tests for which the $p$-values are fractional are recorded. In case of the tests having more than one $p$-values, we record the average of these values along with the ratio of occurrence of fractional $p$-values within parenthesis.

We also test the $32$-bit PRNG for $n=75, 101, 125, 151$ etc. It is seen that, for all such $n$, the results are similar. Likewise, the $64$-bit PRNG (that is, $40$ length window) is tested for $n=51, 101,151,201$ etc., and it is found that, for $n=101$ onwards, the randomness quality of the PRNG is similar. 

	\begin{sidewaystable}[hbtp]
		
		\setlength{\tabcolsep}{1.3pt}
		\scriptsize
		\renewcommand{\arraystretch}{1.30}
		\centering
		\small
		\caption{Empirical test results for the Proposed PRNG using Diehard and TestU01}
		{
			\resizebox{1.00\textwidth}{!}{
				\begin{tabular}{|c|c|c|c|c|c|}
					\hline
					\multirow{2}{*}{\theadfont{Name of the Tests}} & \multicolumn{5}{c|}{\theadfont{$p$-value of tests for random seed of window (Seeds shown in next row)}} \\
					\cline{2-6}
					&  \thead{$1121020001002110001$} &  \thead{$0221022202021120222$} &  \thead{$110010120222201222$} &  \thead{$2011011002112021122$} &  \thead{$21110002102202020221$} \\
					\hline
					\begin{tabular}{c} 	\thead{Rank of} \thead{$31 \times 31$}\thead{Matrices} \end{tabular} & $0.014$ & $0.275$& $0.161$& $0.225$ & $0.200$\\        
					\hline
					\begin{tabular}{c}	\thead{Rank of}  \thead{$32 \times 32$}  	\thead{Matrices}   \end{tabular} & $0.33$ & $0.29$ & $0.37$ & $0.86$ & $0.72$  \\        
					\hline	
					\begin{tabular}{c} \thead{Rank of}   \thead{$320 \times 320$} 	\thead{Matrices}   \end{tabular}& $0.41$ & $0.43$ & $0.77$ & $0.83$ & $0.95$  \\        
					\hline          	 
					\thead{ Monkey Test OPSO} & $0.1380084783 (8/23)$ & $0.1297198261 (8/23)$ & $0.1373205652(7/23) $ & $0.1212254348 (8/23)$ & $0.1781348696 (8/23)$ \\        
					\hline
					\thead{ Monkey Test OQSO} & $0.1558082857 (13/28)$ & $0.1522809286 (11/28)$ & $0.1797470714 (12/28)$& $0.1091943214 (12/28)$ & $0.149736 (13/28)$ \\        
					\hline
					\thead{ Monkey Test DNA}  & $0.3008551935 (19/31)$  & $0.2597885806 (19/31)$ & $0.3132697419 (20/31)$ & $0.2775887419 (19/31)$ & $0.224547 (19/31)$\\        
					\hline
					\begin{tabular}{c}	  \thead{Count the  $1$s} \thead{ in specific} \thead{bytes}   \end{tabular}	& $0.27989148 (17/25)$ & $0.26223256 (17/25)$ & $0.26211536 (17/25)$ & $0.36773772 (17/25)$ & $0.34711684 (17/25)$ \\        
					\hline
					\begin{tabular}{c}\thead{Overlapping} \thead{ Sum test} \end{tabular}    & $0.256689$ & $0.1093805$ & $0.000048$ & $0.025495$ & $0.000184$\\        
					\hline
					\thead{Runs Up Test} & $0.544416$  & $0.790334$ & $0.062062$ & $0.009546$ & $0.813583$ \\        
					\hline
					\thead{Runs Down Test}  & $0.011215$  & $0.006693$ & $ 0.005898$ & $0.000005$ & $0.028227$\\      
					\hline
					\begin{tabular}{c}  \thead{Close pair bit matching Test} \\($t=2$ dimension) \end{tabular}   & $0.04$ & $0.04$ & $0.16$ & $0.94$& $0.04$\\    
					\hline   
					\begin{tabular}{c}  \thead{Close pair bit matching Test} \\($t=4$ dimension) \end{tabular} 
					& $0.51$ & $0.51$ & $0.51$ & $0.04$ & $0.04$ \\        
					\hline  
					\thead{Multinomial overlapping bits test} & $7.0 \times {10}^{-3}$ & $0.19$& $3.4\times {10}^{-3}$& $8.0\times {10}^{-3}$ & $1.8\times {10}^{-3}$\\        
					\hline
					\thead{Appearance spacings test} & $0.94$ & $0.96$ & $0.59$& $0.9989$ & $0.44$ \\        
					\hline
					\begin{tabular}{c} \thead{Linear complexity Test} \\ (Jump complexity \& jump size) \end{tabular}& $0.87$ ~\&~ $0.80$  & $0.30 $ ~\&~ $ 0.27$ & $0.92 $ ~\&~ $ 0.82$ & $0.37 $ ~\&~ $ 0.17$ & $0.27 $ ~\&~ $ 0.03$\\        
					\hline
					\thead{Spectral Test (Fourier Transform)} & $0.17$ & $0.07$ & $0.29$ & $0.10$ & $0.04$\\        
					
					\hline
					\thead{Longest head run test} & $0.91$  & $0.70$ & $ 0.91$ & $0.91$ & $0.91$\\        
					\hline
					\begin{tabular}{c}  \thead{Random walk test}\\ (length $L = 128$, Statistics $R$) \end{tabular} & $0.03$ & $0.04$ & $2.9\times {10}^{-3}$ & $0.27$ & $0.02$ \\    
					\hline
					\begin{tabular}{c} \thead{Random walk test} \\ (length $L = 128$, Statistics $C$) \end{tabular}& $0.02$ & $0.06$ & $0.13$ & $0.01$ & $0.41$\\   
					\hline 
				\end{tabular}}
				\label{battery}
			}
		\end{sidewaystable}

\subsubsection{Space-time Diagram}
\begin{figure}[!h]
\centering
	\vspace{-1.0em}
	\subfloat[$w=20$\label{statespacewin1}]{%
		\includegraphics[width=0.125\textwidth, height = 9.0cm]{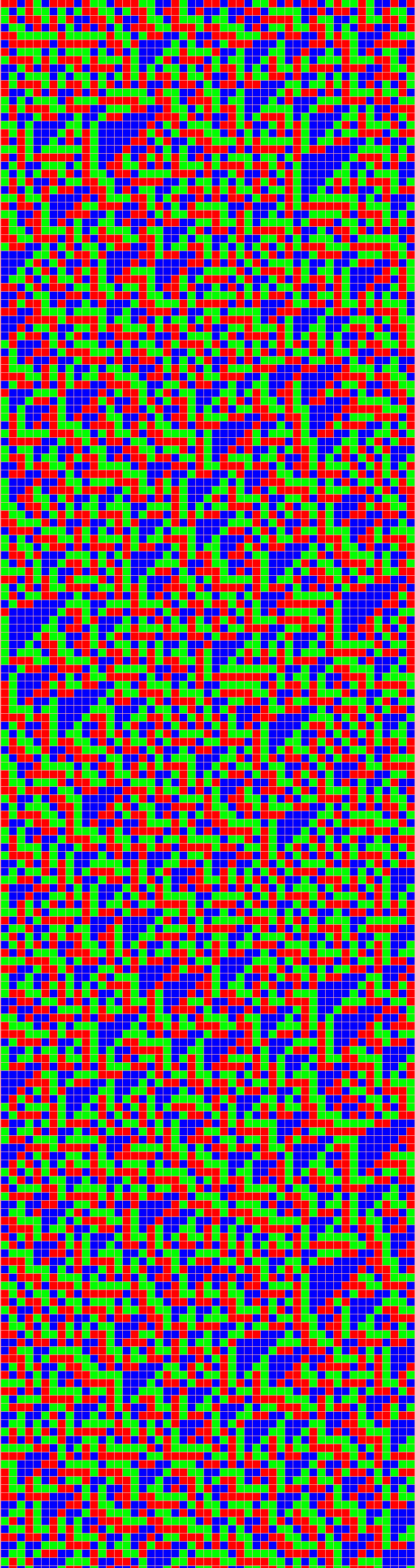}
	}
	\subfloat[$w=40$ \label{statespacewin2}]{%
		\includegraphics[width=0.25\textwidth, height = 9.0cm]{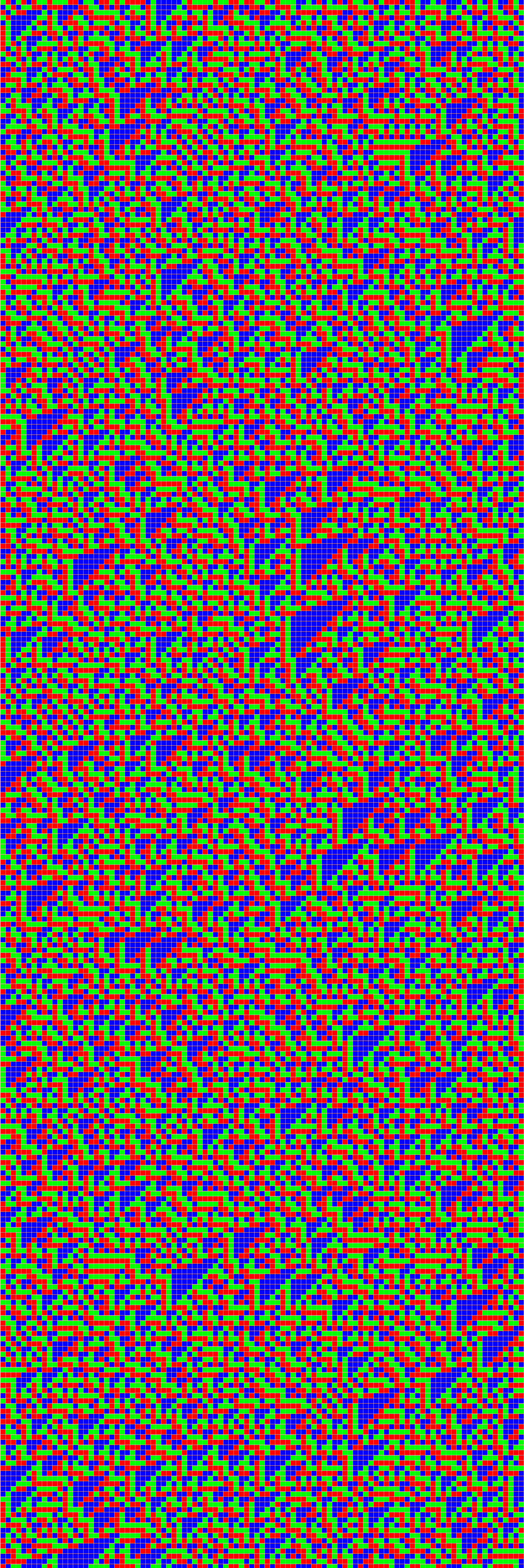}
	}
	\subfloat[$w=80$ \label{statespacewin3}]{%
		\includegraphics[width=0.625\textwidth, height = 9.0cm]{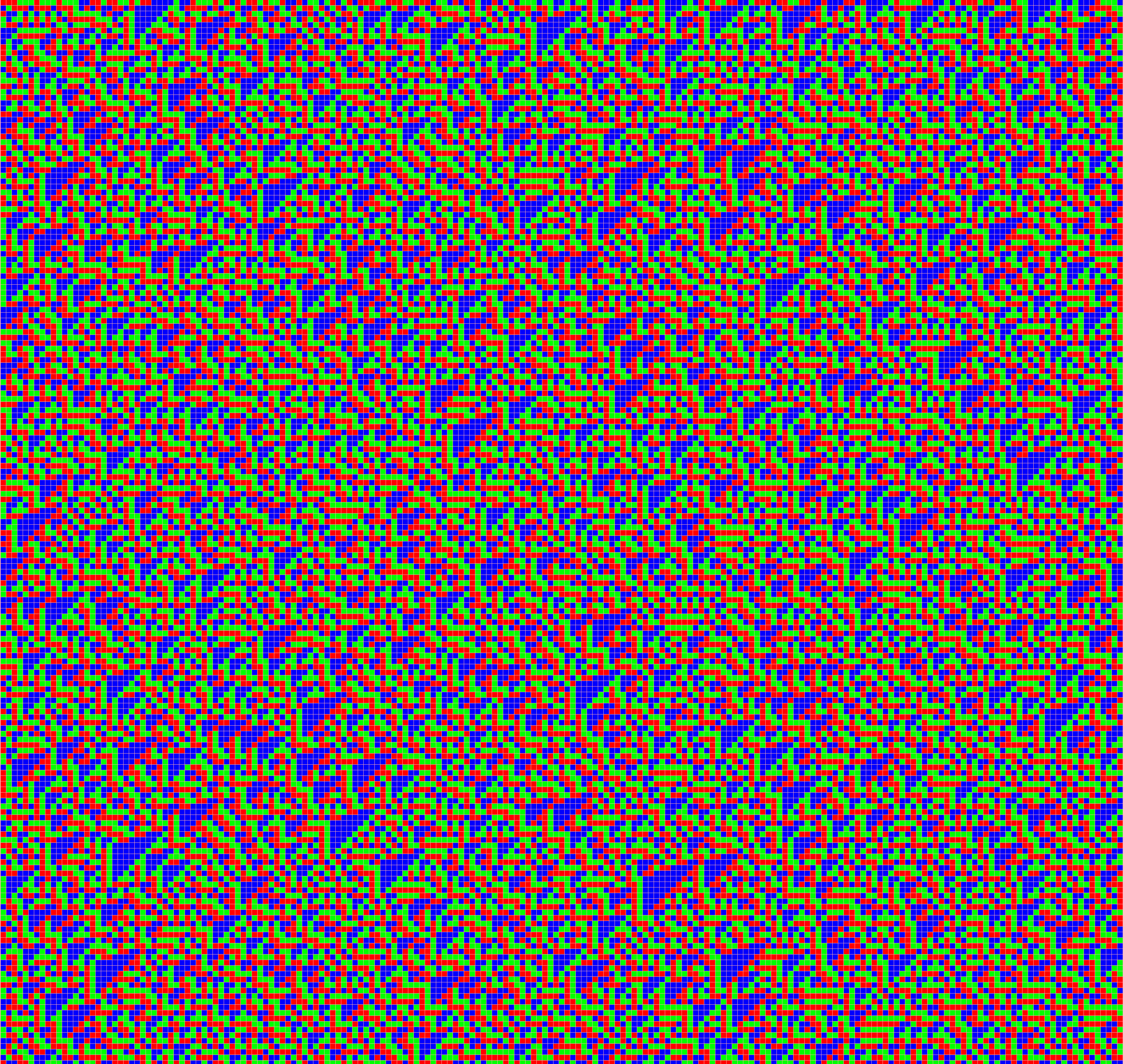}
	}
	\caption{Space-time diagram of the window for PRNG using CA  $\mathbf{\mathscr{R}}$. For Figure~\ref{statespacewin1}, Figure~\ref{statespacewin2} and Figure~\ref{statespacewin3}, CAs sizes are $51$, $101$ and $201$ respectively with seed for window $12120011001120202112$, $2110212120120101210201120012212220001001$ and $01202221000220001020112112120$\\$121202002020111211111220000120010001011122202012222$ respectively. Here, red implies state $2$, green is $1$ and blue implies state $0$}
	\label{Chap:randomness_survey:fig:space-time-window} %
\vspace{-1.0em}
\end{figure}

To test the PRNG on space-time diagram, we take a window of length, say $w$, initialized arbitrarily and set the remaining $n-w$ cells as $0^{n-w-1}1$.
The evolution for the cells of window is observed in Figure~\ref{Chap:randomness_survey:fig:space-time-window}. There is clear lack of repeating patterns and apparent randomness in the space-time diagram which proves the prospect of this window-based scheme with the CA as a good PRNG.

Now, we want to compare and find the actual position of this PRNG among the existing well-known PRNGs. So, in the next section, a brief survey of the existing PRNGs is reported.

\section{The Existing PRNGs: A Brief Survey}\label{chap:randomness_survey:sec:class}
Apart from the CAs, there are other techniques to design PRNGs. The earliest PRNGs were based on linear recurrences modulo a prime number, popularly called \emph{linear congruential sequence}. Introduced by Lehmer \cite{harvard1951annals}, such a PRNG is named \emph{linear congruential generator} (LCG). Many of the existing PRNGs are variants of it.
 
However, another type of linear recurrences, where the modulo operator is $2$, soon became popular due to their ease of implementation and efficiency in computer's binary arithmetic. These types of recurrences work mostly based on a \emph{linear feedback shift register} (LFSR). Introduced by Tausworthe \cite{Tausworthe}, this scheme has instigated many researchers to implement their PRNGs. For example, the celebrated PRNG \emph{Mersenne Twister} \cite{Matsumoto:1998:MTE:272991.272995} is implemented using a variation of this technology. Therefore, we can classify the PRNGs in three main categories -- 1) LCG based, 2) LFSR based and 3) CAs based.

\subsection{LCG based PRNGs}\label{chap:randomness_survey:sec:lcg}
One of the most popular random number generation technique is based on linear recursions on modular arithmetic. These generators are specialization on the linear congruential sequences, represented by 
\begin{equation}
\label{lcg-form}
 x_{n+1}=(ax_n +c) \pmod{m}, ~~n \geq 0
\end{equation}
Here $m>0$ is the modulus, $a$ is the multiplier, $c$ is the increment and $x_0$ is the starting value or seed;  $0 \leq a < m$, $0 \le c < m$, $0\le x_0 < m$ \cite{Knuth2}. The sequence $(x_i)_{i\ge 0}$ is considered as the desired sequence, and the output is $u_i = \frac{x_i}{m}$, if anybody wants to see the numbers from $[0,1)$. However, not all choices of $m,a,c,x_0$ generate a random sequence. For example, if $a=c=1$, the sequence is not random. Similarly, if $a=0$, the case is even worse. Therefore, selection of these magic numbers is crucial for getting a random sequence of numbers.

We can observe that, maximum period possible for an LCG is $m$. However, to get a maximum-period LCG, the following conditions need to be satisfied \cite{Knuth2}: 
\begin{enumerate}
\item $c$ is relatively prime to $m$;
\item if $m $ is multiple of $4$, $a-1$ is also multiple of $4$;
\item for every prime divisor $p$ of $m$, $a-1$ is multiple of $p$.
\end{enumerate}

A good maximum-period LCG is Knuth's LCG \verb MMIX \cite{Knuth2} where $a=6364136223846793005,~ m = 2^{64}$ and $c=1442695040888963407$. The PRNGs used in computer programming are mainly LCGs having maximum period. Some popular examples are \verb rand ~of GNU C Library \cite{GCC} where $a=1103515245$, $c= 12345$ and $m=2^{31}$, and \verb lrand48 ~of same library where $a=25214903917$, $c= 11$ and $m=2^{48}$. Another well-known LCG is \verb drand48 ~of GNU C Library which is similar to \verb lrand48, ~except it produces normalized numbers. Borland LCG is also a well-liked PRNG having $a=22695477$, $c=1$ and $m=2^{32}$.

Many variations of LCGs were proposed. For example, if we take the increment $c=0$, then the generator is called \emph{multiplicative ({\em or,} mixed) congruential generator} (MCG):
\begin{equation}\label{mcg_eq}
 x_{n+1}=ax_{n} \pmod{m}, ~~n \geq 0
\end{equation}
Although generation of numbers is slightly faster in this case, but the maximum period length of $m$ is not achievable. Because, here $x_n=0$ can never appear unless the sequence deteriorates to zero. When $x_n$ is relatively prime to $m$ for all $n$, the length of the period is limited to the number of integers between $0$ and $m$ that are relatively prime to $m$ \cite{Knuth2}. Now, if $m=p^e$, where $p$ is a prime number and $e\in \mathbb{N}$, Equation~\ref{mcg_eq} reduces to: \[x_n = a^nx_0 \pmod{p^e}\]
Taking $a$ as relatively prime to $p$, the period of the MCG is the smallest integer $\lambda$ such that, \[x_0 = a^{\lambda}x_0 \pmod{p^e}\]
Let $p^f$ be the gcd of $x_0$ and $p^e$, then this condition turns down to \[a^{\lambda} = 1 \pmod{p^{e-f}}\]
When $a$ is relatively prime to $m$, the smallest integer $\lambda$ for which $a^{\lambda} = 1 \pmod{p^{e-f}}$ is called the \emph{order of $a \text{ modulo }{m}$}. Any value of $a$ with maximum possible order modulo $m$ is called a \emph{primitive element modulo $m$}. Therefore, the maximum achievable period for MCGs is the order of a primitive element. It can be at maximum $m-1$ \cite{Knuth2}, when 
\begin{enumerate} [topsep=1pt,itemsep=1ex]
\item $m$ is prime;
\item $a$ is a primitive element modulo $m$;
\item $x_0$ is relatively prime to $m$.
\end{enumerate}

Some MCGs with large period are reported in \cite{5388354,fishman1990multiplicative,doi:10.1137/0907002}.  C++11's \verb minstd_rand \cite{Park:1988:RNG:63039.63042,Park:1993}, a good PRNG, is an MCG where $a=48271$ and $m=2^{31}-1$. However, the MCGs perform unsatisfactorily in spectral tests \cite{Knuth2}. So, higher order linear recurrences are proposed of the form 
\begin{equation}\label{mrg-form}
x_{n} = a_1x_{n-1}+\cdots +a_kx_{n-k} \pmod{m}
\end{equation}
where $k\geq 1$ is the order.
Here, $x_0, \cdots, x_{k-1}$ are arbitrary but not all zero. For these recurrences, the best result can be derived when $m$ is a large prime. In this case, according to the theory of finite fields, multipliers $a_1,\cdots, a_k$ exist, such that, the sequence of Equation~\ref{mrg-form} has period of length $p^k-1$, if and only if the polynomial
\begin{equation}\label{mrg-polynomial}
P(z)=z^k-a_1z^{k-1}-\cdots - a_k
\end{equation}
is a \emph{primitive polynomial modulo $p$} \cite{Knuth2}. That is, if and only if, the root of $P(z)$ is a primitive element of the Galois field with $p^k$ elements\footnote{A nonzero polynomial $P(z)$ is \emph{irreducible} if it cannot be factored into two non-constant polynomials over the same field. The straightforward criterion for a polynomial $P(z)$ of degree $k$ over Galois Field $\mathbb{F}(m)$ to be irreducible is -- (1) it divides the polynomial $z^{m^k}-z$ and (2) for all divisors $d$ of $k$, $P(z)$ and $z^{m^d}-z$ are relatively prime. The polynomial $P(z)$ is \emph{primitive}, if it is irreducible and $\min\limits_{n\in \mathbb{N}}\{n | P(z) \text{ divides } z^n-1\}=m^k-1$. In this case, $P(z)$ has a root $\alpha$ in $\mathbb{F}(m^k)$ such that, $\{0,1,\alpha, \alpha^2, \cdots, \alpha^{m^k-2}\}$ is the entire field $\mathbb{F}(m^k)$.}. A generator with such recurrence is called \emph{multiple recursive generator} (MRG) \cite{LEcuyer1990}.

A variant of MRG is the additive \emph{lagged-Fibonacci} generators \cite{brent1994periods}, which take the following form:
 \[x_{n} = (\pm x_{ n-r} \pm x_{ n-s}) \pmod{2^w}\] 
general form of which is a linear recurrence \[q_0x_n+q_1x_{n+1}+\cdots +q_rx_{n+r} = 0 \pmod{2^w}\] defined by a polynomial \[Q(t)= q_0 +q_1t+\cdots + q_rt^r\] with integer coefficients and degree $r>0$. Here, $w$ is an exponent, which may be chosen according to the word length of computer. The desired random sequence is $(x_i)_{i\ge 0}$ where $x_0,\cdots,x_{r-1}$ are initially given and not all even. However, if $Q(t)=q_0 + q_st^s+q_rt^r$ is a primitive trinomial with $r>2$, and if $q_0$ and $q_r$  are chosen as odd, the sequence $(x_i)_{i\ge 0}$ attains the maximal period of length $2^{w-1}(2^r-1)$. The PRNG, proposed in \cite{brent1994periods}, uses this type of trinomials. Some extensions of lagged-Fibonacci generators are the PRNGs named \emph{add-with-carry} (AWC) and \emph{subtract-with-borrow} (SWB) generators \cite{10.2307.2959748}, \emph{Recurring-with-carry} generators \cite{10.2307.3215210}, \emph{multiply-with-carry} (MWC) generators \cite{couture1997distribution} etc. 
 
Another type of generators, named as \emph{inversive congruential generators} (ICGs) are proposed in \cite{Eichenauer1986,EICHENAUERHERRMANN1992345,eichenauer1993statistical}. These generators are defined by the recursion \[x_{n+1}=a{x_{n}}^{-1} + c \pmod{p},~~ n \geq 0\]
where $p$ is a large prime, $x_n$ ranges over the set $\{0,1,\cdots,p-1,\infty\}$ and the $x_n^{-1}$ is the inverse of $x_n$, defined as:  $0^{-1}=\infty$, $\infty^{-1}=0$, otherwise $x^{-1}x \equiv 1 \text{ (modulo }p)$. For the purpose of implementation, one can consider $0^{-1}=0$, as 0 is always followed by $\infty$ and then by $c$ in the sequence. However, for many choices of $a$ and $c$, maximum period length $p+1$ is attainable \cite{Knuth2}. 

To improve the randomness of an LCG, several techniques have been proposed. One important class of PRNGs exists which deals it by combining more than one LCG, see for example \cite{L'Ecuyer:1988:EPC:62959.62969,10.2307.2347988,doi:10.1287.opre.44.5.816,doi:10.1287.opre.47.1.159}. Several combining techniques are suggested in the literature, like addition using integer arithmetic \cite{10.2307.2347988,L'Ecuyer:1988:EPC:62959.62969}, shuffling \cite{nance1978some}, bitwise addition modulo $2$ \cite{bratleyguide} etc. 

In \cite{doi:10.1287.opre.44.5.816}, it is shown that, we can get an MRG equivalent (or approximately equivalent) to the combined generator of two or more component MRGs, where the equivalent MRG has modulus equal to the product of the individual moduli of the component MRGs.
Let us consider $J\geq 2$ component MRGs where each MRG satisfies the recurrence
\begin{equation}\label{cmrg-form}
x_{j,n} = a_{j,1}x_{j,n-1}+\cdots + a_{j,k}x_{j,n-k} \pmod{m_j} \text{,~~~~    $1\leq j \leq J$,}
\end{equation}
having order $k_j$ and coefficients $a_{j,i}$ ($1\le i \le k$). Here, the moduli $m_j$s are pairwise relatively prime and each MRG has period $\rho_j={m_j^{k_j}}-1$.
Now, two combined generators can be defined \cite{doi:10.1287.opre.44.5.816}:
\begin{equation}\label{cmrg_2}
w_{n} = (\sum_{j=1}^{J} \frac{\delta_jx_{j,n}}{m_j}) \pmod{1} 
\end{equation}
\begin{equation}\label{cmrg_1}
z_{n} = (\sum_{j=1}^{J} \delta_jx_{j,n}) \pmod{m_1} \text{;~~~~  $\tilde{u}_n = \frac{z_n}{m_1}$,}
\end{equation}
Here, $\delta_1, \delta_2, \cdots$ are arbitrary integers such that each $\delta_j$ is relatively prime to $m_j$. The MRGs of Equation~\ref{cmrg_2} and \ref{cmrg_1} are approximately equivalent to each other. Further, it can be shown that, the MRGs of equations \ref{cmrg-form} and \ref{cmrg_2} are equivalent to an MRG of Equation~\ref{mrg-form} with $m=\prod\limits_{j=1}^{J} m_j$ and period length = $lcm(\rho_1,\cdots, \rho_J)$.
A well-known example of combined MRG is \verb MRG31k3p \cite{L'Ecuyer:2000:FCM:510378.510476} where two component MRGs of order $3$ are used having the following parameters:  
\[m_1=2^{31}- 1,~ a_{1,1}=0,~ a_{1,2}=2^{22},~ a_{1,3}=2^7+1\]
\[m_2=2^{31}-21069,~ a_{2,1}=2^{15},~ a_{2,2}=0,~ a_{2,3}=2^{15}+1\]
Here, for ease of implementation, each component MRG has two non-zero coefficients of the form $2^q$ and $2^q+1$. The combined MRG folows Equation~\ref{cmrg_1} and its period length is approximately $2^{185}$.

Another technique of improving randomness quality of a generator is to use a randomised algorithm over the outputs of a single LCG. This randomized algorithm is an efficient permutation function or hash function in \cite{o1988pcg}, which introduces the family of generators as \emph{permuted congruential generator} or \emph{PCG}. Here, several operations are performed on the outputs of a fast LCG, like random shifts to drop bits, random rotation of bits, bitwise exclusive-or(XOR)-shift and modular multiplication to perturb the lattice structure inherent to LCGs and improve its randomness quality. A good implementation is \verb PCG-32, ~which produces $32$-bit output and has a period length $2^{64}$. Here, the multiplier is $6364136223846793005$ and increment is taken as $1$.


Sometimes, LCG can be written in a matrix form as 
 \begin{equation}\label{LCG_matrix}
 \mathbf{X_n}=\mathbf{AX_{n-1}+C}\pmod{m}
 \end{equation}
Here, $S=\{\mathbf{X}=(x_1,\cdots,x_k)^T | 0 \leq x_0,\cdots,x_k < m\}$ is the set of $k$-dimensional vectors with elements in $F=\{0,1,\cdots,m-1\}$, $\mathbf{A}=(a_{ij})$ is a $k \times k$ matrix with elements in $F $, $\mathbf{C} \in S$ is a constant vector and $\mathbf{X_0}$ is the seed \cite{LEcuyer1990}. If $k=1$,  the recurrence of Equation~\ref{LCG_matrix} reduces to Equation~\ref{lcg-form}. When $\mathbf{C}=0$, the generator is an MCG:
 \begin{equation}\label{MCG_matrix}
 \mathbf{X_n}=\mathbf{AX_{n-1}}\pmod{m}
 \end{equation}
This form is useful because of its jumping-ahead property. Even for a large $v$, $\mathbf{X}_{i+v}$ can be reached from  $\mathbf{X}_{i}$, by first computing $\mathbf{A}^v \pmod{m}$ in $\mathscr{O}(\log{ v})$ time and applying a matrix-vector multiplication $\mathbf{X}_{i+v} = (\mathbf{A}^v \pmod{m})\mathbf{X}_i \pmod{ m}$ \cite{LEcuyer1990}. Moreover, using this matrix, any LCG of order $k$ can be expressed by an MCG of oder $k+1$: modify $\mathbf{A}$ to add $\mathbf{C}$ as its $(k+1)^{th}$ column and a $(k+1)^{th}$ line containing all $0$s except $1$ in $(k+1)^{th}$ position; modify $\mathbf{X_n}$ to add $1$ as its $(k+1)^{th}$ component. When $m$ is prime and $\mathbf{C}=0$, $F$ and $S$ are equivalent to $\mathbb{F}(m)$ and $\mathbb{F}(m^k)$, where $\mathbb{F}(m^k)$ is the Galois field with $m^k$ elements. In this case, the MCGs have maximal possible period = $m^k-1$ if and only if the characteristic polynomial of $\mathbf{A}$,   
\begin{equation}\label{charac_poly}
 f(x)=|xI-\mathbf{A}|\pmod{m} = (x^k - \sum_{i=1}^{k}a_ix^{k-i})\pmod{m}
 \end{equation}
with coefficients $a_i$ in $\mathbb{F}(m)$ is a primitive modulo $m$. For attaining this period, $\mathbf{A}$ must be nonsingular in modulo $m$ arithmetic. Nevertheless, a polynomial of Equation~\ref{charac_poly} has a companion matrix $\mathbf{A}$:
\begin{equation}\label{a_matrix}
\mathbf{A}={\begin{bmatrix}
 0 & 1 &  \cdots & 0\\
\vdots  & \vdots & \ddots & \vdots \\
 0 & 0  & \cdots & 1 \\
 a_k & a_{k-1} & \cdots & a_1
\end{bmatrix}}
\end{equation}
In this case, by taking $\mathbf{X_n}=(x_n,\cdots,x_{n-k+1})^T$, MCG of Equation~\ref{MCG_matrix} is converted to recurrence of MRG (Equation~\ref{mrg-form}), where $\mathbf{X_n}$ obeys the recursion: \[\mathbf{X}_{n} = a_1\mathbf{X}_{n-1}+\cdots a_k\mathbf{X}_{n-k} \pmod{m}\]

\subsection{LFSR based PRNGs}\label{chap:randomness_survey:sec:lfsr} 

Tausworthe generator \cite{Tausworthe} is a linear recurrence of order $k>1$ like Equation~\ref{mrg-form} where $m=2$, defined by the recurrence
 \begin{equation}\label{lfsr_form}
x_n = (a_1x_{n-1} + \cdots + a_kx_{n-k}) \pmod{2}
 \end{equation}
Here, $a_k = 1$ and $a_1,a_2,\cdots, a_{k-1} \in \mathbb{F}_2$. This recurrence can be implemented on a linear feedback shift register (LFSR). However, the random number is represented by 
 \begin{equation}
u_n = \sum_{l=1}^{L}x_{ns+l-1}2^{-l}
 \end{equation}
which is a number with $L$ consecutive bit sequence of Recurrence \ref{lfsr_form}; with successive $u_n$s spaced $s$ bits apart \cite{Tausworthe}. Here $s$ and $L$ are positive integers.
This PRNG can have a maximal period $\rho = 2^k -1$, if and only if, the characteristic polynomial
 \begin{equation}\label{poly}
 P(z)= 1+a_1z + a_2z^2+\cdots + z^k
 \end{equation}
is primitive over $\mathbb{F}(2)$ and $s$ is relatively prime to $2^k-1$. Then the generated sequence is called \emph{maximal-length linearly recurring sequence modulo $2$}. A popular example of this type of PRNG is \verb random ~of GNU C Library, which returns numbers between $0$ to $2147483647$ having period $ \rho \approx 16\times(2^{31}-1)$

Initially LFSR-based Tausworthe generators used primitive trinomials \cite{Tausworthe,Tootill_1973}. In \cite{Tootill:1971:RUP:321650.321655}, it is shown that, any Tausworthe generator that uses primitive trinomials of form 
 \begin{equation}\label{trinomial}
 P(z)=z^p+z^q+1 ~~(1 \leq q \leq (p-1)/2)
 \end{equation}
as the characteristic polynomial can be represented by a simple linear recurrence in $\mathbb{F}(2^p)$. Moreover, likewise combined MRGs, \emph{combined} Tauseworthe generators have been proposed \cite{l1996maximally}. 
It consists of $J\geq 2$ Tausworthe generators with primitive characteristic polynomials $P_j(z)$ of degree $k_j$ with $s_j$ as mutually prime to $2^{k_j}-1$, $1\leq j \leq J$. The sequence is denoted by $x_{j,n}$ (see Equation~\ref{cmrg-form} with modulus $2$) and random number by $u_{j,n} = \sum\limits_{l=1}^{L}x_{j,ns_j+l-1}2^{-l}$. $L$ is usually the word size of the computer. The output of the combined generator is 
 \[u_n = (u_{1,n}\oplus u_{2,n}\oplus\cdots\oplus u_{J,n})\]
where $\oplus$ is the bitwise XOR operation. As discussed before in Page~\pageref{cmrg_1}, this generator has period $\rho = lcm(2^{k_1}-1,2^{k_2}-1,\cdots, 2^{k_J}-1)$, if the polynomials $P_j(z)$ are pairwise relatively prime, that is, every pair of polynomials have no common factor. \verb Tauss88 \cite{l1996maximally} is such a generator where three component PRNGs are used with order $k_1 = 31$, $k_2 = 29$, and $k_3 = 28$ respectively. This PRNG has period length $\rho = (2^{31} -1)(2^{29}-1)(2^{28}-1) \approx 2^{88}$ and 
returns either $32$-bit unsigned integer or its normalized version. The $C$ code for this PRNG is accessible from $https://github.com/LuaDist/gsl/blob/master/rng/taus.c$. There are two other good combined Tausworthe generators, named \verb LFSR113 ~and \verb LFSR258. ~For \verb LFSR113, ~number of component PRNGs $J = 4$, with period length $\rho \approx 2^{113}$. However, for \verb LFSR258, ~$J = 5$ and period length $\rho \approx 2^{258}$. Both these PRNGs return $64$ bit normalized numbers; for \verb LFSR113 ~the numbers are normalized by multiplying the unsigned long integer output of the LFSR with $2.3283064365387\times 10^{-10}$ and for \verb LFSR258, ~the same is done by multiplying the unsigned $64$-bit output of the LFSR with $5.421010862427522170037264\times 10^{-20}$. $C$ codes for these two PRNGs are available at \cite{RNG}.

A generalization of LFSR (GFSR) has been proposed in \cite{Lewis:1973:GFS:321765.321777} to improve the quality of the PRNGs.
A GFSR sequence can be represented in binary as 
\begin{equation}
{X}_n = x_{j_1+n-1}x_{j_2+n-1}\cdots x_{j_k+n-1}
\end{equation}
where $X_n$ is a sequence of $k$-bit integers and $x_i$ is a LFSR sequence of Equation~\ref{lfsr_form}. 
%
However, this generator fails to reach its theoretical upper bound on period (equal to number of possible states) and has large memory requirement. So, another variation, named twisted GFSR (TGFSR), was proposed in \cite{matsumoto1992twisted,Matsumoto:1994:TGG:189443.189445}. This generator is same as GFSR, but, its linear recurrence is 
\begin{equation}
\mathbf{X}_{l+n} = \mathbf{X}_{l+m} \oplus \mathbf{X}_l \mathbf{A},~~ (l=0,1,\cdots)
\end{equation}
where $\mathbf{A}$ is a $w \times w$ matrix over $\mathbb{F}(2)$, $n,m,w$ are positive integers with $n>m$ and $\mathbf{X}_i$s are vectors in $\mathbb{F}(2^w)$. Usually, matrix $\mathbf{A}$ is chosen as Equation~\ref{a_matrix}. The seed is the tuple $(\mathbf{X}_0,\mathbf{X}_1,\cdots,\mathbf{X}_{n-1})$ with at least one non-zero value.

Likewise LCGs, all LFSR based generators can be represented in the following matrix form: 
 \begin{equation}\label{lfsr1}
\mathbf{X_n} = \mathbf{AX}_{n-1}
 \end{equation}
  \begin{equation}\label{lfsr2}
 \mathbf{Y_n}=\mathbf{BX}_n
 \end{equation}
 \begin{equation}
u_n = \sum_{l=1}^{w}y_{n,l-1}2^{-l}
 \end{equation}
Here, $k,w > 0$, $\mathbf{A}$ is a $k \times k$ matrix, called transition matrix, $\mathbf{B}$ is a $w \times k$ matrix, called output transformation matrix and elements of $\mathbf{A},\mathbf{B}$ are in $\mathbb{F}_2$. The $k$-bit state vector at step $n$ is $\mathbf{X}_n = (x_{n,0} , \cdots, x_{n,k-1} )^T$,  the $w$-bit output vector is $\mathbf{Y}_n = (y_{n,0} , \cdots, y_{n,k-1} )^T$ and output at step $i$ is $u_n \in [0, 1)$. All the operations in equations \ref{lfsr1} and \ref{lfsr2} are modulo $2$ operations. The characteristic polynomial of matrix $\mathbf{A}$ is same as Equation~\ref{charac_poly} with modulus $2$:
\begin{equation}
P(z)=det(z\mathbf{I}-\mathbf{A})=(z^k - \sum_{i=1}^{k}a_iz^{k-i})
\end{equation}
where $a_j \in \mathbb{F}_2$ and $\mathbf{I}$ is the identity matrix. If $a_k = 1$, this recurrence is purely periodic with order $k$. The period of $\mathbf{X}_n$ is maximal, that is, $2^k - 1$, if and only if, $P (z)$ is a primitive polynomial in $\mathbb{F}_2$. In this way, these PRNGs can be portrayed as LCGs in polynomials over $\mathbb{F}_2$.

The matrix $\mathbf{B}$ is usually used for tempering to improve equidistribution property of the PRNG by elementary bitwise transformation operations, like XOR, AND and shift \cite{Matsumoto:1994:TGG:189443.189445}. A TGFSR with tempering operations is called \emph{tempered} TGFSR. The well-known PRNG \emph{Mersenne Twister} (MT) is a variation of TGFSR where the linear recurrence is \cite{Matsumoto:1998:MTE:272991.272995}:
\begin{equation}\label{MT_rec}
\mathbf{X}_{k+n} = \mathbf{X}_{k+m} \oplus (\mathbf{X}^u_{k} | \mathbf{X}^l_{k+1})\mathbf{A}
\end{equation}
Here, $r,m,w$ are positive integers with $0\le r \le w-1$, $m$ ($1 \le m \le n$) is middle term and $r $ is separation point of one word. $\mathbf{A}$ is a $w \times w$ matrix (like Equation~\ref{a_matrix}) with entries in $\mathbb{F}(2)$, $|$ denotes bit vector wise concatenation operation, $\mathbf{X}_{k}^{u}$ is the upper $ w-r$ bits of $\mathbf{X}_{k}$, and $\mathbf{X}_{k+1}^{l}$ is the lower $r$ bits of $\mathbf{X}_{k+1}$. 
The state transition is directed by a linear transformation 
\begin{equation}\nonumber
\mathbf{B} = 
{\begin{bmatrix}
 0 & \mathbf{I}_w & 0 & 0 & \cdots&\cdots &\cdots \\
 0 & 0 & \mathbf{I}_w & 0 &\cdots & \cdots & \cdots\\
 \vdots&  \ddots&  & &\\
  0 &  &\ddots & & \\
  \mathbf{I}_w &  & &\ddots & \\
   0 & &  & & \ddots\\
    \vdots&  &  & &  & \ddots\\
   0 & \cdots &\cdots &\cdots & 0 & \mathbf{I}_w & 0 \\
  0 & \cdots& \cdots& \cdots& 0 & 0 & \mathbf{I}_{w-r}\\
\mathbf{S} & \cdots &\cdots & \cdots& 0 & 0 & 0\\
\end{bmatrix}}
\end{equation}
on an array of size $p=nw-r$ (or an $(n\times w-r)$ array with $r$ bits missing at the upper right corner).
Here, $\mathbf{I}_j$ is a $j\times j$ identity matrix, $\mathbf{0}$ is the zero matrix and 
\begin{equation}
\mathbf{S} = 
{\begin{bmatrix}
\mathbf{0} & \mathbf{I}_r\\
\mathbf{I}_{w-r} & \mathbf{0}\\
\end{bmatrix}}\mathbf{A}
\end{equation}
The generated numbers are integers between $0$ and $2^w-1$ provided $p$ is chosen as a Mersenne exponent such that, the characteristic polynomial of $\mathbf{B}$ is primitive and period is a Mersenne prime $2^p-1 = 2^{nw-r}-1$.
A commonly used example of Mersenne Twister is \verb MT19937 \cite{Matsumoto:1998:MTE:272991.272995} ~having a period $\rho = 2^{19937}-1$. There are $32$-bit and $64$-bit word size variations of it. In case of \verb MT19937-32 ~($32$-bit), the associated parameters are $(w,n,m,r)=(32,624,397,31)$, $\mathbf{a}$=0x9908B0DF, $u=11$, $s=7$, $\mathbf{b}$=0x9D2C5680, $t=15$, $\mathbf{c}$=0xEFC60000, $l=18$ and number of terms in the characteristics polynomial is $135$. However, for \verb MT19937-64 ~($64$-bit), $(w, n, m, r) = (64, 312, 156, 31)$, $\mathbf{a}$ = 0xB5026F5AA96619E916, $u=29$, $s = 17$, $\mathbf{b}$ = 0x71D67FFFEDA6000016, $t=37$, $\mathbf{c}$ = 0xFFF7EEE00000000016 and $l = 43$. 

In \cite{Saito2008}, a variation of Mersenne Twister named single instruction multiple data (SIMD)-oriented Fast Mersenne Twister (SFMT) is proposed. It uses all features of MT along with multi-stage pipelines and Single Instruction Multiple Data (SIMD) (like 128-bit integer) operations of today's computer system. A popular implementation is \verb SFMT19937 ~which has same period as \verb MT19937. ~It can generate both $32$-bit and $64$-bit unsigned integer numbers.
Further, there is a variation of SFMT specialized in producing double precision floating point numbers in IEEE 754 format. This PRNG is, in fact, named as double precision floating point SFMT (dSFMT) \cite{Saito2009}. Two versions are available - \verb dSFMT-32 ~and \verb dSFMT-52 \cite{sfmt}. ~For \verb dSFMT-52, ~the output of the PRNG is a sequence of $52$-bit pseudo-random patterns along with $12$ MSBs (sign and exponent) as constant. Here, instead of linear transition in $\mathbb{F}_2$, an affine transition function is adopted which keeps the constant part as 0x3FF.

Another PRNG, named \emph{well-equidistributed long-period linear} generator or WELL is also based on tampered TGFSR \cite{Panneton:2006:ILG:1132973.1132974}. For this PRNG, the characteristic polynomial of matrix $\mathbf{A}$ has degree $k = rw - j$, where $r > 0$ and $0 \le j < w$, and it is primitive over $\mathbb{F}_2$. Two such good generators are \verb WELL512a ~and \verb WELL1024a \cite{RNG}. In case of \verb WELL512a, ~the parameters are $k = 512$, $w = 32$, $n = 16$ and $r = 0$; so expected period is $\rho = 2^{512}-1$. However, for \verb WELL1024a, ~the parameters are $k = 1024, w = 32, n = 32$ and $r = 0$ with period length $\rho = 2^{1024}-1$. The return values for these PRNGs are $32$-bit numbers normalized by multiplying with $2.32830643653869628906 \times 10^{-10}$. WELL generators follows the general equations \ref{lfsr1} and \ref{lfsr2}.

In \cite{marsaglia2003xorshift}, Marsagila has proposed a very fast PRNG, named \emph{xorshift} generator. The basic concept of such generators is $-$ to get a random number, first shift $a$ positions of a block of bits and then apply XOR on the original block with this shifted block. In general, a xorshift generator has the following recurrence relation \cite{brent2004note,Panneton:2005:XRN:1113316.1113319}:
\begin{equation}\label{xorshift_form}
\mathbf{v}_n = \sum_{j=1}^{t}\mathbf{\tilde{A}_j}\mathbf{v}_{n-m_j} \pmod{2} 
\end{equation}
where $t,m_j >0$, for each $n$, $\mathbf{v}_n$ is a $w$-bit vector and $\mathbf{\tilde{A}}_j$ is either $\mathbf{I}$ or product of $v_j$ xorshift matrices for $v_j\ge 0$. At step $n$, the state of the PRNG is $\mathbf{x}_n = (\mathbf{v}_{n-r+1}^T,\cdots, \mathbf{v}_{n}^T)^T$ where $\mathbf{v}_n = (v_{n,0},\cdots, v_{n,w-1})^T$ and output is $u_n=\sum_{l=1}^{w}v_{n,l-1}2^{-l}$. This generator converts into the general LFSR PRNG of equations \ref{lfsr1} and \ref{lfsr2}, if 
\begin{equation}
\mathbf{A} = 
{\begin{bmatrix}
 0 & \mathbf{I} & \cdots & 0\\
 \vdots &  & \ddots & \vdots \\
 0 & 0 & \cdots & \mathbf{I} \\
\mathbf{A}_r & \mathbf{A}_{r-1} &  \cdots & \mathbf{A}_1
\end{bmatrix}}
\end{equation}
where $k=rw$, $\mathbf{y}_n = \mathbf{v}_n$ and $\mathbf{B}$ matrix has $\mathbf{I}$ matrix of size $w \times w$ in upper left corner with zeros elsewhere. Matrix $\mathbf{A}$ has characteristic polynomial of the form \[P(z)=det(z^r\mathbf{I} + \sum_{j=1}^{r}z^{r-j}\mathbf{A}_j)\]
Therefore, the generator has maximal period length of $2^{rw}-1$, if and only if, this polynomial $P(z)$ is primitive. Marsagila's \verb xorshift32 ~generator \cite{marsaglia2003xorshift} uses $3$ xorshift operations $-$ first XOR with left shift of $13$ bits, then with right shift of $17$ bits and finally again XOR with left shift of $15$bits. Here the returned number is a $32$ bit unsigned integer. There are three other good xorshift generators -- \verb xorshift64*, ~\verb xorshift1024*M_8 ~and \verb xorshift128+. ~In \verb xorshift64* ~generator, the returned number is current state perturbed by a non-linear operation, which is multiplication by $2685821657736338717$ \cite{Vigna:2016:EEM:2956571.2845077}. Here also $3$ xorshifts are performed -- left with $12$ bits, right with $25$ bits and again left with $27$ bits. However, in \verb xorshift1024*M_8, ~the multiplier is $1181783497276652981$ and shift parameters are $31, 11$, and $30$. In both cases, the generated numbers are $64$-bit unsigned integers. In 
\verb xorshift128+, the outputs are $64$-bits and two previous output states are added to get the result \cite{VIGNA2017175}. 

Although these generators are linear, many researchers have developed LFSR based PRNGs by combining these with some non-linear operations \cite{l2003combined,Vigna:2016:EEM:2956571.2845077,L'Ecuyer:2007:TCL:1268776.1268777} to scramble the regularity of linear recurrence. For example, in \cite{l2003combined}, two component combined generators are proposed, where the major component is linear (LFSR or LCG), but the second component is distinct (nonlinear or linear). Whereas, in \cite{Vigna:2016:EEM:2956571.2845077}, to remove the flaws of xorshift generators, a non-linear operation is applied to scramble the results. The \emph{xorshift*} PRNGs, discussed above, are a well-known example.
\subsection{Cellular Automata based PRNGs}
\label{chap:randomness_survey:sec:cas}
We have discussed the potential of CAs in designing PRNGs, and a brief survey has been reported in Section~\ref{Chap:surveyOfCA:sec:randomness} of Chapter~\ref{Chap:surveyOfCA} (Page~\pageref{Chap:surveyOfCA:sec:randomness}). For the sake of completeness, we mention here the well-known CAs which have been used to design PRNGs. 
\begin{description}[topsep=1pt,itemsep=1ex]
	\item[$\bullet$ Rule $30$ CA \cite{wolfram86c}:] In \cite{wolfram86c}, the elementary CA rule $30$ is used to generate random sequence. According to the original proposal, only a single bit is collected from each configuration of the CA. To use it as a PRNG, one may run the CA under periodic boundary condition, and collect $32$ bits from $32$ consecutive configurations to report a single $32$-bit integer.
	
	\item[$\bullet$ Hybrid $30-45$ CA \cite{Horte89a}:] This PRNG uses ECAs rules $30$ and $45$ in alternate positions with periodic boundary condition to generate $32$-bit numbers.
	
	\item[$\bullet$ Maximal-length CA \cite{Horte89a}:] Many of the CAs based PRNGs are maximal length CAs, which are non-uniform (see Section~\ref{Chap:surveyOfCA:scn_eca}, Page~\pageref{Chap:surveyOfCA:scn_eca}) and linear. For example, the maximal length CA with rule vector (see Definition~\ref{Def:RuleVector} of Page~\pageref{Def:RuleVector}) $\mathcal{R}=(90,150,90,90,90,150,150,90,90,90,90,90,150,90,90$,$150$,\\$150,90,150,150,150,90,150,150,150,150,90,150,90,150,90,150)$ under null boundary condition is used in \cite{Horte89a}. In this paper, concept of site spacing (output number is collected from cells spaced by $\gamma$ distance) and time spacing (output numbers are taken $\alpha$ configurations apart) are introduced. However, in this dissertation, we have considered two types of site spacing for the PRNG -- $1$ site spacing (that is, $\gamma=1$) and no site spacing ($\gamma=0$). For $1$ site spacing, two consecutive output sequences are concatenated to get one $32$-bit number, whereas, for no site spacing, the whole configuration of the CA at each time instant is treated as a $32$-bit number.
	
	\item[$\bullet$ Non-linear $2$-state CA \cite{tcad/DasS10}:] Non-linear binary CAs are also used to design PRNGs. As described in \cite{tcad/DasS10}, using the given algorithm, a $45$-cell null boundary $2$-state $3$-neighborhood non-uniform CA is generated. For example, one such CA has rule vector $\mathcal{R}=(5,105,90,90,165,150,90,105,\\150,105,90,165,150,150,165,90,165,90,165,150,150,90,165,105,90,165,\\150,90,105,150,165,90,105,105,90,150,90,90,165,150,150,105,90,165,\\20)$. Its output is a $45$-bit number.
	\end{description}

To conclude this short survey, we list out the good PRNGs in Table~\ref{tab:PRNG_list}, which we have already discussed in this section. The proposed CA $\mathbf{\mathscr{R}}$ based PRNG has to compete with these PRNGs. So, we next rank the existing PRNGs with respect to randomness quality.

\begin{table}[!h]
\setlength{\tabcolsep}{1.5pt}
\scriptsize
\renewcommand{\arraystretch}{1.20}
\centering
\caption{List of $28$ good PRNGs}\label{tab:PRNG_list}
 \begin{tabular}{|c|p{6.0cm}|}
 \hline
\theadfont{Class of PRNGs} & \theadfont{Name of the PRNGs}\\
\hline
\multirow{4}{*}{LCGs} & \verb minstd_rand, ~Borland LCG,\\
&  Knuth's LCG \verb MMIX, \\
& \verb rand, ~\verb lrand48, \\
& \verb MRG31k3p, ~\verb PCG-32 \\
\hline
\multirow{8}{*}{LFSRs} & \verb random, ~\verb Tauss88, \\
& \verb LFSR113, ~\verb LFSR258, \\
& \verb WELL512a, ~\verb WELL1024a, \\
& \verb MT19937-32, ~\verb MT19937-64, \\
& \verb SFMT19937-32, ~\verb SFMT19937-64, \\
& \verb dSFMT19937-32, ~\verb dSFMT19937-52, \\
& \verb xorshift32, ~\verb xorshift64*, ~\verb xorshift1024*, \\ & \verb xorshift128+ \\
\hline
\multirow{3}{*}{CAs} & Rule $30$ with CA size $101$, Hybrid CA with Rules $30$ \& $45$, Maximal Length CA with $\gamma=0$\\ 
&   and $\gamma=1$, Non-linear $2$-state CA\\
\hline
 \end{tabular}
 \end{table}

\section{Who is Better than Whom: An Empirical Study}\label{chap:randomness_survey:sec:facts}
A usual claim of a PRNG is, it is better than others! In this section, we verify which are really better than others with respect to randomness quality. To do so, we choose the PRNGs of Table~\ref{tab:PRNG_list} and apply empirical tests of Section~\ref{chap:randomness_survey:sec:empirical}. For all these PRNGs, we have collected the standard implementation and used them to generate numbers.


To maintain uniformity in testing, we have tested the stream of binary numbers generated in sequence by the PRNGs. For each PRNG, binary (\emph{.bin}) files are produced which contain sequence of numbers (in binary form) without any gap between two consecutive numbers in the sequence. 
If the generated numbers are normalized, then, first we convert fractional part of each number into its binary equivalent and then add these bits to the .bin file.

\subsection{Choice of Seeds}
Although a good PRNG should be independent of seeds, but to run a PRNG we need to choose seeds.
So, to compare all PRNGs on same platform, the following greedy approach to select seed is taken:
\begin{enumerate}[leftmargin=1pt]
\item As we have collected the $C$ programs of the PRNGs from their respective websites, each of them has an available seed for a test run. For example, for \verb MT19937, ~the seed was $19650218$. Nevertheless, in most of the cases, this seed is $1234$ or $12345$. We, therefore, have collected all the seeds hard-coded in the $C$ programs of all PRNGs, and used these as the set of seeds for each PRNG. These seeds are $7$, $1234$, $12345$, $19650218$ and $123456789123456789$. We have tested each PRNG with all these seeds.

\item Apart from studying the behavior of PRNGs for fixed seeds, we also want to observe the average case behavior of the PRNGs. For this reason, we have chosen a simple LCG, \verb rand ~to generate seeds for all other PRNGs. This \verb rand ~is initialized with \verb srand(0). ~The next $1000$ numbers of \verb rand ~ are supplied as seeds to each PRNG to test it $1000$ times with these random seeds.
\end{enumerate}

All PRNGs are tested empirically for each of these seeds and the results are compared impartially. Some PRNGs, such as \verb LFSR113, ~\verb LFSR258, ~etc. require more than one (non-zero) seed to initialize its components. However, we supply here the same seed to all its components.

\subsection{Results of Empirical Tests}\label{chap:randomness_survey:sec:empirical_result}
The selected PRNGs (LCG-based, LFSR-based and CA-based) are tested with statistical tests as well as with the graphical tests. Here, summary of the results of these tests are documented. 

\subsubsection{Results of Statistical Tests} Table~\ref{tab:blind_test} shows the summary of results of Diehard battery of tests, battery \emph{rabbit} of TestU01 library and NIST statistical test suite for the fixed seeds. In this table, for each PRNG, result (in terms of numbers of tests passed) of the testbeds per each seed is recorded. 

\begin{table}[!h]
\setlength{\tabcolsep}{1.3pt}
   \renewcommand{\arraystretch}{1.30}
     \centering
   \small
    \caption{Summary of statistical test results for different fixed seeds}
   \label{tab:blind_test}
     \resizebox{1.00\textwidth}{!}{
   \begin{tabular}{|c|c|c|c|c|c|c|c|c|c|c|c|c|c|c|c|c|c|}
   \hline
\multicolumn{2}{|r|}{\theadfont{Seeds $\longrightarrow$}} & \multicolumn{3}{c|}{\thead{$7$}} &  \multicolumn{3}{c|}{\thead{$1234$}} &   \multicolumn{3}{c|}{\thead{$12345$}} &  \multicolumn{3}{c|}{\thead{$19650218$}} &  \multicolumn{3}{c|}{\thead{$123456789123456789$}} & \thead{Ranking}\\
 \cline{1-17}
 \multicolumn{2}{|c|}{\theadfont{ Name of the PRNGs}} &  Diehard & TestU01 & NIST & Diehard & TestU01 & NIST & Diehard & TestU01 & NIST & Diehard & TestU01 & NIST & Diehard & TestU01 & NIST & (First Level)\\
\hline
\multirow{7}{*}{\rotatebox{90}{LCGs}}& MMIX & 6 & 19 & 7 & 5 & 18 & 7 & 6 & 17 & 8 & 4 & 16 & 8 & 5 & 18 & 8 & 8\\
\cline{2-18}
& minstd\_rand & 0 & 1 & 1 & 0 & 1 & 1 & 0 & 1 & 2  & 0 & 1 & 1 & 0 & 2 & 1 & 12\\
\cline{2-18}
& Borland LCG & 1 & 3 &5 & 0 & 3 & 5 & 1 & 3 & 5 & 1 & 3 & 4 & 1 & 3 & 5 & 11\\
\cline{2-18}
& rand & 1 &1 &2 & 1 & 1& 2& 1 & 3 & 2& 1 & 2 & 2& 1 & 2 & 2 & 11\\
\cline{2-18}
& lrand48() & 1 & 3 & 2& 1 & 2 & 2 & 1 & 3 & 2 & 1 & 3 & 2 & 1 & 2 & 2 & 11\\
\cline{2-18}
& MRG31k3p & 1 & 2 & 1 & 1 & 1 & 1 & 1 & 2 & 1 & 1 & 1 & 1 & 0 & 0 & 2 & 12\\
\cline{2-18}
& PCG-32 & 9 & 25 & 15& 9 & 25 &14 & 11 & 25  &14 & 10 & 24 &15 & 9 & 25 & 15 & 2\\
\cline{2-18}
\hline
\multirow{16}{*}{\rotatebox{90}{LFSRs}}& random() & 1 & 1 & 1 & 1& 1 & 1 & 1 & 3 & 1 & 1 & 2 & 1 & 1 & 2 & 1 & 11\\
\cline{2-18}
& Tauss88 & 11 & 21 & 15 & 9 & 23 & 15 & 11 & 23 & 15 & 11 & 23 & 14 & 10 & 23 & 15 & 4\\
\cline{2-18}
& LFSR113 & 5 & 6 & 1& 11 & 23 & 14& 9 & 23 & 15 & 7 & 23 & 14 & 9 & 23 & 15 & 7\\
\cline{2-18}
& LFSR258 & 0 & 0 & 1& 0 & 5 & 2 & 1 & 5 & 2 & 1 & 5 & 2& 1 & 5 & 0 & 12\\
\cline{2-18}
& WELL512a & 9 & 23 &15 & 10 & 23 & 14 & 10 & 23 & 15 & 8 & 23 & 15 & 7 & 23 & 15 & 5\\
\cline{2-18}
& WELL1024a & 9 & 25 & 15 & 10 & 24 & 15 & 9 & 24& 14& 9 & 25 & 15 & 9 & 25 &15 & 3\\
\cline{2-18}
& MT19937-32 & 10 & 25 & 13& 9 & 25 & 13& 9 & 25 &14 & 9 & 25 &15 & 9 & 25 &15 & 3\\
\cline{2-18}
& MT19937-64 & 10 & 25 &15 & 10 & 24 & 15& 8 & 24 & 15& 11 & 25 &15 & 10 & 25 & 15 & 2\\
\cline{2-18}
& SFMT19937-32 & 10 & 25 & 15& 9 & 25 & 15& 10 & 25 &15 & 9 & 25 & 15& 10 & 25 & 15 & 1\\
\cline{2-18}
& SFMT19937-64 & 11 & 25 & 15 & 10 & 25 &15 & 10 & 25 & 15& 9 & 25 &15 & 10 & 25 & 15 & 1\\
\cline{2-18}
& dSFMT-32 & 7 & 25 &15 & 8 & 25 & 15& 11 & 24 & 13& 11 & 25 &15 & 10 & 24 & 15 & 5\\
\cline{2-18}
& dSFMT-52 & 5 & 11 & 3& 5 & 10 & 3& 7 & 11 & 3 & 6 & 10 & 3 & 7 & 9 & 3 & 9\\
\cline{2-18}
&  xorshift32 & 4 & 17 & 4& 4 & 17 & 4& 4 & 17 &2 & 0 & 17 &13 & 4 & 17 & 13 & 9\\
\cline{2-18}
&  xorshift64* & 10 & 25 & 15 & 10 & 25 & 15& 8 & 25 & 15& 7 & 25 & 15 & 8 & 25 & 14 & 5\\
\cline{2-18}
&  xorshift1024* & 7 & 20 & 6& 9 & 21 & 15& 7 & 20 &15 & 8 & 20 &15 & 6 & 21 & 15 & 6\\
\cline{2-18}
&  xorshift128+ & 9 & 25 & 14& 9 & 25 & 14& 10 & 24 &15 & 10 & 25 &15 & 8 & 24 & 15 & 4\\
\hline
\multirow{5}{*}{\rotatebox{90}{CAs}}& Rule $30$ & 11& 25& 15& 10&25 & 15& 9 & 25& 15 & 8 & 25 & 15 & 11 & 24 & 15 & 2\\
\cline{2-18}
& Hybrid CA with Rules $30$ \& $45$ & 3 & 8 & 3& 0& 1 & 0 & 2  & 8 & 1 & 1 & 7 & 2 & 1 & 8& 2 & 11\\
\cline{2-18}
& Maximal Length CA with $\gamma=0$ & 2 & 12 & 10& 0& 12 & 11 & 1 & 12& 11& 1& 12& 11 & 2& 12 & 10 & 10\\
\cline{2-18}
& Maximal Length CA with $\gamma=1$ & 4& 17& 14 & 3 & 16 & 14 & 3  & 17 & 14 & 4 & 15 & 14 & 3 & 16 & 14 & 8\\
\cline{2-18}
& Non-linear $2$-state CA & 6 & 11& 4& 7& 11 & 2& 5 & 11 & 3 & 6& 11& 4& 5& 12& 4 & 9\\
\hline
 \end{tabular}}
 	\vspace{-0.5em}
 \end{table}  

It is observed that, none of the PRNGs can pass all tests of these blind testbeds. We also notice the following:
\begin{itemize}[leftmargin=1pt]
\item Among all, the LCGs \verb minstd_rand, ~Borland's LCG, \verb rand, ~\verb lrand48, ~\verb MRG31k3p \\~and the LFSRs \verb LFSR258 ~and \verb random ~perform very poorly. In case of diehard tests, these PRNGs can pass at most the runs test, except \verb LFSR258, ~which passes the rank test of $31 \times 31$ and $31 \times 32$ matrices and runs down test.

\item The remaining two LCGs based PRNGs, namely Knuth's \verb MMIX ~and \verb PCG-32 ~behave well in comparison to the other LCGs as well as many of the LFSRs. For example, Knuth's \verb MMIX ~is better than \verb LFSR258 ~and \verb xorshift32, ~whereas \verb PCG-32 ~is better than \verb MMIX ~as well as \verb LFSR113, ~Xorshift PRNGs, WELL PRNGs and dSFMTs. In fact, performance of \verb PCG-32 ~is comparable to MTs and SFMTs.

\item Performance of \verb LFSR113 ~is dependent on seed; whereas, \verb WELL512a ~and \verb WELL1024a ~are invariant of seeds.

\item Among the Mersenne Twister and its variants, performance of dSFMTs (especially, \verb dSFMT-52), ~are unexpectedly poor in terms of the blind empirical tests. 

\item Among the CA-based PRNGs, rule $30$ can compete with the elite group of PRNGs like Mersenne Twisters, WELL and PCG. However, other CA-based PRNGs are not so good. For those CAs, unlike ECA rule $30$, the complete (or, a block from a) configuration of the CA is taken as a number. Among these CAs based PRNGs, performance of the max-length CA ($\gamma=1$) is better than the non-linear $2$-state CA, which is better than the max-length CA ($\gamma=0$) and hybrid CA rule $30-45$.

\end{itemize}

Therefore, we can define first level ranking of the PRNGs from these test results for fixed seeds. Last column of Table~\ref{tab:blind_test} shows this ranking. For this ranking, we have considered the overall tests passed in each test-suite. If two PRNGs gives similar results, they have the same rank. It can be observed that, among the LCG-based PRNGs, \verb PCG-32 ~gives the best result and among the LFSR-based PRNGs, performances of SFMTs ($32$ and $64$ bits) are the best. Moreover, among the CA-based PRNGs, Wolfram's rule $30$ gives best result, which is comparable to the results of SFMTs and MTs.
\begin{itemize}[leftmargin=1pt]
\item As SFMTs are best performers, these are ranked as $1$. Similarly, performance of \verb MT19937-64, ~\verb PCG-32 ~and rule $30$ are comparable (ranked $2$), performance of \verb MT19937-32 ~and \verb WELL1024a ~are comparable (ranked $3$), whereas performance of \verb xorshift128+ ~and \verb Tauss88 ~are comparable (ranked $4$). These are the elite group of best performing PRNGs.

\item \verb xorshift64* ~performs well, but it has more dependency on seed than the MTs and SFMTs. For this reason we rank it lower than MT and SFMT. Similarly, \verb LFSR113 ~can perform better than WELLs and non-linear $2$-state CA-based PRNG for some seeds, but as performance of WELL is more invariant of seed, it has higher rank than \verb LFSR113.    ~

\item As performance of \verb WELL512a, ~\verb Xorshift64* ~and \verb dSFMT-32 ~bit are comparable, they are ranked as $5$. In terms of performance in NIST test-suite, \verb xorshift1024* ~(rank $6$) is better than \verb LFSR113 ~(rank $7$), but worse than \verb dSFMT-32. ~

\item Among the other PRNGs, maximal-length CA with $\gamma=1$ and \verb MMIX ~are ranked $8$, the non-linear $2$-state CA based PRNG, \verb dSFMT-52 ~and \verb xorshift32 ~are ranked $9$ and maximal-length CA with $\gamma=0$ is ranked $10$.

\item Rest of the PRNGs form two groups -- \verb rand, ~\verb lrand48, ~Borland's LCG, \verb random ~and rule $30-45$ as rank $11$ and \verb minstd_rand, ~\verb MRG31k3p ~and \verb LFSR258 ~as rank $12$. 
\end{itemize}

\begin{figure}[!h] 
            \centering
            \vspace{-1.0em}
        \subfloat[SFMT19937-64\label{sfmt64_avg}]{%
        	\includegraphics[width=0.3\linewidth, height=3.0cm]{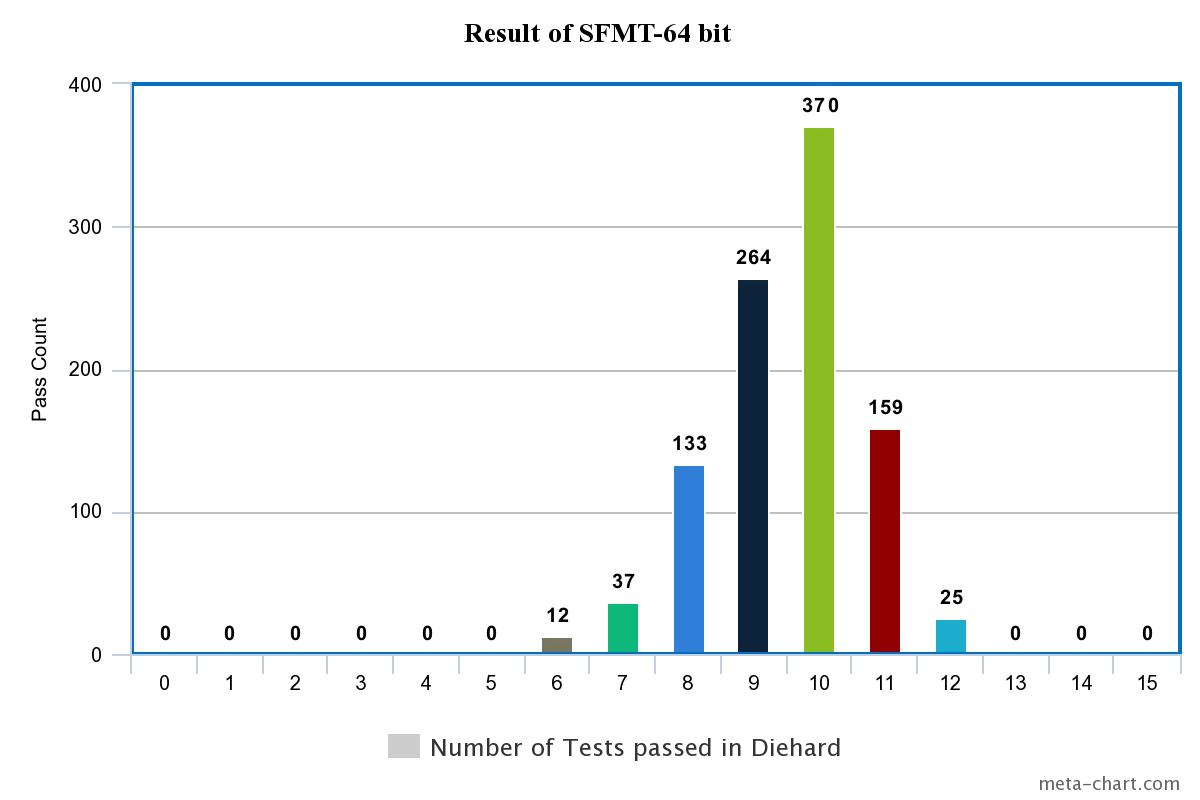}}
        		\hfill
        \subfloat[dSFMT19937-32\label{dsfmt32_avg}]{%
        \includegraphics[width=0.3\linewidth, height=3.0cm]{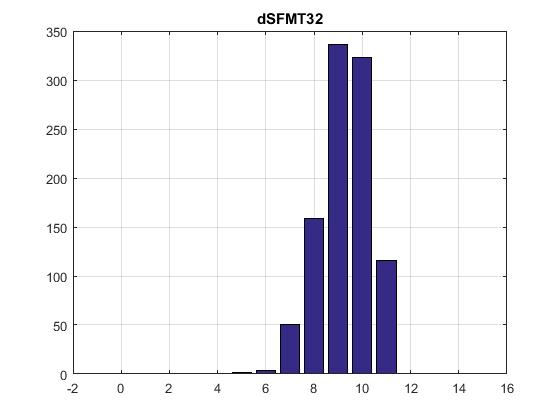}}
        	\hfill
        \subfloat[dSFMT19937-52\label{dsfmt64_avg}]{%
        \includegraphics[width=0.3\linewidth, height=3.0cm]{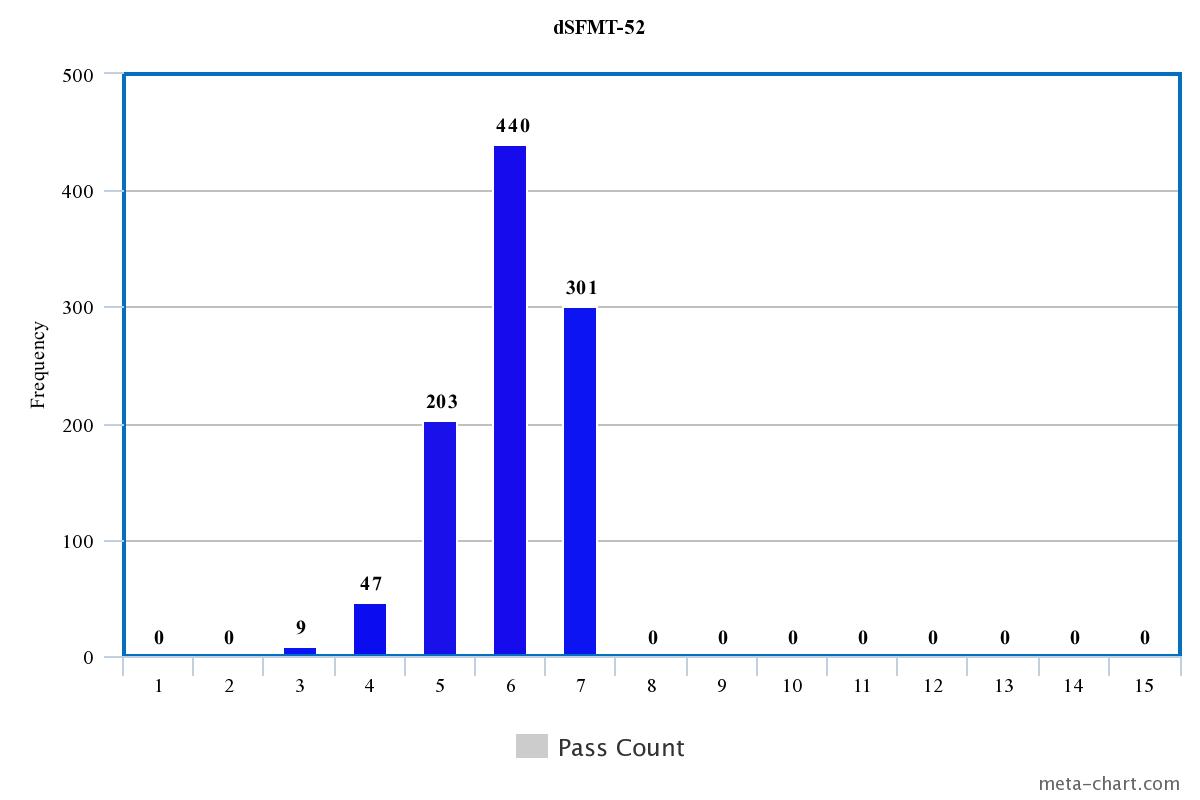}}
        	\hfill
        \subfloat[ xorshift32\label{xor32_avg}]{%
        \includegraphics[width=0.3\linewidth, height=3.0cm]{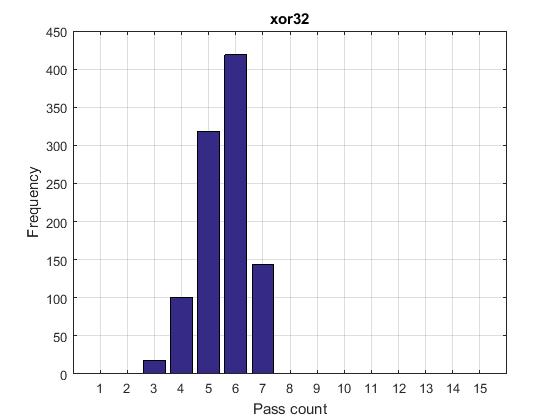}}
       	\hfill
        \subfloat[ xorshift64*\label{xor64_avg}]{%
        \includegraphics[width=0.3\linewidth, height=3.0cm]{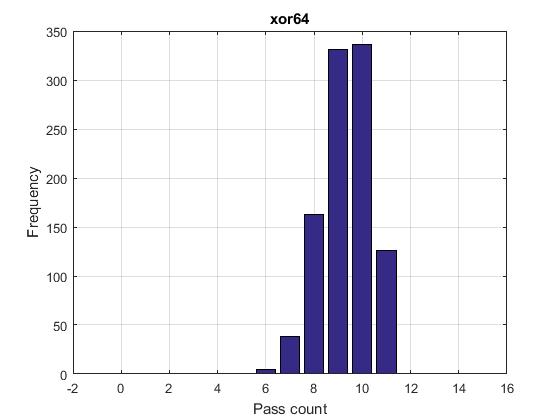}}
        \hfill
    \subfloat[ xorshift1024*\label{xor1024_avg}]{%
          \includegraphics[width=0.3\linewidth, height=3.0cm]{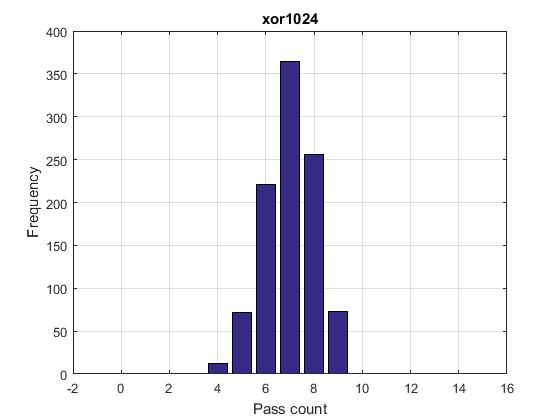}}
          	\hfill
     \subfloat[ xorshift128+\label{xor128}]{%
          \includegraphics[width=0.3\linewidth, height=3.0cm]{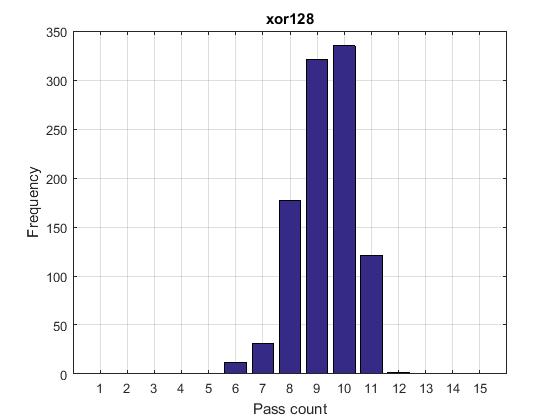}}
          \hfill
     \subfloat[Rule30\label{rule30_avg}]{%
           \includegraphics[width=0.3\linewidth, height=3.0cm]{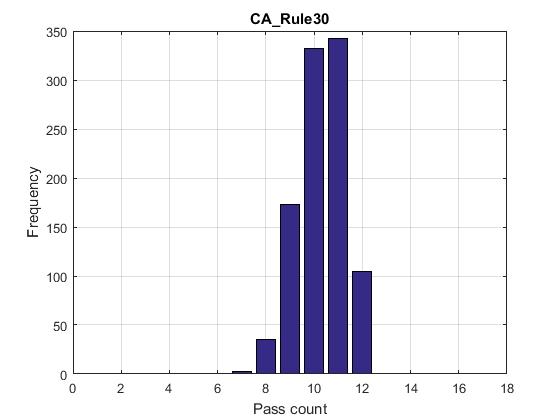}}
           	\hfill
    \subfloat[Rule30-45\label{rule30-45_avg}]{%
           \includegraphics[width=0.3\linewidth, height=3.0cm]{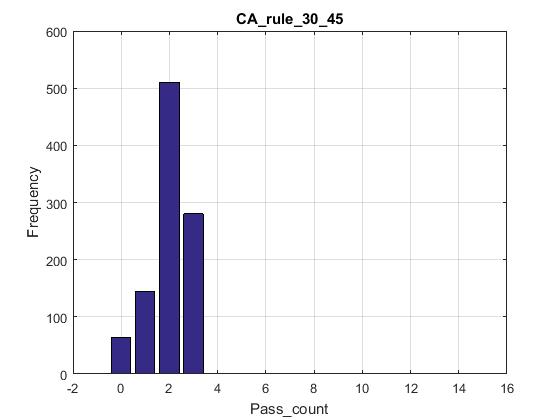}
 }\hfill
    \subfloat[max-length CA with $\gamma =0$\label{maxlength_y=0_avg}]{%
             \includegraphics[width=0.3\linewidth, height=3.0cm]{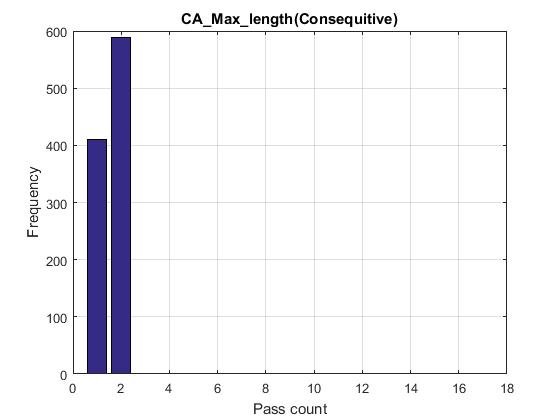}}
             	\hfill
    \subfloat[max-length CA with $\gamma =1$\label{maxlength_y=1_avg}]{%
             \includegraphics[width=0.3\linewidth, height=3.0cm]{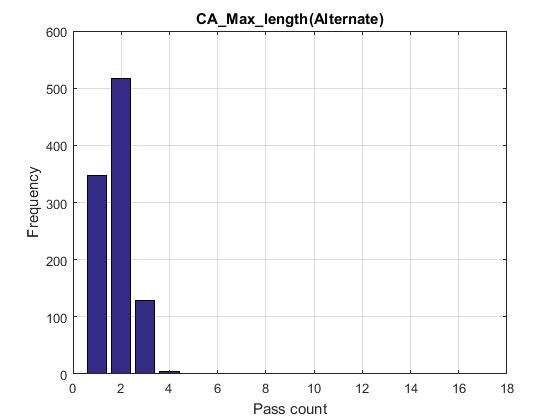}}
             \hfill
     \subfloat[non-linear $2$-state CA\label{nonlinear_avg}]{%
              \includegraphics[width=0.3\linewidth, height=3.0cm]{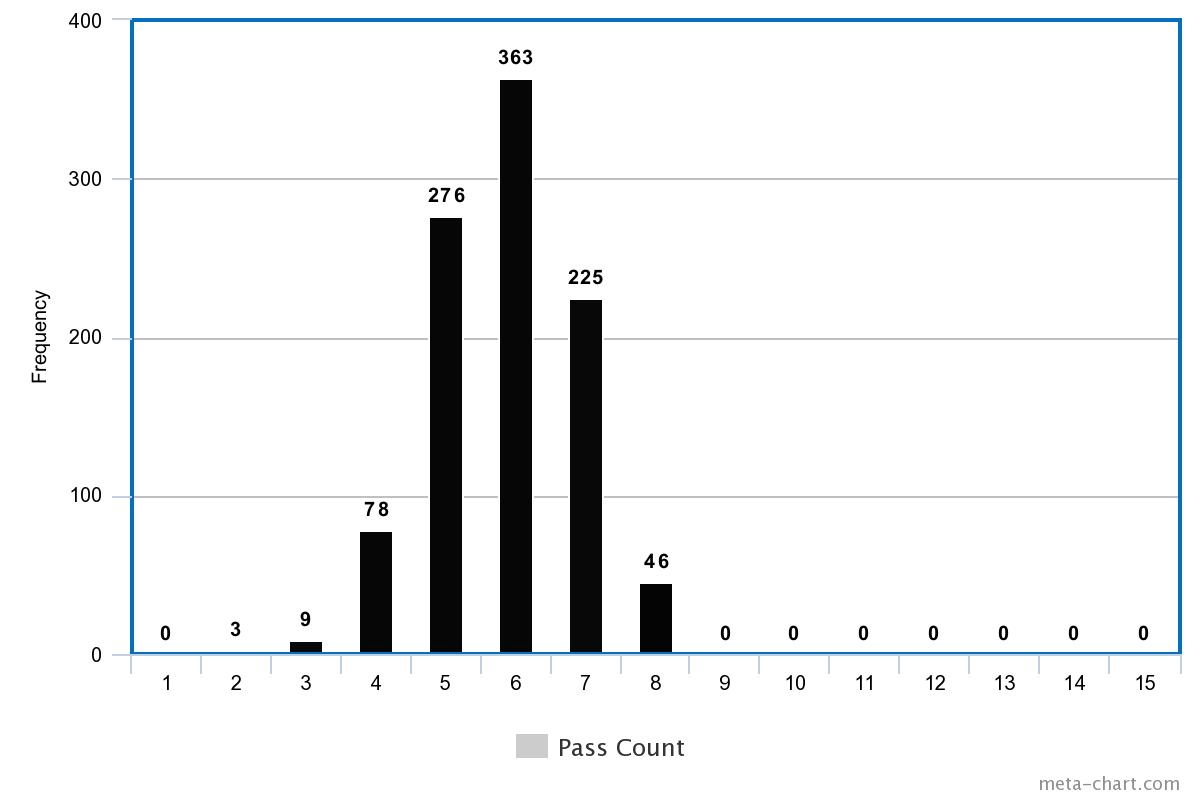}}
        \caption{Average test results of PRNGs for $1000$ seeds with Diehard battery of tests (continued)}
          \label{fig:avg_plot3}
          \end{figure}

In this ranking, however, there are many ties. To improve the ranking, we next use $1000$ seeds, as mentioned before, and observe the average performance of PRNGs in Diehard battery.
      \begin{figure}[!h]
        \centering
        \vspace{-1.0em}
    \subfloat[MMIX\label{mmix_avg}]{%
     \includegraphics[width=0.3\linewidth, height=3.0cm]{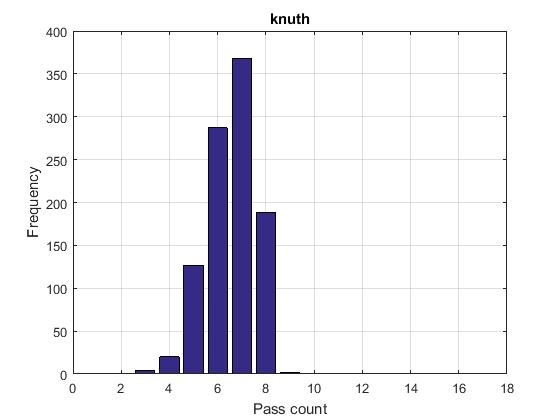}}
     \hfill
      \subfloat[minstd\_rand\label{minstd_avg}]{%
        \includegraphics[width=0.3\linewidth, height=3.0cm]{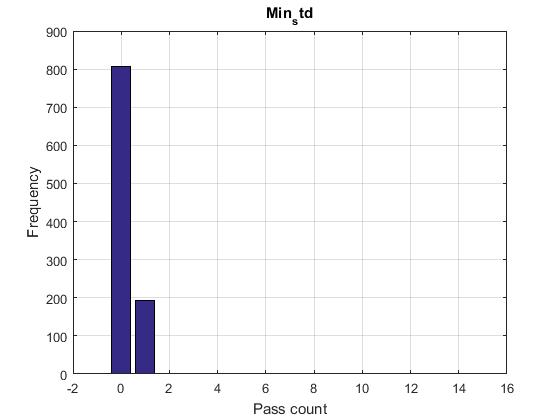}}
        \hfill
      \subfloat[Borland's LCG\label{borland_avg}]{%
 \includegraphics[width=0.3\linewidth, height=3.0cm]{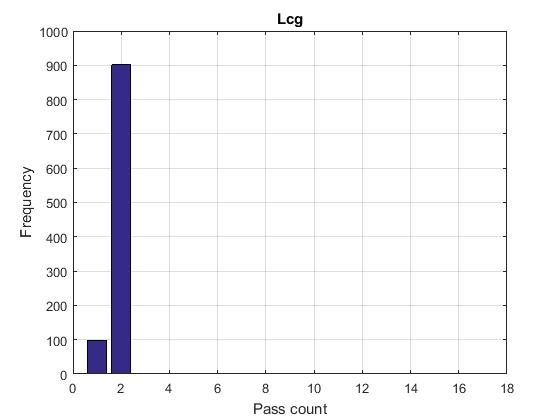}}
        \hfill  
         \subfloat[lrand48()\label{lrand_avg}]{%
           \includegraphics[width=0.3\linewidth, height=3.0cm]{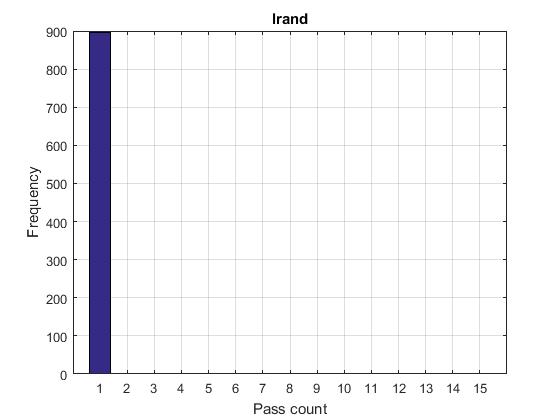}}
           \hfill
        \subfloat[MRG31k3p\label{mrg_avg}]{%
          \includegraphics[width=0.3\linewidth, height=3.0cm]{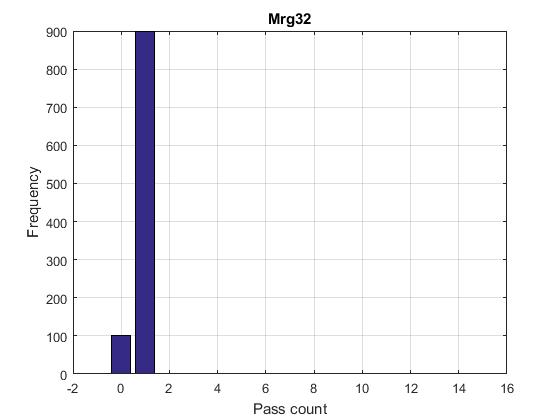}
 }
          \hfill
     \subfloat[PCG-32\label{pcg_avg}]{%
       \includegraphics[width=0.3\linewidth, height=3.0cm]{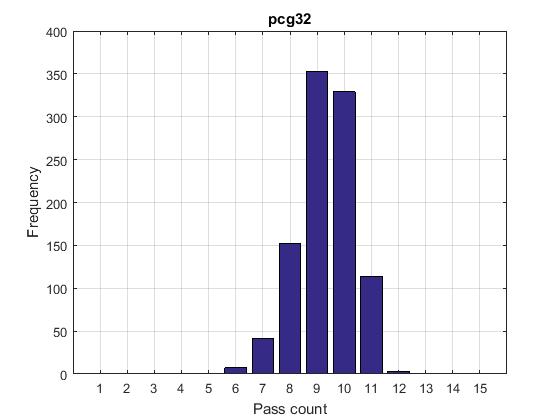}}
       \hfill
        \subfloat[random\label{random_avg}]{%
          \includegraphics[width=0.3\linewidth, height=3.0cm]{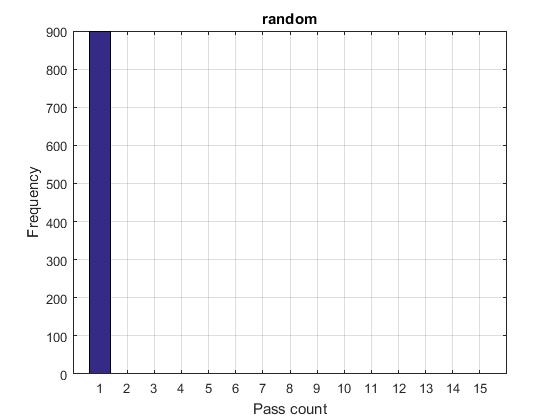}}
          \hfill
       \subfloat[Tauss88\label{tauss_avg}]{%
          \includegraphics[width=0.3\linewidth, height=3.0cm]{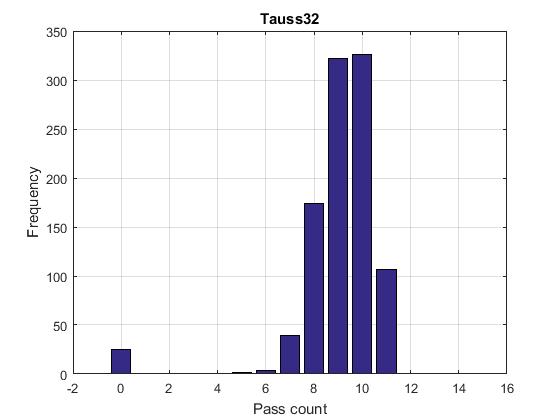}
          }
          \hfill  
           \subfloat[LFSR113\label{lfsr113_avg}]{%
             \includegraphics[width=0.3\linewidth, height=3.0cm]{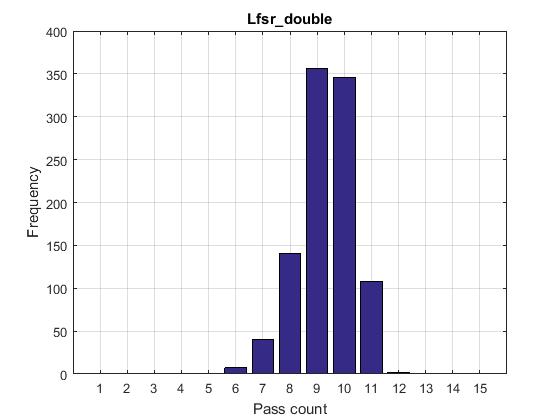}}
             \hfill
          \subfloat[LFSR258\label{lfsr258_avg}]{%
            \includegraphics[width=0.3\linewidth, height=3.0cm]{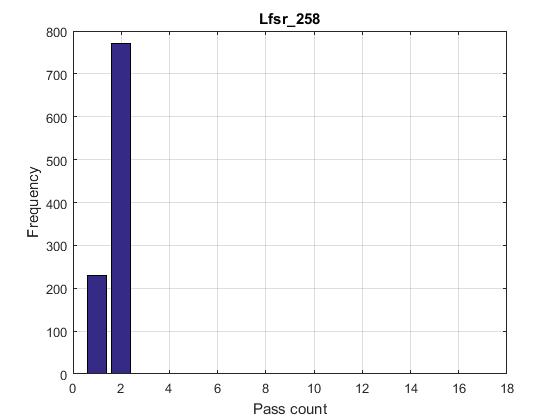}}
            \hfill  
         \subfloat[WELL512a\label{well512_avg}]{%
           	\includegraphics[width=0.3\linewidth, height=3.0cm]{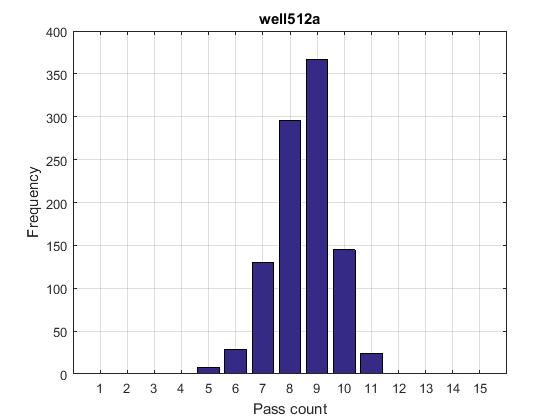}}
           	\hfill
           \subfloat[WELL1024a\label{well1024_avg}]{%
           \includegraphics[width=0.3\linewidth, height=3.0cm]{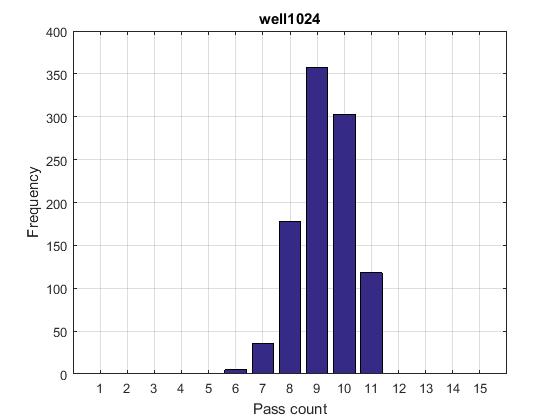}}
           	\hfill
           \subfloat[MT19937-32\label{mt32_avg}]{%
           \includegraphics[width=0.3\linewidth, height=3.0cm]{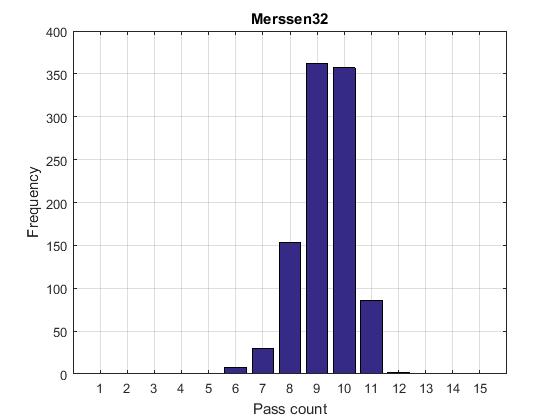}}
           	\hfill
           \subfloat[MT19937-64\label{mt64_avg}]{%
           \includegraphics[width=0.3\linewidth, height=3.0cm]{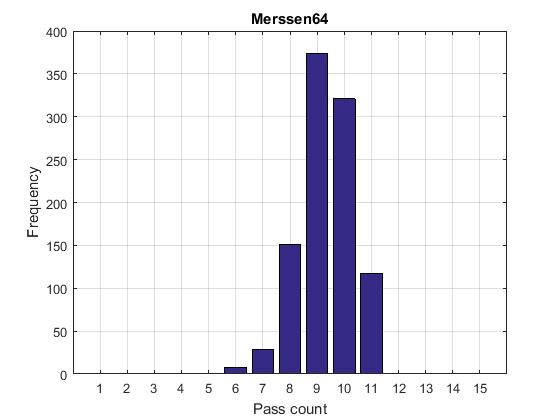}}
           	\hfill
           \subfloat[SFMT19937-32\label{sfmt32_avg}]{%
           \includegraphics[width=0.3\linewidth, height=3.0cm]{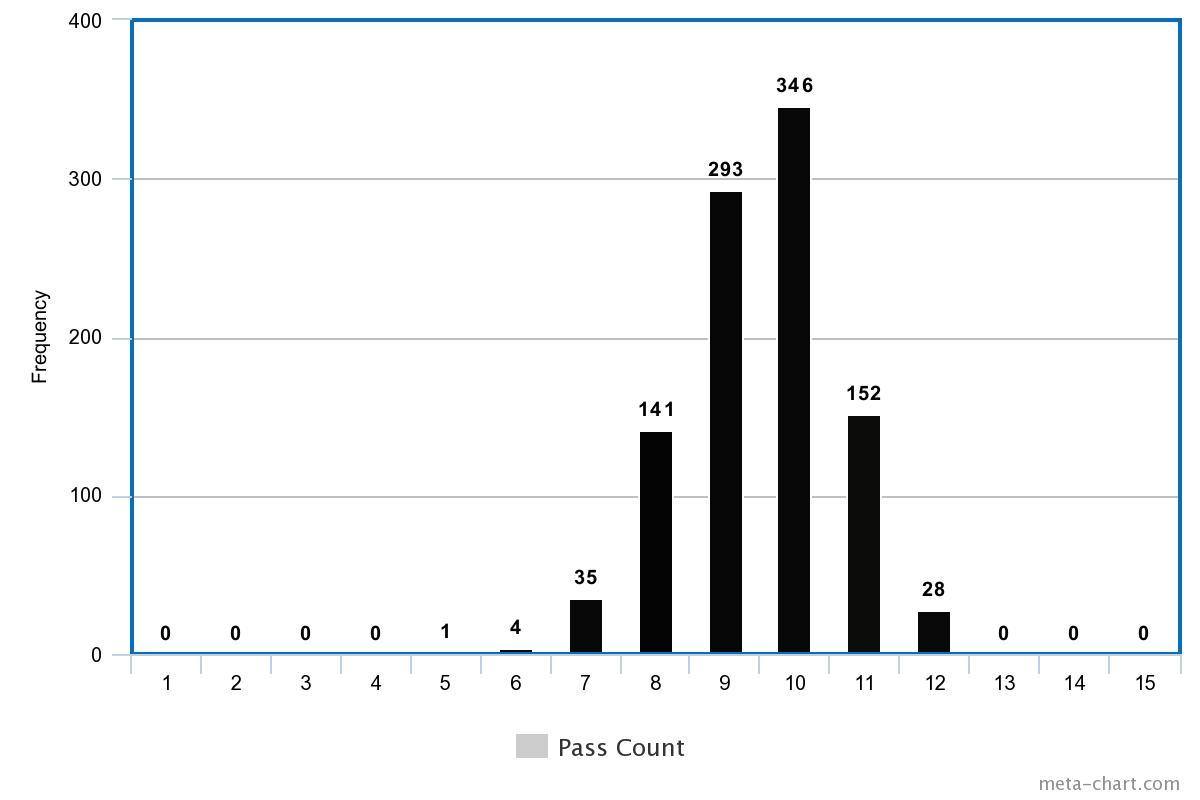}}
 \caption{Average test results of PRNGs for $1000$ seeds with Diehard battery of tests}
          \label{fig:avg_plot2}
          \end{figure}
Figure~\ref{fig:avg_plot2} and  Figure~\ref{fig:avg_plot3} show the plots for all the PRNGs (except \verb rand, ~as \verb rand ~is used to generate the seeds). For each of the figures, $x$ axis represents the number of tests passed by a PRNG and $y$ axis denotes the frequency of passing these tests. By using these plots, we can get a second level of ranking of the PRNGs, as shown in Table~\ref{tab:blind_test_avg}.   
We find average performance by calculating $\frac{\sum\limits_i f_i \times t_i}{1000}$, where $f_i$ is the frequency of a PRNG to pass $t_i$ number of tests for $i^{th}$ seed. The column `Range' in Table~\ref{tab:blind_test_avg} indicates minimum and maximum number of tests passed in the whole experiment for a particular PRNG. The new ranks are discussed below.
\begin{table*}[hbtp]
		\vspace{-0.8em}
\setlength{\tabcolsep}{1.5pt}
   \scriptsize
   \renewcommand{\arraystretch}{1.20}
     \centering
   \small
   \caption{Summary of Statistical test results for different seeds}
     \label{tab:blind_test_avg}
     \resizebox{0.90\textwidth}{5cm}{
   \begin{tabular}{|c|c|c|c|c|c|c|c|c|}
   \hline
   \multicolumn{2}{|c|}{\multirow{2}{*}{\theadfont{Name of the PRNGs}}} & \multicolumn{3}{c|}{\theadfont{Fixed Seeds}} &  \multicolumn{2}{c|}{\theadfont{Random Seeds}} & \multirow{2}{*}{\theadfont{Previous Rank}} & \multirow{2}{*}{\theadfont{$2^{nd}$ level Rank}} \\
   \cline{3-7}
 \multicolumn{2}{|c|}{ } & Diehard & TestU01 & NIST & Average & Range &  & \\
\hline
\multirow{7}{*}{\rotatebox{90}{LCGs}}& MMIX & 4-6 & 16-19 & 7-8 & 6.5 & 2-9  & 8 & 9\\
\cline{2-9}
& minstd\_rand & 0 & 1 & 1-2 &0.38 & 0-1 & 12 & 14\\
\cline{2-9}
& Borland LCG & 1 & 3 & 4-5 & 1.9 & 1-2 & 11 & 12\\
\cline{2-9}
& rand & 1 & 1-3 & 2-3 &  &  & 11 & 13\\
\cline{2-9}
& lrand48() & 1 & 2-3 & 2 & 1 & 1 & 11 & 13\\
\cline{2-9}
& MRG31k3p & 0-1 & 1-2 & 1-2 & 0.9 & 0-1 & 12 & 14\\
\cline{2-9}
& PCG-32 & 9-11 & 24-25 & 14-15 & 9.3 & 6-12 & 2 & 4\\
\cline{2-9}
\hline
\multirow{16}{*}{\rotatebox{90}{LFSRs}}& random() & 1 & 1-3 & 1 & 1 & 1 & 11 & 13\\
\cline{2-9}
& Tauss88 & 9-11 & 21-23 & 14-15 & 9.0 & 0-12 & 4 & 7\\
\cline{2-9}
& LFSR113 & 5-11 & 6-23 & 1-15 & 9.3 & 6-12 & 7 & 7\\
\cline{2-9}
& LFSR258 & 0-1 & 0-5 & 0-2 &  1.8 & 1-2 & 12 & 14\\
\cline{2-9}
& WELL512a & 7-10 & 23 & 14-15 & 8.5 & 5-11 & 5 & 6\\
\cline{2-9}
& WELL1024a & 9-10 & 24-25 & 14-15 & 9.2 & 6-11 & 3 & 4\\
\cline{2-9}
& MT19937-32 & 9-10 & 25 & 13-15 & 9.3 & 6-12 & 3 & 4\\
\cline{2-9}
& MT19937-64 & 8-11 & 24-25 & 15 & 9.4 & 6-11 & 2 & 3\\
\cline{2-9}
& SFMT19937-32 & 9-10 & 25 & 15 & 9.5 & 5-12 & 1 & 1\\
\cline{2-9}
& SFMT19937-64 & 9-11 & 25 & 15 & 9.52 & 6-12 & 1 & 1\\
\cline{2-9}
& dSFMT-32 & 7-11 & 24-25 & 13-15 & 9.3 & 5-11 & 5 & 5\\
\cline{2-9}
& dSFMT-52 & 5-7 & 9-11 & 3 & 5.98 & 3-7 & 9 & 10\\
\cline{2-9}
&  xorshift32 & 2-4 & 17 & 2-13 & 5.5 & 3-7 & 9 & 10\\
\cline{2-9}
&  xorshift64* & 7-10 & 25 & 14-15 & 8.0 & 6-11 & 5 & 6\\
\cline{2-9}
&  xorshift1024* & 6-9 & 20-21 & 6-15 & 7.0 & 4-9 & 6 & 8\\
\cline{2-9}
&  xorshift128+ & 8-10 & 24-25 & 14-15 & 9.4 & 6-12 & 4 & 4\\
\hline
\multirow{5}{*}{\rotatebox{90}{CAs}}& Rule $30$ & 8-11 & 24-25 & 15 & 10.2 & 7-12 & 2 & 2\\
\cline{2-9}
& Hybrid CA with Rules $30$ \& $45$ & 0-3 & 1-8 & 0-3 & 2.0 & 0-3 & 11 & 12\\
\cline{2-9}
& Maximal Length CA with $\gamma=0$ & 0-2 & 12 & 10-11 & 1.6 & 1-2 & 10 & 11\\
\cline{2-9}
& Maximal Length CA with $\gamma=1$ & 3-4 & 15-17 & 14 & 1.8 & 1-4 & 8 & 11\\
\cline{2-9}
& Non-linear $2$-state CA & 5-7 & 11 & 3-4 & 5.85 & 2-8 & 9 & 9\\
\hline
 \end{tabular}}
 	\vspace{-1.0em}
 \end{table*}

\begin{itemize}[leftmargin=1pt]
	\item In terms of average tests passed and range, rule $30$ beats all other PRNGs. However, considering results of NIST and TestU01, SFMTs still hold the first rank, while rule $30$ is ranked $2$. As average of \verb MT19937-64 ~is better than \verb PCG-32, ~it is ranked $3$.
	
	\item Average of \verb PCG-32, ~\verb MT19937-32, ~\verb xorshift128+ ~and \verb WELL1024a ~are similar and in terms of performance in NIST and TestU01 battery of tests, they are alike. So, all these PRNGs are ranked $4$.
	
	\item The next rank holders are \verb dSFMT-32 ~(rank $5$), \verb WELL512a ~and \verb xorshift64* ~(rank $6$). We have observed that, \verb Tauss88 ~sometimes fails to pass any tests of Diehard. So, it is put in the same group (rank $7$) with \verb LFSR113, ~which has a very good average despite having a not-so-good performance for the fixed seeds.
	
	\item \verb xorshift1024a ~has better rank (rank $8$) than non-linear $2$-state CA and \verb MMIX ~(rank $9$), because of performance in NIST and TestU01 library. 
	
	\item The next rank holders are \verb xorshift32 ~and \verb dSFMT-52 ~(rank $10$). 
	
	\item As maximal-length CA with $\gamma=1$ has lower average, its rank is degraded. It is put in the same group as the maximal-length CA with $\gamma=0$ (rank $11$).
	
	\item Rule $30-45$ and Borland's LCG are ranked $12$, whereas other PRNGs previously on the same group, like \verb rand, ~\verb lrand ~and \verb random ~are ranked $13$. 
	
	\item Like previous ranking, \verb minstd_rand, ~\verb MRG31k3p ~and \verb LFSR258 ~are the last rank holders based on their overall performance and the fact that for many seeds, these PRNGs fails to pass any test.
\end{itemize}

In the next section, graphical tests are further incorporated on these PRNGs to verify this ranking as well as to check whether any third level ranking of the intra group PRNGs is possible.
 
\subsubsection{Results of Graphical Tests} As mentioned, we use two types of graphical tests -- lattice tests (2-D and 3-D) and space-time diagram test. Here, each of the PRNGs is tested using only the five fixed seeds, which are renamed as following for ease of presentation:
seed $7$ as $s_1$, $1234$ as $s_2$, $12345$ as $s_3$, $19650218$ as $s_4$ and seed $123456789123456789$ as $s_5$.
The motivation behind these graphical tests are -- 
$(a)$ to understand why some PRNGs perform very poorly, $(b)$ to differentiate the behavior of the PRNGs which perform similarly in the blind empirical tests and $(c)$ to visualize the randomness of the PRNGs. The result of these tests are shown as follows.

\begin{enumerate}[leftmargin=0pt]
\item \textbf{Result of Lattice Tests:} For each of the seeds, the $2$-dimensional and $3$-dimensional lattice tests are performed on every PRNG. As expected, for \verb rand, ~\verb lrand48, ~\verb minstd_rand, ~Borland's LCG, and \verb random, ~the points are either scattered or concentrated on a specific part of the $2$-D and $3$-D planes. However, for the good PRNGs like MTs, SFMTs and WELL, the plots are relatively filled. For example, see Figure~\ref{fig:lattice} for output of \verb MMIX, ~\verb Tauss88, ~\verb WELL1024a ~and rule $30$.
     \begin{figure}[hbtp]
       \centering
  \subfloat[MMIX ($2$-D)\label{2d_knuth}]{%
\includegraphics[width=0.20\linewidth, height=2.0cm]{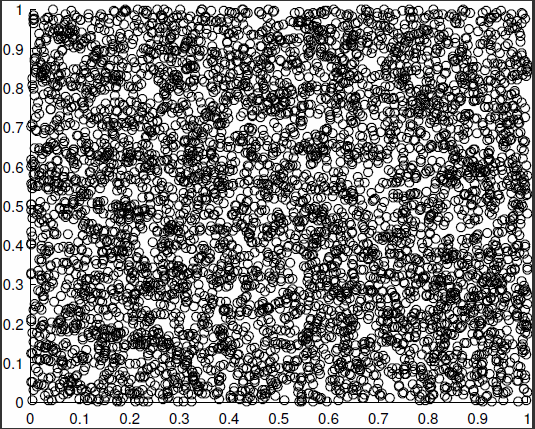}}
\hfill
       \subfloat[MMIX ($3$-D)\label{3d_knuth}]{%
\includegraphics[width=0.20\linewidth, height=2.0cm]{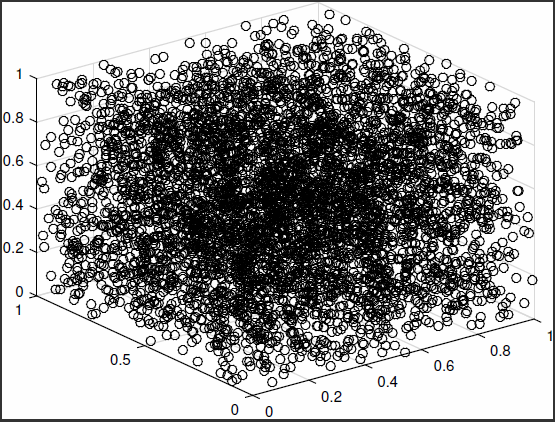}}
\hfill
        \subfloat[Tauss88 ($2$-D)\label{2d_tauss_7-eps-converted-to}]{%
          	\includegraphics[width=0.25\linewidth, height=2.0cm]{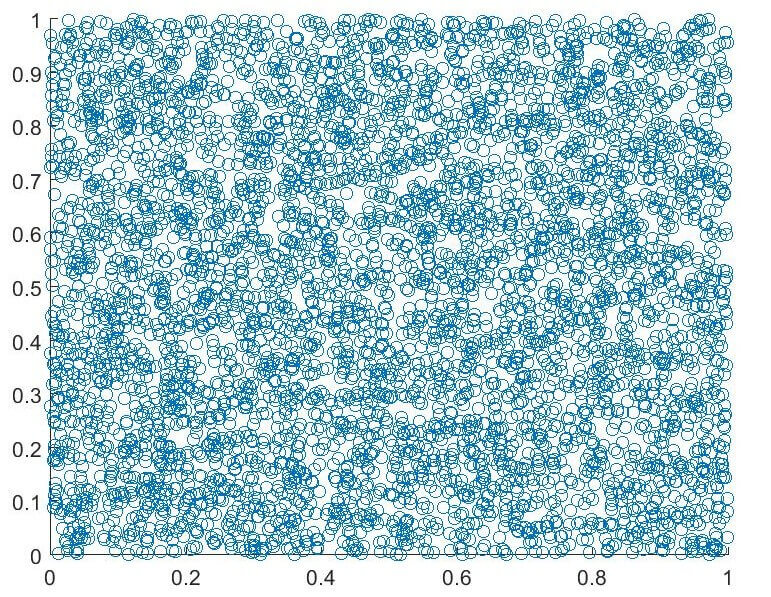}}
          \hfill
          \subfloat[Tauss88 ($3$-D)\label{3d_tauss_7-eps-converted-to}]{%
          	\includegraphics[width=0.20\linewidth, height=2.0cm]{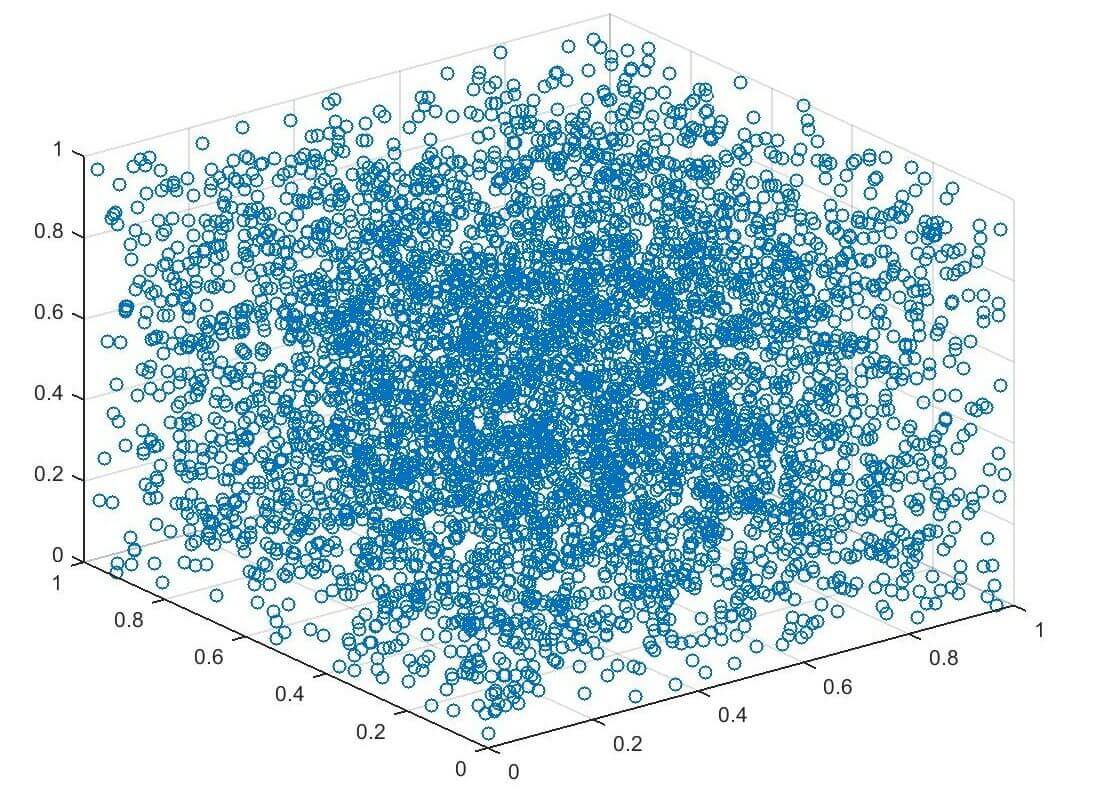}}
          	\hfill\\
      \subfloat[WELL1024a ($2$-D)\label{2d_well1024_7-eps-converted-to}]{%
        	\includegraphics[width=0.20\linewidth, height=2.0cm]{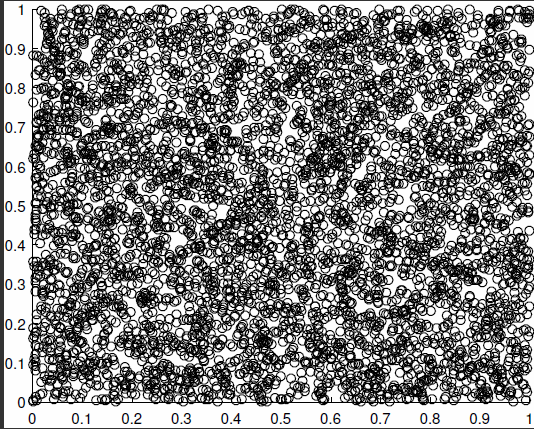}}
  \hfill
    \subfloat[WELL1024a ($3$-D)\label{3d_well1024_7}]{%
     \includegraphics[width=0.20\linewidth, height=2.0cm]{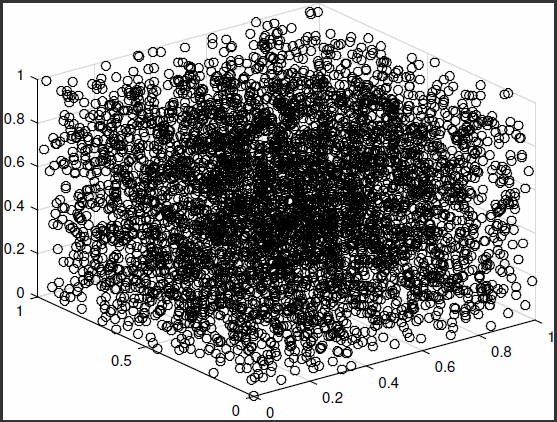}}
\hfill
\subfloat[Rule $30$ ($2$-D)\label{2d_rule30_7}]{%
\includegraphics[width=0.20\linewidth, height=2.0cm]{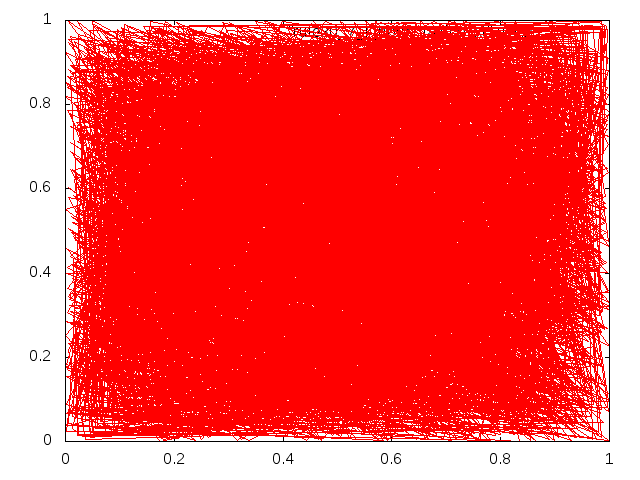}}
\hfill
\subfloat[Rule $30$ ($3$-D)\label{3d_rule30_7}]{%
\includegraphics[width=0.20\linewidth, height=2.0cm]{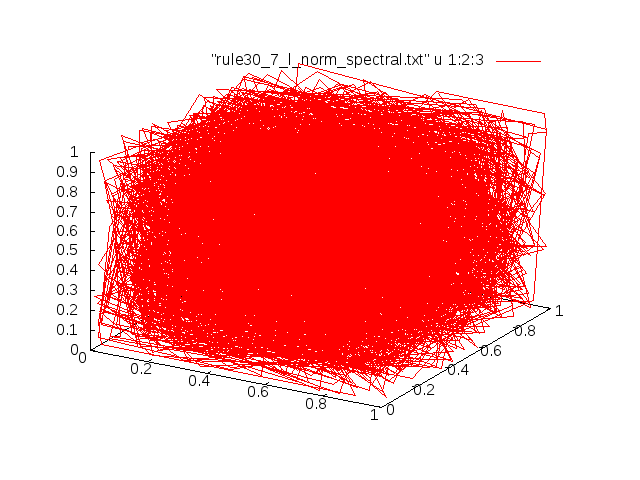}}
      \caption{Lattice Test results for rand, MMIX, Tauss88, WELL1024a, 
      and Rule $30$ with seed $7$}
         \label{fig:lattice}
         \end{figure}
However, this test fails to further enhance or modify the ranking shown in Table~\ref{tab:blind_test_avg}.
So, we avoid supplying all the images of lattice test but move to space-time diagram.

\item \textbf{Result of Space-time Diagram Test:} For space-time diagram, a set of $1000$ numbers are generated from each seed and printed on $X-Y$ plane. The space-time diagrams of $4$ seeds $s_1,s_2,s_3,s_4$ for each PRNG, are shown in Figure~\ref{fig:rand_space-time}
to
\ref{fig:ca_space-time}. From these figures, we can observe the following:

            \begin{figure}[hbtp]
              \centering
              \vspace{-2.0em}
              \subfloat[$s_1$\label{knuth_lcg_7_space}]{%
              	\includegraphics[width=0.2\linewidth, height=5.0cm]{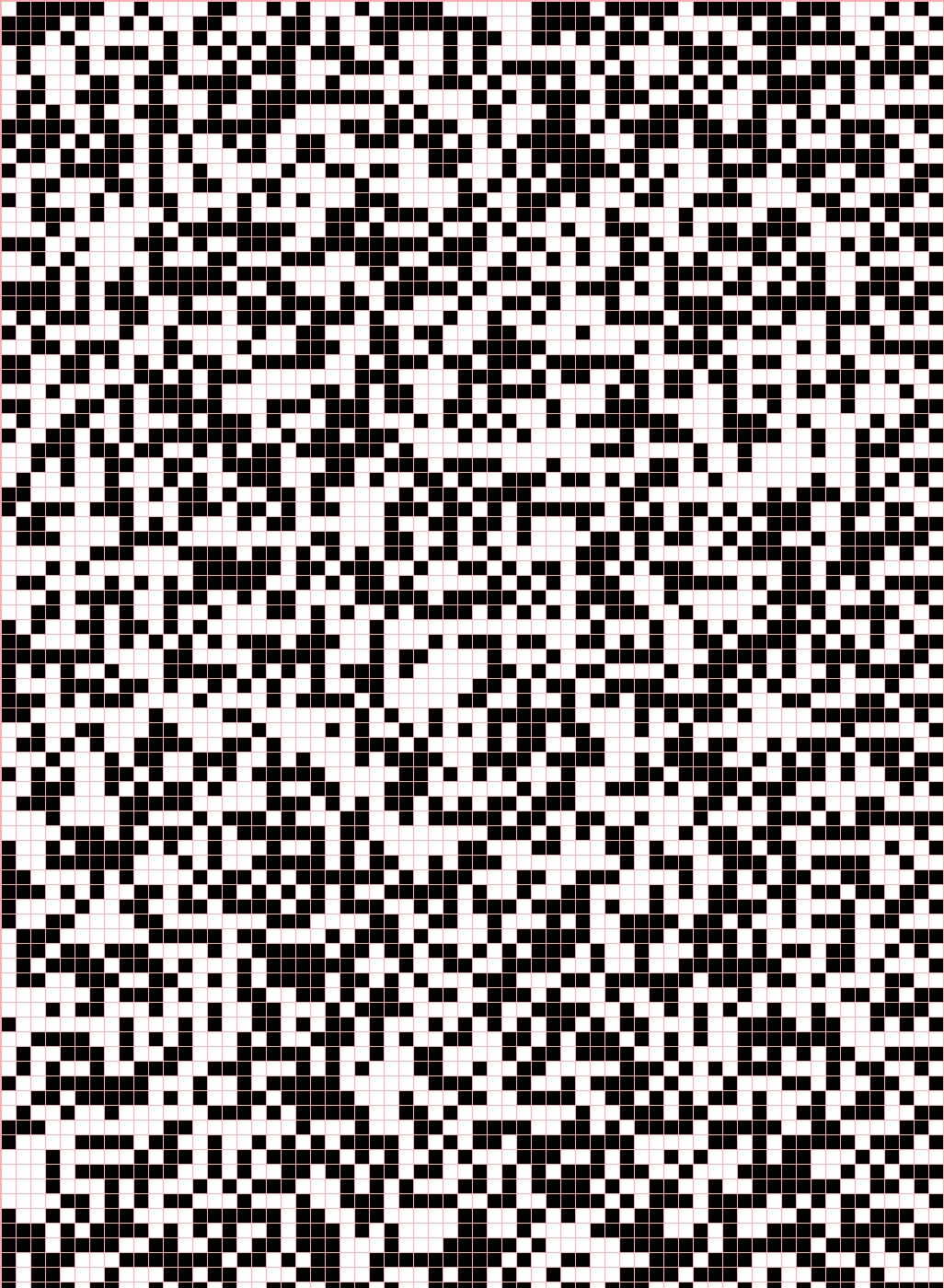}}
              	\hfill
              \subfloat[$s_3$\label{knuth_lcg_12345_space}]{%
              \includegraphics[width=0.2\linewidth, height=5.0cm]{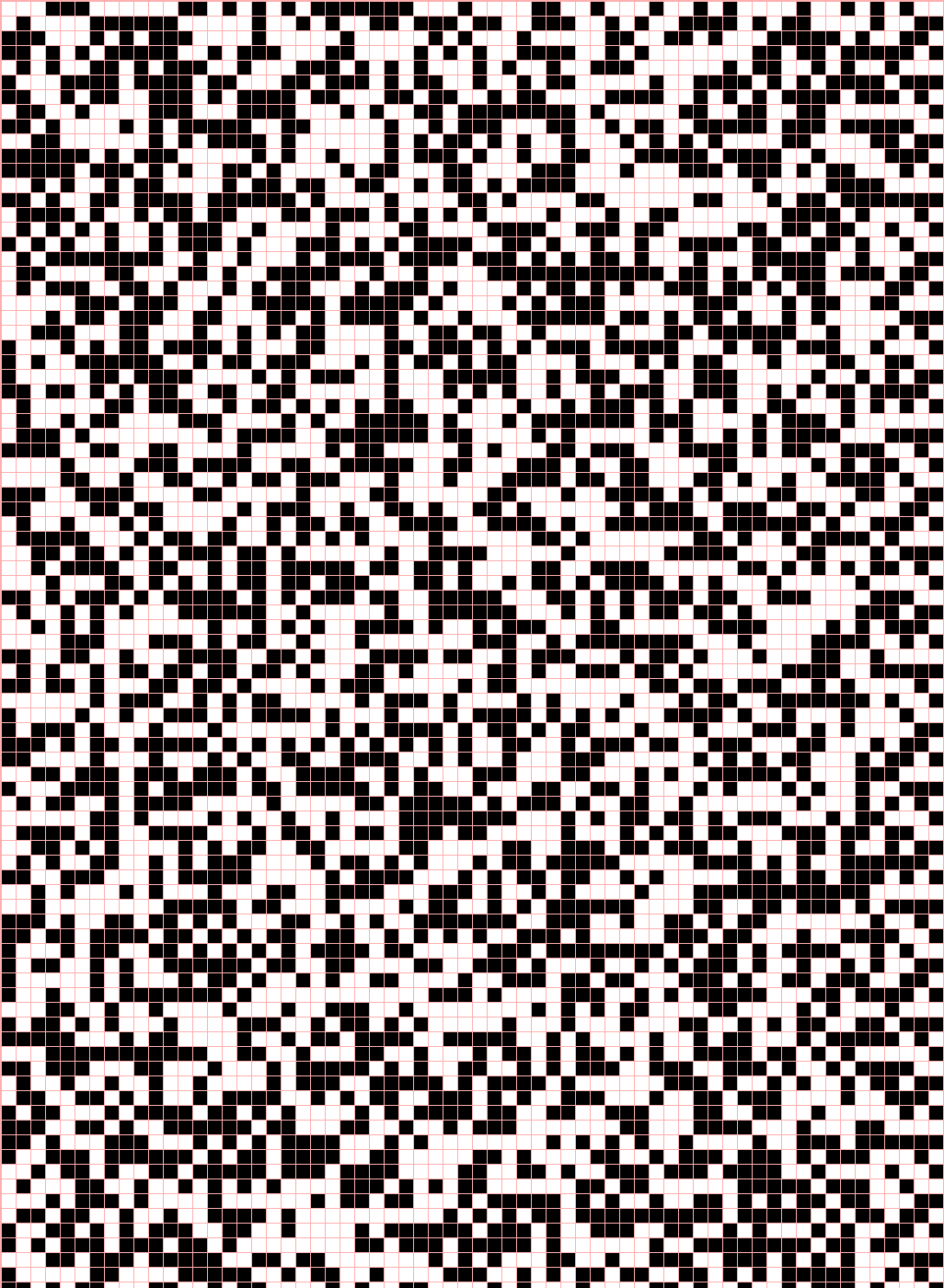}}
              \hfill
              \subfloat[$s_4$\label{knuth_lcg_9650218_space}]{%
              \includegraphics[width=0.2\linewidth, height=5.0cm]{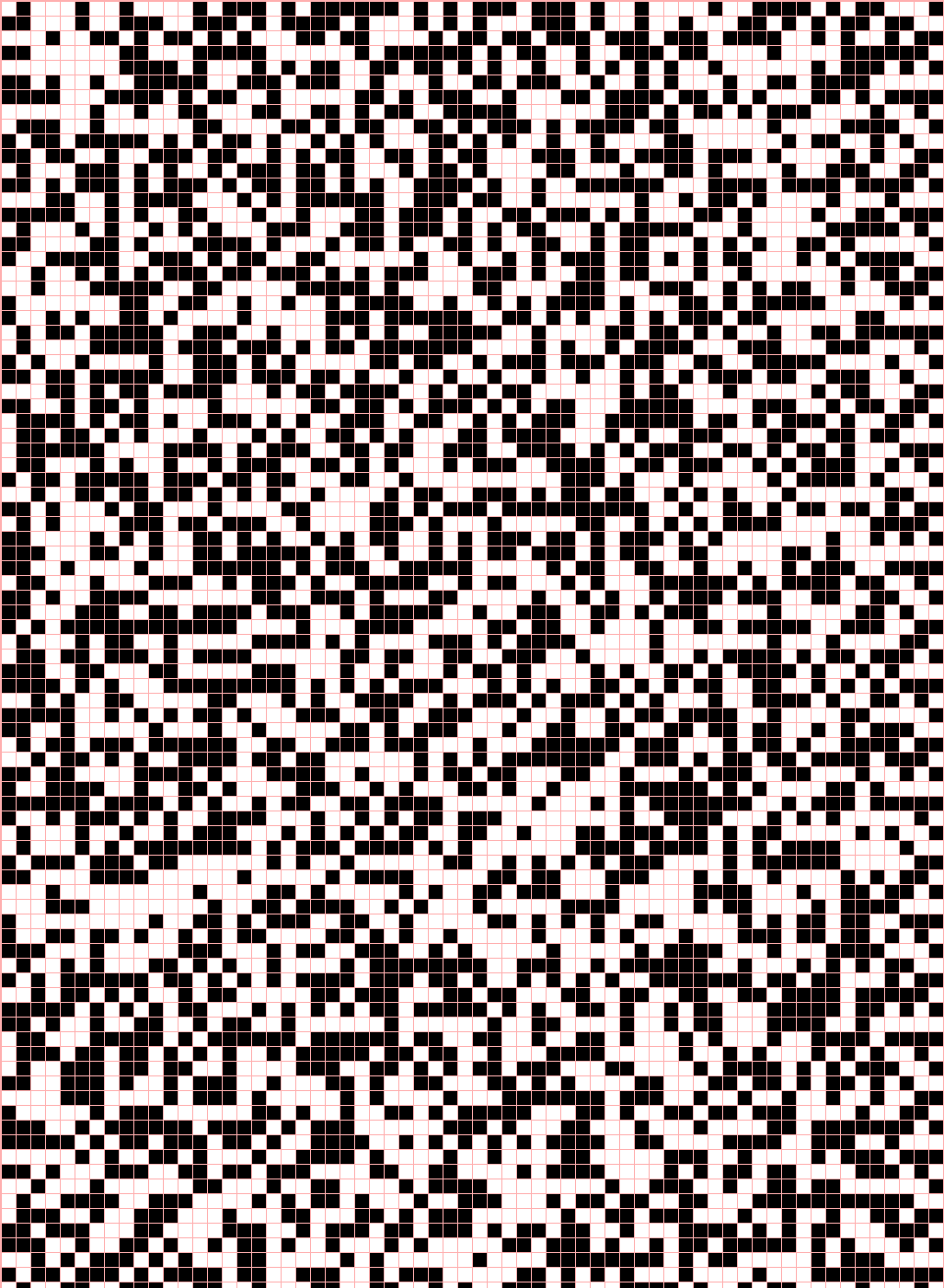}}
              \hfill
              \subfloat[$s_5$\label{knuth_lcg_123456789123456789_space}]{%
              \includegraphics[width=0.2\linewidth, height=5.0cm]{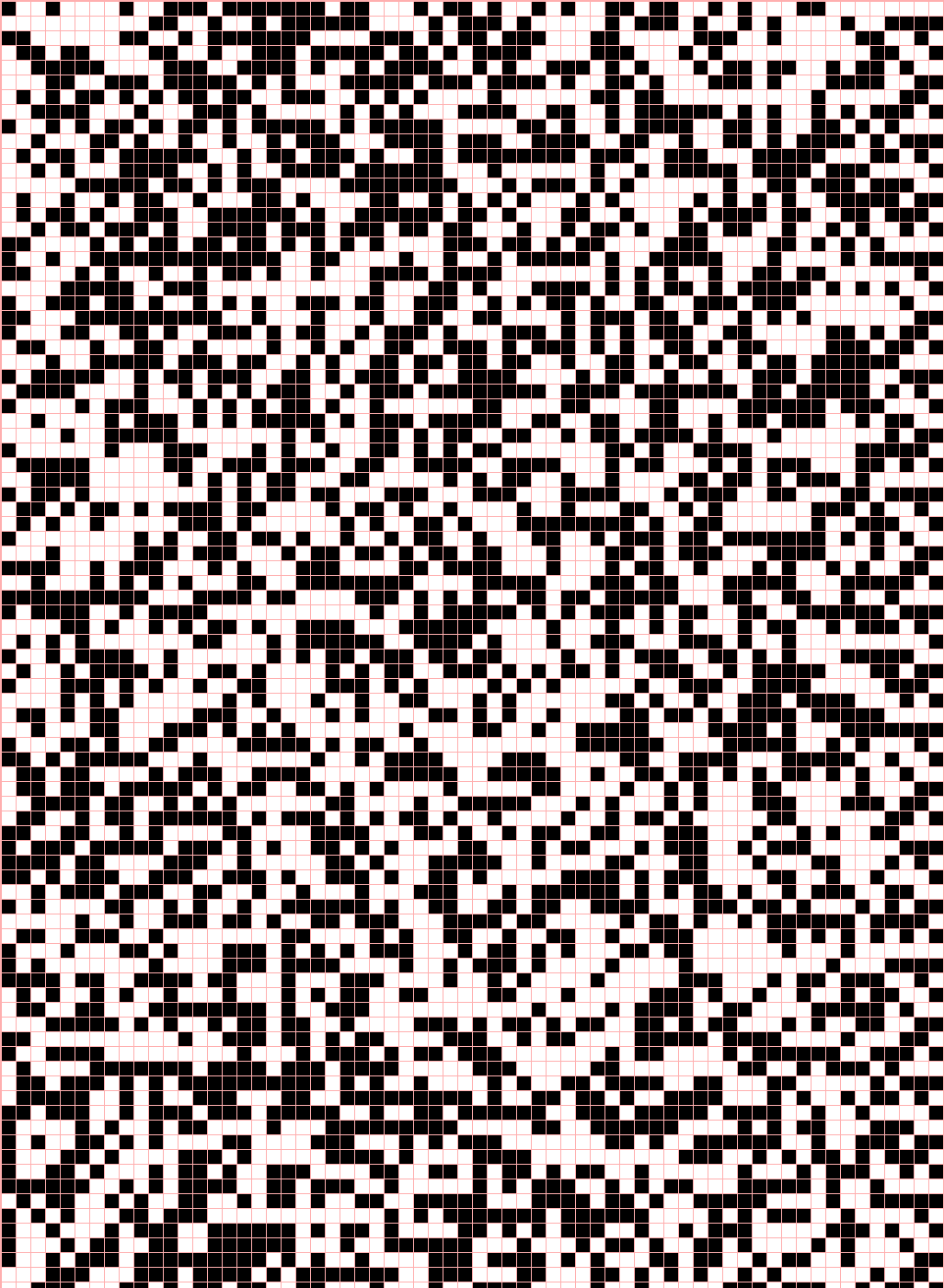}}
%
%
\hfill\\
              \subfloat[$s_1$\label{borland_lcg_7_space}]{%
                           	\includegraphics[width=0.1\linewidth, height=5.0cm]{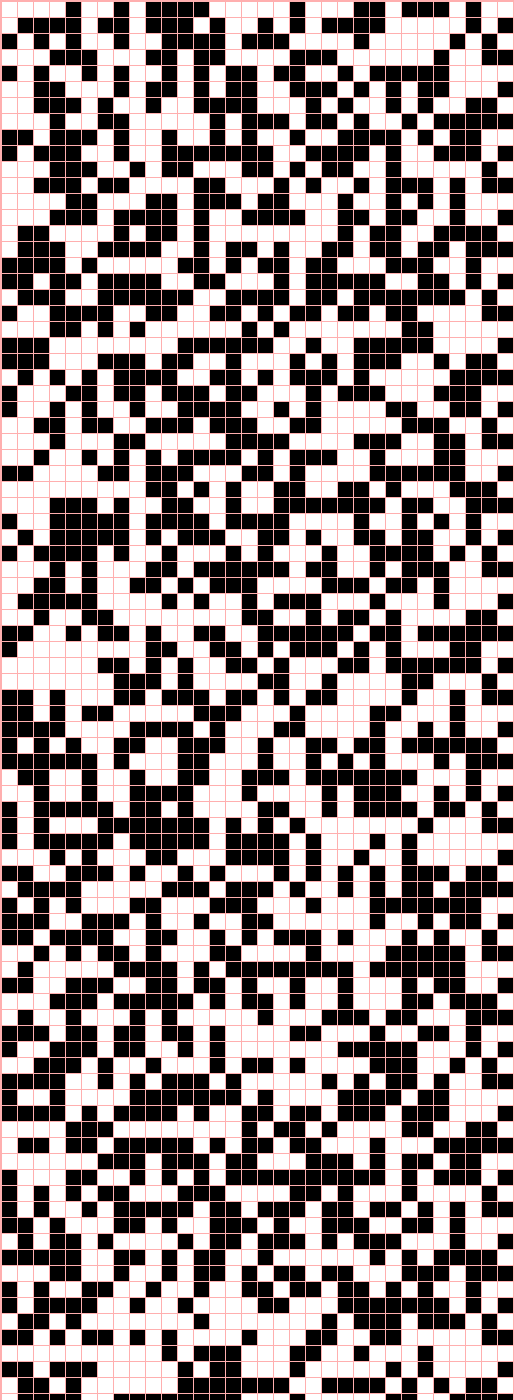}}
                           	\hfill
            \subfloat[$s_3$\label{borland_lcg_12345_space}]{%
          \includegraphics[width=0.1\linewidth, height=5.0cm]{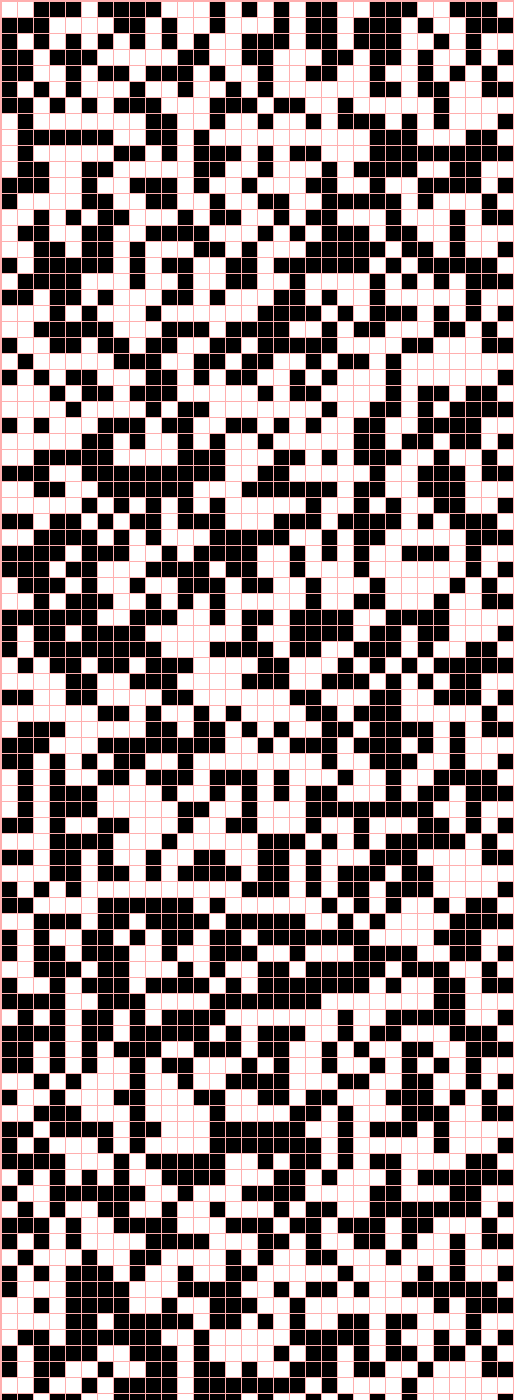}}
                           \hfill
         \subfloat[$s_4$\label{borland_lcg_9650218_space}]{%
          \includegraphics[width=0.1\linewidth, height=5.0cm]{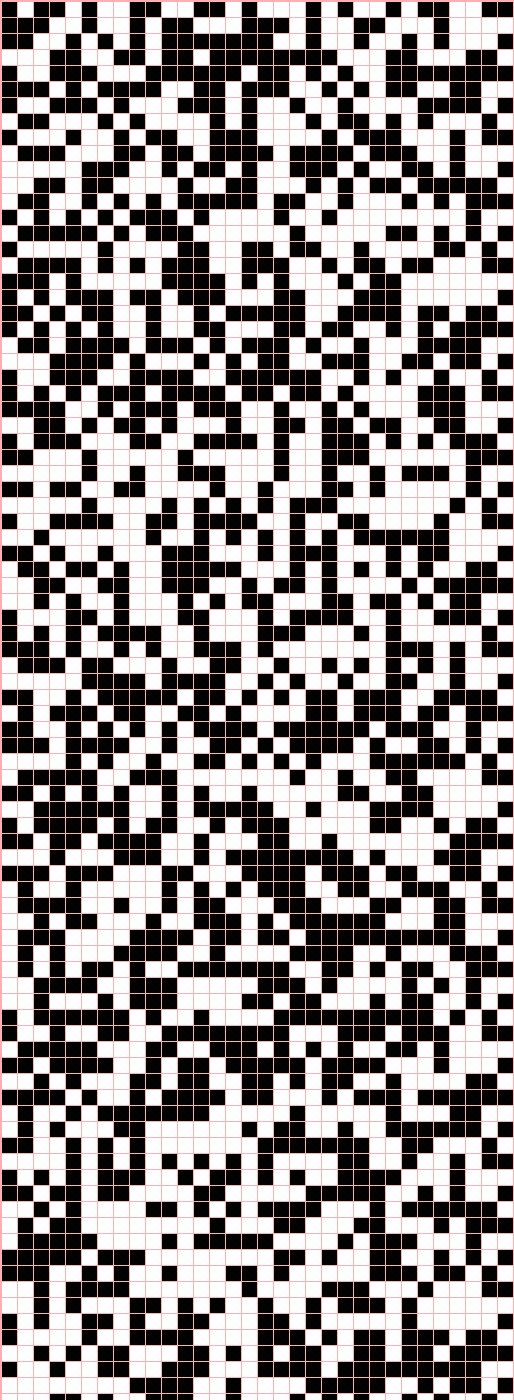}}
                           \hfill
         \subfloat[$s_5$\label{borland_lcg_123456789123456789_spaceo}]{%
           \includegraphics[width=0.1\linewidth, height=5.0cm]{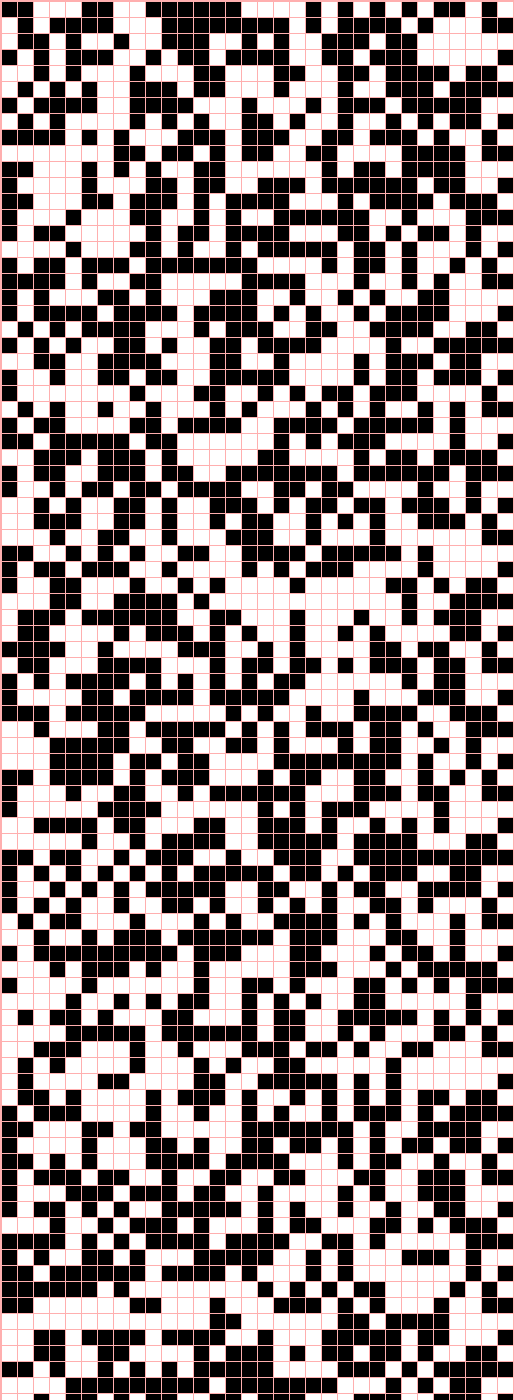}}
             \hfill
                 \subfloat[$s_1$\label{min_std_7_space}]{%
            \includegraphics[width=0.1\linewidth, height=5.0cm]{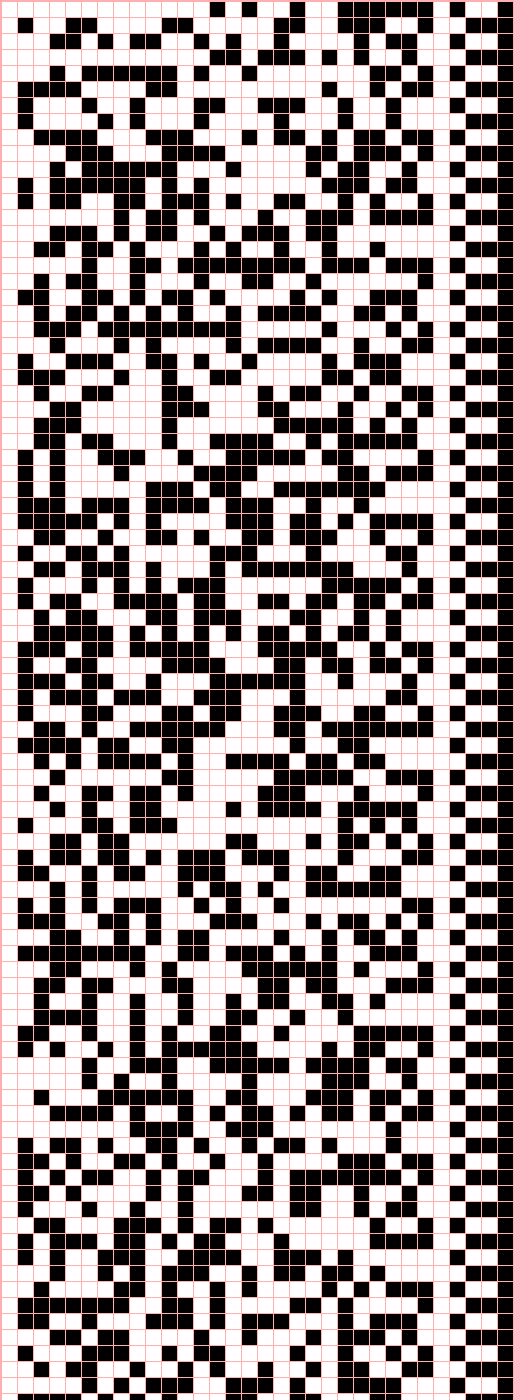}}
                                          	\hfill
                 \subfloat[$s_3$\label{min_std_12345_space}]{%
        \includegraphics[width=0.1\linewidth, height=5.0cm]{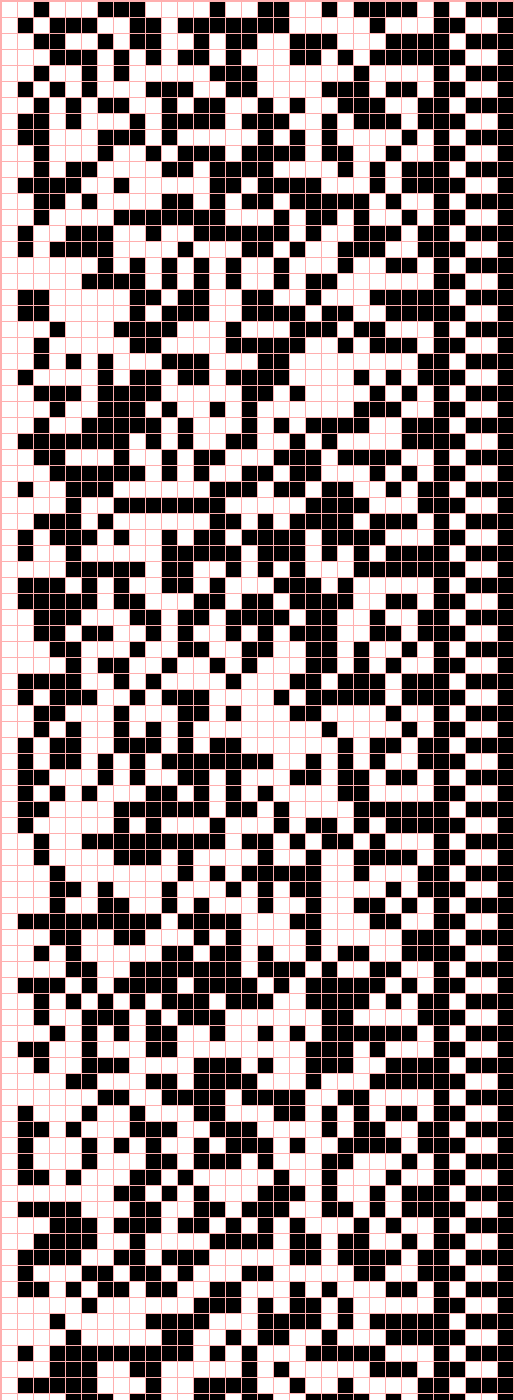}}
                                          \hfill
          \subfloat[$s_4$\label{min_std_9650218_space}]{%
     \includegraphics[width=0.1\linewidth, height=5.0cm]{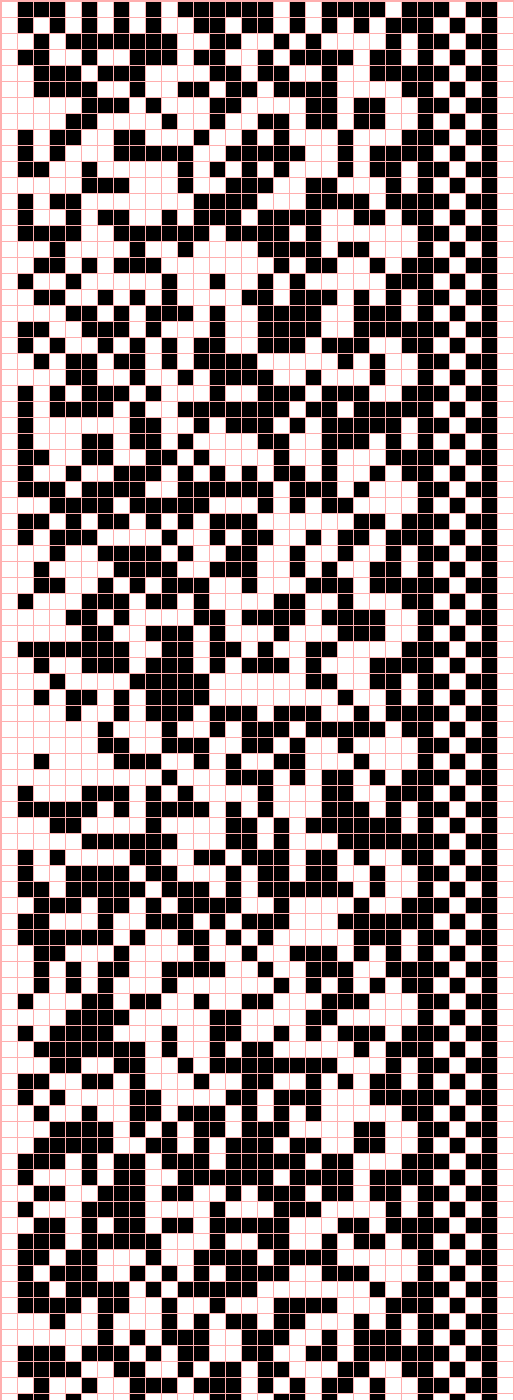}}
                                          \hfill
     \subfloat[$s_5$\label{min_std_123456789123456789_spaceo}]{%
     \includegraphics[width=0.1\linewidth, height=5.0cm]{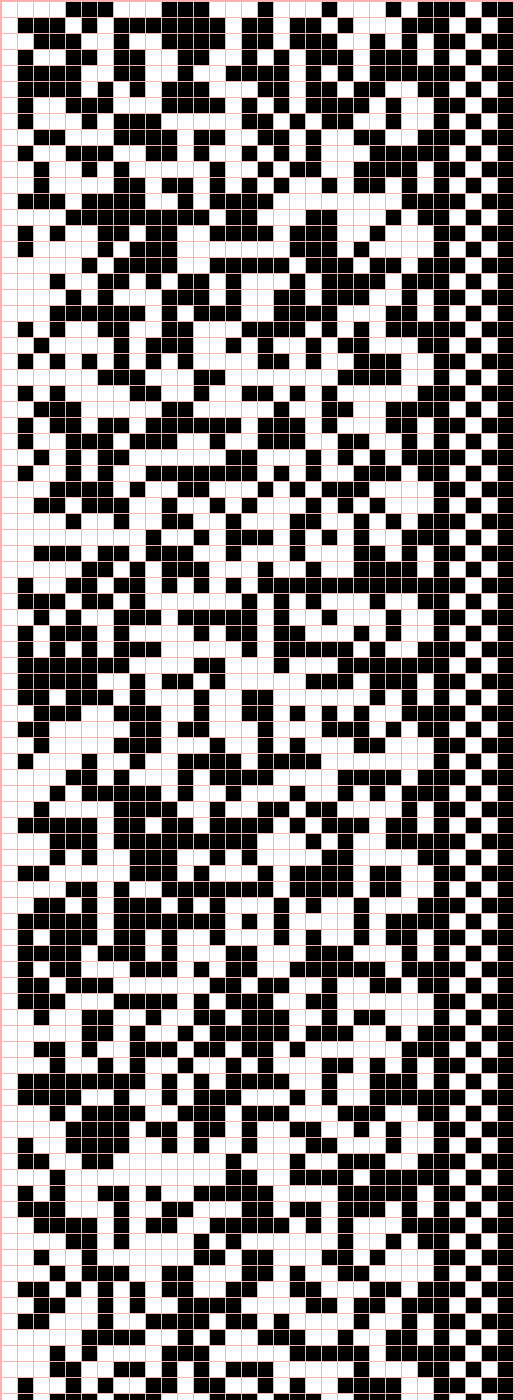}}
    \hfill\\
%
%
\subfloat[$s_1$\label{rand_32_7_space}]{%
\includegraphics[width=0.1\linewidth, height=5.0cm]{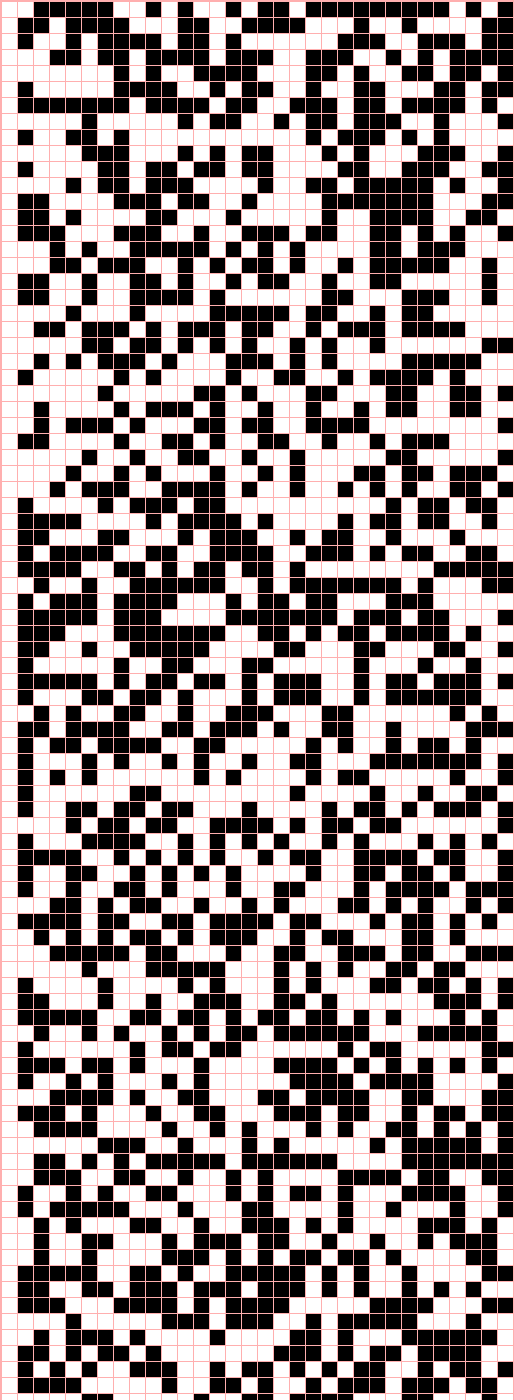}}
\hfill
\subfloat[$s_3$\label{rand_32_12345_space}]{%
 \includegraphics[width=0.1\linewidth, height=5.0cm]{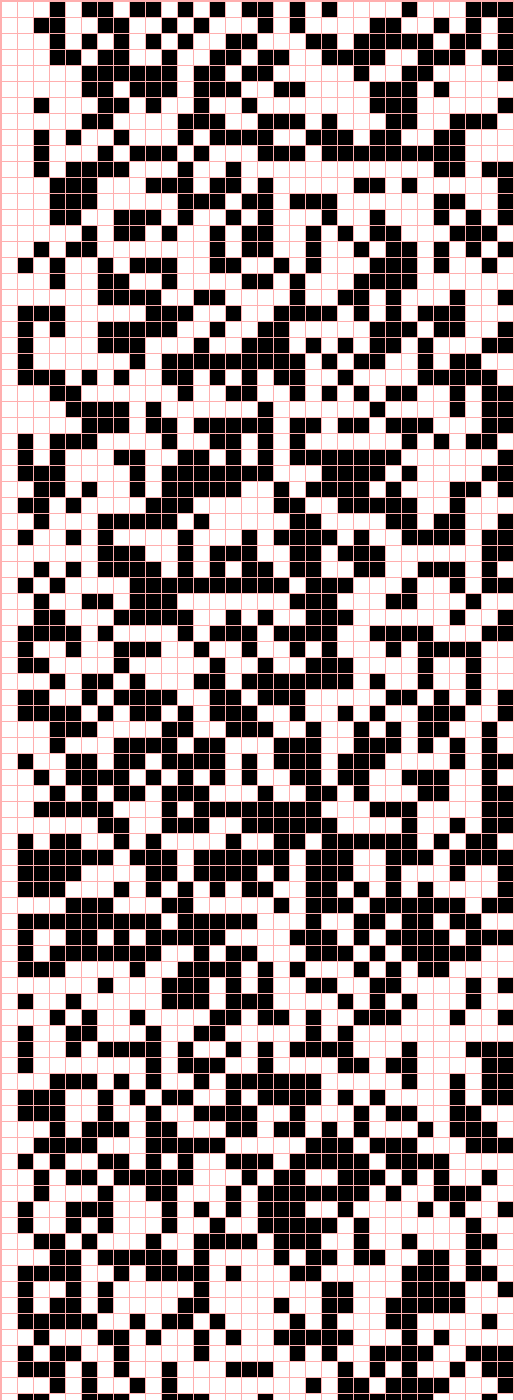}}
\hfill
\subfloat[$s_4$\label{rand_32_9650218_space}]{%
\includegraphics[width=0.1\linewidth, height=5.0cm]{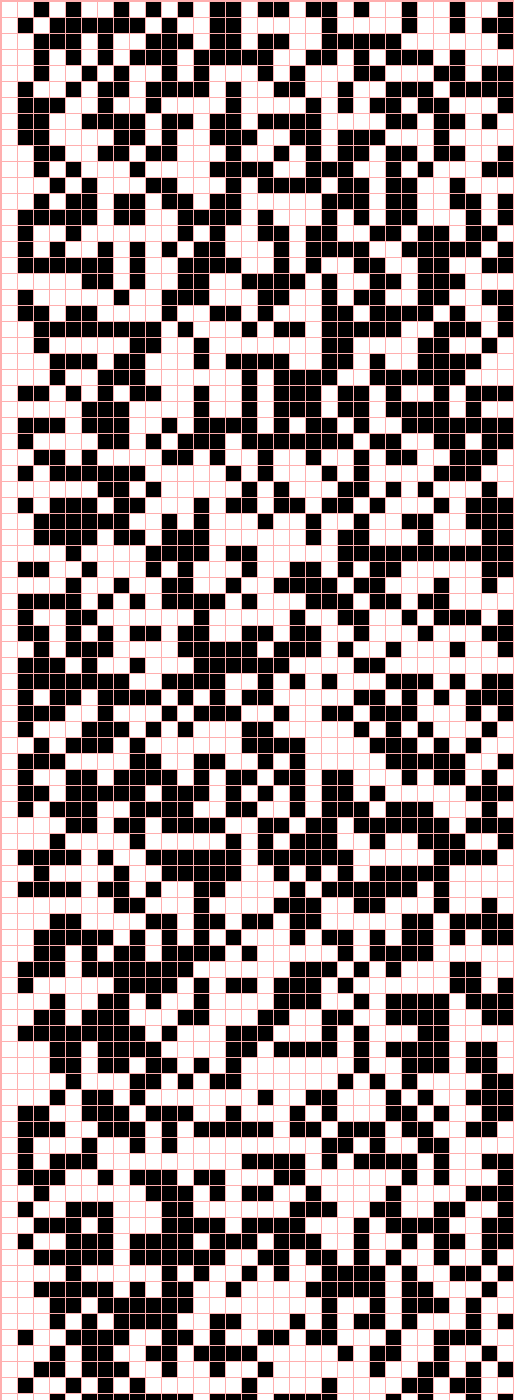}}
\hfill
\subfloat[$s_5$\label{rand_32_123456789123456789_spaceo}]{%
\includegraphics[width=0.1\linewidth, height=5.0cm]{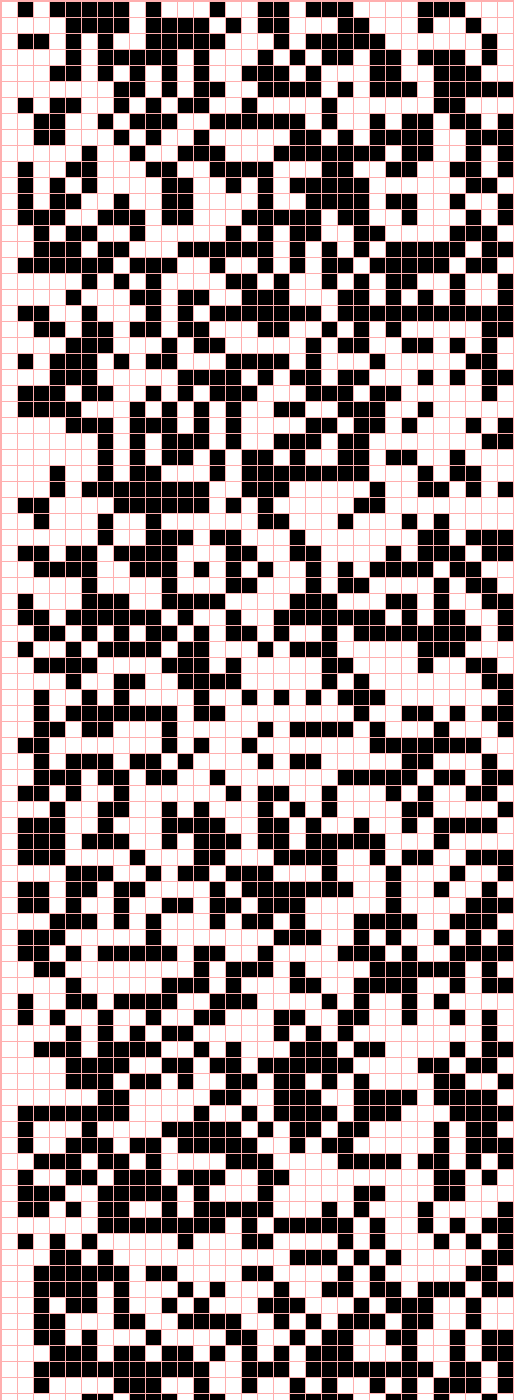}}
\hfill
\subfloat[$s_1$\label{lrand_32_7_space}]{%
\includegraphics[width=0.1\linewidth, height=5.0cm]{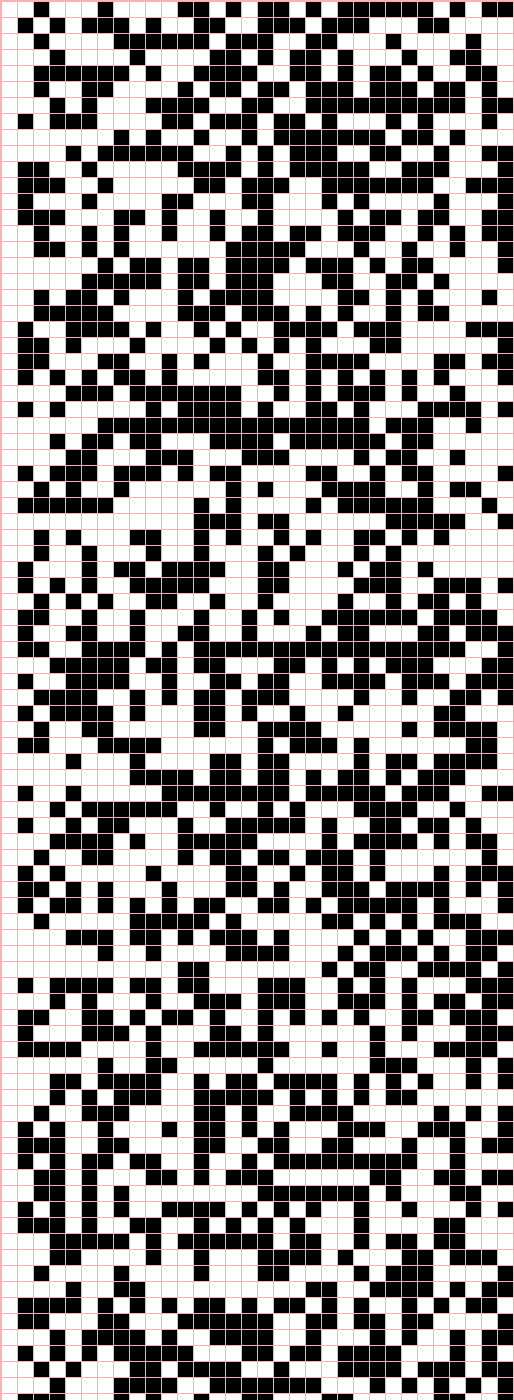}}
\hfill
\subfloat[$s_3$\label{lrand_32_12345_space}]{%
 \includegraphics[width=0.1\linewidth, height=5.0cm]{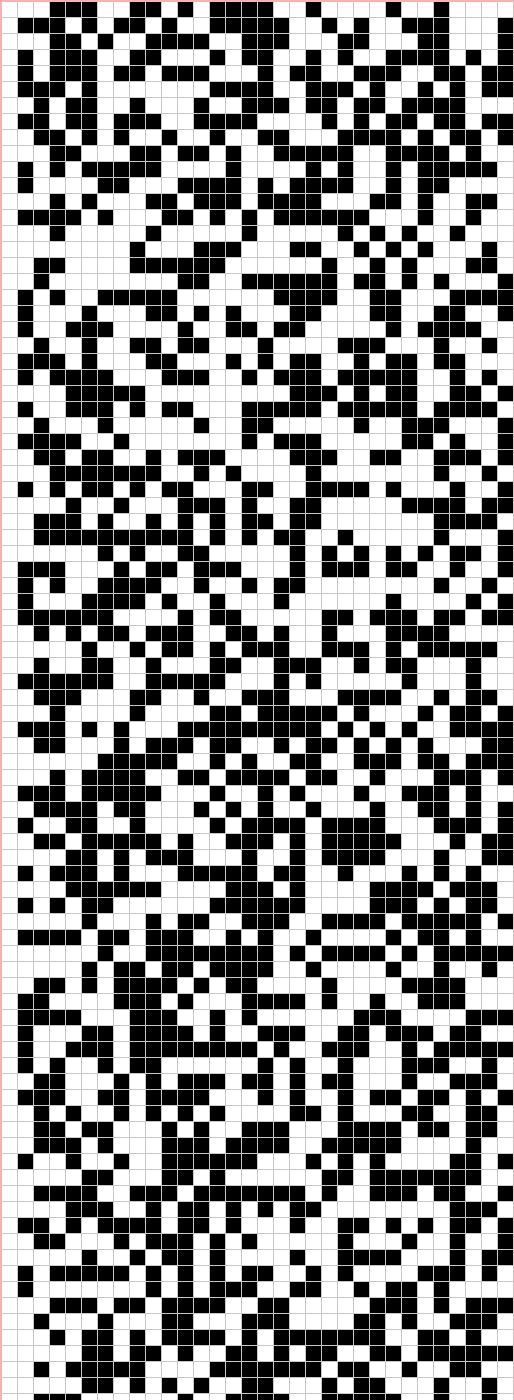}}
\hfill
\subfloat[$s_4$\label{lrand_32_9650218_space}]{%
\includegraphics[width=0.1\linewidth, height=5.0cm]{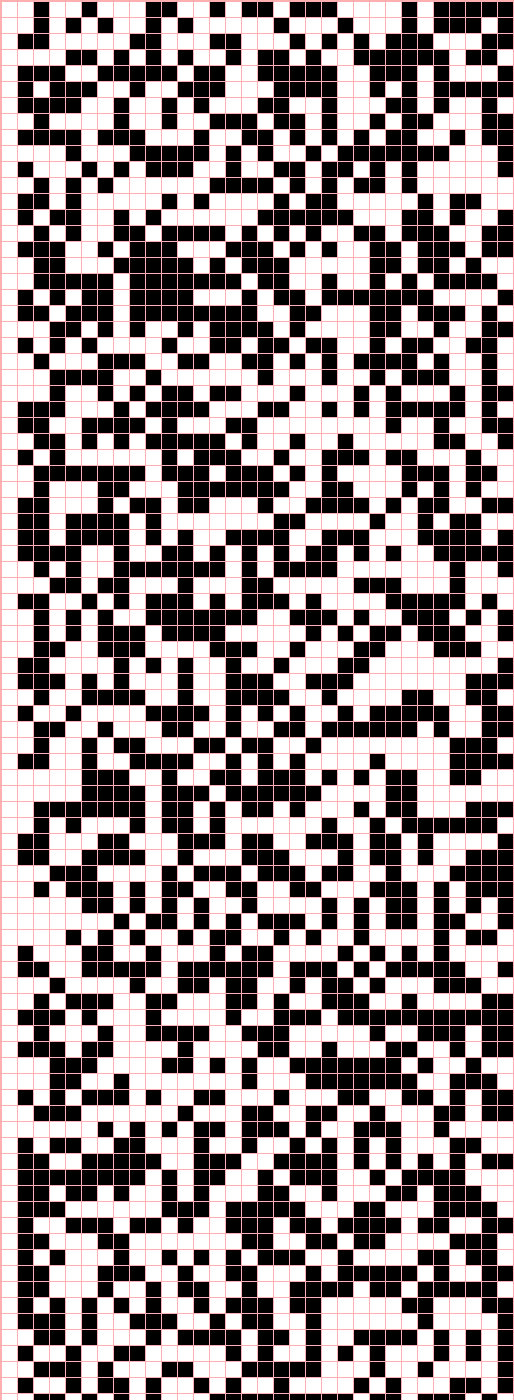}}
\hfill
\subfloat[$s_5$\label{lrand_32_123456789123456789_space}]{%
\includegraphics[width=0.1\linewidth, height=5.0cm]{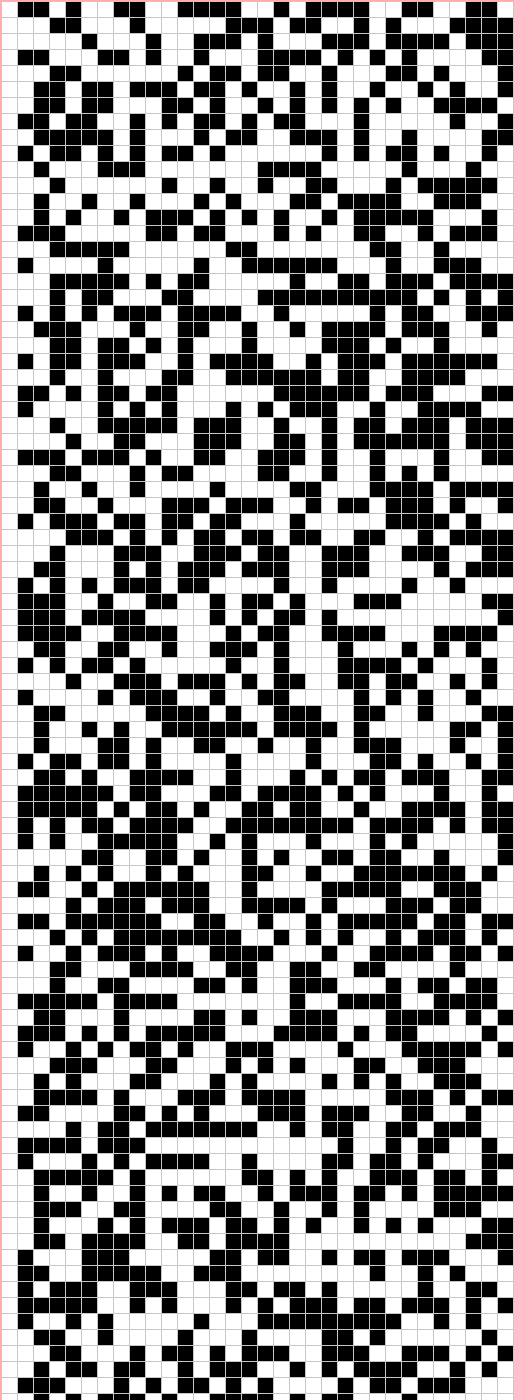}}
\caption{Space-time diagram for Knuth's MMIX (\ref{knuth_lcg_7_space} to \ref{knuth_lcg_123456789123456789_space}), Borland's LCG (\ref{borland_lcg_7_space} to \ref{borland_lcg_123456789123456789_spaceo}) and minstd\_rand (\ref{min_std_7_space} to \ref{min_std_123456789123456789_spaceo}), rand (\ref{rand_32_7_space} to \ref{rand_32_123456789123456789_spaceo}) and lrand (\ref{lrand_32_7_space} to \ref{lrand_32_123456789123456789_space}) of UNIX}
\label{fig:rand_space-time}
\end{figure}          

\begin{figure}[hbtp]
\centering
  \vspace{-2.0em}
\subfloat[$s_1$\label{mrg_7_space}]{%
\includegraphics[width=0.1\linewidth, height=5.0cm]{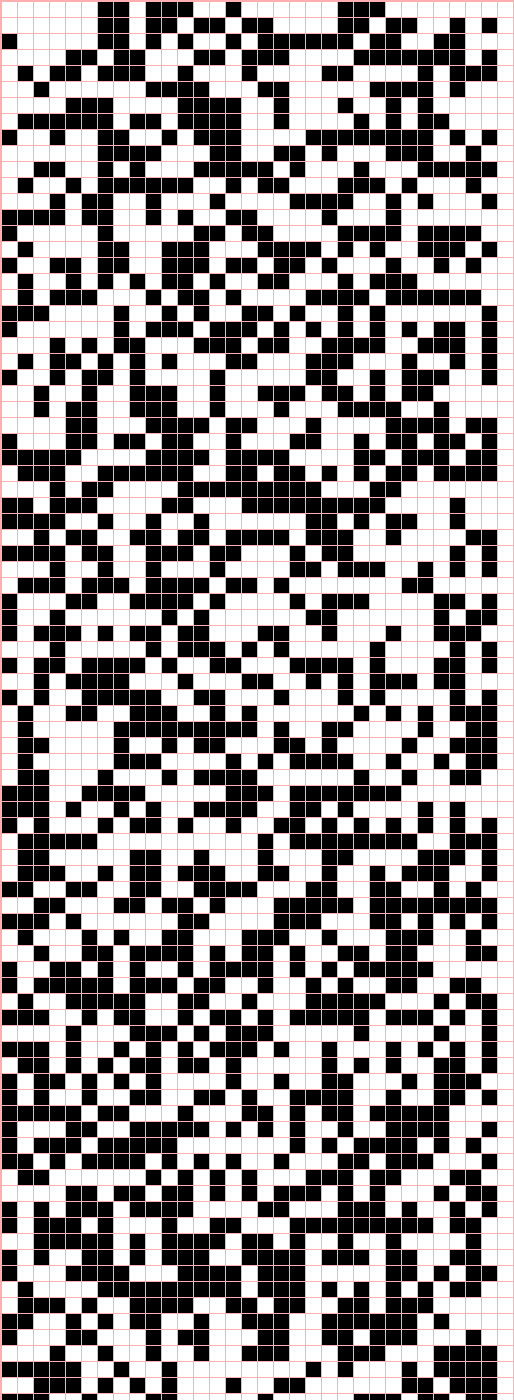}}
\hfill
\subfloat[$s_3$\label{mrg_12345_space}]{%
 \includegraphics[width=0.1\linewidth, height=5.0cm]{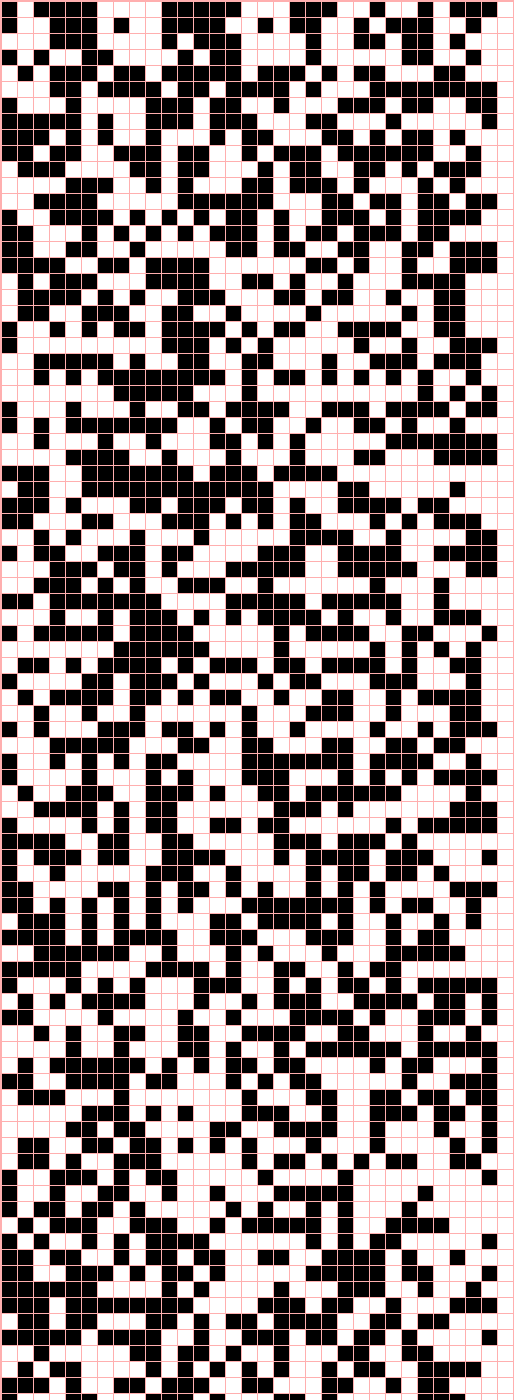}}
\hfill
\subfloat[$s_4$\label{mrg_9650218_space}]{%
\includegraphics[width=0.1\linewidth, height=5.0cm]{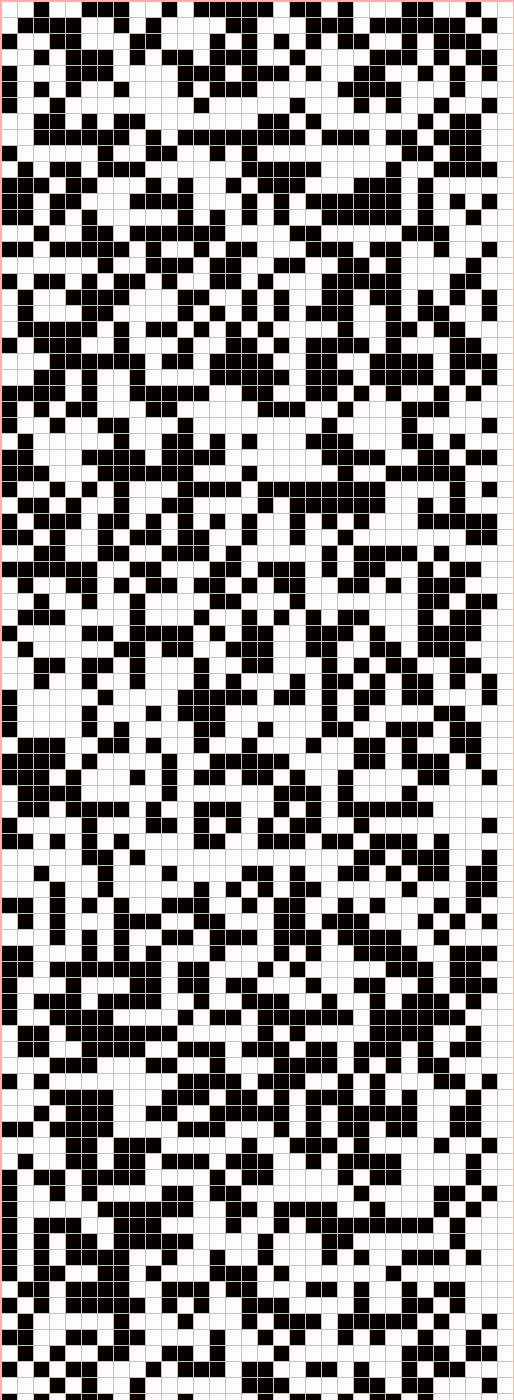}}
\hfill
\subfloat[$s_5$\label{mrg_123456789123456789_spaceo}]{%
\includegraphics[width=0.1\linewidth, height=5.0cm]{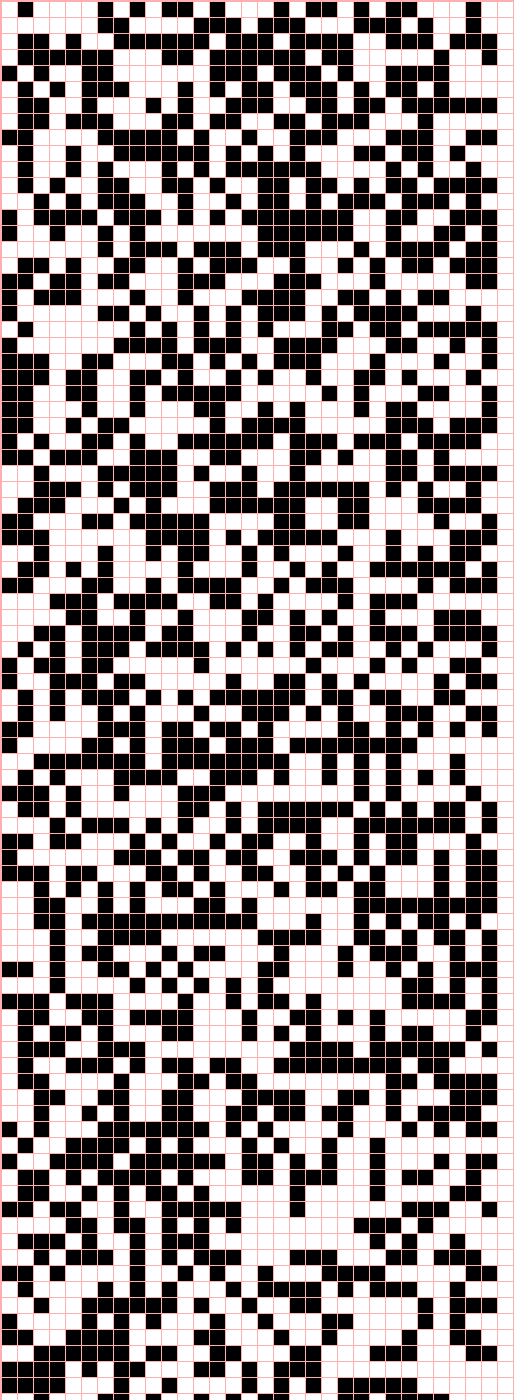}}
\hfill
\subfloat[$s_1$\label{pcg_32_7_space}]{%
\includegraphics[width=0.1\linewidth, height=5.0cm]{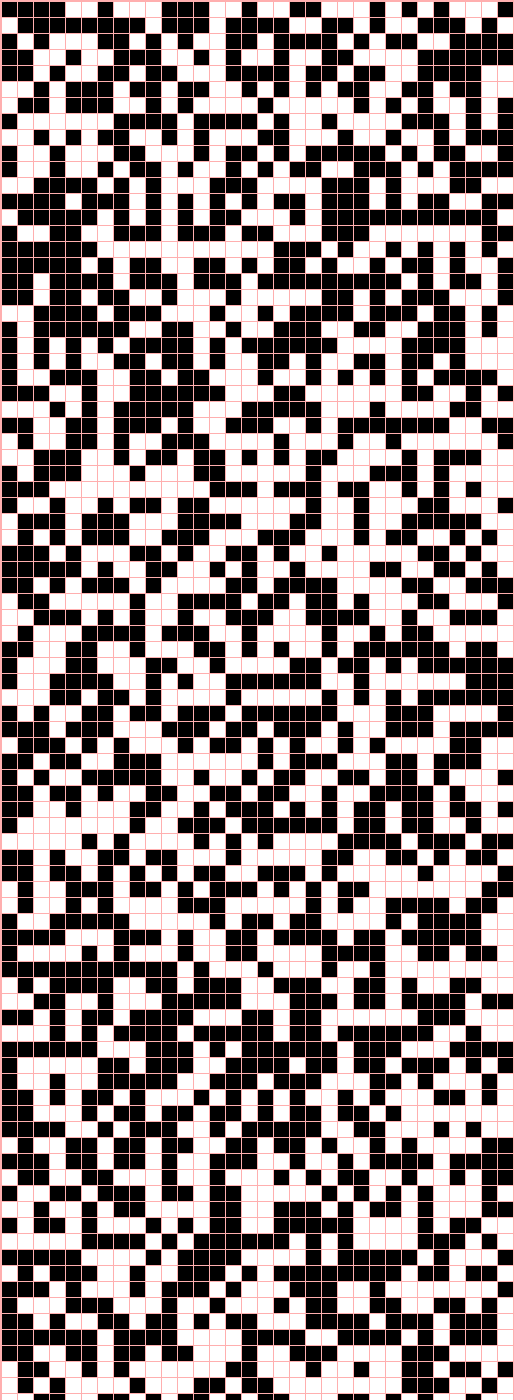}}
\hfill
\subfloat[$s_3$\label{pcg_32_12345_space}]{%
 \includegraphics[width=0.1\linewidth, height=5.0cm]{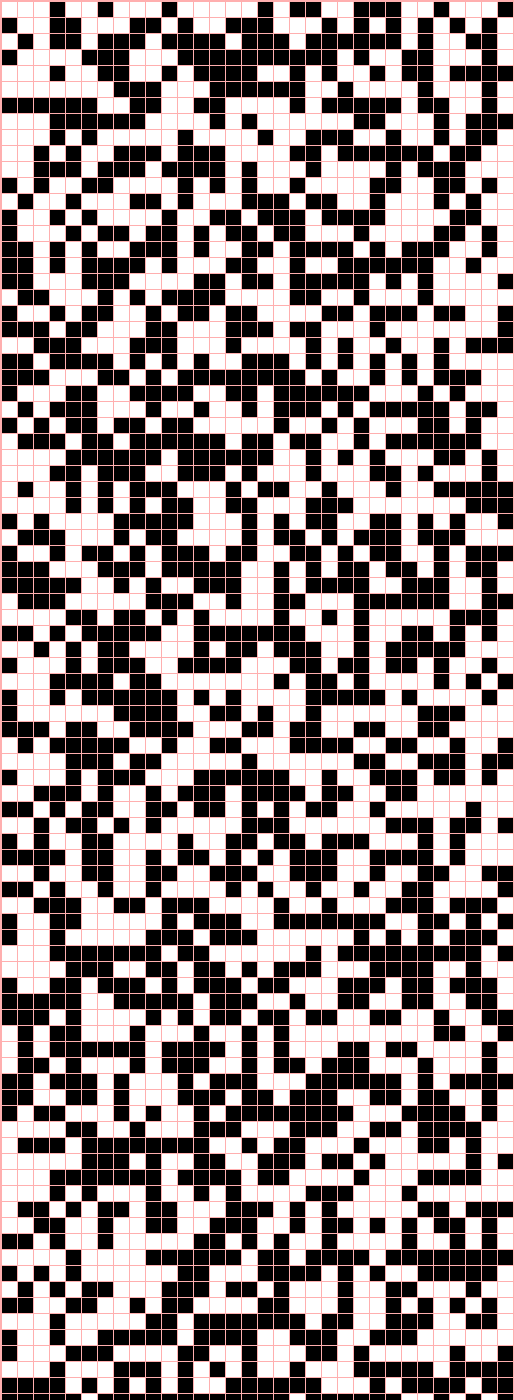}}
\hfill
\subfloat[$s_4$\label{pcg_32_9650218_space}]{%
\includegraphics[width=0.1\linewidth, height=5.0cm]{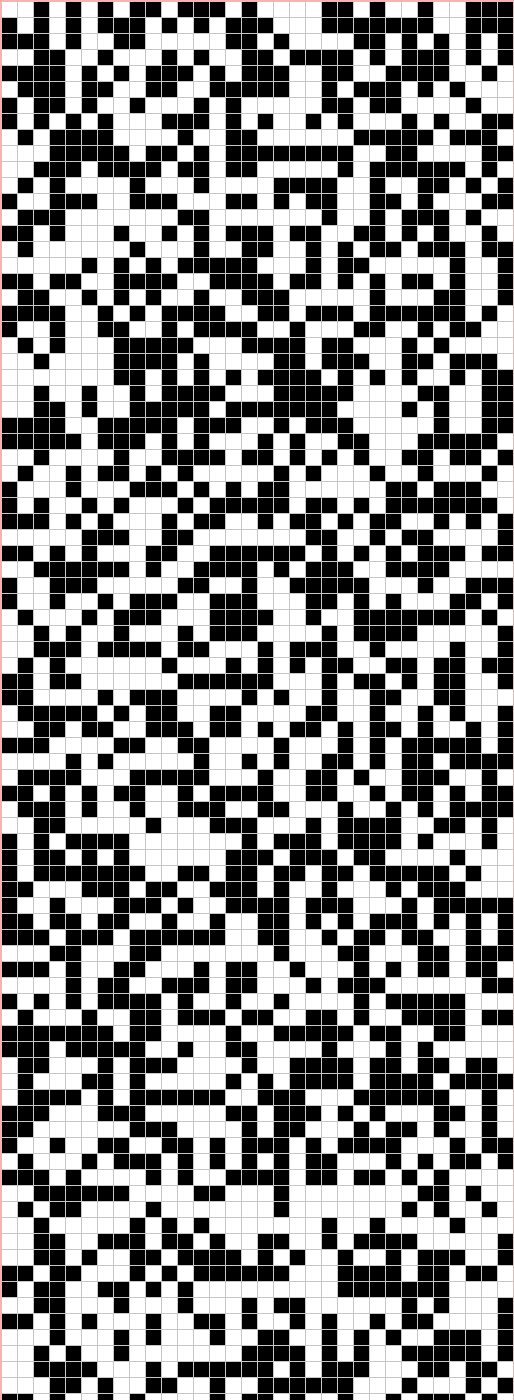}}
\hfill
\subfloat[$s_5$\label{pcg_32_123456789123456789_space}]{%
\includegraphics[width=0.1\linewidth, height=5.0cm]{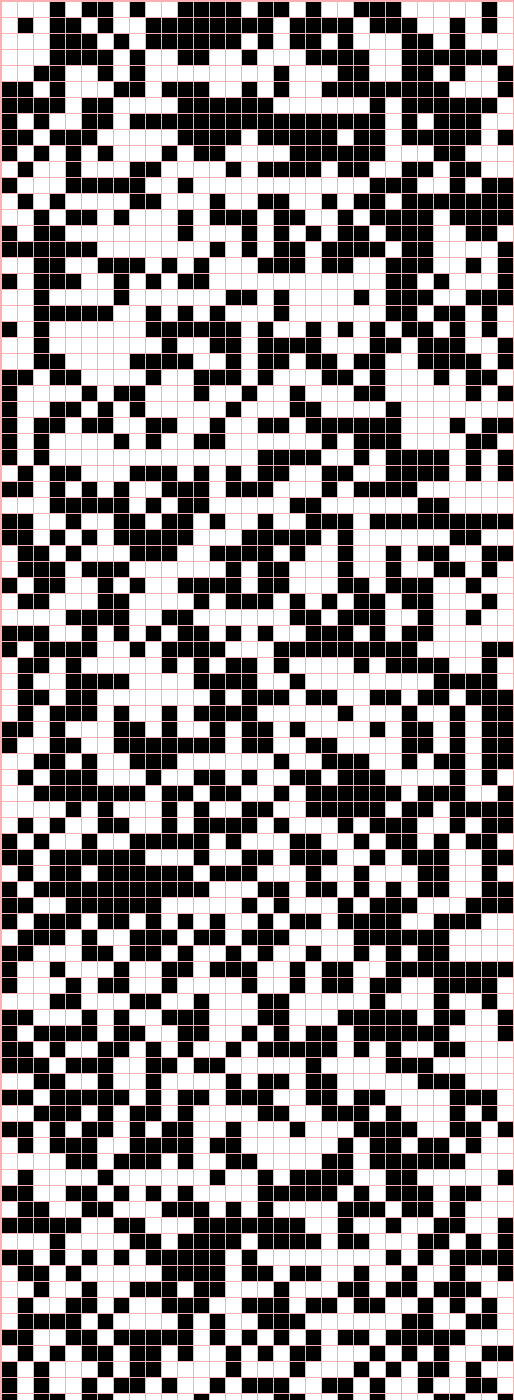}}
%
%
\hfill\\
\subfloat[$s_1$\label{random_32_7_space}]{%
\includegraphics[width=0.1\linewidth, height=5.0cm]{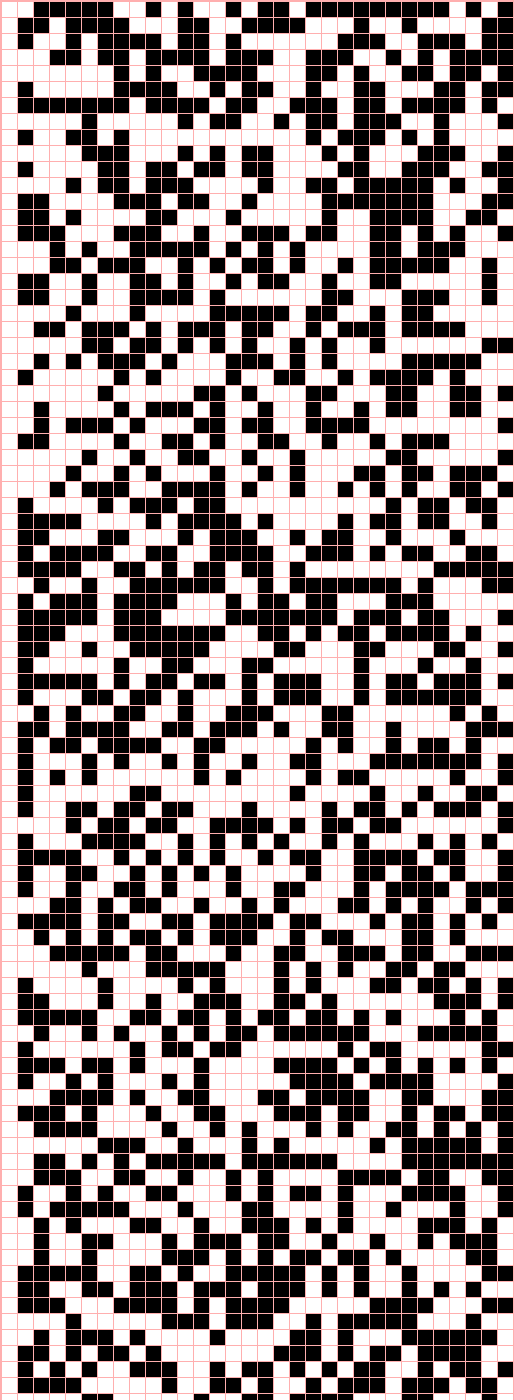}}
\hfill
\subfloat[$s_3$\label{random_32_12345_space}]{%
 \includegraphics[width=0.1\linewidth, height=5.0cm]{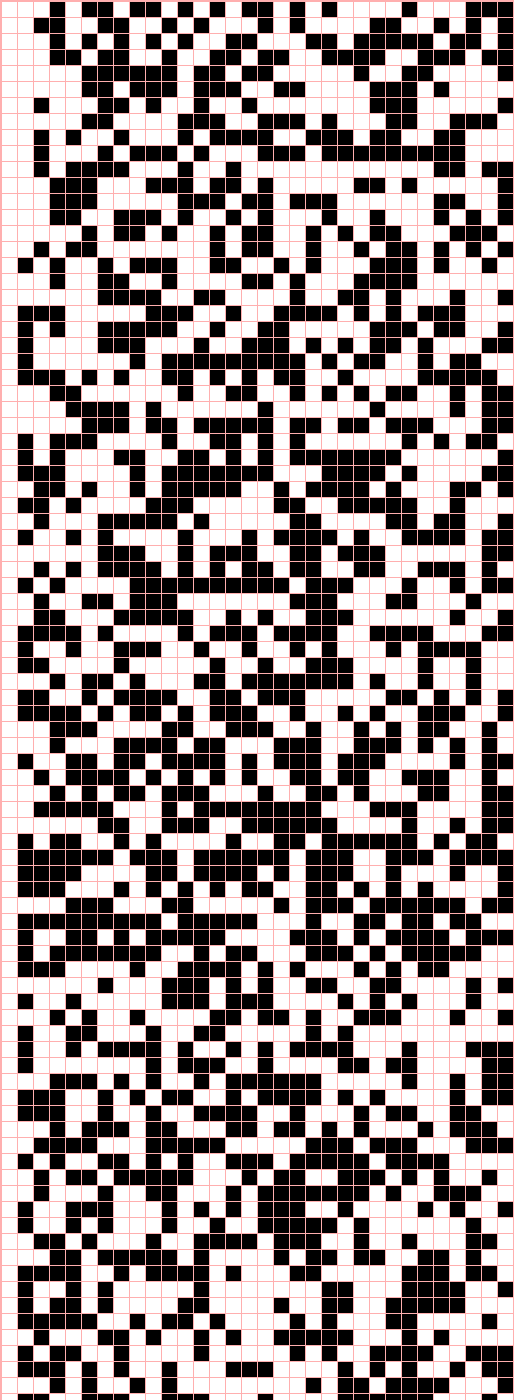}}
\hfill
\subfloat[$s_4$\label{random_32_9650218_space}]{%
\includegraphics[width=0.1\linewidth, height=5.0cm]{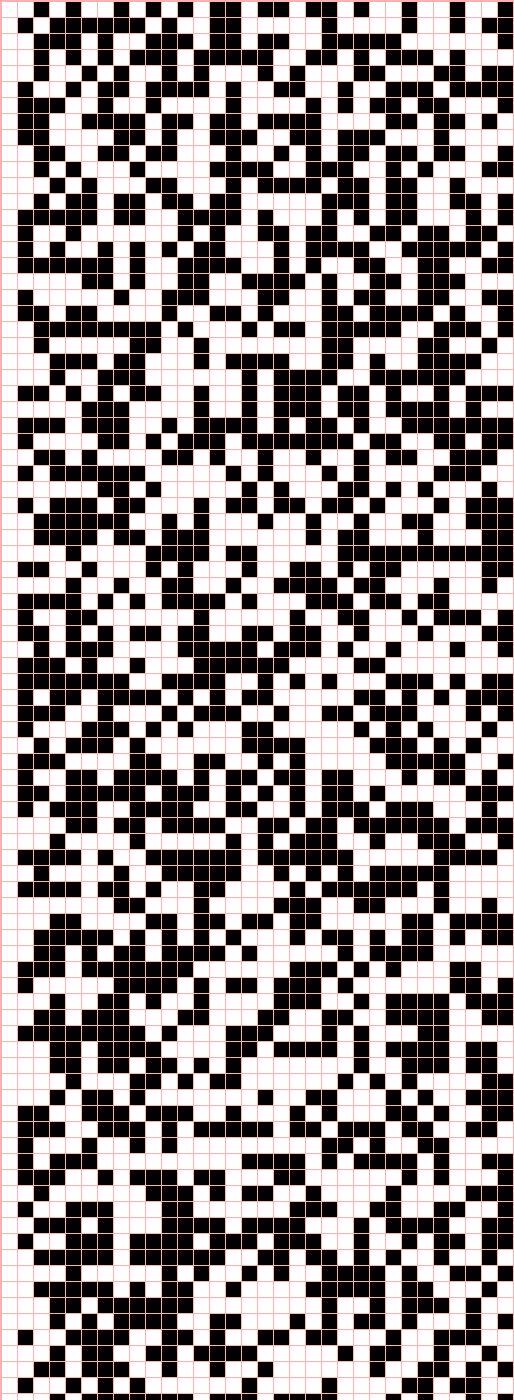}}
\hfill
\subfloat[$s_5$\label{random_32_123456789123456789_spaceo}]{%
\includegraphics[width=0.1\linewidth, height=5.0cm]{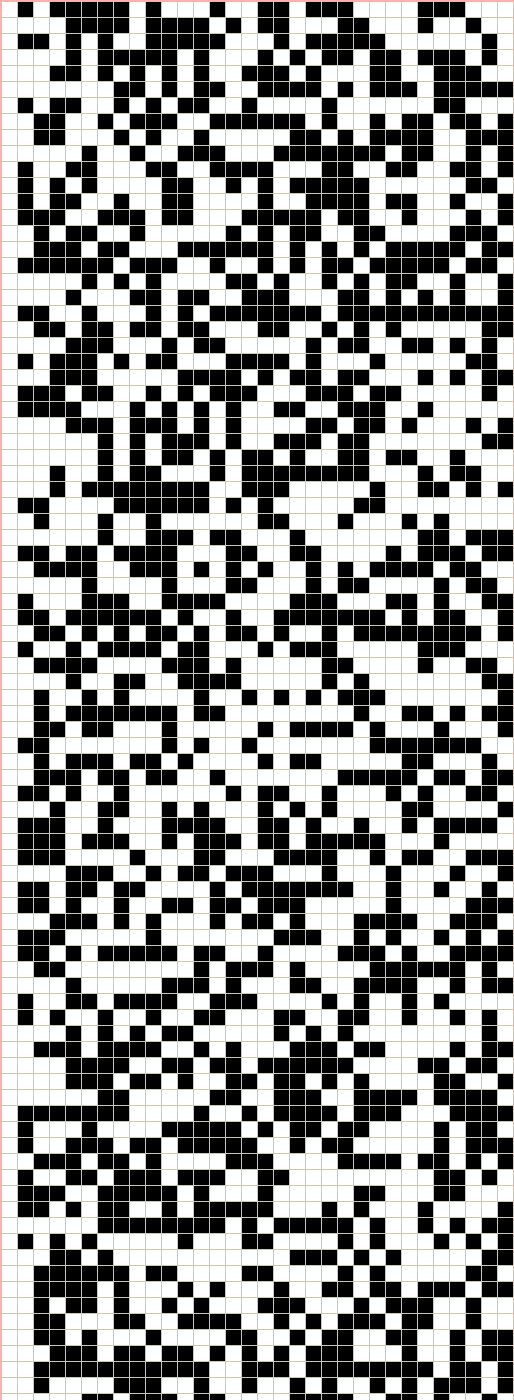}}
\hfill
\subfloat[$s_1$\label{taus_32_7_space}]{%
\includegraphics[width=0.1\linewidth, height=5.0cm]{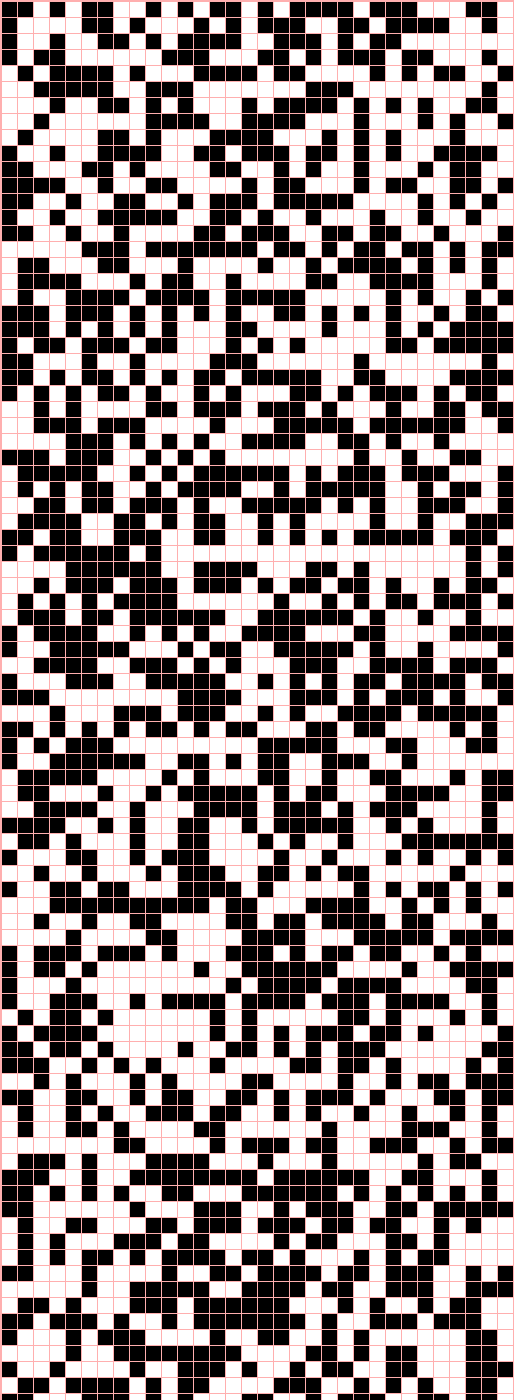}}
\hfill
\subfloat[$s_3$\label{taus_32_12345_space}]{%
 \includegraphics[width=0.1\linewidth, height=5.0cm]{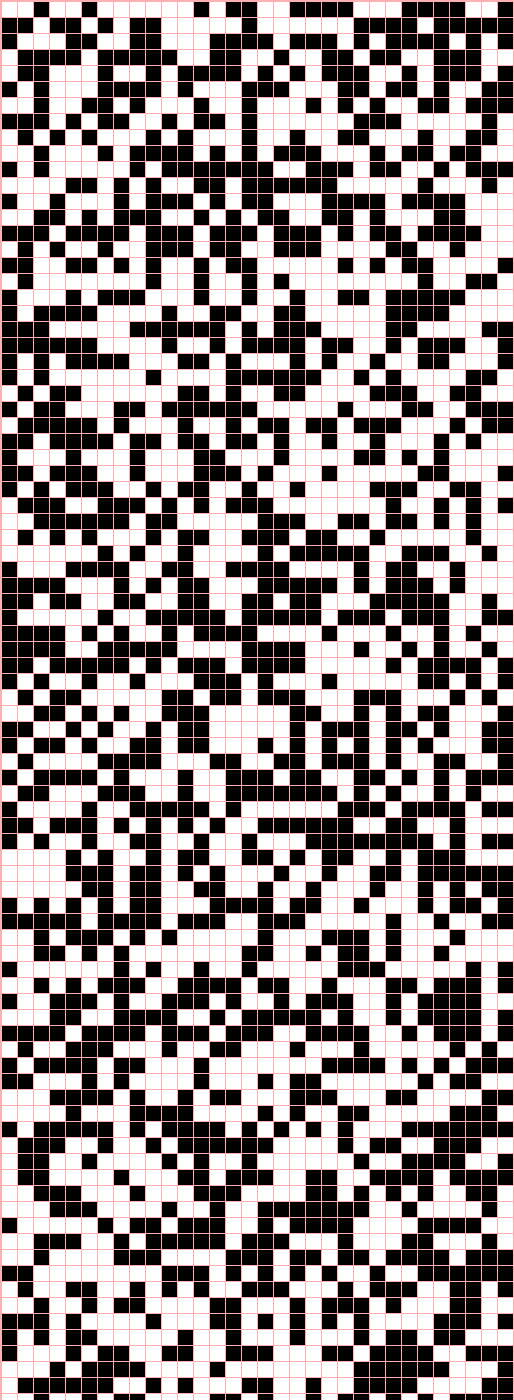}}
\hfill
\subfloat[$s_4$\label{taus_32_9650218_space}]{%
\includegraphics[width=0.1\linewidth, height=5.0cm]{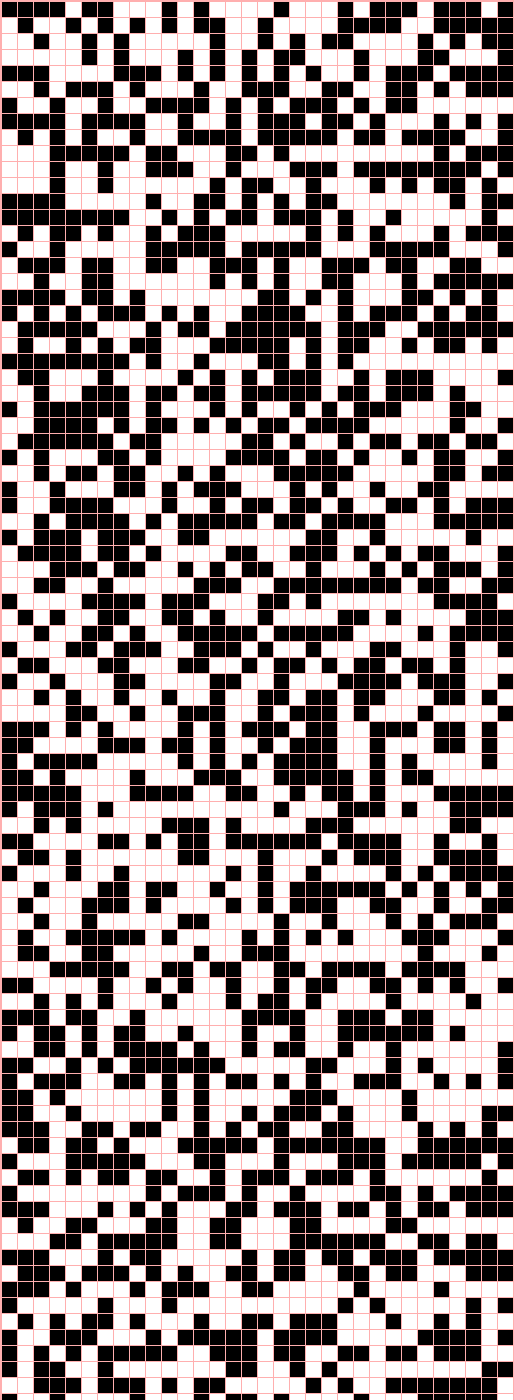}}
\hfill
\subfloat[$s_5$\label{taus_32_123456789123456789_space}]{%
\includegraphics[width=0.1\linewidth, height=5.0cm]{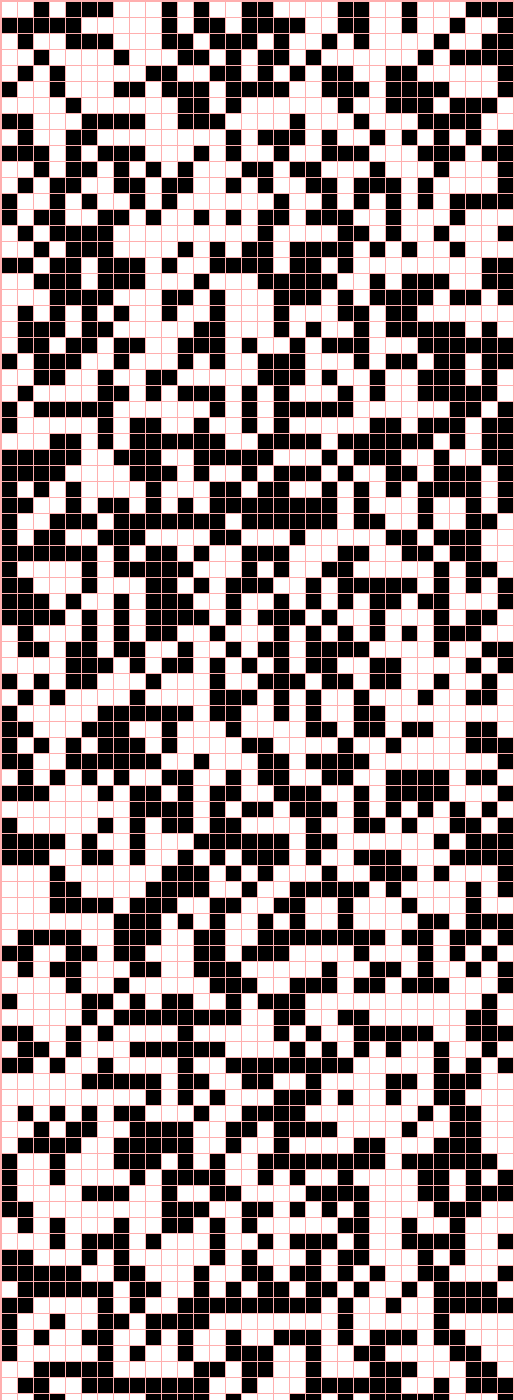}}
%
%
\hfill \\
	\subfloat[$s_1$\label{dsfmt_52_7_space}]{%
		\includegraphics[width=0.2\linewidth, height=5.0cm]{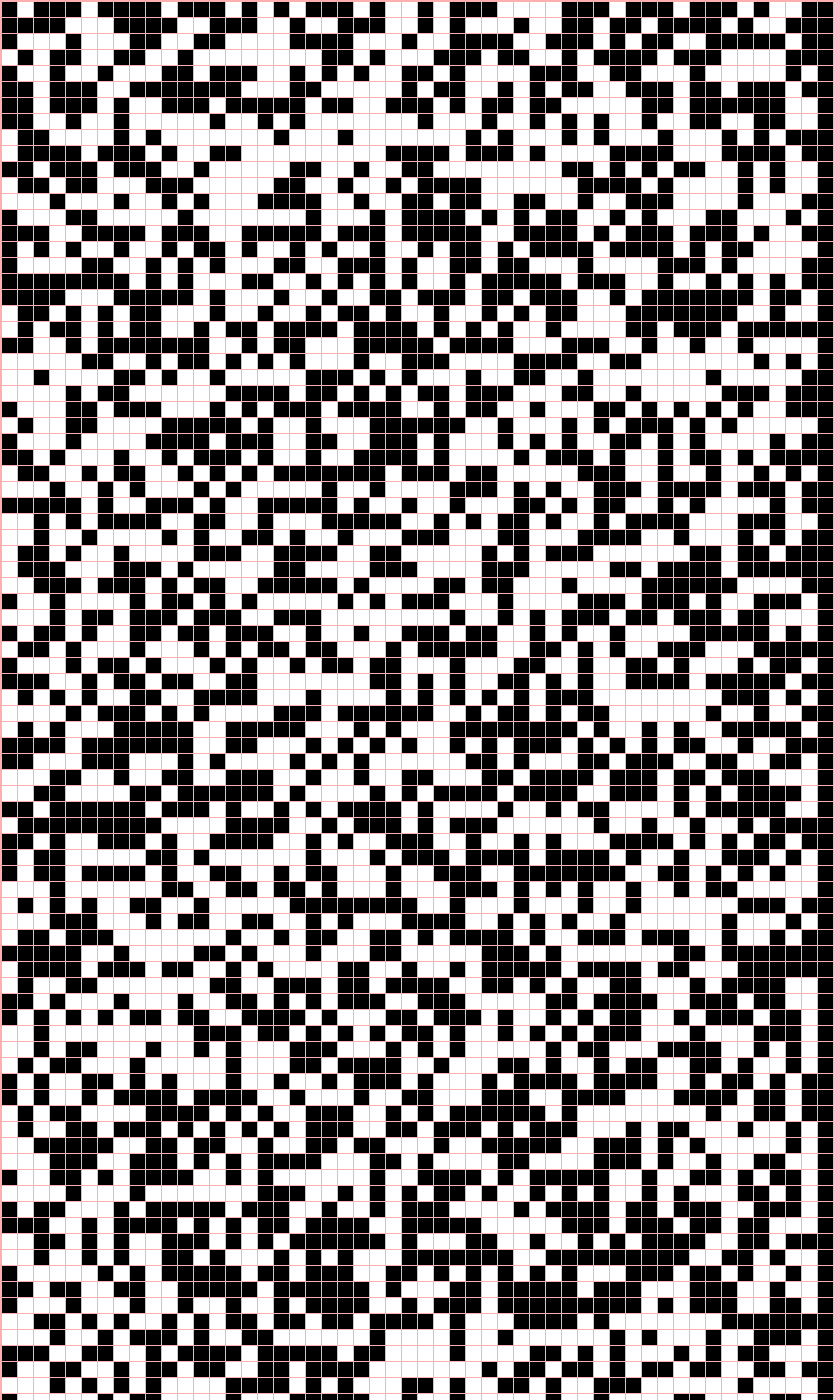}}
	\hfill
	\subfloat[$s_3$\label{dsfmt_52_12345_space}]{%
		\includegraphics[width=0.2\linewidth, height=5.0cm]{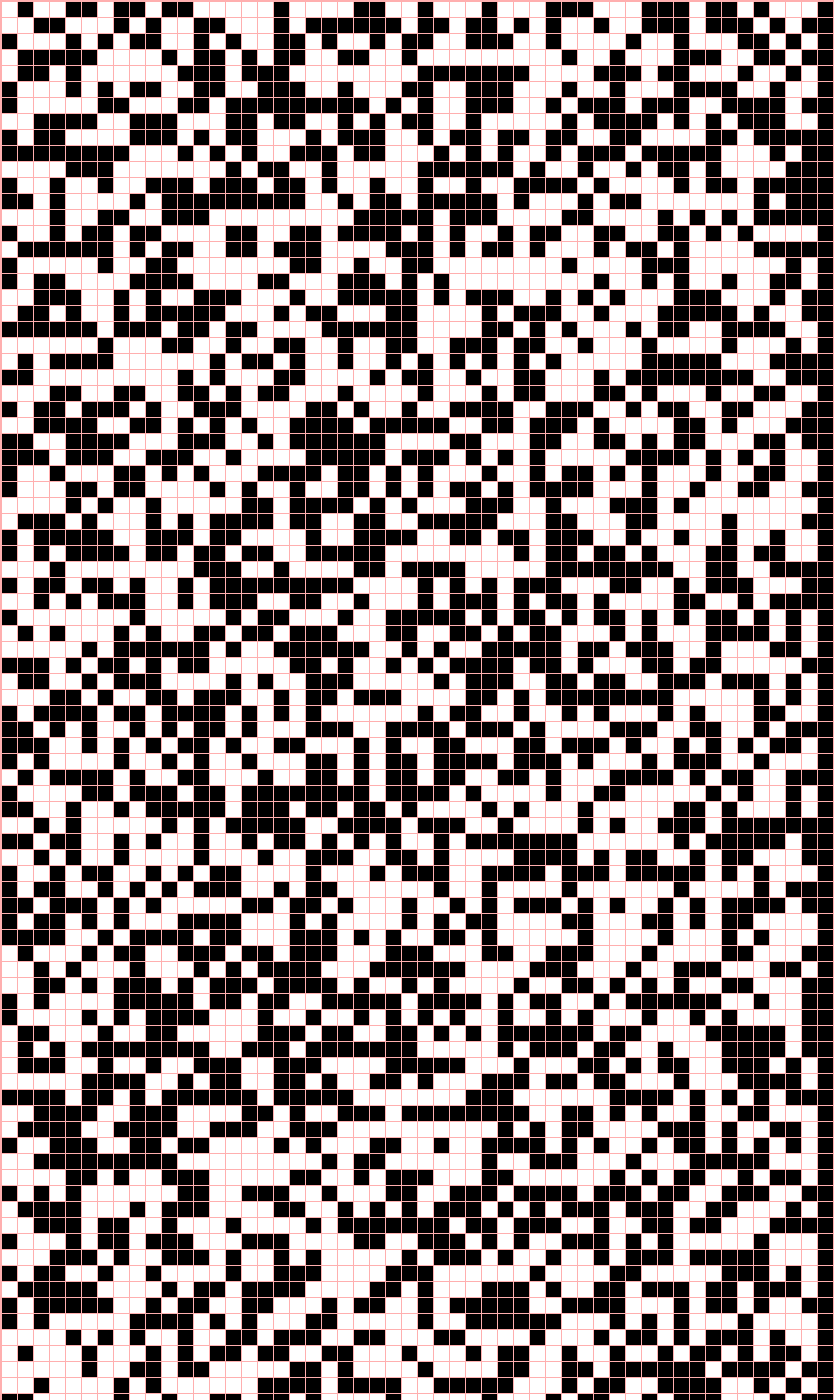}}
	\hfill
	\subfloat[$s_4$\label{dsfmt_52_9650218_space}]{%
		\includegraphics[width=0.2\linewidth, height=5.0cm]{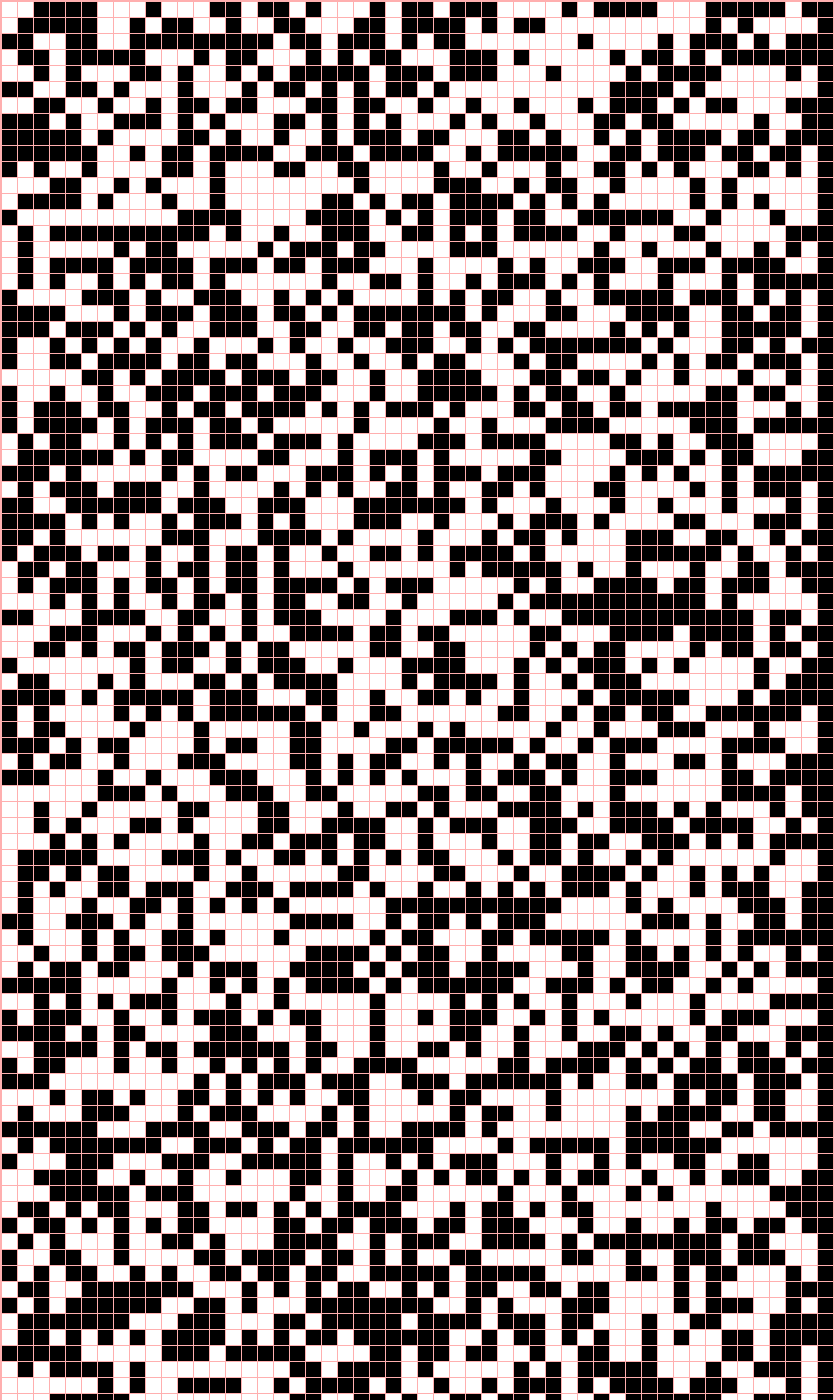}}
	\hfill
	\subfloat[$s_5$\label{dsfmt_52_123456789123456789_space}]{%
		\includegraphics[width=0.2\linewidth, height=5.0cm]{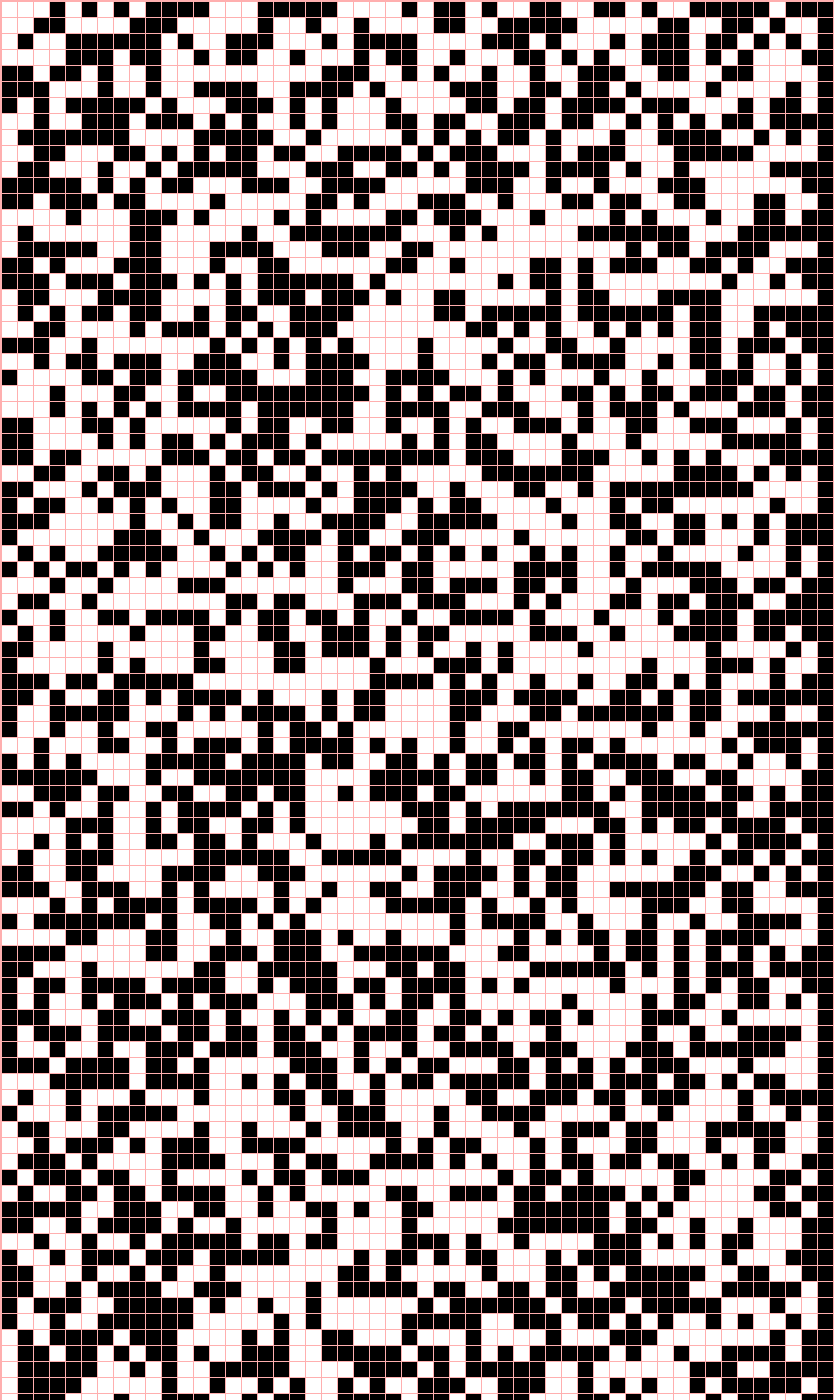}}
		\caption{Space-time diagram for  MRG31k3p (\ref{mrg_7_space} to \ref{mrg_123456789123456789_spaceo}) and PCG $32$-bit (\ref{pcg_32_7_space} to \ref{pcg_32_123456789123456789_space}), random (\ref{random_32_7_space} to \ref{random_32_123456789123456789_spaceo}), Tauss88 (\ref{taus_32_7_space} to \ref{taus_32_123456789123456789_space}) and dSFMT19937 $64$ bit (\ref{dsfmt_52_7_space} to \ref{dsfmt_52_123456789123456789_space})}
		\label{fig:dsfmt32_space-time}
	\end{figure}

\begin{figure}[hbtp]
\centering
  \vspace{-2.0em}
\subfloat[$s_1$\label{lfsr258_7_space}]{%
\includegraphics[width=0.2\linewidth, height=5.0cm]{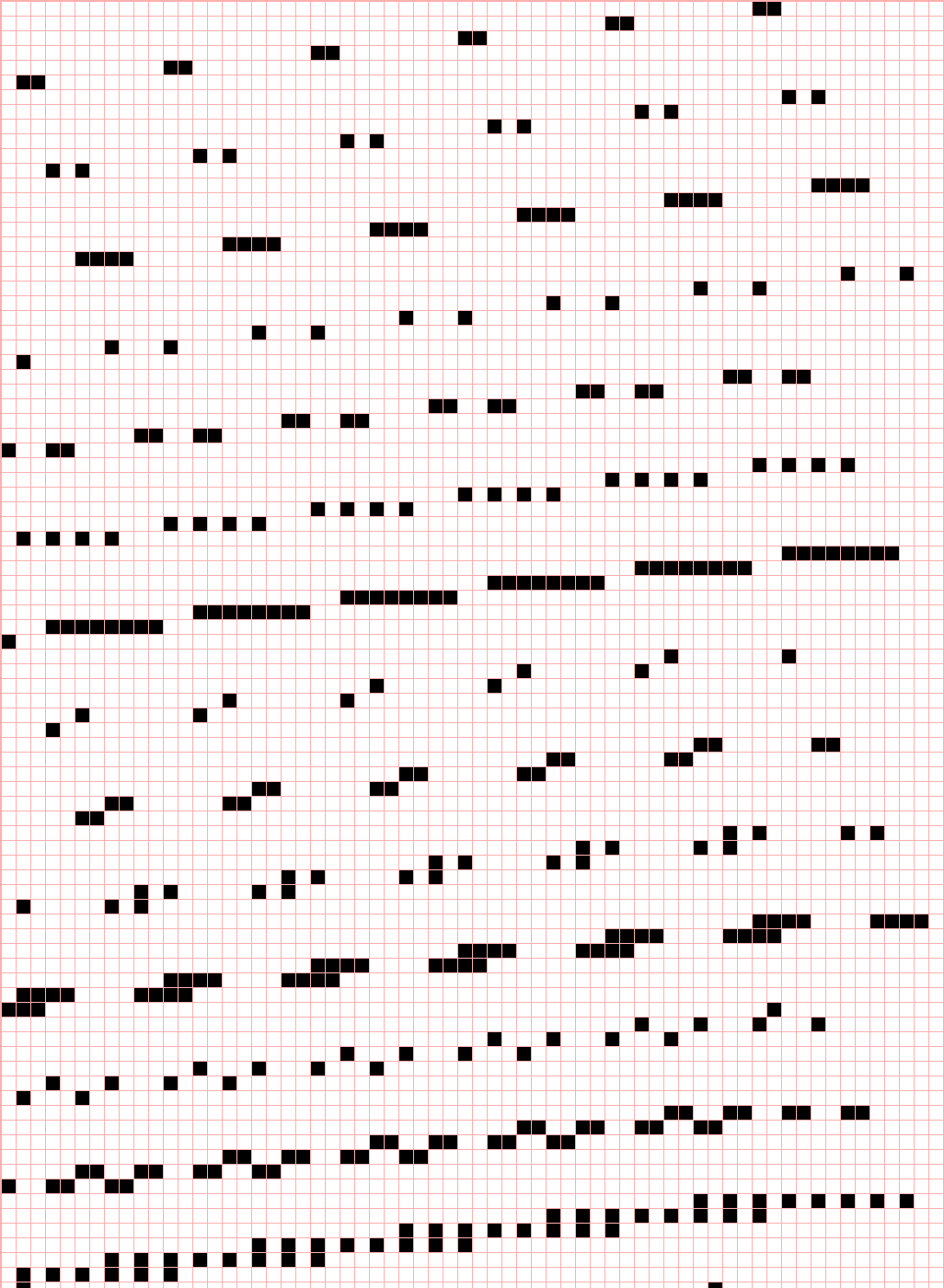}}
\hfill
\subfloat[$s_3$\label{lfsr258_12345_space}]{%
 \includegraphics[width=0.2\linewidth, height=5.0cm]{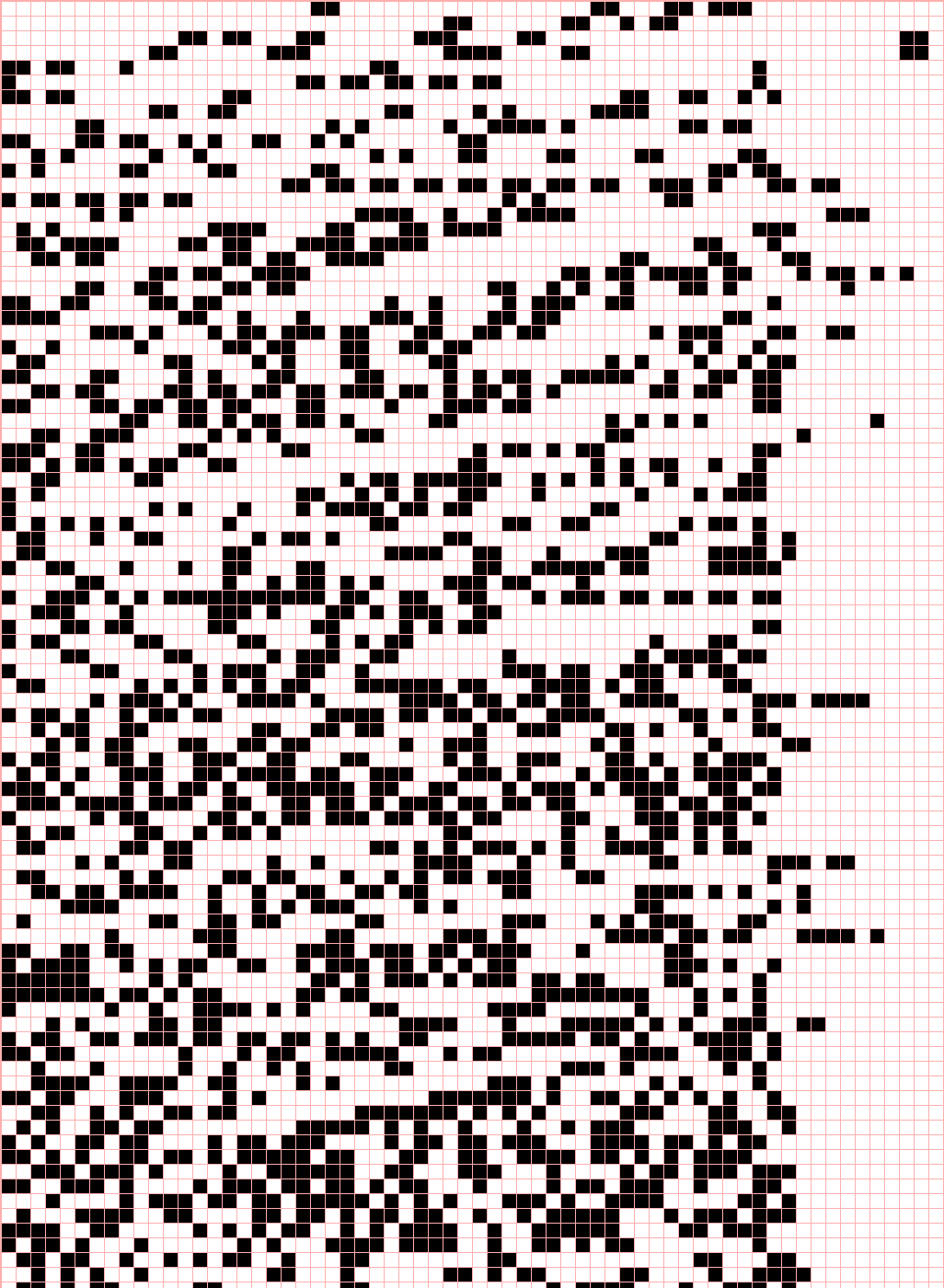}}
\hfill
\subfloat[$s_4$\label{lfsr258_9650218_space}]{%
\includegraphics[width=0.2\linewidth, height=5.0cm]{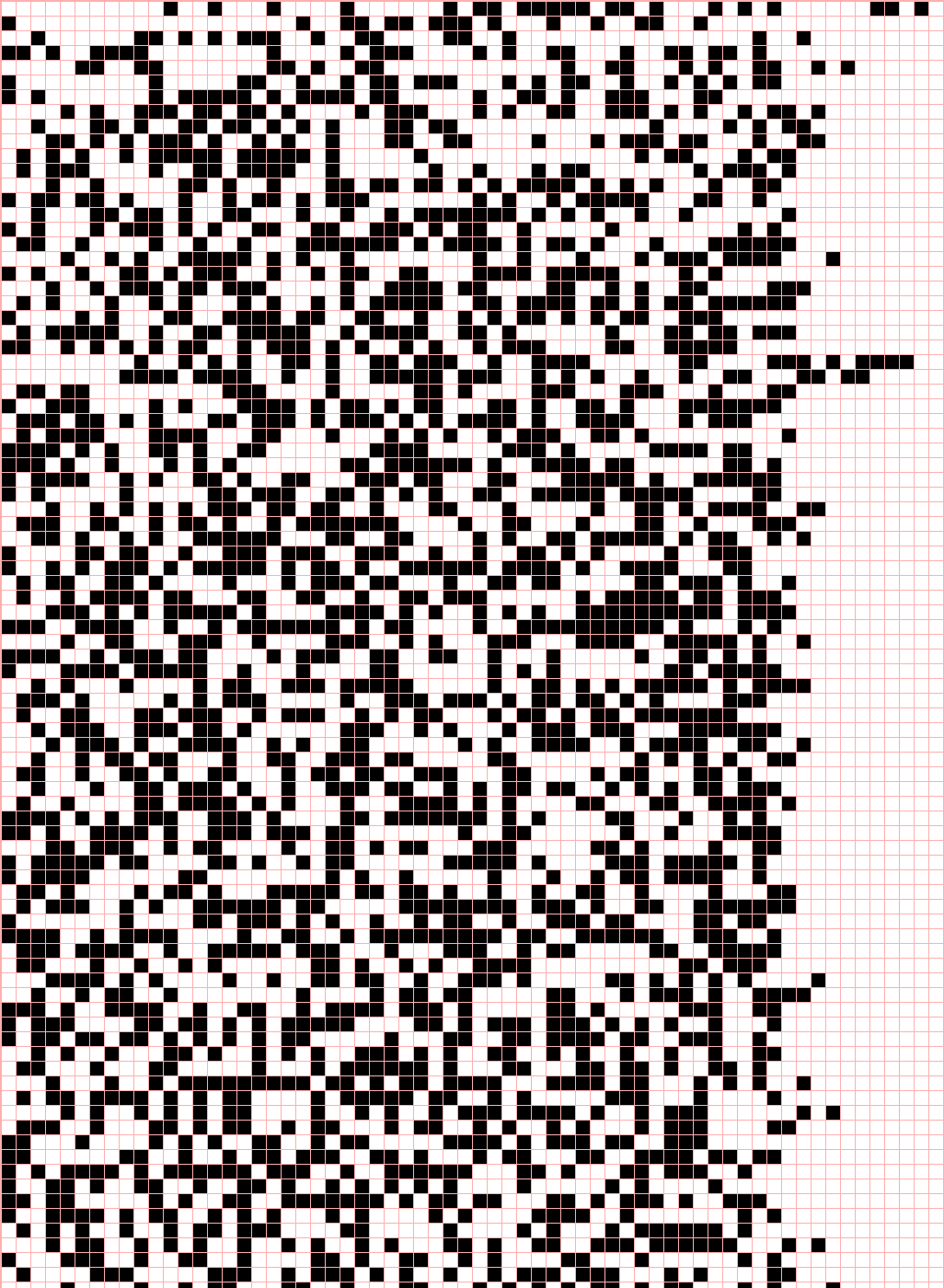}}
\hfill
\subfloat[$s_5$\label{lfsr258_123456789123456789_space}]{%
\includegraphics[width=0.2\linewidth, height=5.0cm]{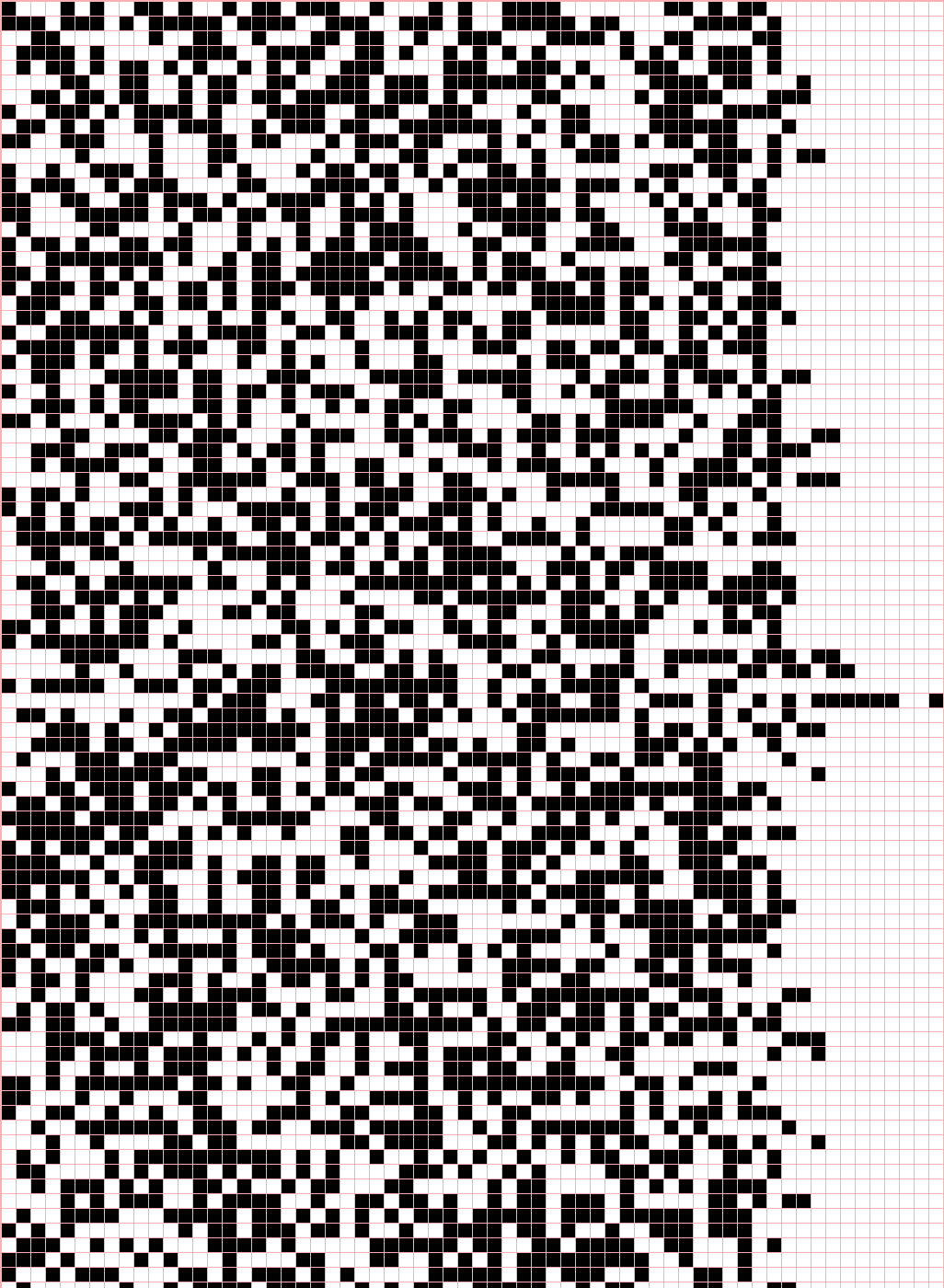}}
%
%
\hfill\\
\subfloat[$s_1$\label{lfsr_double_7_space}]{%
\includegraphics[width=0.1\linewidth, height=5.0cm]{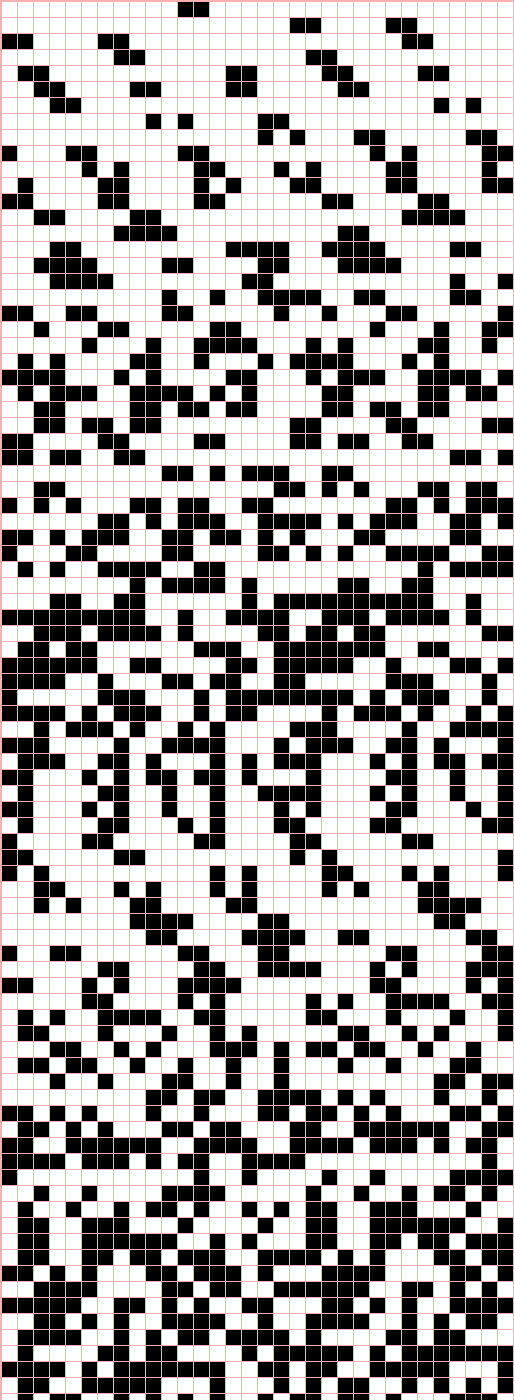}}
\hfill
\subfloat[$s_3$\label{lfsr_double_12345_space}]{%
 \includegraphics[width=0.1\linewidth, height=5.0cm]{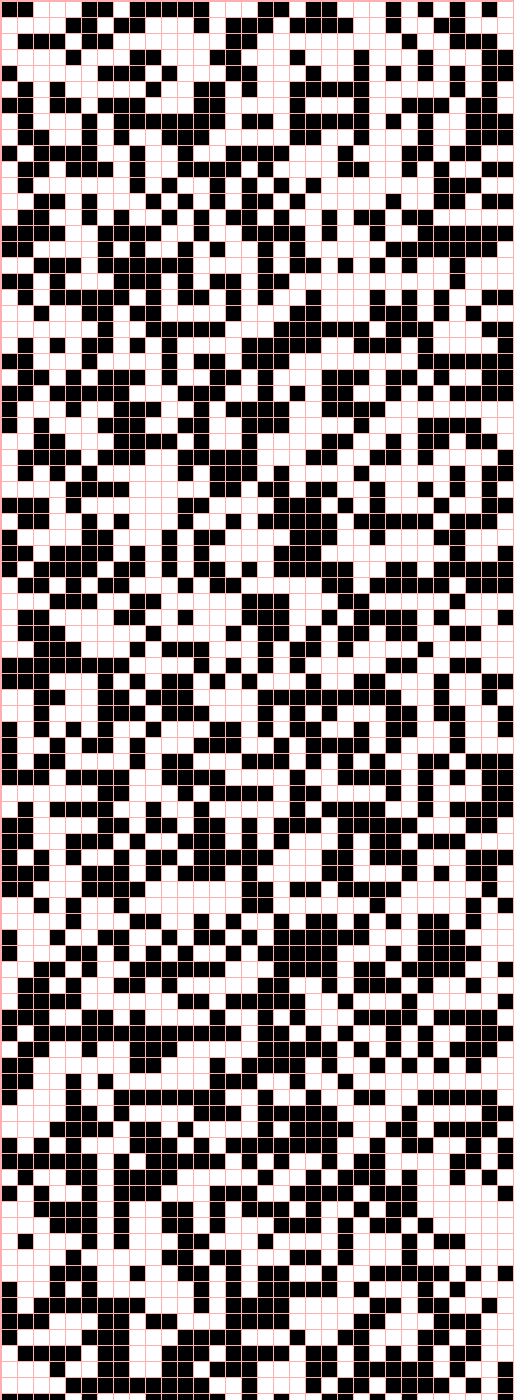}}
\hfill
\subfloat[$s_4$\label{lfsr_double_9650218_space}]{%
\includegraphics[width=0.1\linewidth, height=5.0cm]{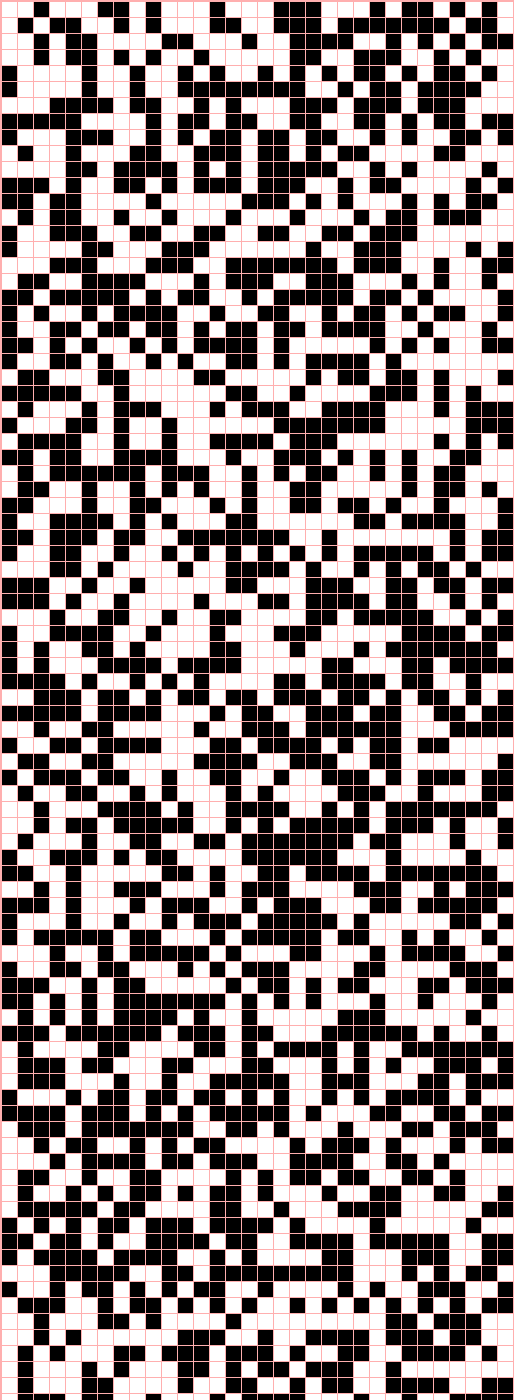}}
\hfill
\subfloat[$s_5$\label{lfsr_double_123456789123456789_spaceo}]{%
\includegraphics[width=0.1\linewidth, height=5.0cm]{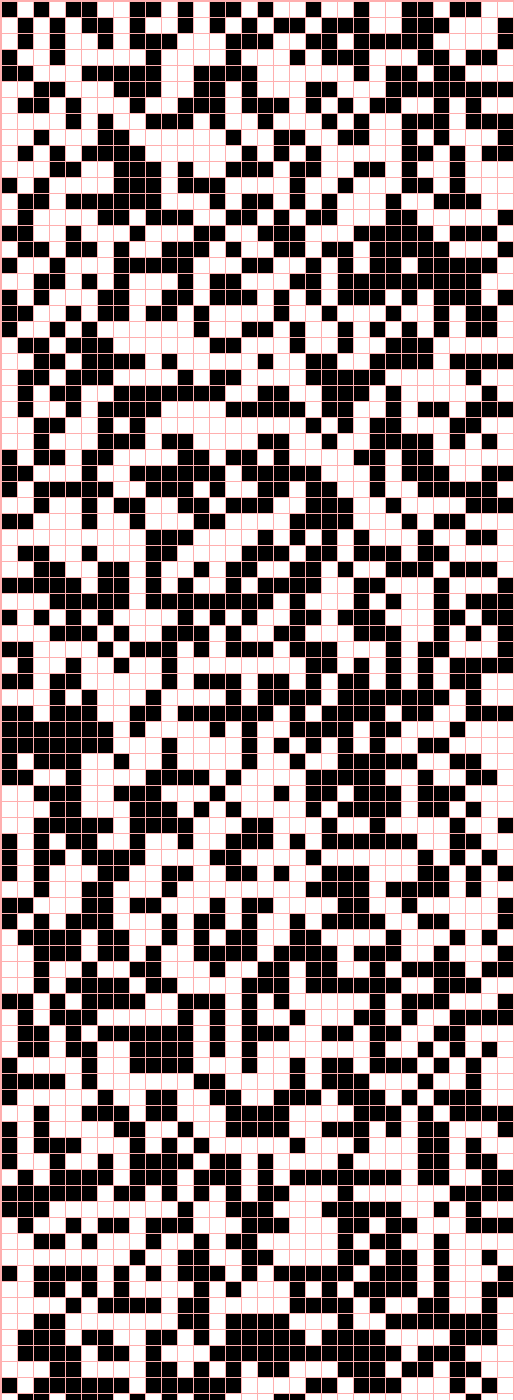}}
\hfill
\subfloat[$s_1$\label{xor32_7_space}]{%
\includegraphics[width=0.1\linewidth, height=5.0cm]{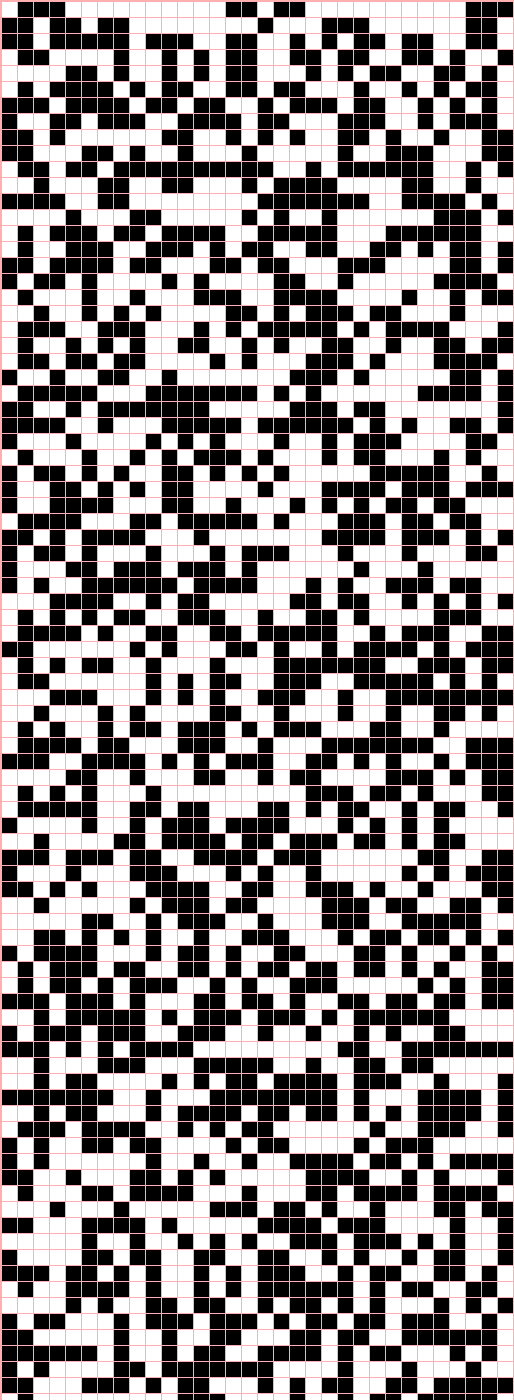}}
\hfill
\subfloat[$s_3$\label{xor32_12345_space}]{%
 \includegraphics[width=0.1\linewidth, height=5.0cm]{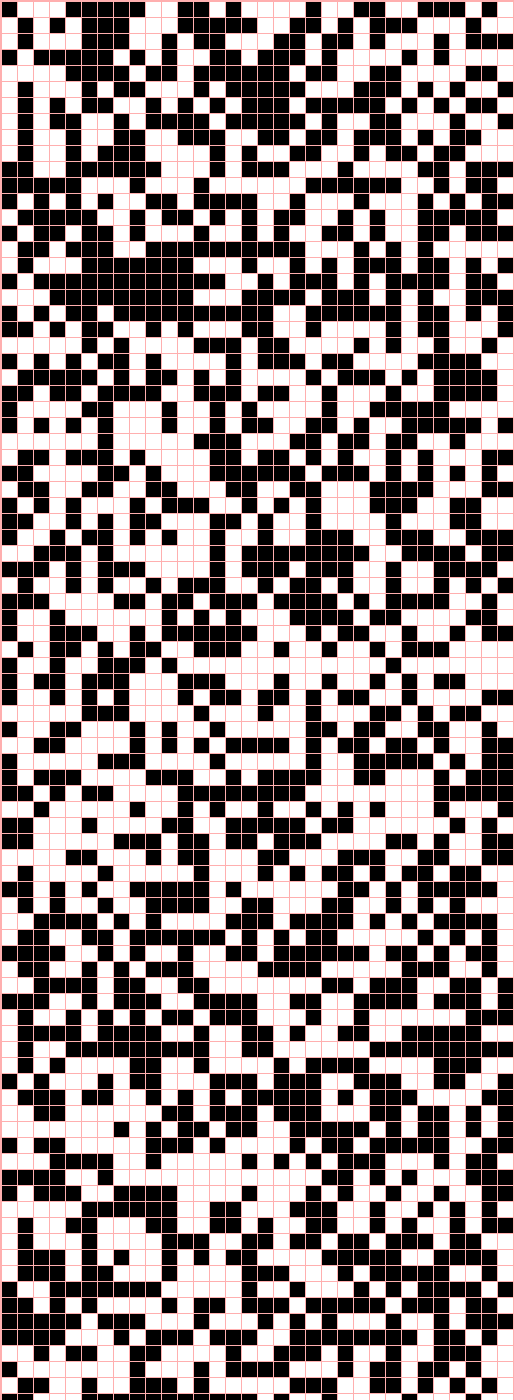}}
\hfill
\subfloat[$s_4$\label{xor32_9650218_space}]{%
\includegraphics[width=0.1\linewidth, height=5.0cm]{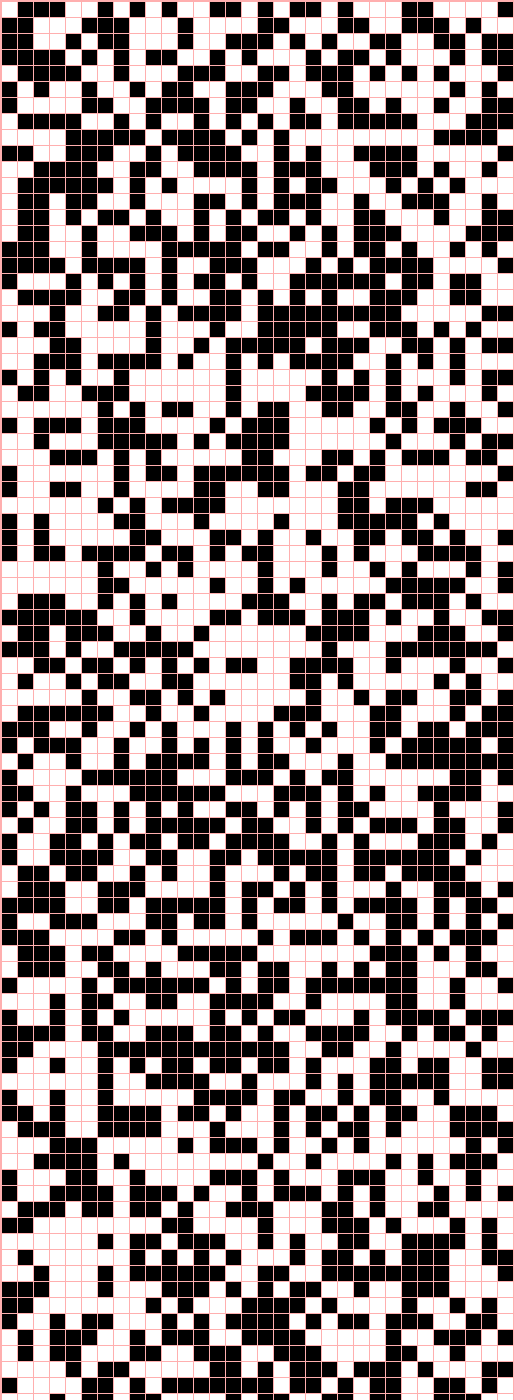}}
\hfill
\subfloat[$s_5$\label{xor32_123456789123456789_spaceo}]{%
\includegraphics[width=0.1\linewidth, height=5.0cm]{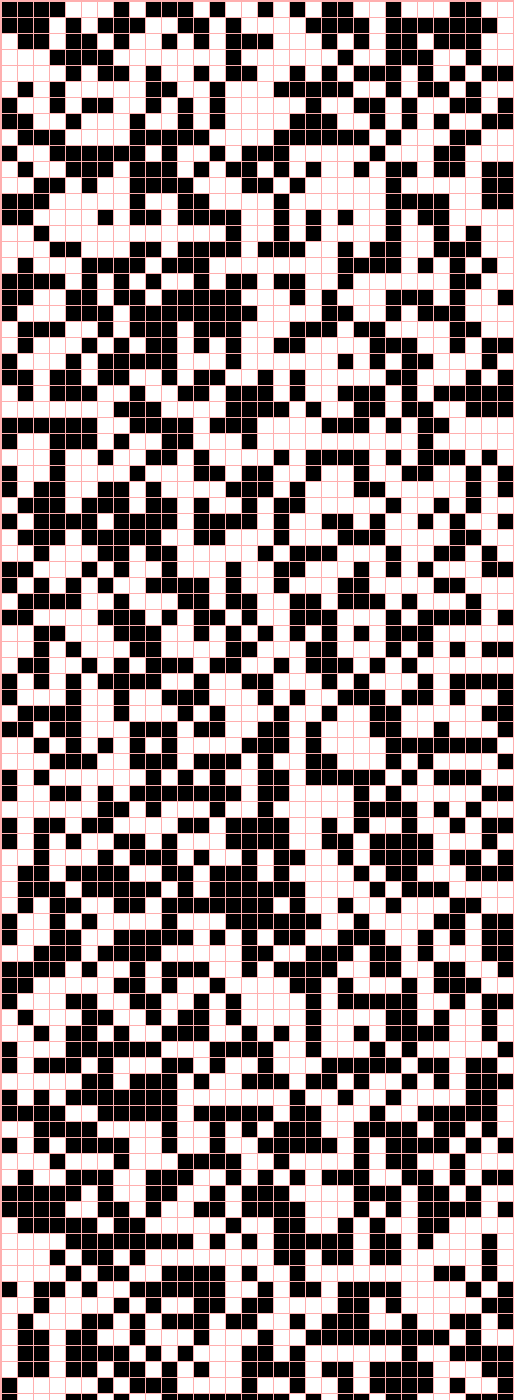}}
%
\hfill\\
	\subfloat[$s_1$\label{well512_7_space}]{%
		\includegraphics[width=0.1\linewidth, height=5.0cm]{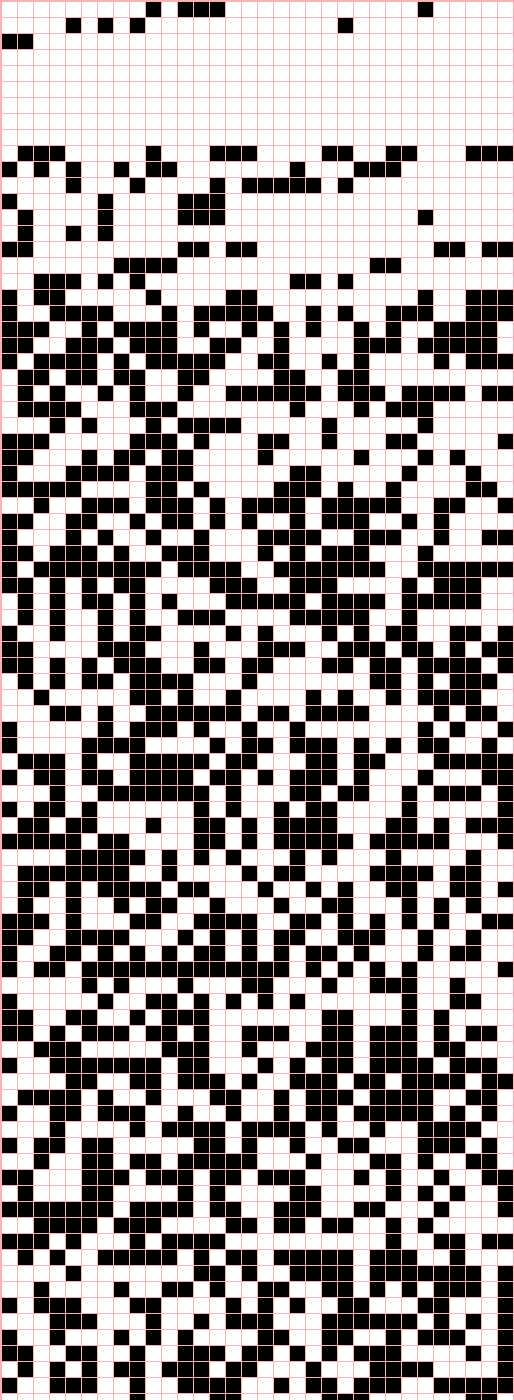}}
	\hfill
	\subfloat[$s_3$\label{well512_12345_space}]{%
		\includegraphics[width=0.1\linewidth, height=5.0cm]{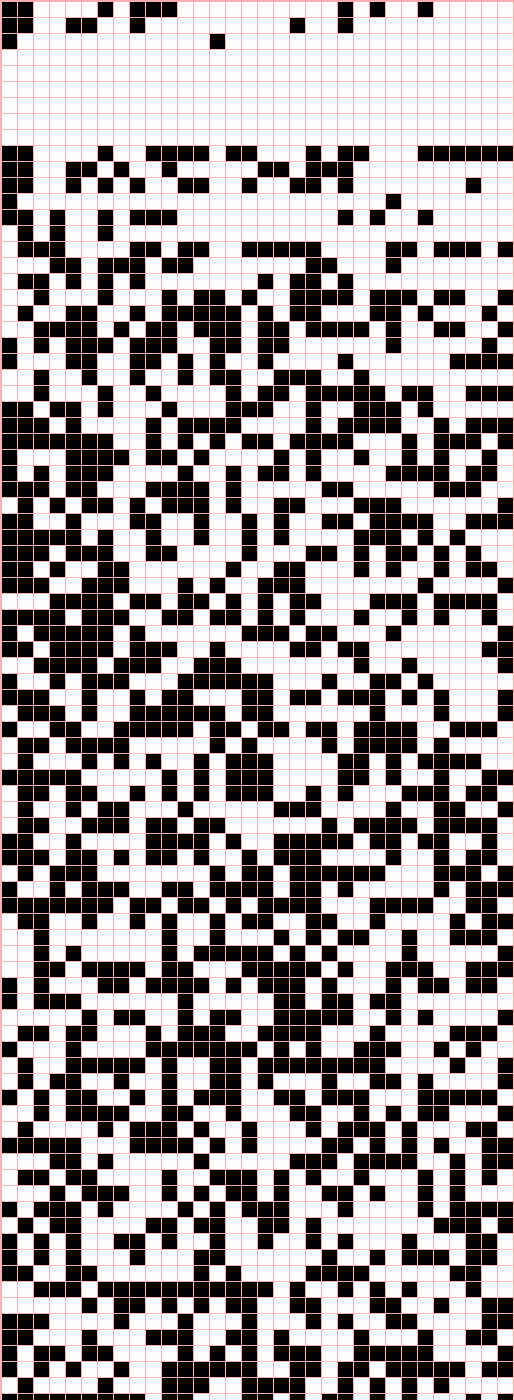}}
	\hfill
	\subfloat[$s_4$\label{well512_9650218_space}]{%
		\includegraphics[width=0.1\linewidth, height=5.0cm]{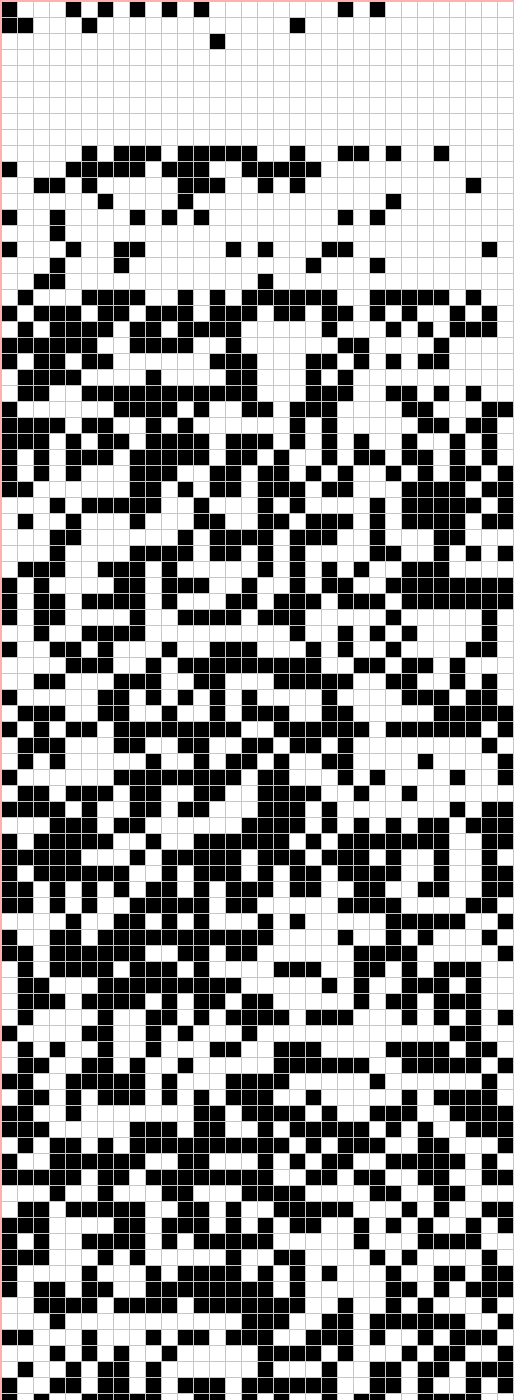}}
	\hfill
	\subfloat[$s_5$\label{well512_123456789123456789_spaceo}]{%
		\includegraphics[width=0.1\linewidth, height=5.0cm]{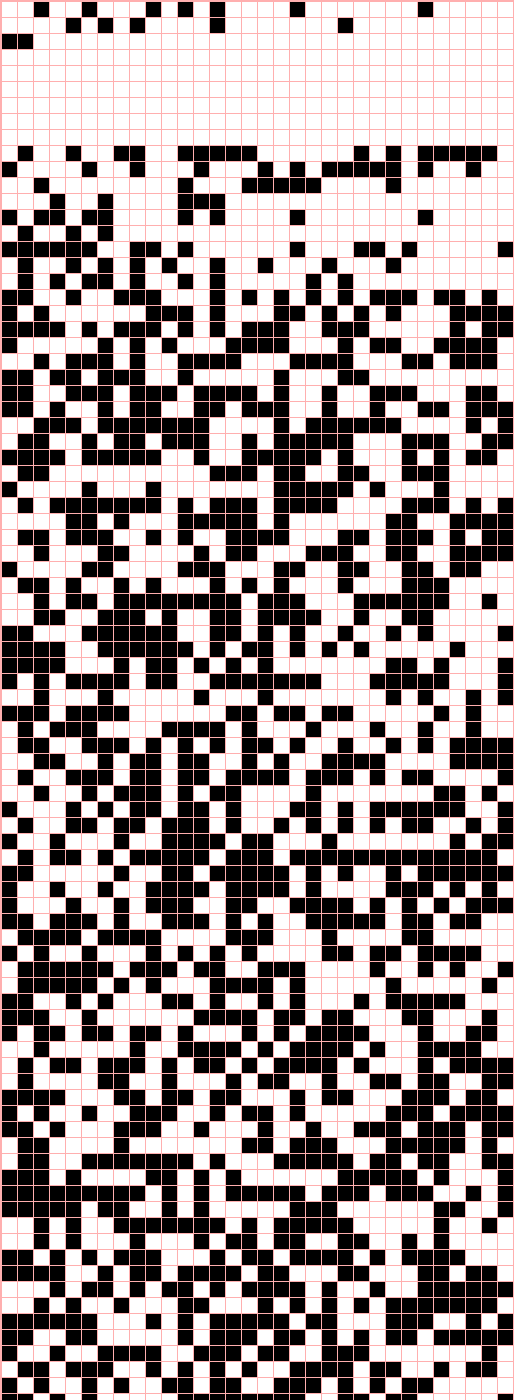}}
	\hfill
	\subfloat[$s_1$\label{well1024_7_space}]{%
		\includegraphics[width=0.1\linewidth, height=5.0cm]{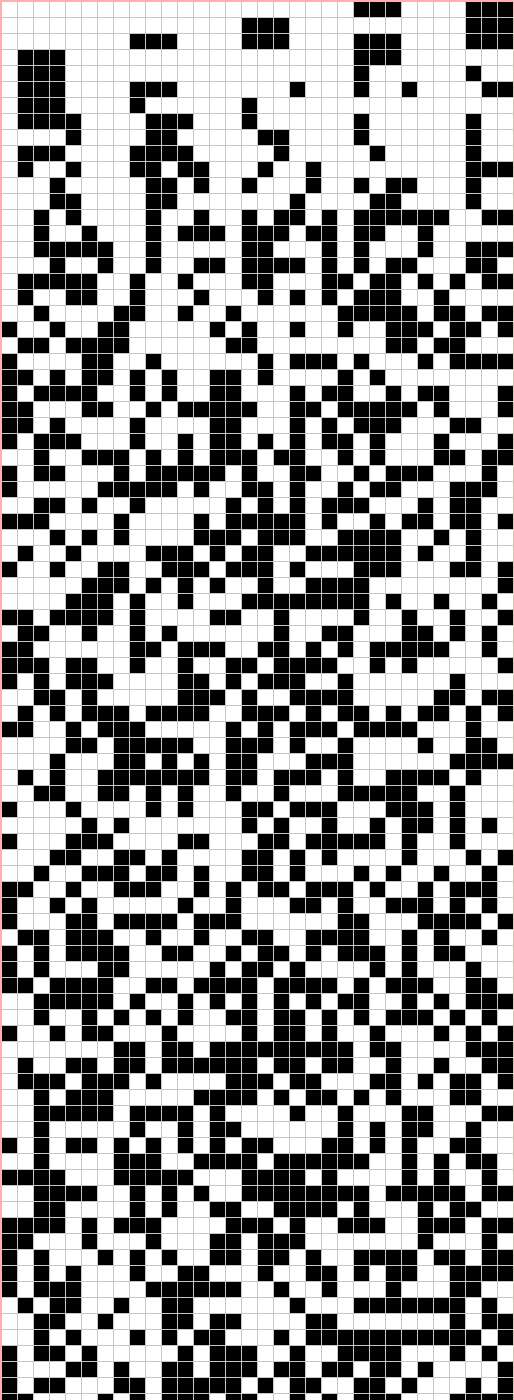}}
	\hfill
	\subfloat[$s_3$\label{well1024_12345_space}]{%
		\includegraphics[width=0.1\linewidth, height=5.0cm]{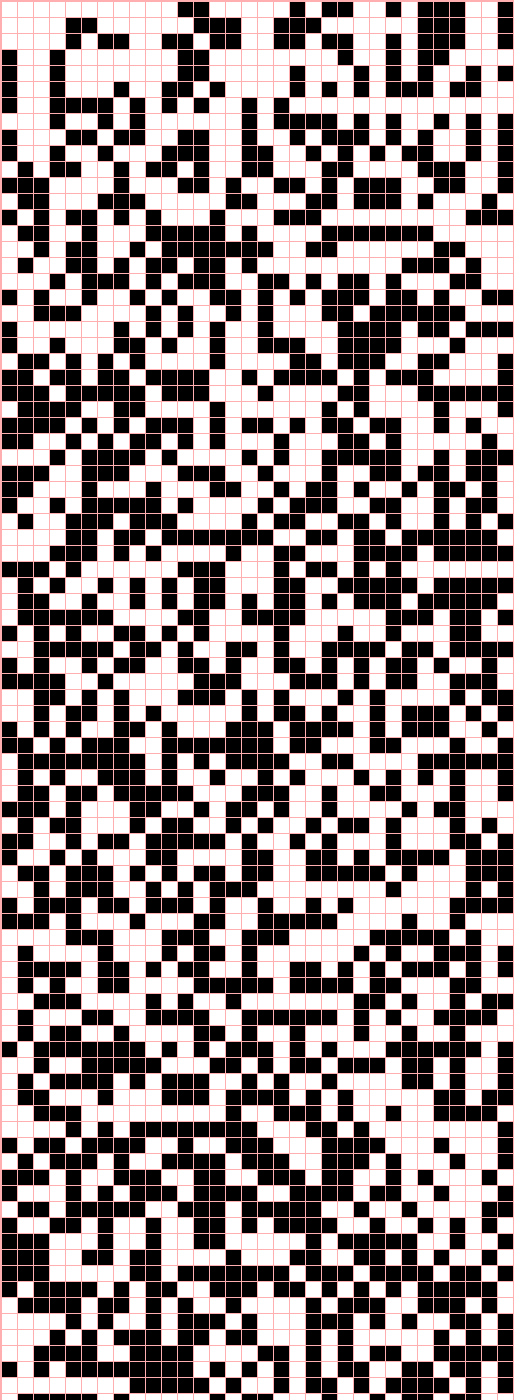}}
	\hfill
	\subfloat[$s_4$\label{well1024_9650218_space}]{%
		\includegraphics[width=0.1\linewidth, height=5.0cm]{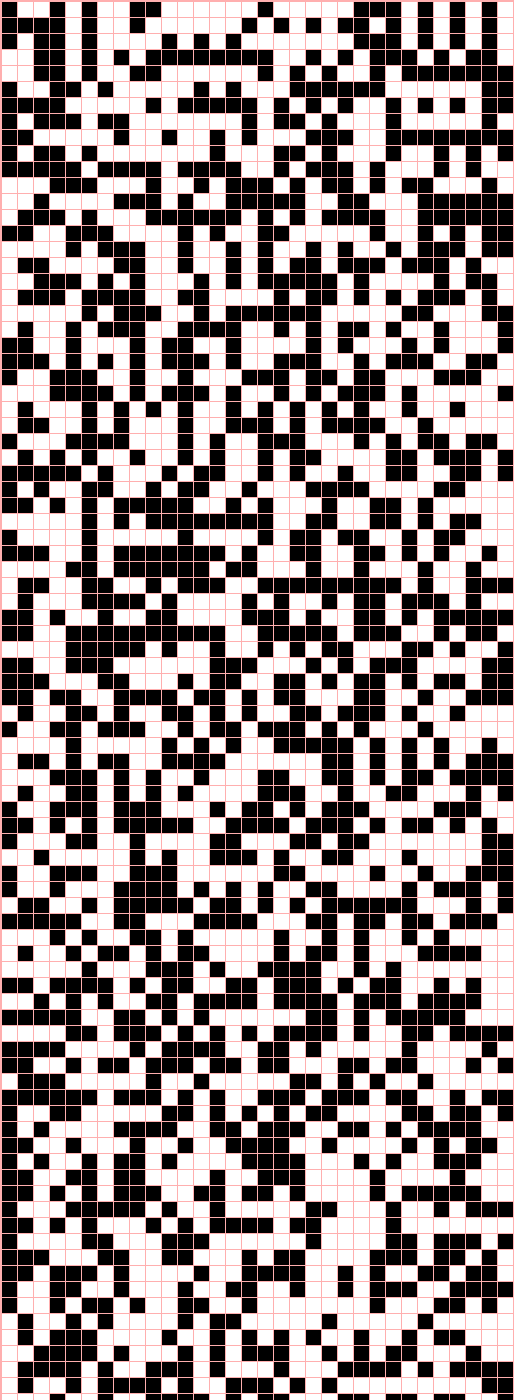}}
	\hfill
	\subfloat[$s_5$\label{well1024_123456789123456789_space}]{%
		\includegraphics[width=0.1\linewidth, height=5.0cm]{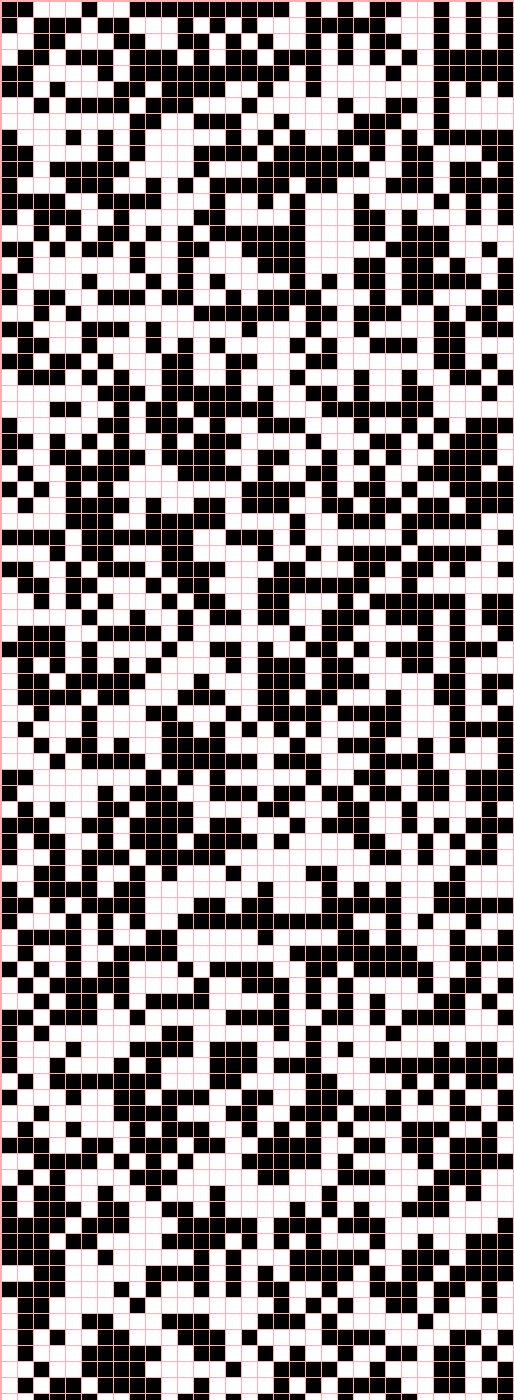}}
	\caption{Space-time diagram for LFSR258 (\ref{lfsr258_7_space} to \ref{lfsr258_123456789123456789_space}), LFSR113 (\ref{lfsr_double_7_space} to \ref{lfsr_double_123456789123456789_spaceo}) and xorshift (\ref{xor32_7_space} to \ref{xor32_123456789123456789_spaceo}), WELL512 (\ref{well512_7_space} to \ref{well512_123456789123456789_spaceo}) and WELL1024a (\ref{well1024_7_space} to \ref{well1024_123456789123456789_space})}
	\label{fig:well_space-time}
\end{figure} 

\begin{figure}[hbtp]
\centering
  \vspace{-2.0em}
\subfloat[$s_1$\label{xorshift64_7_space}]{%
\includegraphics[width=0.2\linewidth, height=5.0cm]{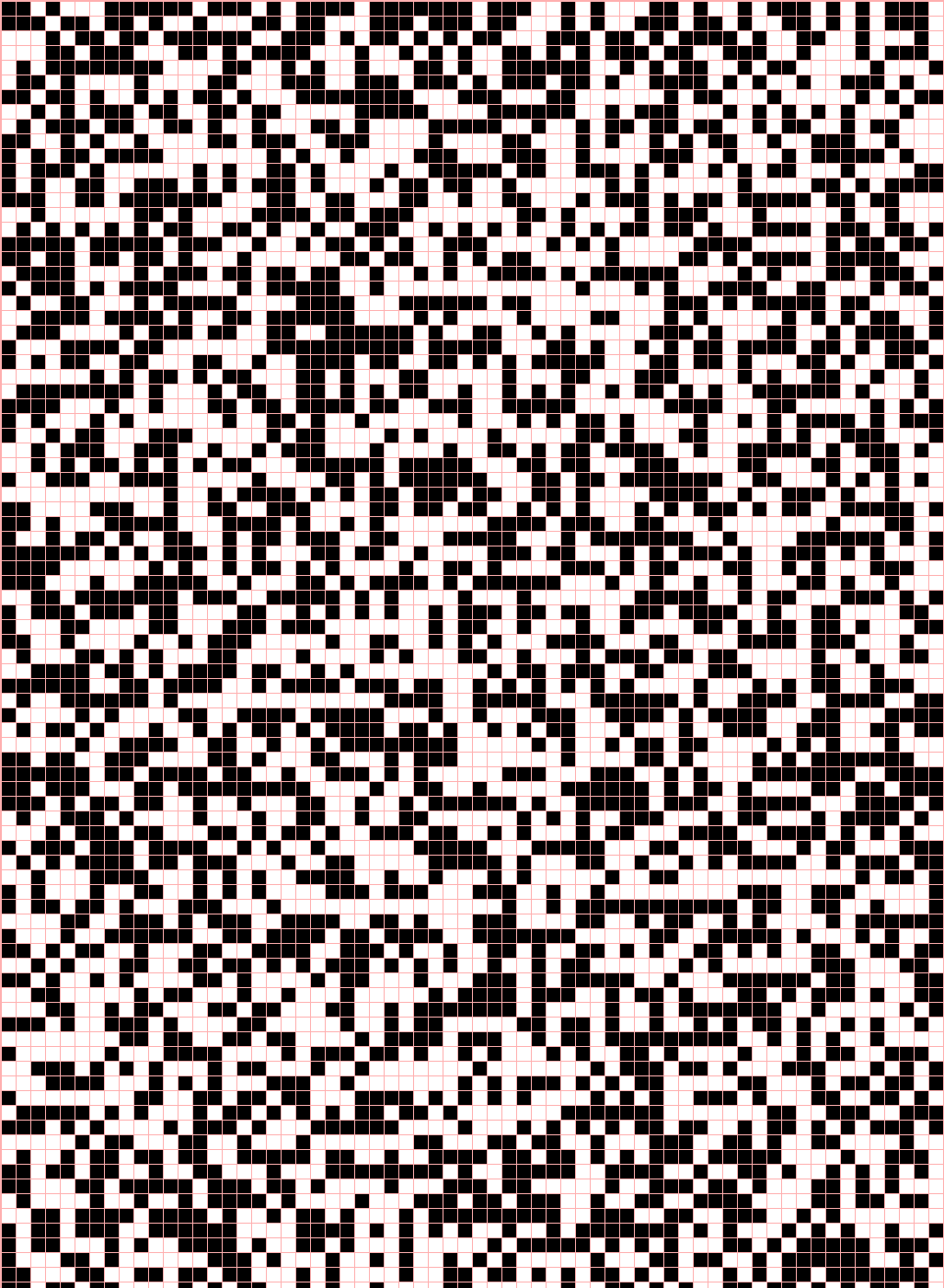}}
\hfill
\subfloat[$s_3$\label{xorshift64_12345_space}]{%
 \includegraphics[width=0.2\linewidth, height=5.0cm]{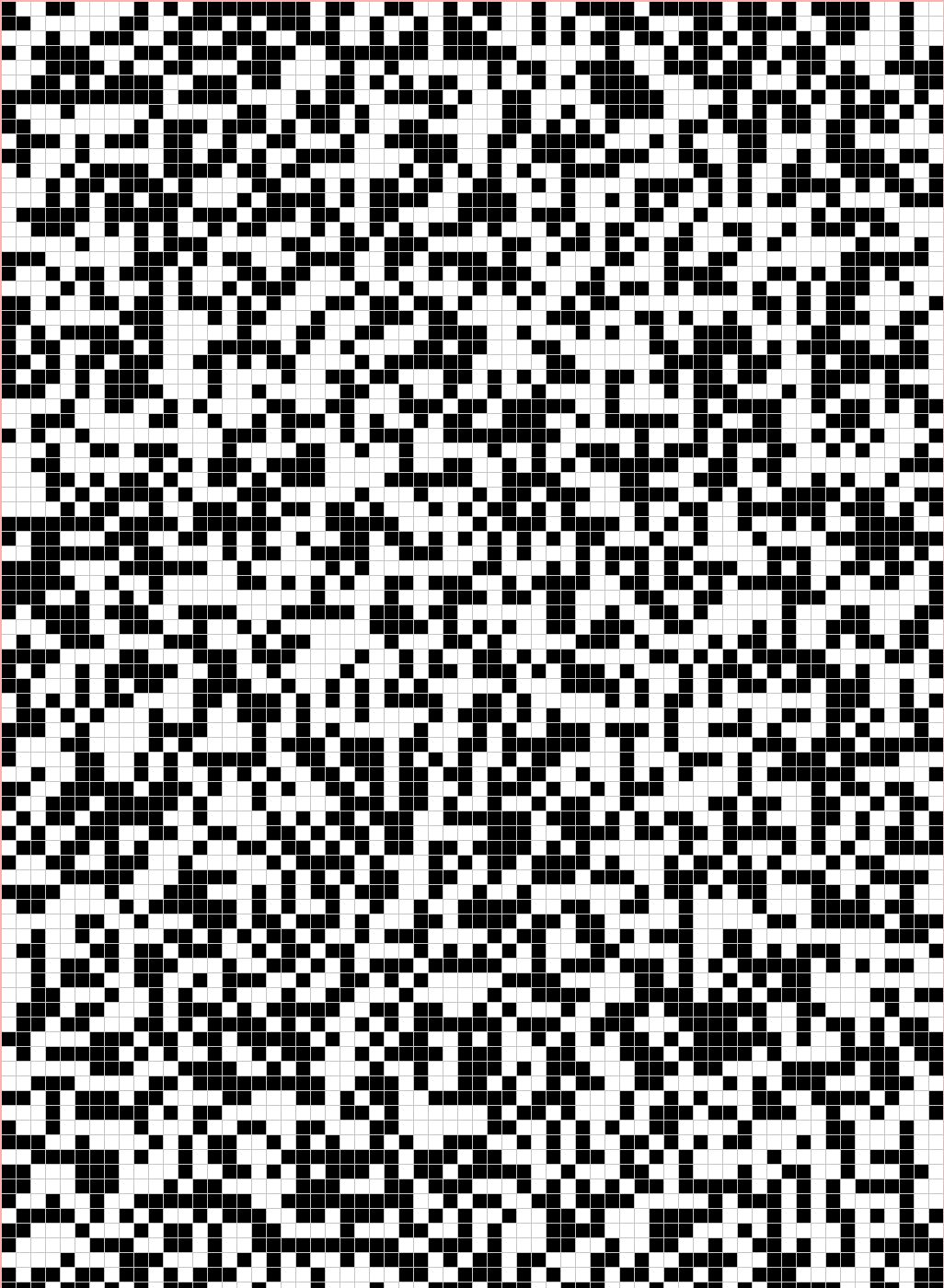}}
\hfill
\subfloat[$s_4$\label{xorshift64_9650218_space}]{%
\includegraphics[width=0.2\linewidth, height=5.0cm]{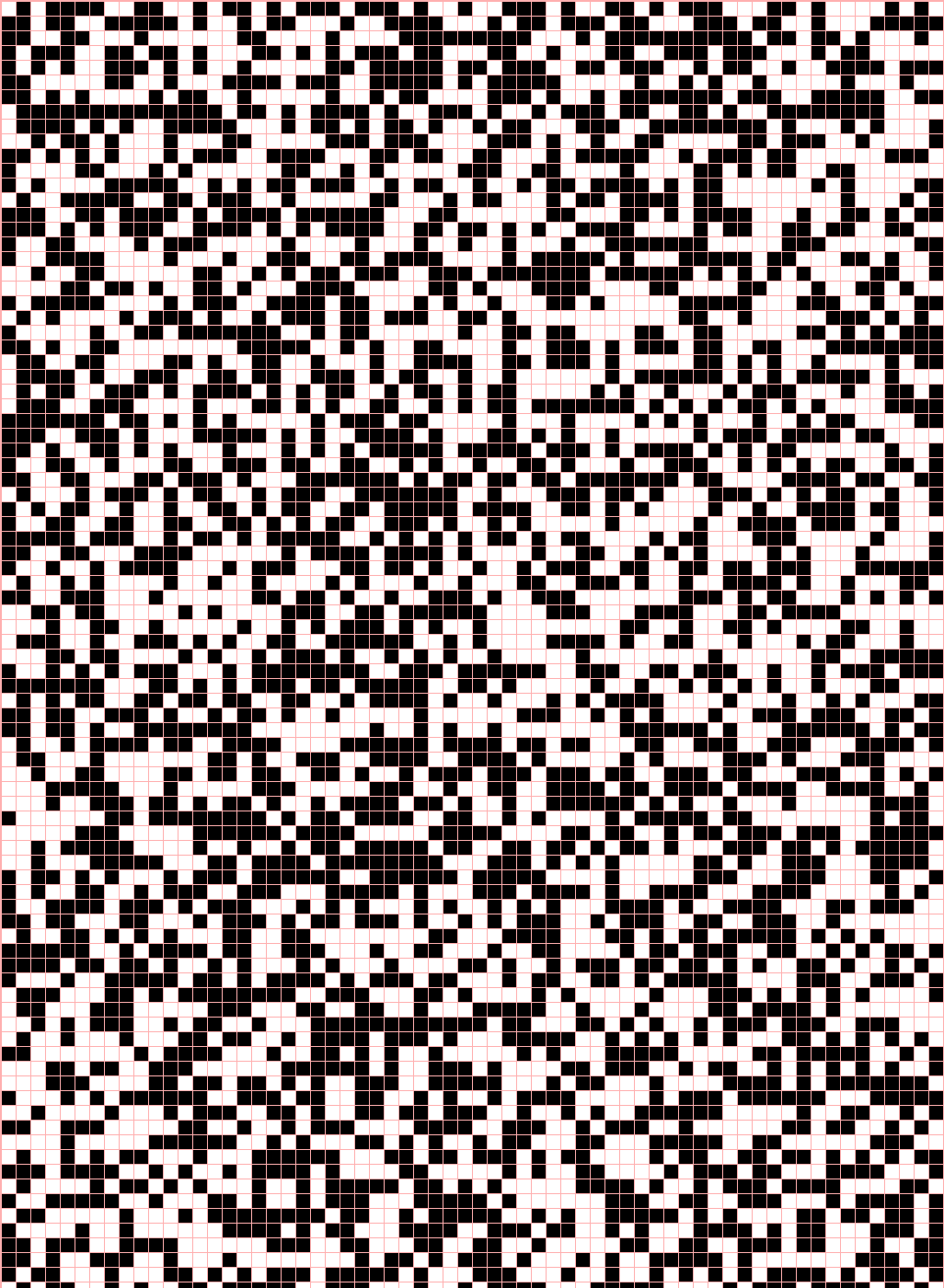}}
\hfill
\subfloat[$s_5$\label{xorshift64_123456789123456789_spaceo}]{%
\includegraphics[width=0.2\linewidth, height=5.0cm]{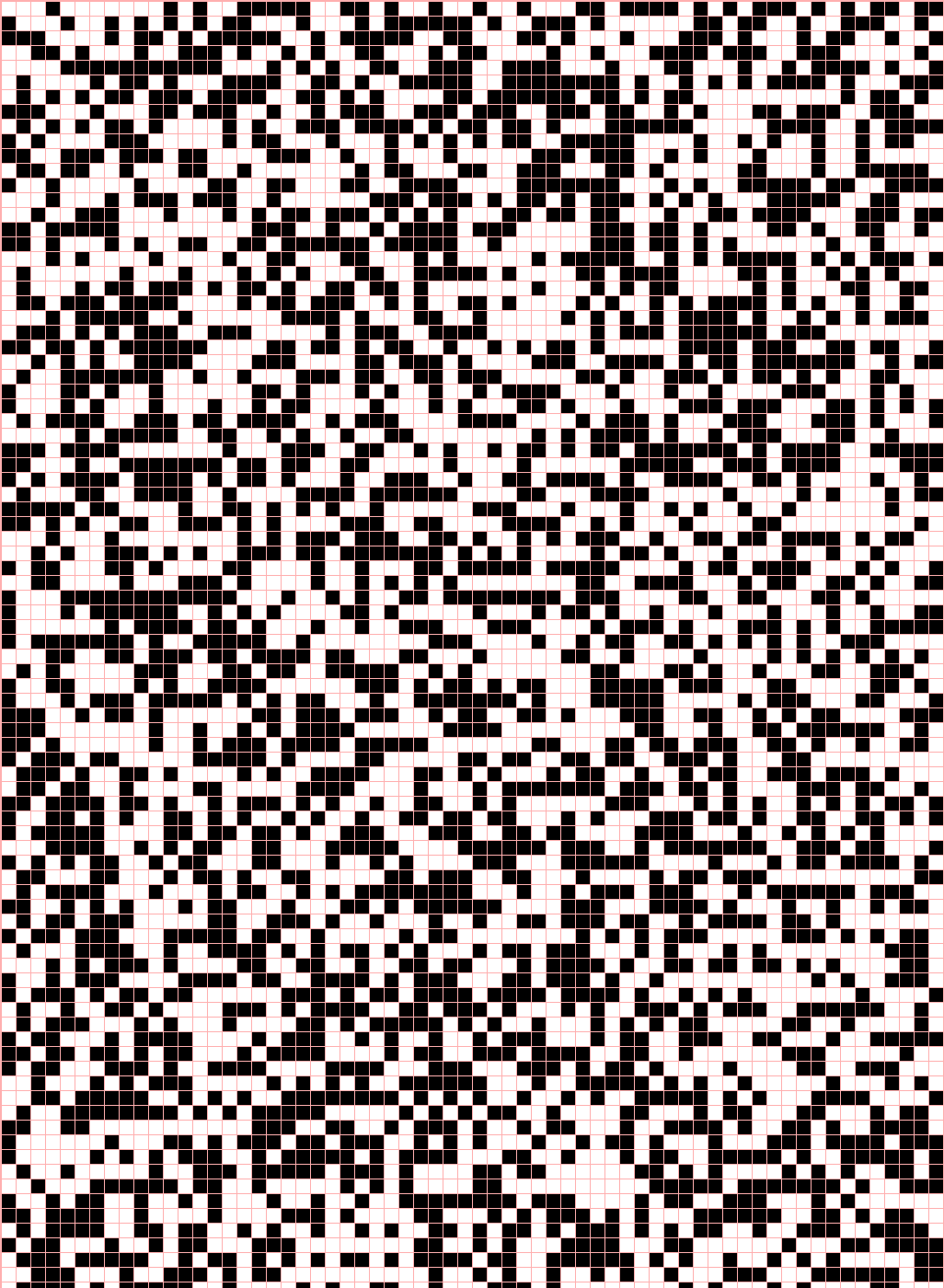}}
%
\hfill\\
\subfloat[$s_1$\label{xorshift1024_7_space}]{%
\includegraphics[width=0.2\linewidth, height=5.0cm]{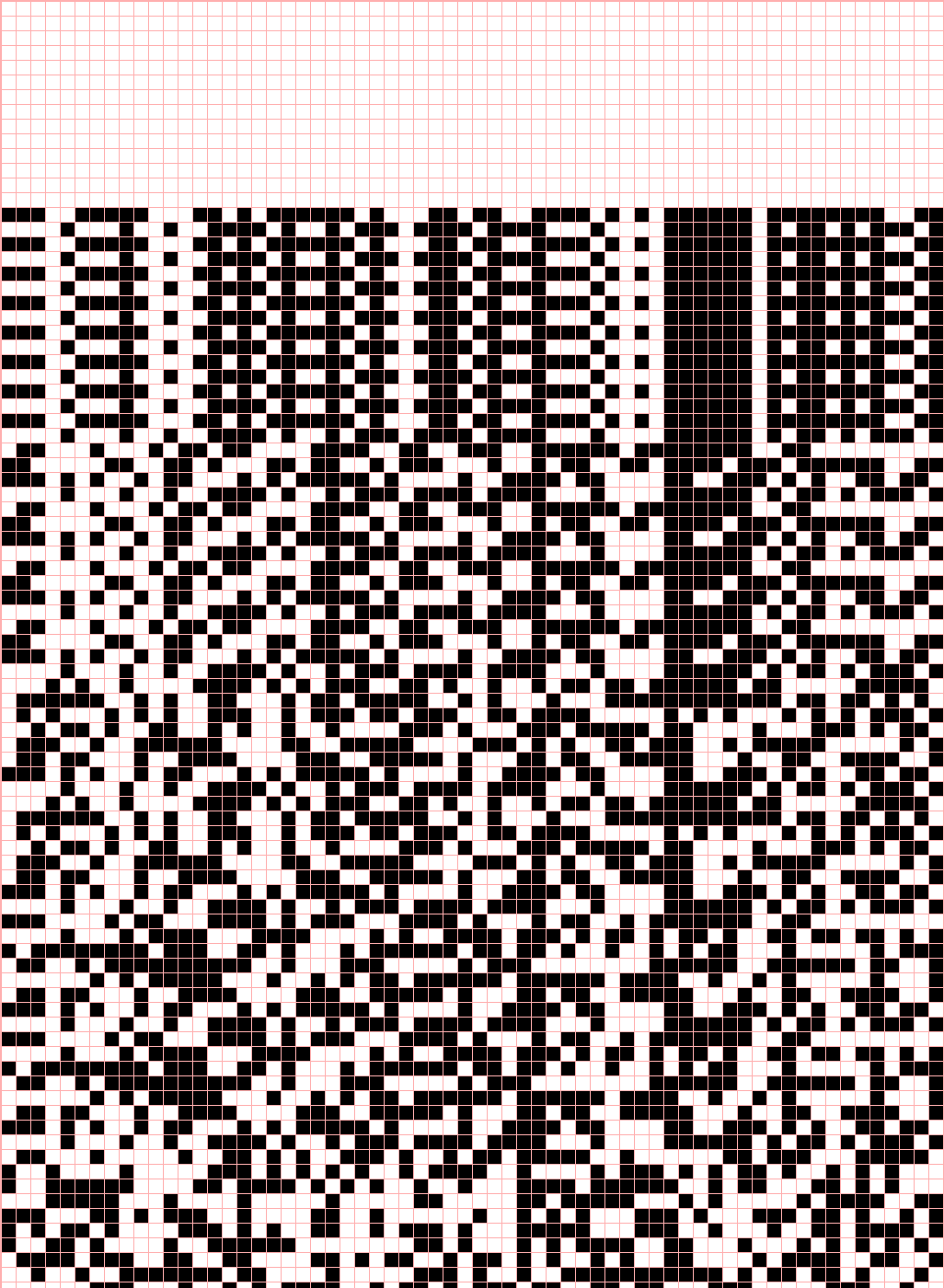}}
\hfill
\subfloat[$s_3$\label{xorshift1024_12345_space}]{%
 \includegraphics[width=0.2\linewidth, height=5.0cm]{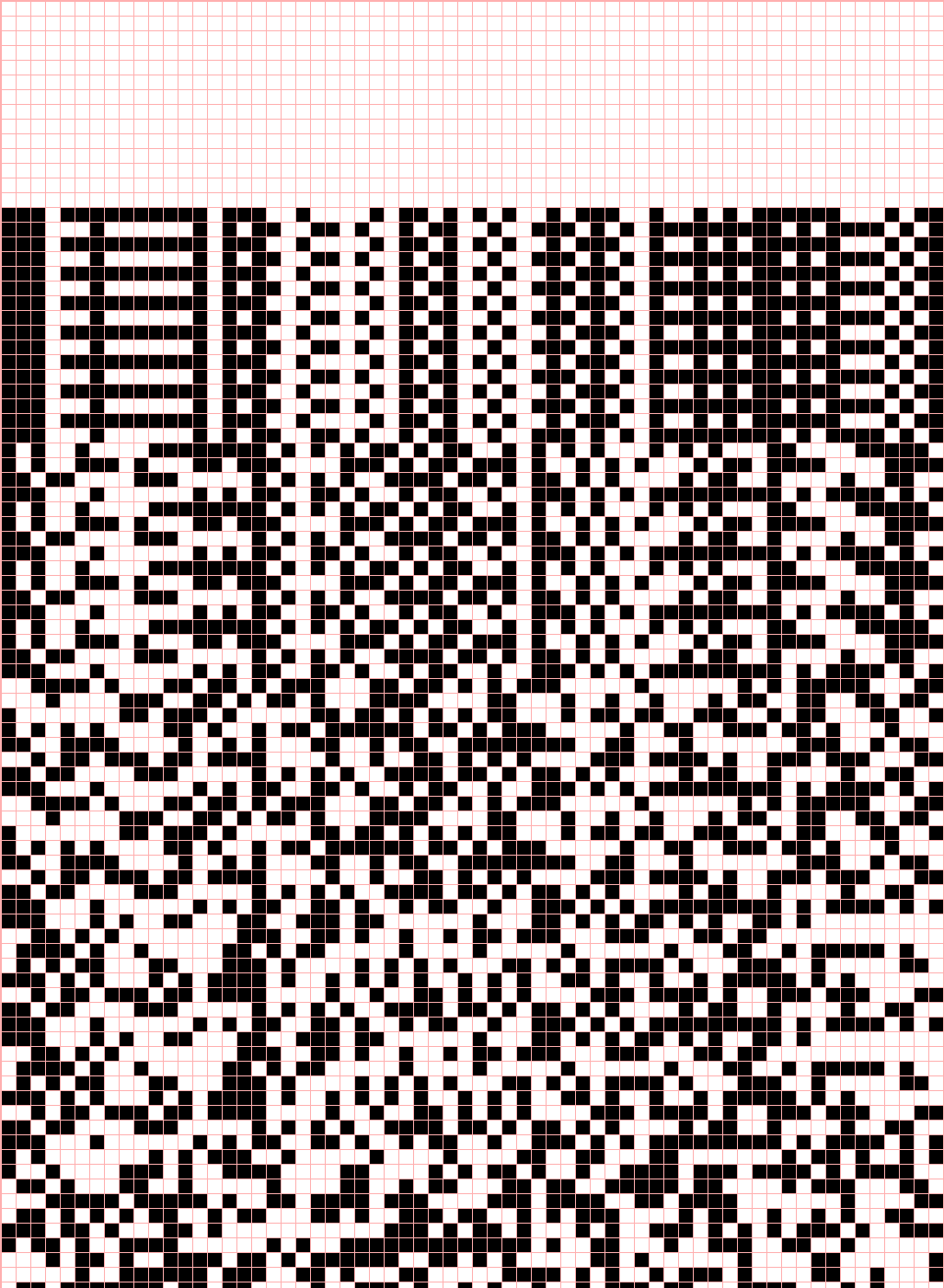}}
\hfill
\subfloat[$s_4$\label{xorshift1024_9650218_space}]{%
\includegraphics[width=0.2\linewidth, height=5.0cm]{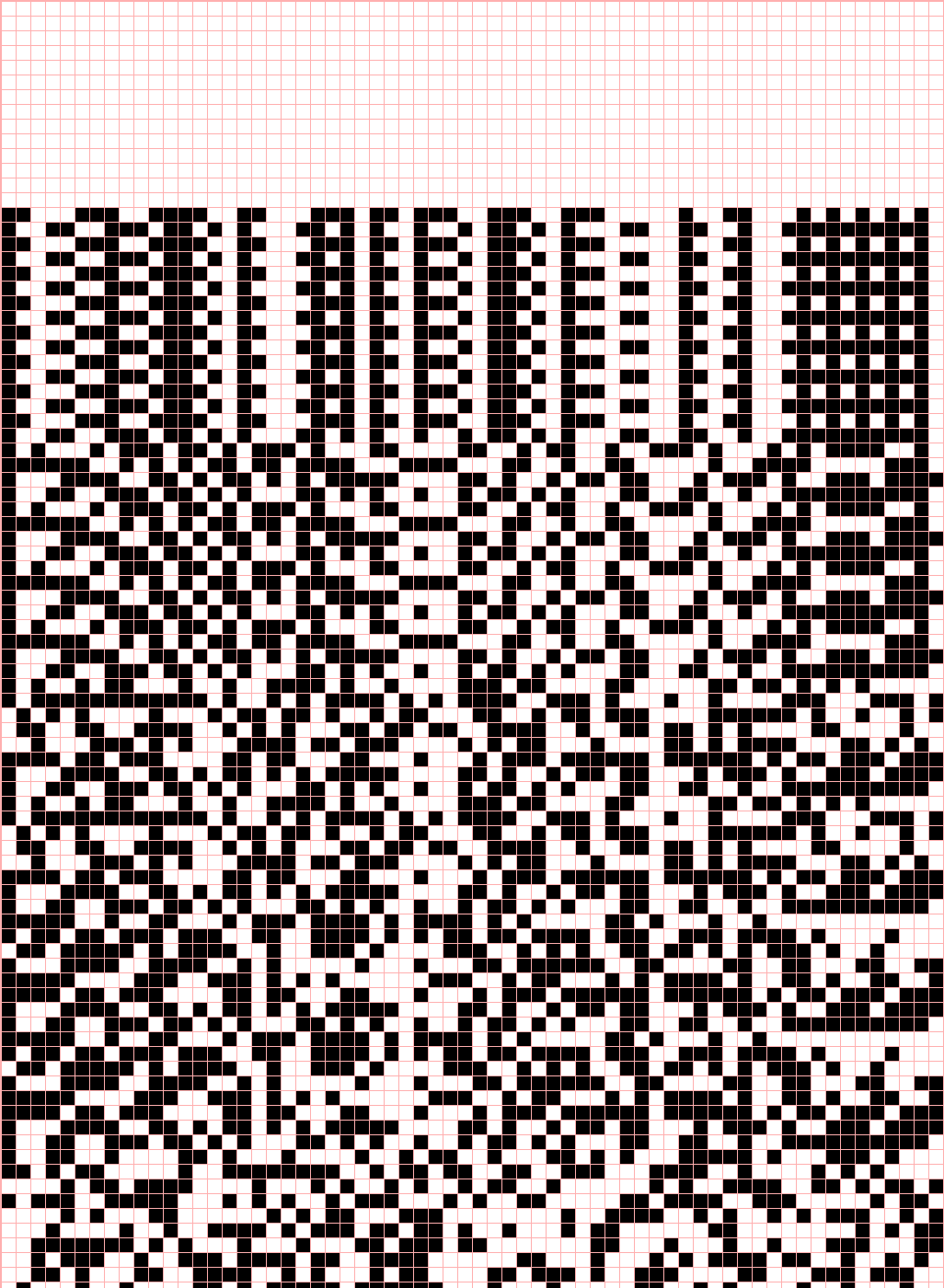}}
\hfill
\subfloat[$s_5$\label{xorshift1024_123456789123456789_space}]{%
\includegraphics[width=0.2\linewidth, height=5.0cm]{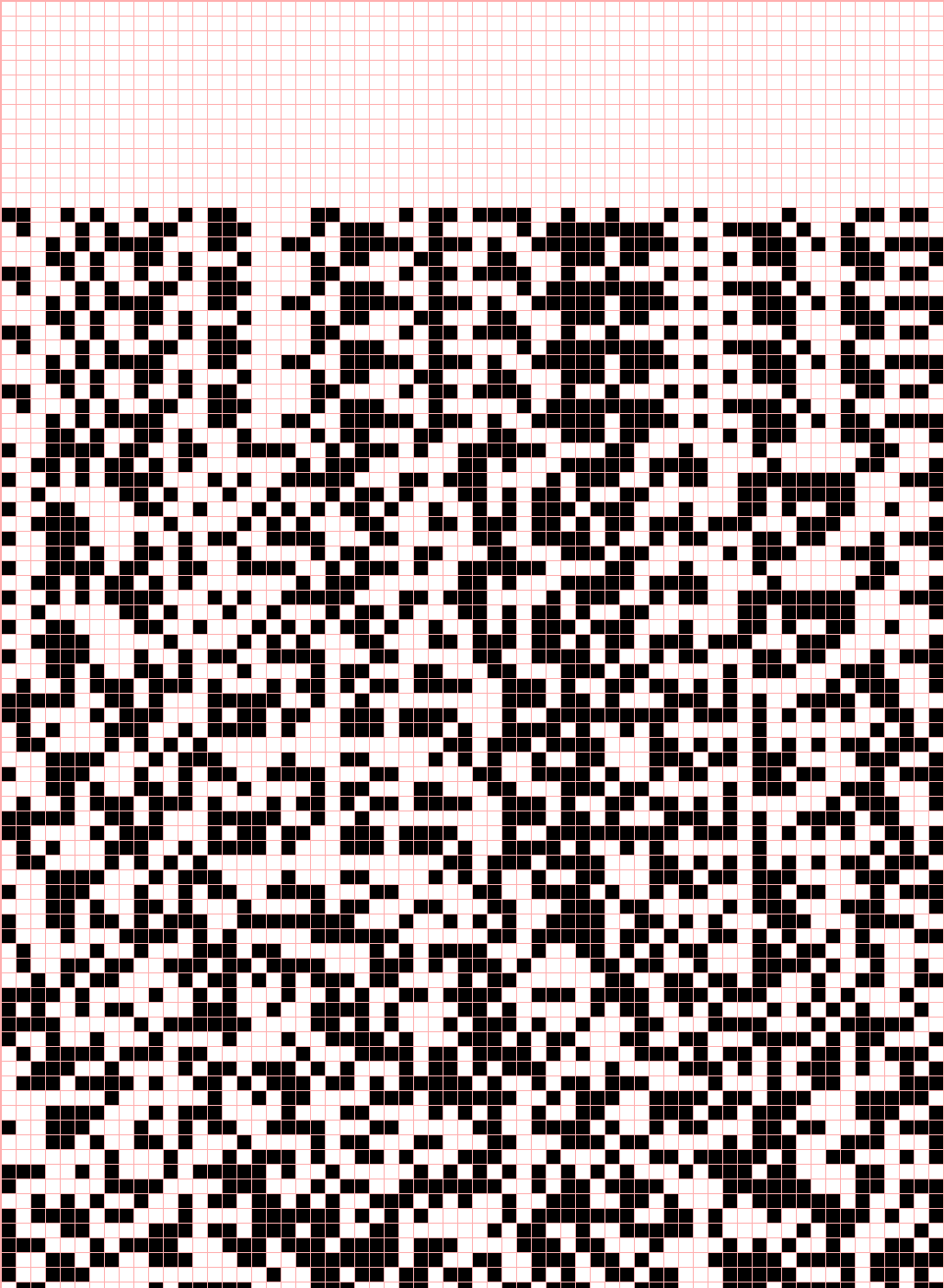}}
%
\hfill\\
\subfloat[$s_1$\label{xorshift_7_space}]{%
\includegraphics[width=0.2\linewidth, height=5.0cm]{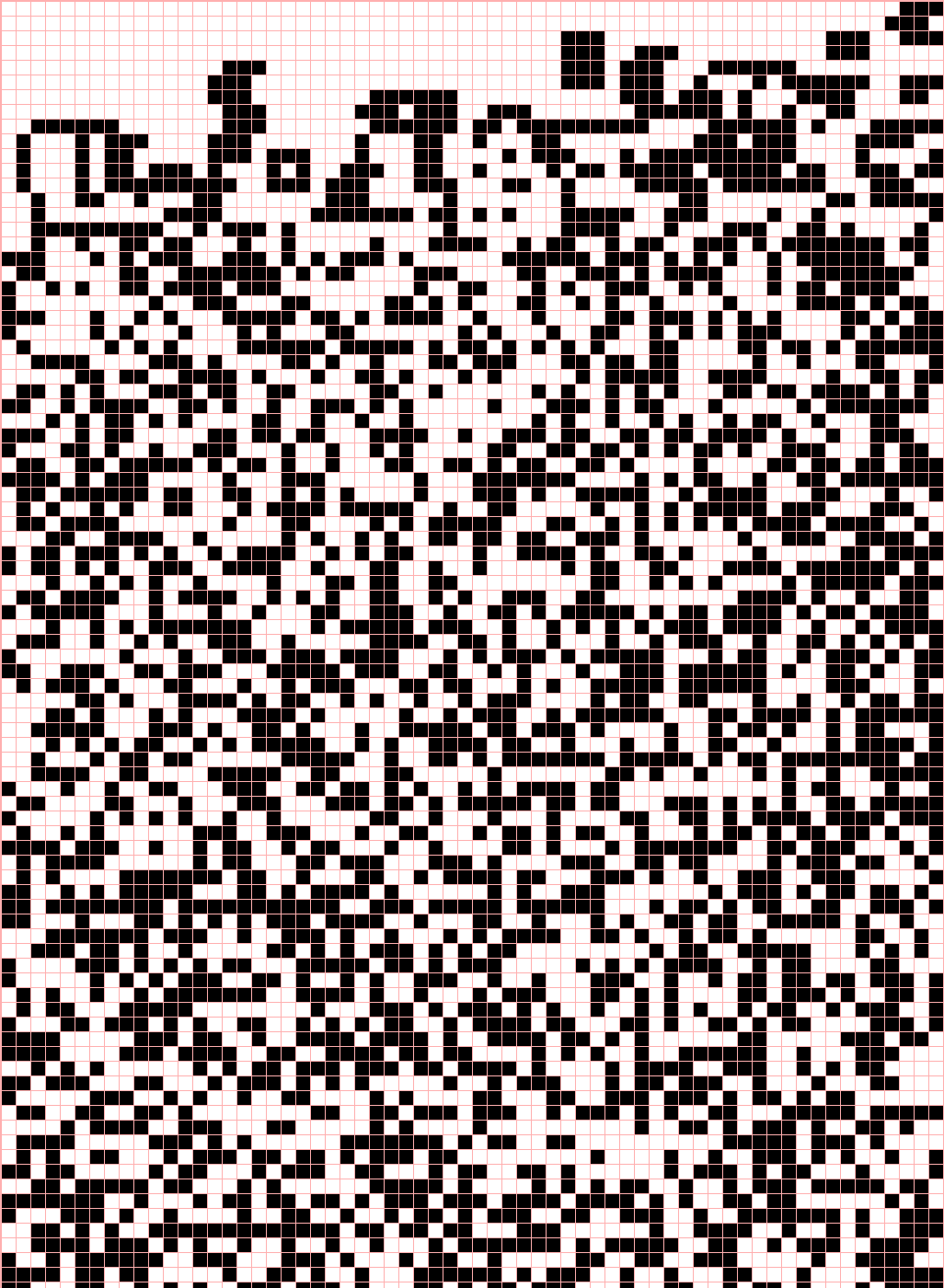}}
\hfill
\subfloat[$s_3$\label{xorshift_12345_space}]{%
 \includegraphics[width=0.2\linewidth, height=5.0cm]{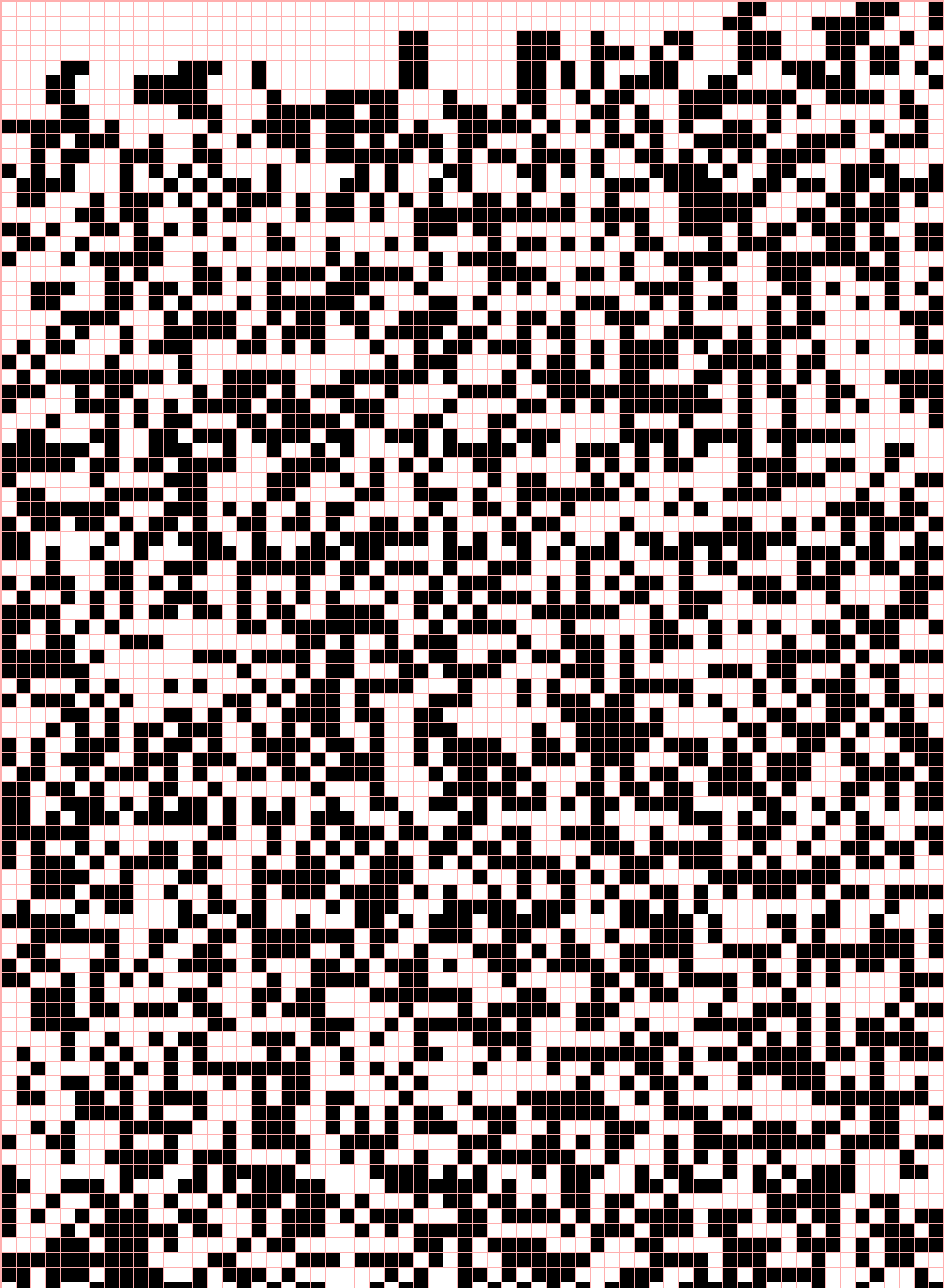}}
\hfill
\subfloat[$s_4$\label{xorshift_9650218_space}]{%
\includegraphics[width=0.2\linewidth, height=5.0cm]{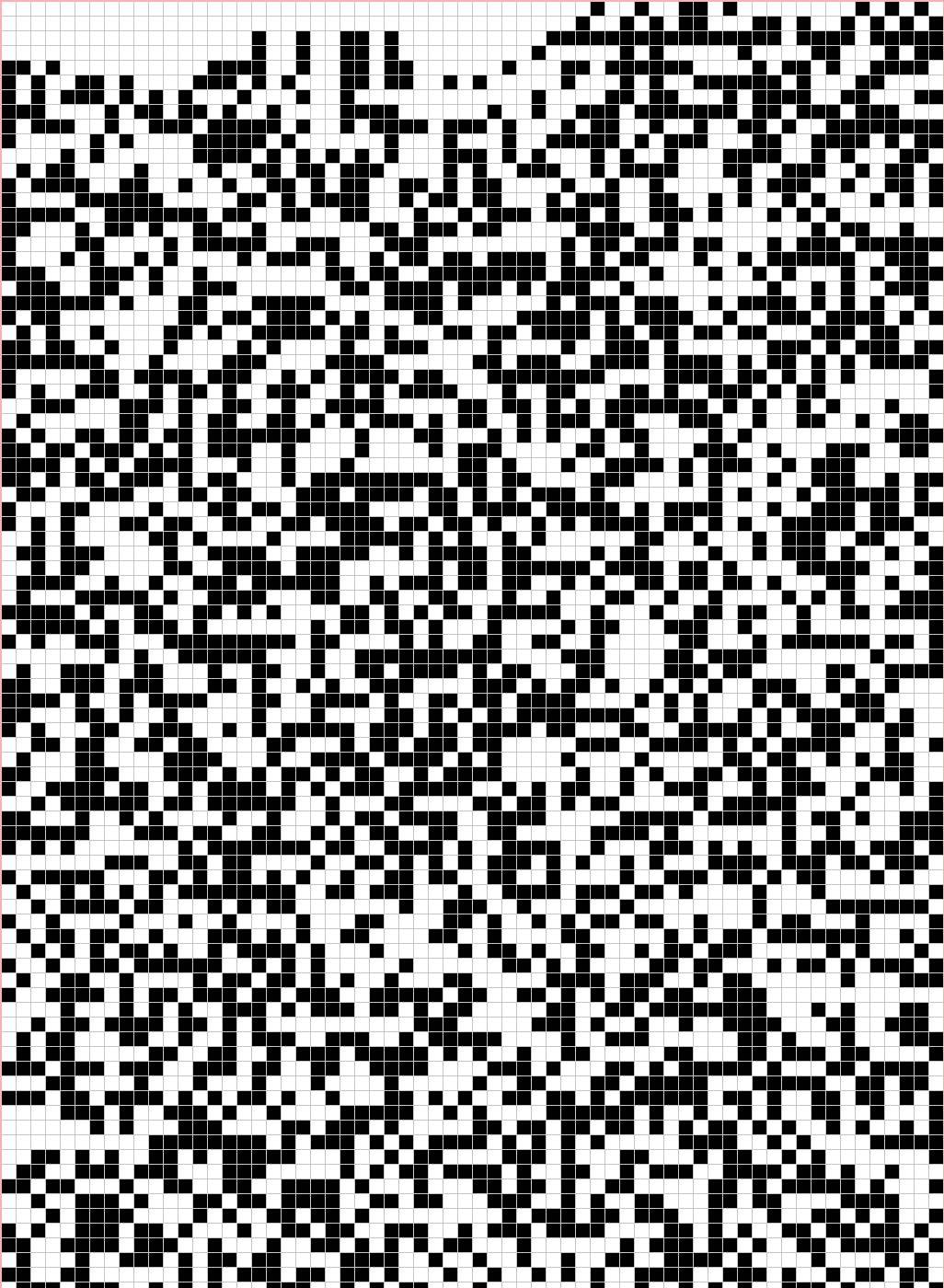}}
\hfill
\subfloat[$s_5$\label{xorshift_123456789123456789_space}]{%
\includegraphics[width=0.2\linewidth, height=5.0cm]{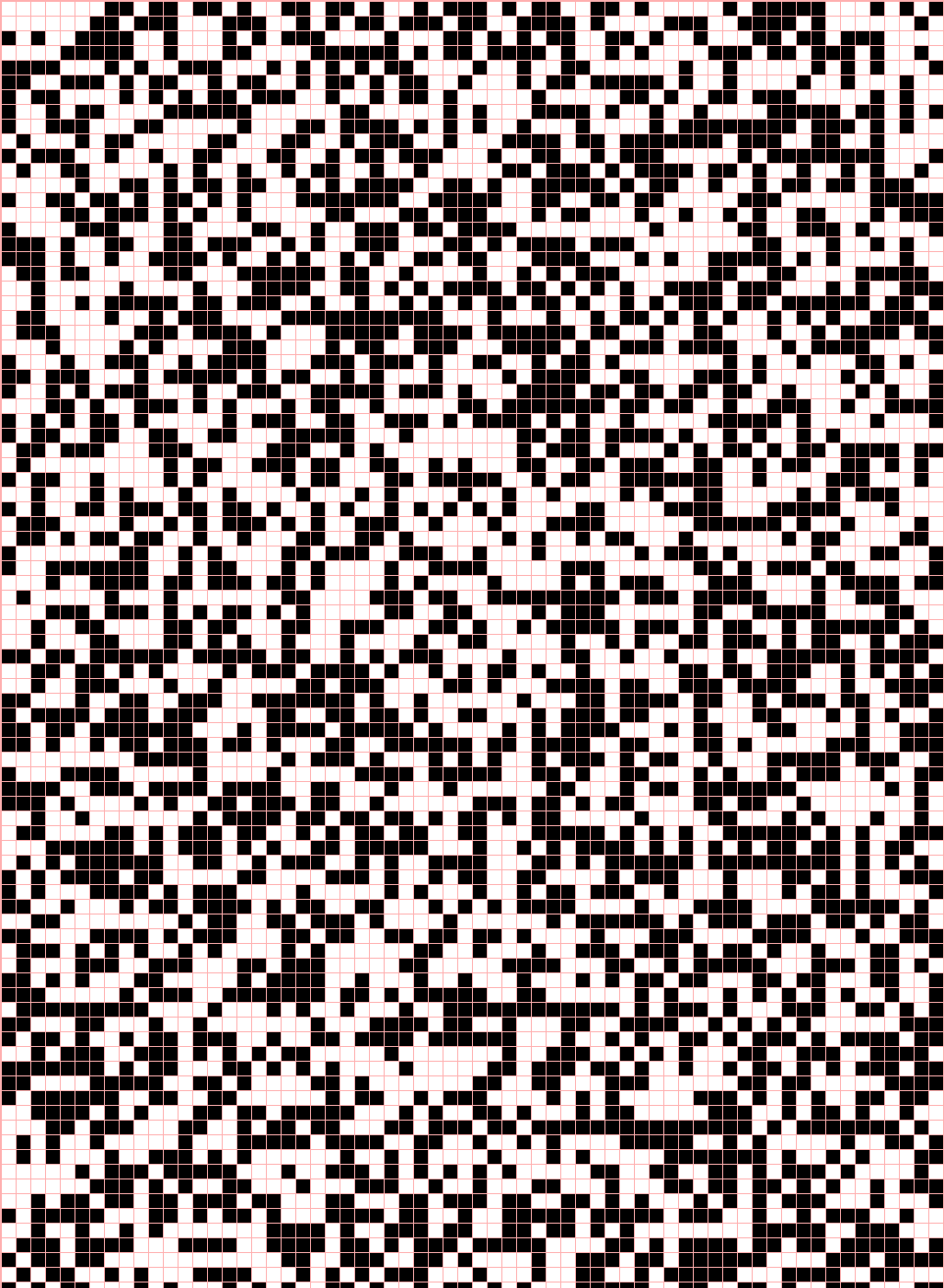}}
\caption{Space-time diagram for xorshift64* (\ref{xorshift64_7_space} to \ref{xorshift64_123456789123456789_spaceo}), xorshift1024* (\ref{xorshift1024_7_space} to \ref{xorshift1024_123456789123456789_space}) and xorshift128+ (\ref{xorshift_7_space} to \ref{xorshift_123456789123456789_space})}
\label{fig:xor_space-time}
\end{figure} 

\begin{figure}[hbtp]
\centering
  \vspace{-2.0em}
\subfloat[$s_1$\label{mt_32_7_space}]{%
\includegraphics[width=0.1\linewidth, height=5.0cm]{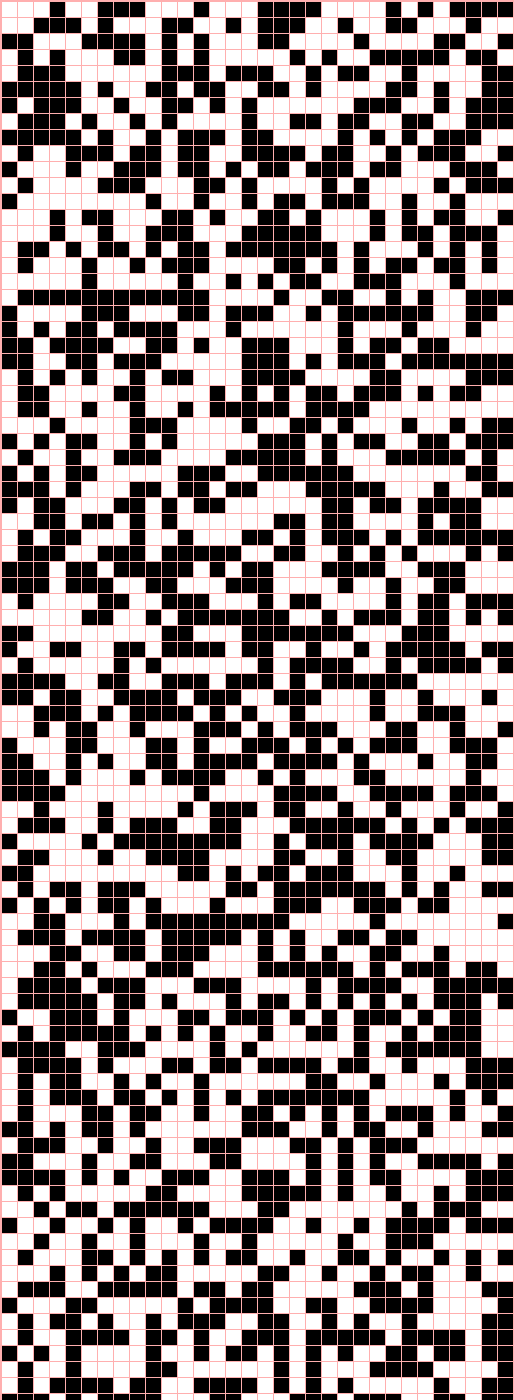}}
\hfill
\subfloat[$s_3$\label{mt_32_12345_space}]{%
 \includegraphics[width=0.1\linewidth, height=5.0cm]{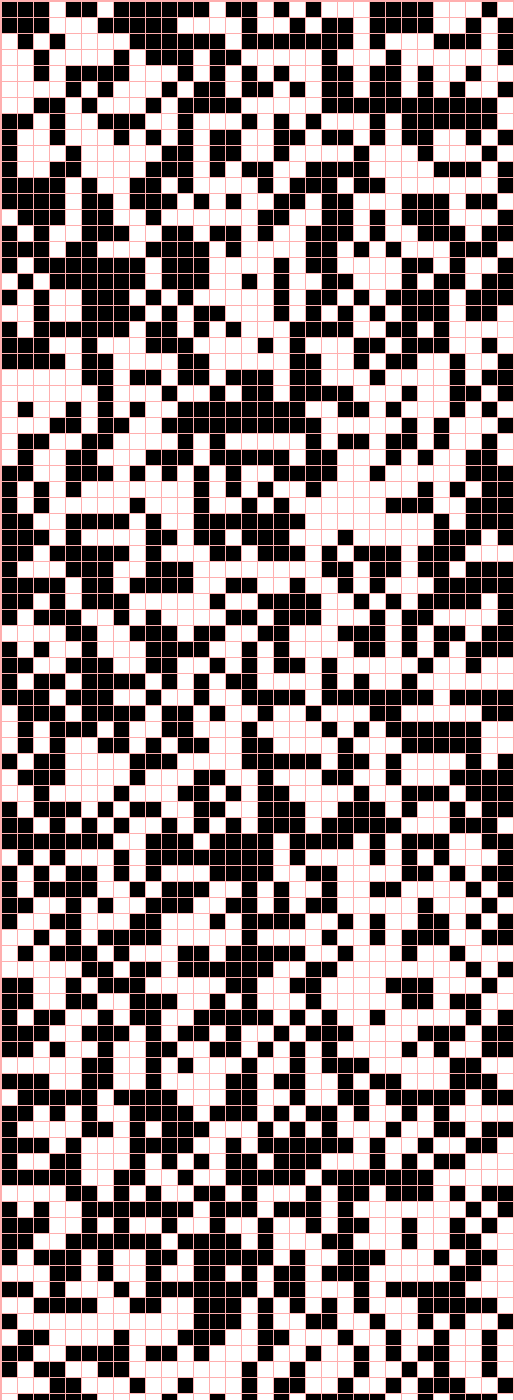}}
\hfill
\subfloat[$s_4$\label{mt_32_9650218_space}]{%
\includegraphics[width=0.1\linewidth, height=5.0cm]{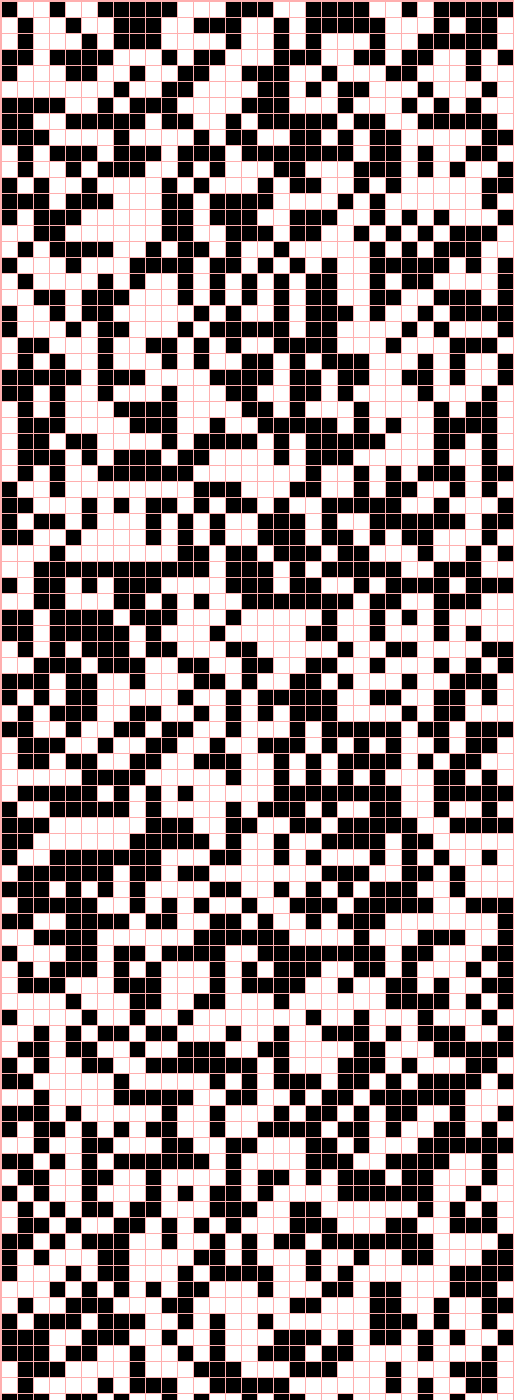}}
\hfill
\subfloat[$s_5$\label{mt_32_123456789123456789_spaceo}]{%
\includegraphics[width=0.1\linewidth, height=5.0cm]{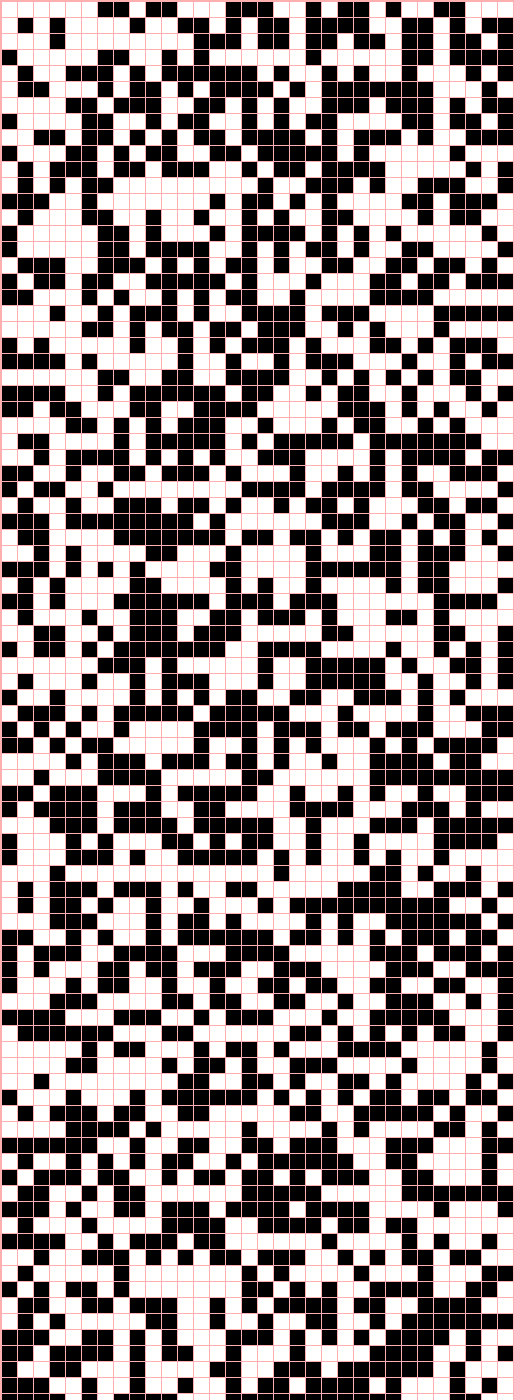}}
\hfill
\subfloat[$s_1$\label{sfmt_32_7_space}]{%
\includegraphics[width=0.1\linewidth, height=5.0cm]{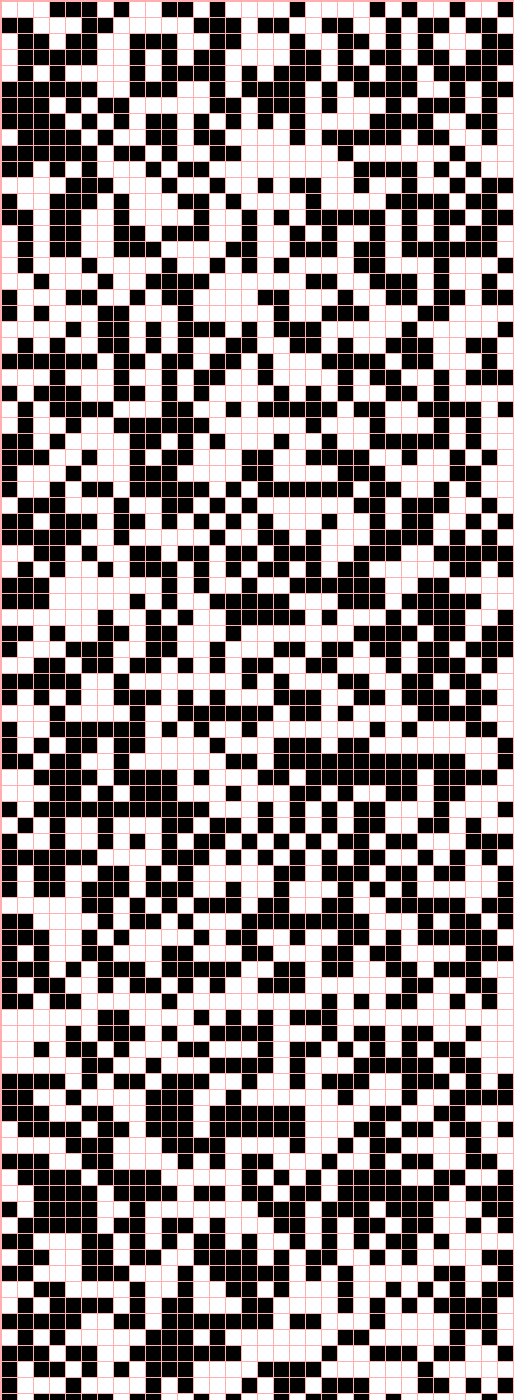}}
\hfill
\subfloat[$s_3$\label{sfmt_32_12345_space}]{%
 \includegraphics[width=0.1\linewidth, height=5.0cm]{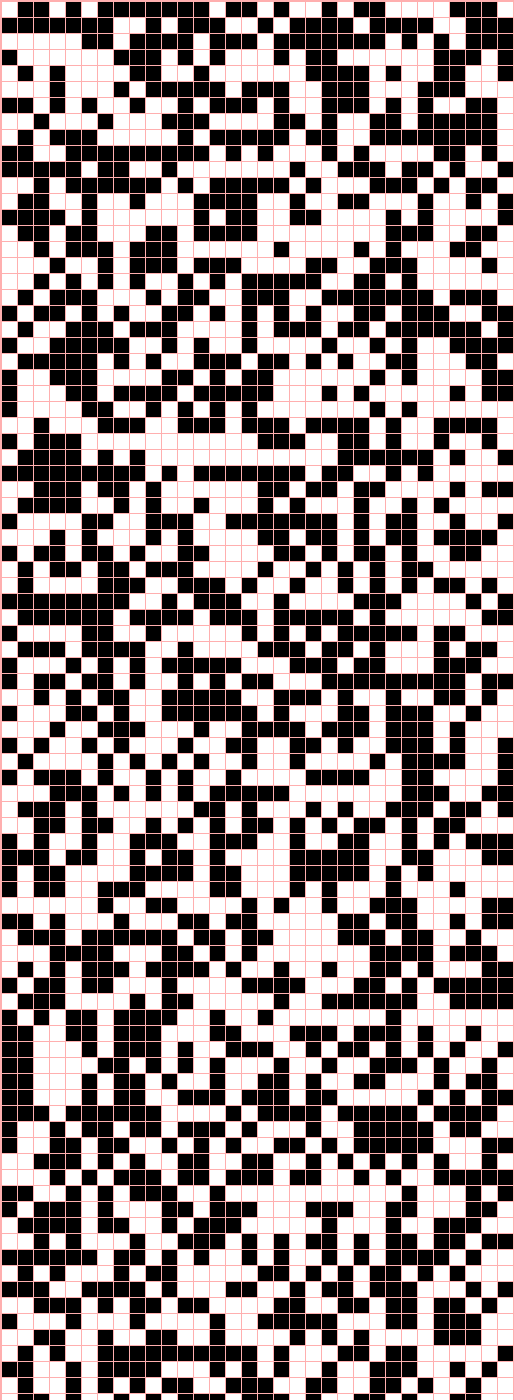}}
\hfill
\subfloat[$s_4$\label{sfmt_32_9650218_space}]{%
\includegraphics[width=0.1\linewidth, height=5.0cm]{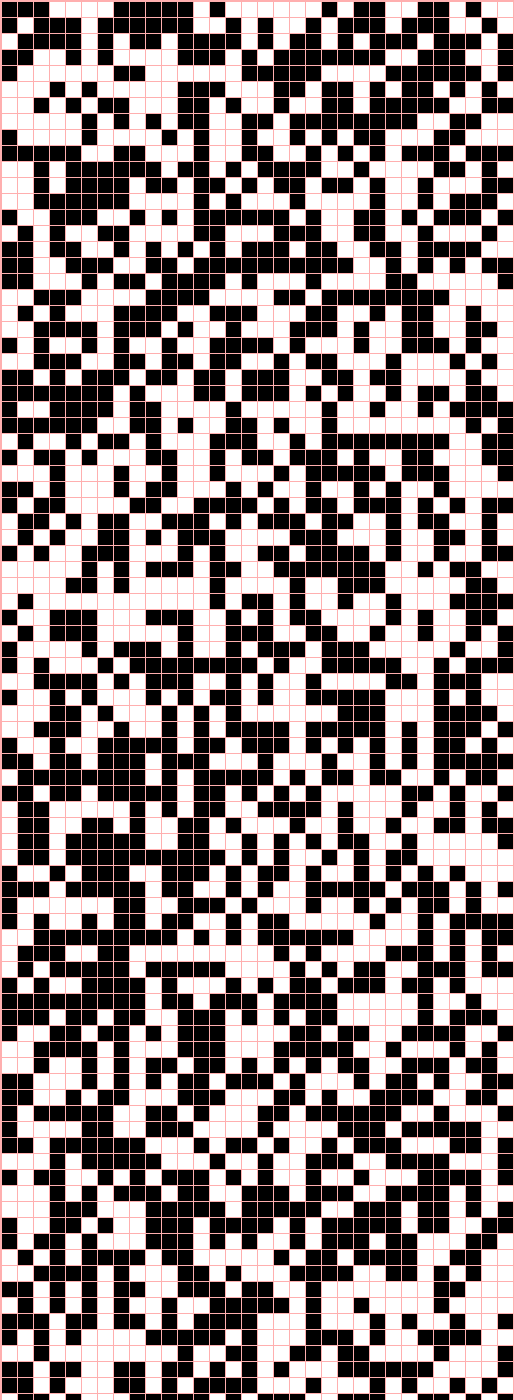}}
\hfill
\subfloat[$s_5$\label{sfmt_32_123456789123456789_spaceo}]{%
\includegraphics[width=0.1\linewidth, height=5.0cm]{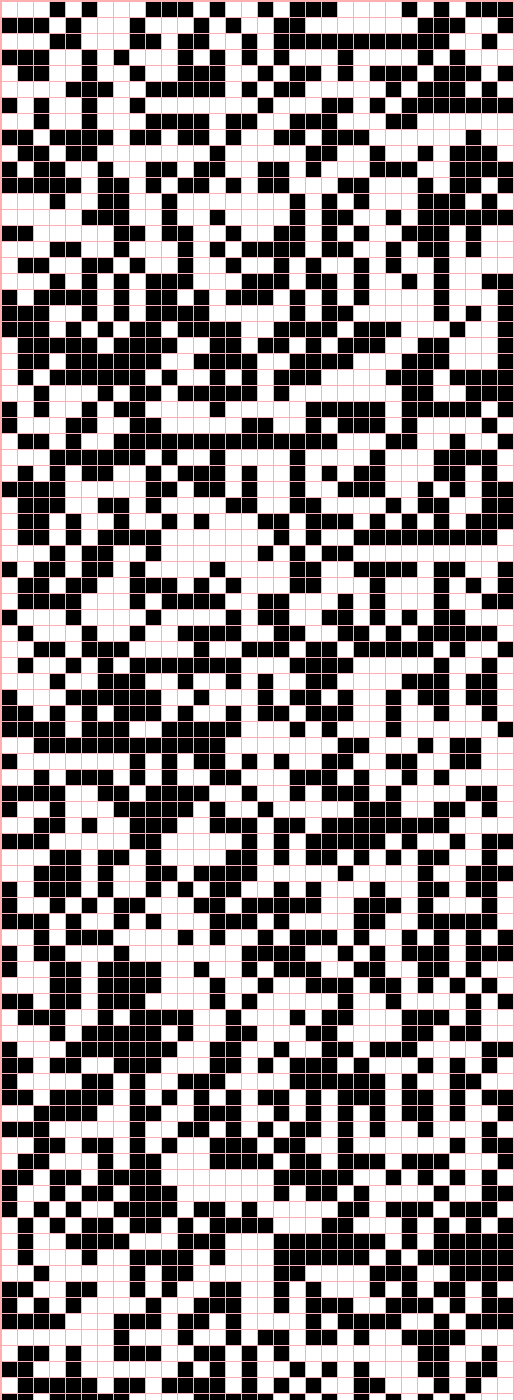}}
\hfill\\
	\subfloat[$s_1$\label{mt_64_7_space}]{%
		\includegraphics[width=0.2\linewidth, height=5.0cm]{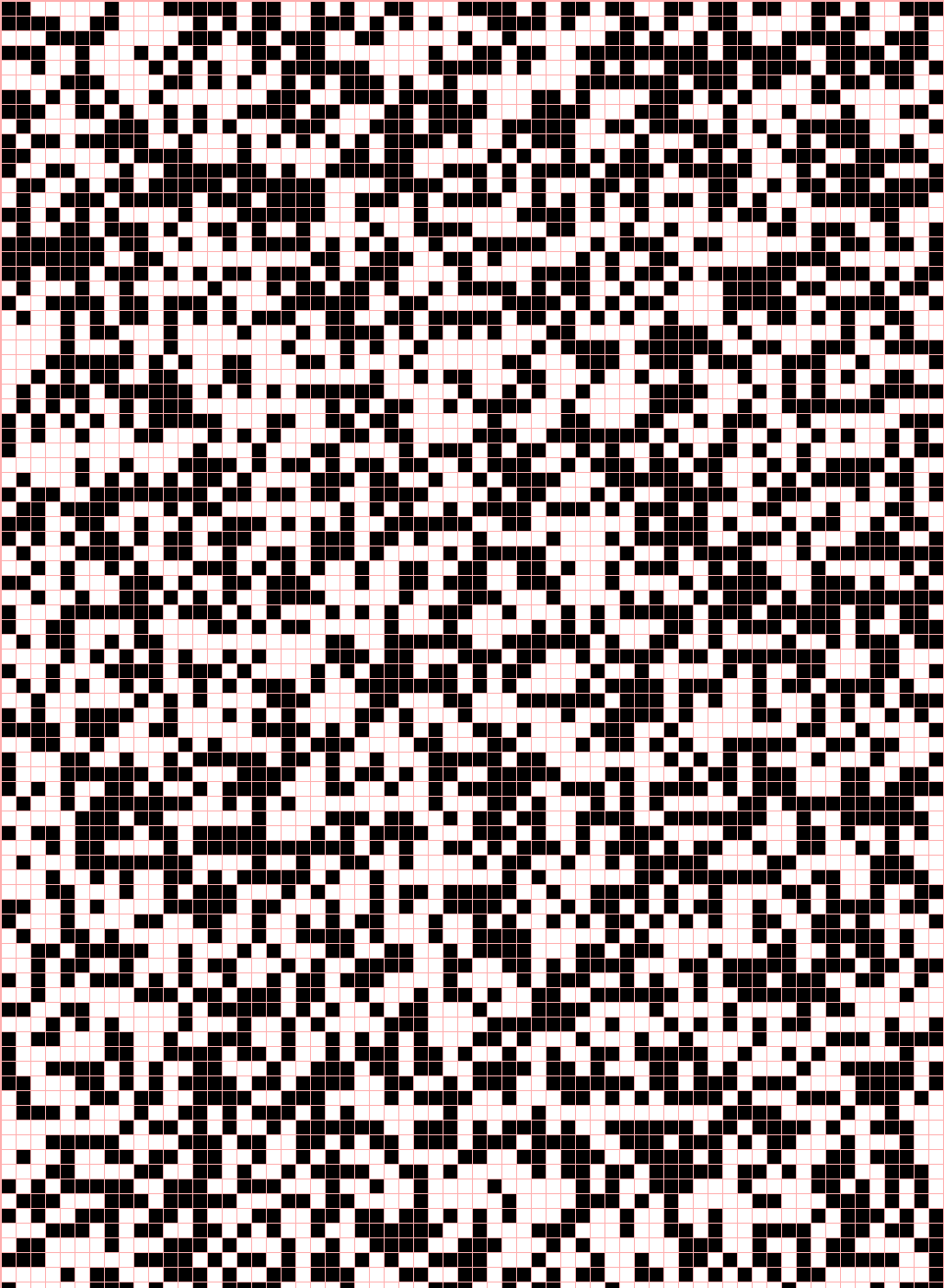}}
	\hfill
	\subfloat[$s_3$\label{mt_64_12345_space}]{%
		\includegraphics[width=0.2\linewidth, height=5.0cm]{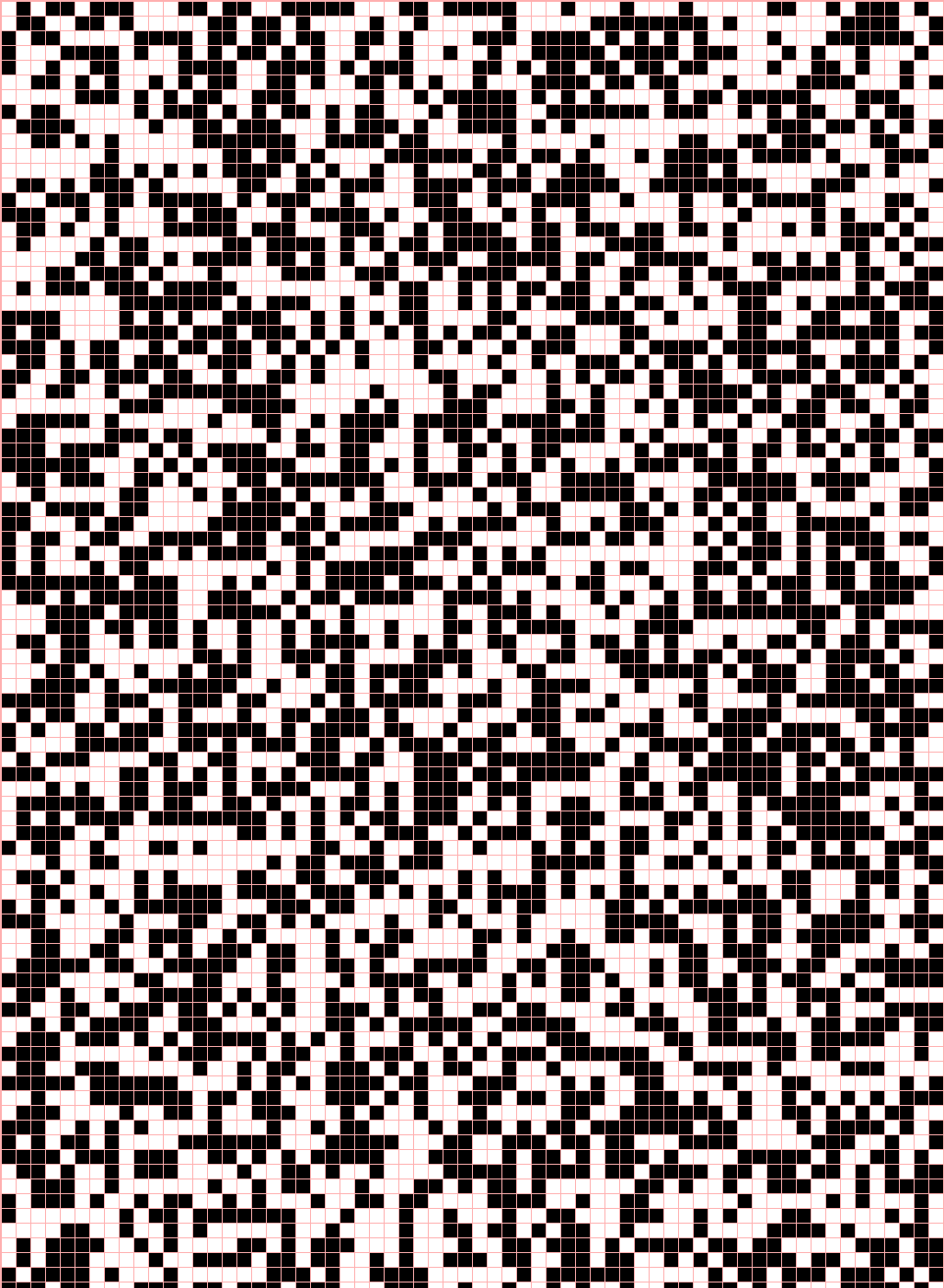}}
	\hfill
	\subfloat[$s_4$\label{mt_64_9650218_space}]{%
		\includegraphics[width=0.2\linewidth, height=5.0cm]{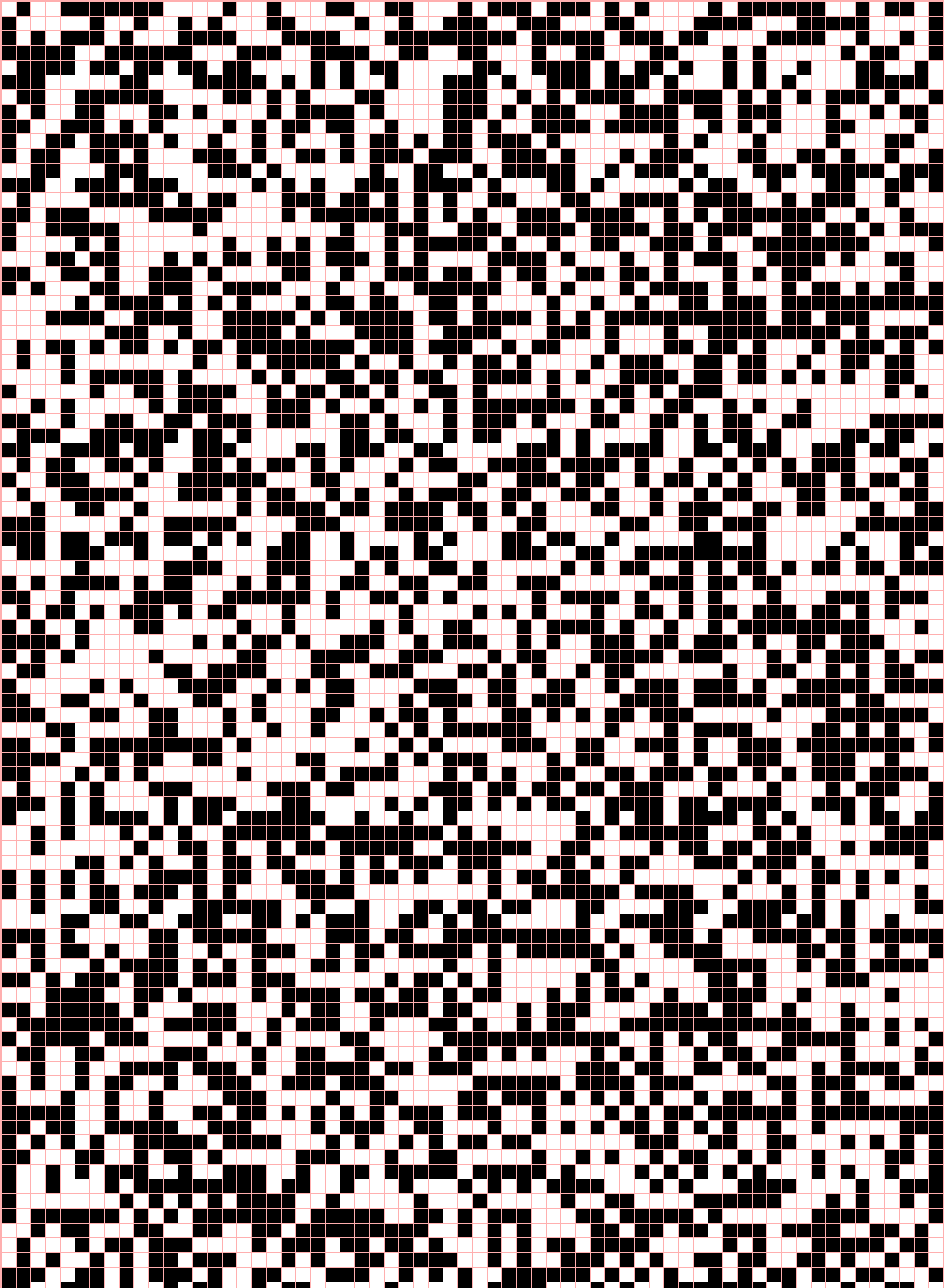}}
	\hfill
	\subfloat[$s_5$\label{mt_64_123456789123456789_space}]{%
		\includegraphics[width=0.2\linewidth, height=5.0cm]{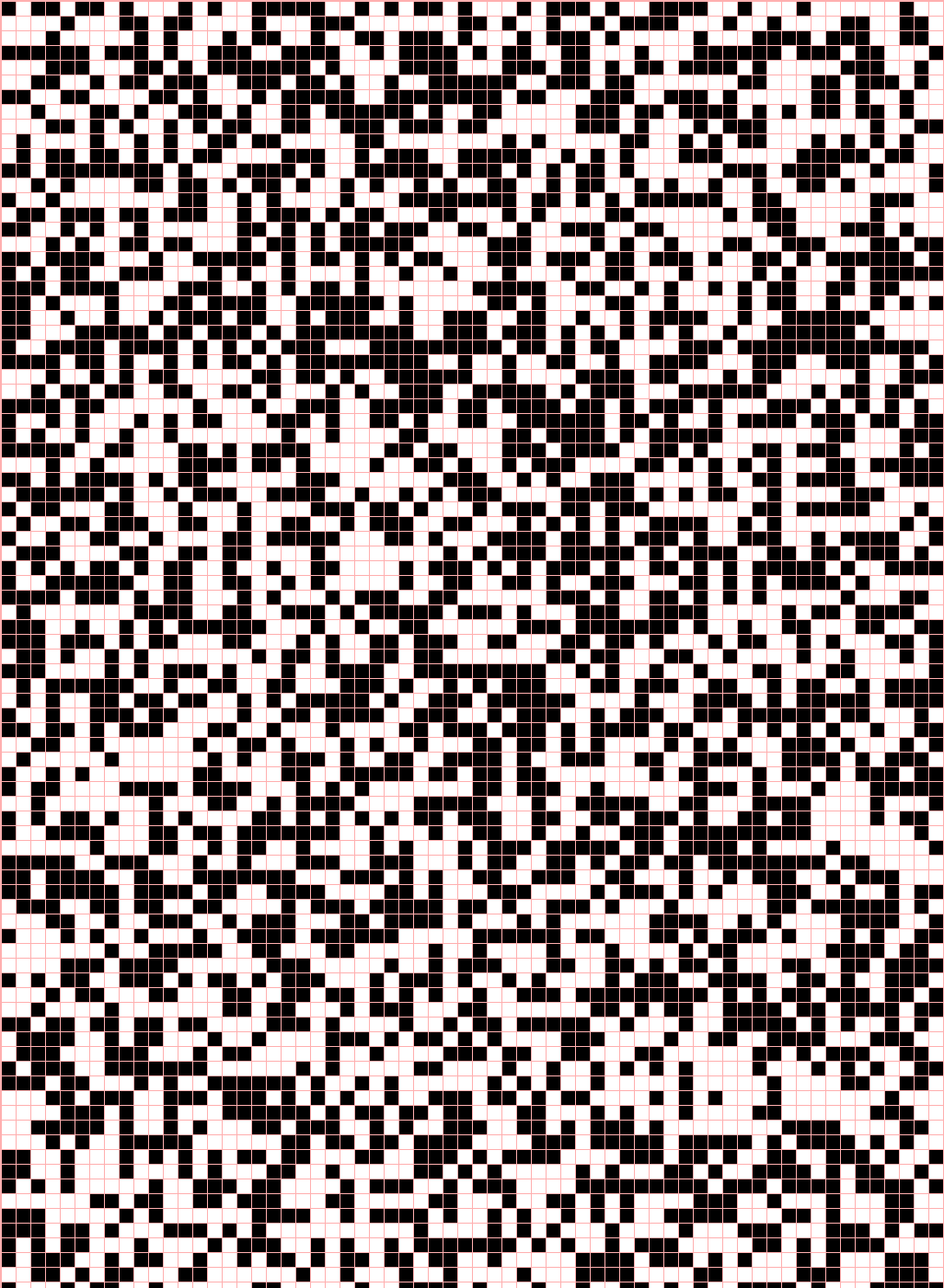}}
%
%
\hfill\\
	\subfloat[$s_1$\label{sfmt_64_7_space}]{%
		\includegraphics[width=0.2\linewidth, height=5.0cm]{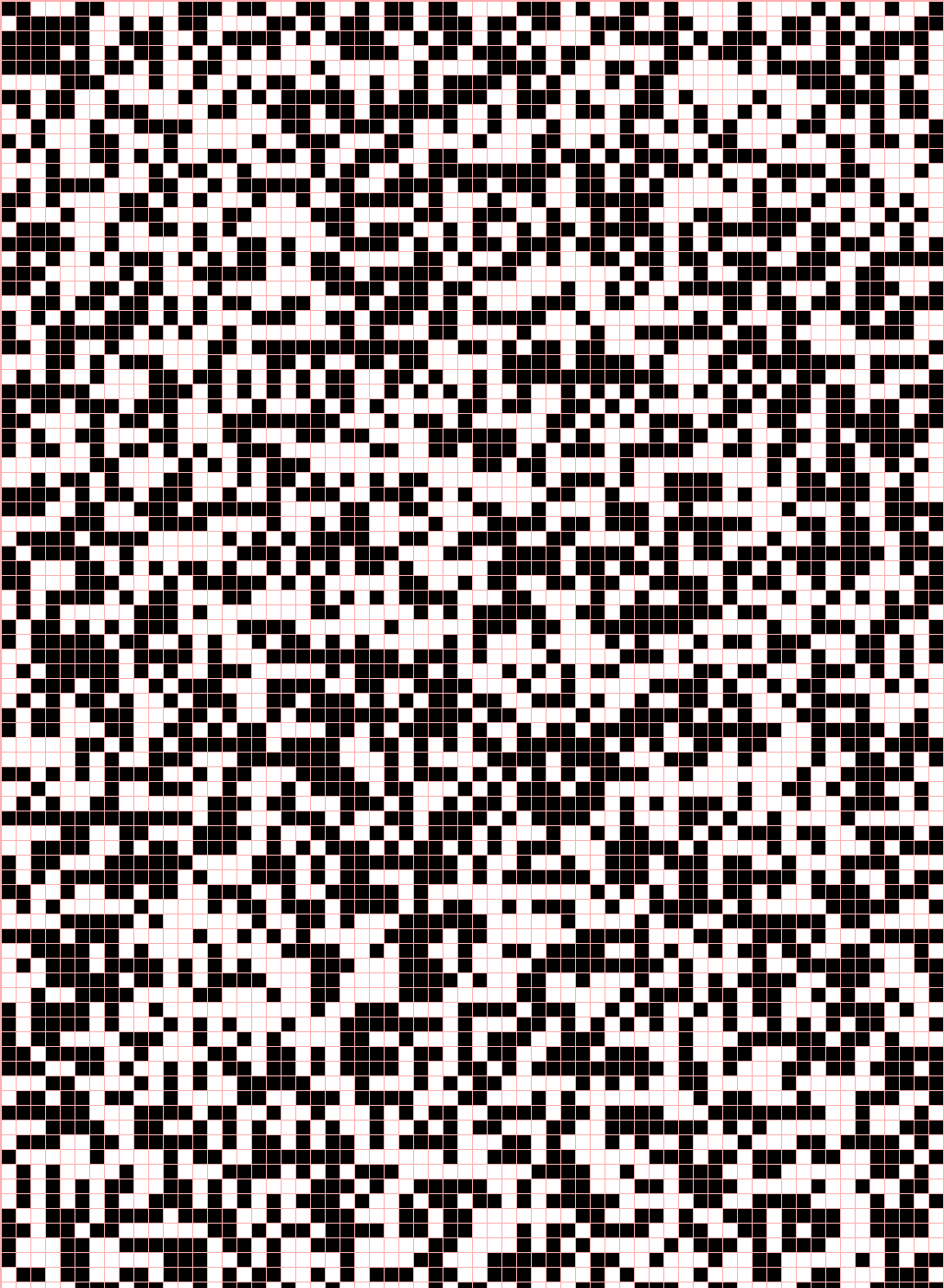}}
	\hfill
	\subfloat[$s_3$\label{sfmt_64_12345_space}]{%
		\includegraphics[width=0.2\linewidth, height=5.0cm]{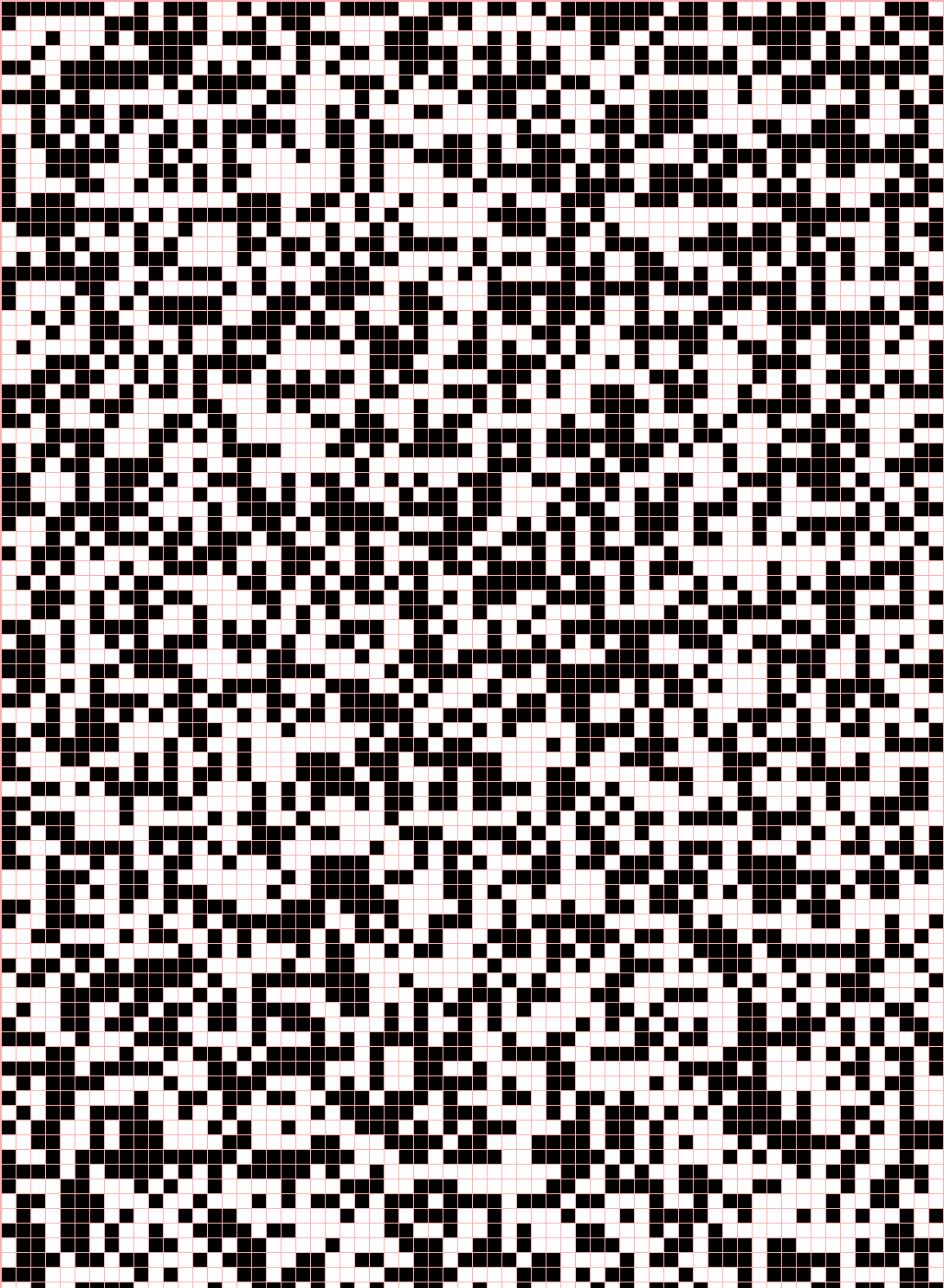}}
	\hfill
	\subfloat[$s_4$\label{sfmt_64_9650218_space}]{%
		\includegraphics[width=0.2\linewidth, height=5.0cm]{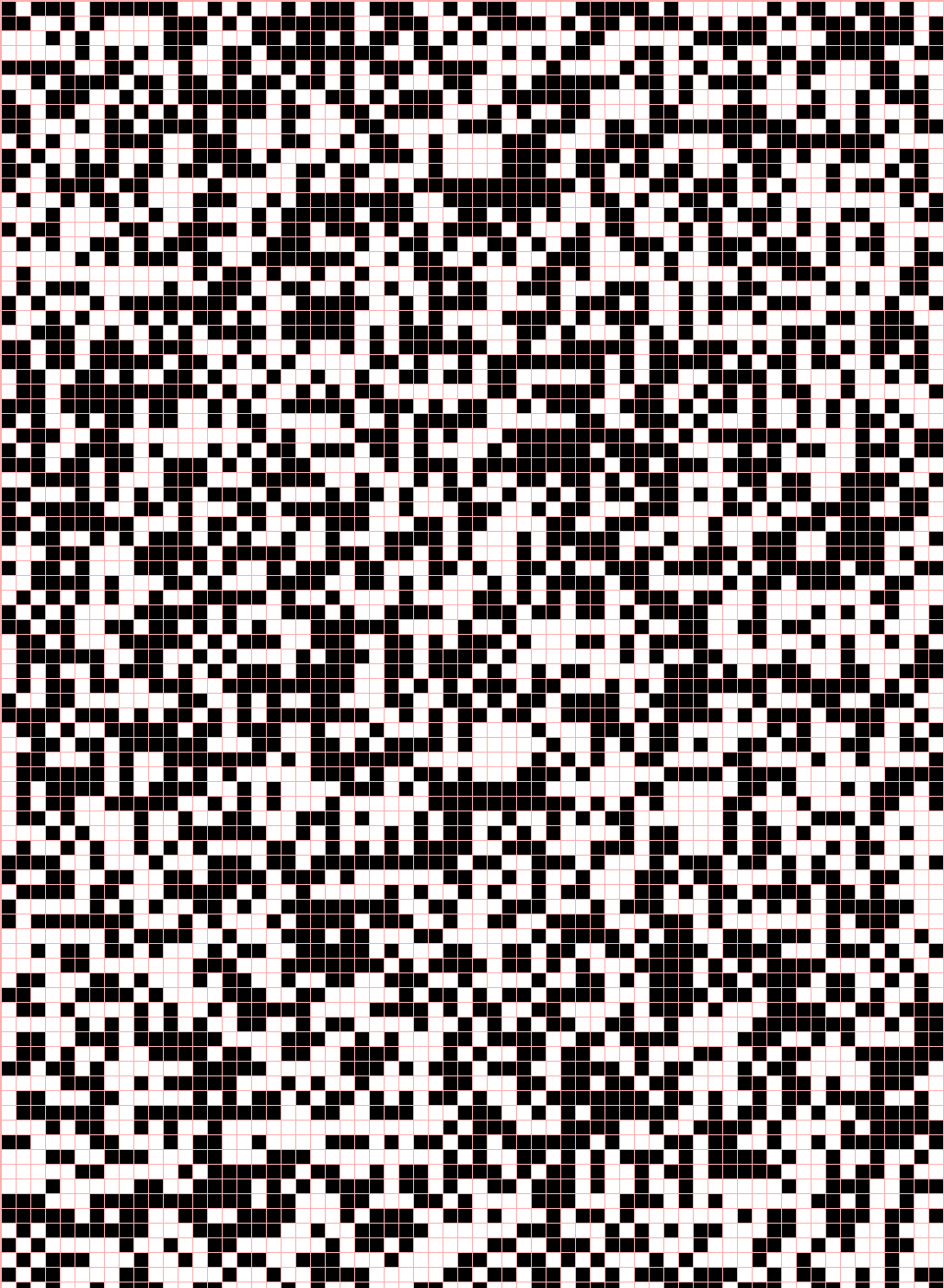}}
	\hfill
	\subfloat[$s_5$\label{sfmt_64_123456789123456789_space}]{%
		\includegraphics[width=0.2\linewidth, height=5.0cm]{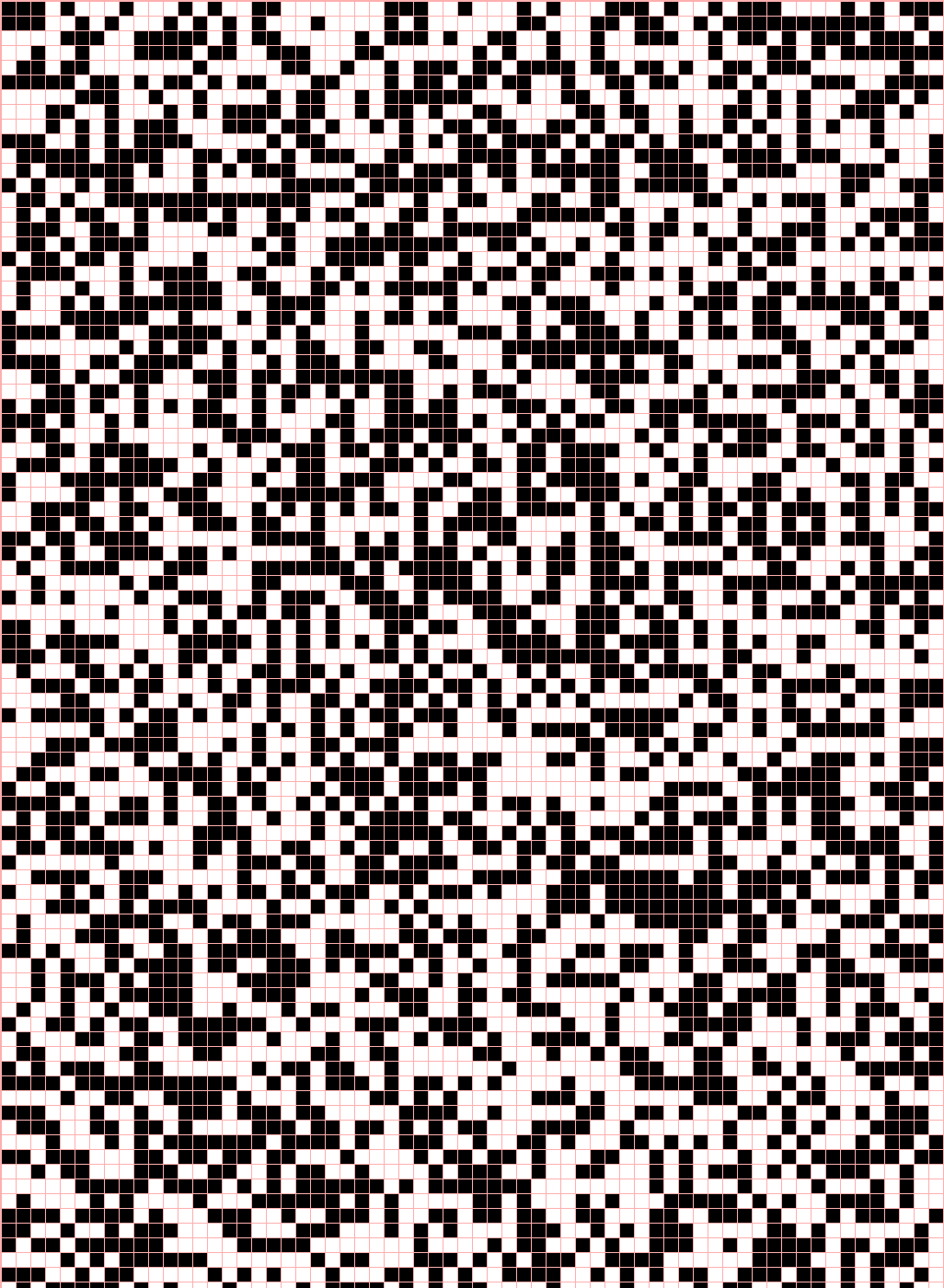}}
	\caption{Space-time diagram for MT19937 $32$ bit (\ref{mt_32_7_space} to \ref{mt_32_123456789123456789_spaceo}), SFMT19937 $32$ bit (\ref{sfmt_32_7_space} to \ref{sfmt_32_123456789123456789_spaceo}), MT19937 $64$ bit (\ref{mt_64_7_space} to \ref{mt_64_123456789123456789_space}) and SFMT19937 $64$ bit (\ref{sfmt_64_7_space} to \ref{sfmt_64_123456789123456789_space})}
	\label{fig:sfmt64_space-time}
\end{figure} 

\begin{figure}[hbtp]
\centering
   	\subfloat[$s_1$\label{dsfmt_7_space}]{%
  		\includegraphics[width=0.1\linewidth, height=5.0cm]{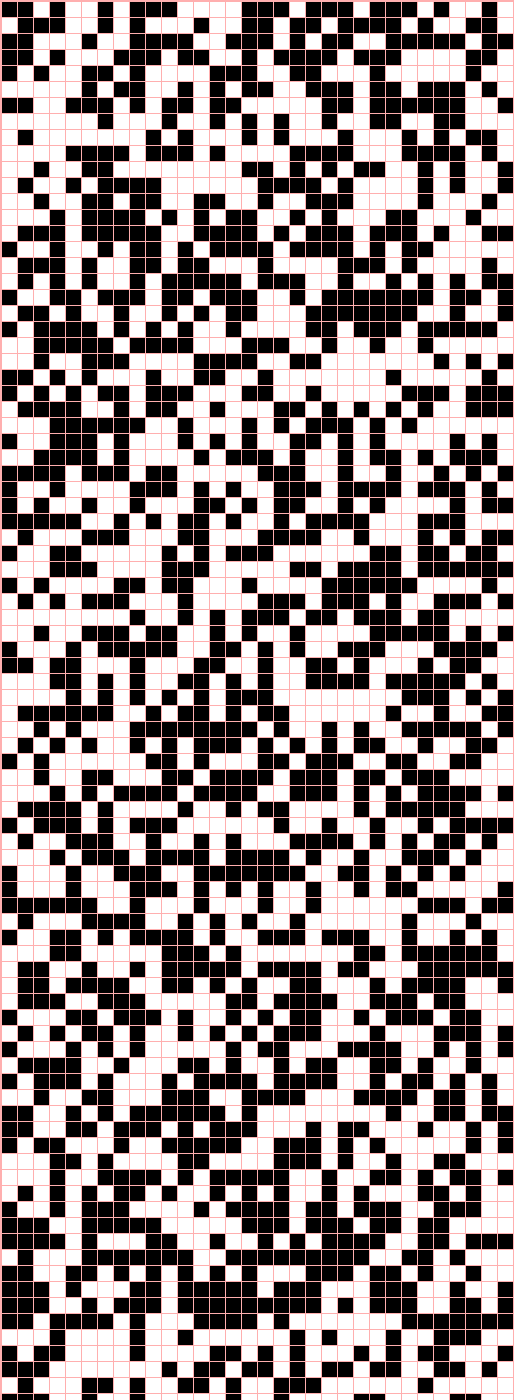}}
  	\hfill
  	\subfloat[$s_3$\label{dsfmt_12345_space}]{%
  		\includegraphics[width=0.1\linewidth, height=5.0cm]{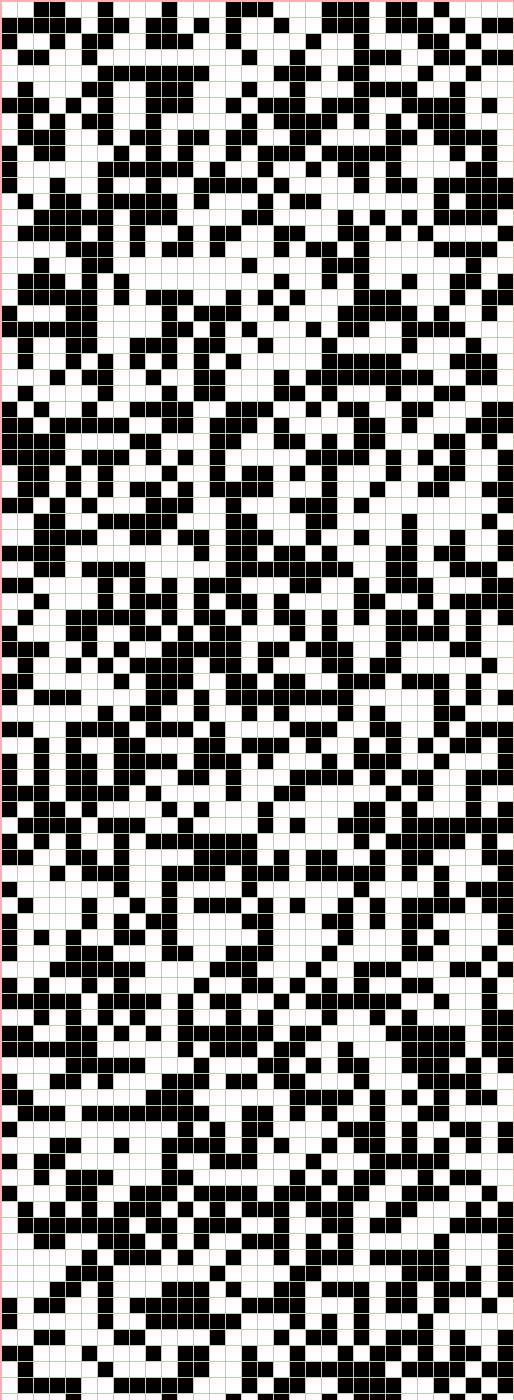}}
  	\hfill
  	\subfloat[$s_4$\label{dsfmt_9650218_space}]{%
  		\includegraphics[width=0.1\linewidth, height=5.0cm]{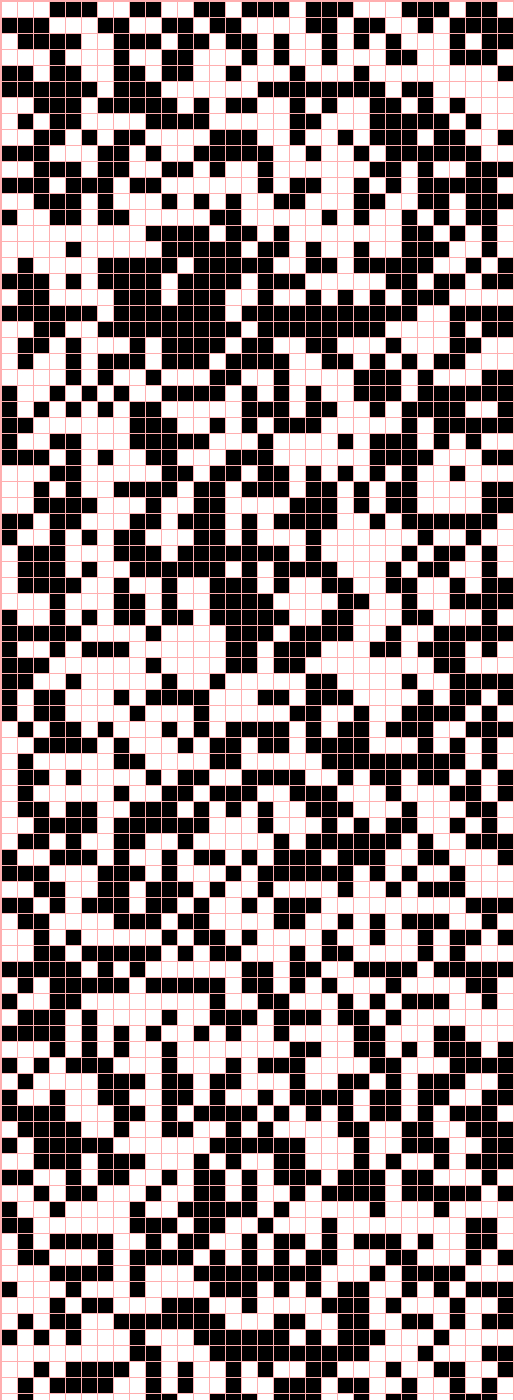}}
  	\hfill
  	\subfloat[$s_5$\label{dsfmt_123456789123456789_spaceo}]{%
  		\includegraphics[width=0.1\linewidth, height=5.0cm]{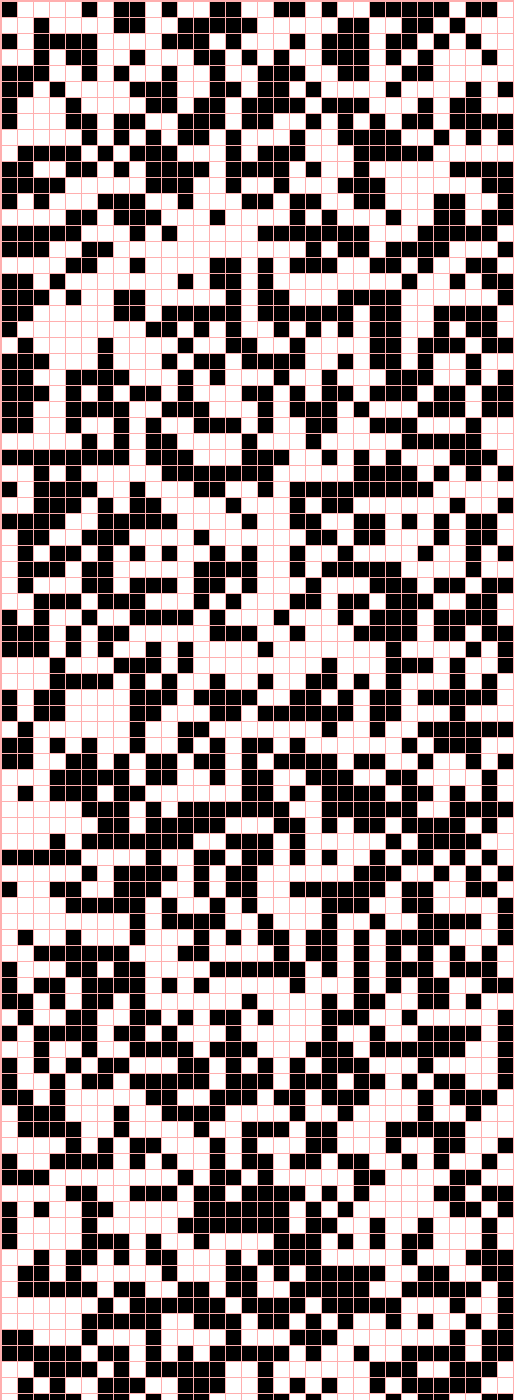}}
  			\hfill
\subfloat[$s_1$\label{rule30_7_space}]{%
\includegraphics[width=0.1\linewidth, height=5.0cm]{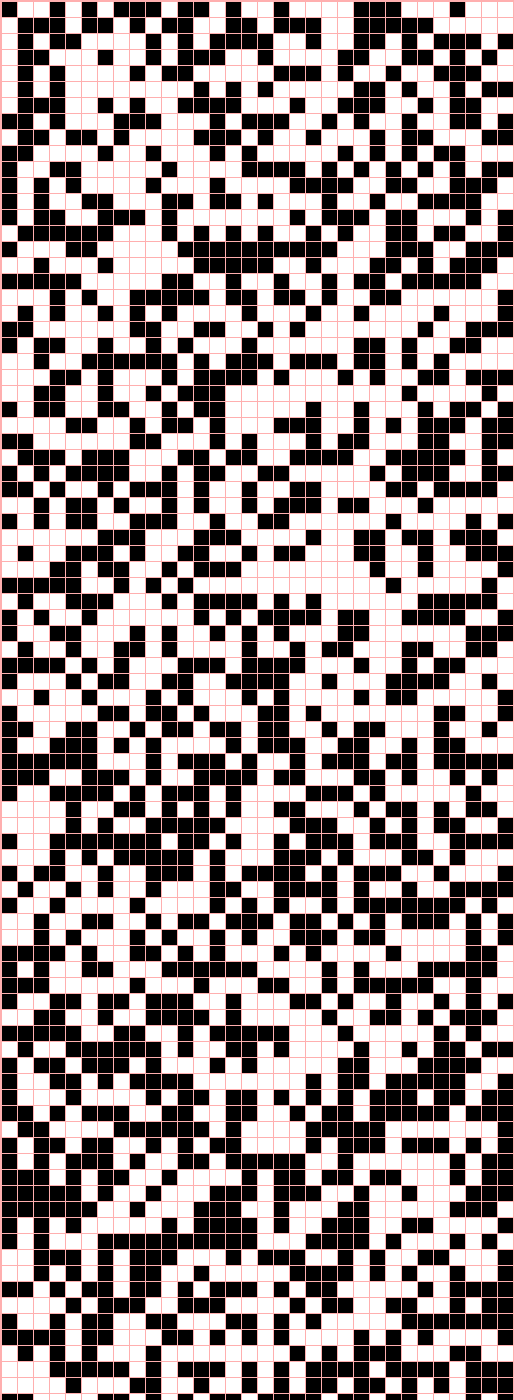}}
\hfill
\subfloat[$s_3$\label{rul30_12345_space}]{%
 \includegraphics[width=0.1\linewidth, height=5.0cm]{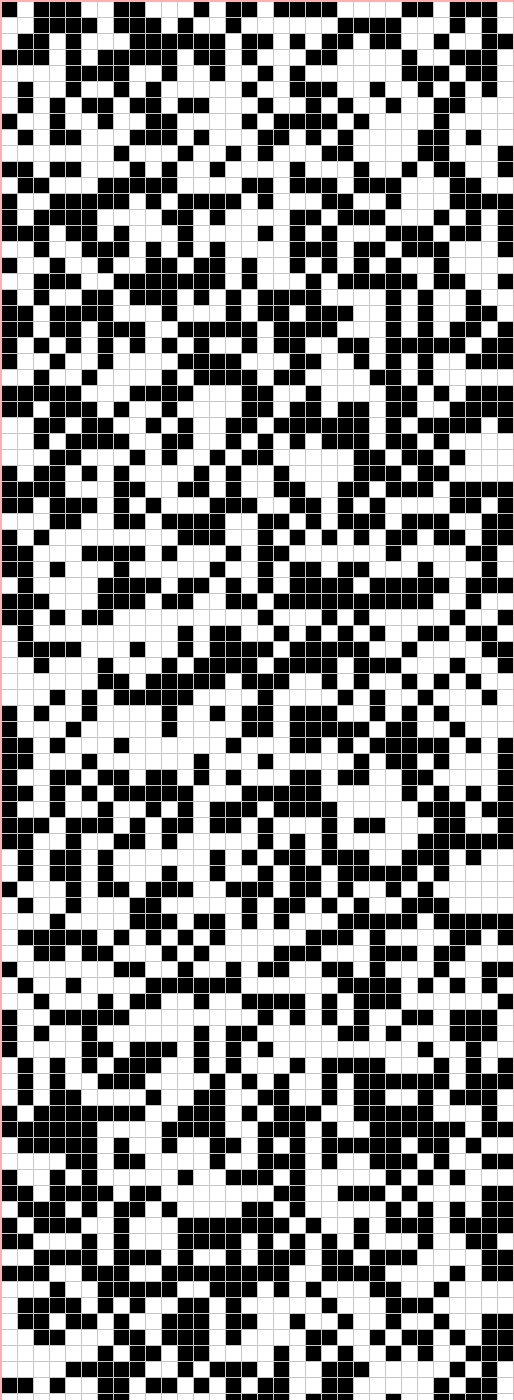}}
\hfill
\subfloat[$s_4$\label{rule30_9650218_space}]{%
\includegraphics[width=0.1\linewidth, height=5.0cm]{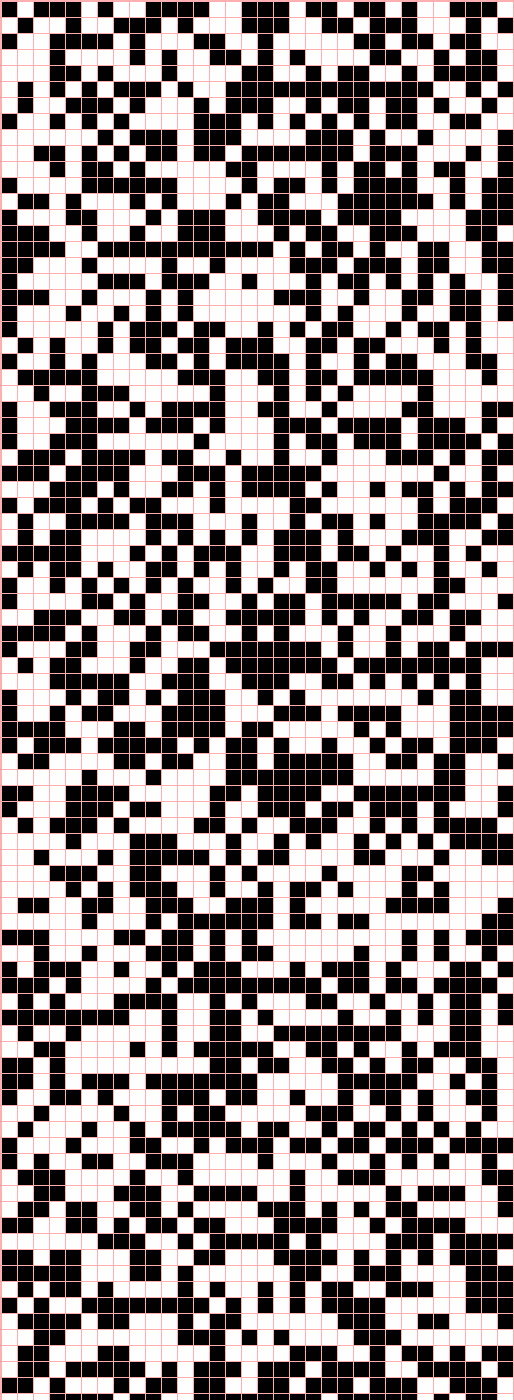}}
\hfill
\subfloat[$s_5$\label{rule30_123456789123456789_spaceo}]{%
\includegraphics[width=0.1\linewidth, height=5.0cm]{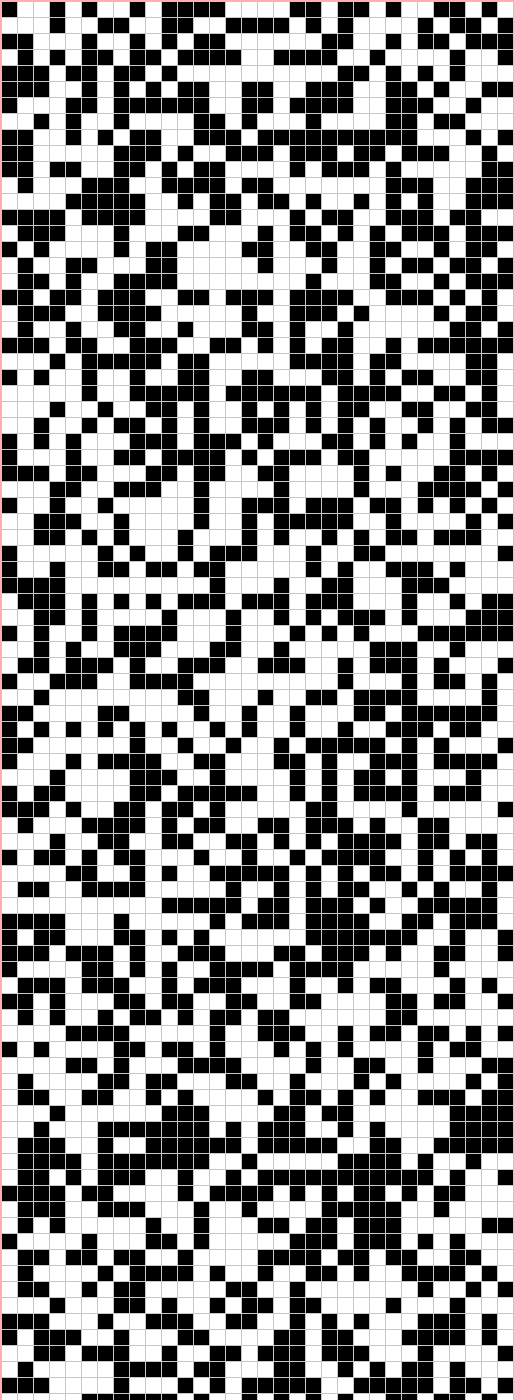}}
\hfill\\
\subfloat[$s_1$\label{rule30-45_7_space}]{%
\includegraphics[width=0.1\linewidth, height=5.0cm]{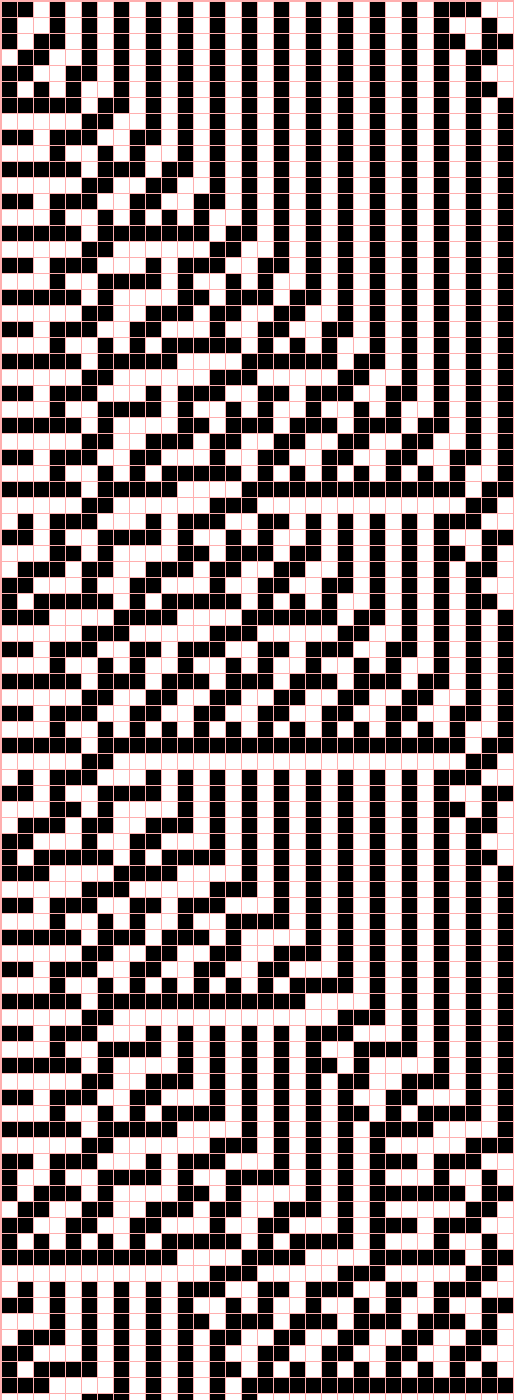}}
\hfill
\subfloat[$s_3$\label{rule30-45_12345_space}]{%
 \includegraphics[width=0.1\linewidth, height=5.0cm]{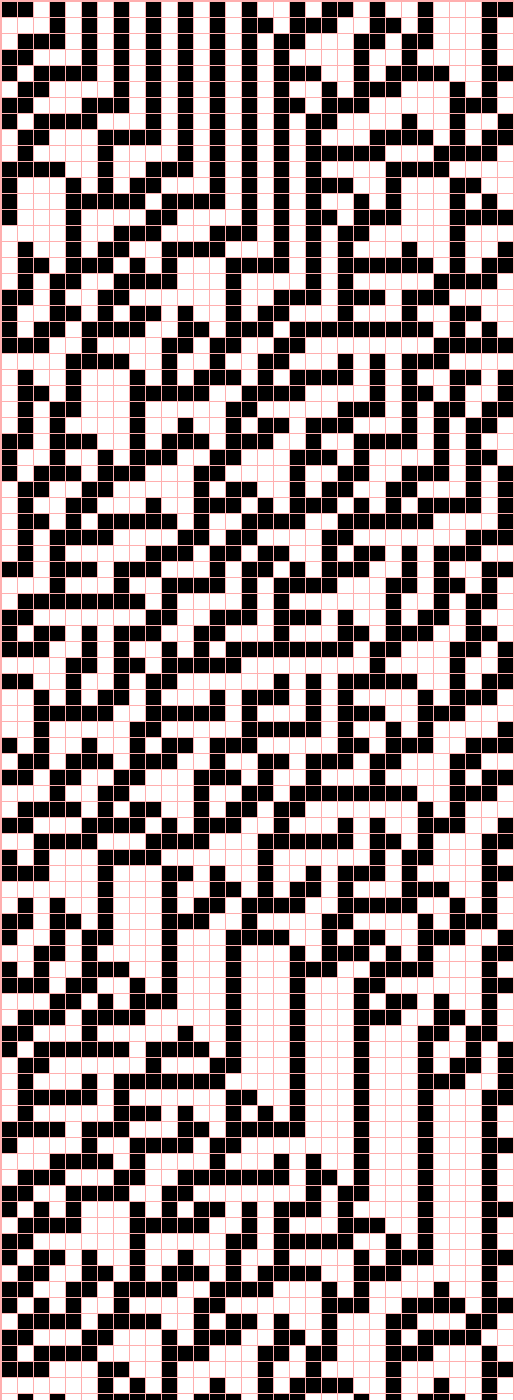}}
\hfill
\subfloat[$s_4$\label{rule30-45_9650218_space}]{%
\includegraphics[width=0.1\linewidth, height=5.0cm]{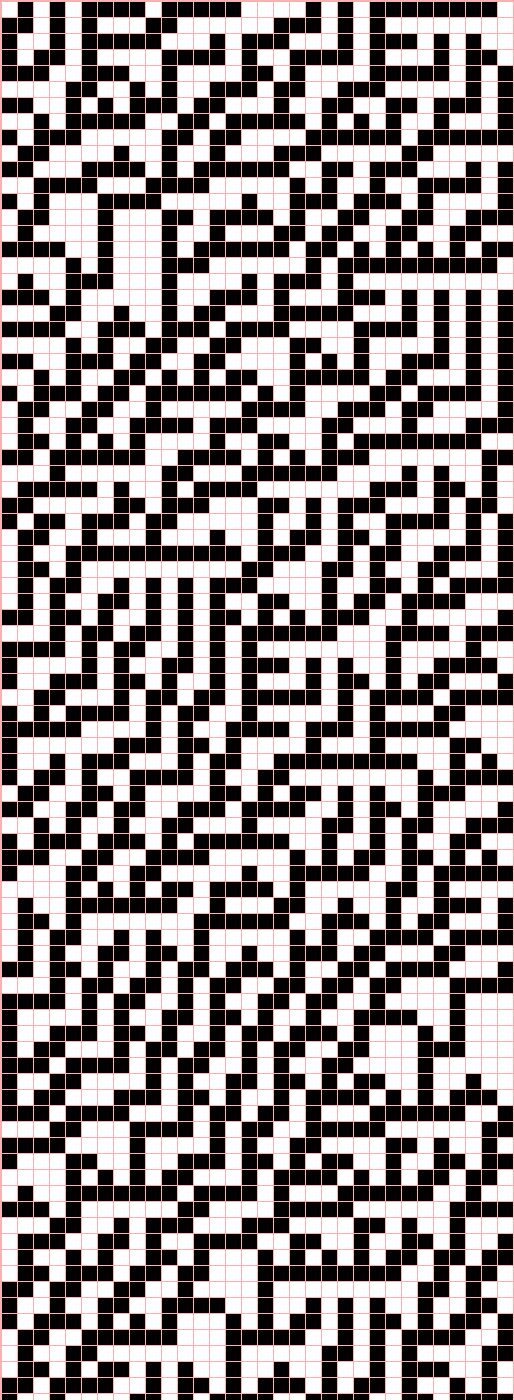}}
\hfill
\subfloat[$s_5$\label{rule30-45_123456789123456789_space}]{%
\includegraphics[width=0.1\linewidth, height=5.0cm]{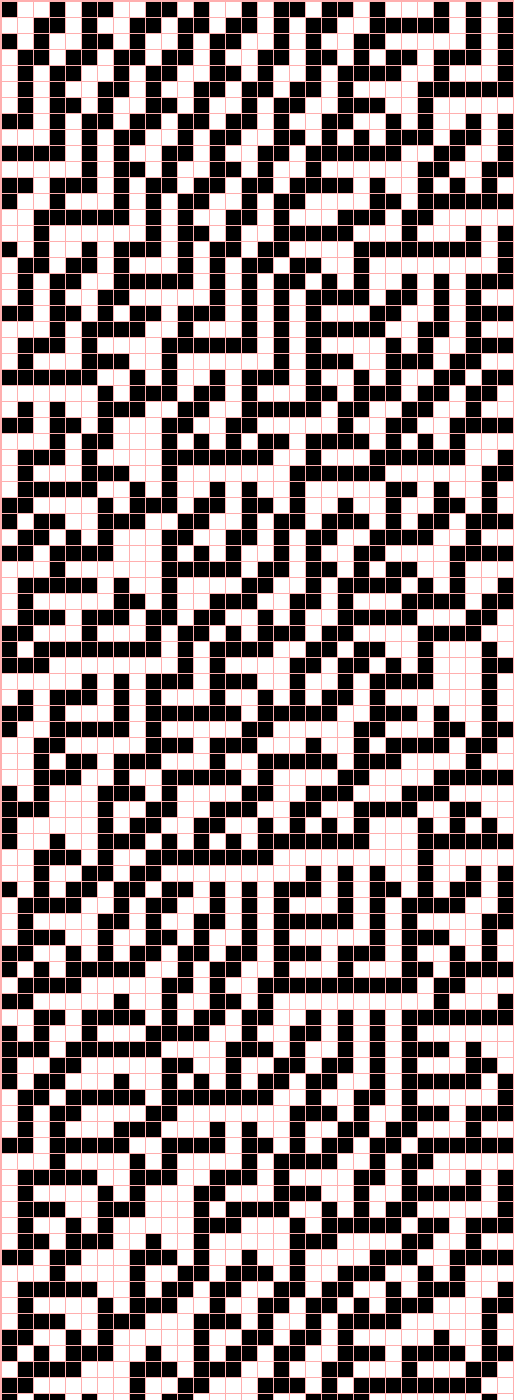}}
\hfill
%
\subfloat[$s_1$\label{maxlength0_7_space}]{%
\includegraphics[width=0.1\linewidth, height=5.0cm]{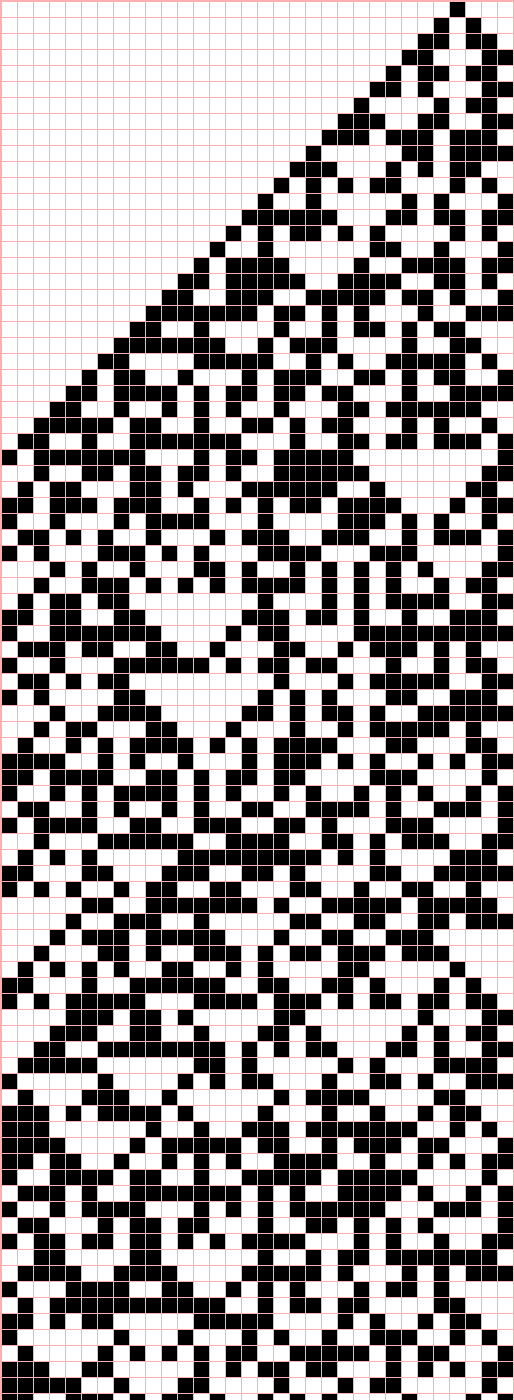}}
\hfill
\subfloat[$s_3$\label{maxlength0_12345_space}]{%
 \includegraphics[width=0.1\linewidth, height=5.0cm]{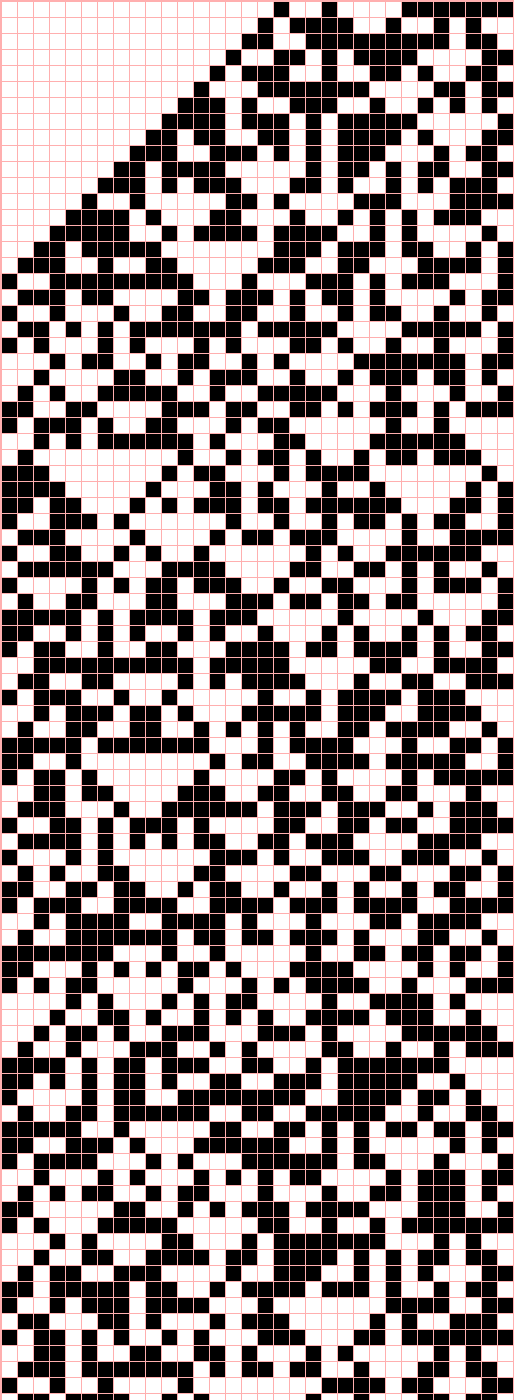}}
\hfill
\subfloat[$s_4$\label{maxlength0_9650218_space}]{%
\includegraphics[width=0.1\linewidth, height=5.0cm]{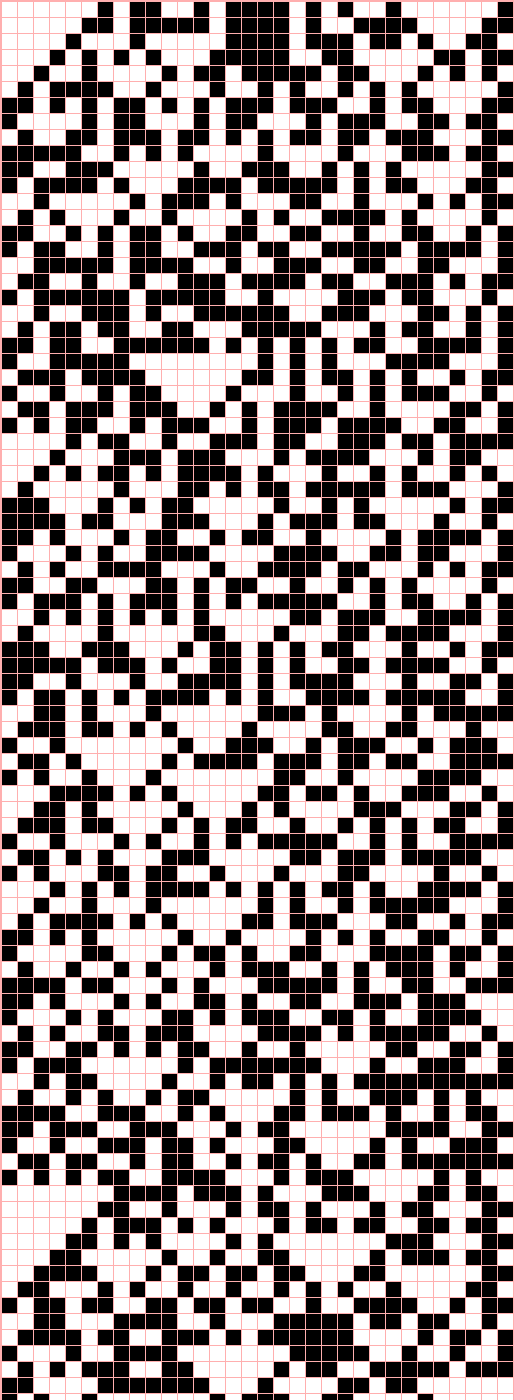}}
\hfill
\subfloat[$s_5$\label{maxlength0_123456789123456789_spaceo}]{%
\includegraphics[width=0.1\linewidth, height=5.0cm]{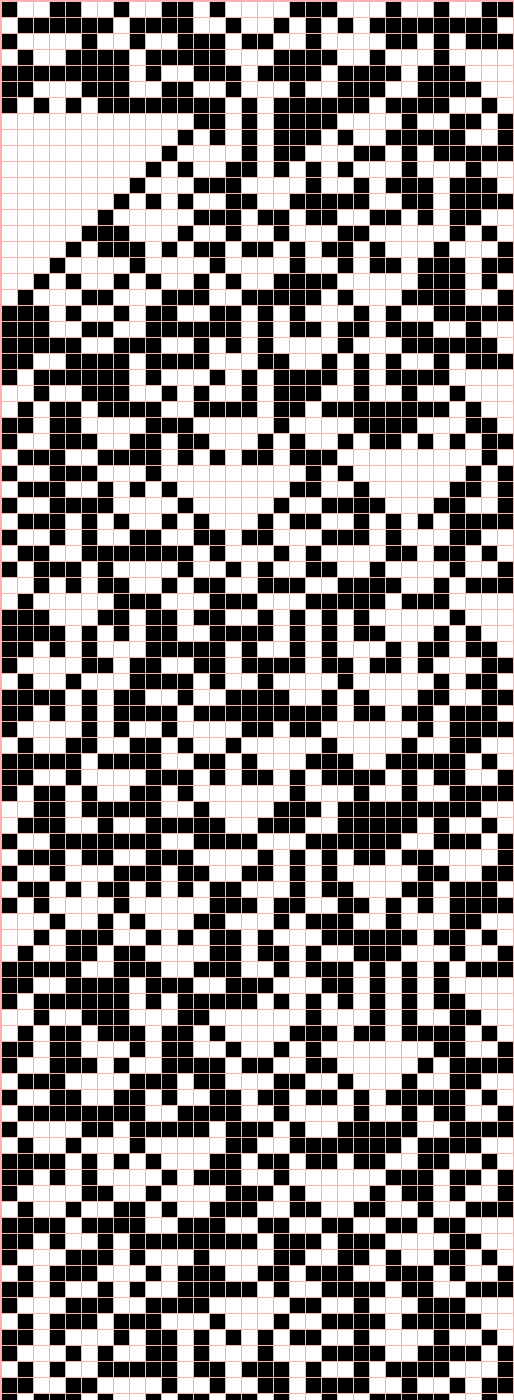}}
\hfill\\
\subfloat[$s_1$\label{maxlength1_7_space}]{%
\includegraphics[width=0.1\linewidth, height=5.0cm]{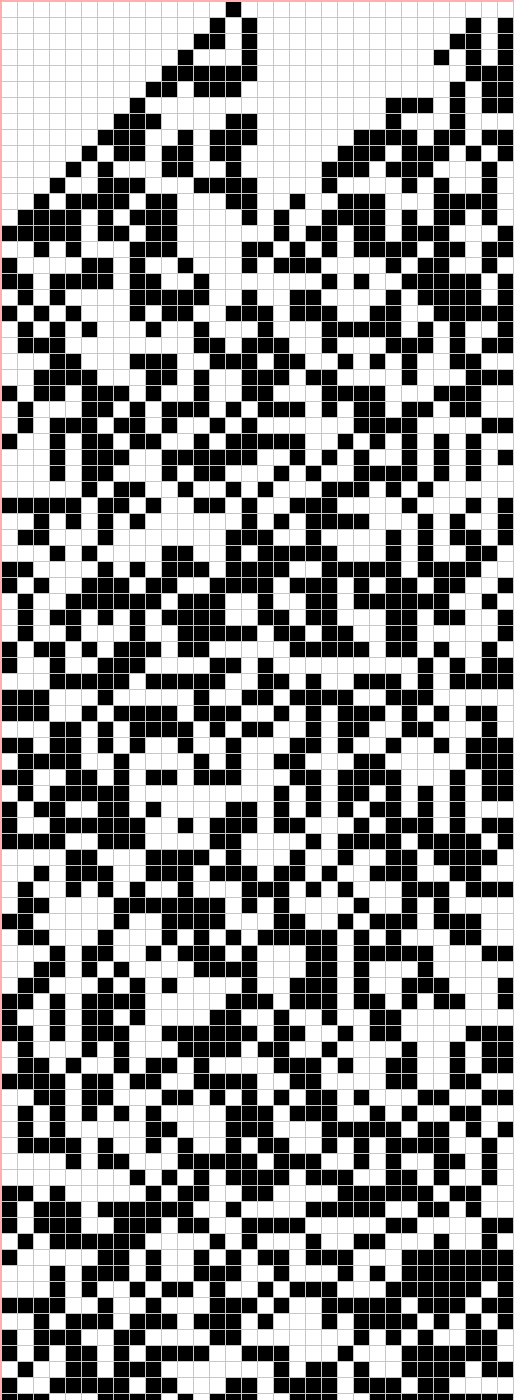}}
\hfill
\subfloat[$s_3$\label{maxlength1_12345_space}]{%
 \includegraphics[width=0.1\linewidth, height=5.0cm]{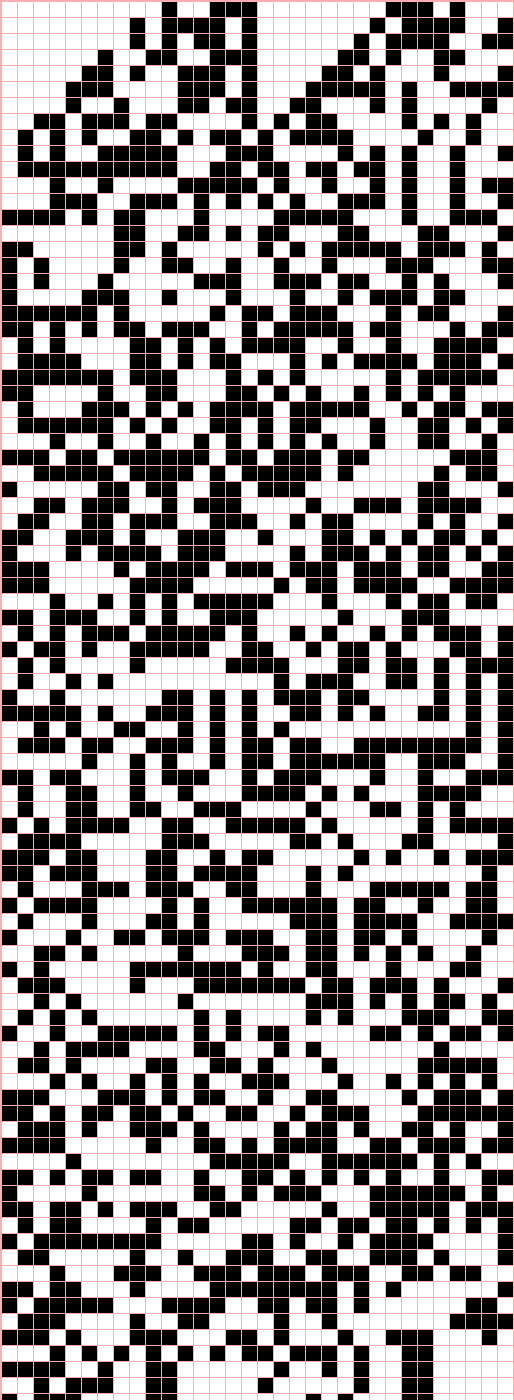}}
\hfill
\subfloat[$s_4$\label{maxlength1_9650218_space}]{%
\includegraphics[width=0.1\linewidth, height=5.0cm]{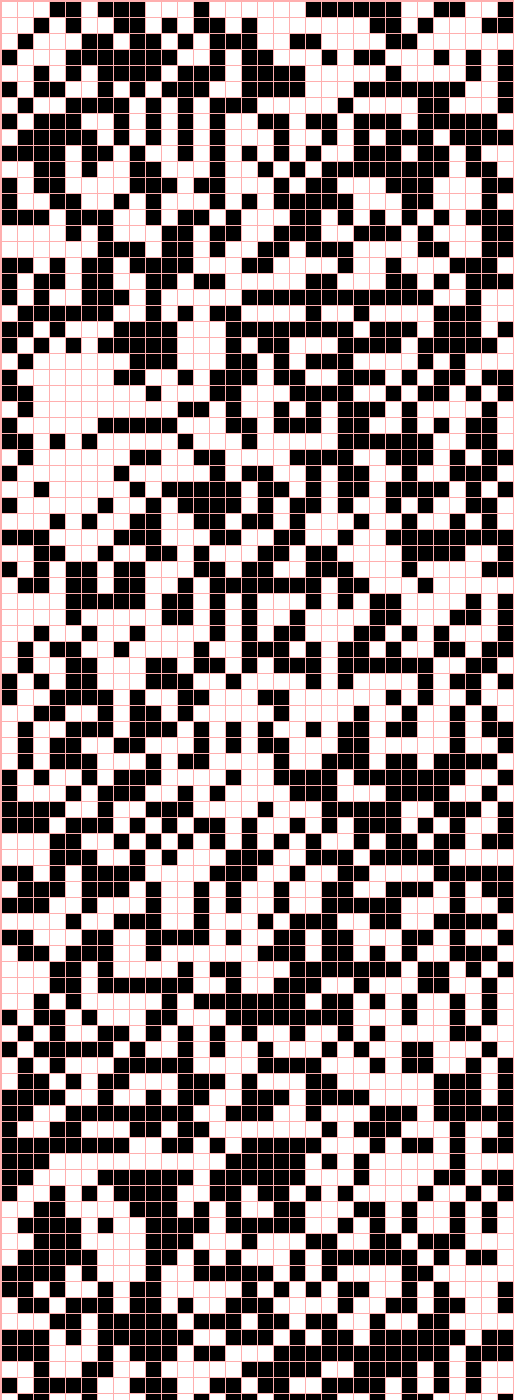}}
\hfill
\subfloat[$s_5$\label{maxlength1_123456789123456789_space}]{%
\includegraphics[width=0.1\linewidth, height=5.0cm]{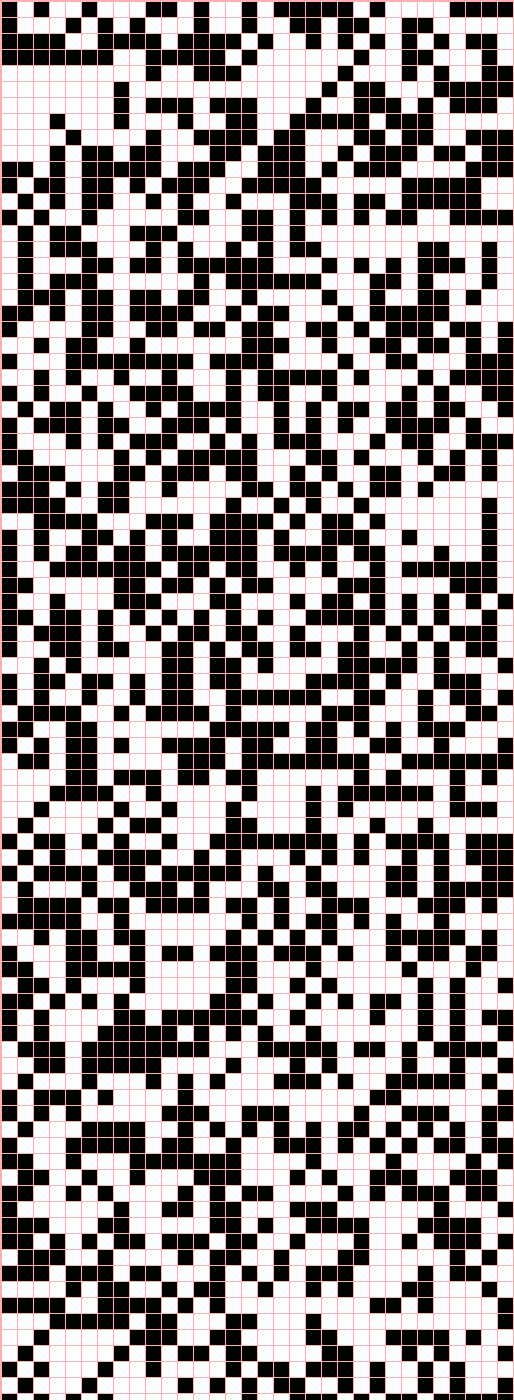}}
\hfill
%
\subfloat[$s_1$\label{nonlinear_7_space}]{%
\includegraphics[width=0.12\linewidth, height=5.0cm]{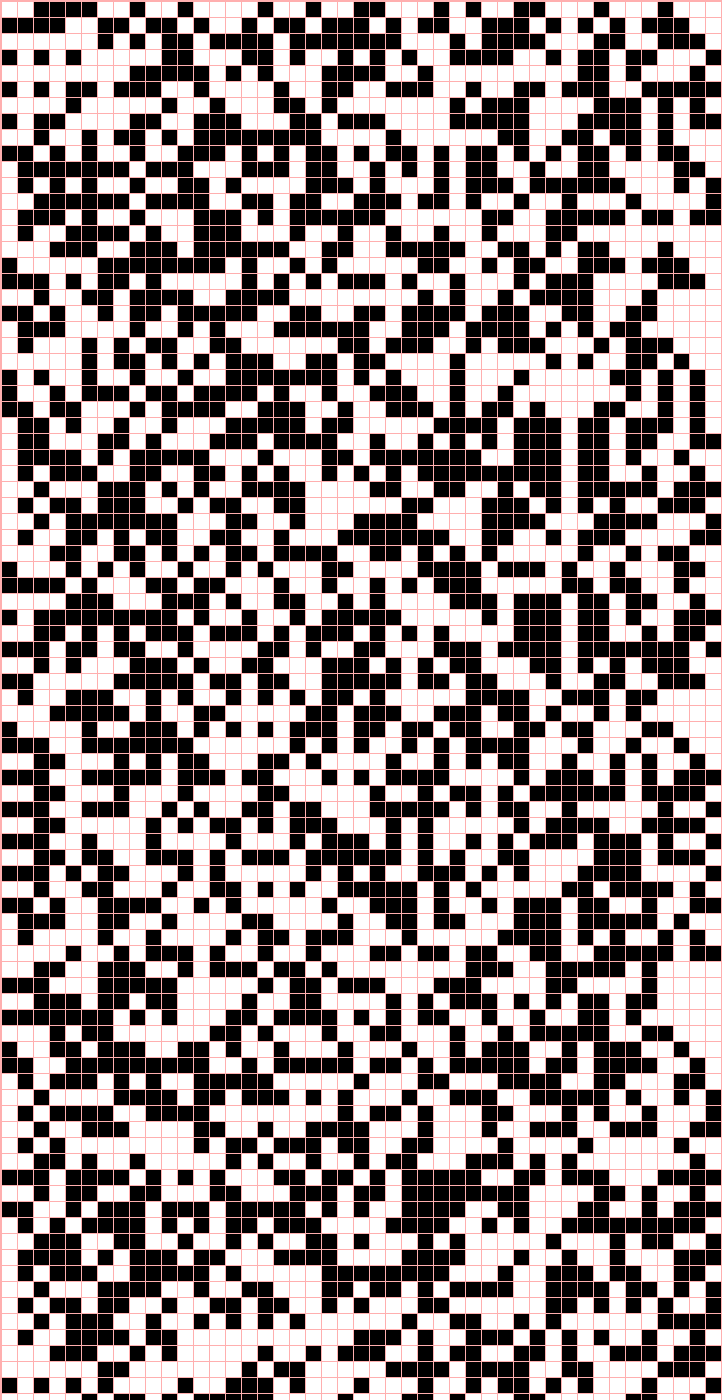}}
\hfill
\subfloat[$s_3$\label{nonlinear_12345_space}]{%
 \includegraphics[width=0.12\linewidth, height=5.0cm]{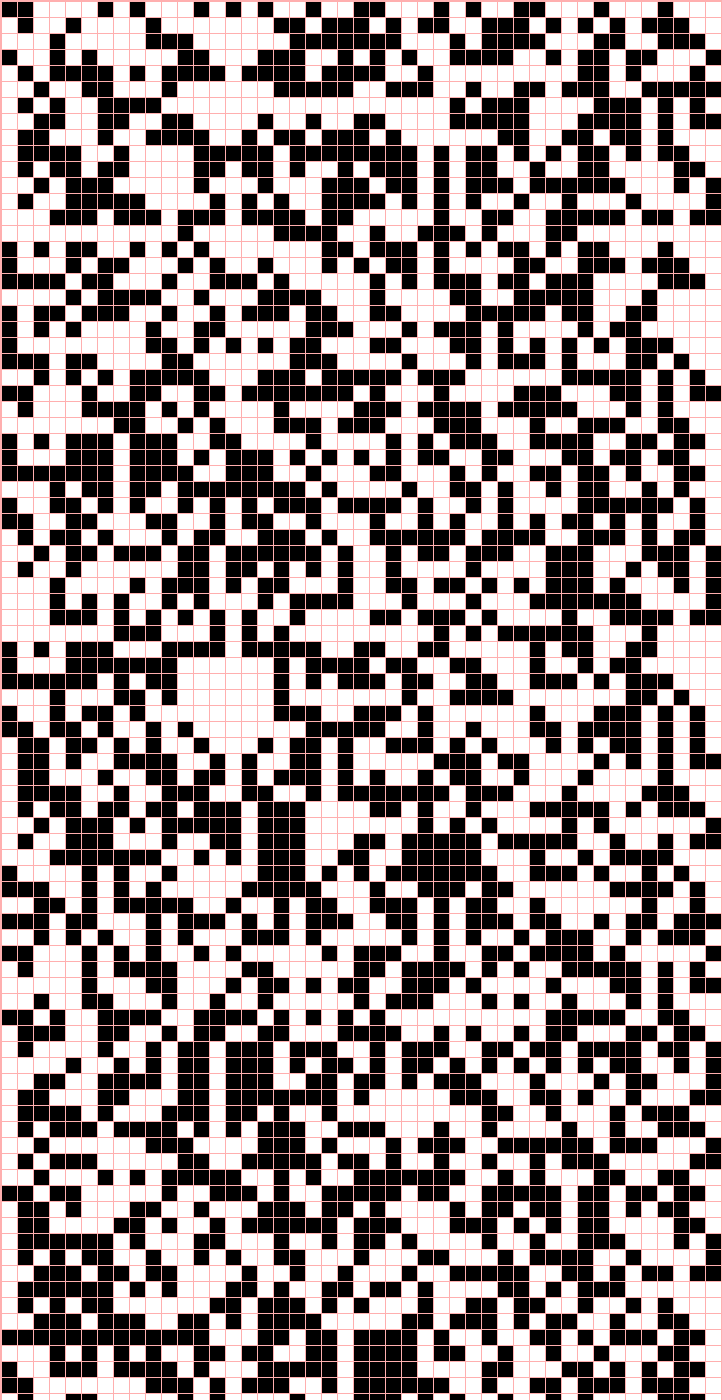}}
\hfill
\subfloat[$s_4$\label{nonlinear_9650218_space}]{%
\includegraphics[width=0.12\linewidth, height=5.0cm]{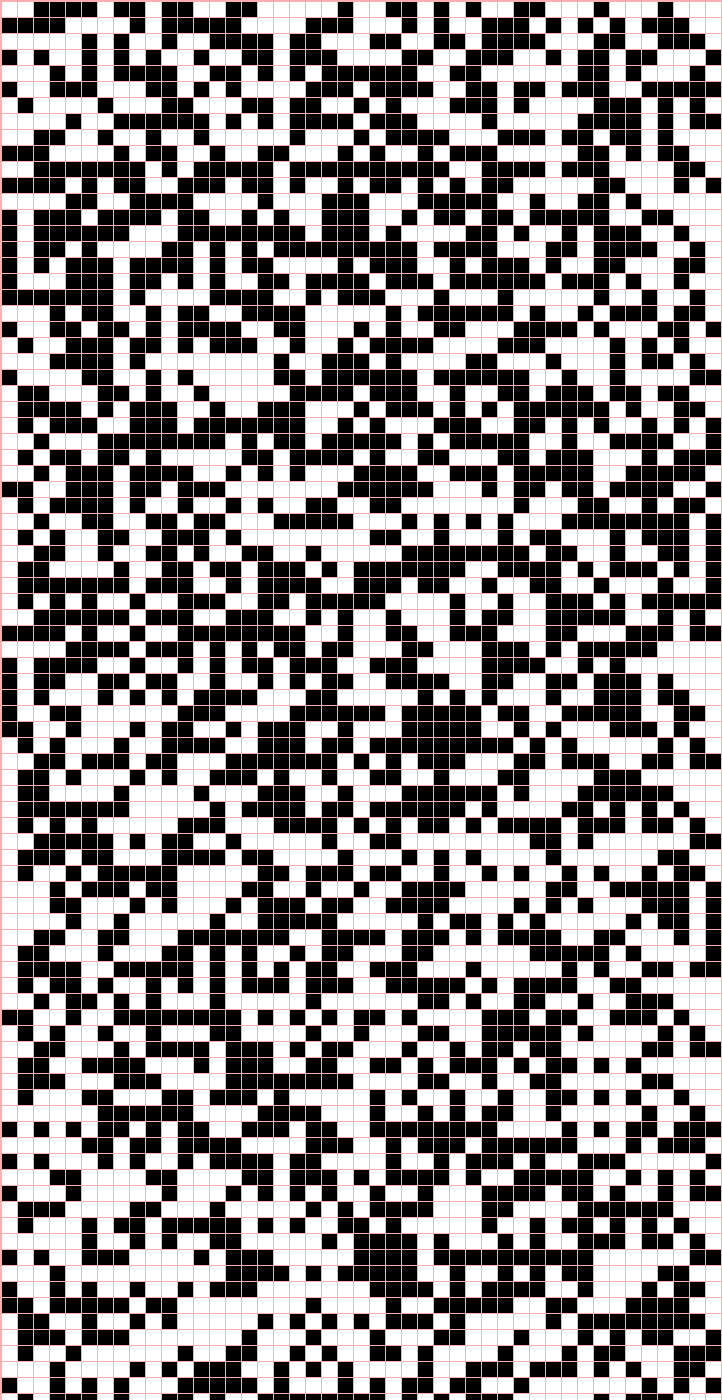}}
\hfill
\subfloat[$s_5$\label{nonlinear_123456789123456789_spaceo}]{%
\includegraphics[width=0.12\linewidth, height=5.0cm]{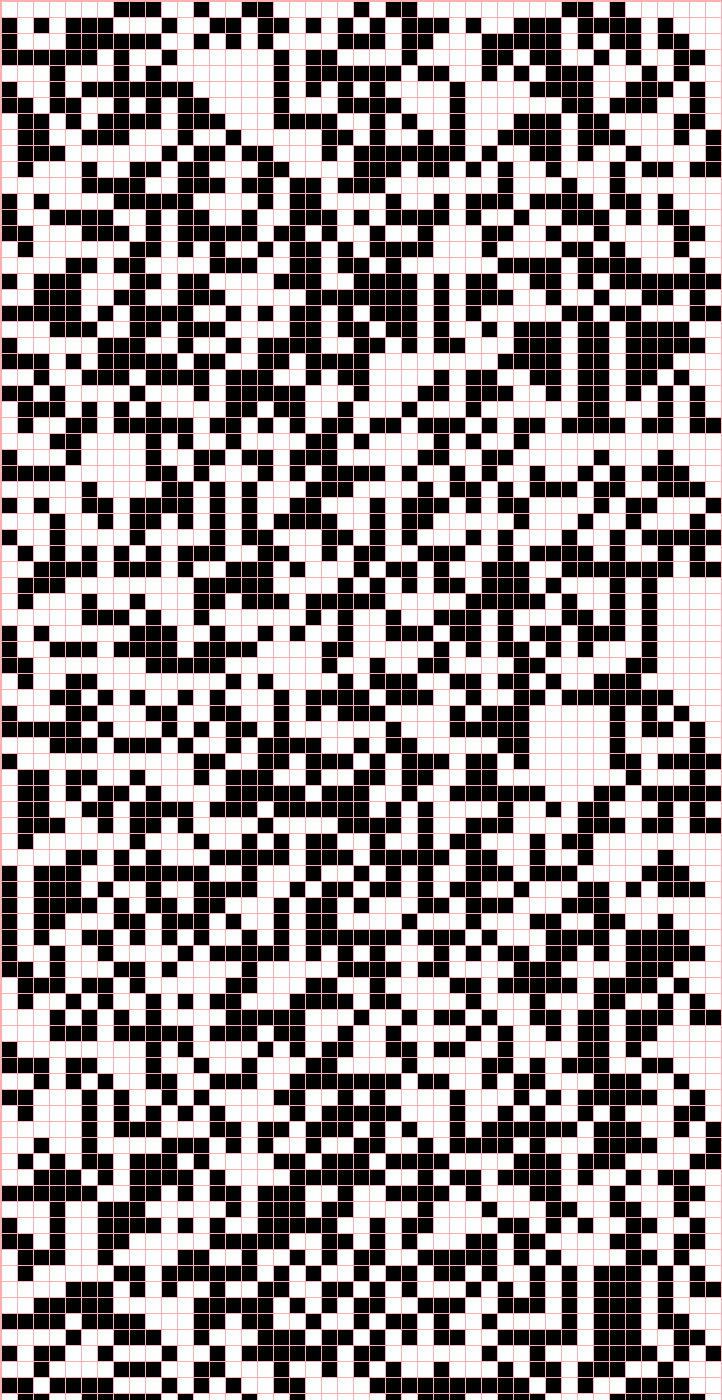}}
\caption{Space-time diagram for dSFMT19937 $32$ bit (\ref{dsfmt_7_space} to \ref{dsfmt_123456789123456789_spaceo}), rule 30 (\ref{rule30_7_space} to \ref{rule30_123456789123456789_spaceo}), rule 30-45 (\ref{rule30-45_7_space} to \ref{rule30-45_123456789123456789_space}), max-length CA with $\gamma=0$ (\ref{maxlength0_7_space} to \ref{maxlength0_123456789123456789_spaceo}), max-length CA with $\gamma =1$ (\ref{maxlength1_7_space} to \ref{maxlength1_123456789123456789_space}) and non-linear $2$-state CA (\ref{nonlinear_7_space} to \ref{nonlinear_123456789123456789_spaceo})
 }
\label{fig:ca_space-time}
\end{figure}

\begin{itemize}[leftmargin=1pt]
\item For \verb minstd_rand, ~the last $6$ bits of the generated numbers are fixed and for Knuth's \verb MMIX, ~last $2$ bits of four consecutive numbers form a pattern.
\item For \verb rand, ~\verb lrand, ~Borland's LCG, \verb MRG31k3p ~and \verb random, ~the percentage of black and white boxes representing $1$s and $0$s are not same. Even for \verb PCG-32, ~there is pattern visible in the diagrams.
\item \verb LFSR113 ~forms pattern for some seeds. For WELL and Xorshift generators, dependency on seed is visible for the initial numbers.
\item For MTs and SFMTs, the dependency on seed is visible for very few levels. For dSFMTs, there is pattern visible in the diagrams.

\item Among the CA-based generators, rule $30-45$ has visible patterns and max-length CAs have dependency on seed up to some initial configurations. However, the figures for rule $30$ CA and non-linear $2$-state CA appear relatively random.
\item For the LCGs, the dependency on seed is less visible than the LFSRs.
\item If observed minutely, every PRNG has some kind of clubbing of white boxes and black boxes, that is, none of the figures is actually free of pattern. However, for the good PRNGs, these patterns are non-repeating.
\end{itemize}

\subsection{Final Ranking and Remark}\label{chap:randomness_survey:sec:final_rank}
Using the space-time diagrams along with the statistical tests, we can further improve the rankings of the PRNGs --
\begin{itemize}
	\item \verb SFMT19937-64 ~holds the first position as it appears more random than \verb SFMT19937-32. ~
	
	\item Rule $30$ holds the $3^{rd}$ rank, whereas \verb MT19937-64 ~holds rank $4$. \verb PCG-32 ~is better than \verb MT19937-32 ~and \verb dSFMT-32. ~So, it holds rank $5$. The next rank holder is \verb MT19937-32. ~
	
	\item \verb dSFMT-32 ~has less dependency on seed than \verb WELL1024a ~and \verb xorshift128+. ~So, it is ranked $7^{th}$ position.
	
	\item \verb xorshift64* ~has no dependency on seed, so it is ranked higher than \verb WELL1024a ~and \verb xorshift128+. ~ 
	
	\item \verb WELL512a ~is ranked lower than \verb WELL1024a ~and \verb xorshift128+, ~as it has more dependency on seed. As \verb Tauss88 ~(rank $11$) sometimes cannot pass any tests, so it is ranked lower than \verb WELL512a ~(rank $10$).
	
	\item \verb dSFMT-52 ~has less dependency on seeds than non-linear $2$-state CA based PRNG and max-length CA with $\gamma=1$. So, it holds rank $12$.
	
	\item Non-linear $2$-state CA based PRNG and max-length CA with $\gamma=1$ based PRNG form the group of $13$ rank holders.
	
	\item Although \verb LFSR113 ~and \verb xorshift1024* ~can perform well for some seeds, but because of its dependency on seeds and visible patterns in the space-time diagram, these are ranked lower than max-length CA with $\gamma=1$. Therefore, \verb LFSR113 ~and \verb xorshift1024* ~downgrade to rank $14$.
	
	
	\item Knuth's \verb MMIX ~and \verb xorshift32 ~generator are in the same group (rank $15$).
	
	\item Max-length CA with $\gamma=0$ has better rank (rank $16$) than rule $30-45$ CA (rank $17$).
	
	\item Among \verb rand, ~\verb lrand, ~\verb minstd_rand, ~Borland's LCG, \verb MRGk13p ~and \verb random, ~the ranking is \verb minstd_rand ~(rank $23$) < \verb MRGk13p ~(rank $22$) < Borland's LCG (rank $21$)< \verb random ~(rank $20$) < \verb rand ~(rank $19$) < \verb lrand ~(rank $18$), where `<' indicates left PRNG has poorer performance than the right one.
	
	\item \verb LFSR258 ~is the worst generator among the selected PRNGs.
\end{itemize}
\end{enumerate}

\begin{table}[!h]
	\centering
	\small
	\caption{Summary of all empirical test results and final ranking}
	\label{tab:final_rank_comparison}
	\resizebox{1.00\textwidth}{5.5cm}{
		\begin{tabular}{|c|c|c|c|c|c|c|c|p{10.2em}|c|c|c|}
			\hline
			\multicolumn{2}{|c|}{\multirow{2}{*}{\theadfont{Name of the PRNGs}}} & \multicolumn{3}{c|}{\theadfont{Fixed Seeds}} &  \multicolumn{2}{c|}{\theadfont{Random Seeds}} & \multirow{2}{*}{\theadfont{Lattice Test}} & \multirow{2}{*}{\theadfont{Space-time Diagram}} & \multicolumn{3}{c|}{\theadfont{Ranking}} \\
			\cline{3-7}\cline{10-12}
			\multicolumn{2}{|c|}{ } & Diehard & TestU01 & NIST & Average & Range &  & & {\theadfont{$1^{st}$ Level}} & \theadfont{$2^{nd}$ Level} &{\theadfont{Final Rank}}\\
			\hline
			\multirow{7}{*}{\rotatebox{90}{LCGs}}& MMIX & 4-6 & 16-19 & 7-8 & 6.5 & 2-9  & Not Filled & Last $2$ bits fixed & 8 & 9 & 15\\
			\cline{2-12}
			& minstd\_rand & 0 & 1 & 1-2 &0.38 & 0-1 &  Not Filled & last $6$ bits fixed & 12 & 14 & 23\\
			\cline{2-12}
			& Borland LCG & 1 & 3 & 4-5 & 1.9 & 1-2 & Not Filled &  Last $2$ bits fixed & 11 & 12 & 21\\
			\cline{2-12}
			& rand & 1 & 1-3 & 2-3 &  &  &  Not Filled & More 0s than 1s & 11 & 13 & 19\\
			\cline{2-12}
			& lrand48() & 1 & 2-3 & 2 & 1 & 1 & Not Filled & More 0s than 1s & 11 & 13 & 18\\
			\cline{2-12}
			& MRG31k3p & 0-1 & 1-2 & 1-2 & 0.9 & 0-1 & Scattered & LSB is 0, block of 0s, dependency on seed & 12 & 14 & 22\\
			\cline{2-12}
			& PCG-32 & 9-11 & 24-25 & 14-15 & 9.3 & 6-12 & Relatively Filled & Independent of seed & 2 & 4 & 5\\
			\cline{2-12}
			\hline
			\multirow{16}{*}{\rotatebox{90}{LFSRs}}& random() & 1 & 1-3 & 1 & 1 & 1 & Not Filled & MSB is 0, blocks of 0s & 11 & 13 & 20\\
			\cline{2-12}
			& Tauss88 & 9-11 & 21-23 & 14-15 & 9.0 & 0-12 & Relatively Filled & Independent of seed, block of $0$s & 4 & 7 & 11\\
			\cline{2-12}
			& LFSR113 & 5-11 & 6-23 & 1-15 & 9.3 & 6-12 & Relatively Filled & Dependency on seed, Block of 0s & 7 & 7 & 14\\
			\cline{2-12}
			& LFSR258 & 0-1 & 0-5 & 0-2 &  1.8 & 1-2 & Scattered & Pattern & 12 & 14 & 24\\
			\cline{2-12}
			& WELL512a & 7-10 & 23 & 14-15 & 8.5 & 5-11 & Relatively filled & First few numbers are fixed with seed dependency & 5 & 6 & 10\\
			\cline{2-12}
			& WELL1024a & 9-10 & 24-25 & 14-15 & 9.2 & 6-11 & Relatively Filled & Dependency on seed & 3 & 4 & 9\\
			\cline{2-12}
			& MT19937-32 & 9-10 & 25 & 13-15 & 9.3 & 6-12 & Relatively Filled & Independent of seed & 3 & 4 & 6\\
			\cline{2-12}
			& MT19937-64 & 8-11 & 24-25 & 15 & 9.4 & 6-11 & Relatively Filled & Independent of seed & 2 & 3 & 4\\
			\cline{2-12}
			& SFMT19937-32 & 9-10 & 25 & 15 & 9.5 & 5-12 & Relatively Filled & Independent of seed & 1 & 1 & 2\\
			\cline{2-12}
			& SFMT19937-64 & 9-11 & 25 & 15 & 9.52 & 6-12 & Relatively Filled & Independent of seed & 1 & 1 & 1\\
			\cline{2-12}
			& dSFMT-32 & 7-11 & 24-25 & 13-15 & 9.3 & 5-11 & Relatively Filled & Independent of seed & 5 & 5 & 7\\
			\cline{2-12}
			& dSFMT-52 & 5-7 & 9-11 & 3 & 5.97 & 3-7 & Relatively Filled & Less dependency on seed & 9 & 10 & 12\\
			\cline{2-12}
			&  xorshift32 & 2-4 & 17 & 2-13 & 5.5 & 3-7 & Not Filled & Blocks of 0s & 9 & 10 & 15\\
			\cline{2-12}
			&  xorshift64* & 7-10 & 25 & 14-15 & 8.0 & 6-11 & Relatively Filled & Independent of seed & 5 & 6 & 8\\
			\cline{2-12}
			&  xorshift1024* & 6-9 & 20-21 & 6-15 & 7.0 & 4-9 & Not Filled & Dependency on seed, Pattern & 6 & 8 & 14\\
			\cline{2-12}
			&  xorshift128+ & 8-10 & 24-25 & 14-15 & 9.4 & 6-12 & Relatively Filled & Dependency on seed for first few numbers & 4 & 4 & 9\\
			\hline
			\multirow{6}{*}{\rotatebox{90}{CAs}}& Rule $30$ & 8-11 & 24-25 & 15 & 10.2 & 7-12 & Relatively Filled & Independent of seed & 2 & 2 & 3\\
			\cline{2-12}
			& Hybrid CA with Rules $30$ \& $45$ & 0-3 & 1-8 & 0-3 & 2.0 & 0-3 & Not Filled & Pattern & 11 & 12 & 17\\
			\cline{2-12}
			& Maximal Length CA with $\gamma=0$ & 0-2 & 12 & 10-11 & 1.6 & 1-2 & Not Filled & Pattern & 10 & 11 & 16\\
			\cline{2-12}
			& Maximal Length CA with $\gamma=1$ & 3-4 & 15-17 & 14 & 1.8 & 1-4 & Relatively Filled & Dependency on seed for first few numbers & 8 & 11 & 13\\
			\cline{2-12}
			& Non-linear $2$-state CA & 5-7 & 11 & 3-4 & 5.85 & 2-8 & Relatively Filled &  Less dependency on seed & 9 & 9 & 13\\
		\cline{2-12}
			& \textbf{Proposed PRNG with CA} $\mathbf{\mathscr{R}}$ & $\mathbf{2-3}$ & $\mathbf{11-12}$ & $\mathbf{4-6}$ & $\mathbf{2.7}$ & $\mathbf{1-4}$ & \textbf{Relatively Filled} & \textbf{Less dependency on seed} & $\mathbf{10}$ & $\mathbf{11}$ & $\mathbf{13}$\\
			\hline
		\end{tabular}}
	\end{table}

Based on the empirical tests, we finally rank the selected $28$ PRNGs into $24$ groups. This final ranking is shown in Table~\ref{tab:final_rank_comparison}.

\section{Rank of Proposed PRNG and Challenges}\label{Chap:randomness_survey:sec:comparison}
To compare our proposed PRNG with the existing $28$ well-known PRNGs, we have tested our PRNG on the same platform with the same seeds. Here, we have used our CA as $32$-bit generator. That is, CA size is taken as $51$ with window length as $20$.

\begin{figure}[!h]
		\centering
		\subfloat[Diehard Test Result\label{fig:3-state_avg}]{\includegraphics[width=0.4\linewidth, height=4.0cm]{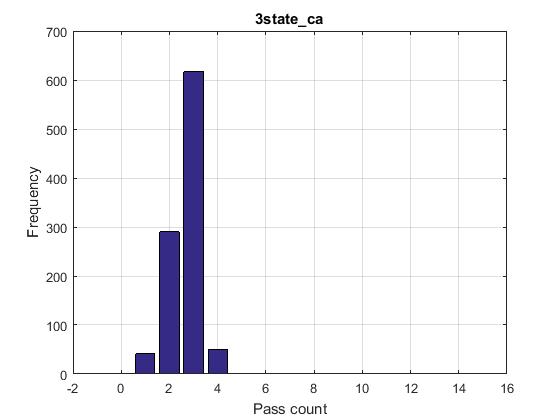}}
		\hfill
\subfloat[$s_1$\label{3state_7_space}]{%
\includegraphics[width=0.1\linewidth, height=5.0cm]{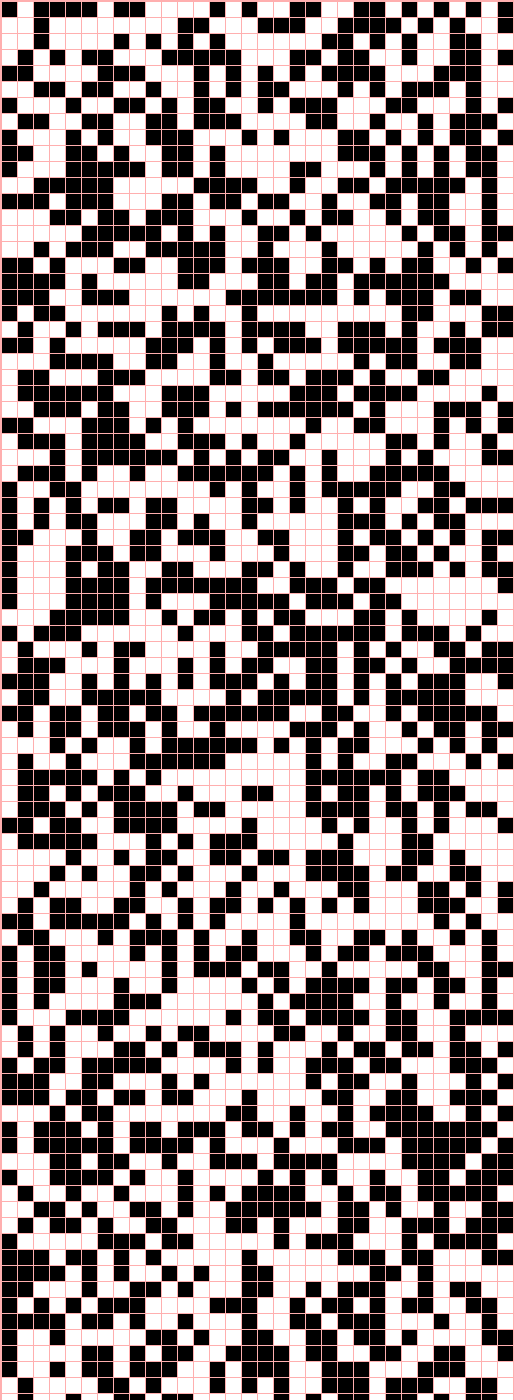}}
\hfill
\subfloat[$s_3$\label{3state_12345_space}]{%
 \includegraphics[width=0.1\linewidth, height=5.0cm]{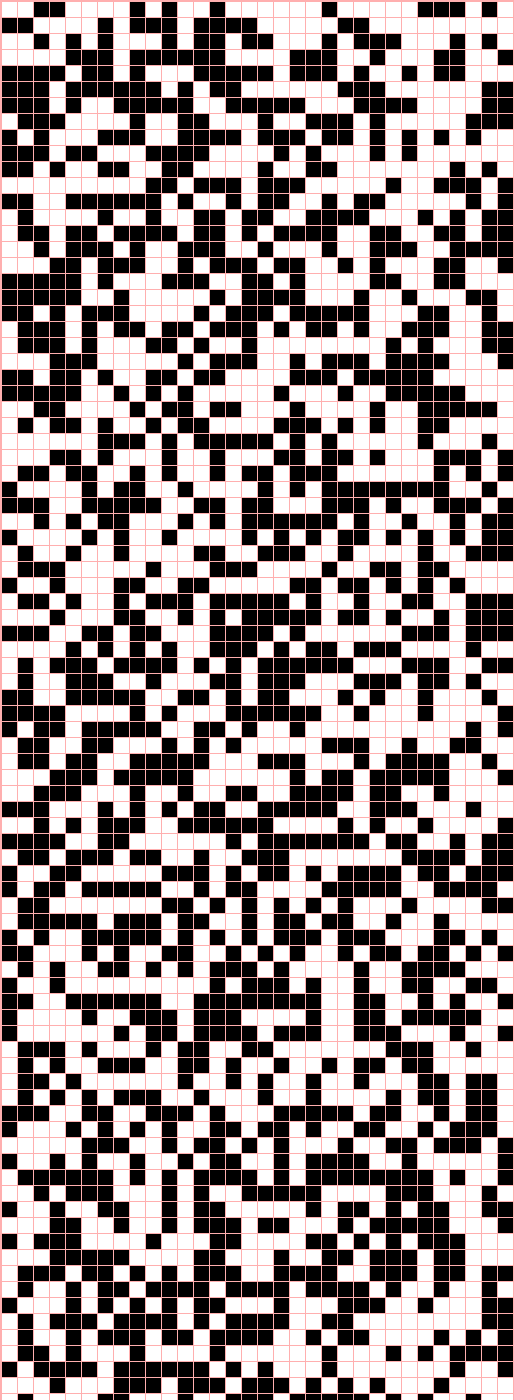}}
\hfill
\subfloat[$s_4$\label{3state_9650218_space}]{%
\includegraphics[width=0.1\linewidth, height=5.0cm]{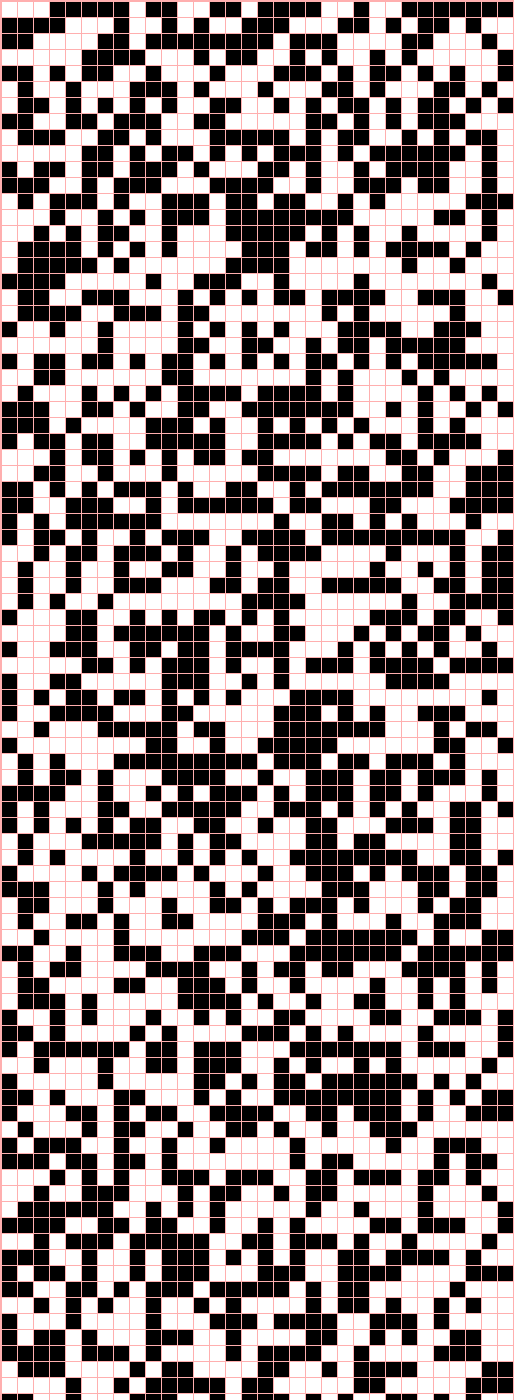}}
\hfill
\subfloat[$s_5$\label{3state_123456789123456789_space}]{%
\includegraphics[width=0.1\linewidth, height=5.0cm]{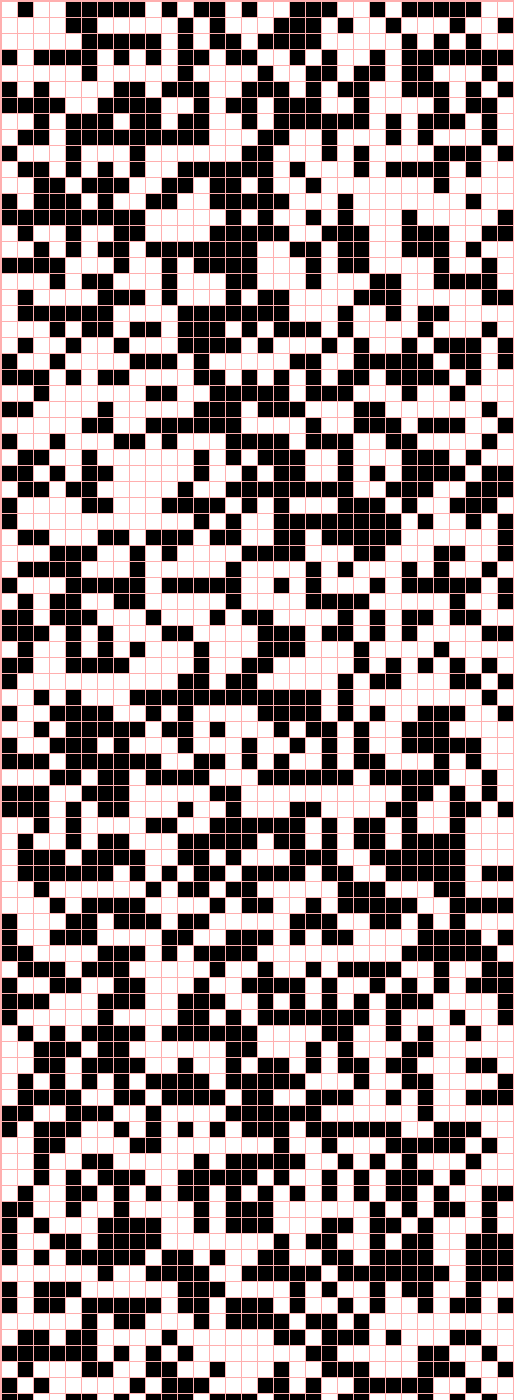}}

	\caption{Average test result for $1000$ seeds and Space-time diagram for the proposed PRNG with CA $\mathbf{\mathscr{R}}$}
	\label{fig:CA_R_space-time}
\end{figure}

However, the proposed PRNG does not perform well for the fixed seeds. It passes $2-3$ tests of Diehard ($2$ for $s_4$ and $3$ for other seeds), $11-12$ tests of battery rabbit of TestU01 library ($11$ for $s_3$, $s_4$ and $s_5$ and $12$ for $s_1$ and $s_2$) and $4-6$ tests of NIST test-suite ($6$ for $s_1$ and $s_2$, $5$ for $s_3$ and $4$ for others) for these fixed seeds. When we test our PRNG with the $1000$ random seeds generated by \verb rand ~with \verb srand(0), ~the average number of tests passed by it is $2.7$ (see Figure~\ref{fig:3-state_avg} for details). So, its performance is comparable to the max-length CA ($\gamma=0$) which has rank $10$ in first level ranking (see Table~\ref{tab:blind_test}) and rank $11$ in second level ranking (see Table~\ref{tab:blind_test_avg}). 

However, when we test the PRNG on lattice tests and space-time diagrams, the figures appear relatively random (Figure~\ref{fig:CA_R_space-time}), like the non-linear $2$-state CA based PRNG and max-length CA with $\gamma=1$ based PRNG (rank $13$ in Table~\ref{tab:final_rank_comparison}). Considering all these facts, we assign an average rank to our own PRNG with respect to its randomness quality.
 However, if we consider performance of the portable PRNGs like \verb rand, ~\verb lrand48, ~\verb MRG31k3p, ~\verb LFSR113, ~\verb LFSR258, ~\verb MMIX, ~Borland's LCG etc., then our PRNG beats all of them. Therefore, this CA is a good candidate to be used in day-to-day applications.

Table~\ref{tab:final_rank_comparison} also indicates that, among the CAs based PRNGs, only ECA rule $30$ falls into the elite category. However, practically building PRNG with rule $30$ following original proposal given in \cite{wolfram86c} is not feasible. Because, here we have to consider huge CA size and at each time-step, only one bit is to be collected from each configuration. So, rule $30$ is to be excluded as a potential PRNG.
This makes the ranking of potential usable CA-based PRNGs around average. 

This scenario puts a challenge in front of us -- can we design a CA based PRNG which can enter to the elite group of PRNGs, and can compete with them in terms of randomness quality? Our minute observation suggests that increment of number of states per cell helps to improve the randomness quality. This is the motivation  to move forward by increasing the number of states per cell. The work in reported in Chapter~\ref{Chap:10-stateCA}.

Apart from the randomness quality, however, there are other issues which are to be cherished by a good PRNG. The $3$-state CA based PRNG fulfills many such criteria, such as ease of implementation, portability, etc. Therefore, for many applications, the proposed PRNG may be more appropriate. So, we continue our research with $3$-state CA and find many more possible CAs in the next chapter that can be good PRNGs. 

%

\section{Conclusion}\label{chap:randomness_survey:sec:conclusion}
This chapter acts as the foundation for the next two chapters. In this chapter, we have discussed about a set of desirable properties for a good PRNG.
We have ascertained some necessary attributes of such a CA which are directly inferred from the context of unpredictability. As example, a $3$-neighborhood $3$-state CA rule $\mathbf{\mathscr{R}}=120021120021021120021021210$ is chosen which satisfies these properties and a window-based PRNG scheme using this CA is proposed to generate random numbers of any length. 

To find the position of this PRNG in terms of its randomness quality, we compare it with $28$ existing widely known \emph{good} PRNGs. We have realized that, the proposed PRNG has average ranking, whereas, the best performing PRNG is \verb SFMT19937-64. ~However, the proposed example PRNG performs better than most of the existing portable PRNGs and binary CAs based PRNGs. Having $3$ states also has some inherent benefits for implementation. Further, the empirical results hint that, by increasing number of states of a CA, its randomness quality can be improved.
In light of these observations, we continue our research with $3$-state CAs, and report the result of the research in the next chapter.

\chapter{A List of $3$-state Cellular Automata as Source of Randomness}\label{Chap:3-stateCA_list}
\begin{center}
\begin{quote}
\emph{I often reflect that had the Ternary instead of the denary Notation been adopted
in the infancy of Society, Machines something like the present would long ere this
been common, as the transition from mental to mechanical calculation would
have been so very obvious and simple}
\end{quote}
\hspace*{3.05in}{\em -- Thomas Fowler, 1841}
\end{center}


\noindent{\small This chapter explores $3$-neighborhood $3$-state cellular automata (CAs) to find a list of potential pseudo-random number generators (PRNGs). As the rule-space is huge ($3^{27}$), we have taken two greedy strategies to select the initial set of rules. These CAs are analyzed theoretically and empirically, to find the CAs with consistently good randomness properties. Finally, we have listed $596$ \emph{good} CAs which qualify as potential PRNGs.}

\section{Introduction}\label{Chap:3-stateCA_list:sec:intro}
{\large\textbf{T}}he cellular automaton (CA), which was first introduced as source of randomness was binary \cite{Wolfram85c}. In fact, most of the works have explored binary CAs to observe pseudo-randomness. When pseudo-randomness is used by real-life applications, such as VLSI test, cryptography, etc., then the binary CAs are the practical options \cite{Horte89a,ppc1,vlsi02a,aspdac04}. Not only the cellular automata based pseudo-randomness, the LFSRs (Linear Feedback Shift Registers) based pseudo-random number generators (PRNGs) are also binary systems. In this era of digital technology, binary systems have become a common choice to work with.



On the contrary, ternary systems remain effective alternative to the binary systems \cite{Hurst1984,5176262}. It has been argued that computing in ternary systems is more cost-effective than in binary systems. In fact, in the last two decades, there has been a renewal of interest in research regarding non-binary or \emph{multi-valued} logic \cite{SERRAN1997533,dunn2012modern,Das2012,6810470,1793814,CHATTOPADHYAY20101014,obiniyi2011arithmetic,gundersen2008aspects,hayes2001third,1498715,connelly2008ternary,nair2015delayOpenAccess,6782156,CHATTOPADHYAY2015123,Mirzaee2696520,1319922,0305-4470-39-24-013,fey2015ternary,1402-4896-2005-T118-025,3297906,0957-4484-17-8-023,LEBARBAJEC20061826,1329407,779713}. This effort may be a reflection of the realization that, in near future, connections will occupy major part of a binary chip. Therefore, due to power consumption and heat dissipation of the transistors and delay associated with the connections, chip size will reach its physical limit (see Moore's law \cite{moore1998cramming}). So, a probable solution is to switch to multi-valued logic where by increasing the density of the logic, more information can be delivered through each connection, hence, lesser memory requirement and operations for a given
length of data. In principle, this can provide an increase in data processing capability per unit chip area and reduction in numbers of gates and connections, with less delay and power consumption.

Among the multi-valued logic systems, computing in $3$-valued or ternary system is considered the most efficient in terms of storage complexity \cite{Hurst1984}, because, it is nearest to the optimal radix -- natural base $e$ \cite{5176262}. In a ternary number system, the unit of information is a ternary digit, termed as \emph{trit}. Each trit can represent $1.58$ bits. Therefore, a $21$-trit word can handle values slightly larger than a $32$-bit word. Moreover, ternary logic can be implemented using \emph{balanced ternary number system}, which is, in words of Donald Knuth ``the prettiest number system'' \cite{Knuth2}. Some advantages of this number system are -- there is no requirement of using explicit sign for any number, there is less frequency of carry generation and propagation, rounding of a number is equivalent to truncation, negation of a number is deduced by exchange of the signed trits, etc. 
Hence, arithmetic computations are more efficient and fast. 
Further, same hardware circuit can implement both addition and subtraction. So, quoting Knuth, ``perhaps the symmetric properties and simple arithmetic of this number system will prove to be quite important some day -- when the ``flip-flop'' is replaced by a ``flip-flap-flop''.''(page 208 of \cite{Knuth2}).
%
%
%

Even if we overlook the above mentioned advantages of ternary systems, they are the natural alternatives if anybody wants to go beyond binary systems. Therefore, $3$-state CA based PRNG is a better choice for ternary computing than others, and it can be used in several applications related to ternary logic. Beyond it, $3$-state CAs can offer better source of randomness and higher degree of flexibility. While we have been identifying the essential properties of CAs to be a good source of randomness, an example $3$-state $3$-neighborhood CA has been explored to design a PRNG.
We have shown that the reported CA has exponentially large cycle length with good theoretical properties, and passes many important empirical tests, like rank tests and spectral tests.
In comparison to the other existing PRNGs, it performs better than all other portable PRNGs, like \verb rand, ~\verb lrand48 ~of UNIX systems, \verb LFSR113 \cite{iLEC16j}, \verb MRG31k3p \cite{iLEC16j}, \verb WELL1024a \cite{iLEC16j}.
These facts suggest that this CA is a good candidate to be used as a portable PRNG for applications like software simulations, randomized algorithms etc. 

However, the following questions arise -- \emph{Is this the only $3$-neighborhood $3$-state CA with such properties? Or, are there other such CAs, which can perform in the similar way?} Incidentally, the answer to the second question is affirmative. In reality, many such CAs exist in the vast rule space of $3$-neighborhood $3$-state CAs with similar randomness quality. These CAs can have different unique unseen properties and may be better applicable for some other purposes. For example, some CAs can be cryptographically secured, while some can be good test pattern generators for hardware circuit testing, etc. This leads to the following question -- \emph{Can we get a list of such CAs with good randomness quality?}

This chapter targets to find such a list of $3$-state finite CAs with $3$-neighborhood (that is, nearest neighbor) condition which are good source of randomness. Before going into the search for such CAs, we have first analyzed the aspects and convenience of employing $3$-state CAs as PRNGs (Section~\ref{Chap:3-stateCA_list:sec:CA_benifits}).


Generally, randomness of a CA is effected by its transition rule, cell size, seed and boundary condition. 
As there are total $3^{3^{3}}=3^{27}=7.625597485\times10^{12}$ $3$-neighborhood $3$-state CAs, it is impossible to individually study all these rules. Here we use two greedy strategies, STRATEGY I and STRATEGY II of Section~\ref{chap:reversibility:Sec:identify} of Chapter~\ref{Chap:reversibility} (Page~\pageref{chap:reversibility:Sec:identify}) to select an initial set of rules. From these rules, the potential CAs are filtered according to some theories developed in Section~\ref{Chap:3-stateCA_list:sec:theory}. (In this chapter, if not otherwise mentioned, by ``CA'', we will mean $1$-dimensional $3$-neighborhood finite CA having $3$ states per cell.) These CAs are, however, further tested for randomness using Diehard battery of tests to select the final set of rules (Section~\ref{Chap:3-stateCA_list:sec:experiment}).

\section{Aspects of $3$-state CA as PRNG}
\label{Chap:3-stateCA_list:sec:CA_benifits}
As discussed, ternary systems are the most potent option among the non-binary systems. They can provide better insight in solving binary problems and offer superior prospect in improving current VLSI circuitry.
For some examples of ternary computational circuits, see \cite{srivastava1996design,6782156,Mirzaee2696520,5248260,4038879,5249619,Santos:1964:AST:1464122.1464168,83076}, which have been developed based on the logic and algorithms \cite{Frieder776392,rine1977computer,Das2012,4037853,4038831,Mouftah:1976:SIT:800111.803598,hubbard1979design} of ternary number system.

Since inception, CAs, with their simplicity, cascadability and modularity nature have gained popularity as technology in the era of VLSI. Therefore, using CAs as the backbone architecture is a good choice for implementing any ternary logic and computation. Further, CA has simple rule with local interaction, hence, CA-based systems are an easy contrivance for practical realization. In terms of randomness quality, tri-state CAs offer improvement over binary CAs (see Table~\ref{tab:final_rank_comparison} of Page~\pageref{tab:final_rank_comparison}). In fact, if one considers the convolution of the LFSR-based PRNGs (see Section~\ref{chap:randomness_survey:sec:lfsr} of Chapter~\ref{Chap:randomness_survey}), then the CA-based PRNGs are more beneficial in terms of lucidity and efficiency.

Moreover, in Chapter~\ref{Chap:randomness_survey} we have observed that, it is possible to design a tri-state CA based PRNG which satisfy the essential attributes (see Section~\ref{Chap:randomness_survey:sec:propertiesOfCA} of Page~\pageref{Chap:randomness_survey:sec:propertiesOfCA}) to be a good source of randomness. Here, the concept of window has been used to generate random ternary numbers of any length by the CA. These tri-state CAs can have exponentially large cycle length (see Table~\ref{Chap:randomness_survey:tab:Tcount} of Page~\pageref{Chap:randomness_survey:tab:Tcount}). So, practically, one can generate sequence of numbers for hours without exhausting the cycle. Further, window-based PRNG scheme offers the additional advantage of unpredictability; here same number can be generated more than once and that has no relation with the completion of cycle. This is because, it may happen that, only the values of window are same for those configurations while the other cells are different to distinguish the configurations. 
This scheme also has some major strengths, as discussed below.



\subsection{Portability} One of the main strengths of this scheme is, it is highly portable, like other portable PRNGs -- \verb rand, ~\verb lrand48 ~of UNIX etc. The PRNG can be implemented using very simple and easy-to-use code by any programming languages. A rudimentary implementation of such an example PRNG for the window-based scheme of Section~\ref{Chap:randomness_survey:sec:prng_R} with CA $\mathbf{\mathscr{R}}=120021120021021120021021210$ (see Page~\pageref{Chap:randomness_survey:sec:prng_R}) using \textit{JAVA} is shown below. 
		


{\small
\begin{lstlisting}[language=JAVA, caption={JAVA code for Window-based PRNG}, linewidth=15.0cm]
int PC[],NC[], seedTri[];	//stores configurations and seed
void srandCA(int size, String seed){ 	//Takes size of output number in bits and a decimal string as seed
 window = (10*size)/16;             //set window size
 n=(int) (window*2.5);            //set CA size
 if(n%2==0)	n=n+1;
 seedTri=toTernary(seed, window);   //converts to ternary array
 for(i=0;i<window;i++)       //initialize window	
	PC[i]=seedTri[i];
 for(i=window;i<n-1;i++) //initialize other values to 00...01
	PC[i]=0; 
 PC[n-1]=1;
for(j=0;j<n;j++){ //leave first n configurations
 	for(i=0;i<n;i++)
 	  NC[i] = Rule[9*PC[(i-1+n)%n]+3*PC[i]+PC[(i+1+n)%n]];
 	for(i=0;i<n;i++)
 	  PC[i] = NC[i];      //update PC[] to use again
 }
}
String randCA(){       //Generates psudo-random number
 for(i=0;i<n;i++)
 	NC[i] = Rule[9*PC[(i-1+n)%n]+3*PC[i]+PC[(i+1+n)%n]];
 for(i=0;i<n;i++)
 	PC[i] = NC[i];    //update PC[] to use again
 return toDecimal(PC,window); //return decimal number corresponding to the window
}
\end{lstlisting}
}
In this code snippet, the method \emph{randCA()} generates the random number as per the requirement of the user. It uses the present configuration of CA as set by the method \emph{srandCA()}. This method takes the length of output number in bits and the seed for window as a decimal string from the user. For any seed, the first $n$ configurations are not used for creating the output random number. 
	
Here, line $3$ sets the window length required for generating ternary numbers corresponding to the size (number of bits) of the output number and line $4$ sets the CA length $n$ for that window. We take CA size $n$ as an odd number at least $2.5$ times of the window length to have sufficient cycle length.
Therefore, practically random number of any length can be generated by this simple scheme. This simplicity makes it a good contender to the existing portable PRNGs. 
		
\subsection{Robustness}The scheme is also highly robust. User can generate random number of any length by giving the desired length and the seed. The size of the window and the CA are set accordingly in the program.
Therefore, practically, there is no limitation on the length of the desired number to be generated. This is an extremely useful property which almost all existing portable PRNGs lack. For example, \verb rand ~has limitation, as the maximum number generated by it, is dependent on implementation, like $16$ bit, $32$ bit etc.

\subsection{Hardware Implementation}
Like binary CAs, $3$-state CAs can also be implemented in hardware in a structured way. To do so, we need ternary logic gates. However, 
several attempts have been made to build logic gates and integrated circuits (ICs) using $3$-valued logic-- e.g. using integrated injection logic ($I^2L$) and emitter-coupled logic (ECL), complementary metal-oxide semiconductor (CMOS) technology, charge-coupled device (CCD) technology, microelectromechanical system (MEMS), Memristors, neural network, optical elements and quantum functional devices etc. \cite{Eichmann86,gundersen2008aspects,CHATTOPADHYAY2015123,fey2015ternary,4038040,4210047,doi:10.1080/00207217508920376,5245386,4038408}. For instance, \cite{doi:10.1080/00207217508920376,doi:10.1080/00207218508939031} record design of ternary storage units (ternary flip-flop) using CMOS technology. Similarly, researchers have worked on the realization of arithmetic and logical operations of ternary systems \cite{5249619,Das2012,BHOWMIK20135561,488807,4038831}. There are many alternatives. For example, minimum and maximum can be replacements of logical AND and OR operation. Similarly, AND (resp. OR) can also be replaced by modulo, or Galois multiplication (resp. addition). So, accordingly combinational circuits can be outlined.
  \begin{figure}[hbtp]
   \vspace{-1.0em}
   \centering
    \includegraphics[width=1.0\textwidth, height = 2.0in]{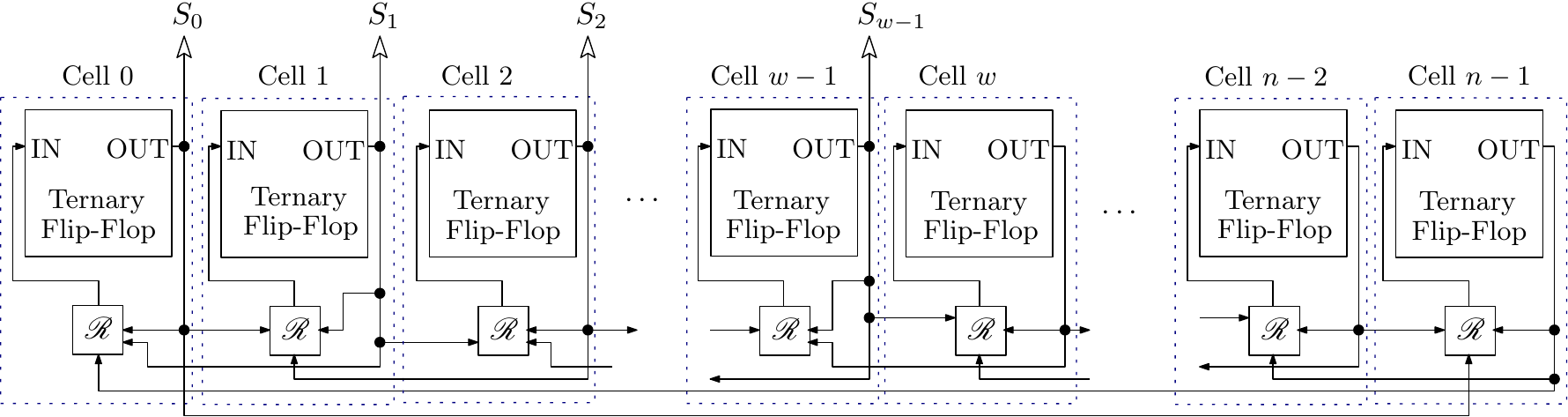}
    \caption{Implementation of a CA based PRNG using $3$-valued Logic Gates}
    \label{Chap:3-stateCA_list:fig:CAarchitechture}
     \vspace{-1.0em}
   \end{figure}

To design hardware for a CA based PRNG, we need to implement each cell of the CA. A cell of a CA is basically a storage unit which can store any of trits $\{0,1,2\}$ along with a rule.
The rule of a CA is a combinational circuit. A schematic hardware implementation for the CA $\mathbf{\mathscr{R}}=120021120021021120021021210$ is shown in Figure~\ref{Chap:3-stateCA_list:fig:CAarchitechture}. Here, $S_0\cdots S_{w-1}$ represents the output of the PRNG (from first $w$ cells) in trits. So, to implement the PRNG, we need $n$ number of cells ($3$VL storage units) and $n$ number of each of the combinational units for rule.

Therefore, using $3$-state CA as PRNG offers strengths like portability, robustness, unpredictability, exponentially large cycle length and non-linearity which may be useful for many purposes. Further, hardware implementation of such a PRNG is easy, which makes it fast and a good choice for testing VLSI circuits using $3$-valued logic. Hence, tri-state CAs based PRNGs can be suitable candidates for day-to-day applications and software simulations, where excellency in randomness quality is not in the forefront, rather efficiency,  implementability, portability, etc. are more important. This motivates us to find a list of CAs with similar qualities.

\section{Greedy Strategies}\label{Chap:3-stateCA_list:sec:CAPRNG}
For $3$-state CAs, total number of rules is $3^{3^3} = 7.625597485\times10^{12}$, which is a huge number for exhaustive testing.
So, we have taken a greedy approach to choose the CAs which fulfill the essential properties for good randomness quality (see Section~\ref{Chap:randomness_survey:sec:propertiesOfCA} of Page~\pageref{Chap:randomness_survey:sec:propertiesOfCA}). Our scheme is to select the balanced rules which have maximum rate of information transmission on at least either of the directions. Reason for this approach is, such a CA is left-permutive or right-permutive (see Definition~\ref{Def:permutivity} of Page~\pageref{Def:permutivity}). According to Devaney's definition \cite{Devaney}, these CAs are chaotic, when defined over infinite lattice \cite{Margara99,kuurka2009topological}. To have a good prospect for randomness, the CAs need to be chaotic (see Section~\ref{Chap:randomness_survey:sec:unpredictability} of Page~\pageref{Chap:randomness_survey:sec:unpredictability}), such that, even a tiny perturbation at any local cell profoundly influence the subsequent configurations. Therefore, our expectation is that, although chaos is the property of CAs having infinite lattice size, for finite sizes also, such CAs will preserve some of its properties and be potential candidates for source of randomness. Success of this scheme, however, remains on how efficiently we are choosing the balanced rules. 
		

\begin{table}[t]
\setlength{\tabcolsep}{1.3pt}
\begin{center}
 \caption{Two rules against STRATEGY I and STRATEGY II}
\label{Chap:3-stateCA_list:tab:rt3}
\resizebox{1.00\textwidth}{!}{
 \begin{tabular}{cccccccccccccccccccccccccccc}
 \toprule
 \thead{P.S.} & \thead{222} & \thead{221} & \thead{220} & \thead{212} & \thead{211} & \thead{210} & \thead{202} & \thead{201} & \thead{200} & \thead{122} & \thead{121} & \thead{120} & \thead{112} & \thead{111} & \thead{110} & \thead{102} & \thead{101} & \thead{100} & \thead{022} & \thead{021} & \thead{020} & \thead{012} & \thead{011} & \thead{010} & \thead{002} & \thead{001} & \thead{000}\\ 
 
 \thead{RMT} & \thead{(26)} & \thead{(25)} & \thead{(24)} & \thead{(23)} & \thead{(22)} & \thead{(21)} & \thead{(20)} & \thead{(19)} & \thead{(18)} & \thead{(17)} & \thead{(16)} & \thead{(15)} & \thead{(14)} & \thead{(13)} & \thead{(12)} & \thead{(11)} & \thead{(10)} & \thead{(9)} & \thead{(8)} & \thead{(7)} & \thead{(6)} & \thead{(5)} & \thead{(4)} & \thead{(3)} & \thead{(2)} & \thead{(1)} & \thead{(0)}\\ 
  \midrule
\multirow{2}{*}{\thead{N.S.}} & 2&1&1&2&1&2&1&1&2&0&2&0&0&0&0&0&2&0&1&0&2&1&2&1&2&0&1\\
 & 1&0&2&0&1&2&1&0&2&0&1&2&1&0&2&1&0&2&0&2&1&0&2&1&0&1&2\\
\bottomrule
\end{tabular}
}
\end{center}
\vspace{-1.5em}
\end{table} 	
\subsection{Maximum Information Transmission on Right Side}
A CA can have maximum information flow on right direction if the RMTs of $Equi_i$, ($0 \leq i \leq 8$) have different next state values (see Section~\ref{Chap:randomness_survey:sec:info_flow} of Page~\pageref{Chap:randomness_survey:sec:info_flow}). So, our first strategy is to choose the CAs for which the equivalent RMTs have different next state values. That is, the CAs are to be synthesized using STRATEGY I (see Section~\ref{chap:reversibility:Sec:identify} of Chapter~\ref{Chap:reversibility}). Here, we reproduce the greedy strategy for $3$-state CAs.  
		
\noindent \textbf{STRATEGY I:} \label{stg1_3}\textit{Pick up the balanced rules in which equivalent RMTs have different next state values, that is, no two RMTs of $Equi_i ~(0 \leq i \leq 8)$ have same next state value.}

\begin{figure}[!h]
			\subfloat[$Equi_5$ of STRATEGY $I$ Rule $211212112020000020102121201$ \label{Chap:3-stateCA_list:fig:equi_5}]{%
				\includegraphics[width=0.40\textwidth, height=1.5cm]{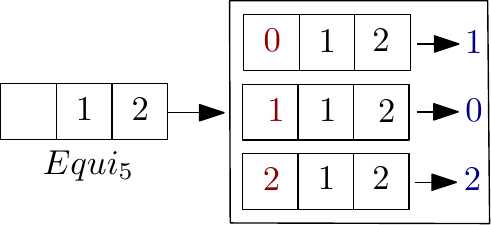}
			}
			\hfill
			\subfloat[$Sibl_5$ of STRATEGY $II$ Rule $102012102012102102021021012$ \label{Chap:3-stateCA_list:fig:sibl_5}]{%
				\includegraphics[width=0.40\textwidth, height=1.5cm]{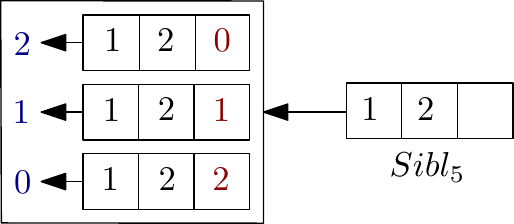}
			}
			\caption{Flow of Information in left and right directions}
			\label{Chap:3-stateCA_list:fig:info_flow}
		\vspace{-1.0em}
		\end{figure}

The first rule of Table~\ref{Chap:3-stateCA_list:tab:rt3} is an example of STRATEGY I CA. Here, for each equivalent RMT set, all equivalent RMTs have different next state values. Obviously, one RMT from each equivalent RMT set is self-replicating. In fact, for any STRATEGY I CA, only one RMT of each $Equi_i$ is self-replicating, because, for any $Equi_i$, the middle cell $y$ is fixed with either of $0/1/2$. Figure~\ref{Chap:3-stateCA_list:fig:equi_5} and Table~\ref{Chap:3-stateCA_list:tab:stgi_right} show the information transmission on right side for the CA. 
Therefore, any STRATEGY I CA has $\frac{18}{27} = 66.667 \%$, that is, the maximum possible information flow on right direction. Such CAs are left-permutive.

	\renewcommand{\arraystretch}{1.2}
	\begin{table}[hbtp]
		\begin{center}
			\caption{RMTs of $Equi_{i} $, $0 \leq i \leq 8$ for CA $211212112020000020102121201$}
			{
				\resizebox{0.9\textwidth}{!}{
					\begin{tabular}{c|ccc|ccc|ccc|c}
						\toprule	
						&	\multicolumn{9}{c|}{\thead{Equivalent RMTs}} & \\
						\thead{\#Set~~} & \thead{P.S.} & \thead{RMT $i$} & \thead{$R[i]$~~} & \thead{P.S.} & \thead{RMT $i$} & \thead{$R[i]$~~}  & \thead{P.S.} & \thead{RMT $i$} & \thead{$R[i]$~~} & \thead{Change of States} \\
						\midrule
						$Equi_0$ & $000$ & $(0)$ & $1$ & $100$ & $(9)$ & $0$ & $200$ & $(18) $ & $ 2$ & $2$ \\ 
						$Equi_1$ & $001$ & $(1)$ & $ 0 $ & $101$ & $(10) $ & $ 2$ & $201$ & $(19) $ & $ 1$ & $2$ \\ 
						$Equi_2$ & $002$ & $(2)$ & $ 2$ & $102$ & $(11) $ & $ 0$ & $202$ & $(20)$ & $1$ & $2$ \\ 
						$Equi_3$ & $010$ & $(3)$ & $1$ & $110$ & $(12)$ & $0$ & $210$ & $(21)$ & $2$ & $2$ \\ 
						$Equi_4$ & $011$ & $(4)$ & $2$ & $111$ & $(13)$ & $0$ & $211$ & $(22)$ & $1$ & $2$ \\ 
						$Equi_5$ & $012$ & $(5)$ & $1$ & $112$ & $(14)$ & $0$ & $212$ & $(23)$ & $2$ & $2$ \\ 
						$Equi_6$ & $020$ & $(6)$ & $2$ & $120$ & $(15)$ & $0$ & $220$ & $(24)$ & $1$ & $2$ \\ 
						$Equi_7$ & $021$ & $(7)$ & $0$ & $121$ & $(16)$ & $2$ & $221$ & $(25) $ & $1$ & $2$ \\ 
						$Equi_8$ & $022$ & $(8)$ & $1$ & $122$ & $(17)$ & $0$ & $222$ & $(26)$ & $2$ & $2$ \\ 
						\midrule
						\multicolumn{10}{c}{\thead{\emph{Total change of $R[i]$} depending on $Equi_i$ = }} & \thead{$18$}\\
						\bottomrule
					\end{tabular}
				}}\label{Chap:3-stateCA_list:tab:stgi_right}
			\end{center}
			\vspace{-1.5em}	
		\end{table} 
	
\subsection{Maximum Information Transmission on Left Side} Similarly, there can be a maximum information flow on left direction, if RMTs of $Sibl_i$ have different next state values. In the next strategy, we choose the CA rules, where the sibling RMTs have different next state values. These CAs follow STRATEGY II (see Section~\ref{chap:reversibility:Sec:identify} of Chapter~\ref{Chap:reversibility}
). Here also, we transcribe the greedy strategy for $3$-state CAs.

\noindent\textbf{STRATEGY II:} \label{stg2_3} \textit{Pick up the balanced rules in which the RMTs of a sibling set have the different next state values, that is, no two RMTs of $Sibl_i ~(0 \leq i \leq 8)$ have same next state value.}

Figure~\ref{Chap:3-stateCA_list:fig:sibl_5} and Table~\ref{Chap:3-stateCA_list:tab:stgii_left} show the information transmission for a STRATEGY II rule $102012102012102102021021012$ (second rule of Table~\ref{Chap:3-stateCA_list:tab:rt3}). For STRATEGY II CAs also, one RMT of each $Sibl_i$ is self-replicating. Hence, $18$ out of $27$ RMTs are affected by the information of their right neighbors; that is, maximum possible rate of information transmission on left direction for STRATEGY II CAs is $66.67\%$. These are right-permutive CAs.		
		
		\renewcommand{\arraystretch}{1.2}
		\begin{table}[!h]
			\begin{center}
				\caption{RMTs of $Sibl_{i} $, $0 \leq i \leq 8$ for CA $102012102012102102021021012$}
				{
					\resizebox{0.9\textwidth}{!}{
						\begin{tabular}{c|ccc|ccc|ccc|c}
							\toprule	
							&	\multicolumn{9}{c|}{\thead{Sibling RMTs}} & \\
							\thead{\#Set~~} & \thead{P.S.} & \thead{RMT $i$} & \thead{$R[i]$~~} & \thead{P.S.} & \thead{RMT $i$} & \thead{$R[i]$~~}  & \thead{P.S.} & \thead{RMT $i$} & \thead{$R[i]$~~} & \thead{Change of States} \\
							\midrule
							$Sibl_0$ & $000$ & ($0$) & $2$ & $001$ & ($1$) & $1$ & $002$ & ($2$) & $ 0 $ & $2$ \\ 
							$Sibl_1$ & $010$ & ($3$) & $1$ & $011$ & ($4$) & $2$ & $012$ & ($5$) & $0$ & $2$\\ 
							$Sibl_2$ & $020$ & ($6$) & $1$ & $021$ & ($7$) & $2$ & $022$ & ($8$) & $0$ & $2$ \\ 
							$Sibl_3$ & $100$ & ($9$) & $2$ & $101$ & ($10$) & $0$ & $102$ & ($11$) & $1$ & $2$ \\ 
							$Sibl_4$ & $110$ & ($12$) & $2$ & $111$ & ($13$) & $0$ & $112$ & ($14$) & $1$ & $2$ \\ 
							$Sibl_5$ & $120$ & ($15$) & $2$ & $121$ & ($16$) & $1$ & $122$ & ($17$) & $0$ & $2$ \\ 
							$Sibl_6$ & $200$ & ($18$) & $2$ & $201$ & ($19$) & $0$ & $202$ & ($20$) & $1$ & $2$ \\ 
							$Sibl_7$ & $210$ & ($21$) & $2$ & $211$ & ($22$) & $1$ & $212$ & ($23$) & $0$ & $2$ \\ 
							$Sibl_8$ & $220$ & ($24$) & $2$ & $221$ & ($25$) & $0$ & $222$ & ($26$) & $1$ & $2$ \\ 
							\midrule
							\multicolumn{10}{c}{\thead{\emph{Total change of $R[i]$} depending on $Sibl_i$ = }} & \thead{$18$}\\
							\bottomrule
						\end{tabular}
					}}\label{Chap:3-stateCA_list:tab:stgii_left}
				\end{center}
			\vspace{-1.0em}	
			\end{table} 
			
		
The CA $\mathbf{\mathscr{R}}=120021120021021120021021210$ used in Chapter~\ref{Chap:randomness_survey} to design an example PRNG is also a STRATEGY II CA.		
There are ${(3!)}^{3^2} = 10077696$ balanced rules that can be selected as candidates following each of STRATEGY $I$ and STRATEGY $II$. These CAs are potential nominees to be good PRNGs. However, we want to ensure that, for rules from each of these strategies, there is information flow in both directions. 

\subsection{Rules with Information Flow in Both Sides}
STRATEGY I and STRATEGY II rules have maximum information flow on right side and left side respectively. Nevertheless, the information flow on the opposite direction for these rules are unknown. For a CA to be a good source of randomness, it is expected that, there is flow of information in both sides.

\renewcommand{\arraystretch}{1.2}
\begin{table}[!h]
\centering
\caption{RMTs of $Sibl_{i} $, $0 \leq i \leq 8$ for CA $211212112020000020102121201$}\label{Chap:3-stateCA_list:tab:stgi_left}
	\resizebox{0.9\textwidth}{!}{
							\begin{tabular}{c|ccc|ccc|ccc|c}
								\toprule	
								&	\multicolumn{9}{c|}{\thead{Sibling RMTs}} & \\
								\thead{\#Set~~} & \thead{P.S.} & \thead{RMT $i$} & \thead{$R[i]$~~} & \thead{P.S.} & \thead{RMT $i$} & \thead{$R[i]$~~}  & \thead{P.S.} & \thead{RMT $i$} & \thead{$R[i]$~~} & \thead{Change of States} \\
								\midrule
								$Sibl_0$ & $000$ & ($0$) & $1$ & $001$ & ($1$) & $0$ & $002$ & ($2$) & $ 2 $ & $2$ \\ 
								$Sibl_1$ & $010$ & ($3$) & $1$ & $011$ & ($4$) & $2$ & $012$ & ($5$) & $1$ & $1$\\ 
								$Sibl_2$ & $020$ & ($6$) & $2$ & $021$ & ($7$) & $0$ & $022$ & ($8$) & $1$ & $2$ \\ 
								$Sibl_3$ & $100$ & ($9$) & $0$ & $101$ & ($10$) & $2$ & $102$ & ($11$) & $0$ & $1$ \\ 
								$Sibl_4$ & $110$ & ($12$) & $0$ & $111$ & ($13$) & $0$ & $112$ & ($14$) & $0$ & $1$ \\ 
								$Sibl_5$ & $120$ & ($15$) & $0$ & $121$ & ($16$) & $2$ & $122$ & ($17$) & $0$ & $1$ \\ 
								$Sibl_6$ & $200$ & ($18$) & $2$ & $201$ & ($19$) & $1$ & $202$ & ($20$) & $1$ & $2$ \\ 
								$Sibl_7$ & $210$ & ($21$) & $2$ & $211$ & ($22$) & $1$ & $212$ & ($23$) & $2$ & $1$ \\ 
								$Sibl_8$ & $220$ & ($24$) & $1$ & $221$ & ($25$) & $1$ & $222$ & ($26$) & $2$ & $1$ \\ 
								\midrule
								\multicolumn{10}{c}{\thead{\emph{Total change of $R[i]$} depending on $Sibl_i$ = }} & \thead{$12$}\\
								\bottomrule
							\end{tabular}
}
\vspace{-1.0em}	
\end{table} 
		
\subsubsection{Flow of Information in Left Side for STRATEGY I} Next state values of sibling RMTs signify the flow of information in left direction. Therefore, we want to select those STRATEGY I CAs, for which there is a significant information flow in left direction also.
		
Table~\ref{Chap:3-stateCA_list:tab:stgi_left} shows the left directional information flow for the rule $211212112020000020102121201$ (Table~\ref{Chap:3-stateCA_list:tab:rt3}) of STRATEGY I. This rule has $\frac{12}{27}=44.44\%$ information transmission on left direction and maximum ($66.67\%$) information flow in right direction.
			

\subsubsection{Flow of Information in Right Side for STRATEGY II} Similarly, different states of equivalent RMTs denotes information flow in left direction. So, the STRATEGY II CAs with significant information flow on right direction are to be chosen.
			
In Table~\ref{Chap:3-stateCA_list:tab:stgii_right}, the information flow on right side for the STRATEGY II CA $102012102012102102021021012$ (see Table~\ref{Chap:3-stateCA_list:tab:rt3}) is shown. It is observed that, for this rule, the rate of information transmission on right direction is $\frac{12}{27}=44.44\%$. So, this rule also has information flow in both directions.
				
\renewcommand{\arraystretch}{1.2}
\begin{table}[hbtp]
\centering
\caption{RMTs of $Equi_{i} $, $0 \leq i \leq 8$ for CA $102012102012102102021021012$}
\label{Chap:3-stateCA_list:tab:stgii_right}
\resizebox{0.9\textwidth}{!}{
								\begin{tabular}{c|ccc|ccc|ccc|c}
									\toprule	
									&	\multicolumn{9}{c|}{\thead{Equivalent RMTs}} & \\
									\thead{\#Set~~} & \thead{P.S.} & \thead{RMT $i$} & \thead{$R[i]$~~} & \thead{P.S.} & \thead{RMT $i$} & \thead{$R[i]$~~}  & \thead{P.S.} & \thead{RMT $i$} & \thead{$R[i]$~~} & \thead{Change of States} \\
									\midrule
									$Equi_0$ & $000$ & $(0)$ & $2$ & $100$ &  $(9)$ & $2$ & $200$ & $(18)$ & $2$ & $1$ \\ 
									$Equi_1$ & $001$ & $(1)$ & $1$ & $101$ & $(10)$ & $0$ & $201$ & $(19)$ & $0$ & $1$ \\ 
									$Equi_2$ & $002$ & $(2)$ & $0$ & $102$ & $(11)$ & $1$ & $202$ & $(20)$ & $1$ & $1$ \\ 
									$Equi_3$ & $010$ & $(3)$ & $1$ & $110$ & $(12)$ & $2$ & $210$ & $(21)$ & $2$ & $1$ \\ 
									$Equi_4$ & $011$ & $(4)$ & $2$ & $111$ & $(13)$ & $0$ & $211$ & $(22)$ & $1$ & $2$ \\ 
									$Equi_5$ & $012$ & $(5)$ & $0$ & $112$ & $(14)$ & $1$ & $212$ & $(23)$ & $0$ & $1$ \\ 
									$Equi_6$ & $020$ & $(6)$ & $1$ & $120$ & $(15)$ & $2$ & $220$ & $(24)$ & $2$ & $1$ \\ 
									$Equi_7$ & $021$ & $(7)$ & $2$ & $121$ & $(16)$ & $1$ & $221$ & $(25)$ & $0$ & $2$ \\ 
									$Equi_8$ & $022$ & $(8)$ & $0$ & $122$ & $(17)$ & $0$ & $222$ & $(26)$ & $1$ & $2$ \\ 
									\midrule
									\multicolumn{10}{c}{\thead{\emph{Total change of $R[i]$} depending on $Equi_i$ = }} & \thead{$12$}\\
									\bottomrule
								\end{tabular}
}
\end{table} 
				
\begin{figure}[!ht]
					\subfloat[Graph for STRATEGY I \label{Chap:3-stateCA_list:fig:infoStgI}]{%
						\includegraphics[width=0.48\textwidth, height = 4.5cm]{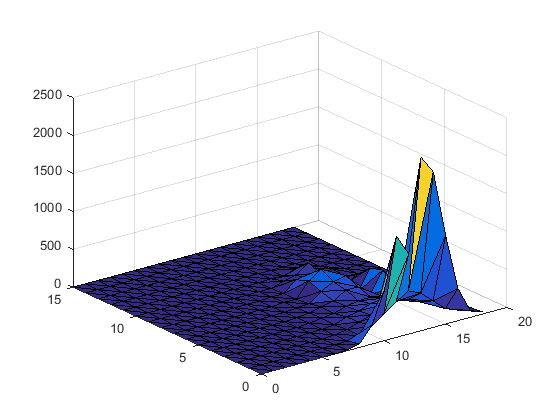}
					}
					\hfill
					\subfloat[Graph for STRATEGY II \label{Chap:3-stateCA_list:fig:infoStgII}]{%
						\includegraphics[width=0.48\textwidth, height = 4.5cm]{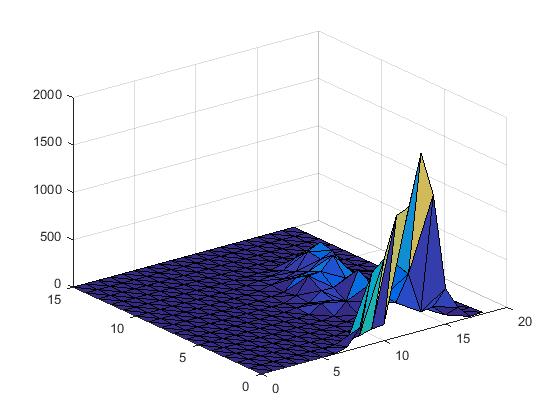}
					}
					\caption{Test Result of Information flow (X-axis) vs Number of Randomness tests passed in Diehard (Y-axis) vs Rule Count (Z-axis) for some arbitrarily selected CAs}         
					\label{Chap:3-stateCA_list:fig:infoFlowStatistic}
				\end{figure}

Our requirement is to take the rules which have a maximum information transmission in one direction, as well as, at least a certain rate of information transmission in the other direction. This is to ensure that, a small ripple in a cell propagates in both directions and because of periodic boundary condition, travels throughout all cells improving the randomness quality of the CAs. To validate our argument, an experiment is conducted, where some rules are arbitrarily chosen from each strategy and tested on Diehard battery of tests; the information flow in reverse direction (that is, left direction for STRATEGY I and right direction for STRATEGY II) for each of these rules is also calculated. For this experiment, $n$ is chosen as $15$ for testing with Diehard (more details on the process of testing is given in Section~\ref{Chap:3-stateCA_list:sec:experiment}).
Figure~\ref{Chap:3-stateCA_list:fig:infoFlowStatistic} shows the plot of these rules. In this figure, X-axis represents information flow in reverse direction in terms of total change in RMT states for sibling RMT sets (STRATEGY I) or for equivalent RMT sets (STRATEGY II). However, the number of randomness tests passed is recorded in Y-axis and the number of rules with any particular rate of transmission value that pass any number of tests, is shown in Z-axis. 
				
This figure clearly shows that, if rate of information transmission in reverse direction is low (total change in RMT states for sibling or equivalent RMT sets $\le 8$), then practically there are insignificant number of rules in each strategy, that passes any randomness tests. So, in this chapter, we have chosen those STRATEGY I and STRATEGY $II$ rules which have reverse directional information transmission rate $\geq \frac{8}{27}\% = 29.63 \%$. There are $10073592$ rules from each strategy that pass this condition.
				
\subsection{Equivalence of CAs of STRATEGY I and STRATEGY II}\label{Chap:3-stateCA_list:sec:equivalence}
There exists the notion of topological equivalence of local rules in CA (defined over infinite lattice). A local rule is \emph{equivalent} to the other rules under the following operations -- (1) reflection, (2) conjugation, and (3) conjugation and reflection together \cite{wolfram84b}. Dynamic behavior of equivalent rules are similar. Hence, they share many similar physical properties such as reversibility, number conservation, chaos, etc.
				
The above three operations are defined as follows \cite{Boccara02}. Let $R_{refl}(x,y,z)$, $R_{conj}(x,y,z)$ and $R_{cr}(x,y,z)$ denote the equivalent rule of $R(x,y,z)$ under reflection, conjugation and conjugation+reflection respectively. Then,
\begin{equation*}
R_{refl}(x,y,z) = R(z,y,x)
\end{equation*}
\begin{equation*}
R_{conj}(x,y,z) = 2-R((2-x),(2-y),(2-z))
\end{equation*}
\begin{equation*}
R_{cr}(x,y,z) = R_{conj}(z,y,x)
\end{equation*}
Here, $R_{cr}(x,y,z)=R_{rc}(x,y,z)$, where $R_{rc}$ is the rule with reflection followed by conjugation operation on $R$. 
				
				
\begin{definition}
\label{Chap:3-stateCA_list:Defi:EquivalentRuleSet}
A set $E$ of rules is said to be \textbf{equivalent rule-set}, if the following conditions are satisfied --
\begin{enumerate}
\item $E$ is closed under each of the three operations (reflection, conjugation and reflection $\&$ conjugation).
\item For each rule $R \in E$, the rules under the above three operations form the set itself.
\end{enumerate}
\end{definition}
Each equivalent rule-set is generally represented by its \emph{minimal rule}, because, from the minimal rule, all other rules of the set can be derived. 
\begin{definition}
A rule $R_{min}$ is called \textbf{minimal} rule of a equivalent rule-set $E$, if 
\begin{enumerate}
\item $R_{min} \in E$,
\item $R_{min}$ is the first rule of the sorted list (on increasing order) containing the rules of $E$.
\end{enumerate}
\end{definition}
				
One can easily observe that, the rules of the two greedy strategies are equivalents to each other. For example, for every rule of STRATEGY II, the rules under reflection operation and conjugation operation are in STRATEGY I, but, the rule under conjugation followed by reflection operation is in STRATEGY II. 

\begin{example}
Table~\ref{Chap:3-stateCA_list:tab:equi_rules} shows two sample rules from each strategy and their equivalent rules. In this table, the rules of $5^{th}$ row and  $8^{th}$ row are the minimal rules for the first equivalent rule-set and the second equivalent rule-set respectively.
				\begin{table}[hbtp]
					\setlength{\tabcolsep}{1.5pt}
\centering
						\caption{Sample Equivalent Rules from STRATEGY I and II}\label{Chap:3-stateCA_list:tab:equi_rules}
						{
							\resizebox{1.00\textwidth}{!}{
								\begin{tabular}{c|ccccccccccccccccccccccccccc}
									\toprule
									\thead{RMT $\rightarrow$} & \thead{222} & \thead{221} & \thead{220} & \thead{212} & \thead{211} & \thead{210} & \thead{202} & \thead{201} & \thead{200} & \thead{122} & \thead{121} & \thead{120} & \thead{112} & \thead{111} & \thead{110} & \thead{102} & \thead{101} & \thead{100} & \thead{022} & \thead{021} & \thead{020} & \thead{012} & \thead{011} & \thead{010} & \thead{002} & \thead{001} & \thead{000}\\ 
									
									\thead{Rules $\downarrow$} & \thead{(26)} & \thead{(25)} & \thead{(24)} & \thead{(23)} & \thead{(22)} & \thead{(21)} & \thead{(20)} & \thead{(19)} & \thead{(18)} & \thead{(17)} & \thead{(16)} & \thead{(15)} & \thead{(14)} & \thead{(13)} & \thead{(12)} & \thead{(11)} & \thead{(10)} & \thead{(9)} & \thead{(8)} & \thead{(7)} & \thead{(6)} & \thead{(5)} & \thead{(4)} & \thead{(3)} & \thead{(2)} & \thead{(1)} & \thead{(0)}\\ 
									\midrule
									$\mathbf{R(x,y,z)}$ & 2 & 1 & 1 & 2 & 1 & 2 & 1 & 1 & 2 & 0 & 2 & 0 & 0 & 0 & 0 & 0 & 2 & 0 & 1 & 0 & 2 & 1 & 2 & 1 & 2 & 0 & 1 \\ 
									$\mathbf{R_{refl}(x,y,z)} $ & 2 & 0 & 1 & 2 & 0 & 1 & 1 & 0 & 2 & 1 & 2 & 0 & 1 & 0 & 2 & 1 & 2 & 0 & 1 & 0 & 2 & 2 & 0 & 1 & 2 & 0 & 1 \\ 
									$\mathbf{R_{conj}(x,y,z)}$ & 1 & 2 & 0 & 1 & 0 & 1 & 0 & 2 & 1 & 2 & 0 & 2 & 2 & 2 & 2 & 2 & 0 & 2 & 0 & 1 & 1 & 0 & 1 & 0 & 1 & 1 & 0 \\ 
									$\mathbf{R_{cr}(x,y,z) }$ & 1 & 2 & 0 & 1 & 2 & 0 & 0 & 2 & 1 & 2 & 0 & 1 & 0 & 2 & 1 & 2 & 0 & 1 & 0 & 2 & 1 & 1 & 2 & 0 & 1 & 2 & 0 \\ 
									\midrule
									$\mathbf{R(x,y,z)}$ &  1 & 2 & 0 & 0 & 2 & 1 & 1 & 2 & 0 & 0 & 2 & 1 & 0 & 2 & 1 & 1 & 2 & 0 & 0 & 2 & 1 & 0 & 2 & 1 & 2 & 1 & 0 \\ 
									$\mathbf{R_{refl}(x,y,z)} $ & 1 & 0 & 0 & 0 & 0 & 0 & 1 & 1 & 2 & 2 & 2 & 2 & 2 & 2 & 2 & 2 & 1 & 0 & 0 & 1 & 1 & 1 & 1 & 1 & 0 & 0 & 0 \\
									$\mathbf{R_{conj}(x,y,z)}$ & 2 & 2 & 2 & 1 & 1 & 1 & 1 & 1 & 2 & 1 & 0 & 0 & 0 & 0 & 0 & 0 & 0 & 0 & 0 & 1 & 1 & 2 & 2 & 2 & 2 & 2 & 1 \\ 
									$\mathbf{R_{cr}(x,y,z) }$ & 2 & 1 & 0 & 1 & 0 & 2 & 1 & 0 & 2 & 2 & 0 & 1 & 1 & 0 & 2 & 1 & 0 & 2 & 2 & 0 & 1 & 1 & 0 & 2 & 2 & 0 & 1 \\ 
									\bottomrule 
								\end{tabular} }}
						\end{table}
\end{example}						
This table indicates that, there is inter-strategy and intra-strategy equivalence among the rules. Therefore, working with the set of minimal rules from the total set of STRATEGY I and STRATEGY II rules may be sufficient for most of the purposes. However, in this chapter, we independently study individual rules from each greedy strategy without concentrating on the minimal rules. 
						%
						%

In the next sections, we define some filtering criteria on these rules and experiment to select the potential PRNGs. Two levels of filtering conditions are applied over the set of rules selected from the greedy strategies. First, theoretical filtering is applied, where inherent structure of the CAs are theoretically explored for improvement in randomness quality. Then, on the theoretically filtered rules, empirical tests are applied repeatedly for different seeds.
						
\section{Theoretical Filtering}	\label{Chap:3-stateCA_list:sec:theory}
It is a common trend to select CAs with at least one quiescent state. Recall that, a state $q$ is quiescent if $R(q,q,q)=q$. This quiescent state signifies existence of a fixed-point attractor in the CA. However, one can observe that, in the configuration transition diagram of CAs, a fixed-point attractor may be associated with long chains of configurations, or with small chains; or it may be isolated like Figure~\ref{fig:config} (Page~\pageref{fig:config}), where most of the configurations are part of a long cycle. Moreover, if a CA has good randomness quality, it usually has a long cycle length. Because, longer cycle length implies more unique configurations are part of the sequence, so lesser the chance of repeating the same sequence; hence, better randomness quality. However, a CA can have long cycle only when it does not have a tendency to converge to the fixed-point attractor(s). Therefore, identifying each fixed-point attractor and its connectivity with other configurations is important for selecting the CAs having potentiality as PRNGs.

Although, only long cycle length of a CA can not ensure its randomness quality; the CA needs to possess the necessary properties (described in Section~\ref{Chap:randomness_survey:sec:propertiesOfCA} of Page~\pageref{Chap:randomness_survey:sec:propertiesOfCA}) and pass important statistical tests (described in Section~\ref{chap:randomness_survey:sec:empirical} of Page~\pageref{chap:randomness_survey:sec:empirical}). Nevertheless, as our selected CAs satisfy the basic properties of potential PRNGs, we can explore the intricate properties of CAs with good randomness quality to find those CAs from this set which satisfy these properties.

\subsection{Identifying Fixed-Point Attractors}\label{Chap:3-stateCA_list:sec:fpg}
Different fixed-point attractors can be associated with a CA of different sizes. Our target is to find all possible fixed point attractors in a CA. For this, we can use a slightly modified de Bruijn graph of the CA (for details on de Bruijn graph representation of CA, please see Section~\ref{Chap:surveyOfCA:dbg} of Page~\pageref{Chap:surveyOfCA:dbg} and Section~\ref{chap:reversibility:sec:intro} of Page~\pageref{chap:reversibility:sec:intro}). In this case, we remove all the edges of the de Bruijn graph which corresponds to non self-replicating RMTs. So, the graph has only the edges with self-replicating RMTs.
						
Recall that, any cycle in a de Bruijn graph corresponds to an RMT sequence. However, if we modify the de Bruijn graph to contain only the edges corresponding to the self-replicating RMTs, any cycle in this graph represents an RMT sequence of self-replicating RMTs. For this RMT sequence, the next configuration is itself, that is a fixed-point attractor. For example, if there is an elementary cycle of length $l$ in this graph, then the RMT sequence corresponding to this cycle portrays a fixed-point attractor for the CA when $n = k\times l$, $k \in \mathbb{N}$.
The cycles of the graph can be detected using many well-known algorithms \cite{tiernan1970efficient,tarjan1973enumeration,ehrenfeucht1973algorithm,johnson1975finding}, even a simple \emph{depth-first-search} (DFS) can detect cycles with a good time-complexity. In this way, every fixed-point attractor in a CA can be identified easily by this graph. 
						
\begin{example}
Take the CA $211212112020000020102121201$ (Table~\ref{Chap:3-stateCA_list:tab:rt3}). Now, the self-replicating RMTs for this CA are $1, 3, 5, 6, 9, 11, 16, 22$ and $26$. So, in the de Bruijn graph, all edges, except the edges corresponding to these self-replicating RMTs, are removed. This graph is shown in Figure~\ref{Chap:3-stateCA_list:fig:fixed}. In this figure, there are two cycles -- one cycle representing RMT sequence $\langle 1,3,9 \rangle$ and another representing $\langle 26 \rangle$. Hence, the CA has a fixed point attractor $(2)^n$ for any length $n$ and another fixed-point attractor $(001)^{\frac{n}{3}}$, when $n=3k$, $k \in \mathbb{N}$.

							\begin{figure}[!ht]
								\subfloat[For CA $211212112020000020102121201$ \label{Chap:3-stateCA_list:fig:fixed}]{%
									\includegraphics[width=0.43\textwidth, height = 3.9cm]{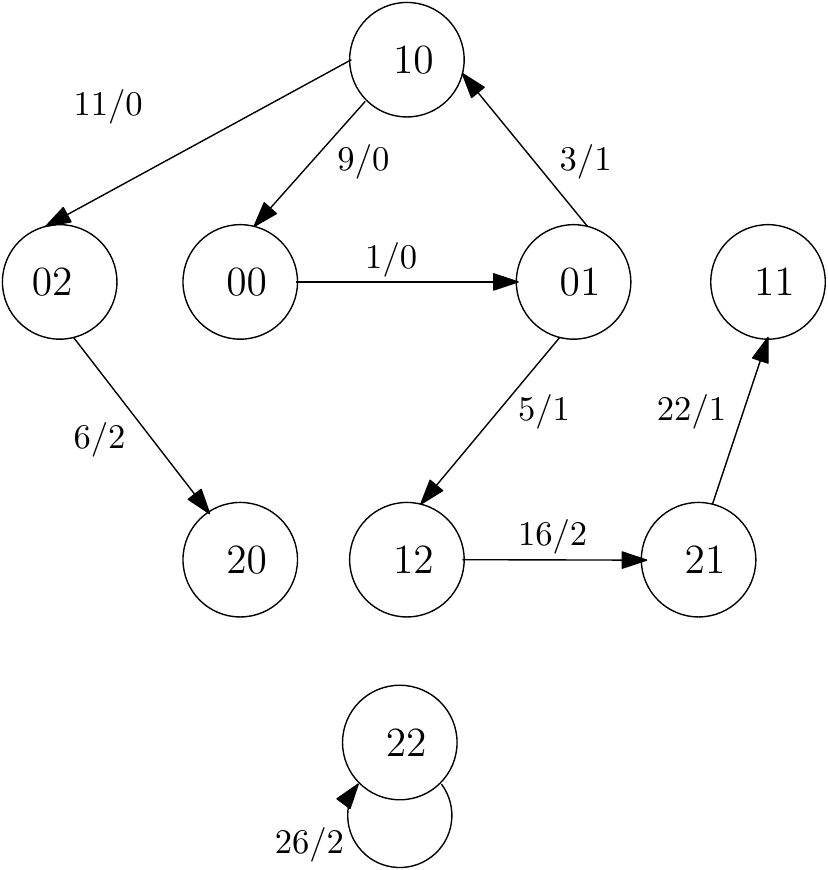}
								}
								\hfill
								\subfloat[For CA $102012102012102102021021012$ \label{Chap:3-stateCA_list:fig:fixed2}]{%
									\includegraphics[width=0.43\textwidth, height = 3.9cm]{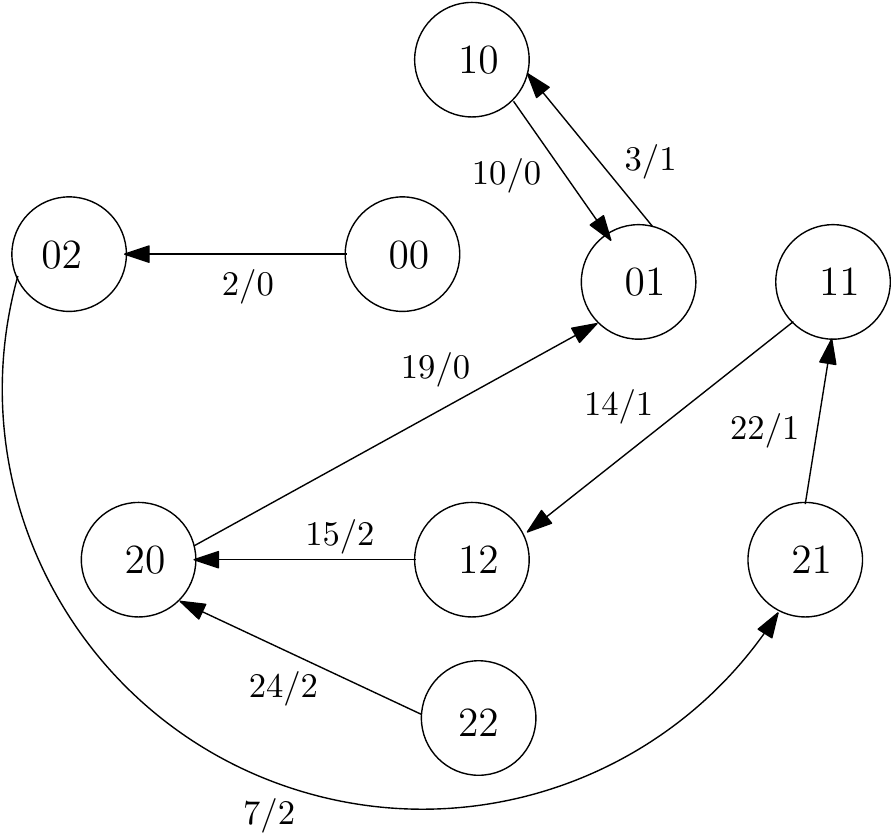}
								}
								\caption{de Bruijn graphs for finding fixed-point attractors of CAs}
								\label{Chap:3-stateCA_list:fig:filter2}
							\end{figure}

Similarly, for the CA $102012102012102102021021012$ (Table~\ref{Chap:3-stateCA_list:tab:rt3}), the self-replicating RMTs are $2,3,7,10,14,15,19,22$ and $24$. Figure~\ref{Chap:3-stateCA_list:fig:fixed2} shows the de Bruijn graph with self-replicating RMTs for this CA. In this graph, there is only one cycle corresponding to the RMT sequence $\langle 3,10 \rangle$. Hence, the only fixed-point attractor in the CA is $(01)^{\frac{n}{2}}$, when $n$ is multiple of $2$.  
\end{example}

						
\subsection{Filtering Conditions}\label{Chap:3-stateCA_list:sec:filter} Our goal is to select those CAs for empirical filtering which have one and only one fixed-point attractor for any CA size. To achieve this, the following two conditions are applied on the CAs.

\subsubsection{Only $1$ quiescent state}\label{Chap:3-stateCA_list:sec:cond1} We take those CAs which have only one quiescent state. For this quiescent state, a fixed-point attractor is formed in the CA of any size $n$. For our CAs, if next state value of RMT $0 (000)$ is $0$, then $0$ is a quiescent state, next state value of RMT $13 (111)$ is $1$, then $1$ is a quiescent state, and next state value of RMT $26 (222)$ is $2$, then $2$ is a quiescent state.
So, the criteria for getting only $1$ quiescent state is --
\begin{itemize}[leftmargin=*,topsep=0pt,itemsep=0ex,partopsep=2ex,parsep=2ex]
\item $R[0]=0$, $R[13] \in \{0,2\}$ and $R[26] \in \{0,1\}$ for $0$ as the only quiescent state,
							
\item $R[13] = 1$, $R[0]\in \{1,2\}$ and $R[26] \in \{0,1\}$ for $1$ as the only quiescent state, or
							
\item $R[26] = 2$, $R[0] \in \{1,2\}$ and $R[13] \in \{0,2\}$ for $2$ as the only quiescent state.
\end{itemize}
where $R[i]$ implies RMT $i$ of rule $R$. Table~\ref{Chap:3-stateCA_list:tab:filter1} records some example CAs satisfying this condition.
						
\begin{table}[h]
\centering
\caption{Some CAs having only one fixed-point}\label{Chap:3-stateCA_list:tab:filter1}
									\resizebox{0.8\textwidth}{1.5cm}{
										\begin{tabular}{|c|c|c|}
											\hline Fixed-point & Rules & Condition \\ 
											\hline $0$ &~102012210120021021012102120 ~& ~$R[0]=0$, $R[13]=2$, $R[26]=1$ \\ 
											\hline $0$ & ~012012120012021012012012210 ~& ~$R[0]=0$, $R[13]=2$, $R[26]=0$\\ 
											\hline $1$ & ~012012102201012210012012102 ~& ~$R[0]=2$, $R[13]=1$, $R[26]=0$\\ 
											\hline $1$ & ~012201201210012201021012201 ~& ~$R[0]=1$, $R[13]=1$, $R[26]=0$ \\ 
											\hline $2$ & ~201210021102102201210012012 ~& ~$R[0]=2$, $R[13]=0$, $R[26]=2$ \\ 
											\hline $2$ & ~210012102210201021210201021 ~& ~$R[0]=1$, $R[13]=0$, $R[26]=2$ \\ 
											\hline 
										\end{tabular} }
									\vspace{-0.9em}
\end{table}
								
\subsubsection{No other fixed-point attractor} We want those CAs for which the fixed-point attractor due to the quiescent state is the only fixed-point attractor in the CA. In Section~\ref{Chap:3-stateCA_list:sec:fpg}, a way of finding all fixed-point attractors in a CA using de Bruijn graph is described. Recall that, any cycle in that graph represents a fixed-point attractor for the CA. However, if a CA fulfills the first condition, then the de Bruijn graph with self-replicating RMTs has a self-loop for either of the RMTs $0$, $13$ or $26$. Our requirement is, there is no other cycle in the graph apart from this self-loop. Otherwise, the CA is rejected.

\begin{example}
In Figure~\ref{Chap:3-stateCA_list:fig:fixed}, for the CA $211212112020000020102121201$, there are two fixed-point attractors -- $2^n$ for the quiescent state $2$ as well as $(001)^{\frac{n}{3}}$. However, for the CA $102012102012102102021021012$, there is no quiescent state. So, none of these CAs satisfy the filtering conditions. Hence, both of these CAs are rejected.
\end{example}
								
For each of the strategies, $1118664$ out of the $10073592$ CAs satisfy these theoretical filtering conditions. However, there is a greater chance of getting long cycle, if the trivial configurations ($0^n$, $1^n$ or $2^n$) are isolated from other non-trivial configurations.

\section{Isolated Trivial Configuration}\label{Chap:3-stateCA_list:sec:trivial}
The two conditions of Section~\ref{Chap:3-stateCA_list:sec:filter} ensure that, one of the trivial configuration ($0^n$, $1^n$ or $2^n$) is a fixed-point attractor and the other two trivial configurations form a cycle. However, from the $1118664$ rules of each strategy, many rules can be screened out if we consider reachability of the trivial configurations ($0^n$, $1^n$ or $2^n$) from a non-trivial configuration.
								
If any of the trivial configurations ($0^n$, $1^n$ or $2^n$) is reachable from other non-trivial configurations, then it implies that the CA converges to trivial configurations for a set of seeds. In this case, the system becomes stable after some time steps. This is an undesirable situation which weakens randomness property of the corresponding CA. Therefore, our target is to select those CAs, for which this situation does not arise. Here, two approaches can be taken -- 
\begin{enumerate}[leftmargin=*,topsep=0pt,itemsep=0ex,partopsep=2ex,parsep=2ex]
\item Discard those rules for which the trivial configurations are reachable from a standard non-trivial configuration;
\item Discard all rules for which the trivial configurations are reachable from \emph{any} non-trivial configuration.
\end{enumerate}
\subsection{First Approach - Nonreachability from ${0^{n-1}2}$}\label{Chap:3-stateCA_list:sec:approach1}
In this approach, we examine if a CA converges to any of the trivial configurations from a {\em special} non-trivial configuration. As the special configuration, we choose here $0^{n-1}2$ (all $0$, except one cell with the highest state value $2$), likewise Wolfram's work \cite{wolfram84b,Wolfr83}, where $n$ is the size of the CA. The rationale behind the choice of this seed is the following: The presence of 2 (may be considered as {\em noise}) in almost homogeneous seed will affect the other cells during evolution of the CA, because of the {\em information propagation} in both directions for the selected CAs. Therefore, it is highly probable that a huge number of configurations will be observed during evolution of a CA. Within a reasonable amount of time, if a trivial configuration is reached, then it implies that, for a good number of seeds, the CA converges to trial configuration. This is an undesirable situation, and we exclude those CAs.
								
Therefore, we search for a set of rules from the $1118664$ rules, which we have already selected in Section~\ref{Chap:3-stateCA_list:sec:filter}. We explore each of the $1118664$ rules from each strategy to see whether it converges to any trivial configuration from $0^{n-1}2$ within a reasonable amount of time. We find that there are $542985$ and $479602$ rules from STRATEGY I and STRATEGY II respectively, which do not converge to trivial configurations from the seed $0^{n-1}2$. Hence, we get a rule set of size $542985+479602=1122587$ using this filtering scheme.
								
\subsection{Second Approach - Isolation from all Non-trivial Configurations}\label{Chap:3-stateCA_list:sec:approach2} 
In this approach, we are more conservative than the previous. Here, we want to choose the CAs for which the trivial configurations are isolated from {\em every} non-trivial configurations. To do that, we again use the de Bruijn graph of the CA. 
								
To find whether a trivial configuration $0^n$ is reachable or not, we modify the de Bruijn graph so that the edges with RMT $r$, where $R[r]\ne 0$, for all $r$, are removed. So, this de Bruijn graph has only the edges corresponding to the RMTs with next state value $0$.
As any cycle in de Bruijn graph represents an RMT sequence, so, any cycle in this de Bruijn graph represents an RMT sequence for which the next configuration is $0^n$. Hence, if there exists any cycle in this graph except the self-loop in node $00$, then this implies that, this trivial configuration $0^n$ is reachable from a non-trivial configuration. In this case, the rule is discarded. Similarly, for the configurations $1^n$ and $2^n$ also, we can draw the de Bruijn graphs having only the corresponding edges to find whether the trivial configurations are reachable from some non-trivial configuration.
								
\begin{example}
The de Bruijn graphs of the CA $102012210120021021012102120$ for finding $0^n$, $1^n$ and $2^n$ are shown in Figure~\ref{Chap:3-stateCA_list:fig:dbg0}, Figure~\ref{Chap:3-stateCA_list:fig:dbg1} and Figure~\ref{Chap:3-stateCA_list:fig:dbg2} respectively.
									
\begin{figure}[hbtp]
\centering										
\subfloat[$0^n$ reachability \label{Chap:3-stateCA_list:fig:dbg0}]{%
		\includegraphics[width=0.30\textwidth, height = 3.2cm]{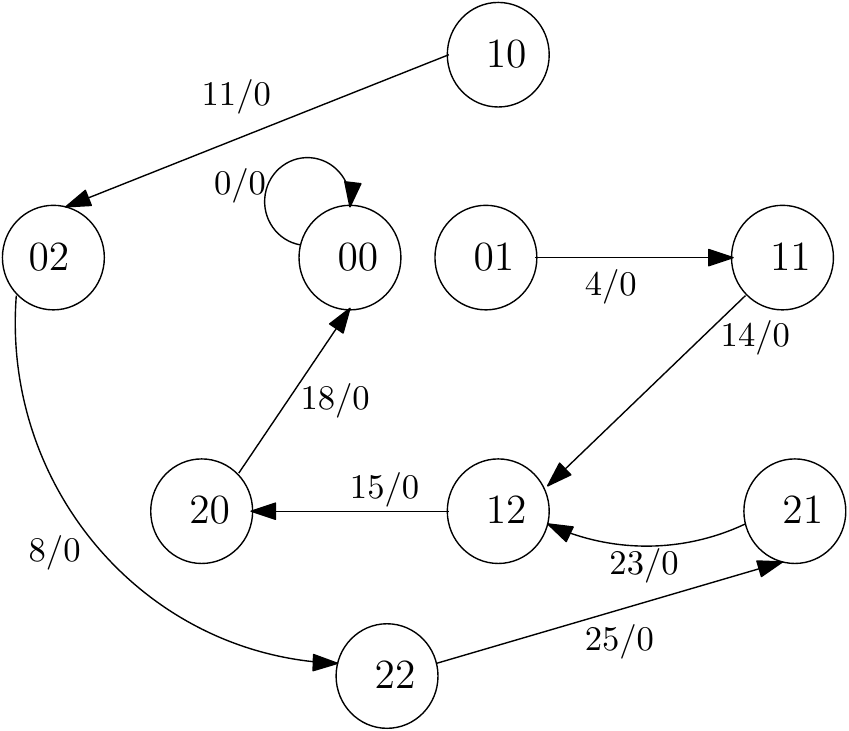}
}
\hfill
\subfloat[$1^n$ reachability \label{Chap:3-stateCA_list:fig:dbg1}]{%
\includegraphics[width=0.30\textwidth, height = 3.2cm]{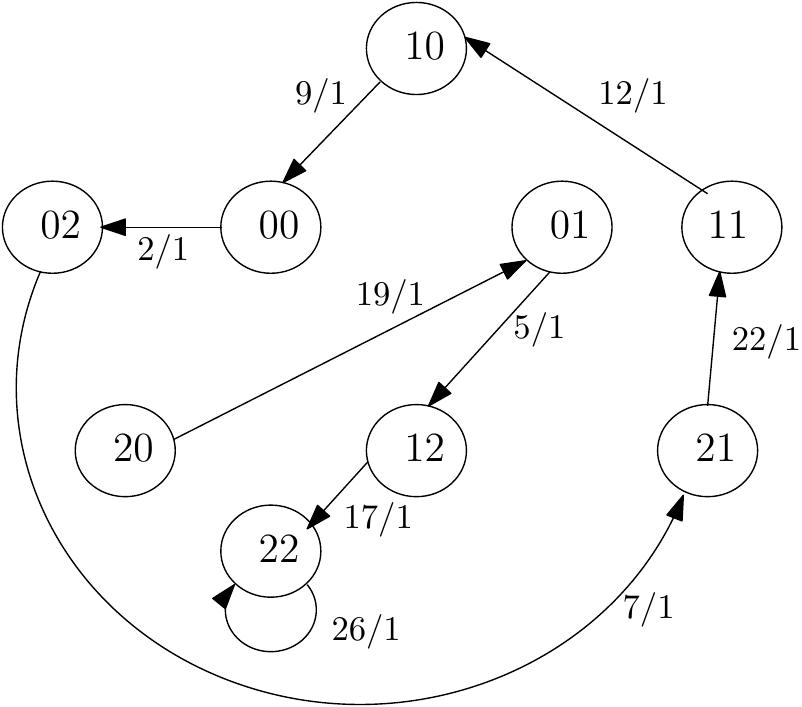}
}
\hfill
\subfloat[$2^n$ reachability\label{Chap:3-stateCA_list:fig:dbg2}]{%
	\includegraphics[width=0.30\textwidth, height = 3.2cm]{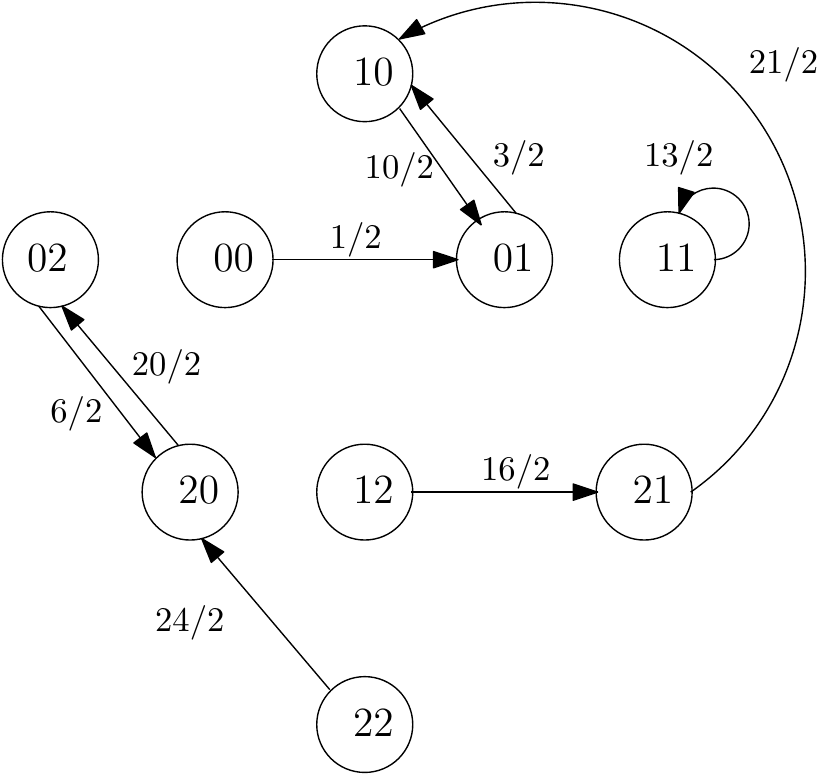}
}
\caption{The de Bruijn graphs of CA $102012210120021021012102120$ for finding reachability of $0^n$, $1^n$ and $2^n$}\label{Chap:3-stateCA_list:fig:filter3}
\end{figure}
									
From these figures it can be observed that, the configuration $0^n$ is not reachable from any non-trivial configuration. However, the configuration $1^n$ is reachable from $2^n$ and $(21100)^{\frac{n}{5}}$ ( that is, RMT sequence $\langle7, 22, 12, 9, 2 \rangle^{\frac{n}{5}}$). Similarly, $2^n$ is reachable from $1^n$ as well as from $(10)^{\frac{n}{2}}$ (RMT sequence $\langle 3,10 \rangle^{\frac{n}{2}}$) and $(20)^{\frac{n}{2}}$ (RMT sequence $\langle 6,20 \rangle^{\frac{n}{2}}$). Hence, this rule is discarded.
\end{example}

In this way, for each rule satisfying the conditions of Section~\ref{Chap:3-stateCA_list:sec:filter}, we can check the reachability of trivial configurations. Huge number of rules are discarded from each strategy by applying this filtering criterion. Finally, for this approach, we get the working rule set of size $5478$ from each of the two strategies. It is expected that, these rules have large cycle lengths for all $n$, as these rules have only one trivial fixed-point attractor and isolated trivial configurations which can only hop among each other. It may be noted here that all of these rules ($5478\times 2$) are included for the obvious reason in the rule set of the first approach.
								

\section{Empirical Filtering}\label{Chap:3-stateCA_list:sec:experiment}
The cellular automata rules, we have selected by exploring the above filters, are expected to be a good source of randomness. In this section, we use statistical tests to identify a {\em better} set of rules from the set of already-selected rules. Like before, we use Diehard battery of tests as the primary empirical testbed.
								
As mentioned in Section~\ref{sec:BTest} (Page~\pageref{sec:diehard}), to test in Diehard, we need to supply a binary file of size $10-12$ MB which contains the sequence of random numbers. There are following issues when we prepare this file.
\begin{enumerate}
\item The CAs under study generate ternary sequences, which are to be converted to equivalent binary sequences. Here, to encode an $n$-length ternary number, we first convert it to equivalent decimal number, then convert the decimal number to binary number.
\item Whole configuration of an $n$-cell CA is considered as an $n$-trit random number. However, there are many other alternatives to extract random numbers from configurations of CAs. Wolfram has taken a single bit from each configuration as a random bit \cite{wolfram86c}. In Section~\ref{Chap:randomness_survey:sec:prng_R} of Chapter~\ref{Chap:randomness_survey} (Page~\pageref{Chap:randomness_survey:sec:prng_R}), we have considered the content of a small window, placed over the configurations, as random numbers. Some other techniques of extracting random numbers from the configurations are also reported\cite{Horte89a}. Using any of these techniques, we could improve the quality of random numbers. However, in this section, we want to identify some rules which are {\em empirically better} than others. So for each CA, we have considered whole configuration during evolution as random number. However, any of the above mentioned techniques can be used to extract the random numbers from configurations when the selected CAs will really be used as PRNGs.

\item As size of a CA, we take $n=15$. This $n$ is sufficiently large to produce a binary file of size $10-12$ MB. Generally, the configuration space of a CA is partitioned into a number of subspaces (see Figure~\ref{fig:config} of Page~\pageref{fig:config} as example). If the number of subspaces is less, then it is expected that the lengths of cycles of the CA are high. In other word, if there are a good number of cycles, then expected lengths of the cycles are less. We want to choose those CAs as good PRNGs, which are having large cycles in their configuration space. For a CA with small-sized cycles, while preparing a binary file of size $10-12$ MB, the configurations may get exhausted before getting the binary file for some seeds. Diehard may indicate this fact by showing bad empirical results. However, if we take large $n$, then relatively small cycles of a CA can be of good lengths, which may lead Diehard to show good results. Considering these facts, we settle the CA size to 15.
	\end{enumerate}

\begin{table}[!h]
\vspace{-1.0em}	
\caption{Sample of candidate rules of Section~\ref{Chap:3-stateCA_list:sec:approach1} (seed = $0^{k}20^{k}$, $k = \floor{\frac{15}{2}}$)}
\label{result1}
\centering
\resizebox{1.00\textwidth}{!}{
\small
\begin{tabular}{|c|c|c||c|c|c|}
\hline 
$3$-state CA Rules & \begin{tabular}{c}
\#Tests passed\\
in Diehard\\
($n = 15$)
\end{tabular}   & \begin{tabular}{c}
\#Tests passed\\
in TestU01\\
($n = 15$)
\end{tabular} & $3$-state CA Rules & \begin{tabular}{c}
\#Tests passed\\
in Diehard\\
($n = 15$)
\end{tabular}   & \begin{tabular}{c}
\#Tests passed\\
in TestU01\\
($n = 15$)
\end{tabular}\\ 
\hline 
222211111111122200000000022 & 10 & 13 & 222211112001000000110122221 & 9 & 13\\
222211112111100200000022021 & 9 & 10 & 222211121101000200010122012 & 8 & 9 \\
222211122001000010110122201 & 8 & 10 & 222211220000102012111020101 & 8 & 12\\
222211222111002110000120001 & 8 & 9 & 222212000000000212111121121 & 9 & 12\\
222212001011121110100000222 & 9 & 12 & 222212002000000120111121211 & 9 & 13\\
222212122000101000111020211 & 9 & 12 & 222212122011101010100020201 & 8 & 12 \\
222212222010001100101120011 & 8 & 13 & 222220000000002112111111221 & 10 & 11\\
222220000000002122111111211 & 10 & 13 & 222220000000102222111011111 & 9 & 12\\
222220000001002222110111111 & 10 & 12 & 222220000001102112110011221 & 7 & 12\\
222220000100002222011111111 & 10 & 12 & 222220101001002220110111012 & 9 & 12 \\
222221122000100000111012211 & 10 & 11 & 222221212000102000111010121 & 9 & 12\\
222221222010000010101112101 & 9 & 13 & 222222000100000122011111211 & 10 & 10\\
222222001000000120111111212 & 9 & 10 & 222222001000100220111011112 & 9 & 15\\
222020200111202021000111112 & 9 & 10 & 222020201110201010001112122 & 8 & 12 \\
222021000000202122111110211 & 8 & 10 & 222021002001102220110210111 & 8 & 14\\
222021020000102202111210111 & 8 & 13 & 222021022011202100100110211 & 9 & 11\\
222021222000100000111212111 & 10 & 9 & 222021222111102000000210111 & 9 & 11\\
222022000000201212111110121 & 8 & 11 & 222022001111100220000211112 & 8 & 11\\
222022002110200110001111221 & 8 & 12 & 222022100111100022000211211 & 8 & 11 \\
~201102102210120201210120021~ & 9 & 13 & ~012021012201012201021210201~ & 8 & 14\\
~012102201201210120201210201~ & 8 & 8 & ~012021021021210201021210021~ & 8 & 11\\
~012021120012102012012021210~ & 8 & 12 & ~012021120012120120021021210~ & 8 & 12\\ 
~012021120120120120120201120~ & 8 & 11 & ~012021210012021021012210120~ & 8 & 12 \\ 

~012102210210102012102102210~ & 8 & 10 & ~012120120012120021120021120~ & 8 & 11 \\ 
~012120201012201120021201120~ & 8 & 12 & ~012201012210210102021210201~ & 8 & 9\\ 
~012201102012012012012210102~ & 8 & 9 & ~012102210201012120210012102~ & 8 & 9 \\ 

~012201201021201210012201210 ~ & 8 & 12 & ~012201210201102102201102120~ & 8 & 12\\ 
~012210012210012201120210012~ & 8 & 11 & ~012210120021012102120120102~ & 8 & 8\\ 
~012210120210210012012210102~ & 8 & 11 & ~012210201021210120012210012~ & 8 & 12 \\ 

~021012012021012012201012102~ & 8 & 12 & ~021012012210012102201012201~ & 8 & 11\\  
~201102012102201120201210012~ & 7 & 9 & ~102210120120210021120210012~ & 7 & 12 \\ 
~102210120012120120102102120~ & 7 & 9 & ~102210120102102120102210210~ & 7 & 11\\ 
~201021201201021021210210102~ & 7 & 12 & ~012021012012012102021012201~ & 7 & 12\\
\hline 
\end{tabular} 
}
\end{table}

\begin{table}[!h]
\caption{Sample of candidate rules of Section~\ref{Chap:3-stateCA_list:sec:approach2} (seed = $0^{k}20^{k}$, $k = \floor{\frac{15}{2}}$)}
\label{result2}
\centering
\resizebox{1.00\textwidth}{!}{
\small
\begin{tabular}{|c|c|c||c|c|c|}
\hline 
$3$-state CA Rules & \begin{tabular}{c}
\#Tests passed\\
in Diehard\\
($n = 15$)
\end{tabular}   & \begin{tabular}{c}
\#Tests passed\\
in TestU01\\
($n = 15$)
\end{tabular} & $3$-state CA Rules & \begin{tabular}{c}
\#Tests passed\\
in Diehard\\
($n = 15$)
\end{tabular}   & \begin{tabular}{c}
\#Tests passed\\
in TestU01\\
($n = 15$)
\end{tabular}\\ 
\hline
222111202000002010111220121 & 9  & 14 & 222111202100000010011222121 & 9  & 13 \\
222111212000000000111222121 & 9  & 11 & 222111212000200000111022121 & 10  & 13 \\
222111222100200000011022111 & 8  & 12 & 222112002000000120111221211 & 8  & 11 \\
222121000000002222111210111 & 7  & 11 & 222121200000002012111210121 & 8  & 11 \\
222121220000200002111012111 & 7  & 10 & 222121222000200000111012111 & 9  & 12 \\
222121222000200010111012101 & 9  & 12 & 222121222101000000010212111 & 8  & 14 \\
222122200101000022010211111 & 8  & 12 & 222211122101000000010122211 & 10  & 12 \\
222211212100000000011122121 & 8  & 12 & 222211222001000000110122111 & 10  & 12 \\
222211222010000100101122011 & 8  & 10 & 222211222101000000010122111 & 10  & 12 \\
222211222110000000001122111 & 10  & 12 & 222212000000001222111120111 & 7  & 14 \\
222212010000000222111121101 & 10  & 13 & 222220000000002122111111211 & 10  & 12  \\
222220000000002222111111111 & 9  & 14 & 222220200001002022110111111 & 8  & 12 \\
012012210210012102021012012 & 9  & 11 & 012021102012012012012012102 & 7  & 13 \\
012021102012012201012012012 & 8  & 11 & 012201102012012012012210102 & 10  & 14 \\
012201102012012210012012012 & 7  & 11 & 012210102012012012012012012 & 7  & 11 \\
012210102012012012012012102 & 8  & 11 & 012210102012012120120201012 & 8  & 10 \\
012210102012012210012012102 & 9  & 12 & 012210102012210012012012102 & 9  & 12 \\
012210102012210012210012102 & 9  & 12 & 012210102012210102012012012 & 9  & 12 \\
012210102012210102012012102 & 8  & 12 & 012210102012210201012012102 & 9  & 12 \\
021012012012012012021102012 & 8  & 11 & 021012012012012012021120012 & 10  & 13 \\
021012012012210021021210012 & 9  & 13 & 021012012021012012201012102 & 9  & 12 \\
021012012021210012021210012 & 8  & 12 & 021012012021210012201210012 & 8  & 12 \\
021012012120210012021012012 & 7  & 11 & 021012021021012012021012012 & 9  & 12 \\
021012021210012012021012012 & 9  & 13 & 021012102012012021012210012 & 8  & 12  \\
021012102012012102210012102 & 9  & 12 & 102201120102120120102120120 & 9  & 11 \\
\hline
\end{tabular}
}
\end{table}
								
								
As mentioned in Section~\ref{sec:BTest} (Page~\pageref{sec:diehard}), Diehard is having 15 different tests. For the screening purpose, we consider a CA as {\em bad} if it passes less than 7 tests for a seed. To identify a set of {\em good} rules, we consider each of the already-selected rules, and then repeat this experiment for several times. At each time, every CA is initialized with a large number of arbitrary seeds and tested for randomness. The CAs which passes the minimum tests (that is, at least $7$ tests) in each of these seeds are collected and put to test again. We have used sixty $64$-bit Intel Core $i5-3570$ CPUs (with $3.40GHz \times 4$ frequency and $4$ GB RAM) for this experiment. 
For a rule, if we get that the CA always passes at least 7 tests in all these runs, then we consider the rule as a {\em good} candidate for randomness.

Although the CAs of STRATEGY I and STRATEGY II are equivalent, we separately experiment with the CAs of each strategy. In fact, we experiment with all rules of an equivalent rule-set (see Definition~\ref{Chap:3-stateCA_list:Defi:EquivalentRuleSet}). If all the rules of such set are {\em good}, then we put their minimal rule in our final list. Rationale behind this style is, a rule along with its equivalents is better exposed to the issue of non-randomness.
								
By Section~\ref{Chap:3-stateCA_list:sec:trivial}, however, one can get two different sets of rules using two approaches. Supposedly, the rules of second approach (Section~\ref{Chap:3-stateCA_list:sec:approach2}) are better than the rules of the other set. Recall that we have identified 542985 rules of STRATEGY I, and 479602 rules of STRATEGY II using first approach. Whereas, 5478 rules of each strategy are chosen using second approach. However, all the rules, obtained by second approach, are the members of the rule set of first approach. To get more candidates, we apply empirical filter on all the selected rules.

For rules from each of the approaches, the experiment is repeated many times. The seed for the tests are taken to be random along with the fixed seed $0^{k}20^{k}$, where $k = \floor{\frac{15}{2}}$. The rules, which have consistently passed at least $7$ tests in all these runs, are selected to be potential pseudo-random number generator. For the first approach, we have got $937$ rules from STRATEGY I and $805$ rules from STRATEGY II which perform well irrespective of any initial condition. Table~\ref{result1} gives some of these rules. Similarly, for the second approach, we have found $230$ rules of STRATEGY I and $264$ rules of STRATEGY II that pass at least $7$ tests. Some of these rules are listed in Table~\ref{result2}.

However, it is impossible to test every rule for all possible seeds, so, we impose a stricter condition on these \emph{good} rules. We only choose those rules for the final set of \emph{better} rules for which all of their equivalent rules are also in the set of good rules. So, finally we get a set of $596$ minimal rules after using all filtering schemes. We claim that these are the good candidates to act as PRNGs. These rules are listed in Table~\ref{Chap:3-stateCA_list:tab:ruleset_second} and Table~\ref{Chap:3-stateCA_list:tab:ruleset1_first}. Table~\ref{Chap:3-stateCA_list:tab:ruleset_second} shows the rules which were selected by second approach (Section~\ref{Chap:3-stateCA_list:sec:approach2}), whereas Table~\ref{Chap:3-stateCA_list:tab:ruleset1_first} notes the rules which were selected by first approach (Section~\ref{Chap:3-stateCA_list:sec:approach1}), but filtered out by second approach. 
								\begin{center}
									\begin{tiny}
									\setlength\tabcolsep{2pt}
\begin{longtable}{|c|c||c|c||c|c|}
											\caption{List of minimal CAs selected as good PRNGs from Section~\ref{Chap:3-stateCA_list:sec:approach2}}
											\label{Chap:3-stateCA_list:tab:ruleset_second}\\
											\hline
											\textbf{SL No.} & \textbf{Rule} & \textbf{SL No.} & \textbf{Rule}  & \textbf{SL No.} & \textbf{Rule} \\
											\hline
											\endfirsthead
											\multicolumn{6}{c}%
											{\tablename\ \thetable\ -- \textit{Continued from previous page}} \\
											\hline
											\textbf{SL No.} & \textbf{Rule} & \textbf{SL No.} & \textbf{Rule}  & \textbf{SL No.} & \textbf{Rule} \\
											\hline
											\endhead
											\hline \multicolumn{6}{r}{\textit{Continued on next page}} \\
											\endfoot
											\hline
											\endlastfoot
 1~ & ~000000000122111211211222122~ & ~2~ & ~000000000222111211111222122~ & ~3~ & ~000000101111211010222122222~ \\ 
								 4~ & ~000000111211111200122222022~ & ~5~ & ~000000120111111001222222212~ & ~6~ & ~000000120211211011122122202~ \\ 
								 7~ & ~000000120221111001112222212~ & ~8~ & ~000000200212111111121222022~ & ~9~ & ~000000200222111011111222122~ \\ 
								 10~ & ~000001000212112111121220222~ & ~11~ & ~000002000112111111221220222~ & ~12~ & ~000002000212111111121220222~ \\ 
								 13~ & ~000002001212111120121220212~ & ~14~ & ~000002100211111021122220212~ & ~15~ & ~000002101121111210212220022~ \\ 
								 16~ & ~000020020212211101121102212~ & ~17~ & ~000020100211211011122102222~ & ~18~ & ~000020121111111000222202212~ \\ 
								 19~ & ~000020200212111111121202022~ & ~20~ & ~000020200222111011111202122~ & ~21~ & ~000022000112111111221200222~ \\ 
								 22~ & ~000022000122111111211200222~ & ~23~ & ~000022000212111111121200222~ & ~24~ & ~000022000212111121121200212~ \\
\hline
										\end{longtable}
									\end{tiny}
								\end{center}

								\begin{center}
									\begin{tiny}\setlength\tabcolsep{2pt}
										\begin{longtable}{|c|c||c|c||c|c|}
											\caption{List of minimal CAs selected as good PRNGs from Section~\ref{Chap:3-stateCA_list:sec:approach1}}
											\label{Chap:3-stateCA_list:tab:ruleset1_first}\\
											\hline
											\textbf{SL No.} & \textbf{Rule} & \textbf{SL No.} & \textbf{Rule} & \textbf{SL No.} & \textbf{Rule} \\
											\hline
											\endfirsthead
											\multicolumn{6}{c}%
											{\tablename\ \thetable\ -- \textit{Continued from previous page}} \\
											\hline
											\textbf{SL No.} & \textbf{Rule} & \textbf{SL No.} & \textbf{Rule}  & \textbf{SL No.} & \textbf{Rule} \\
											\hline
											\endhead
											\hline \multicolumn{6}{r}{\textit{Continued on next page}} \\
											\endfoot
											\hline
											\endlastfoot
	1~ & ~000000011212222222121111100~ & ~2~ & ~000000012112211100221122221~ & ~3~ & ~000000121111111210222222002~ \\ 
											4~ & ~000000200112111121221222012~ & ~5~ & ~000000200222111121111222012~ & ~6~ & ~000000202121211010212122121~ \\ 
											7~ & ~000000210111111002222222121~ & ~8~ & ~000000211111111120222222002~ & ~9~ & ~000001100221110022112222211~ \\ 
											10~ & ~000002000112111221221220112~ & ~11~ & ~000002201111221122222110010~ & ~12~ & ~000002211111111020222220102~ \\ 
											13~ & ~000002212222221021111110100~ & ~14~ & ~000010102111202211222121020~ & ~15~ & ~000010112112222221221101000~ \\ 
											16~ & ~000010122121221011212102200~ & ~17~ & ~000010222112222111221101000~ & ~18~ & ~000011101112122212221200020~ \\ 
											19~ & ~000020112112101221221212000~ & ~20~ & ~000020211112102122221211000~ & ~21~ & ~000021100221110022112202211~ \\ 
											22~ & ~000022121212201212121110000~ & ~23~ & ~000022220222211001111100112~ & ~24~ & ~000022222211100111122211000~ \\
\hline
										\end{longtable}
									\end{tiny}
									\vspace{-1.5em}
								\end{center}
								
We have found $474$ rules from the first approach (Table~\ref{Chap:3-stateCA_list:tab:ruleset1_first}) and $122$ rules from the second approach (Table~\ref{Chap:3-stateCA_list:tab:ruleset_second}). These are the minimal rules selected as potential candidates for PRNGs.
However, there are other $664$ rules from first approach and $166$ rules from second approach, which also perform well for every seed, but all of their equivalents does not conduct similarly, hence not included in the set of good rules. So, they are not included in the final $596$ rules. As, these rules individually are good source of randomness, they are listed here in Table \ref{extraRules_first} and \ref{extraRules_second}).
 
\begin{center}
\begin{tiny}\setlength\tabcolsep{2pt}
\begin{longtable}{|c|c||c|c||c|c|}
\caption{Remaining rules from Section~\ref{Chap:3-stateCA_list:sec:approach1}}
 \label{extraRules_first}\\
	\hline
\textbf{SL No.} & \textbf{Rule} & \textbf{SL No.} & \textbf{Rule} & \textbf{SL No.} & \textbf{Rule} \\
	\hline
	\endfirsthead
\multicolumn{6}{c}%
{\tablename\ \thetable\ -- \textit{Continued from previous page}} \\
											\hline
\textbf{SL No.} & \textbf{Rule} & \textbf{SL No.} & \textbf{Rule}  & \textbf{SL No.} & \textbf{Rule} \\
	\hline
	\endhead
	\hline \multicolumn{6}{r}{\textit{Continued on next page}} \\
	\endfoot
	\hline
	\endlastfoot
1~ & ~112000100001111021220222212~ & ~2~ & ~112000100001211011220122222~ & ~3~ & ~112000101021111210200222022~ \\ 
4~ & ~112000111001111200220222022~ & ~5~ & ~112000111001111220220222002~ & ~6~ & ~112000111020211200201122022~ \\ 
7~ & ~112000111021111200200222022~ & ~8~ & ~112000112000211000221122221~ & ~9~ & ~112000112001111200220222021~ \\ 
10~ & ~112000112001111220220222001~ & ~11~ & ~112000112020211200201122021~ & ~12~ & ~112000120021111001200222212~ \\ 
13~ & ~112000121020211010201122202~ & ~14~ & ~112000121021111000200222212~ & ~15~ & ~112000121021111210200222002~ \\ 
16~ & ~112000121021211010200122202~ & ~17~ & ~112000122001211010220122201~ & ~18~ & ~112001100021112011200220222~ \\ 
19~ & ~112001100021112012200220221~ & ~20~ & ~112001101020112220201220012~ & ~21~ & ~112001101021112020200220212~ \\ 
22~ & ~112002101000211220221120012~ & ~23~ & ~112002101021111210200220022~ & ~24~ & ~112002110001111021220220202~ \\
\hline
\end{longtable}
\end{tiny}
\vspace{-1.0em}
\end{center}
 
  \begin{center}
  \begin{tiny}\setlength\tabcolsep{2pt}
  \begin{longtable}{|c|c||c|c||c|c|}
\caption{Remaining rules from Section~\ref{Chap:3-stateCA_list:sec:approach2}}
   \label{extraRules_second}\\
  	\hline
  \textbf{SL No.} & \textbf{Rule} & \textbf{SL No.} & \textbf{Rule} & \textbf{SL No.} & \textbf{Rule} \\
  	\hline
  	\endfirsthead
  \multicolumn{6}{c}%
  {\tablename\ \thetable\ -- \textit{Continued from previous page}} \\
  											\hline
  \textbf{SL No.} & \textbf{Rule} & \textbf{SL No.} & \textbf{Rule}  & \textbf{SL No.} & \textbf{Rule} \\
  	\hline
  	\endhead
  	\hline \multicolumn{6}{r}{\textit{Continued on next page}} \\
  	\endfoot
  	\hline
  	\endlastfoot
 1~ & ~000000100111211011222122222~ & ~2~ & ~000002101111211010222120222~ & ~3~ & ~000002200122111011211220122~ \\ 
 4~ & ~000020120121111011212202202~ & ~5~ & ~000021101121110210212202022~ & ~6~ & ~000022101121111010212200222~ \\ 
 7~ & ~000022200122111111211200022~ & ~8~ & ~000200011112111100221022222~ & ~9~ & ~000201021212110110121022202~ \\ 
 10~ & ~000220010212111101121002222~ & ~11~ & ~000221001112110110221002222~ & ~12~ & ~001000211112111000220222122~ \\ 
 13~ & ~001020121110111000222202212~ & ~14~ & ~001220121110111010222002202~ & ~15~ & ~001220121112011010220102202~ \\ 
 16~ & ~002000120111111011220222202~ & ~17~ & ~002000121111111000220222212~ & ~18~ & ~002001001210110110121222222~ \\ 
 19~ & ~002220110111111001220002222~ & ~20~ & ~010000001122111110201222222~ & ~21~ & ~010200001202111110121022222~ \\ 
 22~ & ~010222010202111121121000202~ & ~23~ & ~020000200112111111201222022~ & ~24~ & ~020200021112011110201122202~ \\
\hline
 \end{longtable}
 \end{tiny}
 \vspace{-1.0em}
 \end{center}

\begin{figure}[!h]
    \centering
      \includegraphics[width=0.95\textwidth,height =11.5cm]{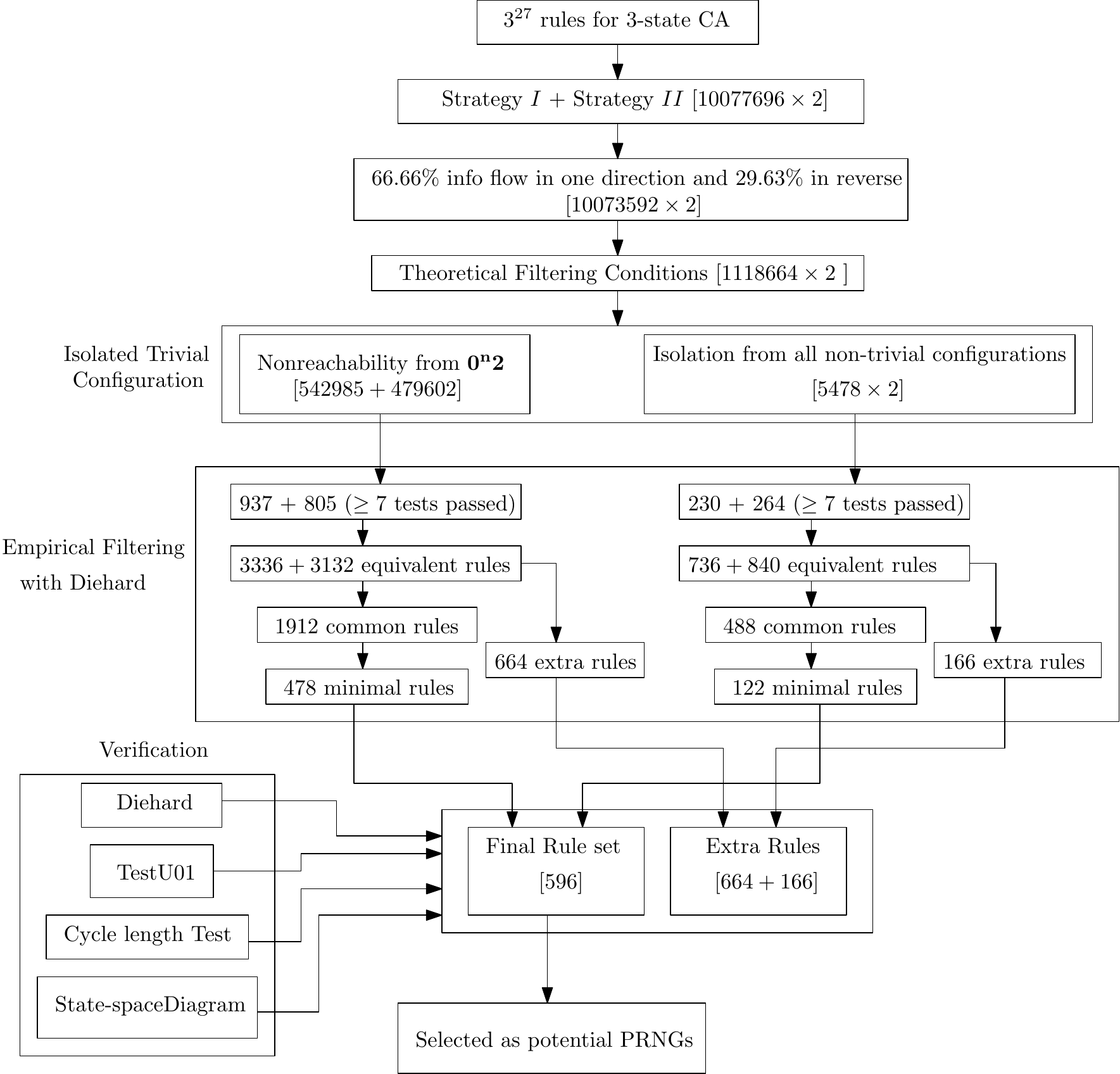}
      \caption{Summarized procedure of selecting $3$-state CAs as potential PRNGs}          
                    \label{fig:flowchart}
    \end{figure}

 The total process of selecting the final set of rules is shown in Figure~\ref{fig:flowchart}.
 
\section{Verification of Result}\label{Chap:3-stateCA_list:sec:results}
Let us now verify the competence  of finally-selected rules as good source of randomness. To do this, the CAs are tested with more stringent TestU01 library and space-time diagram for these CAs are drawn to visualize the dynamic behavior of the CAs. Further, the cycle lengths of those CAs for different values of $n$ are observed. In short, we apply following three tests on the selected rules to verify our result:
\begin{enumerate}
	\item Test with TestU01 library,
	\item Cycle length test
	\item Space-time diagram test
\end{enumerate} 								

\subsection{Test with TestU01 Library}\label{Chap:3-stateCA_list:sec:Testu01}
As TestU01 library offers many more stringent battery of tests, we have tested our selected rule set with TestU01 library. Like before, the battery \emph{rabbit} (bbattery\_RabbitFile()) of the library is selected which takes a binary file as input and contains $25$ stringent tests (see Section~\ref{sec:BTest} of Page~\pageref{sec:BTest}). The process of generating the binary files from the ternary output of the PRNGs is similar to Section~\ref{Chap:3-stateCA_list:sec:experiment}. Each of the rules of the final set, when tested with the battery rabbit, passes $12-15$ tests for any arbitrary initial configuration. Some of these test results for fixed seed ($0^{n-1}2$, $n=15$) is shown in Table~\ref{result1} and \ref{result2}.

\subsection{Cycle Length Test}
If our process of selection is correct, then the CAs selected as potential PRNGs must satisfy the properties that ensure their candidature. One such important property is existence of a very large cycle in their configuration transition diagram, which grows exponentially with $n$ (see Section~\ref{Chap:randomness_survey:sec:cycleLength} of Page~\pageref{Chap:randomness_survey:sec:cycleLength}). Because, in this case, more unique configurations are part of the same cycle, the generated numbers are more random and the chance of the PRNGs to exhaust the period length is lesser. Similarly, longer cycle length intuitively implies less number of cycles in the CA, that is, more configurations are part of one cycle. Therefore, we have to ensure that the selected CAs have extremely large cycle lengths which grow exponentially with $n$. As it is practically impossible to exhaustively derive length of each cycle for a large $n$, so, we conduct an experiment to calculate the cycle length of these rules for some random seeds for all $n\leq 15$.
It is observed that, although cycle length varies with CA size, but each of our selected CAs have very large cycle length for most values of $n$ and the highest recorded cycle length for each $n$ grows exponentially with $n$. This strengthens our selection process. 
								
\begin{example}
Table~\ref{Chap:3-stateCA_list:tab:cycle} gives cycle lengths for a sample run with fixed seed $0^{n-1}2$, $5\leq n \leq 15$, for some rules of Table~\ref{Chap:3-stateCA_list:tab:ruleset_second} and Table~\ref{Chap:3-stateCA_list:tab:ruleset1_first}. Note that, although for most of the CAs, this seed does not give the maximum cycle length for each $n$, but even then, for all these CAs, the cycle lengths with this seed for many CA sizes are sufficiently large.
									
\begin{table}[hbtp]
\caption{Cycle lengths from seed $0^{n-1}2$ for some CAs of Table~\ref{Chap:3-stateCA_list:tab:ruleset_second} and Table~\ref{Chap:3-stateCA_list:tab:ruleset1_first}}
										{
											\centering
											\resizebox{0.95\textwidth}{!}{
												\small
												\begin{tabular}{|c|c|c|c|c|c|c|c|c|c|c|c|}
													\hline 
													Rule & $n= 5$ & $n = 6$ & $n = 7$ & $n = 8$ & $n = 9$ & $n = 10$ & $n = 11$ & $n = 12$ & $n = 13$ & $n = 14$ & $n = 15$\\
													\hline 
													000000000122111211211222122 & 13 & 161 & 1077 & 4415 & 13427 & 21314 & 148081 & 1787 & 863238 & 1510599 & 10339049 \\ 
													000000000222111211111222122 & 154 & 395 & 377 & 679 & 178 & 31079 & 149225 & 483479 & 1001805 & 517145 & 11114804 \\ 
													000000120111111001222222212 & 89 & 308 & 125 & 631 & 15182 & 32959 & 50951 & 661 & 7188 & 1413558 & 1650834 \\ 
													101000111222222222010111000 & 154 & 395 & 107 & 679 & 8234 & 4677 & 149225 & 483479 & 1001805 & 517145 & 11114804 \\ 
													101002111222220222010111000 & 169 & 326 & 314 & 127 & 15164 & 27189 & 120570 & 11264 & 909700 & 3308325 & 867664 \\ 
													101010111222222222010101000 & 139 & 161 & 776 & 4415 & 13427 & 204 & 148081 & 157047 & 456260 & 655731 & 10339049 \\ 
													102102120102120120102120120 & 154 & 395 & 107 & 679 & 8234 & 4677 & 149225 & 483479 & 1001805 & 517145 & 11114804 \\ 
													110001212222222121001110000 & 79 & 104 & 153 & 751 & 10160 & 664 & 52722 & 28247 & 196143 & 56629 & 14282639 \\ 
													110111111001022222222200000 & 169 & 326 & 314 & 127 & 15164 & 27189 & 120570 & 11264 & 909700 & 3308325 & 867664 \\ 
													110111111201222222022000000 & 89 & 308 & 1049 & 27 & 15182 & 32959 & 50951 & 10250 & 1057939 & 1352693 & 1650834 \\ 
													111012111222220222000101000 & 89 & 308 & 1049 & 27 & 15182 & 32959 & 50951 & 10250 & 1057939 & 1352693 & 1650834 \\ 
													111111111200022222022200000 & 139 & 161 & 776 & 4415 & 13427 & 204 & 148081 & 157047 & 456260 & 655731 & 10339049 \\
													000000211111111120222222002 & 224 & 158 & 1805 & 1047 & 2020 & 1729 & 72951 & 23671 & 1501083 & 41411 & 4134219 \\
													000200200211011021122122112 & 149 & 71 & 1112 & 14 & 5177 & 2549 & 88747 & 50 & 870583 & 4277 & 8224979 \\ 
													001000221120111000212222112 & 15 & 119 & 377 & 639 & 251 & 219 & 3101 & 640 & 16665 & 12683 & 147674 \\ 
													001020000222111221110202112 & 18 & 14 & 2106 & 351 & 989 & 149 & 60642 & 71 & 765192 & 45807 & 4755489 \\ 
													012111211200022122121200000 & 134 & 23 & 244 & 39 & 773 & 134 & 703 & 1298 & 3288 & 34229 & 109364 \\ 
													100222000212111221021000112 & 23 & 29 & 88 & 287 & 1223 & 63 & 7435 & 1192 & 18550 & 11078 & 124764 \\ 
													101110222010222111222001000 & 10 & 35 & 1189 & 91 & 8315 & 194 & 83566 & 27 & 687673 & 5242 & 1942229 \\ 
													101111112010020221222202000 & 69 & 14 & 2106 & 351 & 17234 & 18 & 21790 & 71 & 743716 & 45807 & 4755489 \\ 
													102000222220222111011111000 & 10 & 173 & 356 & 87 & 1610 & 4589 & 3145 & 6869 & 74762 & 13502 & 143024 \\ 
													102020111221202222010111000 & 154 & 0 & 146 & 117 & 4643 & 889 & 1979 & 839 & 30640 & 57273 & 473414 \\ 
													102102210102120120102120120 & 39 & 13 & 923 & 1 & 1664 & 13419 & 29160 & 13 & 1448823 & 274987 & 1251059 \\ 
													111201111222020222000112000 & 169 & 182 & 265 & 119 & 2105 & 10529 & 1850 & 182 & 992419 & 43861 & 7776224 \\ 
													120000121212222212001111000 & 234 & 80 & 1147 & 121 & 4499 & 1299 & 166627 & 25619 & 55333 & 147937 & 379199 \\  
													\hline 
												\end{tabular} 
											}} \label{Chap:3-stateCA_list:tab:cycle}
										\end{table}
									\end{example}

\subsection{Space-time Diagram Test} Space-time diagram is an important theoretical tool that has long been used to observe and predict the behavior and evolution of a CA (see Section~\ref{Chap:surveyOfCA:Sec:space-time} of Page~\ref{Chap:surveyOfCA:Sec:space-time} and Section~\ref{Chap:randomness_survey:sec:space-time_R} for details).
 If there is a repeating pattern among the consecutive configurations, then it is visible from this diagram. If there is no such pattern, and the colors appear noisy, the CA has good randomness quality. Therefore, by observing these diagrams for arbitrary seeds, clear idea about the randomness properties of a CA can be developed.
									
\begin{figure}[!h]
										\subfloat[\label{Chap:3-stateCA_list:fig:statespace1}]{%
											\includegraphics[width=0.25\textwidth, height = 12.0cm]{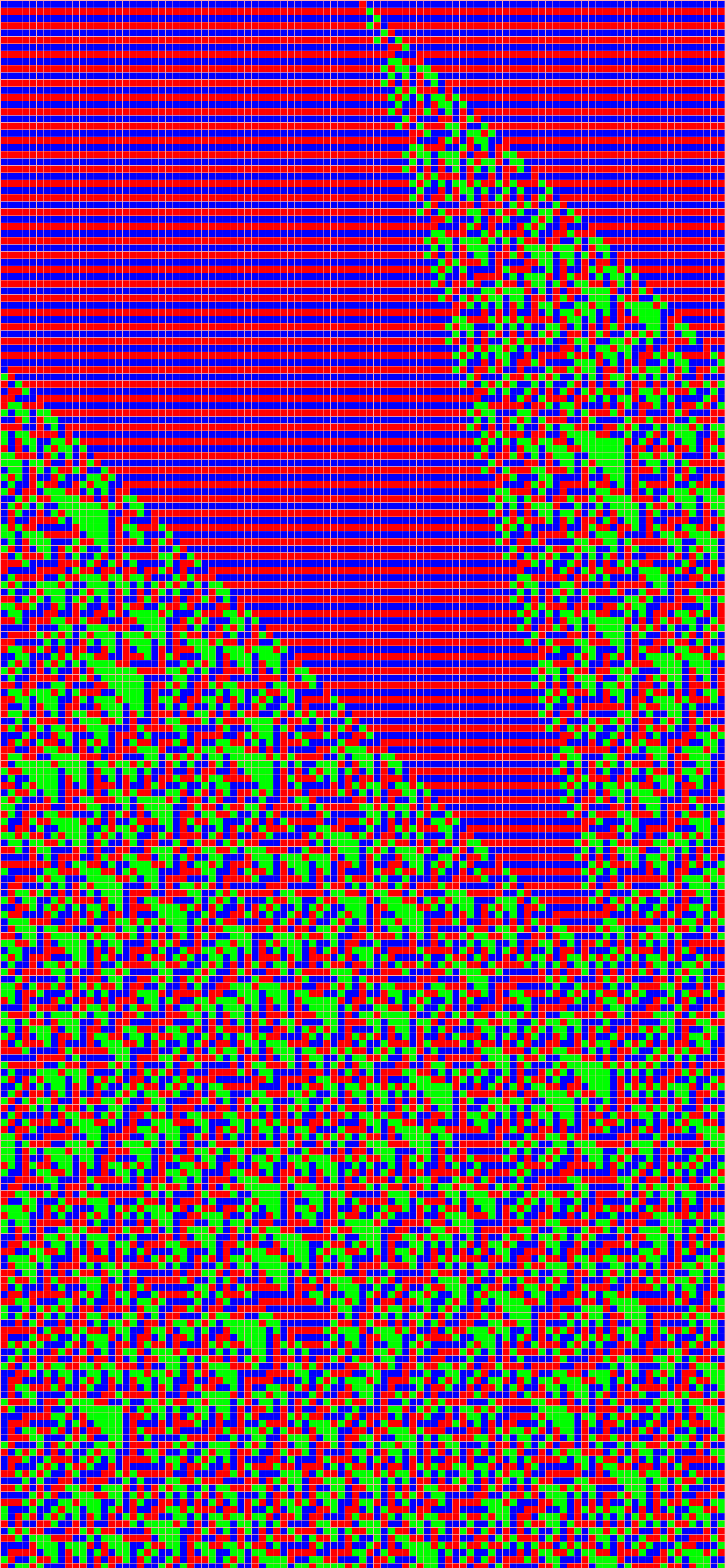}
										}
										\subfloat[ \label{Chap:3-stateCA_list:fig:statespace2}]{%
											\includegraphics[width=0.25\textwidth, height = 12.0cm]{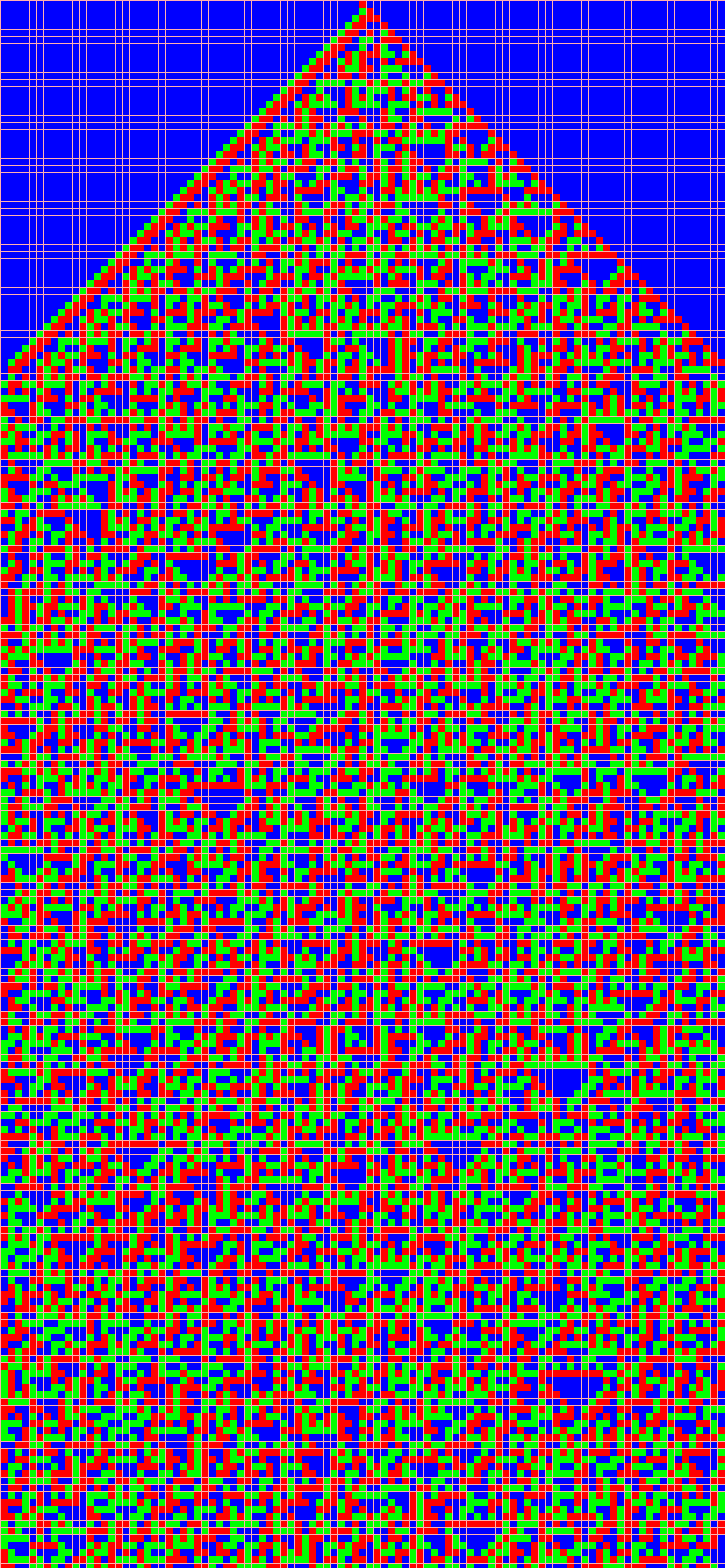}
										}   
										\subfloat[\label{Chap:3-stateCA_list:fig:statespace3}]{%
											\includegraphics[width=0.25\textwidth, height = 12.0cm]{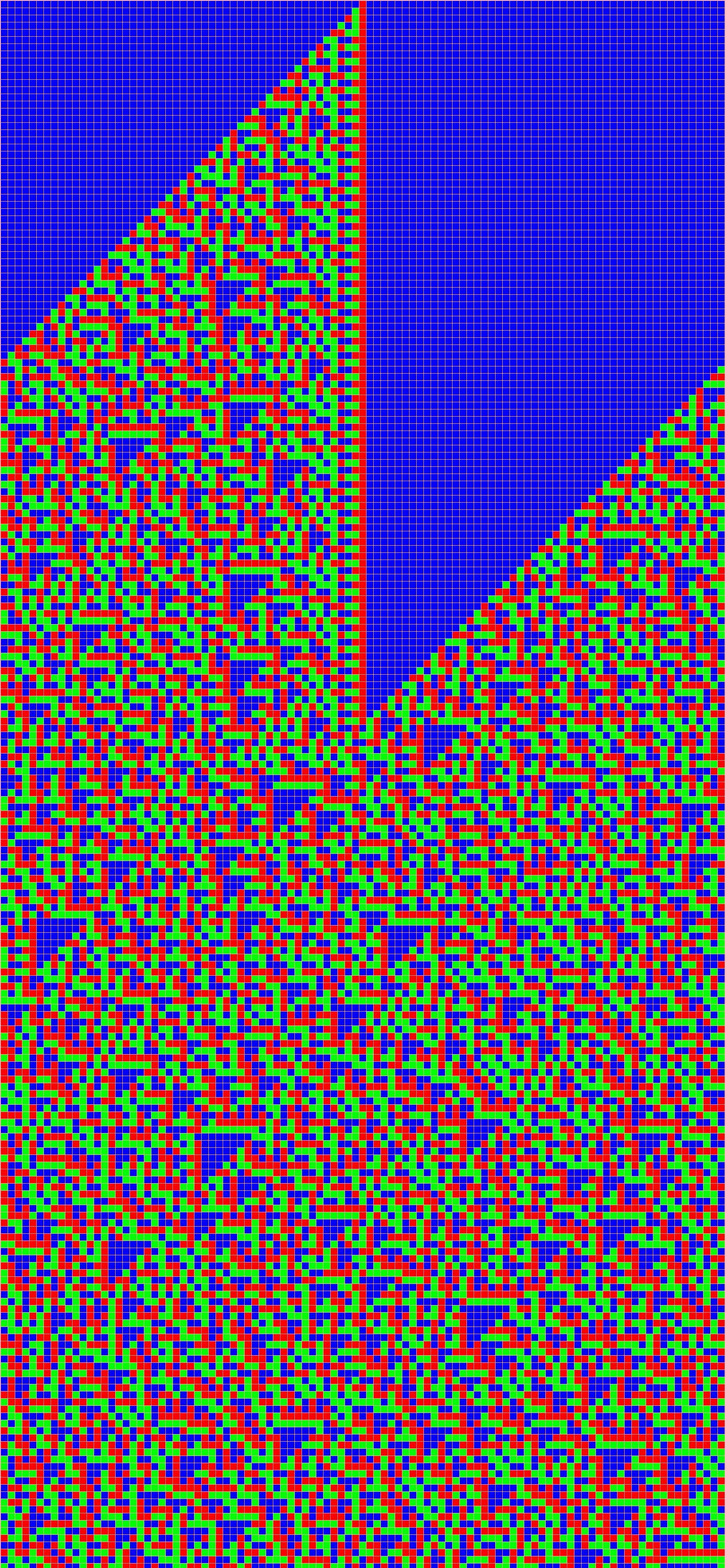}
										}
										\subfloat[\label{Chap:3-stateCA_list:fig:statespace4}]{%
											\includegraphics[width=0.25\textwidth, height = 12.0cm]{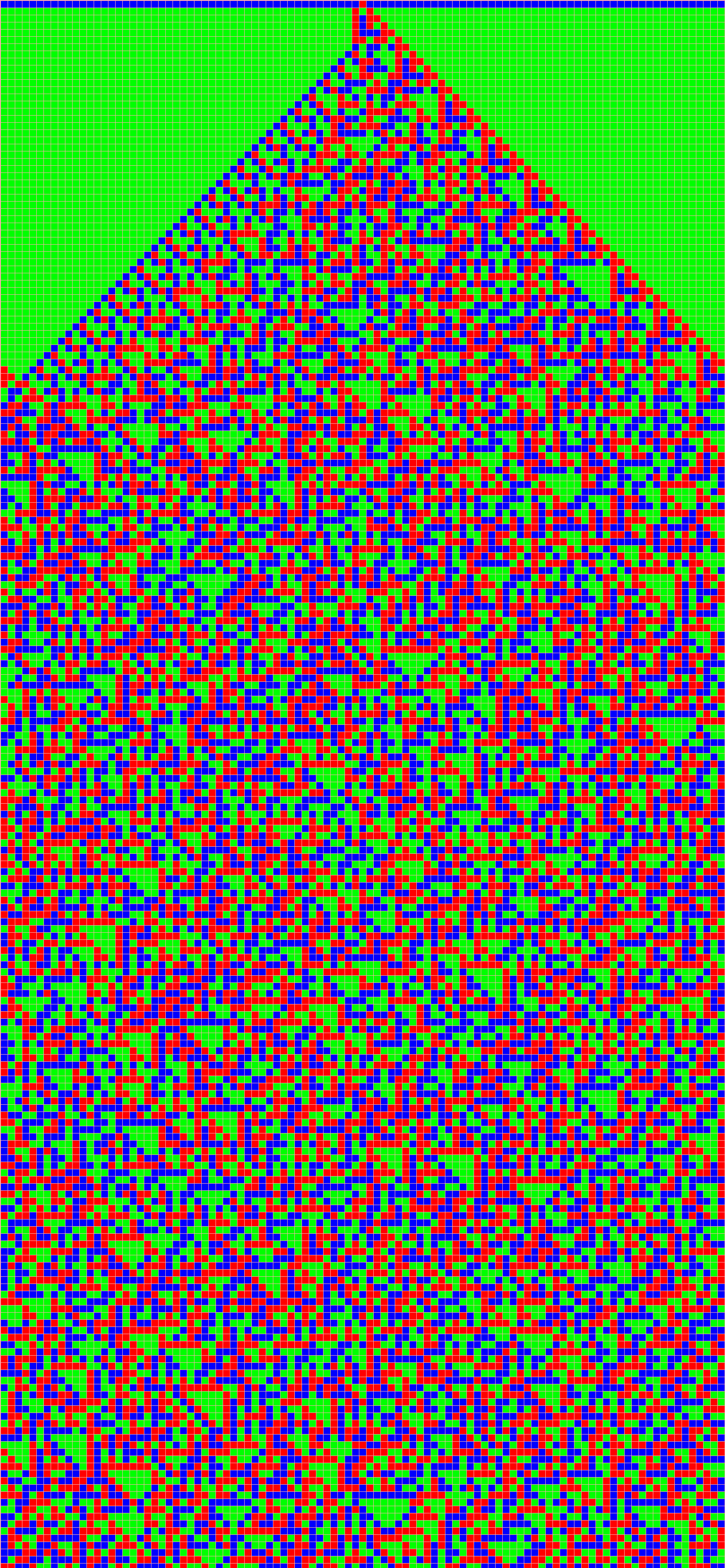}
										}
										\caption{State-space diagram for sample \emph{good} rules. Here $n = 101$ and the evolution has started from $0^{50}20^{50}$. The colors red, green and blue indicate states $2$, $1$ and $0$ respectively. The rules are $001000121110111000222222212$ (Rule $38$ of Table~\ref{Chap:3-stateCA_list:tab:ruleset_second}), $000121102212000221121212010$ (Rule $32$ of Table~\ref{Chap:3-stateCA_list:tab:ruleset1_first}), $102102210102201012102102120$ (Rule $358$ of Table~\ref{Chap:3-stateCA_list:tab:ruleset1_first}) and $110022002022110120201201211$ (Rule $425$ of Table~\ref{Chap:3-stateCA_list:tab:ruleset1_first}) for figures~\ref{Chap:3-stateCA_list:fig:statespace1}, \ref{Chap:3-stateCA_list:fig:statespace2}, \ref{Chap:3-stateCA_list:fig:statespace3} and \ref{Chap:3-stateCA_list:fig:statespace4} respectively}
										\label{Chap:3-stateCA_list:fig:state-space diagram1}
									\end{figure}

Now, for each of the CAs of the final set, space-time diagram is drawn for some fixed seeds (for example, $0^{n-1}2$, $0^{n-1}1$, $1^{n-1}2$ , $1^{n-1}0$, $2^{n-1}0$, $2^{n-1}1$ etc). For $3$-state CAs, three colors are used -- red for state $2$, green for $1$ and blue for state $0$. Some of these diagrams are shown in Figure~\ref{Chap:3-stateCA_list:fig:state-space diagram1}.

It is observed from these figures that, each rule has a dominant tendency of flowing information in at least one direction. Also, there is clear lack of pattern and apparent randomness in the state-space diagrams for these CAs which prove the prospect of these CAs as good source of randomness.  \\

Th results, reported in this section, finally strengthen our claim that the chosen list of \emph{good} CAs are as good as the proposed $3$-state CA based PRNG of Section~\ref{Chap:randomness_survey:sec:prng_R} (Page~\pageref{Chap:randomness_survey:sec:prng_R}). One may further observe that, some CAs are even better than the CA $\mathbf{\mathscr{R}} = 120021120021021120021021210$ (see Table~\ref{result1} and Table~\ref{result2} along with Table~\ref{battery} for comparison). Although, we have considered the whole configuration of the CAs as a number, but, in case of practical implementation, it is more rational to use only a small part of a configuration as the number. This scheme, if properly applied, can greatly improve the randomness quality of the generated numbers by these selected good CAs.

\section{Conclusion}\label{Chap:3-stateCA_list:sec:conclusion}
As ternary systems are good substitutes to the existing binary systems, ternary computation has a bright prospect as an alternative to the current digital computation. Hence, tri-state CA based PRNG has a promise for the future multi-valued computing circuitry. In Chapter~\ref{Chap:randomness_survey}, we have designed an example PRNG using a $3$-neighborhood $3$-state CA. Such PRNGs offer portability and more flexibility in terms of length of the number to be generated. Further, it can also be implemented using ternary logic gates. So, in this chapter, our target has been to find a list of $3$-neighborhood $3$-state CAs having similar properties. We have taken two greedy strategies to select the initial set of rules. Although, rules of these two strategies are topologically equivalent, we have studied these rules individually. Theoretical properties of these rules are then explored to filter the rules with good randomness quality. On theoretically filtered rules, empirical filtering is applied. 
Finally, we have found $596$ rules, which have performed well irrespective of seeds. 
								
All these rules are good source of randomness and can be used for several applications like simulations, randomized algorithms, circuit testing etc. In fact, each of the selected CA rules is unique and may have its own advantage in its suitable domain of application, if explored properly. This intricate task of proper usage is left to the future. However, the tri-state CAs, although performs better than most of the binary CAs, has average performance in empirical testbeds. They can not compete with the elite group of PRNGs. So, in the next chapter, we further increase the number of states to improve the randomness quality of the CA-based PRNGs.
%

\chapter{Random Number Generation using Decimal Cellular Automata}\label{Chap:10-stateCA}
\begin{center}
\begin{quote}
{\em Quand une regle est fort composée, ce qui luy est conforme, passe pour irrégulier.
(When a rule is extremely complex, that which conforms to it passes for irregular (random).)
}
\end{quote}
\hspace*{2.9in}{\em -- Gottfried Leibniz, 1686}
\end{center}

\noindent{\small This chapter illustrates the potentiality of decimal cellular automata (CAs) as pseudo-random number generators (PRNGs). Here, some additional properties for a CA to be a good PRNG are identified. Greedy strategies are utilized to select CAs satisfying these properties. Finally, two heuristic algorithms are given to synthesize such CAs. To generate a number from a configuration, we have used the concept of window. It is observed that, our PRNGs are at least as good as the best known PRNGs.}

\section{Introduction}\label{sec:Introduction}
{\large\textbf{I}}n the last two chapters (chapters~\ref{Chap:randomness_survey} and \ref{Chap:3-stateCA_list}), we have explored cellular automata (CAs) as a source of randomness. We have identified some properties of a cellular automaton (CA) to be a good pseudo-random number generator (PRNG). An example PRNG is developed using a $3$-state CA satisfying these properties, which offers many benefits like portability, ease of implementation, robustness etc. However, while comparing the randomness quality of this PRNG with other existing well-known PRNGs in terms of the performance in empirical testbeds, we have seen that, this PRNG has average ranking. In terms of randomness quality, this PRNG can not compete with the best performing PRNG \verb SFMT19937-64 ~(see Table~\ref{tab:final_rank_comparison}, Page~\pageref{tab:final_rank_comparison}). 

Nevertheless, we have also observed that, increasing number of states of a CA, randomness quality can be improved. Therefore, in this chapter, our target is to further improve randomness quality of a CA by increasing the number of states. We want to probe whether a $d$-state ($d>3$) CA based system, where global behavior is reflection of locally interactive computation, can really be a random system. One way of ascertaining this is to make a PRNG which can over-perform all existing \emph{elite} group of PRNGs. Therefore, in this chapter, we also target to develop PRNG(s) using $d$-state ($d>3$) CAs, which can beat \verb SFMT19937-64. ~

While increasing the number of states of the CA, there are several options. However, as human beings' intuitive training is to compute in decimal number system, the most natural option is $d=10$. Therefore, in this chapter, we choose $1$-dimensional $3$-neighborhood CAs with number of states per cell as $10$. In this chapter, if not otherwise mentioned, by `CAs', we will mean $1$-dimensional $3$-neighborhood $10$-state CAs.



In Section~\ref{Chap:randomness_survey:sec:propertiesOfCA} (Page~\pageref{Chap:randomness_survey:sec:propertiesOfCA}), we identify a list of necessary properties, which a CA needs to fulfill to be a good PRNG. However, as the number of possible rules for $3$-neighborhood $10$-state CAs is gigantic ($10^{10^3}$), innumerable number of rules exist which follow these properties. Hence, throughout this chapter, our goal is to reduce the rule-space to select some rules which follow these properties. To embark on this objective, first, we introduce another stricter property for a CA to be \emph{unpredictable} and hence, a candidate to be a good PRNG (Section~\ref{sec:CA_properties}). Then, in Section~\ref{sec:CA_selection}, greedy strategies are imposed to select some candidate CAs. Rules which follows these criteria are \emph{good} source of randomness and have potential as good PRNGs. Nevertheless, instead of searching exhaustively in the huge rule-space, we opt for synthesis schemes that implement these greedy strategies (Section~\ref{sec:CA_prng}). In Section~\ref{sec:CA_synthesis}, first, a heuristic algorithm is given to synthesize candidate CAs which follow these conditions. Two schemes of using these CAs as PRNGs are reported in Section~\ref{sec:CA_generator_schemes}. Similar to the $3$-state CAs based PRNGs, these PRNGs also offer several benefits like portability, unpredictability, flexibility in length of the generated numbers, etc. (Section~\ref{sec:strength_analysis}). 

However, representing these candidate CAs on pen-and-paper is an issue. In Section~\ref{sec:ca_representation}, we discover a decent way to write some of these rules which also follow the properties of Section~\ref{sec:CA_properties} and the greedy strategies of Section~\ref{sec:CA_selection}. Further, an algorithm for generating such CAs is reported. In a sense, this section is an essence of all the theories developed in the previous sections.

Finally, in Section~\ref{sec:CA_verification}, randomness quality of the proposed PRNGs are empirically verified and compared with existing well-known PRNGs. It is observed that, although the existing best PRNGs can pass all tests of NIST or TestU01 library, but none of the existing PRNGs can actually pass all tests of Diehard.
Further, we show that, performance of our proposed decimal PRNGs are at par with the existing \emph{best} PRNGs (Section~\ref{sec:comparison}). Moreover, in terms of lucidity in the schemes, our PRNGs have potentiality to be better than all their kins.

\section{Properties of Cellular Automata to be Good Source of Randomness}\label{sec:CA_properties}
Like every PRNG, CAs are also completely deterministic. By applying the rule on the initial configuration, next configurations can be deduced. However, in a good PRNG based on CAs, the chosen CAs are \emph{autoplectic} (see Section~\ref{Chap:randomness_survey:sec:unpredictability} of Page~\pageref{Chap:randomness_survey:sec:unpredictability}). During evolution, these CAs produce configurations, whose feature extraction requires more sophisticated computations than the original evolution. In such a random system based on CA, even a tiny change at a local cell should eventually spread to the whole global system. To ensure this, the candidate CAs are expected to have some necessary properties (Discussed in Section~\ref{Chap:randomness_survey:sec:propertiesOfCA} of Page~\pageref{Chap:randomness_survey:sec:propertiesOfCA}). For the sake of completeness, we first briefly recall those properties.

\subsection{Necessary Properties of a CA}\label{sec:necessary_properties} The essential properties a finite CA needs to have to act as a PRNG are listed as following. All these properties are reflection of unpredictability in the CA.
\begin{description}
\item[Balancedness:] To remove bias towards any particular state, the CA rule needs to be balanced. However, there are $ \frac{10^3!}{(10^2!)^{10} }$ number of balanced $10$-state CAs.

\item[Non-linearity:] Non-linear system is said to be more applicable to some applications like cryptography than the linear ones. Therefore, non-linearity is a desirable property for our CAs.

\item[Flow of Information:]  For a CA-based PRNG, the requirement of independence demands that the cells of the CA are dependent on their neighbors. That is, if the state of neighboring cells are changed, the cells update their states. In this case, even a tiny addition of information on any cell affects other cells and eventually propagates throughout the configuration. This situation makes the CAs unpredictable and desirable candidates as PRNGs. 

Our CAs have $3$-neighborhood dependency. So, change in state of the left neighbor (respectively right neighbor) can affect the next state of the cell. This can be estimated by observing the change of states in the sibling (resp. equivalent) RMT set. Therefore, if we consider any change in state of a cell as an information, then this information can flow in two directions -- left and right.
For any CA, the left (resp. right) directional information flow can be calculated as the cumulative sum of the change of states in each sibling (resp. equivalent) RMT set (except the self-replicating RMTs), divided by the total number of RMTs.

\item[Large Cycle Length:] In a CA-based PRNG, it is expected that, the configuration space of the CA is not partitioned into many sub-spaces resulting existence of large cycles. Also, the average cycle length of the CA has to be very large which grows exponentially with CA size $n$. 
\end{description}


However, to make the configurations of a CA more unpredictable, we further explore the structure of CAs for imposing some additional, more constricted properties. 

\subsection{Asymmetric Configuration-space}\label{sec:asymmetricConfig}
\begin{figure}
  \vspace{-1.0em}
  \centering
   \includegraphics[width=1.5in, height = 2.0in]{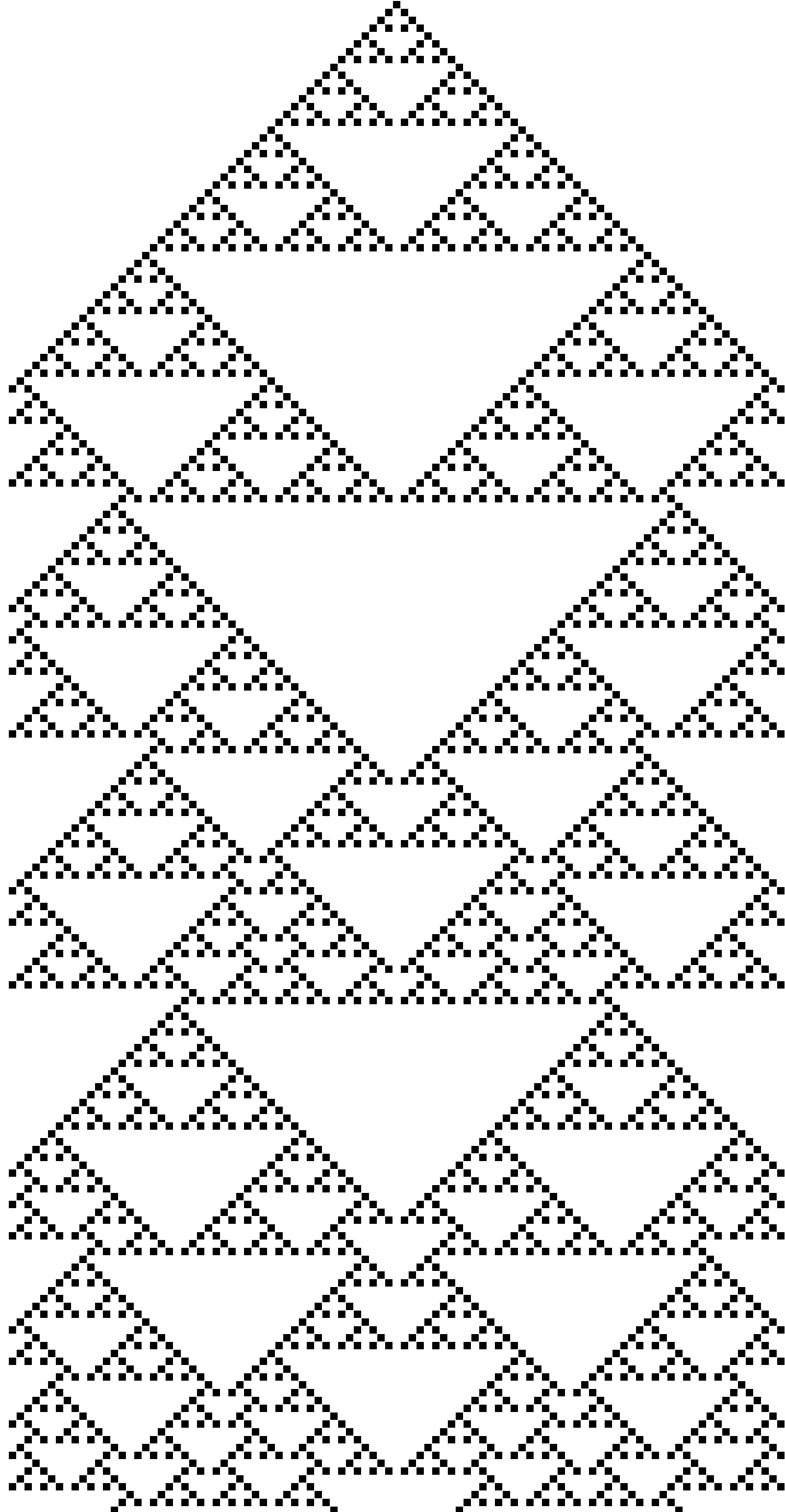}
     \caption{Space-time diagram for rule $90$ of Table~\ref{tab:ruleECA} (Page~\pageref{tab:ruleECA}). Here, number of cells = $100$ and evolution is started from $0^{49}10^{50}$. The color \emph{black} depicts state $1$}        
                   \label{fig:space-time-diagram_ECA90}
                     \vspace{-1.0em}
  \end{figure}

Many of the CAs have \emph{self-organizing} nature \cite{Wolfram:1982gu}. That is, even if the evolution starts from a completely disordered configuration, during evolution, the CAs have tendency to generate structured patterns. An important feature of this self-organizing nature is \emph{self-similarity}, also called \emph{fractal} (see Section~\ref{Chap:randomness_survey:sec:unpredictability} of Page~\pageref{Chap:randomness_survey:sec:unpredictability}). In this case, when magnified, portion of the pattern is indistinguishable from the original pattern. These self-similar patterns are symmetrical and invariant under rescaling. Another striking regularity characteristic that appears in the process of self-organization is existence of a large number of consecutive cells in a configuration with same value. During evolution, these cells can be corrupted by the flow of information from the terminal cells only. Therefore, the length of the sequence of cells with same value progressively reduces with time, producing a uniform triangular pattern. See, for example, Figure~\ref{fig:space-time-diagram_ECA90}. In this figure, a space-time-diagram for rule $90$ (Table~\ref{tab:ruleECA} of Page~\pageref{tab:ruleECA}) is shown, when the initial configuration is $0^{49}10^{50}$. 


This diagram depict the self-similarity property of rule $90$ -- this rule can generate \emph{Sierpinski triangles} in its configuration-space. We can further observe that, for rule $90$, information flow in both directions are equal. So, whenever, a configuration has a sequence of consecutive cells with state $0$, that is, a sub-configuration $0^l$, $l \in \mathbb{N}$, this sub-configuration is altered uniformly from both sides in the successive time steps, generating an equilateral triangular structure. Such triangles are symmetrical to its base and can be of any sizes. However, for a good PRNG, existence of any pattern in the space-time diagram is not desirable.
Therefore, our target is to select the CAs such that, these self-organizing symmetrical regular patterns are (as far as possible) absent in the corresponding space-time diagrams for the configuration-spaces.



\subsubsection{Asymmetric Information Flow}\label{sec:asymmetricFlow}
Having information flow in the CA is essential for a CA to be used as a PRNG. However, if information flow in both directions are symmetrical, then, for a sequence of cells in a configuration with same value, there is a tendency of generating equilateral triangles in the configuration-space for that sub-configuration. See, for instance, Figure~\ref{fig:space-time-diagram_ECA90}. In this figure, every triangle is equilateral and symmetrical. Such symmetrical patterns are undesirable for designing PRNGs. Therefore, to break this symmetry in the triangular patterns, it is desirable that our CAs have asymmetrical information flow in both directions.

\subsubsection{Non-determinism in Equivalent and Sibling RMT sets}\label{sec:asymmetricEquiSiblSet}
Equivalent and sibling RMTs are the building blocks of any rule. If for any $i$, all RMTs of $Equi_i$ (resp. $Sibl_i$) have same next-state values, then this implies, there is less dependency of the rule on its left (resp. right) neighbor. In this case, change of value(s) of left (resp. right) neighbor(s) may not immediately create change in the next-configuration. Because of its symmetrical nature, this assignment of values is also predictable and disagreeable for a good PRNG. Hence, to avoid this unpleasant situation, it is expected that, in the candidate CAs, none of the equivalent and sibling RMT sets have all RMTs with same next state value. This restriction in the equivalent (resp. sibling) RMT set makes the assignment unpredictable for an observer and creates an impression of non-determinism in the CA.

\subsubsection{No large sub-sequence of self-replicating RMTs or RMTs with same next state value}\label{sec:noLargeSubSequence}
If in a CA, a sub-configuration of say, length $l$, exists consisting of only self-replicating RMTs, then for this sub-configuration, the next sub-configuration is itself. During the evolution of the CA, this phenomenon continues to repeat. Because, in each of the next time steps, depending on the information flow, at most the two terminal RMTs of the sub-configuration can be changed. Other RMTs being still self-replicating, the situation continues for at least for the next $\frac{l}{2}$ time steps. In the space-time diagram, this corresponds to a triangle-like pattern.

Similarly, if there exists a sub-configuration of length $k$ in which all RMTs have same next-state value $s\in \mathcal{S}$, then also, in the next time-step, the corresponding sub-configuration has $k$ number of consequtive cells with state $s$, that is a sub-configuration $s^{k}$ of trivial RMTs. Hence, in the next time-step, all RMTs (except at most the terminal RMTs) of the sub-configuration are same and this situation repeats for at least next $\frac{k}{2}$ time-steps. Therefore, in the space-time diagram, again, a triangle-like pattern is generated.

Both of these situations are undesirable for a good PRNG. Therefore, for the candidate CAs, it is expected that, in the configuration-space of the CAs, there is no large sub-sequence of self-replicating RMTs or RMTs with same next state value.\\

\noindent The CAs which satisfy the above properties are our desirable candidates as PRNGs. All these CAs are good source of randomness. However, huge number of such rules exist with similar properties. So, in the next section, we have applied some greedy methods to select some potential CAs which satisfy these properties.

\section{Selection of CA rules}\label{sec:CA_selection}
From the previous section, we find that, to be a good PRNG, the CA rules need to be balanced where the cells are dependent on their neighbors, that is, have a high rate of information flow. This information flow also needs to be asymmetric in directions and there should not be any long chain of self-replicating RMTs or RMTs with same next state value in the configuration-space. The CAs also need to have large cycle lengths that grow exponentially with CA size. However, as the rule-space is gigantic ($10^{10^3}$), it is impossible to test and find all rules which follow these properties. Hence, likewise previous chapter (see Section~\ref{Chap:3-stateCA_list:sec:CAPRNG} of Page~\pageref{Chap:3-stateCA_list:sec:CAPRNG}), here also, we use greedy strategies satisfying these properties to reduce the rule-space.

\subsection{Maximum Information Flow on One Side and Asymmetric Information Flow on the Other}\label{sec:greedy_strategy}
Recall that, information flow in the CA can be measured by calculating the total change of next state values in the sibling or equivalent RMT sets with respect to the maximum possible change of next state values in these RMT sets. Different next state values in the equivalent RMTs (resp. sibling RMTs) implies information flow on right (resp. left) side. So, our greedy strategy is to choose the CA rules such that there is maximum information flow on either direction. This can be ensured by setting RMTs of $Equi_i$ (for right side) or $Sibl_i$ (for left side) different next state values. Recall that, such CAs follow STRATEGY I and STRATEGY II respectively (see Section~\ref{chap:reversibility:Sec:identify} of Chapter~\ref{Chap:reversibility} for details). Here, we reproduce these strategies for the sake of completeness.

\noindent \textbf{STRATEGY I:} \label{stg1_10}\textit{Pick up the balanced rules in which equivalent RMTs have different next state values, that is, no two RMTs of $Equi_i ~(0 \leq i \leq 99)$ have same next state value.}

\noindent\textbf{STRATEGY II:} \label{stg2_10} \textit{Pick up the balanced rules in which the RMTs of a sibling set have the different next state values, that is, no two RMTs of $Sibl_i ~(0 \leq i \leq 99)$ have same next state value.}

Each of these strategies fortifies balancedness property and are left-permutive (STRATEGY I) or, right-permutive (STRATEGY II) (see Definition~\ref{Def:permutivity} of Page~\pageref{Def:permutivity}). However, the rules from each strategy are topologically equivalent of rules from the another strategy (see Section~\ref{Chap:3-stateCA_list:sec:equivalence} of Page~\pageref{Chap:3-stateCA_list:sec:equivalence}). Hence, selecting rules from any one strategy may suffice. In this chapter, we have chosen \emph{STRATEGY II} rules as the first greedy approach. There are ${(10!)}^{10^2}$ balanced rules that follow this strategy. 

%


Now, to ensure asymmetric information flow, we select those CAs from this strategy for which there is non-maximum information flow on right side. Recall that, information flow on right side is dependent on equivalent RMTs. So, to satisfy the property of Section~\ref{sec:asymmetricFlow}, the CAs need to have equivalent RMT sets where not all RMTs of the set have different next state values. Further, to maintain the property of Section~\ref{sec:asymmetricEquiSiblSet}, all equivalent RMT sets should not have all RMTs with same next state value.

These rules satisfy most of the properties of Sections~\ref{sec:necessary_properties} and of Section~\ref{sec:asymmetricFlow} and \ref{sec:asymmetricEquiSiblSet}. However, some of these rules can still be of self-organizing nature; to filter out such rules, we apply the following criterion.
\subsection{Asymmetric Configuration Space}
Although rules satisfying the previous greedy strategy have less probability of generating equilateral triangles in their space-time diagrams, but they may not be completely devoid of regular structures or triangle-like patterns. Further, whether a CA is self-organizing or not can not be predicted beforehand from its rule without observing the space-time diagram of its configuration-space. Nevertheless, if the CA satisfies the property of Section~\ref{sec:noLargeSubSequence}, then its configuration-space has lack of regular structures and patterns. 

However, it is difficult to ascertain that a CA has no large sub-sequence of self-replicating RMTs or RMTs with same next state value in its evolution. Because, firstly, the initial configuration is dependent on user input, hence may not be known in advance. Secondly, even if the initial configuration is known, whether an arbitrary configuration, where consecutive cells have same value or the corresponding RMTs are self-replicating, can be reached from the seed is a PSPACE-complete problem \cite{SUTNER199587}. For a CA of size $n$, there are $10^n$ configurations. Hence, it is also not feasible to individually check reachability of every configuration and its adherence to the property of Section~\ref{sec:noLargeSubSequence}. In this scenario, we again use \emph{primary RMT sets} (see Section~\ref{Chap:semireversible:sec:prim} of Page~\pageref{Chap:semireversible:sec:prim}) which assists us to solve this problem. By informed usage of this tool, we can ensure that, for any RMT sequence in the CA, there is no large sub-sequence of RMTs with same next-state value or of self-replicating RMTs.

\subsubsection{Primary RMT sets and Homogeneous configurations}\label{sec:prim}
Here, we briefly recollect the primary RMT sets for $3$-neighborhood $10$-state CAs. Recall that, if RMT $r$ is part of any RMT sequence, then the next RMT in the sequence must be from $Sibl_k = \{10r \pmod{1000}, 10r+1 \pmod{1000}, \cdots, 10r+9 \pmod{1000}\}$, where $r \equiv k \pmod {100}$. These RMT sequences correspond to some cycles in the de Bruijn graph $B(2,S)$ of a CA. So, the elementary cycles in the graph represent the smallest possible RMT-sequences. The RMT sets corresponding to the RMTs of each elementary cycle is named as \emph{primary RMT sets}


%
%
%

Since de Bruijn graph is a Hamiltonian graph, maximum possible cardinality of any primary RMT set $P$ is $100$, that is, the number of nodes in the graph. However, minimum cardinality is $1$. There are $10$ singleton primary RMT sets $\{0\}$, $\{111\}$, $\{222\}$, $\{333\}$, $\{444\}$, $\{555\}$, $\{666\}$, $\{777\}$, $\{888\}$ and $\{999\}$. Similarly, for each $l \in \{1, 2, \cdots, 100\}$, there exist one or more primary RMT sets with cardinality $l$. For instance, for a $10$-state CA, the number of primary RMT sets of cardinality $2$, $3$ and $4$ are $45$, $90$ and $648$ respectively.


A configuration, RMT sequence of which is formed with the RMTs of a primary RMT set only, is called  \textit{homogeneous configuration}. Length of a homogeneous configuration depends on the size of the CA. However, we represent a homogeneous configuration in terms of its minimal length. For example, $9^n$, where $n \in \mathbb{N}$, is the homogeneous configuration formed by RMT $999$ corresponding to the primary RMT set $\{999\}$.

For any configuration $x$ and a set of primary RMT sets $X$, if $x$ uses all the RMTs of each primary RMT set $P\in X$, it is said to be \emph{followed} from $X$. We describe this by $X \vdash x$.
%
%
For instance, if $X=\{\{202,020\}$,$\{206,063,639,397,971,715,152,52\}\}$, then $X \vdash 063971520202$, where $063971520202$ is the configuration corresponding to the RMT sequence $\langle 206,063,639,397,971,715,152,520,202,020,202,020 \rangle$. If $X$ is a singleton, any $x$, where $X\vdash x$, is a homogeneous configuration.

As any edge of de Bruijn graph can be shared by more than one elementary cycle, an RMT may be part of more than one primary RMT set. But, no two RMTs of $Sibl_i$, $0\leq i\leq 100$, can belong to the same primary RMT set.
 

\subsubsection{Asymmetric distribution of next state values for RMTs of Primary RMT sets} \label{sec:uniforPrim}
Primary RMT sets are building blocks of any RMT sequence. Every RMT sequence is formed with RMTs from one or more primary RMT sets. Therefore, imposing restriction on the next state values of RMTs of each primary RMT set invariably means constraint on the RMT sequences. By assigning asymmetric distribution of next state values for RMTs of Primary RMT sets, we can, in a way, restrict symmetry in the configuration space and uphold property of Section~\ref{sec:noLargeSubSequence}.


Our target is to assign next state values to the RMTs of each primary RMT set such that, in the RMT sequences generated from the RMTs of primary RMT sets, there is no large sub-sequence of RMTs with same next-state value or of self-replicating RMTs. Further, presence of fixed-point attractor in the CA means the CA has a tendency to converge to the fixed-point attractor. For such CAs, the configuration space also gets partitioned, resulting smaller cycles. So, we constrain our CAs to have no non-trivial fixed-point attractor. As we allow appearance of trivial fixed-point attractors (if exists) in the CAs, we need to make certain that, the trivial configurations are non-reachable from every non-trivial configuration. Hence, the CAs can not converge to the trivial fixed-point attractors.

However, to detect the fixed-point attractors as well as reachable trivial configurations of a CA, we can take help of the de Bruijn graph corresponding to the CA. Recall that, if the de Bruijn graph is modified to contain only the self-replicating RMTs, then any cycle in this graph represents a fixed-point attractor in the CA. Therefore, using this graph, all possible fixed-point attractors in a CA can be detected (see Section~\ref{Chap:3-stateCA_list:sec:fpg} of Page~\pageref{Chap:3-stateCA_list:sec:fpg} for more details).  
Similarly, for each $s \in S$, if the de Bruijn graph is altered to contain only the edges with next state value $s$, then a cycle in this graph represents a homogeneous configuration whose next configuration is $s^n$, a trivial configuration. If this graph has any cycle corresponding to a non-trivial configuration, then that implies, the trivial configuration is reachable from a non-trivial configuration (Section~\ref{Chap:3-stateCA_list:sec:approach2} of Page~\pageref{Chap:3-stateCA_list:sec:approach2} reports further illustration).

We already know that each primary RMT set represents the RMTs of an elementary cycle of de Bruijn graph. Therefore, by assigning next state values to the RMTs of primary RMT sets, we can ensure that, there is no non-trivial homogeneous configuration which is a fixed-point attractor, or, for which the next configuration is a trivial one ($s^n$, $s\in \mathcal{S}$, $n \in \mathbb{N}$). All these restraints can be done by allocating the next-state values of the RMTs of primary RMT sets $P_1, P_2, \cdots P_k$ in the following way:
\begin{enumerate}
\item All RMTs of any primary RMT set $\{P_i\}$ having multiple elements do not have same next-state value. Hence, there is no cycle in the de Bruijn graph for the CA representing a non-trivial configuration from which a trivial configuration is reachable.

\item All RMTs of a primary RMT set (having multiple elements) can not be self-replicating. Hence, no cycle in the de bruijn graph exists depicting a non-trivial fixed-point attractor in the CA.

\item If $\{P\}\vdash x$, then number of consecutive RMTs in $\tilde{x}$
with same next-state value is restricted to a limit $l$. Hence, the length of the sub-configurations having same next state value or self-replicating RMTs is limited to $l$. This $l$ is desirable to be very small to establish asymmetric configuration space and reduce possibility of triangles.
\end{enumerate}

This strategy, if applied along with the greedy strategy of Section~\ref{sec:greedy_strategy}, can remove appearances of self-similar regularity and triangle-like patterns in the space-time diagram for most of the seeds. These CAs have no non-trivial fixed-point attractor and all trivial configurations are isolated from the non-trivial configurations. Hence, it is expected that, for the CAs, the configuration-space is less partitioned, resulting small number of cycles and have large cycle lengths. Also, the CAs do not have any inclination towards converging to a configuration. Therefore, these CAs satisfy all the properties of Section~\ref{sec:CA_properties} and are good candidates to be PRNGs.

\section{Cellular Automata as Decimal PRNGs}\label{sec:CA_prng}
In this section, we implement all the aforementioned greedy strategies of Section~\ref{sec:CA_selection} to synthesize CA rules which are good candidates to be PRNGs. 
We use window-based portable scheme (likewise Section~\ref{Chap:randomness_survey:sec:prng_R} of Page~\pageref{Chap:randomness_survey:sec:prng_R}) to generate numbers from the configurations of the CAs. The strengths of our PRNGs are explored in details. Finally, some example rules are shown. 

\subsection{Synthesis of CA}\label{sec:CA_synthesis}
The greedy strategies of the previous section suggest to find the CAs for which the following conditions are satisfied --
\begin{enumerate}
\item \label{cond_1} Each of the sibling RMT sets are balanced, indicating maximum information flow on left side.

\item \label{cond_2} Neither all of the equivalent RMT sets are balanced, nor they have all RMTs with same next state value (asymmetric information flow on right side and non-determinism in equivalent RMT sets).

\item \label{cond_3} None of the primary RMT sets of the CA has all RMTs with same next state value (reachable trivial configuration(s)). Neither are all RMTs of a primary RMT set self-replicating (no non-trivial fixed-point attractor(s)).

\item \label{cond_4} The RMTs of each primary RMT set are assigned values such that number of consecutive RMTs in an RMT sequence with same next state value is less than $l$ (asymmetric configuration-space).
\end{enumerate}

Depending on each values of $l$, we can find many CAs which satisfy these properties among the possible ${(10!)}^{100}$ rule-space. Each of these CAs can be candidate to be a good PRNG. However, because of the gigantic nature of the rule-space, we can not extensively investigate each CA. Hence, we concentrate on heuristically synthesizing some CAs which follow these properties. The synthesis scheme given here is only one of the many possible kinds. One may choose any other scheme to synthesize CAs which follow the aforementioned properties to find good PRNGs.

In this heuristic scheme, we randomly assign values to the RMTs of primary RMT sets of increasing cardinality such that the generated CAs follow the greedy strategy of Section~\ref{sec:greedy_strategy}. The values are allotted in accordance with the conditions \ref{cond_3} and \ref{cond_4}. To avoid backtracking, we stop when all RMTs are assigned values. Here, we have considered $l$ as small as $3$. The step wise synthesis process is given below as following.
\begin{enumerate}
\item[Step 1:]\label{step_1} Randomly assign values to the singleton primary RMT sets $\{0\}$, $\{111\}$, $\{222\}$, $\{333\}$, $\{444\}$, $\{555\}$, $\{666\}$, $\{777\}$, $\{888\}$ and $\{999\}$.

\item[Step 2:]\label{step_2} Set $i=2$. We intend to assign RMTs to the primary RMT sets of cardinality $i$ in the next step.

\item[Step 3:]\label{step_3} Select the primary RMT sets of cardinality $i$. For each set, not all RMTs of the set are to be assigned the same value or can be self-replicating. This assignment to each RMT $r$ is chosen randomly from the possible values while satisfying the conditions \ref{cond_1}, \ref{cond_2}, \ref{cond_3} and \ref{cond_4}. That means, for each $r$, the corresponding sibling RMT set ($Sibl_{\floor{\frac{r}{10}}}$) and equivalent RMT set ($Equi_{k}$, where $r \equiv k \pmod{100}$) are checked to find the possible values that are in accordance with our greedy strategy of Section~\ref{sec:greedy_strategy}. Now, these possible values are further scanned to find whether any RMT sequence can be formed with this RMT where number of consecutive RMTs with same next state value $\geq 3$. If any such value is found, it can not be chosen as a possible assignment. Among the remaining values, we randomly choose an assignment to RMT $r$.

However, if all possible assignments result in an RMT sequence with number of consecutive RMTs with same next state value $\geq 3$, we proceed with the optimal assignment. Here, by optimal assignment, we mean that assignment for which the number of consecutive RMTs with same next state value in the RMT sequences is least. 

To avoid backtracking, we always assign values to unassigned RMTs. That is, once an RMT is allocated a value, it is never reset. 

\item[Step 4:]\label{step_4} If all $1000$ RMTs of the rule are assigned values, go to Step $6$.

\item[Step 5:]\label{step_5} When all RMTs of primary RMT sets of cardinality $i$ are filled, increment $i$ and go to Step $3$.

\item[Step 6:]\label{step_6} As all RMTs are assigned values, we stop further processing and get a rule that preserves the properties of Section~\ref{sec:greedy_strategy}. However, primary RMT sets of all cardinalities up to size $100$ are not needed to be scanned to fill these $1000$ RMTs. In fact, assigning RMTs of primary RMT sets up to cardinality $4$ fills up all the $1000$ RMTs. Hence, the generated rule may still have primary RMT sets whose every RMT has same next state value or every RMT is self-replicating. For these rules some trivial configurations can be reachable from a non-trivial one. There may also be non-trivial fixed-point attractor(s) in the CA for some CA sizes. If it happens, we discard the rule and repeat the process from Step $1$ to synthesize another.

To inspect whether the generated rule is devoid of any non-trivial fixed-point attractor and all trivial configurations are isolated from the non-trivial configurations, we take help of the de Bruijn graph. For fixed-point attractors, we check that, whether the de Bruijn graph with only the edges having self-replicating RMTs have any cycle of length $\ge 2$. Similarly, for reachability of each trivial configuration $s^n$, $s \in \mathcal{S}$, also, we use the de Bruijn graph. We examine that the graph having only the edges corresponding to next state value $s$ has any multi-length cycle. 

Therefore, each of the generated rules are tested using de Bruijn graph to verify whether the CA does not have any non-trivial fixed-point attractor for any $n$ and whether all trivial configurations are isolated. If the randomly synthesized rule passes this test, it is to be used as a good PRNG. 


\end{enumerate}

\begin{Walgo}[!h]{1.5cm}
	\scriptsize
	\SetKw{Fn}{Procedure}
	\SetKwFunction{setRMTs}{setRMTs}
	\SetKwFunction{randPermutation}{randPermutation}
	\SetKwFunction{verifyRule}{verifyRule}
	\SetKwInOut{Input}{Input}
	\SetKwInOut{Output}{Output}

	\Input{$ruleCount$ (Number of rules to be generated)}
	\Output{$ruleCount$ number of Candidate CAs}
	
	\rule[4pt]{0.95\textwidth}{0.95pt}\\
	\hspace{0.04\textwidth} \nlset{Step 1} \label{st1_syn} Initialize $d\leftarrow10$, $noRMTs \leftarrow d^3$, $rC\leftarrow0$, $srand(time(NULL))$ \;
	\hspace{0.04\textwidth}	\nlset{Step 2}\label{st2_syn} 
	\While{$rC<ruleCount$}{
		\lFor {$i = 0$ to $noRMTs-1$}{ 
			Initialize $rule[i]\leftarrow-1$
		}
		\tcc{For Primary RMT sets of cardinality $1$}
		\lFor {$i = 0$ to $d-1$}{ 
			Set $rule[i\times d^2+i\times d +i]\leftarrow rand()\%d$
		}
		\tcc{For Primary RMT Sets of cardinality $2$: $\{iji,jij\}$}
		\For {$i= 0$ to $d-2$}{ 
			\For {$j=i+1$ to $d-1$}{
				$rmtArray.clear()$; \tcp{Clear contents of $rmtArray$}
				$rmtArray.add(i\times d^2+j\times d + i, j\times d^2+i\times d + j)$; \tcp{Add RMT $iji$ and  $jij$ to $rmtArray$}
				\setRMTs{$rmtArray$}; \tcp{Assign values to the RMTs}
			}
		}
		\tcc{For Primary RMT Sets of cardinality $3$: $\{kij, ijk,jki\}$}
		Set $k=0$\;
		\For {$i= 0$ to $d-1$}{ 
			\For {$j=1$ to $d-1$}{
				$rmtArray.clear()$\;
				$rmtArray.add(k\times d^2+i\times d + j, i\times d^2+j\times d + k, j\times d^2+k\times d + i)$\;
				\setRMTs{$rmtArray$}\;
			}
		}
		\tcc{For Primary RMT Sets of cardinality $4$: $\{lij, ijk, jkl, kli\}$}
		\For {$l= 0$ to $d-1$}{ 
			\For {$i=1$ to $d-1$}{
				\For {$j= 0$ to $d-1$}{ 
					\For {$k=1$ to $d-1$}{
						\lIf{$(l=i=j)$ OR $(i=j=k)$ OR $(j=k=l)$ OR $(k=l=i)$}{
							$continue$
						}
						\Else{
							$rmtArray.clear()$\;
							$rmtArray.add(l\times d^2+i\times d + j, i\times d^2+j\times d + k, j\times d^2+k\times d + l, k\times d^2+l\times d + i)$\;
							\lIf{($rmtArray.size()<4$) OR ($\forall p \in rmtArray$, $rule[p]!=-1$)}{
								$continue$
							}
							\lElse{
								\setRMTs{$rmtArray$}
							}
						}
						
					}
				}
			}
		}
		
		\tcp{verify non rechability of trivial configurations and no non-trivial fixed-point attractor}
		\lIf{\verifyRule{rule}=$false$}{
			$continue$
		}
		\tcp{Store each candidate CA as potential PRNG}
		\lFor {$i= noRMTs-1$ to $0$}{ 
			Print $rule[i]$ 
		}
	}
	\caption{\emph{DecimalCASynthesis}}
	\label{algo:CA_synthesis}
\end{Walgo}
A pseudo-code for this heuristic process to generate candidate decimal CAs as PRNGs is shown in Algorithm~\ref{algo:CA_synthesis}. This algorithm (\emph{DecimalCASynthesis}) generates random rules that satisfy the properties of Section~\ref{sec:CA_properties} following greedy strategies of Section~\ref{sec:CA_selection}. Here, primary RMT sets up to cardinality $4$ are generated and the RMTs of each set are assigned values by the procedure \emph{setRMTs}. The algorithm uses an array \emph{rmtArray} which stores the RMTs of each primary RMT set. This \emph{rmtArray} is actually a set, so, does not allow duplicates. As input, it (Algorithm~\ref{algo:CA_synthesis}) takes the number of rules to be generated and outputs the candidate rules. It also uses the following procedures-- \begin{description}
	\item[\textbf{\emph{setRMTs()}}] This procedure assigns values to the unassigned RMTs of each primary RMT set as mentioned in Step $3$. 

	\item[\textbf{\emph{verifyRule()}}] It returns $true$, if there is no non-trivial fixed-point attractor in the de Bruijn graph of the CA and the trivial configurations are non-reachable from any non-trivial configurations. Otherwise, it returns $false$.
\end{description}

Using Algorithm~\ref{algo:CA_synthesis}, any number of CAs can be generated based on the greedy strategies of Section~\ref{sec:CA_selection} that fulfill the properties of Section~\ref{sec:CA_properties}. All these CAs are candidates to be good PRNGs. In the next section, the scheme of using these CAs as PRNGs is described.

\subsection{Scheme of PRNG}\label{sec:CA_generator_schemes}
Like Section~\ref{Chap:randomness_survey:sec:prng_R} (Page~\pageref{Chap:randomness_survey:sec:prng_R}), here also, we use a window of length $w<n$ from the cells of size $n$, where the numbers are generated from the window. The window size $w$ is a variable; an user may choose its size as per requirement. However, in the design of the PRNG, the cell length $n$ is fixed to a value large enough for that $w$. The value of the window works as the seed of the PRNG. 

Although for any seed, one can collect the numbers from the initial configuration onwards, but, to impose more independence on seed, we leave the first $n$ configurations and let the flow of information to spread all cells of the CA.
Therefore, for each seed, the output numbers of the PRNGs are extracted from $n+1st$ configuration onwards. The seed for the window is taken as a user-input. But to assure that the period of the PRNG is extremely large, other $n-w$ cells of the initial configuration are fixed to a non-homogeneous configuration. In this scheme, we choose $0^{n-w-1}1$ as the fixed seed for the $n-w$ cells. One may choose any other non-homogeneous configuration as this seed as long as the resulting cycle length is very large. 

While designing PRNGs, two schemes are contemplated. Firstly, the user may need to generate decimal numbers of $w$ digits, where $w$ can be any variable. This requirement is often observed as an usage of portable PRNGs, mostly LCGs, like \verb rand, ~\verb drand48 ~etc. of UNIX Systems, where user has to use modular operation. However, as configurations of our CAs are decimal strings, unlike these LCGs, user can directly generate numbers of required digits from the PRNG depending on the variable window size. Moreover, there may be requirement of generating binary numbers of some specific bits, mostly multiple of computer word size. Many of the well-known PRNGs commonly in use, dwell on this scheme. In this case also, we can use our CAs to generate binary numbers as per requirement. Therefore, in this chapter, we present the following two schemes for generating numbers --\\

$(1)$ For decimal numbers of $w$ digits,

$(2)$ For binary numbers of $32\times b$ bits, where $b$ is a variable.
%
\subsubsection{Generation of $w$-digit Decimal numbers} To generate random numbers of $w$ digits, the window size is taken as $w$. For example, if an user wants to generate random numbers in the range $0$ to $999$, the input window size is to be given as $3$. Depending on this input variable window size $w$, the cell size $n$ also varies in the design. However, for each specific $w$, $n$ is a fixed value. We recommend taking this $n$ as at least $10$ times larger than $w$ and $\geq 100$. 
A sample code for implementing this scheme using \emph{JAVA} is given as below. In this scheme, we specify the CA size $n$ as follows-- for window $w< 10\times i$, the CA size is fixed as $n=100\times i+1$. For instance, if $w=3$, $n=101$.

 
{\small
\begin{lstlisting}[language=JAVA, caption={JAVA code for Generating Decimal Random Number}, linewidth=14.5cm]
int PC[],NC[],seedDeci[];	// to store configurations and seed
int[] Rule; 			//The rule to be used as PRNG
void srandCA(int window, String seed){ //Takes size of output number in digits as window, a decimal string as seed
 n=(int) ((window/10+1)*100+1);   //set CA size
 seedDeci=splitDigits(seed, window); //To array of digits
 for(i=0;i<window;i++)           //initialize window	
 	PC[i]=seedDeci[i];
 for(i=window;i<n-1;i++) //initialize other values to 0...01
 	PC[i]=0;
 PC[n-1]=1;
 for(j=0;j<n;j++){ //leave first n configurations
 	for(i=0;i<n;i++)
 	  NC[i] = Rule[100*PC[(i-1+n)%n]+10*PC[i]+PC[(i+1+n)%n]];
 	for(i=0;i<n;i++)
 	  PC[i] = NC[i];      //update PC[] to use again
 	}
 }
String randCA(){   /**Generates psudo-random number**/
 for(i=0;i<n;i++)
	NC[i] = Rule[100*PC[(i-1+n)%n]+10*PC[i]+PC[(i+1+n)%n]];
 for(i=0;i<n;i++)
	PC[i] = NC[i];    //update PC[] to use again
 return extractNumber(PC,window); //return decimal number corresponding to the window
}
 	\end{lstlisting}\label{listing1}}
 
The method \emph{randCA()} generates the random number as per the requirement of the user. It uses the present configuration of CA as set by the method \emph{srandCA()}. This method \emph{srandCA()} takes the size of window as the number of digits of output number, the seed for window as a decimal string from the user. The output number is extracted from the current configuration using the method \emph{extractNumber()}. Therefore, using this code, random decimal numbers of any range can be generated.

\subsubsection{Generation of $32\times b$-bit Binary numbers}\label{sec:32-bitPRNG} In this scheme, decimal strings from the configurations of our CAs are used to generate random binary numbers of size as multiples of word size of a $32$-bit computer. To generate $32\times b$ bit numbers, we take window size as $w=14\times b$. CA size $n$ is considered as $100\times b+1$. That is, $b$ is a variable taken as user input. Depending on this value, the size of the generated numbers and number of cells of the CA are fixed inside the design. For example, CA sizes for generating $32$-bit, $64$-bit and $128$-bit numbers are $101$, $201$ and $401$ respectively, whereas user input $b$ is to be given as $1$, $2$ or $4$ for window size $14$, $28$ or $56$ respectively.

Here, for improving randomness of the PRNGs, we use two level extraction of values. In the first level, values of the window $w=14\times b$ cells are educed from the $n$-cell configurations. Then, in the second level, we use \emph{modular} operation with $2^{32 \times b}$, like the LCGs, to generate the $32\times b$ bit number. For example, we can generate $32$-bit numbers by using modular $2^{32}$ operation on the $14$ digit decimal number extracted from the window. This implies, the output is the binary number corresponding to the decimal number of the window transformed under modular operation. By this method, we can generate all numbers within the range $0$ to $2^{32 \times b}-1$. A sample code snippet for implementation is \emph{JAVA} is shown as follows.
 
 {\small
 \begin{lstlisting}[language=JAVA, caption={JAVA code for Window-based PRNG}, linewidth=14.5cm]
int PC[],NC[],seedTri[];   //To store configurations and seed
int[] Rule; 			//The rule to be used as PRNG
void srandCABin(int b, String seed){ //Takes size of output number as 32*b bits, a decimal string as seed
 window = (14*b);          //set window size
 n=(int) (100*b+1);        //set CA size
 seedDeci=splitDigits(seed, window);    //To array of digits
 for(i=0;i<window;i++)        //initialize window	
	PC[i]=seedDeci[i];
 for(i=window;i<n-1;i++)    //initialize other values to 0...01
	PC[i]=0; 
 PC[n-1]=1;
 for(j=0;j<n;j++){ //leave first n configurations
	for(i=0;i<n;i++)
	  NC[i] = Rule[100*PC[(i-1+n)%n]+10*PC[i]+PC[(i+1+n)%n]];
	for(i=0;i<n;i++)
	  PC[i] = NC[i];      //update PC[] to use again
 }
}
String randCABin(){     /**Generates psudo-random number**/
 for(i=0;i<n;i++)
 	NC[i] = Rule[100*PC[(i-1+n)%n]+10*PC[i]+PC[(i+1+n)%n]];
 for(i=0;i<n;i++)
   	PC[i] = NC[i];      //update PC[] to use again
 Number=extractNumber(PC,window); //extract decimal number for window
 return (Number % Math.pow(2,32*b)); //return the output number
}
\end{lstlisting}}

The method \emph{randCABin()} generates the $32\times b$ binary number as per the requirement of the user. For initialization, it uses the method \emph{srandCABin()}. This method takes $b$ and the seed for window as a decimal string from the user. The other cells are initialized with $0^{n-w-1}1$. It returns a $32\times b$ bit number morphed by using modular operation as mentioned above. Hence, using this scheme, binary numbers of any range can also be generated by our CAs. 

\subsection{Strength Analysis}\label{sec:strength_analysis}
As we are using window-based PRNG scheme, likewise Section~\ref{Chap:3-stateCA_list:sec:CA_benifits} (Page~\pageref{Chap:3-stateCA_list:sec:CA_benifits}), our proposed PRNGs also have the similar advantages, as discussed here.
\begin{description}
\item[Portability:] This PRNG is highly portable and can be implemented using very simple and easy-to-use codes. 
\item[Robustness:] Because of variable window size, the PRNGs are also highly robust. One can generate random number of any desirable length by giving proper input. 

\item[Large Cycle Length:] As our CAs satisfy the properties of Section~\ref{sec:CA_properties}, they have very large cycle lengths. So, practically, even generation of numbers for hours can not exhaust the period of the PRNG(s).
We have experimentally observed the cycle lengths of some CAs of Table~\ref{tab:exampleRules} having size $n=5,6,\cdots,9$. For that purpose, the initial configuration has been taken as $0^{n-1}1$. The evolution of a CA is stopped when an already visited configuration is repeated. It is observed that, even for this fixed seed, the cycle length is very large for each $n$. Therefore, for the substantial $n$ taken as CA size in the schemes for PRNGs, it is quite obvious that the period of the PRNGs are prodigious.
 
 \item[Unpredictability:] One of the disadvantages of a PRNG is, if the same number is repeated, the cycle will be repeated in the same sequence. That means, if an adversary can keep track of all numbers in a cycle, she can predict the generation algorithm. However, because of window-based PRNG schemes, same number can be generated more than once, but that does not have any relation with the completion of cycle.
  This scenario makes the generated numbers unpredictable and more suitable for cryptographic purposes. 
\end{description}

\subsection{Representation of Rules}\label{sec:ca_representation} 
From the above discussion, it is evident that, all of our CAs are outstanding as PRNGs. However, our CAs have $1000$ RMTs. Although, this is not an issue for programming purpose, in case of representation on pen-and-paper, it is cumbersome to write in a tabular form or as a string. Therefore, our target has been to represent these good CAs in a simpler form. We have observed that, not all of the rules synthesized using Algorithm~\ref{algo:CA_synthesis} can be written in a simplified structure. However, as the rule-space is enormous, we can find many rules which follow the properties of Section~\ref{sec:CA_properties}, can be synthesized using the heuristic scheme of Algorithm~\ref{algo:CA_synthesis} and can be represented by a simple expression.

To represent these CAs, we explore an interesting property of \emph{STRATEGY II} CAs. Recall that, for these CAs, the RMTs of each of the sibling RMT sets are balanced; that is, each sibling RMT set contains one RMT per every possible state. Therefore, if we take next state values of each sibling RMT set as a string, it is a permutation of $``0123456789$''. There are total $10!=3628800$ possible permutations. So, a STRATEGY II CA is a concatenation of $100$ such permutations corresponding to the $100$ sibling RMT sets.

All our greedy strategies of Section~\ref{sec:CA_selection} and the synthesis scheme of Algorithm~\ref{algo:CA_synthesis} are based on STRATEGY II CAs. Hence, all these rules can be viewed as a collection of the permutations. Moreover, some of our proposed CAs have sibling RMT sets which are cyclic permutations of one another. Therefore, these CAs can be constituted by a single permutation only, provided we know the offset for the cyclic permutations for each sibling RMT set. Here, we always take cyclic permutation in right direction only.

For ease of presentation, we choose those CAs which can be represented by cyclic permutation of $Sibl_0$ and follow the properties of Section~\ref{sec:CA_properties} and the greedy strategies of Section~\ref{sec:CA_selection}. These CAs can also be synthesized using Algorithm~\ref{algo:CA_synthesis}. Each rule is embodied by the chosen permutation of $Sibl_0$, where the other sibling RMT sets $Sibl_i$, $1\le i \le 99$ are generated by cyclic permutation of $Sibl_0$. The offset of these cyclic permutations are to be chosen wisely such that each rule satisfies the  conditions \ref{cond_1}, \ref{cond_2}, \ref{cond_3} and \ref{cond_4}. A simple scheme which ensures most of these properties is given in Algorithm~\ref{algo:CA_generation}.

\begin{Walgo}[!h] {1.0cm}
	\scriptsize
	\BlankLine
	\SetKwInOut{Input}{Input}
	\SetKwInOut{Output}{Output}
	
	\Input{A permutation of $`0123456789$' ($initPerm$)}
	\Output{A CA rule}
	\BlankLine
	\rule[4pt]{0.95\textwidth}{0.99pt}\\
	\BlankLine
		\hspace{0.04\textwidth} \nlset{Step 1}\label{st1}  Set $Sibl_0 \leftarrow initPerm$; \tcp{Initialize $Sibl_0$ with the given permutation} 
		\hspace{0.04\textwidth}	\nlset{Step 2}\label{st2}
			 \For {$i = 1$ to $9$}{ 
					Set	$Sibl_i \leftarrow Sibl_{0} >> i$ ; \tcp{Initialize next $9$ $Sibl_i$ with $i$ times right shift of $Sibl_0$} 
				}
		\hspace{0.04\textwidth}	\nlset{Step 3}\label{st3}
			 \For {$j = 10$ to $99$}{ 
			 	Set	$Sibl_j \leftarrow Sibl_{j \pmod{10}} >> (j/10)$ ; \tcp{Initialize next sibling RMT sets such that equivalent RMT sets are balanced} 
			}
		\hspace{0.04\textwidth}	\nlset{Step 4}\label{st4} 
		\For {$i = 0$ to $8$}{ 
			\For {$j=i+1$ to $9$}{
			\tcp{Find sibling RMT set numbers for RMTs of primary RMT sets of cardinality $2$ and assign one set to the another}
			Set $Sibl_{j*10+i} \leftarrow Sibl_{i*10+j}$ \;
		 }
		}	
\BlankLine
		\caption{\emph{GenerateRuleFromPermutation}}
	\label{algo:CA_generation}
\end{Walgo}

In this algorithm, from a random permutation of $``0123456789$" assigned to $Sibl_0$, a candidate CA is generated.  First, we initialize each sibling RMT set $Sibl_i$, $1\le i \le 9$ by cyclic permutation of $Sibl_0$ with offset $1$ (\ref{st1}). The remaining sibling RMT sets $Sibl_j$, $10\leq j \leq 99$, are then initialized by an offset of $\floor{\frac{j}{10}}$ of $Sibl_{k}$, where $j\equiv k \pmod{10}$ (\ref{st2}). This initialization ensures that, for each equivalent RMT set, the RMTs have different next state values. However, our requirement is to make some of these equivalent RMT sets unbalanced and the primary RMT sets balanced. We can observe that, for each primary RMT sets of cardinality $2$, the RMTs of the set are of different positions in their corresponding sibling RMT sets. Therefore, the easiest way to ensure balancedness among any two RMTs belonging to two different sibling RMT sets is, assign the same permutation to those sets. This is done in \ref{st3}. By this reassignment of some sibling RMT sets, non-determinism in equivalent RMT sets and asymmetric information flow on left side are ensured. For keeping the process simple, we stop at the primary RMT sets of cardinality $2$.

In this way, using Algorithm~\ref{algo:CA_generation}, we can easily generate rule from a random permutation which satisfies most of the conditions of Section~\ref{sec:CA_selection}. Success of this algorithm as a good PRNG, however depends on how efficiently we are choosing the initial permutation. Whether a rule, generated by this algorithm for a random permutation, is a good candidate can be recognized if we test this rule by the method \emph{verifyRule()} of Algorithm~\ref{algo:CA_synthesis}. For many input permutations, the generated CAs satisfy all the properties identified in Section~\ref{sec:CA_properties} and are obviously candidates to be good PRNGs. Some sample candidate CAs are recorded in Table~\ref{tab:exampleRules}. For each CA in this table, only the permutation for $Sibl_0$ is shown.

\begin{table}[h]
\setlength{\tabcolsep}{1.9pt}
\renewcommand{\arraystretch}{1.45}
\begin{center}
\vspace{-0.5em}	
 \caption{Some rules of $10$-state CAs. Here, rules are represented by $Sibl_0$}
\label{tab:exampleRules}
\resizebox{0.85\textwidth}{!}{
\begin{tabular}{cc|cc|cc|cc}
 \toprule

 	 \theadfont{SL No.} & \theadfont{Rule} & \theadfont{SL No.} & \theadfont{Rule} & \theadfont{SL No.} & \theadfont{Rule} & \theadfont{SL No.} & \theadfont{Rule}\\
\midrule
1 ~~&~~ 8572036419 ~~&~~ 2 ~~&~~ 3154968072 ~~&~~ 3 ~~&~~ 1632405789 ~~&~~ 4 ~~&~~ 5102847369\\
5 ~~&~~ 1973502846 ~~&~~ 6 ~~&~~ 7028415369 ~~&~~ 7 ~~&~~ 1592407368 ~~&~~ 8 ~~&~~ 2469587301\\
9 ~~&~~ 8135940672 ~~&~~ 10 ~~&~~ 0271584936 ~~&~~ 11 ~~&~~ 9821354706 ~~&~~ 12 ~~&~~ 6924135087 \\
13 ~~&~~ 5983076412 ~~&~~ 14 ~~&~~ 4319256807 ~~&~~ 15 ~~&~~ 9837205146 ~~&~~  ~~&~~  \\ 
\bottomrule
\end{tabular}
}
\end{center}
\vspace{-1.5em}
\end{table}

\section{Verification of the PRNGs}\label{sec:CA_verification} In this section, randomness of our proposed CA-based PRNGs of Section~\ref{sec:CA_prng} are verified. We have used two methods for this verification -- $(1)$ using space-time diagram of the CAs, $(2)$ using empirical testbeds (e.g. Diehard, TestU01 and NIST). Finally, our proposed PRNGs are compared with the existing PRNGs.

\subsection{Space-time diagram}

\begin{figure}[!h]
\centering
	\subfloat[Rule $8572036419$\label{statespace1}]{%
		\includegraphics[width=0.30\textwidth, height = 17.0cm]{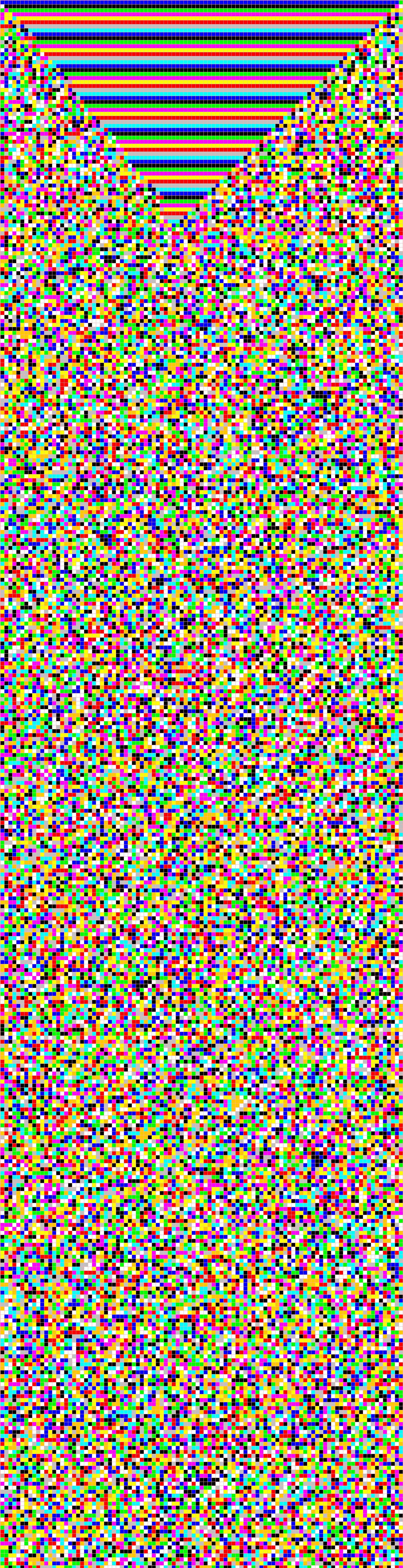}
	}
	\hfill
	\subfloat[Rule $1973502846$\label{statespace2}]{%
		\includegraphics[width=0.30\textwidth, height = 17.0cm]{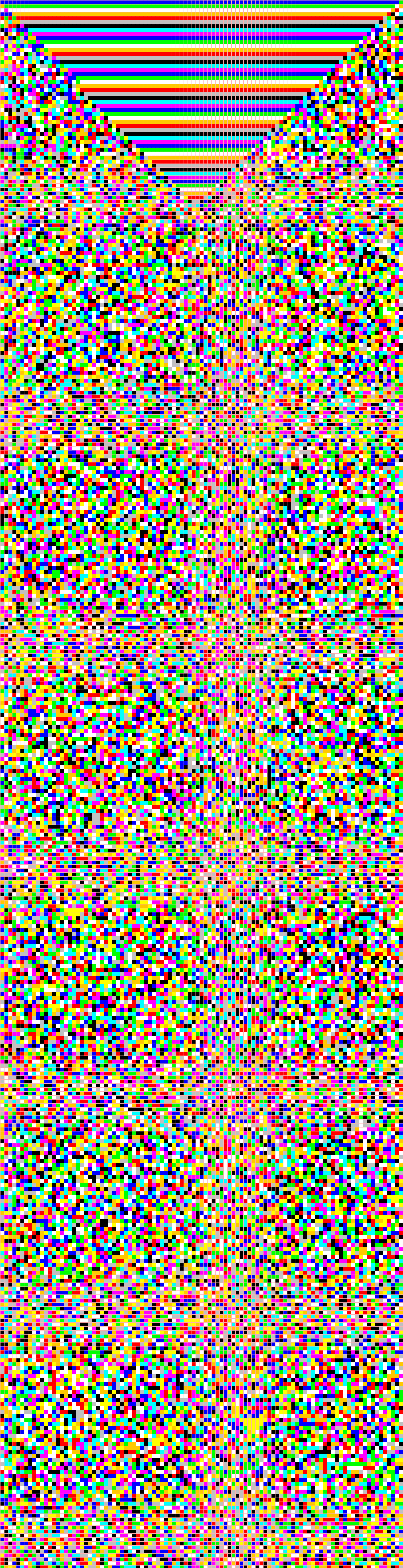}
	}
	\hfill
	\subfloat[Rule $9837205146$\label{statespace3}]{%
		\includegraphics[width=0.30\textwidth, height = 17.0cm]{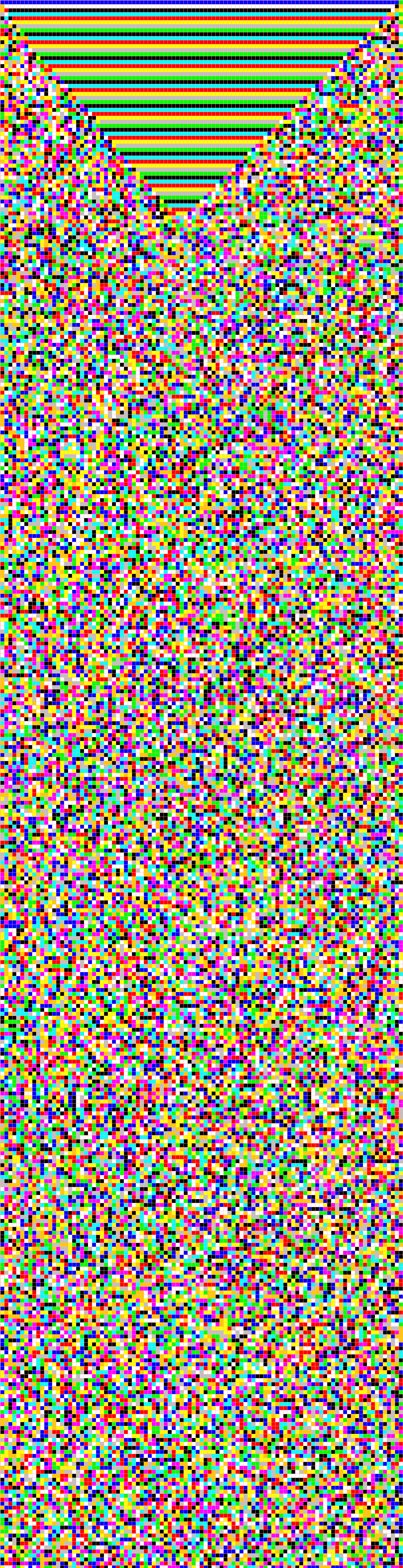}
	}
	\caption{State-space diagrams for three CAs of Table~\ref{tab:exampleRules} with seed $0^{100}1$. For each seed, $5000$ numbers are generated. The rules are represented by $Sibl_0$}
	\label{fig:state-space diagram}
	\vspace{-1.5em}
\end{figure}

We have drawn space-time diagrams for many candidate CAs generated using algorithms~\ref{algo:CA_synthesis} and \ref{algo:CA_generation}. For our CAs, the colors blue, green, red, yellow, cyan, magenta, orange, light gray, black and white respectively represent the states in increasing order. Sample diagrams for some CAs of Table~\ref{tab:exampleRules} are shown in Figure~\ref{fig:state-space diagram}. From this figure, it is evident that, these CAs are free from any self-similar regular structures and triangle like patterns. Hence, the CAs selected by our heuristic algorithms are good PRNGs. 

\subsection{Empirical Test Results}
To verify the randomness quality of the our candidate CAs as PRNGs, we have used the well-known empirical testbeds Diehard, TestU01 and NIST (discussed in Section~\ref{chap:randomness_survey:sec:empirical}). Our target has been to compare these empirical test results with those of the existing well-known PRNGs. Therefore, to be on the same platform, we have mainly tested the CAs as $32$-bit binary number generators (see Section~\ref{sec:32-bitPRNG}).

A large number of CAs generated using algorithms~\ref{algo:CA_synthesis} and \ref{algo:CA_generation} are tested by these battery of tests for many seeds. Recall that, to test a generator for a seed, a binary file (size $10-12$ MB for Diehard and \emph{bbattery\_RabbitFile()} of TestU01 and $125$ MB for NIST) is generated containing sequence of random numbers. This file is then tested by the individual testbeds.

We have observed that, for random seeds, our CAs based PRNGs pass all tests of battery \emph{rabbit} of TestU01 library and all tests of NIST statistical test-suite. This proves that, the randomness quality of the PRNGs are excellent and these PRNGs can be suitable for cryptographic applications. We have also observed that, all the candidate CAs perform almost similarly. Hence, each of the CAs generated by our algorithms are good PRNGs. This result, in turn, concord with our claim that, all the CAs following properties of Section~\ref{sec:CA_properties} or greedy strategies of Section~\ref{sec:CA_selection} are good PRNGs. Table~\ref{tab:result} gives sample results for some rules of Table~\ref{tab:exampleRules}. Detailed results for two rules (rule $8572036419$ and $8135940672$ of Table~\ref{tab:exampleRules}) are shown in Table~\ref{Chap:10-stateCA:tab:detail_rule1} and Table~\ref{Chap:10-stateCA:tab:detail_rule9} respectively. In this table, if a particular test returns a single $p$-value, that value is recorded. Similarly, if a test has some modules and individual modules returns specific $p$-values, then those $p$-values (up to $5$ modules per test) are documented in the table. However, if a test has more than $5$ $p$-values, then, in the table we are indicating it by writing the number of $p$-values in the acceptable range out of the total $p$-values returned by the test. Recall that, a PRNG \emph{passes} a test when each of the $p$-values of the test is within the acceptable range.

\begin{table*}[!h]
	\setlength{\tabcolsep}{1.3pt}
	\scriptsize
	\renewcommand{\arraystretch}{1.30}
	\centering
	\caption{Result of empirical tests on some CAs of Table~\ref{tab:exampleRules} for $4$ arbitrary seeds}
	\label{tab:result}
	\resizebox{0.89\textwidth}{2.3cm}{
		\begin{tabular}{|c|c|c|c|c|c|c|c|c|c|c|c|c|}
			\hline
			{\theadfont{Seeds $\longrightarrow$}} & \multicolumn{3}{c|}{\thead{$75922897361293$}} &  \multicolumn{3}{c|}{\thead{$36062618792023$}} &   \multicolumn{3}{c|}{\thead{$36753562912709$}} &  \multicolumn{3}{c|}{\thead{$19478450361063$}} \\
			\hline
			{\theadfont{$Sibl_0$ for the CA}} &  Diehard & TestU01 & NIST & Diehard & TestU01 & NIST & Diehard & TestU01 & NIST & Diehard & TestU01 & NIST \\
			\hline
			8572036419 & {10}& {25}& {15} &{11} &{25} &{15} & {9} & {25} & {15} & {10} & {25} & {15} \\
			3154968072 & {13}& {25}& {15} &{9} &{25} &{15} & {9} & {25} & {15} & {9} & {25} & {15} \\
			1632405789 & {10}& {25}& {15} &{10} &{25} &{15} & {11} & {25} & {15} & {9} & {25} & {15} \\
			5102847369 & {10}& {25}& {15} &{10} &{25} &{15} & {9} & {25} & {15} & {10} & {25} & {15} \\
			1973502846 & {10}& {25}& {15} &{10} &{25} &{15} & {10} & {25} & {15} & {9} & {25} & {15} \\
			1592407368 & {9}& {25}& {15} &{9} &{25} &{15} & {9} & {25} & {15} & {11} & {25} & {15} \\
			2469587301 & {10}& {25}& {15} &{10} &{25} &{15} & {9} & {25} & {15} & {10} & {25} & {15} \\
			8135940672 & {9}& {25}& {15} &{11} &{25} &{15} & {10} & {25} & {15} & {10} & {25} & {15} \\
			0271584936 & {11}& {25}& {15} &{12} &{25} &{15} & {11} & {25} & {15} & {10} & {25} & {15} \\
			\hline
		\end{tabular}
	}
\end{table*}

\begin{table}[hbtp]
\setlength{\tabcolsep}{1.3pt}
\scriptsize
\renewcommand{\arraystretch}{1.30}
\centering
\small
\caption{Empirical test results for PRNG based on CA $8572036419$ of Table~\ref{tab:exampleRules}}
\label{Chap:10-stateCA:tab:detail_rule1}
\resizebox{1.00\textwidth}{10.5cm}{
\begin{tabular}{|c|c|c|c|c|c|c|}
\hline
\multicolumn{2}{|c|}{\multirow{2}{*}{\theadfont{Name of the Tests}}} & \multicolumn{5}{c|}{\theadfont{$p$-value of tests for fixed seeds(Seeds shown in next row)}} \\
\cline{3-7}
\multicolumn{2}{|c|}{}	&  \thead{$7$} &  \thead{$1234$} &  \thead{$12345$} &  \thead{$19650218$} &  \thead{$123456789123456789$}\\
\hline
\multirow{15}{*}{\rotatebox{90}{\theadfont{Diehard}}} & \begin{tabular}{c} 	\thead{Birthday spacings}\end{tabular} & $0.167889$ & $0.111253$ & $0.518874$ & $0.198537$ & $0.466659$ \\        
\cline{2-7}
& \begin{tabular}{c}	\thead{Overlapping permutations}   \end{tabular} & -nan & -nan & -nan & -nan & -nan \\        
\cline{2-7}
& \begin{tabular}{c} \thead{Rank of} \thead{$31 \times 31$ and $32 \times 32$} 	\thead{Matrices}   \end{tabular}& $ 0.864626, 0.728301$ & $0.772509, 0.594911$ & $0.736078, 0.274377$ & $0.738911, 0.732214$ & $0.826499, 709313$ \\        
\cline{2-7}      	 
& \thead{Ranks of $6\times8$ matrices} & $0.098736$ & $0.280243$ & $0.840722$ & $0.128395$ & $0.828469$ \\        
\cline{2-7}
& \thead{Monkey tests on $20$-bit Words} & $20/20$ & $20/20$ & $20/20$& $19/20$ & $18/20$ \\        
\cline{2-7}
& \thead{Monkey tests: OPSO, OQSO and DNA}  & $21/23, 26/28, 28/31$  & $22/23, 25/28, 31/31$ & $20/23, 25/28, 29/31$ & $23/23, 26/28, 30/31$ & $23/23, 26/28, 28/31$ \\        
\cline{2-7}
& \begin{tabular}{c}  \thead{Count the $1$s} \thead{in a stream of bytes}  \end{tabular}	& $0.184827$ & $0.025716$ & $0.240414$ & $0.561768$ & $0.891073$ \\        
\cline{2-7}
& \begin{tabular}{c}  \thead{Count the $1$s} \thead{ in specific} \thead{bytes}   \end{tabular}	& $24/25$ & $22/25$ & $23/25$ & $23/25$ & $23/25$ \\        
\cline{2-7}
& \thead{Parking lot Test} & $0.0$  & $0.0$ & $0.0$ & $0.0$ & $0.0$ \\        
\cline{2-7}
& \thead{Minimum distance Test} & $0.856481$  & $0.052007$ & $0.698626$ & $0.061421$ & $0.691974$ \\        
\cline{2-7}
& \thead{Random spheres Test} & $ 0.017546$  & $0.893265$ & $0.825793$ & $0.271041$ & $0.526544$ \\        
\cline{2-7}
& \thead{Squeeze Test} & $ 0.3983336$  & $0.290256$ & $0.920403$ & $0.550350$ & $0.215773$ \\        
\cline{2-7}
& \begin{tabular}{c}\thead{Overlapping} \thead{ Sum Test} \end{tabular} & $0.002281$ & $0.699830$ & $0.3919863$ & $0.641299$ & $0.243911$\\        
\cline{2-7}
& \thead{Runs Up \& Runs Down Test}  & $0.234826, 0.779649$  & $ 0.369253, 0.419243$ & $0.571309, 0.885434$ & $0.472121, 0.134062$ & $0.393852, 0.439961$ \\      
\cline{2-7}
& \begin{tabular}{c} \thead{Craps Test}\end{tabular}& $0.845926, 0.177776$  & $0.725311, 0.797186 $ & $0.356666, 0.994164$ & $0.512236, 0.363162$ & $0.774946, 0.253801$ \\        
\hline
				
\multirow{25}{*}{\rotatebox{90}{\theadfont{TestU01 \emph{rabbit}}}} & \begin{tabular}{c} \thead{MultinomialBitsOver Test} \end{tabular}   & $0.76$ & $0.03$ & $0.07$ & $0.54$& $0.9942$ \\    
\cline{2-7}
& \begin{tabular}{c}  \thead{ClosePairsBitMatch Test ($t=2$)} \end{tabular} & $0.16$ & $0.01$ & $0.16$ & $0.16$ & $0.16$ \\  
\cline{2-7}
& \begin{tabular}{c} \thead{ClosePairsBitMatch Test ($t=4$)} \end{tabular} & $0.51$ & $0.04$ & $0.51$ & $0.04$ & $0.04$ \\        
\cline{2-7}
& \thead{AppearanceSpacings Test} & $0.69$ & $0.79$ & $0.04$& $0.89$ & $0.48$ \\        
\cline{2-7}
& \begin{tabular}{c} \thead{LinearComplexity Test} \end{tabular}& $0.48, 0.36$  & $0.36, 0.75$ & $0.35, 0.18$ & $0.62, 0.62$ & $0.26, 0.71$ \\        
\cline{2-7}
& \begin{tabular}{c} \thead{LempelZiv Test} \end{tabular}& $0.72$  & $0.64 $ & $0.83$ & $0.54$ & $0.9956$ \\        
\cline{2-7}
& \thead{Spectral Test (Fourier1)} & $0.42$ & $0.12$ & $0.81$ & $0.2$ & $0.88$ \\        
\cline{2-7}
& \thead{Spectral Test (Fourier3)} & $0.84, 0.28, 0.72$ & $0.33, 0.96, 0.29$ & $0.6, 0.87, 0.9918$ & $0.6, 0.62, 0.98$ & $0.97, 0.17, 0.15$ \\        
\cline{2-7}
& \thead{LongestHeadRun Test} & $0.6, 0.45$  & $0.73, 0.45$ & $ 0.35, 0.7$ & $0.55, 0.26$ & $0.68, 0.7$ \\        
\cline{2-7}
& \thead{PeriodsInStrings Test} & $0.89$  & $0.45$ & $ 0.87$ & $0.8$ & $0.37$\\        
\cline{2-7}
& \thead{HammingWeight Test ($L = 32$)} & $0.08$  & $0.26$ & $ 0.94$ & $0.96$ & $0.12$ \\        
\cline{2-7}
& \thead{HammingCorrelation Test ($L = 32$)} & $0.04$  & $0.61$ & $ 0.87$ & $0.22$ & $0.46$ \\        
\cline{2-7}
& \thead{HammingCorrelation Test ($L = 64$)} & $0.02$  & $0.05$ & $ 0.61$ & $0.88$ & $0.24$ \\        
\cline{2-7}
& \thead{HammingCorrelation Test ($L = 128$)} & $0.08$  & $0.0087$ & $ 0.12$ & $0.34$ & $0.53$ \\        
\cline{2-7}
&  \begin{tabular}{c}\thead{HammingIndependence Test} \theadfont{($L = 16$)}\end{tabular} & $0.35$  & $0.32$ & $ 0.46$ & $0.93$ & $0.29$ \\        
\cline{2-7}
& \begin{tabular}{c}\thead{HammingIndependence Test} \theadfont{($L = 32$)}\end{tabular} & $0.52$  & $0.32$ & $ 0.95$ & $0.53$ & $0.16$ \\        
\cline{2-7}
& \begin{tabular}{c}\thead{HammingIndependence Test} \theadfont{($L = 64$)} \end{tabular}& $0.82$  & $0.64$ & $ 0.95$ & $0.08$ & $0.1$ \\        
\cline{2-7}
& \thead{AutoCorrelation Test ($d=1$)} & $0.86$  & $0.11$ & $ 0.35$ & $0.3$ & $0.9902$ \\        
\cline{2-7}
& \thead{AutoCorrelation Test ($d=2$)} & $0.86$  & $0.35$ & $ 0.07$ & $0.41$ & $0.92$ \\        
\cline{2-7}
& \thead{Run Test} & $0.38, 0.14$  & $0.46, 0.94$ & $ 0.48, 0.54$ & $0.42, 0.72$ & $0.03, 0.08$ \\        
\cline{2-7}
& \thead{MatrixRank Test ($32 \times 32$)} & $0.21$  & $0.65$ & $0.07$ & $0.52$ & $0.7$ \\        
\cline{2-7}
& \thead{MatrixRank Test ($320 \times 320$)} & $0.1$  & $0.32$ & $ 0.84$ & $0.78$ & $0.97$ \\        
\cline{2-7}	
& \begin{tabular}{c}  \thead{RandomWalk1 Test}\\ ($L = 128$) \end{tabular} & \begin{tabular}{c}$0.34, 0.19, 0.03,$\\ $0.36, 0.72$\end{tabular} & \begin{tabular}{c}$0.16, 0.66, 0.82,$\\ $ 0.14, 0.61$\end{tabular} & \begin{tabular}{c}$0.88, 0.12,0.03,$\\ $ 0.72, 0.31$\end{tabular} & \begin{tabular}{c}$0.88, 0.14, 0.95,$\\ $ 0.17, 0.56$\end{tabular} & \begin{tabular}{c}$0.77, 0.35, 0.36,$\\ $ 0.51, 0.05$\end{tabular} \\    
\cline{2-7}
& \begin{tabular}{c} \thead{RandomWalk1 Test} \\ ($L = 1024$) \end{tabular}& \begin{tabular}{c}$0.49, 0.08, 0.24,$\\ $0.17, 0.55$\end{tabular} & \begin{tabular}{c}$0.2, 0.49, 0.59,$\\ $0.79, 0.19$\end{tabular} & \begin{tabular}{c}$0.69, 0.81, 0.07,$ \\ $0.84, 0.92$\end{tabular} & \begin{tabular}{c}$0.45, 0.07, 0.25,$\\ $0.48, 0.93$\end{tabular} & \begin{tabular}{c}$0.9, 0.75, 0.91,$\\ $ 0.4, 0.16$\end{tabular} \\ 
\cline{2-7}
& \begin{tabular}{c} \thead{RandomWalk1 Test} \\ ($L = 10016$) \end{tabular}& \begin{tabular}{c}$0.22, 0.17, 0.29,$\\ $0.98, 0.86$\end{tabular} & \begin{tabular}{c}$0.79, 0.12, 0.45,$\\ $0.39, 0.27$\end{tabular} & \begin{tabular}{c}$0.36, 0.9913, 0.99,$\\ $0.55, 0.44$\end{tabular} & \begin{tabular}{c}$0.43, 0.29, 0.06,$\\ $0.5, 0.6$\end{tabular} & \begin{tabular}{c}$0.74, 0.24, 0.49,$\\ $0.8, 0.61$\end{tabular} \\   

\hline 
\multirow{15}{*}{\rotatebox{90}{\theadfont{NIST}}} & \begin{tabular}{c}	\thead{Frequency Test} \end{tabular} & $0.854708$ & $0.044220$& $0.977480$& $0.007750$ & $0.285427$ \\        
\cline{2-7}
& \begin{tabular}{c}\thead{BlockFrequency Test}   \end{tabular} & $0.528111$ & $0.020689$ & $0.170922$ & $0.494392$ & $0.064015$ \\        
\cline{2-7}
& \thead{CumulativeSums Test} & $0.272977, 0.313041$ & $0.132640,0.534146$& $0.773405, 0.881662$& $0.000856, 0.212184$ & $0.383827, 0.979226$ \\        
\cline{2-7}
& \begin{tabular}{c} \thead{Runs Test}   \end{tabular}& $0.614226$ & $0.087162$ & $0.308561$ & $0.624627$ & $0.520102$ \\        
\cline{2-7}      	 
& \thead{LongestRun Test} & $0.931185$ & $0.167184$ & $0.155499$ & $0.927677$ & $0.378705$ \\        
\cline{2-7}
& \thead{Rank Test} & $0.110083$ & $0.508172$ & $0.325206$& $0.657933$ & $0.975012$ \\        
\cline{2-7}
& \thead{FFT Test}  & $0.344048$  & $0.192724$ & $0.103138$ & $0.514124$ & $0.027313$\\        
\cline{2-7}
& \begin{tabular}{c} \thead{NonOverlappingTemplate Test} \end{tabular}	& $148/148$ & $148/148$ & $148/148$ & $148/148$ & $148/148$ \\        
\cline{2-7}
& \begin{tabular}{c}\thead{OverlappingTemplate Test} \end{tabular}    & $0.186566$ & $0.893482$ & $0.649612$ & $0.593478$ & $0.041169$ \\        
\cline{2-7}
& \thead{Universal Test} & $0.959347$  & $0.390721$ & $0.820143$ & $0.614226$ & $0.037566$ \\        
\cline{2-7}
& \begin{tabular}{c}  \thead{ApproximateEntropy Test} \end{tabular} 
& $0.332970$ & $0.643366$ & $0.870856$ & $0.947308$ & $0.856359$ \\ 
\cline{2-7}
& \thead{RandomExcursions Test} & $8/8$ & $8/8$ & $8/8$& $8/8$ & $8/8$ \\        
\cline{2-7}
& \begin{tabular}{c} \thead{RandomExcursionsVariant Test} \end{tabular}& $18/18$  & $18/18$ & $18/18$ & $18/18$ & $18/18$ \\
\cline{2-7}
& \begin{tabular}{c}  \thead{Serial Test}\end{tabular}   & $0.915317, 0.231956$ & $0.496351, 0.680755$ & $0.295391, 0.157251$ & $0.046269, 0.293952 $& $0.089843, 0.183547$ \\       
\cline{2-7}
& \thead{LinearComplexity Test}  & $0.480771$  & $0.903338$ & $ 0.433590$ & $0.019857$ & $0.599693$ \\      
\hline 					
\end{tabular}
}
\end{table} 

\begin{table}[hbtp]
	\setlength{\tabcolsep}{1.3pt}
	\scriptsize
	\renewcommand{\arraystretch}{1.30}
	\centering
	\small
	\caption{Empirical test results for PRNG based on CA $8135940672$ of Table~\ref{tab:exampleRules}}
	\label{Chap:10-stateCA:tab:detail_rule9}
	\resizebox{1.00\textwidth}{10.5cm}{
\begin{tabular}{|c|c|c|c|c|c|c|}
\hline
\multicolumn{2}{|c|}{\multirow{2}{*}{\theadfont{Name of the Tests}}} & \multicolumn{5}{c|}{\theadfont{$p$-value of tests for fixed seeds(Seeds shown in next row)}} \\
\cline{3-7}
\multicolumn{2}{|c|}{}	&  \thead{$7$} &  \thead{$1234$} &  \thead{$12345$} &  \thead{$19650218$} &  \thead{$123456789123456789$}\\
\hline
\multirow{15}{*}{\rotatebox{90}{\theadfont{Diehard}}} & \begin{tabular}{c} 	\thead{Birthday spacings}\end{tabular} & $0.592408$ & $0.821591$ & $0.308180$ & $0.796304$ & $0.284138$ \\        
\cline{2-7}
& \begin{tabular}{c}	\thead{Overlapping permutations}   \end{tabular} & -nan & -nan & -nan & -nan & -nan \\        
\cline{2-7}
& \begin{tabular}{c} \thead{Rank of} \thead{$31 \times 31$ and $32 \times 32$} 	\thead{Matrices}   \end{tabular}& $ 0.501055, 0.9434391$ & $0.171841, 0.940631$ & $0.152404, 0.139550$ & $0.089720, 0.089720$ & $0.442147, 0.120516$ \\        
\cline{2-7}      	 
& \thead{Ranks of $6\times8$ matrices} & $0.380331$ & $0.441550$ & $0.577535$ & $0.866641$ & $0.252487$ \\        
\cline{2-7}
& \thead{Monkey tests on $20$-bit Words} & $18/20$ & $20/20$ & $18/20$& $20/20$ & $20/20$ \\        
\cline{2-7}
& \thead{Monkey tests: OPSO, OQSO and DNA}  & $22/23, 25/28, 31/31$  & $23/23, 28/28, 25/31$ & $22/23, 28/28, 30/31 $ & $22/23, 26/28, 30/31$ & $22/23, 26/28, 30/31$ \\        
\cline{2-7}
& \begin{tabular}{c}  \thead{Count the $1$s} \thead{in a stream of bytes}  \end{tabular}	& $0.862859$ & $0.898851$ & $0.396091$ & $0.356230$ & $0.354165$ \\        
\cline{2-7}
& \begin{tabular}{c}  \thead{Count the $1$s} \thead{ in specific} \thead{bytes}   \end{tabular}	& $25/25$ & $24/25$ & $25/25$ & $23/25$ & $23/25$ \\        
\cline{2-7}
& \thead{Parking lot Test} & $0.0$  & $0.0$ & $0.0$ & $0.0$ & $0.0$ \\        
\cline{2-7}
& \thead{Minimum distance Test} & $0.414339$  & $0.169058$ & $0.317138$ & $0.176496$ & $0.517575$ \\        
\cline{2-7}
& \thead{Random spheres Test} & $ 0.585376$  & $0.952433$ & $0.537112$ & $0.243585$ & $0.681059$ \\        
\cline{2-7}
& \thead{Squeeze Test} & $ 0.976537$  & $0.639014$ & $0.571032$ & $0.754493$ & $0.974104$ \\        
\cline{2-7}
& \begin{tabular}{c}\thead{Overlapping} \thead{ Sum Test} \end{tabular} & $0.004237$ & $0.972409$ & $0.095085$ & $0.033195$ & $0.320996$\\        
\cline{2-7}
& \thead{Runs Up \& Runs Down Test}  & $0.410629, 0.604856$  & $ 0.366584, 0.836334$ & $0.386436, 0.880923$ & $0.301850, 0.850938$ & $0.116072, 0.651308$ \\      
\cline{2-7}
& \begin{tabular}{c} \thead{Craps Test}\end{tabular}& $0.338492, 0.617232$  & $0.393970, 0.013012 $ & $0.971003, 0.836949$ & $0.214099, 0.150841,
$ & $0.054105, 0.022516$ \\        
\hline
				
\multirow{25}{*}{\rotatebox{90}{\theadfont{TestU01 \emph{rabbit}}}} & \begin{tabular}{c} \thead{MultinomialBitsOver Test} \end{tabular}   & $0.007$ & $0.77$ & $0.01$ & $0.54$& $0.1$ \\    
\cline{2-7}
& \begin{tabular}{c}  \thead{ClosePairsBitMatch Test ($t=2$)} \end{tabular} & $0.51$ & $0.16$ & $0.04$ & $0.94$ & $0.51$ \\  
\cline{2-7}
& \begin{tabular}{c} \thead{ClosePairsBitMatch Test ($t=4$)} \end{tabular} & $0.51$ & $0.04$ & $0.04$ & $0.51$ & $0.51$ \\        
\cline{2-7}
& \thead{AppearanceSpacings Test} & $0.83$ & $0.63$ & $0.12$& $0.17$ & $0.13$ \\        
\cline{2-7}
& \begin{tabular}{c} \thead{LinearComplexity Test} \end{tabular}& $0.94, 0.18$  & $0.74, 0.68$ & $0.74, 0.82$ & $0.71, 0.64,$ & $0.22, 0.15$ \\        
\cline{2-7}
& \begin{tabular}{c} \thead{LempelZiv Test} \end{tabular}& $0.3$  & $0.3 $ & $0.15$ & $0.81$ & $0.72$ \\        
\cline{2-7}
& \thead{Spectral Test (Fourier1)} & $0.19$ & $0.27$ & $0.66$ & $0.75$ & $0.78$ \\        
\cline{2-7}
& \thead{Spectral Test (Fourier3)} & $0.54, 0.38, 0.58$ & $0.14, 0.2, 0.09$ & $0.44, 0.98, 0.64$ & $0.15, 0.71, 0.32$ & $0.77, 0.12, 0.19$ \\        
\cline{2-7}
& \thead{LongestHeadRun Test} & $0.8, 0.26$  & $0.79, 0.14$ & $0.82, 0.5$ & $0.93, 0.9915$ & $0.66, 0.7$ \\        
\cline{2-7}
& \thead{PeriodsInStrings Test} & $0.13$  & $0.74$ & $ 0.53$ & $0.06$ & $0.23$\\        
\cline{2-7}
& \thead{HammingWeight Test ($L = 32$)} & $0.37$  & $0.44$ & $ 0.98$ & $0.51$ & $0.52$ \\        
\cline{2-7}
& \thead{HammingCorrelation Test ($L = 32$)} & $0.26$  & $0.92$ & $ 0.57$ & $0.59$ & $0.14$ \\        
\cline{2-7}
& \thead{HammingCorrelation Test ($L = 64$)} & $0.67$  & $0.82$ & $ 0.25$ & $0.32$ & $0.07$ \\        
\cline{2-7}
& \thead{HammingCorrelation Test ($L = 128$)} & $0.84$  & $0.6$ & $ 0.11$ & $0.19$ & $0.65$ \\        
\cline{2-7}
&  \begin{tabular}{c}\thead{HammingIndependence Test} \theadfont{($L = 16$)}\end{tabular} & $0.81$  & $0.6$ & $ 0.84$ & $0.83$ & $0.22$ \\        
\cline{2-7}
& \begin{tabular}{c}\thead{HammingIndependence Test} \theadfont{($L = 32$)}\end{tabular} & $0.92$  & $0.63$ & $ 0.51$ & $0.74$ & $0.32$ \\        
\cline{2-7}
& \begin{tabular}{c}\thead{HammingIndependence Test} \theadfont{($L = 64$)} \end{tabular}& $0.03$  & $0.71$ & $ 0.63$ & $0.05$ & $0.0064$ \\        
\cline{2-7}
& \thead{AutoCorrelation Test ($d=1$)} & $0.59$  & $0.55$ & $ 0.15$ & $0.03$ & $0.99$ \\        
\cline{2-7}
& \thead{AutoCorrelation Test ($d=2$)} & $0.31$  & $0.26$ & $ 0.88$ & $0.41$ & $0.78$ \\        
\cline{2-7}
& \thead{Run Test} & $0.04, 0.56$  & $0.55, 0.49$ & $0.86, 0.79$ & $0.15, 0.76$ & $0.06, 0.02$ \\        
\cline{2-7}
& \thead{MatrixRank Test ($32 \times 32$)} & $0.6$  & $0.27$ & $0.72$ & $0.24$ & $0.49$ \\        
\cline{2-7}
& \thead{MatrixRank Test ($320 \times 320$)} & $0.19$  & $0.26$ & $ 0.83$ & $0.91$ & $0.29$ \\        
\cline{2-7}	
& \begin{tabular}{c}  \thead{RandomWalk1 Test}\\ ($L = 128$) \end{tabular} & \begin{tabular}{c}$0.42, 0.23, 0.45, $\\ $0.91, 0.36$\end{tabular} & \begin{tabular}{c}$0.67, 0.52, 0.22,$\\ $ 0.62, 0.66$\end{tabular} & \begin{tabular}{c}$0.22, 0.52, 0.23, $\\ $ 0.04, 0.69$\end{tabular} & \begin{tabular}{c}$0.13, 0.07, 0.22, $\\ $0.1, 0.11$\end{tabular} & \begin{tabular}{c}$0.01, 0.01, 0.54, $\\ $ 0.23, 0.46$\end{tabular} \\    
\cline{2-7}
& \begin{tabular}{c} \thead{RandomWalk1 Test} \\ ($L = 1024$) \end{tabular}& \begin{tabular}{c}$0.04, 0.43, 0.87, $\\ $0.99, 0.45$\end{tabular} & \begin{tabular}{c}$0.43, 0.41, 0.29,$\\ $ 0.43, 0.28$\end{tabular} & \begin{tabular}{c}$0.47, 0.84, 0.96,$ \\ $ 0.7, 0.25$\end{tabular} & \begin{tabular}{c}$0.45, 0.96, 0.82,$\\ $ 0.61, 0.64$\end{tabular} & \begin{tabular}{c}$0.57, 0.75, 0.98,$\\ $ 0.32, 0.87$\end{tabular} \\ 
\cline{2-7}
& \begin{tabular}{c} \thead{RandomWalk1 Test} \\ ($L = 10016$) \end{tabular}& \begin{tabular}{c}$0.45, 0.91, 0.02, $\\ $0.38, 0.73$\end{tabular} & \begin{tabular}{c}$ 0.49, 0.83, 0.22,$\\ $ 0.26, 0.18$\end{tabular} & \begin{tabular}{c}$0.41, 0.66, 0.9, $\\ $0.95, 0.7$\end{tabular} & \begin{tabular}{c}$0.29, 0.58, 0.08, $\\ $0.28, 0.57$\end{tabular} & \begin{tabular}{c}$0.43, 0.79, 0.0065, $\\ $ 0.9925, 0.52$\end{tabular} \\   

\hline 
\multirow{15}{*}{\rotatebox{90}{\theadfont{NIST}}} & \begin{tabular}{c}	\thead{Frequency Test} \end{tabular} & $0.328297$ & $0.872425$& $0.928857$& $0.114712$ & $0.029205$ \\        
\cline{2-7}
& \begin{tabular}{c}\thead{BlockFrequency Test}   \end{tabular} & $0.337688$ & $0.072514$ & $0.965860$ & $0.585209$ & $0.542228$ \\        
\cline{2-7}
& \thead{CumulativeSums Test} & $0.041981, 0.161703$ & $0.794391, 0.856359$& $0.931185, 0.336111$& $0.170922, 0.859637$ & $0.428095, 0.030399$ \\        
\cline{2-7}
& \begin{tabular}{c} \thead{Runs Test}   \end{tabular}& $0.825505$ & $0.657933$ & $0.699313$ & $0.292519$ & $0.988677$ \\        
\cline{2-7}      	 
& \thead{LongestRun Test} & $0.883171$ & $0.268917$ & $0.054314 $ & $0.278461$ & $0.295391$ \\        
\cline{2-7}
& \thead{Rank Test} & $0.818343$ & $0.568739$ & $0.368587$& $0.382115$ & $0.060492$ \\        
\cline{2-7}
& \thead{FFT Test}  & $0.922855$  & $0.085068$ & $0.006425$ & $0.707513$ & $0.284024$\\        
\cline{2-7}
& \begin{tabular}{c} \thead{NonOverlappingTemplate Test} \end{tabular}	& $148/148$ & $148/148$ & $148/148$ & $148/148$ & $148/148$ \\        
\cline{2-7}
& \begin{tabular}{c}\thead{OverlappingTemplate Test} \end{tabular}    & $0.877083$ & $0.371941$ & $0.893482$ & $0.540204$ & $0.927677$ \\        
\cline{2-7}
& \thead{Universal Test} & $0.818343$  & $0.735908$ & $0.317565$ & $0.089301$ & $0.877083$ \\        
\cline{2-7}
& \begin{tabular}{c}  \thead{ApproximateEntropy Test} \end{tabular} 
& $0.807412$ & $0.230755$ & $0.761719$ & $0.234373$ & $0.488534$ \\ 
\cline{2-7}
& \thead{RandomExcursions Test} & $8/8$ & $8/8$ & $8/8$& $8/8$ & $8/8$ \\        
\cline{2-7}
& \begin{tabular}{c} \thead{RandomExcursionsVariant Test} \end{tabular}& $18/18$  & $18/18$ & $18/18$ & $18/18$ & $18/18$ \\
\cline{2-7}
& \begin{tabular}{c}  \thead{Serial Test}\end{tabular}   & $0.668321, 0.873987$ & $0.240501, 0.331408$ & $0.420827, 0.721777$ & $0.457825, 0.193767$& $0.336111, 0.940080$ \\       
\cline{2-7}
& \thead{LinearComplexity Test}  & $0.889118$  & $0.800005$ & $ 0.908760$ & $0.388990$ & $0.454053$ \\      
\hline 					
\end{tabular}
}
\end{table} 
		
From these tables, we can observe that, our CAs fail to pass only the \emph{overlapping permutation} and \emph{parking lot} tests of Diehard. 
In fact, from the empirical test results of Section~\ref{chap:randomness_survey:sec:empirical_result}, it is noticed that, none of the existing PRNGs can pass these two tests of Diehard for any seeds (see Table~\ref{tab:comparison}).

\subsubsection{Comparison with existing PRNGs}\label{sec:comparison} We now compare our proposed PRNGs with the existing PRNGs. For comparison, we have chosen the well-known PRNGs, which are considered as \emph{good} (see Table~\ref{tab:PRNG_list} of Page~\pageref{tab:PRNG_list}), like Knuth's \verb MMIX, ~\verb rand, ~\verb lrand48 ~\& \verb random ~in GNU C Library, ~\verb MRG31k3p, ~\verb PCG-32, ~\verb Taus88, ~\verb LFSR113, ~\verb LFSR258, ~\verb WELL512a ~\& \verb WELL1024a, ~\verb xorshift32, \verb xorshift64*, \verb xorshift1024*M_8, ~\verb xorshift128+, \\~\verb MT19937-32, ~\verb MT19937-64, ~\verb SMFT19937-32 ~\verb SFMT19937-64, \verb dSFMT19937-32 ~\& \verb dSFMT19937-64 ~along with the CA-based generators using maximal-length CAs with $1$ cell spacing ($\gamma=1$) and no cell spacing ($\gamma=0$), non-linear $2$-state CA and the $3$-state CA-based PRNG of Section~\ref{Chap:randomness_survey:sec:prng_R} (Page~\pageref{Chap:randomness_survey:sec:prng_R}) etc. For more details on the selected PRNGs, please see Section~\ref{chap:randomness_survey:sec:class}.

Recall that, to test those PRNGs, the $C$ programs of the PRNGs are collected from their respective websites, where each PRNG has an available seed for normal usage. For example, for MT$19937$, the seed is $19650218$, whereas, for most of the others, this seed is $1234$ or $12345$. We have collected all these seeds hard-coded in the $C$ programs of all PRNGs, and used these as the set of seeds for each PRNG. These seeds are $7$, $1234$, $12345$, $19650218$ and $123456789123456789$. Section~\ref{chap:randomness_survey:sec:empirical_result} reports a detailed discussion about these empirical test results (see Table~ \ref{tab:blind_test} of Page~\pageref{tab:blind_test}, \ref{tab:blind_test_avg} of Page~\pageref{tab:blind_test_avg} and \ref{tab:final_rank_comparison} of Page~\pageref{tab:final_rank_comparison}). To be on the same platform, we have tested our proposed PRNGs with all these seeds. Some sample results for these fixed seeds for two rules are depicted in Table~\ref{Chap:10-stateCA:tab:detail_rule1} and Table~\ref{Chap:10-stateCA:tab:detail_rule9}. Table~\ref{tab:comparison} reports a comparison of the existing PRNGs (like Table~\ref{tab:blind_test} of Page~\pageref{tab:blind_test}) with one of our proposed PRNGs (rule $8135940672$ of Table~\ref{tab:exampleRules}) using the fixed seeds.



\begin{table}[!h]
\setlength{\tabcolsep}{1.3pt}
   \renewcommand{\arraystretch}{1.30}
     \centering
   \small
 \caption{Comparison of existing PRNGs with our proposed PRNG(s) using Diehard, TestU01 and NIST for the fixed seeds 
 	}
 	\label{tab:comparison}
     \resizebox{1.00\textwidth}{!}{
   \begin{tabular}{|c|c|c|c|c|c|c|c|c|c|c|c|c|c|c|c|c|c|}
   \hline
\multicolumn{2}{|r|}{\theadfont{Seeds $\longrightarrow$}} & \multicolumn{3}{c|}{\thead{$7$}} &  \multicolumn{3}{c|}{\thead{$1234$}} &   \multicolumn{3}{c|}{\thead{$12345$}} &  \multicolumn{3}{c|}{\thead{$19650218$}} &  \multicolumn{3}{c|}{\thead{$123456789123456789$}} & \thead{Ranking}\\
 \cline{1-17}
 \multicolumn{2}{|c|}{\theadfont{ Name of the PRNGs}} &  Diehard & TestU01 & NIST & Diehard & TestU01 & NIST & Diehard & TestU01 & NIST & Diehard & TestU01 & NIST & Diehard & TestU01 & NIST & (First Level)\\
\hline
\multirow{7}{*}{\rotatebox{90}{LCGs}}& MMIX & 6 & 19 & 7 & 5 & 18 & 7 & 6 & 17 & 8 & 4 & 16 & 8 & 5 & 18 & 8 & 8\\
\cline{2-18}
& minstd\_rand & 0 & 1 & 1 & 0 & 1 & 1 & 0 & 1 & 2  & 0 & 1 & 1 & 0 & 2 & 1 & 12\\
\cline{2-18}
& Borland LCG & 1 & 3 &5 & 0 & 3 & 5 & 1 & 3 & 5 & 1 & 3 & 4 & 1 & 3 & 5 & 11\\
\cline{2-18}
& rand & 1 &1 &2 & 1 & 1& 2& 1 & 3 & 2& 1 & 2 & 2& 1 & 2 & 2 & 11\\
\cline{2-18}
& lrand48() & 1 & 3 & 2& 1 & 2 & 2 & 1 & 3 & 2 & 1 & 3 & 2 & 1 & 2 & 2 & 11\\
\cline{2-18}
& MRG31k3p & 1 & 2 & 1 & 1 & 1 & 1 & 1 & 2 & 1 & 1 & 1 & 1 & 0 & 0 & 2 & 12\\
\cline{2-18}
& PCG-32 & 9 & 25 & 15& 9 & 25 &14 & 11 & 25  &14 & 10 & 24 &15 & 9 & 25 & 15 & 2\\
\cline{2-18}
\hline
\multirow{16}{*}{\rotatebox{90}{LFSRs}}& random() & 1 & 1 & 1 & 1& 1 & 1 & 1 & 3 & 1 & 1 & 2 & 1 & 1 & 2 & 1 & 11\\
\cline{2-18}
& Tauss88 & 11 & 21 & 15 & 9 & 23 & 15 & 11 & 23 & 15 & 11 & 23 & 14 & 10 & 23 & 15 & 4\\
\cline{2-18}
& LFSR113 & 5 & 6 & 1& 11 & 23 & 14& 9 & 23 & 15 & 7 & 23 & 14 & 9 & 23 & 15 & 7\\
\cline{2-18}
& LFSR258 & 0 & 0 & 1& 0 & 5 & 2 & 1 & 5 & 2 & 1 & 5 & 2& 1 & 5 & 0 & 12\\
\cline{2-18}
& WELL512a & 9 & 23 &15 & 10 & 23 & 14 & 10 & 23 & 15 & 8 & 23 & 15 & 7 & 23 & 15 & 5\\
\cline{2-18}
& WELL1024a & 9 & 25 & 15 & 10 & 24 & 15 & 9 & 24& 14& 9 & 25 & 15 & 9 & 25 &15 & 3\\
\cline{2-18}
& MT19937-32 & 10 & 25 & 13& 9 & 25 & 13& 9 & 25 &14 & 9 & 25 &15 & 9 & 25 &15 & 3\\
\cline{2-18}
& MT19937-64 & 10 & 25 &15 & 10 & 24 & 15& 8 & 24 & 15& 11 & 25 &15 & 10 & 25 & 15 & 2\\
\cline{2-18}
& SFMT19937-32 & 10 & 25 & 15& 9 & 25 & 15& 10 & 25 &15 & 9 & 25 & 15& 10 & 25 & 15 & 1\\
\cline{2-18}
& SFMT19937-64 & 11 & 25 & 15 & 10 & 25 &15 & 10 & 25 & 15& 9 & 25 &15 & 10 & 25 & 15 & 1\\
\cline{2-18}
& dSFMT-32 & 7 & 25 &15 & 8 & 25 & 15& 11 & 24 & 13& 11 & 25 &15 & 10 & 24 & 15 & 5\\
\cline{2-18}
& dSFMT-52 & 5 & 11 & 3& 5 & 10 & 3& 7 & 11 & 3 & 6 & 10 & 3 & 7 & 9 & 3 & 9\\
\cline{2-18}
&  xorshift32 & 4 & 17 & 4& 4 & 17 & 4& 4 & 17 &2 & 0 & 17 &13 & 4 & 17 & 13 & 9\\
\cline{2-18}
&  xorshift64* & 10 & 25 & 15 & 10 & 25 & 15& 8 & 25 & 15& 7 & 25 & 15 & 8 & 25 & 14 & 5\\
\cline{2-18}
&  xorshift1024* & 7 & 20 & 6& 9 & 21 & 15& 7 & 20 &15 & 8 & 20 &15 & 6 & 21 & 15 & 6\\
\cline{2-18}
&  xorshift128+ & 9 & 25 & 14& 9 & 25 & 14& 10 & 24 &15 & 10 & 25 &15 & 8 & 24 & 15 & 4\\
\hline
\multirow{7}{*}{\rotatebox{90}{CAs}}& Rule $30$ & 11& 25& 15& 10&25 & 15& 9 & 25& 15 & 8 & 25 & 15 & 11 & 24 & 15 & 2\\
\cline{2-18}
& Hybrid CA with Rules $30$ \& $45$ & 3 & 8 & 3& 0& 1 & 0 & 2  & 8 & 1 & 1 & 7 & 2 & 1 & 8& 2 & 11\\
\cline{2-18}
& Maximal Length CA with $\gamma=0$ & 2 & 12 & 10& 0& 12 & 11 & 1 & 12& 11& 1& 12& 11 & 2& 12 & 10 & 10\\
\cline{2-18}
& Maximal Length CA with $\gamma=1$ & 4& 17& 14 & 3 & 16 & 14 & 3  & 17 & 14 & 4 & 15 & 14 & 3 & 16 & 14 & 8\\
\cline{2-18}
& Non-linear $2$-state CA & 6 & 11& 4& 8& 10 & 2& 5 & 12 & 3 & 5& 12& 4& 7& 12& 4 & 9\\
\cline{2-18}
& {$3$-state CA} ${\mathscr{R}}$ & ${3}$& ${12}$& ${6}$ & ${3}$ & ${12}$ & ${6}$ & ${3}$ & $\mathbf{11}$ & ${5}$ & ${2}$ & ${11}$ & ${4}$ & ${3}$ & ${11}$ & ${4}$ & ${10}$\\
\cline{2-18}
& \thead{Rule $\mathbf{8135940672}$ of Table~\ref{tab:exampleRules}} & \textbf{9}& \textbf{25}& \textbf{15} &\textbf{10} &\textbf{25} &\textbf{15} & \textbf{11} & \textbf{25} & \textbf{15} & \textbf{11} & \textbf{25} & \textbf{15} & \textbf{10} & \textbf{25} & \textbf{15} & \textbf{1}\\
			\hline
 \end{tabular}}
 	\vspace{-0.5em}
 \end{table}

\begin{table}[!h]
	\centering
	\small
	\caption{Summary of all empirical test results and final ranking}
	\label{tab:final_rank_update}
	\resizebox{1.00\textwidth}{6.0cm}{
		\begin{tabular}{|c|c|c|c|c|c|c|c|p{10.2em}|c|c|c|}
			\hline
			\multicolumn{2}{|c|}{\multirow{2}{*}{\theadfont{Name of the PRNGs}}} & \multicolumn{3}{c|}{\theadfont{Fixed Seeds}} &  \multicolumn{2}{c|}{\theadfont{Random Seeds}} & \multirow{2}{*}{\theadfont{Lattice Test}} & \multirow{2}{*}{\theadfont{Space-time Diagram}} & \multicolumn{3}{c|}{\theadfont{Ranking}} \\
			\cline{3-7}\cline{10-12}
			\multicolumn{2}{|c|}{ } & Diehard & TestU01 & NIST & Average & Range &  & & {\theadfont{$1^{st}$ Level}} & \theadfont{$2^{nd}$ Level} &{\theadfont{Final Rank}}\\
			\hline
			\multirow{7}{*}{\rotatebox{90}{LCGs}}& MMIX & 4-6 & 16-19 & 7-8 & 6.5 & 2-9  & Not Filled & Last $2$ bits fixed & 8 & 9 & 15\\
			\cline{2-12}
			& minstd\_rand & 0 & 1 & 1-2 &0.38 & 0-1 &  Not Filled & last $6$ bits fixed & 12 & 14 & 23\\
			\cline{2-12}
			& Borland LCG & 1 & 3 & 4-5 & 1.9 & 1-2 & Not Filled &  Last $2$ bits fixed & 11 & 12 & 21\\
			\cline{2-12}
			& rand & 1 & 1-3 & 2-3 &  &  &  Not Filled & More 0s than 1s & 11 & 13 & 19\\
			\cline{2-12}
			& lrand48() & 1 & 2-3 & 2 & 1 & 1 & Not Filled & More 0s than 1s & 11 & 13 & 18\\
			\cline{2-12}
			& MRG31k3p & 0-1 & 1-2 & 1-2 & 0.9 & 0-1 & Scattered & LSB is 0, block of 0s, dependency on seed & 12 & 14 & 22\\
			\cline{2-12}
			& PCG-32 & 9-11 & 24-25 & 14-15 & 9.3 & 6-12 & Relatively Filled & Independent of seed & 2 & 4 & 5\\
			\cline{2-12}
			\hline
			\multirow{16}{*}{\rotatebox{90}{LFSRs}}& random() & 1 & 1-3 & 1 & 1 & 1 & Not Filled & MSB is 0, blocks of 0s & 11 & 13 & 20\\
			\cline{2-12}
			& Tauss88 & 9-11 & 21-23 & 14-15 & 9.0 & 0-12 & Relatively Filled & Independent of seed, block of $0$s & 4 & 7 & 11\\
			\cline{2-12}
			& LFSR113 & 5-11 & 6-23 & 1-15 & 9.3 & 6-12 & Relatively Filled & Dependency on seed, Block of 0s & 7 & 7 & 14\\
			\cline{2-12}
			& LFSR258 & 0-1 & 0-5 & 0-2 &  1.8 & 1-2 & Scattered & Pattern & 12 & 14 & 24\\
			\cline{2-12}
			& WELL512a & 7-10 & 23 & 14-15 & 8.5 & 5-11 & Relatively filled & First few numbers are fixed with seed dependency & 5 & 6 & 10\\
			\cline{2-12}
			& WELL1024a & 9-10 & 24-25 & 14-15 & 9.2 & 6-11 & Relatively Filled & Dependency on seed & 3 & 4 & 9\\
			\cline{2-12}
			& MT19937-32 & 9-10 & 25 & 13-15 & 9.3 & 6-12 & Relatively Filled & Independent of seed & 3 & 4 & 6\\
			\cline{2-12}
			& MT19937-64 & 8-11 & 24-25 & 15 & 9.4 & 6-11 & Relatively Filled & Independent of seed & 2 & 3 & 4\\
			\cline{2-12}
			& SFMT19937-32 & 9-10 & 25 & 15 & 9.5 & 5-12 & Relatively Filled & Independent of seed & 1 & 1 & 2\\
			\cline{2-12}
			& SFMT19937-64 & 9-11 & 25 & 15 & 9.52 & 6-12 & Relatively Filled & Independent of seed & 1 & 1 & 1\\
			\cline{2-12}
			& dSFMT-32 & 7-11 & 24-25 & 13-15 & 9.3 & 5-11 & Relatively Filled & Independent of seed & 5 & 5 & 7\\
			\cline{2-12}
			& dSFMT-52 & 5-7 & 9-11 & 3 & 5.97 & 3-7 & Relatively Filled & Less dependency on seed & 9 & 10 & 12\\
			\cline{2-12}
			&  xorshift32 & 2-4 & 17 & 2-13 & 5.5 & 3-7 & Not Filled & Blocks of 0s & 9 & 10 & 15\\
			\cline{2-12}
			&  xorshift64* & 7-10 & 25 & 14-15 & 8.0 & 6-11 & Relatively Filled & Independent of seed & 5 & 6 & 8\\
			\cline{2-12}
			&  xorshift1024* & 6-9 & 20-21 & 6-15 & 7.0 & 4-9 & Not Filled & Dependency on seed, Pattern & 6 & 8 & 14\\
			\cline{2-12}
			&  xorshift128+ & 8-10 & 24-25 & 14-15 & 9.4 & 6-12 & Relatively Filled & Dependency on seed for first few numbers & 4 & 4 & 9\\
			\hline
			\multirow{7}{*}{\rotatebox{90}{CAs}}& Rule $30$ & 8-11 & 24-25 & 15 & 10.2 & 7-12 & Relatively Filled & Independent of seed & 2 & 2 & 3\\
			\cline{2-12}
			& Hybrid CA with Rules $30$ \& $45$ & 0-3 & 1-8 & 0-3 & 2.0 & 0-3 & Not Filled & Pattern & 11 & 12 & 17\\
			\cline{2-12}
			& Maximal Length CA with $\gamma=0$ & 0-2 & 12 & 10-11 & 1.6 & 1-2 & Not Filled & Pattern & 10 & 11 & 16\\
			\cline{2-12}
			& Maximal Length CA with $\gamma=1$ & 3-4 & 15-17 & 14 & 1.8 & 1-4 & Relatively Filled & Dependency on seed for first few numbers & 8 & 11 & 13\\
			\cline{2-12}
			& Non-linear $2$-state CA & 5-8 & 10-12 & 3-4 & 5.85 & 2-8 & Relatively Filled &  Less dependency on seed & 9 & 9 & 13\\
		\cline{2-12}
			& {$3$-state CA} ${\mathscr{R}}$ & ${2-3}$ & ${11-12}$ & ${4-6}$ & ${2.7}$ & ${1-4}$ & {Relatively Filled} & {Less dependency on seed} & ${10}$ & ${11}$ & ${13}$\\
			\cline{2-12}
			& \thead{Rule $\mathbf{1632405789}$ of Table~\ref{tab:exampleRules}} & \textbf{9-11}& \textbf{25}& \textbf{15} &\textbf{9.588} & \textbf{6-12} & \textbf{Relatively Filled} & \textbf{Independent of seed} & \textbf{1} & \textbf{1} & \textbf{1}\\
			
			\hline
		\end{tabular}}
	\end{table}  

\begin{figure}[!h]
	\subfloat[SFMT\label{testu01_sfmt64}]{%
		\includegraphics[width=0.45\textwidth, height = 5.0cm]{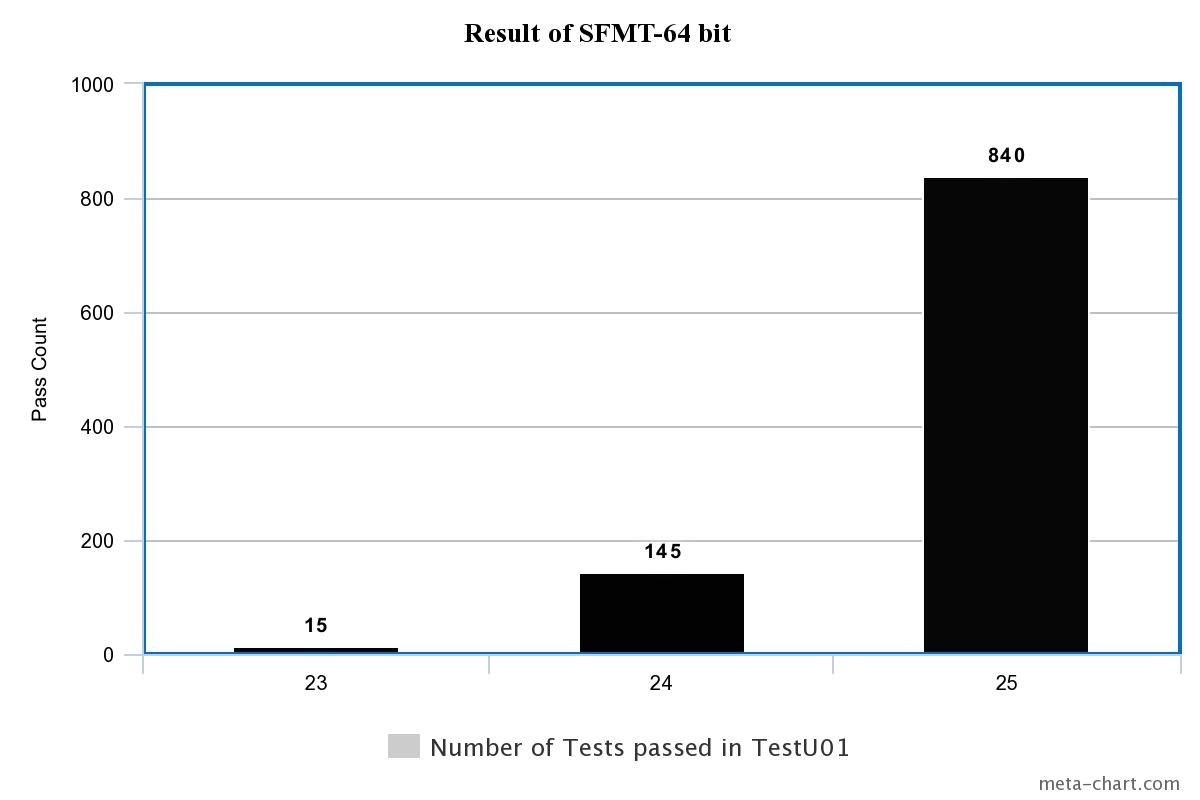}
	}
	\hfill
	\subfloat[Rule $1973502846$ of Table~\ref{tab:exampleRules}\label{testu01_proposedPRNG}]{%
		\includegraphics[width=0.45\textwidth, height = 5.0cm]{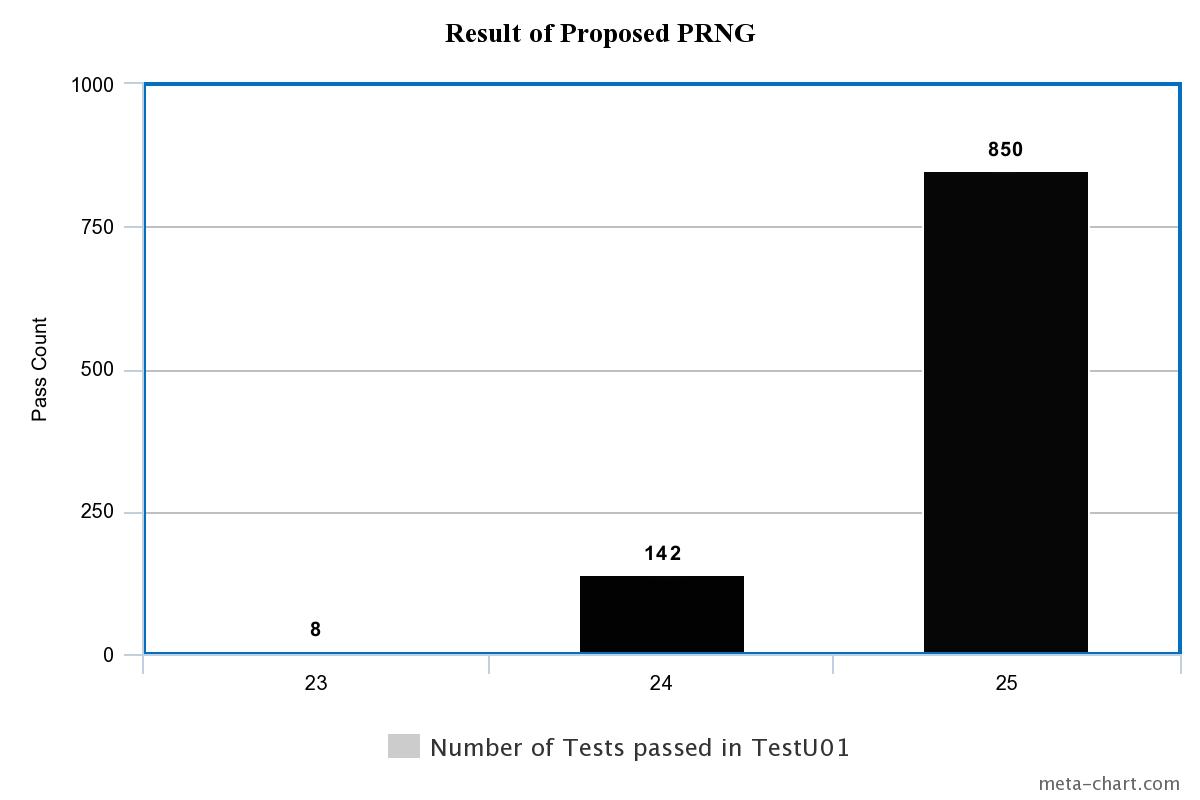}
	}
	\caption{TestU01 test results with $1000$ seeds for SFMT ($64$-bit) and a proposed PRNG. The seeds are generated by consecutive calls of \emph{rand()} initialized with \emph{srand(0)}}
	\label{fig:testu01_rand_result}
\end{figure}

\begin{figure}[!h]
	\subfloat[SFMT\label{diehard_sfmt64}]{%
		\includegraphics[width=0.45\textwidth, height = 6.0cm]{./10-stateCA/Diehard_SFMT64.jpeg}
	}
	\hfill
	\subfloat[Rule $1632405789$ of Table~\ref{tab:exampleRules}\label{diehard_proposedPRNG}]{%
		\includegraphics[width=0.45\textwidth, height = 6.0cm]{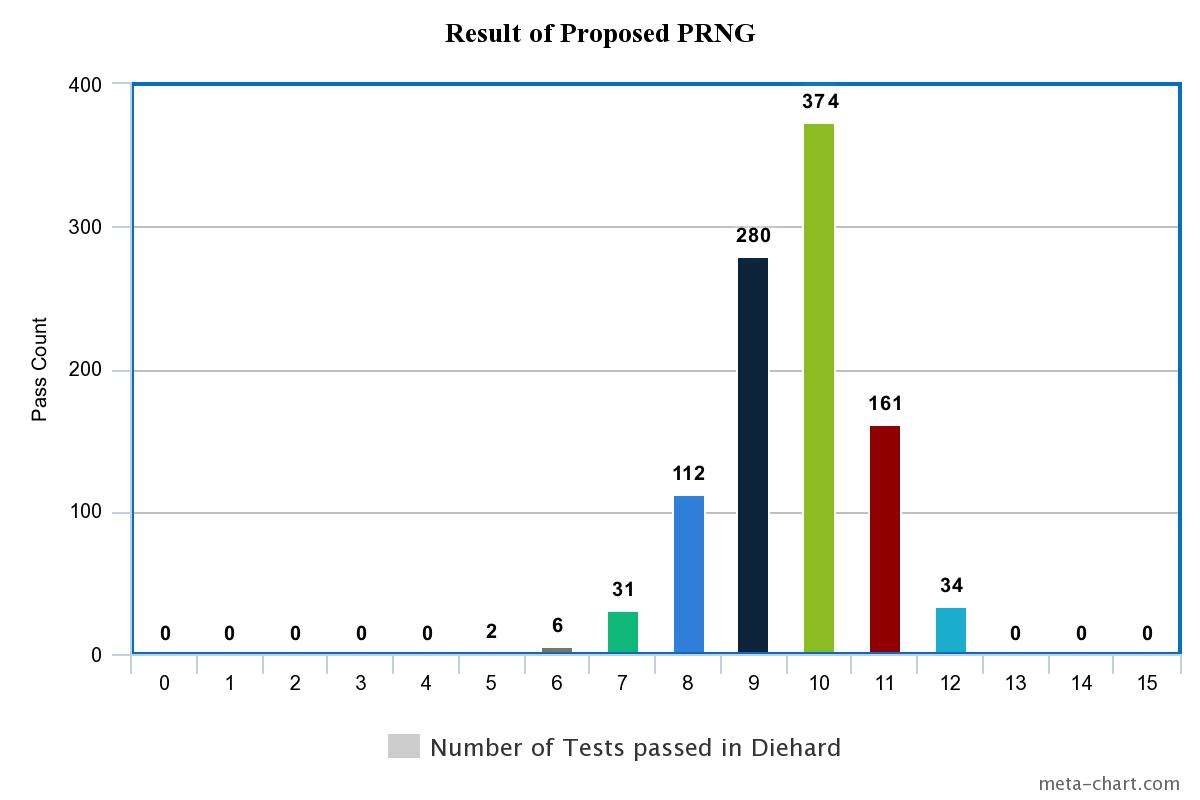}
	}
	\caption{Diehard test results with $1000$ seeds for SFMT ($64$-bit) and a proposed PRNG. The seeds are generated by consecutive calls of \emph{rand()} initialized with \emph{srand(0)}}
	\label{fig:diehard_rand_result}
\end{figure}

From this table, we can observe that, 
results of our proposed CAs are at par with the \verb SFMT19937-64 ~(which is the best performer among all existing PRNGs in Table~\ref{tab:final_rank_comparison} of Page~\pageref{tab:final_rank_comparison}). To confirm this claim, we again test our proposed PRNGs and \verb SFMT19937-64 ~with TestU01 library and our PRNGs with Diehard battery of tests for $1000$ random seeds. Like Chapter~\ref{Chap:randomness_survey}, here also, for each PRNG, these random seeds are generated by \verb rand, ~which is initialized with \verb srand(0). ~
Some results of this experiment are shown in figures~\ref{fig:testu01_rand_result} and \ref{fig:diehard_rand_result} respectively. 

From these figures, we can clearly see that, average of our CA is $9.588$ in Diehard battery of tests (Figure~\ref{diehard_proposedPRNG}) and $24.842$ in TestU01 library (Figure~\ref{testu01_proposedPRNG}) in comparison to average $9.52$ in Diehard (Figure~\ref{diehard_sfmt64}) and $24.825$ in TestU01 (Figure~\ref{testu01_sfmt64}) for \verb SFMT19937-64. ~So, our PRNGs have slightly better average than even \verb SFMT19937-64. ~These PRNGs also have no pattern in the space-time diagrams or in lattice tests and are independent of seed. Therefore, our CAs hold at least the same rank as \verb SFMT19937-64. ~Here, we reproduce Table~\ref{tab:final_rank_comparison} (Page~\pageref{tab:final_rank_comparison}) in Table~\ref{tab:final_rank_update} to exhibit the final ranking of all the PRNGs.

These results further strengthen our claim that, our proposed PRNGs are at least one of the best PRNGs existing today. Further, in terms of simplicity, portability, robustness, unpredictability and ease of implementation, we claim that, our CA-based decimal number generators can be better than all their kins.
 
\section{Conclusion}\label{sec:Conclusion}
In Chapter~\ref{Chap:randomness_survey}, we have observed that, ranking of existing CAs based PRNGs is average. So, in this chapter, our target has been to improve randomness quality of CAs and develop PRNGs that can be at par with the best PRNGs. With that objective, we have explored the prospect of decimal CAs as pseudo-random number generators. Two heuristic synthesis schemes of generating candidate CAs have been reported which ensure that the CAs are unpredictable and devoid of self-similar and triangular patterns.
We have proposed two window-based schemes to use these CAs as PRNGs. By comparing our PRNGs with the existing well-known $28$ PRNGs on same platform, we have finally shown that, randomness quality of these PRNGs in terms of performance in empirical testbeds are at par with the best PRNG \verb SFMT19937-64. ~Further, considering simplicity and other strengths of the scheme, we claim that, our CA-based generators are at least one of the best PRNGs today. This result also confirms our hypothesis that, if we increase the number of states, than, locally interactive computational model, like CA, can also be an excellent source of randomness.



\chapter{Conclusion}\label{Chap:conclusion}
{\small This chapter abridges the dissertation by designating the main contributions of the research reported in each chapter. Further, it records some open problems related to this work, which can be explored in future.}

\section{Key Contributions}
{\large\textbf{C}}ellular automaton (CA), a discrete mathematical model, is an elementary tool in addressing many fundamental problems in theoretical computer science and physics. In this dissertation, our paramount goal has been to study two global properties of cellular automata (CAs), namely, reversibility and randomness. Both these issues are related to the study of nature and are well-known in the domain of physics and mathematics. We have explored $1$-dimensional $d$-state CAs under periodic boundary condition to study these characteristics.

Before delving into the actual research, in Chapter~\ref{Chap:surveyOfCA}, we have surveyed the journey of cellular automaton from its rudimentary form to the current conformation where CAs are used as technological solutions to many real-life problems. However, the definitions and terminologies we have used for our work are described in Chapter~\ref{Chap:reversibility}. This chapter has also addressed the reversibility of finite $1$-dimensional $d$-state CAs. To analyze such CAs, we have developed a tool, named reachability tree, which can efficiently characterize reversible finite CAs. Using this tool, a number of theorems have been developed which constitute the conditions, a reachability tree of a reversible CA of size $n$ needs to fulfill. Further, to restrict the growth of the reachability tree with $n$, we have introduced minimized reachability tree which conserves all informations of a CA of size $n$ in its fixed height. Based on this minimized reachability tree, a decision algorithm has been proposed which inspects whether a $1$-dimensional $3$-neighborhood $d$-state CA is reversible for a size $n$. Finally, three greedy strategies have been reported to recognize a large set of CAs reversible for a size $n$.

We have observed that, reversibility of finite CAs is dependent not only on the rule but also on the size of a CA. Whereas, if a CA is defined over infinite lattice, reversibility is decided based on the rule only. Therefore, there is an apparent disagreement between two cases. In this scenario, Chapter~\ref{Chap:semireversible} has targeted to establish a connection between these two cases. To do this, we have redefined the notion of reversibility and introduced the idea of semi-reversibility. As a result, three classes of CAs have been observed-- (1) reversible CAs, (2) semi-reversible CAs and (3) strictly irreversible CAs. To decide reversibility/semi-reversibility, an algorithm has been proposed by exploiting the minimized reachability tree. The height of the minimized tree for a CA rule is bound to a small size $n_0$ and for any $n\ge n_0$, its height does not change. Consequently, using the minimized tree, we have conferred about the reversibility of the CA when it is defined over infinite lattice. Hence, eventually, the reversibility of CAs for finite and infinite cases have been related. 

The next three chapters have investigated cellular automata as source of randomness.
In Chapter~\ref{Chap:randomness_survey}, a set of necessary properties have been identified for a CA to be a good source of randomness. Then a sample pseudo-random number generator (PRNG) has been developed using a $3$-neighborhood $3$-state CA which satisfies these properties. The numbers in the PRNG have been generated from the output of a variable window out of the total $n$-cell configuration. While comparing this PRNG with the existing well-known PRNGs, we have observed that, performance of this CA in the empirical tests is average. It can not compete with the best PRNG like \verb SFMT19937-64, ~although it has better performance than the existing portable PRNGs and many of the $2$-state CAs based PRNGs. This scheme has also offered portability, robustness, ease of implementation in hardware, etc.
So, in Chapter~\ref{Chap:3-stateCA_list}, we have further explored the rule space of $3$-neighborhood $3$-state CAs and reported a list of $596$ CAs with similar quality. To find such rules, we have employed theoretical and empirical filtering on an initial set of CAs selected by using greedy approach. All these CAs can be used in applications like randomized algorithms, simulations etc., where ease of implementability and portability predominates randomness quality.

Our intuitive understanding is, by increasing number of states of CAs, randomness quality can be improved. On this speculation, in Chapter~\ref{Chap:10-stateCA}, we have taken the number of states per cell as $10$ and utilized these decimal CAs as source of randomness. 
Here, a stricter property has been imposed on the CAs to be unpredictable and again greedy strategies are used for identifying CAs which are promised to be good source of randomness. There have been many possible options to implement these greedy strategies. We have reported two such heuristic algorithms to generate the candidate CAs. Two window-based portable PRNG schemes have been instigated using these CAs -- (1) as decimal number generators and as (2) binary number generators. Finally, we have compared the proposed PRNGs on the same platform with the empirical testbeds. The results have clearly indicated that, our CAs perform at least at par with \verb SFMT19937-64. ~Hence, the decimal CAs based PRNGs are one of the best PRNGs today. Further, it has also validated our hypothesis that, by increasing number of states, a CA based system, where local computation is the signature characteristic, can also be an excellent source of randomness.
\section{Future Directions}

In future, the reported works can be extended to the following directions:
\begin{itemize}

\item In Chapter~\ref{Chap:reversibility}, we have observed in Algorithm~\ref{chap:reversibility:algo:rev_algo} (Page~\pageref{chap:reversibility:algo:rev_algo}) (which tests reversibility of a finite CA of length $n$) that, not much unique nodes in the reachability tree of a given $n$-cell CA are generated. In fact, after certain number of levels, no new nodes are generated. Since, number of levels and number of cells are related, this observation raises the following question-- 
what is the tight upper bound of the necessary and sufficient lattice size $(n_0)$ to decide reversibility of a finite CA with any $n$? Efforts may be taken to answer this question. Further, the algorithm for deciding reversibility of finite CAs may be improved.


\item The greedy strategies of Chapter~\ref{Chap:reversibility} select a set of CAs, candidates to be reversible. However, some rules may exist which does not follow these strategies but yet are reversible for some $n$. Discovering appropriate strategy to identify such rules is a task of future. 

\item Many of the works on finite CAs consider open, particularly null, boundary condition, whereas we have used here periodic boundary condition. In future, all the works, related to reversibility and randomness, reported here can be extended for open boundary condition.

\item In Chapter~\ref{Chap:semireversible}, we have observed that, for the trivial semi-reversible CAs, we need not to complete the construction of minimized reachability tree. Hence, the question arises, is there any sufficient lattice size or any other criteria for identifying such CAs without constructing the minimized tree?

\item Our observation is that semi-reversible CAs offer more complex global behavior than other CAs. For example, a semi-reversible CA, in general, has better randomness quality than a reversible CA. However, exploring these CAs and developing necessary theories behind their behavior is a task of future research. 
For example, it can be explored whether there exists any relation between semi-reversibility, reversibility and chaos.

\item We have noticed that, among the list of $596$ CAs of Chapter~\ref{Chap:3-stateCA_list}, many CAs have better performance in empirical tests than the $3$-state CA of Chapter~\ref{Chap:randomness_survey}. These CAs may also offer other advantages in appropriate application areas. Utilizing these CAs properly is a task of future research.

\item Ternary CAs can also be used for implementing ternary computational circuitry (like arithmetic logic unit etc.) based on balanced-ternary number system.

\item Chapter~\ref{Chap:10-stateCA} has indicated that, decimal CAs based PRNGs are one of the best PRNGs. However, still two questions remain unanswered-- (1) Can we build a CA based PRNG which can fool all empirical tests and beat every other PRNG? (2) Are these CAs based PRNGs cryptographically secured?
\end{itemize}

\cleardoublepage
\phantomsection
\addcontentsline{toc}{chapter}{Author's Statement}
\chapter*{Author's Statement}
\begin{enumerate}
\small{
\item { {Kamalika Bhattacharjee} and Sukanta Das. 
\newblock Random Number Generation using Decimal Cellular Automata.
\newblock \emph{Communications in Nonlinear Science and Numerical Simulation}, 78: 104878, 2019.}

\item { {Kamalika Bhattacharjee} and Sukanta Das. 
\newblock On Finite $1$-Dimensional Cellular Automata: Reversibility and Semi-reversibility
\newblock \emph{arXiv preprint arXiv:1903.0601}, 2019.}

\item {Kamalika Bhattacharjee} and Sukanta Das. 
\newblock A List of Tri-state Cellular Automata which are Potential Pseudo-random Number Generators.
\newblock \emph{International Journal of Modern Physics C}, 29(09): 1850088, 2018.

\item {Kamalika Bhattacharjee}, Nazma Naskar, Souvik Roy, and Sukanta Das. 
\newblock A survey of cellular automata: types, dynamics, non-uniformity and applications. 
\newblock \emph{Natural Computing} (2018). DOI:http://dx.doi.org/10.1007/s11047-018-9696-8

\item {Kamalika Bhattacharjee}.
\newblock Design of an Adder for Ternary Numbers using a $3$- state Cellular Automaton
\newblock {\em In Proceedings of EAIT 2018, Kolkata}. pages 1--4. IEEE, 2018.

\item {Kamalika Bhattacharjee}, Krishnendu Maity and Sukanta Das.
\newblock A Search for Good Pseudo-random Number Generators: Survey and Empirical Studies.
\newblock {\em arXiv preprint arXiv:1811.04035}, 2018.

\item {Kamalika Bhattacharjee}, Dipanjyoti Paul and Sukanta Das.
\newblock Pseudo-random Number Generation using a $3$-state Cellular Automaton.
\newblock {\em International Journal of Modern Physics C}, 28(06): 1750078, 2017.

\item {Kamalika Bhattacharjee} and Sukanta Das.
\newblock Reversibility of $d$-state Finite Cellular Automata.
\newblock {\em Journal of Cellular Automata}, 11(2-3): 213--245, 2016.

\item {Kamalika Bhattacharjee}, Dipanjyoti Paul and Sukanta Das.
\newblock Pseudorandom Pattern Generation using $3$-state Cellular Automata.
\newblock {\em In Proceedings of ACRI 2016, Morocco}, pages 3 -- 13. Springer International
Publishing, 2016.


}
\end{enumerate}

\newpage \thispagestyle{empty} \mbox{}

\pagestyle{plain}
\bibliographystyle{IEEEtran}

\cleardoublepage
\phantomsection
\addcontentsline{toc}{chapter}{Bibliography}
\bibliography{References_thesis}

 \newcommand{\noop}[1]{}
\begin{thebibliography}{100}
\providecommand{\url}[1]{#1}
\csname url@samestyle\endcsname
\providecommand{\newblock}{\relax}
\providecommand{\bibinfo}[2]{#2}
\providecommand{\BIBentrySTDinterwordspacing}{\spaceskip=0pt\relax}
\providecommand{\BIBentryALTinterwordstretchfactor}{4}
\providecommand{\BIBentryALTinterwordspacing}{\spaceskip=\fontdimen2\font plus
\BIBentryALTinterwordstretchfactor\fontdimen3\font minus
  \fontdimen4\font\relax}
\providecommand{\BIBforeignlanguage}[2]{{%
\expandafter\ifx\csname l@#1\endcsname\relax
\typeout{** WARNING: IEEEtran.bst: No hyphenation pattern has been}%
\typeout{** loaded for the language `#1'. Using the pattern for}%
\typeout{** the default language instead.}%
\else
\language=\csname l@#1\endcsname
\fi
#2}}
\providecommand{\BIBdecl}{\relax}
\BIBdecl

\bibitem{morita2016universality}
K.~Morita, ``Universality of 8-state reversible and conservative triangular
  partitioned cellular automata,'' in \emph{International Conference on
  Cellular Automata}.\hskip 1em plus 0.5em minus 0.4em\relax Springer
  International Publishing, 2016, pp. 45--54.

\bibitem{wolfram86}
S.~Wolfram, \emph{Theory and applications of cellular automata}.\hskip 1em plus
  0.5em minus 0.4em\relax Singapore: World Scientific Publishing Co. Ltd.,
  1986.

\bibitem{hilbert19281928}
D.~Hilbert and W.~Ackermann, ``Grundz{\"u}ge der theoretischen logik,''
  \emph{Die Grundlehren der Mathematischen Wissenschaften in
  Einzeldarstellungen}, vol.~27, 1928.

\bibitem{PhysRevLett.88.237901}
S.~Lloyd, ``Computational capacity of the universe,'' \emph{Physical Review
  Letters}, vol.~88, p. 237901, 2002.

\bibitem{Zuse1982}
K.~Zuse, ``The computing universe,'' \emph{International Journal of Theoretical
  Physics}, vol.~21, no. 6--7, pp. 589--600, 1982.

\bibitem{Wolframbook1}
S.~Wolfram, \emph{A New kind of Science}.\hskip 1em plus 0.5em minus
  0.4em\relax Wolfram-Media, 2002.

\bibitem{Hoekstra:2010:SCS:1855006}
A.~G. Hoekstra, J.~Kroc, and P.~M. Sloot, \emph{Simulating Complex Systems by
  Cellular Automata}, 1st~ed.\hskip 1em plus 0.5em minus 0.4em\relax Springer
  Publishing Company, Incorporated, 2010.

\bibitem{Neuma66}
J.~von Neumann, \emph{Theory of Self-Reproducing Automata}, A.~W. Burks,
  Ed.\hskip 1em plus 0.5em minus 0.4em\relax USA: University of Illinois Press,
  1966.

\bibitem{ulam1952random}
S.~Ulam, ``Random processes and transformations,'' in \emph{Proceedings of the
  International Congress on Mathematics}, vol.~2, 1952, pp. 264--275.

\bibitem{Burks}
A.~W. Burks, ``Essays on {C}ellular {A}utomata,'' \emph{Technical Report,
  University of Illinois, Urbana}, 1970.

\bibitem{Thatcher}
J.~W. Thatcher, ``Universality in the von {N}eumann cellular model,'' DTIC
  Document, Tech. Rep., 1964.

\bibitem{Gardner71}
M.~Gardner, ``On cellular automata self-reproduction, the {G}arden of {E}den
  and the {G}ame of `{L}ife','' \emph{Scientific American}, vol. 224, no.~2,
  pp. 112--118, 1971.

\bibitem{Wolfr83}
S.~Wolfram, ``Statistical mechanics of cellular automata,'' \emph{Reviews of
  Modern Physics}, vol.~55, no.~3, pp. 601--644, 1983.

\bibitem{Horte89a}
P.~D. Hortensius, R.~D. McLeod, W.~Pries, D.~M. Miller, and H.~C. Card,
  ``Cellular automata-based pseudorandom number generators for built-in
  self-test,'' \emph{IEEE Transactions on Computer-Aided Design of Integrated
  Circuits and Systems}, vol.~8, no.~8, pp. 842--859, 1989.

\bibitem{ppc1}
P.~Pal~Chaudhuri, D.~Roy~Chowdhury, S.~Nandi, and S.~Chattopadhyay,
  \emph{Additive Cellular Automata -- Theory and Applications}.\hskip 1em plus
  0.5em minus 0.4em\relax IEEE Computer Society Press, USA, ISBN 0-8186-7717-1,
  1997, vol.~1.

\bibitem{SukantaTH}
S.~Das, ``{T}heory and {A}pplications of {N}onlinear {C}ellular {A}utomata {I}n
  {VLSI} {D}esign,'' Ph.D. dissertation, {B}engal {E}ngineering and {S}cience
  {U}niversity, {S}hibpur, {I}ndia, 2007.

\bibitem{Nandi94a}
S.~Nandi, B.~K.~Kar, and P.~Pal~Chaudhuri, ``{T}heory and {A}pplication of
  {C}ellular {A}utomata in {C}ryptography,'' \emph{IEEE Transactions on
  Computers}, vol.~43, no.~12, pp. 1346--1357, 1994.

\bibitem{DBLP:journals/jca/DasR11}
S.~Das and D.~Roy~Chowdhury, ``Cryptographically suitable maximum length
  cellular automata,'' \emph{Journal of Cellular Automata}, vol.~6, no.~6, pp.
  439--459, 2011.

\bibitem{SMITH1972233}
A.~R. Smith, III, ``Real-time language recognition by one-dimensional cellular
  automata,'' \emph{Journal of Computer and System Sciences}, vol.~6, no.~3,
  pp. 233 -- 253, 1972.

\bibitem{Maji05}
P.~Maji and P.~Pal~Chaudhuri, ``Fuzzy cellular automata for modeling pattern
  classifier,'' \emph{IEICE Transactions on Information and Systems}, vol.
  E88-D, no.~4, pp. 691--702, 2005.

\bibitem{maji2007rbffca}
P.~Maji and P.~P. Chaudhuri, ``{RBFFCA}: {A} hybrid pattern classifier using
  radial basis function and fuzzy cellular automata,'' \emph{Fundamenta
  Informaticae}, vol.~78, no.~3, pp. 369--396, 2007.

\bibitem{jen1986invariant}
E.~Jen, ``Invariant strings and pattern recognizing properties of 1{D} {CA},''
  \emph{Journal of statistical physics}, vol.~43, pp. 243--265, 1986.

\bibitem{RAGHAVAN1993145}
R.~Raghavan, ``Cellular automata in pattern recognition,'' \emph{Information
  Sciences}, vol.~70, no.~1, pp. 145 -- 177, 1993.

\bibitem{wolfram86c}
S.~Wolfram, ``Random sequence generation by cellular automata,'' \emph{Advances
  in applied mathematics}, vol.~7, no.~2, pp. 123--169, 1986.

\bibitem{khan98}
A.~R. Khan, ``Replacement of some {G}raphics {R}outines with the help of {2D}
  {C}ellular {A}utomata {A}lgorithms for {F}aster {G}raphics {O}perations,''
  Ph.D. dissertation, University of Kashmir, 1998.

\bibitem{Rosin06}
P.~L. Rosin, ``Training cellular automata for image processing,'' \emph{IEEE
  Transactions on Image Proceedings}, vol.~15, no.~7, pp. 2076--2087, 2006.

\bibitem{MitraDCN96}
S.~Mitra, S.~Das, P.~Pal~Chaudhuri, and S.~Nandi, ``Architecture of a {VLSI}
  chip for modeling amino acid sequence in proteins,'' in \emph{Proceedings of
  9th International Conference on {VLSI} Design}, 1996, pp. 316--317.

\bibitem{DBLP:conf/iicai/GhoshLMC07}
S.~Ghosh, N.~Laskar, S.~Mahapatra, and P.~Pal~Chaudhuri, ``Probabilistic
  cellular automata model for identification of {CpG} island in {DNA} string,''
  in \emph{Proceedings of Indian International Conference on Artificial
  Intelligence}, 2007, pp. 1490--1509.

\bibitem{iet.cp.2013.2283}
J.~Li, Z.~Chen, and T.~Qin, ``\BIBforeignlanguage{English}{Using cellular
  automata to model evolutionary dynamics of social network},''
  \emph{\BIBforeignlanguage{English}{IET Conference Proceedings}}, pp.
  200--205, 2013.

\bibitem{10.1007/978-3-642-40495-5_3}
K.~Ma{\l}ecki, J.~Jankowski, and M.~Rokita, ``Application of graph cellular
  automata in social network based recommender system,'' in \emph{Computational
  Collective Intelligence. Technologies and Applications}.\hskip 1em plus 0.5em
  minus 0.4em\relax Springer Berlin Heidelberg, 2013, pp. 21--29.

\bibitem{6108515}
R.~Hunt, E.~Mendi, and C.~Bayrak, ``Using cellular automata to model social
  networking behavior,'' in \emph{12th International Symposium on Computational
  Intelligence and Informatics}, 2011, pp. 287--290.

\bibitem{ZIMBRES2009157}
R.~A. Zimbres and P.~P.~B. de~Oliveira, ``Dynamics of quality perception in a
  social network: A cellular automaton based model in aesthetics services,''
  \emph{Electronic Notes in Theoretical Computer Science}, vol. 252, pp. 157 --
  180, 2009.

\bibitem{Dabbaghian2012}
V.~Dabbaghian, V.~K. Mago, T.~Wu, C.~Fritz, and A.~Alimadad, ``Social
  interactions of eating behaviour among high school students: a cellular
  automata approach,'' \emph{BMC Medical Research Methodology}, vol.~12, no.~1,
  p. 155, 2012.

\bibitem{doi:10.1080/13658810210157769}
F.~Wu, ``Calibration of stochastic cellular automata: the application to
  rural-urban land conversions,'' \emph{International Journal of Geographical
  Information Science}, vol.~16, no.~8, pp. 795--818, 2002.

\bibitem{goldenberg2001using}
J.~Goldenberg, B.~Libai, and E.~Muller, ``Using complex systems analysis to
  advance marketing theory development: Modeling heterogeneity effects on new
  product growth through stochastic cellular automata,'' \emph{Academy of
  Marketing Science Review}, vol.~9, no.~3, pp. 1--18, 2001.

\bibitem{doi:10.1002}
J.~Klüver and J.~Schmidt, ``Control parameters in {B}oolean networks and
  cellular automata revisited from a logical and a sociological point of
  view,'' \emph{Complexity}, vol.~5, no.~1, pp. 45--52, 1999.

\bibitem{doi:10.1080/136588198241617}
K.~C. Clarke and L.~J. Gaydos, ``Loose-coupling a cellular automaton model and
  {GIS}: long-term urban growth prediction for {S}an {F}rancisco and
  {W}ashington/{B}altimore,'' \emph{International Journal of Geographical
  Information Science}, vol.~12, no.~7, pp. 699--714, 1998.

\bibitem{Chopard}
B.~Chopard and M.~Droz, \emph{{C}ellular {A}utomata {M}odelling of {P}hysical
  {S}ystems}.\hskip 1em plus 0.5em minus 0.4em\relax {C}ambridge {U}niversity
  {P}ress, 1998.

\bibitem{nagel1992cellular}
K.~Nagel and M.~Schreckenberg, ``A cellular automaton model for freeway
  traffic,'' \emph{Journal de physique I}, vol.~2, no.~12, pp. 2221--2229,
  1992.

\bibitem{PhysRevLett-35}
J.~Matsukidaira and K.~Nishinari, ``Euler-{L}agrange correspondence of cellular
  automaton for traffic-flow models,'' \emph{Physical Review Letters}, vol.~90,
  pp. 088\,701(1--4), 2003.

\bibitem{hartman90}
H.~Hartman and P.~Tamayo, ``Reversible cellular automata and chemical
  turbulence,'' \emph{Physica D: Nonlinear Phenomena}, vol.~45, no.~1, pp. 293
  -- 306, 1990.

\bibitem{Kari2012180}
J.~Kari, ``Universal pattern generation by cellular automata,''
  \emph{Theoretical Computer Science}, vol. 429, pp. 180 -- 184, 2012.

\bibitem{aspdac04}
S.~Das, D.~Dey, S.~Sen, B.~K. Sikdar, and P.~P. Chaudhuri, ``An {E}fficient
  {D}esign of {N}on-linear {CA} {B}ased {PRPG} for {VLSI} {C}ircuit
  {T}esting,'' in \emph{Proceedings of the Asia and South Pacific Design
  Automation Conference}, 2004, pp. 110--112.

\bibitem{Kutrib20081142}
M.~Kutrib and A.~Malcher, ``Fast reversible language recognition using cellular
  automata,'' \emph{Information and Computation}, vol. 206, no. 9-10, pp. 1142
  -- 1151, 2008, {S}pecial {I}ssue: 1st International Conference on Language
  and Automata Theory and Applications (LATA 2007).

\bibitem{Kirschenmann1972}
P.~Kirschenmann, ``Concepts of randomness,'' \emph{Journal of Philosophical
  Logic}, vol.~1, no. 3-4, pp. 395--414, 1972.

\bibitem{doi:10.1093/bjps/axi138}
A.~Eagle, ``Randomness is unpredictability,'' \emph{The British Journal for the
  Philosophy of Science}, vol.~56, no.~4, pp. 749--790, 2005.

\bibitem{0034-4885-80-12-124001}
M.~N. Bera, A.~Acín, M.~Kuś, M.~W. Mitchell, and M.~Lewenstein, ``Randomness
  in quantum mechanics: philosophy, physics and technology,'' \emph{Reports on
  Progress in Physics}, vol.~80, no.~12, p. 124001, 2017.

\bibitem{hedlund69}
G.~A. Hedlund, ``Endomorphisms and automorphisms of the shift dynamical
  system,'' \emph{Mathematical Systems Theory}, vol.~3, no.~4, pp. 320--375,
  1969.

\bibitem{Richa72}
D.~Richardson, ``Tessellations with local transformations,'' \emph{Journal of
  Computer and System Sciences}, vol.~6, pp. 373--388, 1972.

\bibitem{Amoroso72}
S.~Amoroso and Y.~N. Patt, ``Decision procedures for surjectivity and
  injectivity of parallel maps for tesselation structures,'' \emph{Journal of
  Computer and System Sciences}, vol.~6, pp. 448--464, 1972.

\bibitem{nasu1977local}
M.~Nasu, ``Local maps inducing surjective global maps of one-dimensional
  tessellation automata,'' \emph{Mathematical Systems Theory}, vol.~11, no.~1,
  pp. 327--351, 1977.

\bibitem{Maruoka197947}
A.~Maruoka and M.~Kimura, ``Injectivity and surjectivity of parallel maps for
  cellular automata,'' \emph{Journal of Computer and System Sciences}, vol.~18,
  no.~1, pp. 47 -- 64, 1979.

\bibitem{Maruoka1982269}
------, ``Strong surjectivity is equivalent to c-injectivity,''
  \emph{Theoretical Computer Science}, vol.~18, no.~3, pp. 269 -- 277, 1982.

\bibitem{sato77}
T.~Sato and N.~Honda, ``Certain relations between properties of maps of
  tessellation automata,'' \emph{Journal of Computer and System Sciences},
  vol.~15, no.~2, pp. 121 -- 145, 1977.

\bibitem{di1975reversibility}
S.~Di~Gregorio and G.~Trautteur, ``On reversibility in cellular automata,''
  \emph{Journal of Computer and System Sciences}, vol.~11, no.~3, pp. 382--391,
  1975.

\bibitem{culik1987invertible}
K.~Culik, ``On invertible cellular automata,'' \emph{Complex Systems}, vol.~1,
  no.~6, pp. 1035--1044, 1987.

\bibitem{suttner91}
K.~Sutner, ``De bruijin graphs and linear cellular automata,'' \emph{Complex
  Systems}, vol.~5, no.~1, pp. 19--30, 1991.

\bibitem{zubeyir11}
Z.~Cinkir, H.~Akin, and I.~Siap, ``Reversibilty of 1d cellular automata with
  periodic boundary over finite fields $ \mathbb{Z}_{p}$,'' \emph{Journal of
  Statistical Physics}, vol. 143, pp. 807--823, 2011.

\bibitem{entcs/DasS09}
S.~Das and B.~K. Sikdar, ``Characterization of 1-d periodic boundary reversible
  {CA},'' \emph{Electronic Notes in Theoretical Computer Science}, vol. 252,
  pp. 205--227, 2009.

\bibitem{marti2011reversibility}
A.~Mart{\i}n~del Rey and G.~Rodr{\i}guez~S{\i}nchez, ``Reversibility of linear
  cellular automata,'' \emph{Applied Mathematics and Computation}, vol. 217,
  no.~21, pp. 8360--8366, 2011.

\bibitem{Ino05}
S.~Inokuchi, K.~Honda, H.~Y. Lee, T.~Sato, Y.~Mizoguchi, and Y.~Kawahara, ``On
  reversible cellular automata with finite cell array,'' in
  \emph{Unconventional Computation}.\hskip 1em plus 0.5em minus 0.4em\relax
  Springer, 2005, pp. 130--141.

\bibitem{Soumya2011}
S.~Ghosh, N.~S.~Maiti, P.~Pal~Chaudhuri, and B.~K.~Sikdar, ``On invertible
  three neighborhood null-boundary uniform cellular automata,'' \emph{Complex
  Systems}, vol.~20, pp. 47--65, 2011.

\bibitem{Soumya2010}
N.~S.~Maiti, S.~Ghosh, S.~Munshi, and P.~Pal~Chaudhuri, ``Linear time algorithm
  for identifying the invertibility of null-boundary three neighborhood
  cellular automata,'' \emph{Complex Systems}, vol.~19, no.~1, pp. 89--113,
  2010.

\bibitem{alonso2009elementary}
R.~Alonso-Sanz and L.~Bull, ``Elementary cellular automata with minimal memory
  and random number generation,'' \emph{Complex Systems}, vol.~18, no.~2, pp.
  195 -- 213, 2009.

\bibitem{tcad/DasS10}
S.~Das and B.~K. Sikdar, ``A scalable test structure for multicore chip,''
  \emph{IEEE Transactions on Computer-Aided Design of Integrated Circuits and
  Systems}, vol.~29, no.~1, pp. 127--137, 2010.

\bibitem{114093}
P.~H. Bardell, ``Analysis of cellular automata used as pseudorandom pattern
  generators,'' in \emph{Proceedings International Test Conference}, 1990, pp.
  762--768.

\bibitem{Tomassini96}
M.~Sipper and M.~Tomassini, ``Generating parallel random number generators by
  cellular programming,'' \emph{International Journal of Modern Physics C},
  vol.~7, no.~2, pp. 180--190, 1996.

\bibitem{Marco00}
M.~Tomassini, M.~Sipper, and M.~Perrenoud, ``On the generation of high-quality
  random numbers by two-dimensional cellular automata,'' \emph{IEEE
  Transactions on Computers}, vol.~49, no.~10, pp. 1146--1151, 2000.

\bibitem{Guan04a}
S.~Guan, S.~Zhang, and T.~Quieta, ``2-{D} {CA} {V}ariation {W}ith {A}symetric
  {N}eighborship for {P}sedorandom {N}umber {G}eneration,'' \emph{IEEE
  Transactions on Computer-Aided Design of Integrated Circuits and Systems},
  vol.~23, no.~3, pp. 378--388, 2004.

\bibitem{Guan04}
S.~Guan and S.~K. Tan, ``Pseudorandom {N}umber {G}eneration {W}ith
  {S}elf-{P}rogrammable {C}ellular {A}utomata,'' \emph{IEEE Transactions on
  Computer-Aided Design of Integrated Circuits and Systems}, vol.~23, no.~7,
  pp. 1095--1101, 2004.

\bibitem{Matsumoto:1998:MTE:272991.272995}
M.~Matsumoto and T.~Nishimura, ``Mersenne {T}wister: A 623-dimensionally
  equidistributed uniform pseudo-random number generator,'' \emph{ACM
  Transactions on Modeling and Computer Simulation}, vol.~8, no.~1, pp. 3--30,
  1998.

\bibitem{Saito2008}
M.~Saito and M.~Matsumoto, ``{SIMD}-{O}riented {F}ast {M}ersenne {T}wister: a
  128-bit {P}seudorandom {N}umber {G}enerator,'' in \emph{7th International
  Conference on Monte Carlo and Quasi-Monte Carlo Methods in Scientific
  Computing}.\hskip 1em plus 0.5em minus 0.4em\relax Springer Berlin
  Heidelberg, 2008, pp. 607--622.

\bibitem{Saito2009}
------, ``A {PRNG} specialized in double precision floating point numbers using
  an affine transition,'' in \emph{8th International Conference on Monte Carlo
  and Quasi-Monte Carlo Methods in Scientific Computing}.\hskip 1em plus 0.5em
  minus 0.4em\relax Springer Berlin Heidelberg, 2009, pp. 589--602.

\bibitem{Smith71}
A.~R. Smith, III, ``Cellular automata complexity trade-offs,''
  \emph{Information and Control}, vol.~18, no.~5, pp. 466--482, 1971.

\bibitem{bohm2002wholeness}
D.~Bohm, \emph{Wholeness and the implicate order}.\hskip 1em plus 0.5em minus
  0.4em\relax Routledge, ISBN 0-203-99515-5, 1980.

\bibitem{moreno1932application}
J.~L. Moreno, E.~S. Whitin, and H.~H. Jennings, \emph{Application of the group
  method to classification}.\hskip 1em plus 0.5em minus 0.4em\relax National
  committee on prisons and prison labor, New York. Reprinted in The first book
  on group psychotherapy, 3d ed., 1957 [pp. 3 - 103], 1932.

\bibitem{McCulloch1943}
W.~S. McCulloch and W.~Pitts, ``A logical calculus of the ideas immanent in
  nervous activity,'' \emph{The bulletin of mathematical biophysics}, vol.~5,
  no.~4, pp. 115--133, 1943.

\bibitem{banks1970universality}
E.~R. Banks, ``Universality in cellular automata,'' in \emph{Proceedings of
  IEEE Conference Record of 11th Annual Symposium on Switching and Automata
  Theory}, 1970, pp. 194--215.

\bibitem{Toffo87}
T.~Toffoli and N.~Margolus, \emph{Cellular Automata Machines: A New Environment
  for Modeling}.\hskip 1em plus 0.5em minus 0.4em\relax Cambridge, MA, USA: MIT
  Press, 1987.

\bibitem{moore1962machine}
E.~F. Moore, ``Machine models of self-reproduction,'' in \emph{Proceedings of
  Symposia in Applied Mathematics}, vol.~14, 1962, pp. 17--33.

\bibitem{refId0}
K.~Morita, M.~Margenstern, and K.~Imai, ``Universality of reversible hexagonal
  cellular automata,'' \emph{RAIRO - Theoretical Informatics and Applications},
  vol.~33, no.~6, pp. 535--550, 1999.

\bibitem{Siap2d}
I.~Siap, H.~Akin, and S.~U\u{g}uz, ``Structure and reversibility of {2D}
  hexagonal cellular automata,'' \emph{Computers $\&$ Mathematics with
  Applications}, vol.~62, no.~11, pp. 4161--4169, 2011.

\bibitem{IMAI2000181}
K.~Imai and K.~Morita, ``A computation-universal two-dimensional 8-state
  triangular reversible cellular automaton,'' \emph{Theoretical Computer
  Science}, vol. 231, no.~2, pp. 181 -- 191, 2000.

\bibitem{margenstern1999polynomial}
M.~Margenstern and K.~Morita, ``A polynomial solution for 3-{SAT} in the space
  of cellular automata in the hyperbolic plane,'' \emph{Journal of Universal
  Computer Science}, vol.~5, no.~9, pp. 563--573, 1999.

\bibitem{Margenstern200199}
------, ``{NP} problems are tractable in the space of cellular automata in the
  hyperbolic plane,'' \emph{Theoretical Computer Science}, vol. 259, no. 1-2,
  pp. 99--128, 2001.

\bibitem{Marr20}
C.~Marr and M.~T. H\"{u}tt, ``Outer-totalistic cellular automata on graphs,''
  \emph{Physics Letters A}, vol. 373, no.~5, pp. 546 -- 549, 2009.

\bibitem{tomassini29}
M.~Tomassini, M.~Giacobini, and C.~Darabos, ``Evolution and dynamics of
  small-world cellular automata,'' \emph{Complex Systems}, vol.~15, pp.
  261--284, 2005.

\bibitem{Darabos7}
C.~Darabos, M.~Giacobini, and M.~Tomassini, ``Performance and robustness of
  cellular automata computation on irregular networks.'' \emph{Advances in
  Complex Systems}, vol.~10, pp. 85--110, 2007.

\bibitem{Jump74}
J.~R. Jump and J.~S. Kirtane, ``On the interconnection structure of cellular
  networks,'' \emph{Information Control}, vol.~24, pp. 74--91, 1974.

\bibitem{boccara-1998-31}
N.~Boccara and H.~Fuk\'{s}, ``Cellular automaton rules conserving the number of
  active sites,'' \emph{Journal of Physics A: Mathematical and General},
  vol.~31, no.~28, p. 6007, 1998.

\bibitem{Dyer80}
C.~Dyer, ``One-way bounded cellular automata,'' \emph{Information Control},
  vol.~44, pp. 261--281, 1980.

\bibitem{Pries86}
W.~Pries, A.~Thanailakis, and H.~C. Card, ``Group properties of cellular
  automata and {VLSI} applications,'' \emph{IEEE Transactions on Computers},
  vol. C-35, no.~12, pp. 1013--1024, 1986.

\bibitem{biplabtcad}
B.~K.~Sikdar, N.~Ganguly, and P.~Pal~Chaudhuri, ``Design of hierarchical
  cellular automata for on-chip test pattern generator,'' \emph{IEEE
  Transactions on Computer-Aided Design of Integrated Circuits and Systems},
  vol.~21, no.~12, pp. 1530--1539, 2002.

\bibitem{Wolframbook}
S.~Wolfram, \emph{{C}ellular {A}utomata and {C}omplexity--Collected
  Papers}.\hskip 1em plus 0.5em minus 0.4em\relax Westview Press, 1994.

\bibitem{Packa85b}
N.~H. Packard and S.~Wolfram, ``Two-dimensional cellular automata,''
  \emph{Journal of {S}tatistical {P}hysics}, vol.~38, no. 5/6, pp. 901--946,
  1985.

\bibitem{Kari90}
J.~Kari, ``Reversibility of {2D} cellular automata is undecidable,''
  \emph{Physica D: Nonlinear Phenomena}, vol.~45, pp. 386--395, 1990.

\bibitem{Durand93}
B.~Durand, ``Global properties of {2D} cellular automata: Some complexity
  results,'' in \emph{Proceedings of 18th International Symposium on
  Mathematical Foundations of Computer Science}.\hskip 1em plus 0.5em minus
  0.4em\relax Springer Berlin Heidelberg, 1993, pp. 433--441.

\bibitem{terrier2004two}
V.~Terrier, ``Two-dimensional cellular automata and their neighborhoods,''
  \emph{Theoretical computer science}, vol. 312, no. 2-3, pp. 203--222, 2004.

\bibitem{deOliveira20061}
G.~M.~B. de~Oliveira and S.~R.~C. Siqueira, ``Parameter characterization of
  two-dimensional cellular automata rule space,'' \emph{Physica D: Nonlinear
  Phenomena}, vol. 217, no.~1, pp. 1 -- 6, 2006.

\bibitem{chr2D}
S.~U{\u{g}}uz, U.~Sahin, H.~Akin, and I.~Siap, ``Self-replicating patterns in
  {2D} linear cellular automata,'' \emph{International Journal of Bifurcation
  and Chaos}, vol.~24, no.~1, p. 1430002, 2014.

\bibitem{gandin19973d}
C.-A. Gandin and M.~Rappaz, ``A {3D} cellular automaton algorithm for the
  prediction of dendritic grain growth,'' \emph{Acta Materialia}, vol.~45,
  no.~5, pp. 2187--2195, 1997.

\bibitem{Palas1}
P.~Sarkar and R.~Barua, ``{M}ulti-dimensional $\sigma$-automata,
  $\pi$-polynomial and generalized $s$-matrices,'' \emph{{T}heoretical
  {C}omputer {S}cience}, vol. 197, no. 1-2, pp. 111--138, 1998.

\bibitem{miller2005two}
D.~B. Miller and E.~Fredkin, ``Two-state, reversible, universal cellular
  automata in three dimensions,'' in \emph{Proceedings of the 2nd conference on
  Computing frontiers}, 2005, pp. 45--51.

\bibitem{Mo201431-3DCA}
Y.~Mo, B.~Ren, W.~Yang, and J.~Shuai, ``The 3-dimensional cellular automata for
  {HIV} infection,'' \emph{Physica A: Statistical Mechanics and its
  Applications}, vol. 399, pp. 31 -- 39, 2014.

\bibitem{DENNUNZIO201440}
A.~Dennunzio, E.~Formenti, and M.~Weiss, ``Multidimensional cellular automata:
  closing property, quasi-expansivity, and (un)decidability issues,''
  \emph{Theoretical Computer Science}, vol. 516, pp. 40 -- 59, 2014.

\bibitem{Codd68}
E.~F. Codd, \emph{Cellular {A}utomata}.\hskip 1em plus 0.5em minus 0.4em\relax
  Academic Press Inc., 1968.

\bibitem{Arbib66}
M.~A. Arbib, ``Simple self-reproducing universal automata,'' \emph{Information
  and {C}ontrol}, vol.~9, pp. 177--189, 1966.

\bibitem{Banks71}
E.~R. Banks, ``Information {P}rocessing and {T}ransmission in {C}ellular
  {A}utomata,'' Ph.D. dissertation, {M}{I}{T}, 1971.

\bibitem{Martin84a}
O.~Martin, A.~M. Odlyzko, and S.~Wolfram, ``Algebraic {P}roperties of
  {C}ellular {A}utomata,'' \emph{Communications in Mathematical Physics},
  vol.~93, pp. 219--258, 1984.

\bibitem{vlsi00d}
B.~K.~Sikdar, K.~Paul, G.~P.~Biswas, V.~Boppana, C.~Yang, S.~Mukherjee, and
  P.~Pal~Chaudhuri, ``Theory and application of {GF}($2^p$) cellular automata
  as on-chip test pattern generator,'' in \emph{Proceedings of 13th
  International Conference on VLSI Design}, 2000, pp. 556--561.

\bibitem{ito1983linear}
M.~It{\^o}, N.~{\^O}sato, and M.~Nasu, ``Linear cellular automata over
  $\mathbb{ Z}_m$,'' \emph{Journal of Computer and System Sciences}, vol.~27,
  no.~1, pp. 125--140, 1983.

\bibitem{CATTANEO2004249}
G.~Cattaneo, A.~Dennunzio, and L.~Margara, ``Solution of some conjectures about
  topological properties of linear cellular automata,'' \emph{Theoretical
  Computer Science}, vol. 325, no.~2, pp. 249 -- 271, 2004.

\bibitem{Sarkar00}
P.~Sarkar, ``A brief history of cellular automata,'' \emph{{ACM} Computing
  Survey}, vol.~32, no.~1, pp. 80--107, 2000.

\bibitem{Tsali91}
P.~Tsalides, T.~A. York, and A.~Thanailakis, ``Pseudo-random {N}umber
  {G}enerators for {VLSI} {S}ystems based on {L}inear {C}ellular {A}utomata,''
  \emph{IEE {P}roceedings {E} - {C}omputers and {D}igital {T}echniques}, vol.
  138, no.~4, pp. 241--249, 1991.

\bibitem{Jin2012538}
J.~Jin and Z.~H. Wu, ``A secret image sharing based on neighborhood
  configurations of 2-d cellular automata,'' \emph{Optics $\&$ Laser
  Technology}, vol.~44, no.~3, pp. 538 -- 548, 2012.

\bibitem{uguz2013reversibility}
S.~U{\u{g}}uz, H.~Akin, and I.~Siap, ``Reversibility algorithms for 3-state
  hexagonal cellular automata with periodic boundaries,'' \emph{International
  Journal of Bifurcation and Chaos}, vol.~23, no.~06, p. 1350101, 2013.

\bibitem{Nandi96}
S.~Nandi and P.~Pal~Chaudhuri, ``Analysis of periodic and intermediate boundary
  90/150 cellular automata,'' \emph{{IEEE} Transactions on Computer}, vol.~45,
  no.~1, pp. 1--12, 1996.

\bibitem{PhysRevE-CKS}
S.~Cheybani, J.~Kert{\'e}sz, and M.~Schreckenberg, ``Stochastic boundary
  conditions in the deterministic {N}agel-{S}chreckenberg traffic model,''
  \emph{Physical Review E}, vol.~63, p. 016107, 2000.

\bibitem{mitcourseware}
J.~Li, E.~Demaine, and M.~Gymrek, ``Es.268 {T}he {M}athematics in {T}oys and
  {G}ames, {S}pring 2010. ({M}assachusetts {I}nstitute of {T}echnology: {MIT}
  opencourseware),'' in \emph{http://ocw.mit.edu (Accessed)}, 2010.

\bibitem{Fates20061}
N.~Fat{\`{e}}s, E.~Thierry, M.~Morvan, and N.~Schabanel, ``Fully asynchronous
  behavior of double-quiescent elementary cellular automata,''
  \emph{Theoretical Computer Science}, vol. 362, no. 1-3, pp. 1--16, 2006.

\bibitem{fates00608485}
N.~Fat{\`e}s, ``{Stochastic Cellular Automata Solutions to the Density
  Classification Problem - When randomness helps computing},'' \emph{{Theory of
  Computing Systems}}, vol.~53, no.~2, pp. 223--242, 2013.

\bibitem{Soto2008}
J.~M.~G. Soto, ``Computation of explicit preimages in one-dimensional cellular
  automata applying the de bruijn diagram,'' \emph{Special Issues on Journal of
  Cellular Automata}, vol.~3, no.~3, pp. 219--230, 2008.

\bibitem{voorhees2008remarks}
B.~Voorhees, ``Remarks on applications of de bruijn diagrams and their
  fragments.'' \emph{Journal of Cellular Automata}, vol.~3, no.~3, pp. 187 --
  204, 2008.

\bibitem{Martinez2008}
G.~J. Martinez, H.~V. McIntosh, J.~C. S.~T. Mora, and S.~V.~C. Vergara,
  ``Determining a regular language by glider-based structures called phases
  $f_{i}\_1$ in rule 110,'' \emph{Special Issues on Journal of Cellular
  Automata}, vol.~3, no.~3, pp. 231--270, 2008.

\bibitem{de2014complete}
G.~J. Mart{\'\i}nez, A.~Adamatzky, and H.~V. McIntosh, ``Complete
  characterization of structure of rule 54,'' \emph{Complex Systems}, vol.~23,
  no.~3, pp. 259--293, 2014.

\bibitem{Mora2008}
S.~T. Mora, J.~Carlos, M.~G. Hern{\'a}ndez, and S.~V. Chapa~Vergara, ``Pair
  diagram and cyclic properties characterizing the inverse of reversible
  automata,'' \emph{Special Issues on Journal of Cellular Automata}, vol.~3,
  no.~3, pp. 205--218, 2008.

\bibitem{Betel2013}
H.~Betel, P.~P.~B. de~Oliveira, and P.~Flocchini, ``Solving the parity problem
  in one-dimensional cellular automata,'' \emph{Natural Computing}, vol.~12,
  no.~3, pp. 323--337, 2013.

\bibitem{Mariot2017}
L.~Mariot, A.~Leporati, A.~Dennunzio, and E.~Formenti, ``Computing the periods
  of preimages in surjective cellular automata,'' \emph{Natural Computing},
  vol.~16, no.~3, pp. 367--381, 2017.

\bibitem{DennunzioFP13}
A.~Dennunzio, E.~Formenti, and J.~Provillard, ``Local rule distributions,
  language complexity and non-uniform cellular automata,'' \emph{Theoretical
  Computer Science}, vol. 504, pp. 38 -- 51, 2013.

\bibitem{Das90c}
A.~K.~Das, A.~Ganguly, A.~Dasgupta, S.~Bhawmik, and P.~Pal~Chaudhuri,
  ``Efficient characterisation of cellular automata,'' \emph{IEE Proceedings E
  -- Computers and Digital Techniques}, vol. 137, no.~1, pp. 81--87, 1990.

\bibitem{biplab}
B.~K.~Sikdar, N.~Ganguly, A.~Karmakar, S.~Chowdhury, and P.~Pal~Chaudhuri,
  ``{M}ultiple {A}ttractor {C}ellular {A}utomata for {H}ierarchical {D}iagnosis
  of {VLSI} {C}ircuits,'' in \emph{Proceedings of {A}sian {T}est {S}ymposium
  ({ATS})}, 2001, pp. 385 -- 390.

\bibitem{Chattopadhyay2d}
P.~Chattopadhyay, P.~Pal~Choudhury, and K.~Dihidar, ``Characterisation of a
  particular hybrid transformation of two-dimensional cellular automata,''
  \emph{Computers $\&$ Mathematics with Applications}, vol.~38, no. 5 - 6, pp.
  207 -- 216, 1999.

\bibitem{Acri04}
S.~Das, B.~K.~Sikdar, and P.~Pal~Chaudhuri, ``{C}haracterization of
  {R}eachable/{N}onreachable {C}ellular {A}utomata {S}tates,'' in
  \emph{Proceedings of International Conference on Cellular Automata, Research
  and Industry (ACRI)}.\hskip 1em plus 0.5em minus 0.4em\relax Springer Berlin
  Heidelberg, 2004, pp. 813--822.

\bibitem{Adak2016OnSO}
S.~Adak, N.~Naskar, P.~Maji, and S.~Das, ``On synthesis of non-uniform cellular
  automata having only point attractors,'' \emph{Journal of Cellular Automata},
  vol.~12, no. 1-2, pp. 81--100, 2016.

\bibitem{langton90}
C.~G. Langton, ``Computation at the edge of chaos,'' \emph{Physica D: Nonlinear
  Phenomena}, vol.~42, pp. 12--37, 1990.

\bibitem{WuenscheRePEc}
A.~Wuensche, ``Complexity in {O}ne-{D} {C}ellular {A}utomata: {G}liders,
  {B}asins of {A}ttraction and the {Z} {P}arameter,'' Santa Fe Institute,
  Working Papers 94-04-025, 1994.

\bibitem{WuenscheI}
------, ``{C}lassifying {C}ellular {A}utomata {A}utomatically,'' \emph{{S}anta
  {F}e {I}nstitute {W}orking {P}aper 98-02-018}, 1998.

\bibitem{voorhees1997some}
B.~Voorhees, ``Some parameters characterizing cellular automata rules,''
  \emph{Complex Systems}, vol.~11, no.~5, pp. 373--385, 1997.

\bibitem{LangtonII}
W.~Li, N.~H. Packard, and C.~G. Langton, ``Transition {P}henomena in {C}ellular
  {A}utomata rule space,'' \emph{Physica D: Nonlinear Phenomena}, vol.~45, pp.
  77--94, 1990.

\bibitem{wolfram84b}
S.~Wolfram, ``Universality and complexity in cellular automata,''
  \emph{Physica}, vol.~10, pp. 1--35, 1984.

\bibitem{DennunzioFP14}
A.~Dennunzio, E.~Formenti, and J.~Provillard, ``Three research directions in
  non-uniform cellular automata,'' \emph{Theoretical Computer Science}, vol.
  559, pp. 73--90, 2014.

\bibitem{FiatNazim}
N.~Fat{\'e}s, ``{FiatLux : cellular automata and discrete complex systems
  simulator},'' \url{http://fiatlux.loria.fr/}, 2017, [Online; accessed on
  August 25, 2017].

\bibitem{wuensche1992global}
A.~Wuensche and M.~Lesser, \emph{Global Dynamics Of Cellular Automata: An Atlas
  Of Basin Of Attraction Fields Of One-dimensional Cellular Automata}.\hskip
  1em plus 0.5em minus 0.4em\relax Andrew Wuensche, 1992, no.~1.

\bibitem{DDLab}
A.~Wuensche, ``{Discrete Dynamics Lab},'' \url{http://www.ddlab.com/}, 2017,
  [Online; accessed on August 25, 2017].

\bibitem{Vichn84}
G.~Y. Vichniac, ``Simulating physics with cellular automata,'' \emph{Physica D:
  Nonlinear Phenomena}, vol.~10, no. 1-2, pp. 96 -- 116, 1984.

\bibitem{Frisc86}
U.~Frisch, B.~Hasslacher, and Y.~Pomeau, ``Lattice gas automata for the
  {N}avier-{S}tokes equation,'' \emph{Physical Review Letters}, vol.~56,
  no.~14, pp. 1505--1508, 1986.

\bibitem{Grins85}
G.~Grinstein, C.~Jayaprakash, and Y.~He, ``Statistical mechanics of
  probabilistic cellular automata,'' \emph{Physical Review Letters}, vol.~55,
  pp. 2527--2530, 1985.

\bibitem{379-440-26}
B.~S. Kerner, ``Three-phase traffic theory and highway capacity,''
  \emph{Physica A: Statistical Mechanics and its Applications}, vol. 333,
  no.~C, pp. 379--440, 2004.

\bibitem{Smith:1971}
A.~R. Smith, III, ``Simple computation-universal cellular spaces,''
  \emph{Journal of the ACM}, vol.~18, no.~3, pp. 339--353, 1971.

\bibitem{Morita89}
K.~Morita and M.~Harao, ``Computation universality of one dimensional
  reversible injective cellular automata,'' \emph{IEICE Transactions}, vol. E
  72, pp. 758--762, 1989.

\bibitem{Dubacq95}
J.~C. Dubacq, ``How to simulate {T}uring machines by invertible one-dimensional
  cellular automata,'' \emph{International Journal of Foundations Computer
  science}, vol.~6, no.~4, pp. 395--402, 1995.

\bibitem{iirgen1987simple}
J.~Albert and K.~Culik~II, ``A simple universal cellular automaton and its
  one-way and totalistic version,'' \emph{Complex Systems}, vol.~1, pp. 1--16,
  1987.

\bibitem{Durand-Lose98}
J.~O. Durand{-}Lose, ``About the universality of the {B}illiard ball model,''
  in \emph{Proceedings International Colloquium Universal Machines and
  Computations}, 1998, pp. 118--132.

\bibitem{lindgren90}
K.~Lindgren and M.~G. Nordahl, ``Universal computation in simple
  one-dimensional cellular automata,'' \emph{Complex Systems}, vol.~4, no.~3,
  pp. 299--318, 1990.

\bibitem{cook2004universality}
M.~Cook, ``Universality in elementary cellular automata,'' \emph{Complex
  systems}, vol.~15, no.~1, pp. 1--40, 2004.

\bibitem{Kari05}
J.~Kari, ``Theory of cellular automata: {A} survey,'' \emph{Theoretical
  Computer Science}, vol. 334, no. 1-3, pp. 3--33, 2005.

\bibitem{ollinger2011four}
N.~Ollinger and G.~Richard, ``Four states are enough!'' \emph{Theoretical
  Computer Science}, vol. 412, pp. 22--32, 2011.

\bibitem{Martin94}
B.~Martin, ``A universal cellular automata in quasi-linear time and its {S}-m-n
  form,'' \emph{Theoritical Computer Science}, vol. 123, no.~2, pp. 199--237,
  1994.

\bibitem{ollinger2002quest}
N.~Ollinger, ``The quest for small universal cellular automata,'' in
  \emph{Proceedings of 29th International Colloquium Automata, Languages and
  Programming}.\hskip 1em plus 0.5em minus 0.4em\relax Springer Berlin
  Heidelberg, 2002, pp. 318--329.

\bibitem{ollinger2003intrinsic}
------, ``The intrinsic universality problem of one-dimensional cellular
  automata,'' in \emph{Proceedings of International Symposium on Theoretical
  Aspects of Computer Science}.\hskip 1em plus 0.5em minus 0.4em\relax
  Springer, 2003, pp. 632--641.

\bibitem{toffoli77}
T.~Toffoli, ``Computation and construction universality of reversible cellular
  automata,'' \emph{Journal of Computer and System Sciences}, vol.~15, pp.
  213--231, 1977.

\bibitem{morita1995reversible}
K.~Morita, ``Reversible simulation of one-dimensional irreversible cellular
  automata,'' \emph{Theoretical Computer Science}, vol. 148, no.~1, pp.
  157--163, 1995.

\bibitem{Kari94}
J.~Kari, ``Reversibility and surjectivity problems of cellular automata,''
  \emph{Journal of Computer and System Sciences}, vol.~48, no.~1, pp. 149--182,
  1994.

\bibitem{Kari2005}
------, ``Reversible cellular automata,'' in \emph{Proceedings of International
  Conference on Developments in Language Theory}.\hskip 1em plus 0.5em minus
  0.4em\relax Springer Berlin Heidelberg, 2005, pp. 57--68.

\bibitem{moraal2000graph}
H.~Moraal, ``Graph-theoretical characterization of invertible cellular
  automata,'' \emph{Physica D: Nonlinear Phenomena}, vol. 141, no. 1--2, pp.
  1--18, 2000.

\bibitem{MoraMM06}
J.~C. S.~T. Mora, G.~J. Mart{\'i}nez, and H.~V. McIntosh, ``The inverse
  behavior of a reversible one-dimensional cellular automaton obtained by a
  single welch diagram,'' \emph{Journal of Cellular Automata}, vol.~1, no.~1,
  pp. 25--39, 2006.

\bibitem{Morita2008101}
K.~Morita, ``Reversible computing and cellular automata - a survey,''
  \emph{Theoretical Computer Science}, vol. 395, no.~1, pp. 101 -- 131, 2008.

\bibitem{Myhill63}
J.~Myhill, ``The converse of {M}oore's {G}arden of {E}den theorem,'' in
  \emph{Proceedings of American Mathematical Society}, vol.~14, 1963, pp.
  685--686.

\bibitem{amoroso1970garden}
S.~Amoroso and G.~Cooper, ``The {G}arden-of-{E}den theorem for finite
  configurations,'' \emph{Proceedings of the American Mathematical Society},
  vol.~26, no.~1, pp. 158--164, 1970.

\bibitem{maruoka1976condition}
A.~Maruoka and M.~Kimura, ``Condition for injectivity of global maps for
  tessellation automata,'' \emph{Information and Control}, vol.~32, no.~2, pp.
  158--162, 1976.

\bibitem{toffoli90}
T.~Toffoli and N.~H. Margolus, ``Invertible cellular automata: A review,''
  \emph{Physica D: Nonlinear Phenomena}, vol.~45, no. 1--3, pp. 229 -- 253,
  1990.

\bibitem{doi:10.1137/0406004}
A.~Mach\`{\i} and F.~Mignosi, ``Garden of {E}den {C}onfigurations for
  {C}ellular {A}utomata on {C}ayley {G}raphs of {G}roups,'' \emph{SIAM Journal
  on Discrete Mathematics}, vol.~6, no.~1, pp. 44--56, 1993.

\bibitem{ceccherini1999amenable}
T.~G. Ceccherini-Silberstein, A.~Machi, and F.~Scarabotti, ``Amenable groups
  and cellular automata,'' \emph{Annales de l'institut Fourier}, vol.~49,
  no.~2, pp. 673--685, 1999.

\bibitem{capobianco2009surjunctivity}
S.~Capobianco, ``Surjunctivity for {C}ellular {A}utomata in {B}esicovitch
  {S}paces,'' \emph{Journal of Cellular Automata}, vol.~4, no.~2, pp. 89--98,
  2009.

\bibitem{margenstern2009garden}
M.~Margenstern, ``About the {G}arden of {E}den theorems for cellular automata
  in the hyperbolic plane,'' \emph{Electronic Notes in Theoretical Computer
  Science}, vol. 252, pp. 93--102, 2009.

\bibitem{Devaney}
R.~L. Devaney, ``An introduction to chaotic dynamical systems,''
  \emph{Addison-Wesley}, 1986.

\bibitem{CM96}
B.~Codenotti and L.~Margara, ``Transitive cellular automata are sensitive,''
  \emph{The American Mathematical Monthly}, vol. 103, no.~1, pp. 58--62, 1996.

\bibitem{DCTMitchell93}
M.~Mitchell, P.~T. Hraber, and J.~P. Crutchfield, ``Revisiting the edge of
  chaos: Evolving cellular automata to perform computations,'' \emph{Complex
  Systems}, vol.~7, pp. 89--130, 1993.

\bibitem{Margara99}
G.~Cattaneo, E.~Formenti, L.~Margara, and G.~Mauri, ``On the dynamical behavior
  of chaotic cellular automata,'' \emph{Theoretical Computer Science}, vol.
  217, no.~1, pp. 31 -- 51, 1999.

\bibitem{Cattaneocht}
G.~Cattaneo, M.~Finelli, and L.~Margara, ``Investigating topological chaos by
  elementary cellular automata dynamics,'' \emph{Theoretical Computer Science},
  vol. 244, no. 1 - 2, pp. 219 -- 241, 2000.

\bibitem{AcerbiDF09}
L.~Acerbi, A.~Dennunzio, and E.~Formenti, ``Conservation of some dynamical
  properties for operations on cellular automata,'' \emph{Theoretical Computer
  Science}, vol. 410, no. 38-40, pp. 3685--3693, 2009.

\bibitem{Kurka97}
P.~Kurka, ``Languages, equicontinuity and attractors in cellular automata,''
  \emph{Ergodic Theory and Dynamical Systems}, vol.~17, no.~02, pp. 417--433,
  1997.

\bibitem{durand2003undecidability}
B.~Durand, E.~Formenti, and G.~Varouchas, ``On undecidability of equicontinuity
  classification for cellular automata,'' \emph{Discere Mathematics and
  Theoretical Computer Science}, vol.~3, pp. 117--128, 2003.

\bibitem{DENNUNZIO20094823}
A.~Dennunzio, P.~D. Lena, E.~Formenti, and L.~Margara, ``On the directional
  dynamics of additive cellular automata,'' \emph{Theoretical Computer
  Science}, vol. 410, no. 47--49, pp. 4823 -- 4833, 2009.

\bibitem{dennunzio2013periodic}
A.~Dennunzio, P.~Di~Lena, E.~Formenti, and L.~Margara, ``Periodic orbits and
  dynamical complexity in cellular automata,'' \emph{Fundamenta Informaticae},
  vol. 126, no. 2-3, pp. 183--199, 2013.

\bibitem{culik88}
K.~Culik and S.~Yu, ``Undecidability of cellular automata classification
  schemes,'' \emph{Complex Systems}, vol.~2, pp. 177--190, 1988.

\bibitem{Culik90}
K.~Culik, L.~P. Hard, and S.~Yu, ``Computation theoretic aspects of cellular
  automata,'' \emph{Physica D: Nonlinear Phenomena}, vol.~45, no. 1-3, pp.
  357--378, 1990.

\bibitem{Kari92a}
J.~Kari, ``The nilpotency problem of one-dimensional cellular automata,''
  \emph{SIAM Journal on Computing}, vol.~21, no.~3, pp. 571--586, 1992.

\bibitem{blanchard1997dynamical}
F.~Blanchard and A.~Maass, ``Dynamical properties of expansive one-sided
  cellular automata,'' \emph{Israel Journal of Mathematics}, vol.~99, no.~1,
  pp. 149--174, 1997.

\bibitem{WARD1994495}
T.~Ward, ``Automorphisms of $\mathbb{Z}^d$-subshifts of finite type,''
  \emph{Indagationes Mathematicae}, vol.~5, no.~4, pp. 495 -- 504, 1994.

\bibitem{Sheresh}
M.~A. Shereshevsky, ``Lyapunov {E}xponents for one-dimensional {C}ellular
  {A}utomata,'' \emph{Journal of Nonlinear Science}, vol.~2, pp. 1--8, 1992.

\bibitem{Finelli199}
M.~Finelli, G.~Manzini, and L.~Margara, ``Lyapunov exponents versus expansivity
  and sensitivity in cellular automata,'' \emph{Journal of Complexity},
  vol.~14, no.~2, pp. 210 -- 233, 1998.

\bibitem{Wolfram85c}
S.~Wolfram, ``Origins of randomness in physical systems,'' \emph{Physics Review
  Letters}, vol.~55, pp. 449--452, 1985.

\bibitem{Makato98}
M.~Matsumoto, ``Simple cellular automata as pseudorandom m-sequence generators
  for built-in self-test,'' \emph{ACM Transactions on Modeling and Computer
  Simulation}, vol.~8, no.~1, pp. 31--42, 1998.

\bibitem{Horte89c}
P.~D. Hortensius, R.~D. McLeod, and H.~C. Card, ``Parallel random number
  generation for {VLSI} systems using cellular automata,'' \emph{IEEE
  Transactions on Computers}, vol. C-38, no.~10, pp. 1466--1473, 1989.

\bibitem{870571}
M.~Saraniti and S.~M. Goodnick, ``Hybrid fullband cellular automaton/monte
  carlo approach for fast simulation of charge transport in semiconductors,''
  \emph{IEEE Transactions on Electron Devices}, vol.~47, no.~10, pp.
  1909--1916, 2000.

\bibitem{comer2012random}
J.~M. Comer, J.~C. Cerda, C.~D. Martinez, and D.~H. Hoe, ``Random number
  generators using cellular automata implemented on {FPGAs},'' in
  \emph{Proceedings of Southeastern Symposium on System Theory}, 2012, pp.
  67--72.

\bibitem{Wolfr86b}
S.~Wolfram, ``Cryptography with {C}ellular {A}utomata,'' \emph{Advances in
  {C}ryptology - {C}rypto'85}, vol. 218, pp. 429--432, 1986.

\bibitem{DBLP:journals/ccds/DasC13}
S.~Das and D.~Roy~Chowdhury, ``{CAR30:} {A} new scalable stream cipher with
  rule 30,'' \emph{Cryptography and Communications}, vol.~5, no.~2, pp.
  137--162, 2013.

\bibitem{Formenti2014}
E.~Formenti, K.~Imai, B.~Martin, and J.-B. Yun{\`e}s, ``Advances on random
  sequence generation by uniform cellular automata,'' in \emph{Computing with
  New Resources: Essays Dedicated to Jozef Gruska on the Occasion of His 80th
  Birthday}.\hskip 1em plus 0.5em minus 0.4em\relax Cham: Springer
  International Publishing, 2014, pp. 56--70.

\bibitem{Leporati2014CryptographicPO}
A.~Leporati and L.~Mariot, ``Cryptographic properties of bipermutive cellular
  automata rules,'' \emph{Journal of Cellular Automata}, vol.~9, pp. 437--475,
  2014.

\bibitem{COMPAGNER1987391}
A.~Compagner and A.~Hoogland, ``Maximum-length sequences, cellular automata,
  and random numbers,'' \emph{Journal of Computational Physics}, vol.~71,
  no.~2, pp. 391 -- 428, 1987.

\bibitem{wang2008generating}
Q.~Wang, S.~Yu, W.~Ding, and M.~Leng, ``Generating high-quality random numbers
  by cellular automata with {PSO},'' in \emph{Proceedings of 4th International
  Conference on Natural Computation}, 2008, pp. 430--433.

\bibitem{Guan03}
S.~Guan and S.~Zhang, ``An {E}volutionary {A}pproach to the {D}esign of
  {C}ontrollable {C}ellular {A}utomata {S}tructure for {R}andom {N}umber
  {G}eneration,'' \emph{IEEE Transactions on Computer-Aided Design of
  Integrated Circuits and Systems}, vol.~7, no.~1, pp. 23--36, 2003.

\bibitem{ats03}
S.~Das, A.~Kundu, S.~Sen, B.~K.~Sikdar, and P.~Pal~Chaudhuri, ``Non-{L}inear
  {C}elluar {A}utomata {B}ased {PRPG} {D}esign ({W}ithout {P}rohibited
  {P}attern {S}et) {I}n {L}inear {T}ime {C}omplexity,'' in \emph{Proceedings of
  Asian Test Symposium ({ATS})}, 2003, pp. 78--83.

\bibitem{fredkin82}
E.~Fredkin and T.~Toffoli, ``Conservative logic,'' \emph{International Journal
  of Theoretical Physics}, vol.~21, pp. 219--253, 1982.

\bibitem{pivato2002-45}
M.~Pivato, ``Conservation laws in cellular automata,'' \emph{Nonlinearity},
  vol.~15, no.~6, p. 1781, 2002.

\bibitem{Boccara02}
N.~Boccara and H.~Fuk\'{s}, ``Number-conserving cellular automaton rules,''
  \emph{Fundamenta Informaticae}, vol.~52, no. 1-3, pp. 1--13, 2002.

\bibitem{pCA18}
H.~Fuk\'{s}, ``Probabilistic cellular automata with conserved quantities,''
  \emph{Nonlinearity}, vol.~17, no.~1, p. 159, 2004.

\bibitem{Durand03}
B.~Durand, E.~Formenti, and Z.~R\'oka, ``Number-conserving cellular automata
  {I}: decidability,'' \emph{Theoretical Computer Science}, vol. 299, no. 1-3,
  pp. 523--535, 2003.

\bibitem{Moreira2003711}
A.~Moreira, ``Universality and decidability of number-conserving cellular
  automata,'' \emph{Theoretical Computer Science}, vol. 292, no.~3, pp.
  711--721, 2003.

\bibitem{Formenti2003269}
E.~Formenti and A.~Grange, ``Number conserving cellular automata {II}:
  dynamics,'' \emph{Theoretical Computer Science}, vol. 304, pp. 269--290,
  2003.

\bibitem{PhysRevE-40}
K.~Nagel, ``Particle hopping models and traffic flow theory,'' \emph{Physical
  Review E}, vol.~53, pp. 4655--4672, 1996.

\bibitem{Kohyama01011989-27}
T.~Kohyama, ``Cellular automata with particle conservation,'' \emph{Progress of
  Theoretical Physics}, vol.~81, no.~1, pp. 47--59, 1989.

\bibitem{Kohyama01011989-28}
------, ``Cluster growth in particle-conserving cellular automata,''
  \emph{Journal of Statistical Physics}, vol.~63, no. 3-4, pp. 637--651, 1991.

\bibitem{Margolus198481}
N.~Margolus, ``Physics-like models of computation,'' \emph{Physica D: Nonlinear
  Phenomena}, vol.~10, no. 1 - 2, pp. 81 -- 95, 1984.

\bibitem{Hattori}
T.~Hattori and S.~Takesue, ``Additive conserved quantities in discrete-time
  lattice dynamical systems,'' \emph{Physica D: Nonlinear Phenomena}, vol.~49,
  no.~3, pp. 295 -- 322, 1991.

\bibitem{takesue1995staggered}
S.~Takesue, ``Staggered invariants in cellular automata,'' \emph{Complex
  Systems}, vol.~9, no.~2, pp. 149--168, 1995.

\bibitem{Morita98}
{Morita, Kenichi} and {Imai, Katsunobu}, ``Number-conserving reversible
  cellular automata and their computation-universality,'' \emph{RAIRO -
  Theoretical Informatics and Applications}, vol.~35, no.~3, pp. 239--258,
  2001.

\bibitem{Morita:99}
K.~{M}orita, Y.~{T}ojima, and K.~{I}mai, ``A simple computer embedded in a
  reversible and number conserving two-dimensional cellular space,'' in
  \emph{Multiple-Valued Logic}, 1999, pp. 483--514.

\bibitem{Kurka:2003}
P.~{K}urka, ``Cellular automata with vanishing particles,'' \emph{Fundamenta
  Informaticae}, vol.~58, no. 3--4, pp. 203--221, 2003.

\bibitem{GOLES20113616}
E.~Goles, A.~Moreira, and I.~Rapaport, ``Communication complexity in
  number-conserving and monotone cellular automata,'' \emph{Theoretical
  Computer Science}, vol. 412, no.~29, pp. 3616 -- 3628, 2011.

\bibitem{das2011characterization}
S.~Das, ``Characterization of non-uniform number conserving cellular
  automata,'' in \emph{Proceedings of International workshop on Cellular
  Automata and Discrete Complex Systems, (AUTOMATA)}, 2011, pp. 17--28.

\bibitem{hazari14}
R.~Hazari and S.~Das, ``Number conservation property of elementary cellular
  automata under asynchronous update,'' \emph{Complex Systems}, vol.~23, pp.
  177--195, 2014.

\bibitem{PhysRevLett.74.5148}
M.~Land and R.~K. Belew, ``No perfect two-state cellular automata for density
  classification exists,'' \emph{Physical Review Letters}, vol.~74, pp.
  5148--5150, 1995.

\bibitem{Kari:2012:MTC:2385073.2385086}
J.~Kari and B.~Le~Gloannec, ``Modified traffic cellular automaton for the
  density classification task,'' \emph{Fundamenta Informaticae}, vol. 116, no.
  1-4, pp. 141--156, 2012.

\bibitem{morales01}
F.~J. Morales, J.~P. Crutchfield, and M.~Mitchell, ``Evolving two-dimensional
  {C}ellular {A}utomata to {P}erform {D}ensity {C}lassification : {A} report on
  work in {P}rogress,'' \emph{Parallel Computing}, vol.~27, pp. 571--585, 2001.

\bibitem{DCTMaiti06}
N.~S.~Maiti, S.~Munshi, and P.~Pal~Chaudhuri, ``An analytical formulation for
  cellular automata {(CA)} based solution of density classification task
  {(DCT)},'' in \emph{Proceedings of International Conference on Cellular
  Automata, Research and Industry (ACRI)}.\hskip 1em plus 0.5em minus
  0.4em\relax Springer International Publishing, 2006, pp. 147--156.

\bibitem{Fuk05}
H.~Fuk\'s, ``Solution of the density classification problem with two cellular
  automata rules,'' \emph{Physics Review}, vol. E 55, pp. R2081--R2084, 1997.

\bibitem{Oliveira13}
P.~P.~B. de~Oliveira, ``Conceptual connections around density determination in
  cellular automata,'' \emph{Proceedings of International workshop on Cellular
  Automata and Discrete Complex Systems (AUTOMATA)}, pp. 1--14, 2013.

\bibitem{FSSPCA2}
K.~Noguchi, ``Simple 8-state minimal time solution to the firing squad
  synchronization problem,'' \emph{Theoretical Computer Science}, vol. 314,
  no.~3, pp. 303 -- 334, 2004.

\bibitem{FSSPCA3}
H.~Umeo, M.~Hisaoka, and T.~Sogabe, ``A survey on optimum-time firing squad
  synchronization algorithms for one-dimensional cellular automata,''
  \emph{International Journal of Unconventional Computing}, vol.~1, no.~4, pp.
  403--426, 2005.

\bibitem{MANZONI2014108}
L.~Manzoni and H.~Umeo, ``The firing squad synchronization problem on {CA} with
  multiple updating cycles,'' \emph{Theoretical Computer Science}, vol. 559,
  pp. 108 -- 117, 2014.

\bibitem{Culik91}
K.~Culik and S.~Dube, ``An efficient solution to the firing mob problem,''
  \emph{Theoretical Computer Science}, vol.~91, pp. 57--69, 1991.

\bibitem{Smith76}
A.~R. Smith, III, ``Introduction to and survey of polyautomata theory,''
  \emph{Automata Languages Development}, pp. 405--422, 1976.

\bibitem{Mazoyer}
J.~Mazoyer, C.~Nichitiu, and E.~R{\'e}mila, ``Compass permits leader
  election,'' in \emph{Proceedings of the Tenth Annual ACM-SIAM Symposium on
  Discrete Algorithms}, 1999, pp. 947--948.

\bibitem{BeckersW01}
A.~Beckers and T.~Worsch, ``A perimeter-time {CA} for the queen bee problem.''
  \emph{Parallel Computing}, vol.~27, no.~5, pp. 555--569, 2001.

\bibitem{Stratmann154}
M.~Stratmann and T.~Worsch, ``Leader election in d-dimensional {CA} in time
  diam log(diam),'' \emph{Future Generation Computer Systems}, vol.~18, no.~7,
  pp. 939--950, 2002.

\bibitem{banda1}
P.~Banda, J.~Caughman, and J.~Pospichal, ``Configuration symmetry and
  performance upper bound of one-dimensional cellular automata for the leader
  election problem,'' \emph{Journal of Cellular Automata}, vol.~10, no. 1-2,
  pp. 1--21, 2015.

\bibitem{rosen}
P.~Rosenstiehl, J.~R. Fiksel, and A.~Holliger, \emph{R.C. Read (Ed.), Graph
  Theory and Computing}.\hskip 1em plus 0.5em minus 0.4em\relax Academic Press,
  New York, 1972, pp. 219--265.

\bibitem{Vollmar2}
R.~Vollmar, ``On two modified problems of synchronization in cellular
  automata,'' \emph{Acta Cybernetica}, vol.~3, no.~4, pp. 293--300, 1977.

\bibitem{Legendi}
T.~Legendi and E.~Katona, ``A 5 state solution of the early bird problem in a
  one dimensional cellular space,'' \emph{Acta Cybernetica}, vol.~5, no.~2, pp.
  173--179, 1981.

\bibitem{Legendi1}
------, ``A solution of the early bird problem in an n-dimensional cellular
  space,'' \emph{Acta Cybernetica}, vol.~7, no.~1, pp. 81--87, 1986.

\bibitem{SRoy2017}
S.~Roy and S.~Das, ``Distributed mutual exclusion problem in cellular
  automata,'' \emph{Journal of Cellular Automata}, vol.~12, no.~6, pp.
  493--512, 2017.

\bibitem{Herman01081973}
G.~T. Herman and W.~H. Liu, ``The daughter of {C}elia, the {F}rench flag, and
  the firing squad,'' \emph{Simulation}, vol.~21, no.~2, pp. 33--41, 1973.

\bibitem{short12}
A.~I. Adamatzky, ``Computation of shortest path in cellular automata,''
  \emph{Mathl. Comput. Modelling}, vol.~23, no.~4, pp. 105--113, 1996.

\bibitem{Delorme1999347}
M.~Delorme, J.~Mazoyer, and L.~Tougne, ``Discrete parabolas and circles on {2D}
  cellular automata,'' \emph{Theoretical Computer Science}, vol. 218, no.~2,
  pp. 347 -- 417, 1999.

\bibitem{naka}
K.~Nakamura, ``Asynchronous cellular automata and their computational
  ability,'' \emph{Systems, Computers, Controls}, vol.~5, no.~5, pp. 58--66,
  1974.

\bibitem{GOLZE1978176}
U.~Golze, ``(a-)synchronous (non-)deterministic cell spaces simulating each
  other,'' \emph{Journal of Computer and System Sciences}, vol.~17, no.~2, pp.
  176 -- 193, 1978.

\bibitem{Nakamura22}
K.~Nakamura, ``Synchronous to asynchronous transformation of polyautomata,''
  \emph{Journal of Computer and System Sciences}, vol.~23, no.~1, pp. 22 -- 37,
  1981.

\bibitem{Hem82}
A.~Hemmerling, ``On the computational equivalence of synchronous and
  asynchronous cellular spaces,'' \emph{Elektronische Informationsverarbeitung
  und Kybernetik}, vol.~18, no. 7/8, pp. 423--434, 1982.

\bibitem{Ingerson84}
T.~E. Ingerson and R.~L. Buvel, ``Structure in asynchronous cellular
  automata,'' \emph{Physica D: Nonlinear Phenomena}, vol.~10, no. 1 - 2, pp. 59
  -- 68, 1984.

\bibitem{LeC89}
G.~Le~Ca\"er, ``Comparison between simultaneous and sequential updating in
  $2^{n+1}-1$ cellular automata,'' \emph{Physica A: Statistical Mechanics and
  its Applications}, vol. 157, no.~2, pp. 669 -- 687, 1989.

\bibitem{Cor}
R.~Cori, Y.~Metivier, and W.~Zielonka, ``Asynchronous mappings and asynchronous
  cellular automata,'' \emph{Information and Computation}, vol. 106, no.~2, pp.
  159 -- 202, 1993.

\bibitem{probing12}
O.~Bour{\'e}, N.~Fat{\`{e}}s, and V.~Chevrier, ``Probing robustness of cellular
  automata through variations of asynchronous updating,'' \emph{Natural
  Computing}, vol.~11, no.~4, pp. 553--564, 2012.

\bibitem{Dennunzio13}
A.~Dennunzio, E.~Formenti, L.~Manzoni, and G.~Mauri, ``m-{A}synchronous
  cellular automata: from fairness to quasi-fairness,'' \emph{Natural
  Computing}, vol.~12, no.~4, pp. 561--572, 2013.

\bibitem{Fates14}
N.~Fat{\`{e}}s, ``A guided tour of asynchronous cellular automata,''
  \emph{Journal of Cellular Automata}, vol.~9, no. 5-6, pp. 387--416, 2014.

\bibitem{Dennunzio16}
A.~Dennunzio, E.~Formenti, and L.~Manzoni, ``Computing {I}ssues of
  {A}synchronous {CA},'' \emph{Fundamenta Informaticae}, vol. 120, no.~2, pp.
  165--180, 2012.

\bibitem{Golze1982121}
U.~Golze and L.~Priese, ``Petri net implementations by a universal cell
  space,'' \emph{Information and Control}, vol.~53, no. 1 - 2, pp. 121 -- 138,
  1982.

\bibitem{Pighizzini1994179}
G.~Pighizzini, ``Asynchronous automata versus asynchronous cellular automata,''
  \emph{Theoretical Computer Science}, vol. 132, no. 1–2, pp. 179 -- 207,
  1994.

\bibitem{Droste20001}
M.~Droste, P.~Gastin, and D.~Kuske, ``Asynchronous cellular automata for
  pomsets,'' \emph{Theoretical Computer Science}, vol. 247, no. 1-2, pp. 1--38,
  2000.

\bibitem{PhysRevE}
H.~J. Blok and B.~Bergersen, ``Synchronous versus asynchronous updating in the
  ``game of life'','' \emph{Physical Review E}, vol.~59, pp. 3876--3879, 1999.

\bibitem{Ruxton}
G.~Ruxton and L.~A. Saravia, ``{The need for biological realism in the updating
  of cellular automata models},'' \emph{Ecological Modelling}, vol. 107, no.
  2-3, pp. 105--112, 1998.

\bibitem{Tomassini02}
M.~Tomassini and M.~Venzi, ``Artificially evolved asynchronous cellular
  automata for the density task,'' in \emph{Proceedings of International
  Conference on Cellular Automata, Research and Industry (ACRI)}.\hskip 1em
  plus 0.5em minus 0.4em\relax Springer Berlin Heidelberg, 2002, pp. 44--55.

\bibitem{Suzudo2004185}
T.~Suzudo, ``Spatial pattern formation in asynchronous cellular automata with
  mass conservation,'' \emph{Physica A: Statistical Mechanics and its
  Applications}, vol. 343, pp. 185--200, 2004.

\bibitem{sarkar2012reversibility}
A.~Sarkar, A.~Mukherjee, and S.~Das, ``Reversibility in asynchronous cellular
  automata,'' \emph{Complex Systems}, vol.~21, no.~1, p.~71, 2012.

\bibitem{SethiD14}
B.~Sethi, N.~Fat{\`e}s, and S.~Das, ``Reversibility of elementary cellular
  automata under fully asynchronous update,'' in \emph{Proceedings of 11th
  Annual Conference Theory and Applications of Models of Computation}.\hskip
  1em plus 0.5em minus 0.4em\relax Springer International Publishing, 2014, pp.
  39--49.

\bibitem{ACASir}
S.~Das, A.~Sarkar, and B.~K. Sikdar, ``Synthesis of reversible asynchronous
  cellular automata for pattern generation with specific hamming distance,'' in
  \emph{Proceedings of International Conference on Cellular Automata, Research
  and Industry, (ACRI)}.\hskip 1em plus 0.5em minus 0.4em\relax Springer Berlin
  Heidelberg, 2012, pp. 643--652.

\bibitem{Sethi2016}
B.~Sethi and S.~Das, ``On the use of asynchronous cellular automata in
  symmetric-key cryptography,'' in \emph{Proceedings of 4th International
  Symposium on Security in Computing and Communications}.\hskip 1em plus 0.5em
  minus 0.4em\relax Springer Singapore, 2016, pp. 30--41.

\bibitem{Mariot2016}
L.~Mariot, ``Asynchrony immune cellular automata,'' in \emph{Proceedings of
  International Conference on Cellular Automata, Research and Industry
  (ACRI)}.\hskip 1em plus 0.5em minus 0.4em\relax Springer International
  Publishing, 2016, pp. 176--181.

\bibitem{Manzoni2012}
L.~Manzoni, ``Asynchronous cellular automata and dynamical properties,''
  \emph{Natural Computing}, vol.~11, no.~2, pp. 269--276, 2012.

\bibitem{Sethi2015}
B.~Sethi and S.~Das, \emph{Recent Advances in Natural Computing: Selected
  Results from the IWNC 7 Symposium}.\hskip 1em plus 0.5em minus 0.4em\relax
  Springer Japan, 2015, ch. Convergence of Asynchronous Cellular Automata
  (Under Null Boundary Condition) and Their Application in Pattern
  Classification, pp. 35--55.

\bibitem{CPLX:CPLX21749}
B.~Sethi, S.~Roy, and S.~Das, ``Asynchronous cellular automata and pattern
  classification,'' \emph{Complexity}, vol.~21, no.~S1, pp. 370--386, 2016.

\bibitem{boccara1994some}
N.~Boccara and M.~Roger, ``Some properties of local and nonlocal site exchange
  deterministic cellular automata,'' \emph{International Journal of Modern
  Physics C}, vol.~5, no.~03, pp. 581--588, 1994.

\bibitem{newman99}
M.~E.~J. Newman and D.~J. Watts, ``Scaling and percolation in the small-world
  network model,'' \emph{Physical Review E}, vol.~60, no.~6, pp. 7332--7342,
  1999.

\bibitem{yang2007}
X.~S. Yang and Y.~Z. Yang, ``Cellular automata networks,'' in \emph{Proceedings
  of Unconventional Computing}, 2007, pp. 280--302.

\bibitem{Adami199529}
C.~Adami, ``Self-organized criticality in living systems,'' \emph{Physics
  Letters A}, vol. 203, no.~1, pp. 29 -- 32, 1995.

\bibitem{watts1998collective}
D.~J. Watts and S.~H. Strogatz, ``Collective dynamics of `small-world'
  networks,'' \emph{Nature}, vol. 393, no. 6684, pp. 440--442, 1998.

\bibitem{Tomassini15}
M.~Tomassini, ``Generalized automata networks,'' in \emph{Proceedings of
  International Conference on Cellular Automata, Research and Industry
  (ACRI)}.\hskip 1em plus 0.5em minus 0.4em\relax Springer Berlin Heidelberg,
  2006, pp. 14--28.

\bibitem{8718492.ch2}
P.~Domosi and C.~L. Nehaniv, \emph{Algebraic theory of automata networks: An
  Introduction}.\hskip 1em plus 0.5em minus 0.4em\relax Society for Industrial
  and Applied Mathematics, ISBN: 978-0-89871-569-9, 2005.

\bibitem{Kayama2011}
Y.~Kayama and Y.~Imamura, ``Network representation of the game of life,''
  \emph{Journal of Artificial Intelligence and Soft Computing Research}, vol.
  Vol. 1, No. 3, pp. 233--240, 2011.

\bibitem{Kayama2012}
Y.~Kayama, ``Network view of binary cellular automata,'' in \emph{Proceedings
  of International Conference on Cellular Automata, Research and Industry
  (ACRI)}.\hskip 1em plus 0.5em minus 0.4em\relax Springer Berlin Heidelberg,
  2012, pp. 224--233.

\bibitem{Das91}
A.~K.~Das, A.~Sanyal, and P.~Pal~Chaudhuri, ``On characterization of cellular
  automata with matrix algebra,'' \emph{Information Sciences}, vol.~61, no.~3,
  pp. 251 -- 277, 1992.

\bibitem{Serra90c}
M.~Serra, T.~Slater, J.~C. Muzio, and D.~M. Miller, ``The analysis of
  one-dimensional linear cellular automata and their aliasing properties,''
  \emph{IEEE Transactions on Computer-Aided Design of Integrated Circuits and
  Systems}, vol.~9, no.~7, pp. 767--778, 1990.

\bibitem{CattaneoDFP09}
G.~Cattaneo, A.~Dennunzio, E.~Formenti, and J.~Provillard, ``Non-uniform
  cellular automata,'' in \emph{Proceedings of 3rd International Conference
  Language and Automata Theory and Applications, {LATA}}, 2009, pp. 302--313.

\bibitem{DennunzioFP12}
A.~Dennunzio, E.~Formenti, and J.~Provillard, ``Non-uniform cellular automata:
  Classes, dynamics, and decidability,'' \emph{Information and Computation},
  vol. 215, pp. 32--46, 2012.

\bibitem{salo2014realization}
V.~Salo, ``Realization problems for nonuniform cellular automata,''
  \emph{Theoretical Computer Science}, vol. 559, pp. 91--107, 2014.

\bibitem{NiloyFI08}
N.~Ganguly, B.~K.~Sikdar, and P.~Pal~Chaudhuri, ``Exploring cycle structures of
  additive cellular automata,'' \emph{Fundamenta Informaticae}, vol.~87, no.~2,
  pp. 137--154, 2008.

\bibitem{Das90b}
A.~K. Das, ``Additive {C}ellular {A}utomata : {T}heory and {A}pplication as a
  {B}uilt-in {S}elf-test {S}tructure,'' Ph.D. dissertation, {IIT}, {K}haragpur,
  {I}ndia, 1990.

\bibitem{Barde90}
P.~H. Bardell, ``Analysis of {C}ellular {A}utomata {U}sed as {P}seudo-random
  {P}attern {G}enerators,'' in \emph{Proceedings of International Test
  Conference}, 1990, pp. 762--768.

\bibitem{cattell1996synthesis}
K.~Cattell and J.~C. Muzio, ``Synthesis of one-dimensional linear hybrid
  cellular automata,'' \emph{IEEE Transactions on Computer-Aided Design of
  Integrated Circuits and Systems}, vol.~15, no.~3, pp. 325--335, 1996.

\bibitem{Jetta95}
K.~Cattell and S.~Zhang, ``Minimal cost one-dimensional linear hybrid cellular
  automata of degree through 500,'' \emph{Journal of Electronic Testing: Theory
  and Applications}, vol.~6, no.~2, pp. 255--258, 1995.

\bibitem{Bhatt95}
S.~Bhattacharjee, J.~Bhattacharya, and P.~Pal~Chaudhuri, ``An {E}fficient
  {D}ata {C}ompression {H}ardware based on {C}ellular {A}utomata,'' in
  \emph{Proceedings of Data Compression Conference}, 1995, p. 472.

\bibitem{Chakr93}
S.~Chakraborty, D.~Roy~Chowdhury, and P.~Pal~Chaudhuri, ``Theory and
  {A}pplication of {N}on-{G}roup {C}ellular {A}utomata for {S}ynthesis of
  {E}asily {T}estable {F}inite {S}tate {M}achines,'' \emph{IEEE Transactions on
  Computers}, vol.~45, no.~7, pp. 769--781, 1996.

\bibitem{santanu00}
S.~Chattopadhyay, S.~Adhikari, S.~Sengupta, and M.~Pal, ``{H}ighly {R}egular,
  {M}odular, and {C}ascadable {D}esign of {C}ellular {A}utomata-{B}ased
  {P}attern {C}lassifier,'' \emph{{IEEE} {T}ransactions on {VLSI} {S}ystems},
  vol.~8, no.~6, pp. 724--735, 2000.

\bibitem{Rappid}
M.~Roncken, K.~Stevens, R.~Pendurkar, S.~Rotem, and P.~Pal~Chaudhuri,
  ``{CA-BIST} for asynchronous circuits: a case study on the {RAPPID}
  asynchronous instruction length decoder,'' in \emph{Proceedings of Advanced
  Research in Asynchronous Circuits and Systems}, 2000, pp. 62--72.

\bibitem{Acri08b}
S.~Das and B.~K. Sikdar, ``Characterization of {N}on-reachable {S}tates in
  {I}rreversible {CA} {S}tate {S}pace,'' in \emph{Proceedings of International
  Conference on Cellular Automata, Research and Industry (ACRI)}.\hskip 1em
  plus 0.5em minus 0.4em\relax Springer Berlin Heidelberg, 2008, pp. 160--167.

\bibitem{maji2003theory}
P.~Maji, C.~Shaw, N.~Ganguly, B.~K. Sikdar, and P.~P. Chaudhuri, ``Theory and
  application of cellular automata for pattern classification,''
  \emph{Fundamenta Informaticae}, vol.~58, no. 3--4, pp. 321--354, 2003.

\bibitem{Chowd92d}
D.~Roy~Chowdhury, ``Theory and {A}pplications of {A}dditive {C}ellular
  {A}utomata for {R}eliable and {T}estable {VLSI} {C}ircuit {D}esign,'' Ph.D.
  dissertation, {IIT}, {K}haragpur, {I}ndia, 1994.

\bibitem{adcom00}
N.~Ganguly, D.~Halder, J.~Deb, B.~K.~Sikdar, and P.~Pal~Chaudhuri, ``Hashing
  {T}hrough {C}ellular {A}utomata,'' in \emph{Proceedings of $8^{th}$
  International Conference of Advanced Computing and Communication}, 2000, pp.
  95 -- 101.

\bibitem{NiloyIV}
N.~Ganguly, P.~Maji, S.~Dhar, B.~K. Sikdar, and P.~Pal~Chaudhuri, ``Evolving
  cellular automata as pattern classifier,'' in \emph{Proceedings of
  International Conference on Cellular Automata, Research and Industry
  (ACRI)}.\hskip 1em plus 0.5em minus 0.4em\relax Springer Berlin Heidelberg,
  2002, pp. 56--68.

\bibitem{Chowd93a}
D.~Roy~Chowdhury, S.~Chakraborty, B.~Vamsi, and P.~Pal~Chaudhuri, ``{C}ellular
  {A}utomata based synthesis of easily and fully testable {FSM}s,'' in
  \emph{Proceedings of International Conference on Computer Aided Design},
  1993, pp. 650--653.

\bibitem{NazmaTh}
N.~Naskar, ``{C}haracterization and {S}ynthesis of {N}on-{U}niform {C}ellular
  {A}utomata with {P}oint {S}tate {A}ttractors,'' Ph.D. dissertation, {I}ndian
  {I}nstitute of {E}ngineering {S}cience and {T}echnology, {S}hibpur, {I}ndia,
  2015.

\bibitem{MajiPhd}
P.~Maji, ``{C}ellular {A}utomata {E}volution for {P}attern {R}ecognition,''
  Ph.D. dissertation, {Jadavpur University}, {Kolkata}, {I}ndia, 2005.

\bibitem{DasMNS09}
S.~Das, S.~Mukherjee, N.~Naskar, and B.~K. Sikdar, ``Characterization of single
  cycle {CA} and its application in pattern classification,'' \emph{Electronic
  Notes in Theoretical Computer Science}, vol. 252, pp. 181--203, 2009.

\bibitem{NiloyV}
N.~Ganguly, P.~Maji, B.~K.~Sikdar, and P.~Pal~Chaudhuri, ``{G}eneralized
  {M}ultiple {A}ttractor {C}ellular {A}utomata ({GMACA}) {F}or {A}ssociative
  {M}emory,'' \emph{{I}nternational {J}ournal of {P}attern {R}ecognition and
  {A}rtificial {I}ntelligence, {S}pecial {I}ssue: {C}omputational
  {I}ntelligence for {P}attern {R}ecognition}, vol.~16, no.~07, pp. 781--795,
  2002.

\bibitem{DASFAA04}
P.~Maji and P.~Pal~Chaudhuri, ``{FMACA}: {A} {F}uzzy {C}ellular {A}utomata
  {B}ased {P}attern {C}lassifier,'' in \emph{Proceedings of 9th International
  Conference on Database Systems for Advanced Applications}.\hskip 1em plus
  0.5em minus 0.4em\relax Springer Berlin Heidelberg, 2004, pp. 494--505.

\bibitem{Atrub65}
A.~J. Atrubin, ``A one-dimensional real-time iterative multiplier,'' \emph{IEEE
  Transactions on Electronic Computers}, vol. EC-14, no.~3, pp. 394--399, 1965.

\bibitem{Cole69}
S.~N. Cole, ``Real time computation by n-dimensional iterative arrays of finite
  state machines,'' \emph{IEEE Transactions on Computers}, vol. C-18, pp. 55 --
  77, 1969.

\bibitem{Manni77}
F.~B. Manning, ``An approach to highly integrated, computer-maintained cellular
  arrays,'' \emph{IEEE Transactions on Computers}, vol. C-26, pp. 536 -- 552,
  1977.

\bibitem{Fisch65}
P.~C. Fischer, ``Generation of primes by a one-dimensional real-time iterative
  array,'' \emph{Journal of {ACM}}, vol.~12, pp. 388--394, 1965.

\bibitem{Nishi81}
H.~Nishio, ``Real time sorting of binary numbers by one-dimensional cellular
  automata,'' Kyoto University, Tech. Rep., 1981.

\bibitem{Nishio75}
H.~Nishio and Y.~Kobuchi, ``Fault tolerant cellular space,'' \emph{Journal of
  Compututer and System Sciences}, vol.~11, pp. 150--170, 1975.

\bibitem{Benjamin97}
S.~C. Benjamin and N.~F. Johnson, ``A {P}ossible {N}anometer-scale {C}omputing
  {D}evice based on an {A}dding {C}ellular {A}utomaton,'' \emph{Applied Physics
  Letters}, vol.~70, no.~17, pp. 2321 -- 2323, 1997.

\bibitem{Tsali90}
P.~Tsalides, ``Cellular {A}utomata based {B}uilt-{I}n {S}elf-{T}est
  {S}tructures for {VLSI} {S}ystems,'' \emph{Electronics Letters}, vol.~26,
  no.~17, pp. 1350--1352, 1990.

\bibitem{DBLP:conf/ats/ChakrabortyC09}
R.~Chakraborty and D.~Roy~Chowdhury, ``A novel seed selection algorithm for
  test time reduction in {BIST},'' in \emph{Proceedings of Asian Test Symposium
  ({ATS})}, 2009, pp. 15--20.

\bibitem{Albic87a}
A.~Albicki and M.~Khare, ``Cellular {A}utomata used for test pattern
  generation,'' in \emph{Proceedings of International Conference Computer
  Design}, 1987, pp. 56--59.

\bibitem{Das89}
A.~K.~Das and P.~Pal~Chaudhuri, ``An {E}fficient {O}n-chip {D}eterministic
  {T}est {P}attern {G}eneration {S}cheme,'' \emph{Microprocessing {\em $\&$}
  Microprogramming}, vol.~26, pp. 195--204, 1989.

\bibitem{vlsi02a}
N.~Ganguly, B.~K.~Sikdar, and P.~Pal~Chaudhuri, ``Design of an {O}n-{C}hip
  {T}est {P}attern {G}enerator {W}ithout {P}rohibitited {P}attern {S}et
  ({PPS}),'' in \emph{Proceedings of Asia and South Pacific Design Automation
  Conference/VLSI Design}, 2002, pp. 689--694.

\bibitem{Das93}
A.~K.~Das and P.~Pal~Chaudhuri, ``Vector space theoretic analysis of additive
  cellular automata and its applications for pseudo-exhaustive test pattern
  generation,'' \emph{IEEE Transactions on Computers}, vol.~42, no.~3, pp.
  340--352, 1993.

\bibitem{ubist}
S.~Das, N.~Ganguly, B.~K.~Sikdar, and P.~Pal~Chaudhuri, ``{D}esign of a
  {U}niversal {BIST} ({UBIST}) {S}tructure,'' in \emph{Proceedings of $16^{th}$
  International Conference on VLSI Design}, 2003, pp. 161--166.

\bibitem{Mitra91b}
B.~Mitra, P.~R.~Panda, and P.~Pal~Chaudhuri, ``A flexible scheme for state
  assignment based on characteristics of the {FSM},'' in \emph{Proceedings of
  International Conference on Computer Aided Design}, 1991, pp. 226--229.

\bibitem{Misra92b}
S.~Misra, B.~Mitra, , S.~Sengupta, and P.~Pal~Chaudhuri, ``Synthesis of
  self-testable sequential logic using programmable cellular automata,'' in
  \emph{Proceedings of International Conference on VLSI Design}, 1992, pp.
  193--198.

\bibitem{Bao04}
F.~Bao, ``Cryptanalysis of a partially known cellular automata cryptosystem,''
  \emph{IEEE Transactions on Computers}, vol.~53, no.~11, pp. 1493--1497, 2004.

\bibitem{Seredynski2004753}
F.~Seredynski, P.~Bouvry, and A.~Y. Zomaya, ``Cellular automata computations
  and secret key cryptography,'' \emph{Parallel Computing}, vol.~30, no. 5 - 6,
  pp. 753 -- 766, 2004.

\bibitem{ref1}
M.~Seredynski and P.~Bouvry, ``Block cipher based on reversible cellular
  automata,'' \emph{New Generation Computing}, vol.~23, no.~3, pp. 245--258,
  2005.

\bibitem{S0129626409000225}
A.~WUENSCHE, ``Cellular automata encryption: the reverse algorithm, z-parameter
  and chain-rules,'' \emph{Parallel Processing Letters}, vol.~19, no.~02, pp.
  283--297, 2009.

\bibitem{Chowd94a}
D.~Roy~Chowdhury, S.~Basu, I.~Sengupta, and P.~Pal~Chaudhuri, ``Design of
  {CAECC} --- {C}ellular automata based error correcting code,'' \emph{IEEE
  Transactions on Computers}, vol.~43, no.~6, pp. 759--764, 1994.

\bibitem{vlsi00b}
K.~Paul and D.~Roy~Chowdhury, ``Application of {GF}(2$^p$) {CA} in {B}urst
  {E}rror {C}orrecting {C}odes,'' in \emph{Proceedings of International
  Conference of {VLSI} Design}, 2000, pp. 562--567.

\bibitem{Horte90b}
P.~D. Hortensius, R.~D. McLeod, and H.~C. Card, ``Cellular automata based
  signature analysis for built-in self-test,'' \emph{IEEE Transactions on
  Computers}, vol.~39, no.~10, pp. 1273--1283, 1990.

\bibitem{Das90e}
A.~K.~Das, D.~Saha, A.~Roy~Chowdhury, S.~Misra, and P.~Pal~Chaudhuri,
  ``Signature analysers based on additive cellular automata,'' in
  \emph{Proceedings of 20th Fault Tolerant Computing Systems}, 1990, pp.
  265--272.

\bibitem{Rosin2010790}
P.~L. Rosin, ``Image processing using 3-state cellular automata,''
  \emph{Computer Vision and Image Understanding}, vol. 114, no.~7, pp. 790 --
  802, 2010.

\bibitem{Okba11}
O.~Kazar and S.~Slatnia, ``Evolutionary cellular automata for image
  segmentation and noise filtering using genetic algorithms,'' \emph{Journal of
  Applied Computer Science $\&$ Mathematics}, vol.~5, no.~10, pp. 33--40, 2011.

\bibitem{Paul99}
K.~Paul, D.~Roy~Chowdhury, and P.~Pal~Chaudhuri, ``Cellular {A}utomata {B}ased
  {T}ransform {C}oding for {I}mage {C}ompression,'' in \emph{Proceedings of
  International Conference on High Performance Computing}, 1999, pp. 269--273.

\bibitem{Wongthanavasu03}
S.~Wongthanavasu and R.~Sadananda, ``A {CA}-based edge operator and its
  performance evaluation,'' \emph{Journal of Visual Communication and Image
  Representation}, vol.~14, no.~2, pp. 83 -- 96, 2003.

\bibitem{Sadeghi12}
S.~Sadeghi, A.~Rezvanian, and E.~Kamrani, ``An efficient method for impulse
  noise reduction from images using fuzzy cellular automata,'' \emph{AEU -
  International Journal of Electronics and Communications}, vol.~66, no.~9, pp.
  772 -- 779, 2012.

\bibitem{Jen86}
E.~Jen, ``Invariant strings and pattern-recognizing properties of
  one-dimensional cellular automata,'' \emph{Journal of Statistical Physics},
  vol.~43, no. 1--2, pp. 243--265, 1986.

\bibitem{Maji2}
P.~Maji, N.~Ganguly, and P.~Pal~Chaudhuri, ``{E}rror {C}orrecting {C}apability
  of {C}ellular {A}utomata {B}ased {A}ssociative {M}emory,'' \emph{{IEEE}
  {T}ransactions on {S}ystems, {M}an and {C}ybernetics, {P}art {A}}, vol.~33,
  no.~4, pp. 466--480, 2003.

\bibitem{Lafe}
O.~Lafe, ``Data compression and encryption using cellular automata
  transforms,'' \emph{Engineering Applications of Artificial Intelligence},
  vol.~10, no.~6, pp. 581 -- 591, 1997.

\bibitem{vlsi00a}
K.~Paul, D.~Roy~Chowdhury, and P.~Pal~Chaudhuri, ``Scalable {P}ipelined
  {M}icro-{A}rchitecture for {W}avelet {T}ransform,'' in \emph{Proceedings of
  International Con. on VLSI Design}, 2000, pp. 144--147.

\bibitem{ShawSM04}
C.~Shaw, B.~K. Sikdar, and N.~C. Maiti, ``{CA} based document compression
  technology,'' in \emph{Proceedings of 11th International Conference on Neural
  Information Processing}, 2004, pp. 679--685.

\bibitem{ShawDS06}
C.~Shaw, S.~Das, and B.~K. Sikdar, ``Cellular automata based encoding technique
  for wavelet transformed data targeting still image compression,'' in
  \emph{Proceedings of International Conference on Cellular Automata, Research
  and Industry (ACRI)}, 2006, pp. 141--146.

\bibitem{lafe2002method}
O.~E. Lafe, ``Method and apparatus for video compression using sequential frame
  cellular automata transforms,'' 2002, {US} {P}atent 6,456,744.

\bibitem{ye2008novel}
R.~Ye and H.~Li, ``A novel image scrambling and watermarking scheme based on
  cellular automata,'' in \emph{Proceedings of International Symposium on
  Electronic Commerce and Security}, 2008, pp. 938--941.

\bibitem{Chandrama}
C.~Shaw, D.~Chatterji, P.~Maji, S.~Sen, B.~N. Roy, and P.~Pal~Chaudhuri, ``A
  pipeline architecture for encompression (encryption + compression)
  technology.'' in \emph{Proceedings of VLSI Design}, 2003, pp. 277--282.

\bibitem{ShawMSSRC04}
C.~Shaw, P.~Maji, S.~Saha, B.~K.~Sikdar, S.~Roy, and P.~Pal~Chaudhuri,
  ``Cellular automata based encompression technology for voice data,'' in
  \emph{Proceedings of International Conference on Cellular Automata, Research
  and Industry (ACRI)}, 2004, pp. 258--267.

\bibitem{Burks1984157}
C.~Burks and D.~Farmer, ``Towards modeling dna sequences as automata,''
  \emph{Physica D: Nonlinear Phenomena}, vol.~10, no. 1-2, pp. 157--167, 1984.

\bibitem{ermentrout1993cellular}
G.~B. Ermentrout and L.~Edelstein-Keshet, ``Cellular automata approaches to
  biological modeling,'' \emph{Journal of Theoretical Biology}, vol. 160,
  no.~1, pp. 97--133, 1993.

\bibitem{zorzenon01}
R.~M.~Z. dos Santos and S.~Coutinho, ``Dynamics of {HIV} {I}nfection : {A}
  {C}ellular {A}utomata {A}pproach,'' \emph{Physical Review Letters}, vol.~87,
  p. 168102, 2001.

\bibitem{MOREIRA02}
J.~Moreira and A.~Deutsch, ``Cellular automaton models of tumor development: A
  critical review,'' \emph{Advances in Complex Systems}, vol.~05, no. 02n03,
  pp. 247--267, 2002.

\bibitem{JCC:JCC20354}
X.~Xiao, S.~Shao, Z.~Huang, and K.~Chou, ``Using pseudo amino acid composition
  to predict protein structural classes: Approached with complexity measure
  factor,'' \emph{Journal of Comutational Chemistry}, vol.~27, no.~4, pp.
  478--482, 2006.

\bibitem{Santos:2013}
J.~Santos, P.~Villot, and M.~Di{\'e}guez, ``Protein folding with cellular
  automata in the {3D} {HP} model,'' in \emph{Proceedings of the 15th Annual
  Conference Companion on Genetic and Evolutionary Computation}, 2013, pp.
  1595--1602.

\bibitem{DBLP:conf/acri/GhoshBMMC10}
S.~Ghosh, T.~Bachhar, N.~S.~Maiti, I.~Mitra, and P.~Pal~Chaudhuri, ``Theory and
  application of equal length cycle cellular automata {(ELCCA)} for enzyme
  classification,'' in \emph{Proceedings of International Conference on
  Cellular Automata, Research and Industry (ACRI)}, 2010, pp. 46--57.

\bibitem{Ghosh2012}
S.~Ghosh, N.~S.~Maiti, and P.~Pal~Chaudhuri, ``Theory and application of
  restricted five neighborhood cellular automata ({R5NCA}) for protein
  structure prediction,'' in \emph{Proceedings of International Conference on
  Cellular Automata, Research and Industry (ACRI)}.\hskip 1em plus 0.5em minus
  0.4em\relax Springer Berlin Heidelberg, 2012, pp. 360--369.

\bibitem{DBLP:conf/acri/GhoshMC14}
------, ``{C}ellular {A}utomata {M}odel for {P}rotein {S}tructure {S}ynthesis
  {(PSS)},'' in \emph{Proceedings of International Conference on Cellular
  Automata, Research and Industry (ACRI)}, 2014, pp. 268--277.

\bibitem{doi:10.1142/S0218001494000280}
K.~Morita and S.~Ueno, ``Parallel generation and parsing of array languages
  using reversible cellular automata,'' \emph{International Journal of Pattern
  Recognition and Artificial Intelligence}, vol.~08, no.~02, pp. 543--561,
  1994.

\bibitem{tome1994necessary}
J.~A.~B. Tome, ``Necessary and sufficient conditions for reversibility in one
  dimensional cellular automata,'' in \emph{Proceedings Workshop on Physics and
  Computation}, 1994, pp. 156--159.

\bibitem{MANZINI199860}
G.~Manzini and L.~Margara, ``Invertible linear cellular automata over ${Z}_m$:
  Algorithmic and dynamical aspects,'' \emph{Journal of Computer and System
  Sciences}, vol.~56, no.~1, pp. 60 -- 67, 1998.

\bibitem{Aso85}
H.~Aso and N.~Honda, ``Dynamical characteristics of {L}inear {C}ellular
  {A}utomata,'' \emph{Journal of Computer and System Sciences}, vol.~30, pp.
  291--317, 1985.

\bibitem{czeizler2007tight}
E.~Czeizler and J.~Kari, ``A tight linear bound on the synchronization delay of
  bijective automata,'' \emph{Theoretical computer science}, vol. 380, no. 1-2,
  pp. 23--36, 2007.

\bibitem{mora4construction}
J.~C. S.~T. Mora, M.~G. Hern{\'a}ndez, H.~V. McIntosh, and S.~V.~C. Vergara,
  ``Construction of reversible cellular automata by amalgamations and
  permutations of states,'' \emph{Journal of Cellular Automata}, vol.~4, pp.
  311--322, 2009.

\bibitem{CHANG2016217}
C.-H. Chang and H.~Chang, ``On the bernoulli automorphism of reversible linear
  cellular automata,'' \emph{Information Sciences}, vol. 345, pp. 217 -- 225,
  2016.

\bibitem{DOW199767}
R.~Dow, ``Additive cellular automata and global injectivity,'' \emph{Physica D:
  Nonlinear Phenomena}, vol. 110, no. 1--2, pp. 67 -- 91, 1997.

\bibitem{0305-4470-37-22-006}
A.~Nobe and F.~Yura, ``On reversibility of cellular automata with periodic
  boundary conditions,'' \emph{Journal of Physics A: Mathematical and General},
  vol.~37, no.~22, p. 5789, 2004.

\bibitem{Sato2009}
T.~Sato, K.~Honda, H.~Lee, and Y.~Kawahara, ``A classification of triplet local
  rules with inverse image sequences,'' in \emph{Natural Computing}, ser.
  Proceedings in Information and Communications Technology.\hskip 1em plus
  0.5em minus 0.4em\relax Springer Japan, 2009, vol.~1, pp. 167--178.

\bibitem{hazariJCA17}
{R}aju {H}azari and {S}ukanta {D}as, ``Number conservation property of binary
  cellular automata under $\alpha$-asynchronous update,'' \emph{Journal of
  Cellular Automata}, vol.~13, no.~3, pp. 247--265, 2018.

\bibitem{tiernan1970efficient}
J.~C. Tiernan, ``An efficient search algorithm to find the elementary circuits
  of a graph,'' \emph{Communications of the ACM}, vol.~13, no.~12, pp.
  722--726, 1970.

\bibitem{tarjan1973enumeration}
R.~Tarjan, ``Enumeration of the elementary circuits of a directed graph,''
  \emph{SIAM Journal on Computing}, vol.~2, no.~3, pp. 211--216, 1973.

\bibitem{ehrenfeucht1973algorithm}
A.~Ehrenfeucht, L.~D. Fosdick, and L.~J. Osterweil, ``An algorithm for finding
  the elementary circuits of a directed graph,'' \emph{University of Colorado
  at Boulder, Department of Computer Science, Report (CU-CS-024--73)}, 1973.

\bibitem{johnson1975finding}
D.~B. Johnson, ``Finding all the elementary circuits of a directed graph,''
  \emph{SIAM Journal on Computing}, vol.~4, no.~1, pp. 77--84, 1975.

\bibitem{galton1890dice}
F.~Galton, ``Dice for statistical experiments,'' \emph{Nature}, vol.~42, no.
  1070, pp. 13--14, 1890.

\bibitem{tippett1927random}
L.~Tippett, \emph{Random sampling numbers}, ser. Tracts for computers.\hskip
  1em plus 0.5em minus 0.4em\relax Cambridge University Press, 1927.

\bibitem{2980655}
M.~G. Kendall and B.~Babington-Smith, ``Randomness and random sampling
  numbers,'' \emph{Journal of the Royal Statistical Society}, vol. 101, no.~1,
  pp. 147--166, 1938.

\bibitem{2983623}
------, ``Second paper on random sampling numbers,'' \emph{Supplement to the
  Journal of the Royal Statistical Society}, vol.~6, no.~1, pp. 51--61, 1939.

\bibitem{von195113}
J.~Von~Neumann, ``13. various techniques used in connection with random
  digits,'' \emph{Applied Mathematics and Computation}, vol.~12, pp. 36--38,
  1951.

\bibitem{l1996maximally}
P.~L'ecuyer, ``Maximally equidistributed combined {T}ausworthe generators,''
  \emph{Mathematics of Computation}, vol.~65, no. 213, pp. 203--213, 1996.

\bibitem{Lewis:1973:GFS:321765.321777}
T.~G. Lewis and W.~H. Payne, ``Generalized feedback shift register pseudorandom
  number algorithm,'' \emph{Journal of the ACM}, vol.~20, no.~3, pp. 456--468,
  1973.

\bibitem{Panneton:2005:XRN:1113316.1113319}
F.~Panneton and P.~L'Ecuyer, ``On the xorshift random number generators,''
  \emph{ACM Transactions on Modeling and Computer Simulation}, vol.~15, no.~4,
  pp. 346--361, 2005.

\bibitem{Panneton:2006:ILG:1132973.1132974}
F.~Panneton, P.~L'Ecuyer, and M.~Matsumoto, ``Improved long-period generators
  based on linear recurrences modulo 2,'' \emph{ACM Transactions on
  Mathematical Software}, vol.~32, no.~1, pp. 1--16, 2006.

\bibitem{Tausworthe}
R.~C. Tausworthe, ``Random numbers generated by linear recurrence modulo two,''
  \emph{Mathematics of Computation}, vol.~19, no.~90, pp. 201--209, 1965.

\bibitem{Tezuka:1987:DGP:31846.31848}
S.~Tezuka, ``On the discrepancy of {GFSR} pseudorandom numbers,'' \emph{Journal
  of the ACM}, vol.~34, no.~4, pp. 939--949, 1987.

\bibitem{Marco99}
M.~Tomassini, M.~Sipper, M.~Zolla, and M.~Perrenoud, ``Generating high-quality
  random numbers in parallel by cellular automata,'' \emph{Future Generation
  Computer Systems}, vol.~16, no. 2-3, pp. 291--305, 1999.

\bibitem{Vigna:2016:EEM:2956571.2845077}
S.~Vigna, ``An experimental exploration of {M}arsaglia's {X}orshift generators,
  scrambled,'' \emph{ACM Transactions on Mathematical Software}, vol.~42,
  no.~4, pp. 30:1--30:23, 2016.

\bibitem{122655}
K.~Cattell and M.~Serra, ``The analysis of one dimensional multiple-valued
  linear cellular automata,'' in \emph{Proceedings of the Twentieth
  International Symposium on Multiple-Valued Logic}, 1990, pp. 402--409.

\bibitem{HOSSEINI2014149}
S.~M. Hosseini, H.~Karimi, and M.~V. Jahan, ``Generating pseudo-random numbers
  by combining two systems with complex behaviors,'' \emph{Journal of
  Information Security and Applications}, vol.~19, no.~2, pp. 149 -- 162, 2014.

\bibitem{LEcuyer1990}
P.~L'Ecuyer, ``Random numbers for simulation,'' \emph{Communications of the
  ACM}, vol.~33, no.~10, pp. 85--97, 1990.

\bibitem{Supreeti_2018_chaos}
S.~Kamilya and S.~Das, ``A study of chaos in cellular automata,''
  \emph{International Journal of Bifurcation and Chaos}, vol.~28, no.~03, p.
  1830008, 2018.

\bibitem{Meier1990}
W.~Meier and O.~Staffelbach, \emph{Nonlinearity Criteria for Cryptographic
  Functions}.\hskip 1em plus 0.5em minus 0.4em\relax Springer Berlin
  Heidelberg, 1990, pp. 549--562.

\bibitem{Knuth2}
D.~E. Knuth, \emph{The {A}rt of {C}omputer {P}rogramming -- {S}eminumerical
  {A}lgorithms}, 3rd~ed.\hskip 1em plus 0.5em minus 0.4em\relax Pearson
  Education, 2000, vol.~2.

\bibitem{L'Ecuyer:2007:TCL:1268776.1268777}
P.~L'Ecuyer and R.~Simard, ``Test{U}01: A {C} library for empirical testing of
  random number generators,'' \emph{ACM Transactions on Mathematical Software},
  vol.~33, no.~4, pp. 22:1--22:40, 2007.

\bibitem{diehard}
G.~Marsaglia, ``{DIEHARD}: {A} battery of tests of randomness,'' in
  \emph{http://stat.fsu.edu/\~~geo/diehard.html}, 1996.

\bibitem{rukhin2001statistical}
A.~Rukhin, J.~Soto, J.~Nechvatal, M.~Smid, and E.~Barker, ``A statistical test
  suite for random and pseudorandom number generators for cryptographic
  applications,'' DTIC Document, Tech. Rep., 2001.

\bibitem{harvard1951annals}
D.~H. Lehmer, ``Mathematical methods in large-scale computing units,''
  \emph{Annals of the Computation Laboratory of Harvard University}, vol.~26,
  pp. 141--146, 1951.

\bibitem{GCC}
R.~McGrath \emph{et~al.}, ``{GNU} {C} {L}ibrary,''
  \url{https://www.gnu.org/software/libc/manual/html_node/Pseudo\_002dRandom-Numbers.html\#index-pseudo\_002drandom-numbers}.

\bibitem{5388354}
P.~A.~W. Lewis, A.~S. Goodman, and J.~M. Miller, ``A pseudo-random number
  generator for the system/360,'' \emph{IBM Systems Journal}, vol.~8, no.~2,
  pp. 136--146, 1969.

\bibitem{fishman1990multiplicative}
G.~S. Fishman, ``Multiplicative congruential random number generators with
  modulus $2^{\beta}$: an exhaustive analysis for $\beta= 32$ and a partial
  analysis for $\beta= 48$,'' \emph{Mathematics of Computation}, vol.~54, no.
  189, pp. 331--344, 1990.

\bibitem{doi:10.1137/0907002}
G.~Fishman and L.~R. Moore, III, ``An exhaustive analysis of multiplicative
  congruential random number generators with modulus $2^{31} - 1$,'' \emph{SIAM
  Journal on Scientific and Statistical Computing}, vol.~7, no.~1, pp. 24--45,
  1986.

\bibitem{Park:1988:RNG:63039.63042}
S.~K. Park and K.~W. Miller, ``Random number generators: Good ones are hard to
  find,'' \emph{Communications of the ACM}, vol.~31, no.~10, pp. 1192--1201,
  1988.

\bibitem{Park:1993}
S.~K. Park, K.~W. Miller, and P.~K. Stockmeyer, ``Technical correspondence:
  Response,'' \emph{Communications of the ACM}, vol.~36, no.~7, pp. 105--110,
  1993.

\bibitem{brent1994periods}
R.~P. Brent, ``On the periods of generalized {F}ibonacci recurrences,''
  \emph{Mathematics of Computation}, vol.~63, no. 207, pp. 389--401, 1994.

\bibitem{10.2307.2959748}
G.~Marsaglia and A.~Zaman, ``A new class of random number generators,''
  \emph{The Annals of Applied Probability}, vol.~1, no.~3, pp. 462--480, 1991.

\bibitem{10.2307.3215210}
C.~Ko\c{c}, ``Recurring-with-carry sequences,'' \emph{Journal of Applied
  Probability}, vol.~32, no.~4, pp. 966--971, 1995.

\bibitem{couture1997distribution}
R.~Couture and P.~L'ecuyer, ``Distribution properties of multiply-with-carry
  random number generators,'' \emph{Mathematics of Computation of the American
  Mathematical Society}, vol.~66, no. 218, pp. 591--607, 1997.

\bibitem{Eichenauer1986}
J.~Eichenauer and J.~Lehn, ``A non-linear congruential pseudo random number
  generator,'' \emph{Statistische Hefte}, vol.~27, no.~1, pp. 315--326, 1986.

\bibitem{EICHENAUERHERRMANN1992345}
J.~Eichenauer-Herrmann, ``Construction of inversive congruential pseudorandom
  number generators with maximal period length,'' \emph{Journal of
  Computational and Applied Mathematics}, vol.~40, no.~3, pp. 345 -- 349, 1992.

\bibitem{eichenauer1993statistical}
------, ``Statistical independence of a new class of inversive congruential
  pseudorandom numbers,'' \emph{Mathematics of Computation}, vol.~60, no. 201,
  pp. 375--384, 1993.

\bibitem{L'Ecuyer:1988:EPC:62959.62969}
P.~L'Ecuyer, ``Efficient and portable combined random number generators,''
  \emph{Communications of the ACM}, vol.~31, no.~6, pp. 742--751, 1988.

\bibitem{10.2307.2347988}
B.~A. Wichmann and I.~D. Hill, ``Algorithm {AS} 183: An efficient and portable
  pseudo-random number generator,'' \emph{Journal of the Royal Statistical
  Society. Series C (Applied Statistics)}, vol.~31, no.~2, pp. 188--190, 1982.

\bibitem{doi:10.1287.opre.44.5.816}
P.~L'Ecuyer, ``Combined multiple recursive random number generators,''
  \emph{Operations Research}, vol.~44, no.~5, pp. 816--822, 1996.

\bibitem{doi:10.1287.opre.47.1.159}
------, ``Good parameters and implementations for combined multiple recursive
  random number generators,'' \emph{Operations Research}, vol.~47, no.~1, pp.
  159--164, 1999.

\bibitem{nance1978some}
R.~E. Nance and C.~Overstreet~Jr, ``Some experimental observations on the
  behavior of composite random number generators,'' \emph{Operations Research},
  vol.~26, no.~5, pp. 915--935, 1978.

\bibitem{bratleyguide}
P.~Bratley, B.~L. Fox, and L.~E. Schrage, \emph{A guide to simulation}.\hskip
  1em plus 0.5em minus 0.4em\relax Springer Science \& Business Media, 1987.

\bibitem{L'Ecuyer:2000:FCM:510378.510476}
P.~L'Ecuyer and R.~Touzin, ``Fast combined multiple recursive generators with
  multipliers of the form $a = \pm 2^q \pm 2^r$,'' in \emph{Proceedings of the
  32nd Conference on Winter Simulation}, 2000, pp. 683--689.

\bibitem{o1988pcg}
M.~E. O'Neill, ``{PCG}: {A} {F}amily of {S}imple {F}ast {S}pace-{E}fficient
  {S}tatistically {G}ood {A}lgorithms for {R}andom {N}umber {G}eneration,'' no.
  HMC-CS-2014-0905, 2014.

\bibitem{Tootill_1973}
J.~P.~R. Tootill, W.~D. Robinson, and D.~J. Eagle, ``An asymptotically random
  {T}ausworthe sequence,'' \emph{Journal of the ACM}, vol.~20, no.~3, pp.
  469--481, 1973.

\bibitem{Tootill:1971:RUP:321650.321655}
J.~P.~R. Tootill, W.~D. Robinson, and A.~G. Adams, ``The {R}uns {U}p-and-{D}own
  {P}erformance of {T}ausworthe {P}seudo-{R}andom {N}umber {G}enerators,''
  \emph{Journal of the ACM}, vol.~18, no.~3, pp. 381--399, 1971.

\bibitem{RNG}
P.~L'Ecuyer, ``Random number generators,'' 2017,
  \url{http://www-labs.iro.umontreal.ca/~simul/rng/}.

\bibitem{matsumoto1992twisted}
M.~Matsumoto and Y.~Kurita, ``Twisted {GFSR} generators,'' \emph{ACM
  Transactions on Modeling and Computer Simulation}, vol.~2, no.~3, pp.
  179--194, 1992.

\bibitem{Matsumoto:1994:TGG:189443.189445}
------, ``Twisted {GFSR} generators {II},'' \emph{ACM Transactions on Modeling
  and Computer Simulation}, vol.~4, no.~3, pp. 254--266, 1994.

\bibitem{sfmt}
M.~Saito and M.~Matsumoto, ``{SIMD}-oriented {F}ast {M}ersenne {T}wister
  ({SFMT}): twice faster than {M}ersenne {T}wister,'' 2017,
  \url{http://www.math.sci.hiroshima-u.ac.jp/\~m-mat/MT/SFMT/\#dSFMT}.

\bibitem{marsaglia2003xorshift}
G.~Marsaglia, ``Xorshift {RNGs},'' \emph{Journal of Statistical Software},
  vol.~8, no.~14, pp. 1--6, 2003.

\bibitem{brent2004note}
R.~P. Brent, ``Note on {M}arsaglia's {X}orshift {R}andom {N}umber
  {G}enerators,'' \emph{Journal of Statistical Software}, vol.~11, no.~5, pp.
  1--5, 2004.

\bibitem{VIGNA2017175}
S.~Vigna, ``Further scramblings of marsaglia's xorshift generators,''
  \emph{Journal of Computational and Applied Mathematics}, vol. 315, no.
  Supplement C, pp. 175 -- 181, 2017.

\bibitem{l2003combined}
P.~L'Ecuyer and J.~Granger-Pich{\'e}, ``Combined generators with components
  from different families,'' \emph{Mathematics and Computers in Simulation},
  vol.~62, no. 3--6, pp. 395--404, 2003.

\bibitem{Hurst1984}
S.~L. Hurst, ``Multiple-valued logic - its status and its future,'' \emph{IEEE
  Transactions on Computers}, vol. C-33, no.~12, pp. 1160 -- 1179, 1984.

\bibitem{5176262}
W.~Alexander, ``The ternary computer,'' \emph{Electronics and Power}, vol.~10,
  no.~2, pp. 36--39, 1964.

\bibitem{SERRAN1997533}
N.~Serran, A.~M. Jorge, and J.~S. Dias, ``A proposal for the implementation of
  ternary digital circuits,'' \emph{Microelectronics Journal}, vol.~28, no.~5,
  pp. 533 -- 541, 1997.

\bibitem{dunn2012modern}
M.~Dunn and G.~Epstein, \emph{Modern Uses of Multiple-Valued Logic: Invited
  Papers from the Fifth International Symposium on Multiple-Valued Logic held
  at Indiana University, Bloomington, Indiana, May 13--16, 1975}.\hskip 1em
  plus 0.5em minus 0.4em\relax Springer Science \& Business Media, 2012,
  vol.~2.

\bibitem{Das2012}
S.~Das, P.~S. Dasgupta, and S.~Sensarma, \emph{Arithmetic Algorithms for
  Ternary Number System}.\hskip 1em plus 0.5em minus 0.4em\relax Springer
  Berlin Heidelberg, 2012, pp. 111--120.

\bibitem{6810470}
B.~Parhami and M.~McKeown, ``Arithmetic with binary-encoded balanced ternary
  numbers,'' in \emph{2013 Asilomar Conference on Signals, Systems and
  Computers}, 2013, pp. 1130--1133.

\bibitem{1793814}
S.~Aguzzoli, A.~Ciabattoni, B.~Gerla, C.~Manara, and V.~Marra, Eds.,
  \emph{Algebraic and Proof-theoretic Aspects of Non-classical Logics: Papers
  in Honor of Daniele Mundici on the Occasion of His 60th Birthday}.\hskip 1em
  plus 0.5em minus 0.4em\relax Berlin, Heidelberg: Springer-Verlag, 2007.

\bibitem{CHATTOPADHYAY20101014}
T.~Chattopadhyay, ``All-optical symmetric ternary logic gate,'' \emph{Optics
  $\&$ Laser Technology}, vol.~42, no.~6, pp. 1014 -- 1021, 2010.

\bibitem{obiniyi2011arithmetic}
A.~A. Obiniyi, E.~E. Absalom, and K.~Adako, ``Arithmetic logic design with
  color-coded ternary for ternary computing,'' \emph{International Journal of
  Computer Applications}, vol.~26, no.~11, pp. 31--37, 2011.

\bibitem{gundersen2008aspects}
H.~Gundersen, ``Aspects of balanced ternary arithmetics implemented using
  {CMOS} recharged semi-floating gate devices,'' Ph.D. dissertation, Institute
  of Informatics, University of Oslo, Norway, 2008.

\bibitem{hayes2001third}
B.~Hayes, ``Third base,'' \emph{American scientist}, vol.~89, no.~6, pp.
  490--494, 2001.

\bibitem{1498715}
M.~Glusker, D.~M. Hogan, and P.~Vass, ``The ternary calculating machine of
  {T}homas {F}owler,'' \emph{IEEE Annals of the History of Computing}, vol.~27,
  no.~3, pp. 4--22, 2005.

\bibitem{connelly2008ternary}
J.~Connelly, ``Ternary computing testbed: 3-trit computer architecture,'' Ph.D.
  dissertation, California Polytechnic State University, 2008.

\bibitem{nair2015delayOpenAccess}
R.~S. Nair, S.~C. Smith, and J.~Di, ``Delay {I}nsensitive {T}ernary {CMOS}
  {L}ogic for {S}ecure {H}ardware,'' \emph{Journal of Low Power Electronics and
  Applications}, vol.~5, no.~3, pp. 183--215, 2015.

\bibitem{6782156}
M.~Eaton, ``Design and construction of a balanced ternary {ALU} with potential
  future cybernetic intelligent systems applications,'' in \emph{IEEE 11th
  International Conference on Cybernetic Intelligent Systems}, 2012, pp.
  30--35.

\bibitem{CHATTOPADHYAY2015123}
T.~Chattopadhyay, ``Design of optical reconfigurable balanced ternary
  arithmetic logic unit using mems based design,'' \emph{Optics
  Communications}, vol. 356, no. Supplement C, pp. 123 -- 135, 2015.

\bibitem{Mirzaee2696520}
R.~F. Mirzaee, K.~Navi, and N.~Bagherzadeh, ``High-efficient circuits for
  ternary addition,'' \emph{VLSI Design}, vol. 2014, pp. 10:10--10:10, 2014.

\bibitem{1319922}
P.~Kerntopf, M.~A. Perkowski, and M.~H.~A. Khan, ``On universality of general
  reversible multiple-valued logic gates,'' in \emph{Proceedings. 34th
  International Symposium on Multiple-Valued Logic}, 2004, pp. 68--73.

\bibitem{0305-4470-39-24-013}
G.~Yang, F.~Xie, X.~Song, and M.~A. Perkowski, ``Universality of 2-qudit
  ternary reversible gates,'' \emph{Journal of Physics A: Mathematical and
  General}, vol.~39, no.~24, p. 7763, 2006.

\bibitem{fey2015ternary}
D.~Fey, ``Ternary arithmetic pipeline architectures using multi-bit
  memristors,'' \emph{Future Computing}, pp. 1--6, 2015.

\bibitem{1402-4896-2005-T118-025}
J.~Yi, H.~Huacan, and L.~Yangtian, ``Ternary optical computer architecture,''
  \emph{Physica Scripta}, vol. 2005, no. T118, p.~98, 2005.

\bibitem{3297906}
M.~Klein, J.~A. Mol, J.~Verduijn, G.~P. Lansbergen, S.~Rogge, R.~D. Levine, and
  F.~Remacle, ``Ternary logic implemented on a single dopant atom field effect
  silicon transistor,'' \emph{Applied Physics Letters}, vol.~96, no.~4, p.
  043107, 2010.

\bibitem{0957-4484-17-8-023}
I.~L. Bajec, N.~Zimic, and M.~Mraz, ``The ternary quantum-dot cell and ternary
  logic,'' \emph{Nanotechnology}, vol.~17, no.~8, p. 1937, 2006.

\bibitem{LEBARBAJEC20061826}
------, ``Towards the bottom-up concept: Extended quantum-dot cellular
  automata,'' \emph{Microelectronic Engineering}, vol.~83, no.~4, pp. 1826 --
  1829, 2006, micro- and Nano-Engineering MNE 2005.

\bibitem{1329407}
H.~Gundersen and Y.~Berg, ``Max and min functions using multiple-valued
  recharged semi-floating gate circuits,'' in \emph{IEEE International
  Symposium on Circuits and Systems}, vol.~2, 2004, pp. II--857--860.

\bibitem{779713}
A.~Herrfeld and S.~Hentschke, ``Ternary multiplication circuits using 4-input
  adder cells and carry look-ahead,'' in \emph{Proceedings 29th IEEE
  International Symposium on Multiple-Valued Logic}, 1999, pp. 174--179.

\bibitem{moore1998cramming}
G.~E. Moore, ``Cramming more components onto integrated circuits,''
  \emph{Proceedings of the IEEE}, vol.~86, no.~1, pp. 82--85, 1998.

\bibitem{iLEC16j}
P.~L'Ecuyer, ``{SSJ}: Stochastic simulation in {Java}, software library,''
  2016, \url{http://simul.iro.umontreal.ca/ssj/}.

\bibitem{srivastava1996design}
A.~Srivastava and K.~Venkatapathy, ``Design and implementation of a low power
  ternary full adder,'' \emph{VLSI Design}, vol.~4, no.~1, pp. 75--81, 1996.

\bibitem{5248260}
I.~Halpern and M.~Yoeli, ``Ternary arithmetic unit,'' \emph{Proceedings of the
  Institution of Electrical Engineers}, vol. 115, no.~10, pp. 1385--1388, 1968.

\bibitem{4038879}
A.~Mukhopadhyay, ``Symmetric ternary switching functions,'' \emph{IEEE
  Transactions on Electronic Computers}, vol. EC-15, no.~5, pp. 731--739, 1966.

\bibitem{5249619}
D.~I. Porat, ``Three-valued digital systems,'' \emph{Proceedings of the
  Institution of Electrical Engineers}, vol. 116, no.~6, pp. 947--954, 1969.

\bibitem{Santos:1964:AST:1464122.1464168}
J.~Santos and H.~Arango, ``On the analysis and synthesis of three-valued
  digital systems,'' in \emph{Proceedings of the Spring Joint Computer
  Conference}.\hskip 1em plus 0.5em minus 0.4em\relax ACM, 1964, pp. 463--475.

\bibitem{83076}
J.~A.~C. Webb and S.~J. Laverty, ``Extension of the {T}-gate concept to a
  hybrid {U}-gate architecture for ternary and higher order logic systems,''
  \emph{Electronics Letters}, vol.~26, no.~19, pp. 1629--1630, 1990.

\bibitem{Frieder776392}
G.~Frieder, ``Ternary computers: Part {I}: Motivation for ternary computers,''
  in \emph{Conference Record of the 5th Annual Workshop on Microprogramming},
  1972, pp. 83--86.

\bibitem{rine1977computer}
D.~Rine, \emph{Computer Science and Multiple Valued Logic: Theory and
  Applications}.\hskip 1em plus 0.5em minus 0.4em\relax North-Holland
  Publishing Company, 1977.

\bibitem{4037853}
W.~H. Hanson, ``Ternary threshold logic,'' \emph{IEEE Transactions on
  Electronic Computers}, vol. EC-12, no.~3, pp. 191--197, 1963.

\bibitem{4038831}
R.~D. Merrill, ``A tabular minimization procedure for ternary switching
  functions,'' \emph{IEEE Transactions on Electronic Computers}, vol. EC-15,
  no.~4, pp. 578--585, 1966.

\bibitem{Mouftah:1976:SIT:800111.803598}
H.~T. Mouftah, ``A study on the implementation of three-valued logic,'' in
  \emph{Proceedings of the Sixth International Symposium on Multiple-valued
  Logic}, 1976, pp. 123--126.

\bibitem{hubbard1979design}
C.~J. Hubbard, ``A design of a ternary arithmetic logic unit,'' Ph.D.
  dissertation, Texas Tech University, 1979.

\bibitem{Eichmann86}
G.~Eichmann, Y.~Li, and R.~R. Alfano, ``Optical binary coded ternary arithmetic
  and logic,'' \emph{Applied Optics}, vol.~25, no.~18, pp. 3113--3121, 1986.

\bibitem{4038040}
D.~J. Anderson and D.~L. Dietmeyer, ``A magnetic ternary device,'' \emph{IEEE
  Transactions on Electronic Computers}, vol. EC-12, no.~6, pp. 911--914, 1963.

\bibitem{4210047}
W.~D. Ryan and H.~Madany, ``Transistor-tunnel-diode ternary-logic circuits,''
  \emph{Electronics Letters}, vol.~4, no.~7, pp. 127--128, 1968.

\bibitem{doi:10.1080/00207217508920376}
S.~S. Rath, ``A ternary flip-flop circuit,'' \emph{International Journal of
  Electronics}, vol.~38, no.~1, pp. 41--47, 1975.

\bibitem{5245386}
R.~P. Hallworth and F.~G. Heath, ``Semiconductor circuits for ternary logic,''
  \emph{Proceedings of the IEE - Part C: Monographs}, vol. 109, no.~15, pp.
  219--225, 1962.

\bibitem{4038408}
J.~Santos, H.~Arango, and M.~Pascual, ``A ternary storage element using a
  conventional ferrite core,'' \emph{IEEE Transactions on Electronic
  Computers}, vol. EC-14, no.~2, pp. 248--248, 1965.

\bibitem{doi:10.1080/00207218508939031}
S.~A.~H. Shalash, M.~A. Al-Zubaldy, J.~A. Rahim, S.~I. Adhem, and M.~A.~H.
  Abdul-karim, ``{CMOS} {J-K} triflop,'' \emph{International Journal of
  Electronics}, vol.~58, no.~2, pp. 365--371, 1985.

\bibitem{BHOWMIK20135561}
P.~Bhowmik, T.~Chattopadhyay, and J.~N. Roy, ``Ternary galois field arithmetic
  operations with optical nonlinear material ({OPNLM}) switch,'' \emph{Optik -
  International Journal for Light and Electron Optics}, vol. 124, no.~22, pp.
  5561 -- 5566, 2013.

\bibitem{488807}
M.~A.~H. Ali, F.~J. Hassan-Alshiroofi, and H.~G. Rotithor, ``A framework for
  design of multivalued logic functions and its application using {CMOS}
  ternary switches,'' \emph{IEEE Transactions on Circuits and Systems I:
  Fundamental Theory and Applications}, vol.~43, no.~4, pp. 279--289, 1996.

\bibitem{kuurka2009topological}
P.~K{\r{u}}rka, \emph{Topological Dynamics of Cellular Automata}.\hskip 1em
  plus 0.5em minus 0.4em\relax Springer New York, 2009, pp. 9246--9268.

\bibitem{Wolfram:1982gu}
S.~Wolfram, ``{Cellular automata as simple self-organizing systems},''
  California Institute of Technology, Pasadena, CA, Tech. Rep. CALT-68-938,
  1982.

\bibitem{SUTNER199587}
K.~Sutner, ``On the computational complexity of finite cellular automata,''
  \emph{Journal of Computer and System Sciences}, vol.~50, no.~1, pp. 87 -- 97,
  1995.

\end{thebibliography}
\end{document}